\documentclass[12pt]{article}
\usepackage{amsmath,amssymb}
\usepackage[english]{babel}
\usepackage[cp866]{inputenc}
\usepackage{hyperref}

\numberwithin{equation}{section}
\numberwithin{equation}{subsection}
\newcommand{\bq}{\begin{eqnarray}}
\newcommand{\eq}{\end{eqnarray}}

\newcommand{\bbq}{\begin{equation*}}
\newcommand{\eeq}{\end{equation*}}

\newcommand{\ra}{\rightarrow}
\newcommand{\w}{{\mathcal W}}
\newcommand{\cw}{{\cal W}}
\newcommand{\gd}{ \gamma_q}
\newcommand{\ov}{\overline}
\newcommand{\nd}{{\ov N}_c}
\newcommand{\lym}{{\Lambda}_{SYM}}

\newcommand{\la}{ \Lambda_Q }

\newcommand{\te}{d^2\theta}
\newcommand{\ote}{d^2\ov \theta}

\newcommand{\mph}{\mu_{\Phi}}

\newcommand{\bo}{\rm{b_o}}
\newcommand{\bd}{\rm {\ov b}_o}

\newcommand{\ql}{Q^{\it l}}
\newcommand{\oql}{{\ov Q}_{\it  l}}
\newcommand{\qh}{Q^{\it h}}
\newcommand{\oqh}{{\ov Q}_{\it  h}}
\newcommand{\dl}{{q}_{\it l}}
\newcommand{\odl}{{\ov q}^{\it  l}}
\newcommand{\hd}{q_{\it h}}
\newcommand{\ohd}{{\ov q}^{\it h}}
\newcommand{\mh}{m_{\it h}}
\newcommand{\ml}{m_{\it l}}
\newcommand{\hml}{\hat{m}_{\it l}}

\newcommand{\Qo}{({\ov Q}Q)_1}
\newcommand{\Qt}{({\ov Q}Q)_2}

\newcommand{\mo}{m_{1}}

\newcommand{\mql}{m^{\rm pole}_{Q,\,L}}

\newcommand{\mgl}{\mu_{{\rm gl},\,L}}
\newcommand{\mgs}{\mu_{{\rm gl},\,S}}

\newcommand{\bt}{{\rm b}_2}

\newcommand{\mut}{{\ov\mu}_{{\rm gl},\,2}}

\newcommand{\mgt}{\mu_{\rm gl, 2}}
\newcommand{\mgo}{\mu_{\rm gl, 1}}

\newcommand{\mtq}{\langle m_Q^{\rm tot}\rangle}
\newcommand{\qma}{\langle m_{Q,1}^{\rm tot}\rangle}
\newcommand{\qmb}{\langle m_{Q,2}^{\rm tot}\rangle}
\newcommand{\qtp}{\mu^{\rm pole}_{q,2}}
\newcommand{\muo}{ {\ov\mu}_{{\rm gl},\,1} }

\newcommand{\oqt}{\ov{\textsf{Q}}_2}

\newcommand{\sqt}{\textsf{Q}^2}

\newcommand{\tq}{\textsf{q}}
\newcommand{\otq}{\ov{\textsf{q}}}

\newcommand{\sq}{\textsf{Q}}
\newcommand{\dq}{\textsf{q}}
\newcommand{\odq}{\textsf{q}}
\newcommand{\oq}{\ov{\textsf{Q}}}

\newcommand{\hh}{\mathbf{Q}^H_H}

\newcommand{\qop}{\mu^{\rm pole}_{q,1}}
\newcommand{\mgh}{\mu_{\it gl,\,h}}

\newcommand{\lm}{\Lambda_2}
\newcommand{\mx}{\mu_{\rm x}}
\newcommand{\bb}{2N_c-N_F}
\newcommand{\lt}{\tilde\Lambda}
\newcommand{\mos}{\mu_o^{\rm str}}

\newcommand{\mhp}{m^{\rm pole}_{\it h}}
\newcommand{\mlp}{m^{\rm pole}_{\it l}}
\newcommand{\nl}{ N_{\it l}}
\newcommand{\nh}{ N_{\it h}}
\newcommand{\mulp}{\mu^{\rm pole}_{q,\,\it l}}
\newcommand{\hlp}{{\hat\mu}^{\rm pole}_{q,\,\it l}}
\newcommand{\olp}{{\ov\mu}^{\rm pole}_{q,\,\it l}}
\newcommand{\hhp}{{\hat\mu}^{\rm pole}_{q,\,\it h}}
\newcommand{\muhp}{\mu^{\rm pole}_{q,\,\it h}}

\newcommand{\mpQ}{ m^{\rm pole}_Q}
\newcommand{\ogh}{{\ov \mu}_{gl,\,\it h}}

\newcommand{\ma}{m^{\rm pole}_{\sq,2} }
\newcommand{\mg}{\mu_{\rm gl}}
\newcommand{\wt}{\widetilde}
\newcommand{\qo}{({\ov q}q)_1}
\newcommand{\qt}{{(\ov q}q)_2}
\newcommand{\qq}{ ({\ov q}q) }
\newcommand{\QQ}{ ({\ov Q}Q) }
\newcommand{\mi}{m_{Q,i}}

\newcommand{\tp}{M}
\newcommand{\tm}{\widetilde m}

\newcommand{\wmu}{\widetilde{\mu}_{\rm x}}

\newcommand{\no}{{\rm n}_1}
\newcommand{\nt}{{\rm n}_2}

\newcommand{\mtl}{\langle m_Q^{\rm tot}\rangle_L}

\newcommand{\omp}{\mu^{\rm pole}_{q,1}}
\newcommand{\tmp}{\mu^{\rm pole}_{q,2}}

\newcommand{\wh}{\widehat}

\newcommand{\ha}{\hat m}

\begin{document}

\begin{center}
\bf \Large Mass Spectra in $\mathbf{{\cal N}=1}$ SQCD, 
in $\mathbf{{\cal N}=1}$  SQCD-type~ theory \\
and in softly broken $\mathbf{{\cal N}=2\ra {\cal N}=1}$ SQCD.
\end{center}
\begin{center} 
{ \vspace*{ - 2mm}{\bf \Large And problems with the 
$\mathbf{{\cal N}=1}$ Seiberg duality}}
\end{center}

\vspace{4mm}

\begin{center}
\bf \large Victor L. Chernyak 
\end{center}

\begin{center} 
e-mail: victorchernyak41@gmail.com 
\end{center}
\begin{center} 
Novosibirsk State University, 630090 Novosibirsk,
Pirogova str.2, Russia 
\end{center}
\begin{center} 
Budker Institute of Nuclear Physics, RAS, \\
630090 Novosibirsk-90, Lavrentev ave.11, Russia
\end{center}
\vspace{2mm}

\begin{center}
{\bf \large Abstract}
\end{center}

{Mass spectra are calculated for ${\cal N}=1$ SQCD and ${\cal N}=1$ SQCD-type 
theories  and  for  softly broken   ${\cal N}=2$ SQCD in vacua with unbroken
$Z_{(2N_c-N_F)\geq 2}$ symmetry. It is shown that  the Seiberg ${\cal N}=1$
duality works, at best, for massless quarks within the conformal
window only.

Besides:\, a)\, the gauge invariant order parameter for scalar quarks 
is introduced which distinguishes confinement and higgs phases,\, 
b)\, the phase transition are described in  ${\cal N}=1$ SQCD-type  and
broken  ${\cal N}=2$ SQCD theories between confinement and higgs phases,\,
c)\, the confinement mechanism is proposed for the ordinary YM and 
 ${\cal N}=1$ SYM theories}.
 
\tableofcontents
\addcontentsline{toc}{section}
{\bf \Large  Part I. \,\,Standard ${\cal N}=1$ SQCD and its Seiberg's dual}. 

{\section {\bf \large {The overall  introduction and summary}} }

{\hspace {0.5 cm}} The dynamics of 4-dimensional strongly coupled
non-Abelian gauge theories is complicated. It is well known that
supersymmetry (SUSY) leads to some simplifications in comparison with
the ordinary (i.e. non-SUSY) theories. Besides, it is widely believed
that SUSY is relevant to the real world. In any case, it is of great
interest to study the dynamics of nearest SUSY-relative of 
ordinary QCD, i.e. the ${\cal N}=1$ SQCD with $SU(N_c)$ colors 
and  quarks   with  $N_F$  flavors and masses~$m_{Q,i}$.

Because supersymmetric (SUSY) theories are much more constrained 
by  SUSY than the ordinary ones, it is easier for theory to deal with
them. So, they can serve, at least, as useful models for elucidating
the complicated strong coupling gauge dynamics (not even speaking
about their potential relevance to a real world).  And  the ${\cal N}=1$ 
SQCD  was considered in many papers, see e.g. 
the reviews  \cite{AKM, IS, SV-r}.   

This review paper  deals with 
${\cal N}=1$ and ${\cal N}=2$ SQCD and  with the ${\cal N}=1$  
SQCD-type theories, all mainly in the strong  coupling regimes.

But even with ${\cal N}=1$ SQCD, there is currently no proven physical 
picture of even main nontrivial characteristic features of its dynamics in 
the strong coupling region. In particular, it remained unclear how to 
calculate  even the main characteristic features of mass spectra.\\

The best previously  known proposal was the duality proposed by 
N.Seiberg, i.e. the  lower  energy dual theory at scales $\mu < \la$ 
\cite{S1,S2}  which, in a  number of  cases,   {\it is weakly coupled when 
the direct one is strongly   coupled,   and \it vice versa}.

The Seiberg duality passed a number non-trivial checks (mainly the 't
Hooft triangles and the behavior in the conformal regime), but  no proof 
has been given up to  now that the direct and dual theories are (or are 
not) really equivalent at low energies $\mu < \la$. In particular, { \it no 
way was shown up to now how to obtain at $N_c+1 < N_F < 3N_c/2$ solitons
with quantum numbers of dual quarks in the direct theory, and vice
versa}. Therefore, {\it the Seiberg proposal of equivalence of direct and
dual theories at $\mu < \la$ remains a hypothesis up to now}. 
 
The reason is that  a proof of equivalence of direct and dual theories  
needs real understanding of and the control over the dynamics of both 
theories in the strong coupling   regimes. Therefore, in the framework 
of gauge theories, the Seiberg  proposal remains a hypothesis.
 
We quote here three papers only. {\bf 1)}\, \cite{AR} (with N.
Seiberg among the authors): "By now many examples of such dualities
have been found, and a lot of evidence has been collected for their 
validity. However, there is still no general understanding of the origin 
of  these dualities, nor a prescription to find the dual for a given
gauge theory". \,{\bf 2)}\, \cite{G}:\, "The most established example
of such a duality is the Seiberg duality in ${\cal N}= 1$ SQCD... The
validity of this duality is still a conjecture but it have anyway
remarkably passed numerous checks..."\, {\bf 3)}\,\,\cite{A}:\, " All
of these dualities do not have any rigorous derivation so far, but
they pass many consistency checks..."

As for the string theory, we quote here two papers. {\bf 4)}\, The
review \cite{K}:\, "Specifically, we have seen using branes that the
quantum moduli spaces of vacua and quantum chiral rings of the
electric and magnetic SQCD theories coincide. This leaves open the
question whether Seiberg's duality extends to an equivalence of the
full infrared theory, since in general the chiral ring does not fully
specify the infrared conformal field theory. It is believed that in
gauge theory the answer is yes, and to prove it in brane theory will
require an understanding of the smoothness of the transition when
fivebranes cross. It is important to emphasize that the question
cannot be addressed using any currently available tools". {\bf 5)}\,
The most detailed discussion of implicit dynamical assumptions and
weak points in attempts to derive Seiberg's duality with a help of
moving branes has been given in \cite{V}, whose authors discussed
their results with N. Seiberg. Reinforcing conclusions from \cite{K},
the authors \cite{V} emphasize  that, {\it within the approach  with 
moving branes, no reasons are seen for the equivalence of the direct
and dual ${\cal N}=1$ theories with massive quarks at all values of
$N_F/N_c$, and for massless quarks outside the conformal window}.\\

 The purpose of this review is to introduce and to use the definite
dynamical scenario \cite{ch3} which {\it allows to calculate the main
characteristic properties of  mass spectra in ${\cal N}=1$
SQCD-type theories}. The only dynamical assumption of this scenario is
that {\it quarks in such ${\cal N}=1$ theories can be in two different 
standard phases only}\,: these are either the HQ (heavy quark)  phase 
where they are not higgsed but may be confined, or the Higgs phase 
where they form nonzero coherent vacuum condensate breaking the 
color   symmetry (and so, such quarks are not confined), at the 
appropriate   values of the  Lagrangian parameters.

The word "standard" implies also that, unlike the very special ${\cal
N}=2$ SQCD with its additional elementary colored  adjoint  scalar 
superfields  $X^{adj}$ and enhanced supersymmetry, 
{\it no extra, i.e. in addition to the standard set of (pseudo)
Nambu-Goldstone particles described below in this paper, 
parametrically light solitons (e.g. magnetic monopoles or dyons)
are formed at those lower scales where the effectively  massless 
regime  is  broken explicitly by nonzero particle masses
in such ${\cal N}=1$  SQCD-type  theories}.  These may be 
the  quark pole masses below which they decouple as heavy,  or 
masses of gluon multiplets  due to higgsed quarks. It is worth noting 
also  that  the appearance of  such {\it  additional} light  solitons will 
influence   the  't Hooft triangles of these  ${\cal N}=1$ theories.  

Let us note that e.g. in the direct ${\cal N}=1$ SQCD with   
light quarks with $m_Q\ll\la$, in most considered cases in  the strong 
coupling  regimes  the quarks decouple  as heavy  or are higgsed 
at scales  parametrically larger than $\lym=(\la^{3N_c-N_F}\det m_Q)
^{1/3N_c}$.  And in such cases,   when not the whole $SU(N_c)$  group is 
higgsed,  all or a part of gluons remain  effectively massless at these scales. 
Then instantons are not  operating at such scales $\mu\gg\lym$ due to 
remained gluino zero modes.  The nonperturbative  instanton  
contributions  become then relevant  at the scale $\sim\lym$ only. 

{\it And  this justifies  the  main assumption of the  proposed dynamical  
scenario in \cite{ch3} that no additional parametrically light  solitons are 
formed due to nonperturbative effects at those scales $\mu \gg\lym$ 
where  quarks are  decoupled as heavy  or are higgsed}. 

Let us emphasize also that within this scenario:\, 1)\, {\it no
contradictions with proven results have been found,\, 2)\, no
internal inconsistencies  have been found,\, 3) this dynamical
scenario  satisfies all those tests which were used as checks of the
Seiberg hypothesis about equivalence of the direct and dual
theories at low energies} $\mu <\la$. This shows, in particular, that
{\it all these tests, although necessary, may well be insufficient}. \\

According to Seiberg's view \cite{S2, IS} of e.g. the standard
direct ${\cal N}=1$ SQCD with $N_c+1<N_F<3N_c/2$,  
the scale  factor $\la$ of $SU(N_c)$ gauge
coupling  and direct quarks with $m_Q\ra 0$ (or $0 < m_Q\ll\la$),
the regime of the direct theory at $\mu<\la$ is in this case: {\it
\,''confinement without chiral symmetry breaking''}\, (as far as small
$m_Q\neq 0$ can be neglected). {\it  And  the dual ``magnetic''\,  theory   
with dual  $SU(\nd=N_F-N_c)$ colors and with dual  quarks and 
elementary $M$-fields 
is considered in this case  as the lower energy form at scales 
$\mu < \la$ of  the direct ``electric''\, theory}.\\

It is written in \cite{S2}.\,  {\it "The quarks and gluons of one theory 
can be interpreted as  solitons (non-Abelian magnetic monopoles) 
of the elementary fields of the  other  theory...  When one 
of the theories is Higgsed by an expectation value of a  squark, 
the other theory is confined. Massless glueballs, baryons and
Abelian magnetic monopoles in the confining description are the
weakly  coupled elementary quarks (i.e. solitons of the confined 
quarks)  in the dual Higgs description''}.

This means in the case $N_c+1 < N_F < 3N_c/2$ that, with the
effectively unbroken chiral symmetry (and R-charge symmetry),  
{\it all direct  ``electric'' quarks remained massless (or light), 
but hadrons made from these massless (or light) quarks 
and direct  gluons  acquired  large masses $\sim\la$ due to 
mysterious strong confinement with the string  tension 
$\sigma^{1/2}\sim\la$,  and decoupled as heavy at $\mu<\la$. 
Instead  of them, there   mysteriously appeared 
``magnetic'' massless (or light) composite
solitons. These last are particles of the dual theory}.\\

This viewpoint was criticized in section 7 of \cite{ch1} for the following
reasons. There is no confinement in Yukawa-type theories without
gauge interactions.   The confinement of original ``electric'' \, particles  
originates   in ${\cal N}=1$  SQCD-type theories  {\bf only} from the pure
${\cal N}=1$ SYM unbroken by  (possibly) higgsed quarks.  And because 
${\cal N}=1$ SYM has only one
dimensional parameter $\langle\lym\rangle$,  the string tension  is 
 $\sigma^{1/2}\sim\langle\lym\rangle$. But in the  standard ${\cal N}=1$ 
SQCD the {\it universal value} of $\langle\lym\rangle$  is well known: 
$\langle\lym\rangle=\langle S\rangle^{1/3}\sim (\Lambda_Q^{3N_c-N_F}
\det  m_Q)^{1/3N_c}\ll\la$ at $m_{Q,i}\ll\la$. Therefore, {\it such SYM can not 
produce so strong  confinement with the string tension $\sim\Lambda_Q$, 
but only with  $\sigma^{1/2}\sim\langle\lym\rangle\ll \la$} (and there is no
 confinement at all if the mass of even one quark tends to~zero).\\
 
And besides,  {\it nobody succeeded  up to now to  obtain 
dual quarks as solitons of the direct theory, and vice versa}. 
(The attempt made in \cite{APS} was erroneous in a number of 
respects. At least, it is sufficient to say that quarks from the  
$SU(N_F-N_c)$ subgroup of $SU(N_c)$  in  br2-vacua (the baryonic 
branch) of broken ${\cal N}=2$ SQCD  are not the magnetic ones but 
the original pure electric quarks. See  section 41 in Part III). \\

Additionally, as was argued in details in section 7 of \cite{ch1}, the
problem  with this Seiberg's view is that it is impossible to write in 
the direct theory at  the scale  $\mu\sim\la$ the effective   Lagrangian 
of  massive flavored  ``electric'' hadrons   with masses $\sim\la$, 
made of  strongly   confined 
effectively massless  flavored ``electric''  quarks with $m_Q\ll\la$, 
which will be  nonsingular at $m_Q\ra 0$  and preserving 
both the R-symmetry and  chiral symmetry $SU(N_F)_L\times SU
(N_F)_R\,$. Recall that these arguments in \cite{ch1} about the effective 
hadron Lagrangian at the scale  $\sim\la$ do not  use any assumptions 
about possible dynamics.
This is similar to our ordinary QCD in the chiral limit
$m_Q\ra 0$. If the chiral symmetry were not broken spontaneously, it
would be impossible to have massive nucleons (and other fermions)
with masses $\sim\Lambda_{QCD}$ because the term $\sim
\Lambda_{QCD}{\ov N}N$ in the potential is incompatible with the unbroken 
chiral symmetry. The situation in ${\cal N}=1$ SQCD with the unbroken chiral
symmetry at $N_c<N_F<3N_c/2$ and at $m_Q\ra 0$ is even more restrictive
due to unbroken R-symmetry and holomorphic dependence of the 
superpotential on its chiral parameters. \\

Besides, clearly, within the  conformal window $3N_c/2 < N_F < 3N_c$, 
the direct theory with massless (or light, $m_Q\ll\la$) quarks 
{\it enters smoothly the perturbative  conformal regime at $\mu < \la$, 
with   all its quarks and gluons remaining effectively massless}. 

In this case, the dual Seiberg's theory has to be understood not as the
low energy form of the direct one, but as possibly {\it equivalent alternative 
``magnetic''\,  description of the direct  ``electric''\,  theory  at $\mu < \la$}.\\

For reasons described above, the approach in e.g. \cite{ch3,ch19,ch21,ch23} 
was to  consider the direct and Seiberg's dual theories {\it as independent
theories, to calculate their mass spectra and to compare}. This is
then the direct {\it additional check} of possible equivalence of
these two theories at $\mu < \la$.

Then, in addition to above described differences between the direct 
and dual  theories at $N_c < N_F < 3N_c/2$, 
as was shown in particular in \cite{ch3,ch19} within the dynamical
scenario \cite{ch3}, the mass spectra of the direct and Seiberg's dual theories
are {\it parametrically different} at both ends of the conformal window, i.e. either 
at   $0 < (3N_c-N_F)/N_F\ll 1$ or at $0 < (2N_F-3N_c)\ll 1$. There are no reasons 
then to expect that  these two theories become equavalent in the other part of 
the conformal winow.  (See also \cite{ch23}   and  Appendix B in \cite{Session}).\\

Let us note that {\it  the results for the mass spectra of all considered  
in this review ${\cal N}=1$ direct and dual theories, i.e. the standars SQCD 
or SQCD-type ones,  fully confirm  the conclusions of  \cite{V,K},  see above}. 
Moreover,  our results {\it  strengthen} the results \cite{V,K}. I.e., {\it  the direct 
${\cal N}=1$ SQCD and Seiberg's  dual one are  equivalent, at best, only  
for massless quarks with $m_Q\ra 0$ at  $3N_c/2 < N_F < 3N_c$,  i.e.  
within the conformal window}. \\

And, in particular,   see eqs. (7.2),(7.3)  in \cite{ch1},  at  $N_c < N_F  < 3N_c/2$ 
and light quarks with $m_Q\ll\la$   the exact values of two-point correlators  
of conserved  $SU(N_F)_{left}\times  SU(N_F)_{right}$  currents  {\it are 
parametrically  different} at $m^{\rm pole}_Q\ll\mu\ll\la$  when all particles  
are effectively massless  in both the  direct and dual 
theories. To show this exact inequality,  no
additional statements  about possible dynamical  properties of the direct 
and dual  theories were used, except those one presented above :\, that 
confinement  originates from the SYM theory only. \\

On the whole, the main results presented in  this review. - \\

{\it 1) The dynamical scenario introduced in \cite{ch3}, which  satisfies all 
needed requirements,  was checked on internal self  consistency through calculations 
of  mass spectra in vacua of the standard $\mathbf{ {\cal N}=1}$ SQCD and 
$\mathbf{ {\cal N}=1}$ SQCD-type theory at different values of $1 \leq N_F < 3N_c$ 
and different quark masses. This not only confirmed the conclusions of 
papers  \cite{V,K}, but strengthened them in that the Seiberg duality works, 
at best, in one case only : for massless quarks with $3N_c/2 < N_F < 3N_c$, 
i.e. within  the conformal window.   It was shown that in other considered cases 
the mass spectra of the direct  and Seiberg's  dual theories are parametrically 
different. See Parts I and II.

{\bf 2)}\, The main characteristic features of mass spectra in vacua of the  
standard  $\mathbf{ {\cal N}=1}$ SQCD and $\mathbf{ {\cal N}=1}$  SQCD-type 
theory  were calculated at different values of $1 \leq N_F < 3N_c$ and different quark 
masses. 

In particular, it was shown in Appendix B in \cite{Session}  for $N_F=N_c+1$ and 
in  \cite{ch23} for $N_F=N_c$  that  proposed by N. Seiberg so called ``S-confinement''\,  
\cite{S1} with  the tension of the confining string $\sigma^{1/2}\sim\la$ in the direct 
theory is   erroneous, see  section 12.

{\bf 3)}\, Independently, the mass spectra were calculated quantitatively (i.e. the main 
terms  and first nonperturbative power corrections)  in numerous vacua with  unbroken 
$Z_{(2N_c-N_F)\geq 2}$ symmetry of softly broken $\mathbf{ {\cal N}=2}$ SQCD.
See Part III.

In comparison e.g. with corresponding results from related papers
\cite{SY1,SY2,SY6} of M.Shifman and A.Yung (and from a number of their \, 
other numerous papers on this subject), our results are essentially \, 
different. In  addition to critical remarks given in Introduction to Part III 
and  in sections 41.1, 41.2, 45.2 and 45.4, an extended criticism of a number of
results from \cite{SY1,SY2} is given in section 8 of \cite{ch13}\,.

In particular, the arguments in \cite{SY3} are based on considering 
the  unequal mass quarks with $\Delta m\neq 0$, the
``outside''\, region $\Delta m\gg\lm$ and ``inside''\, region 
$\Delta m\ll\lm$ and the path between them going through 
Argyres-Douglas point.  

We would like to emphasize that these ``outside''\,  and ``inside''\, 
regions are completely separated by so-called ``wall crossing curves''.
And there is no such path in the complex plane $\Delta m/\lm$
which connects the ``outside''\,  and ``inside''\,  regions and does 
not cross this ``wall crossing curve''.  And the path going through
Argyres-Douglas point is not the exclusion.  

The quantum numbers  of light particles jump generically when
the generic path crosses the ``wall crossing curve'' and have to be 
carefully considered separately in the ``outside''\,  and ``inside''\, 
regions.  

And besides, as for the example considered in \cite{SY3} in support of 
crossover between the confinement and higgs regimes (and 
proposal of  ``instead of confinement''\,  regime),  i.e.   $N_c=3,\, 
N_F=5,\, 2N_c-N_F=1$. It is the example  with the trivial 
$Z_{2N_c-N_F=1}$ symmetry which gives no  restrictions. It belongs 
to another type of vacua, see section 45.5.
And this  clearly  invalidates  all statements of M. Shifman and
A.Yung about quantum numbers of light particles, about crossover 
and  about ``instead of confinement'' regime in vacua with 
non-trivial unbroken $Z_{2N_c-N_F\geq 2}$ symmetry.

{\bf 4)}\, The colored but gauge invariant quark field ${\hat Q}^i_\beta$  was 
 introduced  in \cite{ch21}, see sections 10 and 11.     Its mean values in various 
 vacua  determine the gauge invariant order 
parameter  $\rho,\,\, \langle {\hat Q}^{i}_{\beta}\rangle=\delta^{i}_{\beta}\,\rho$. This 
last  behaves non-analytically\, :\, $\rho_{\rm higgs}\neq 0$ in the higgs phase, while 
$\rho_{HQ}=0$ in   the HQ (heavy quark) phase where quarks are confined.
This is the explicit counter example to the widely spread opinion that the 
gauge  invariant order parameter does not exist for scalar quarks.

{\bf 5)}.\, It was obtained by E. Fradkin and S.H. Shenker \cite{FS} that the transition
between the confinement and Higgs regimes is the analytic crossover, not the phase 
transitoion. It was shown in \cite{ch21,ch23} that  this result is a consequence of unphysical
QCD-model used in \cite{FS}.  Instead of QCD with dynamical scalar quarks, it was 
the Stueckelberg pure YM theory without dynamical quarks and with fixed by hands
nonzero masses of all electric gluons.  Really, in the QCD with normal dynamical  scalar 
quarks the  thansition between these regimes  is the phase transition, 
see sections 10.3 and 12.5.

{\bf 6)}.\, The arguments presented in \cite{IS} by K. Intriligator and N. Seiberg for the 
standard direct $SU(N_c),\,\, N_F=N_c$ $\,\,\,{\cal N}=1$ SQCD theory  in support of 
conclusions \cite{FS} about crossover between  the confinement and higgs regimes,  
are  criticized in \cite{ch23} as erroneous, see section 12.5.

{\bf 7)}.\, It was shown in \cite{ch21} for $1\leq  N_F \leq N_c-1$ in the standard ${\cal N}=1$ 
SQCD that,  at small fixed  quark mass $m_Q\ll\la$ and increased $N_c$,  there  is the phase 
transition from the higgs phase to confinement one. And there is the phase 
transition at  fixed $N_c$ and decreased $m_Q$ from the HQ (heavy quark)  phase  
at $m_Q\gg\la$ where quarks  are confined to the higgs phase at $m_Q\ll\la$ 
where they are higgsed,  see sections 10 and 11.

{\bf 8)}.\,  The confinement mechanism was  proposed in \cite{ch23} for the ordinary 
(i.e. non SUSY) YM and  for ${\cal N}=1$  SYM, see section 12.5}.\\

To the best   of our knowledge, all results obtained  look 
self-consistent,  satisfy a large number of independent checks,
and do not contradict  any proven results.\\

\section{Direct theory}

\subsection{ Definition and some examples}

\numberwithin{equation}{subsection}

Let us recall first in short some properties of the standard ${\cal
N}=1$ SQCD with $SU(N_c)$ colors and $1\leq N_F <3 N_c$ flavors of
{\it light} equal mass quarks $Q^i_\beta,\,\,{\ov Q}^\beta_i,\,\,
i=1...N_F,\,\,\beta=1...N_c$,\,\, with $m_Q\ll\la$, see e.g.
section 2 in \cite{ch1}. It is convenient to start e.g. with $3 N_c/2 < N_F <
3N_c$ and the scale $\mu=\la$. The Lagrangian at $\mu=\la$ looks as
\footnote{\,
The gluon exponents are implied in Kahler terms.  \label{(f1)}
}
\bq
K={\rm Tr}\,\Bigl (Q^\dagger Q+ (Q\ra {\ov Q}) \Bigr )\,, \quad {\cal
W}=\frac{2\pi}{\alpha(\mu=\la)} S+
 {\rm Tr}\,(m_Q {\ov Q} Q)\,.   \label{(2.1.1)}
\eq
$(m_Q)^i_j=\delta^i_j \,m_Q(\mu=\la)$ is the mass parameter
(it is taken as real positive), $S=\sum_{A,\beta}
W^{A,\,\beta}W^{A}_{\beta}/32\pi^2$, where $W^A_{\beta}$ is the 
gauge field strength, $A=1...N_c^2-1,\, \beta=1,2$,\, $a(\mu)=N_c
g^2(\mu)/8\pi^2=N_c\alpha(\mu)/2\pi$ is the gauge coupling
with its scale factor $\la$. Let us take now $m_Q\ra 0$ and evolve to
the UV Pauli-Villars (PV) scale $\mu_{PV}$ to define {\it the parent
UV theory}. The only change in comparison with \eqref{(2.1.1)} will
be the appearance of the corresponding logarithmic renormalization
factor $z_Q(\la, \mu_{PV})=z^{-1})(\mu_{PV},\la)\gg 1$ in the Kahler
term for {\it massless} quarks and the logarithmic evolution of the 
gauge coupling: $\alpha(\mu=\la)
\ra\alpha(\mu=\mu_{PV})\ll \alpha(\mu=\la)$, while the scale factor
$\la$ of the gauge coupling remains the same. Now, we continue the
parameter $m_Q$
from zero to some nonzero value, e.g. $0 < m_Q\ll\la$.
 And this will be {\it  a definition of our parent UV theory}.

The Konishi anomaly \cite{Konishi} for this theory looks as 
\bbq
\hspace*{-2mm} m_Q(\mu)=z^{-1}_Q(\la, \mu)m_Q=z(\mu,\la)
m_Q\,,\,\,\,m_Q\equiv
m_Q(\mu=\la)\,,\,\, M\equiv M (\mu=\la)\,,
\eeq
\bq
\hspace*{-3mm} m_Q(\mu)\langle M(\mu)\rangle= \langle S\rangle\,,\,
M(\mu)= z_Q(\la, \mu)M,\,\,\sum_{\beta=1}^{N_c}\langle{\ov Q}^\beta_j
Q^i_\beta\rangle=\delta^i_j\langle
M\rangle=\delta^i_j M,\,\,M=\sum_{\beta=1}^{N_c}\langle{\ov
Q}^\beta_1Q^1_\beta\rangle. \,\,\, \label{(2.1.2)}
\eq

Let is evolve now this theory from the scale $\mu_{PV}$ to lower
energies. It is in the weak coupling logarithmic perturbative regime
at scales $\la < \mu < \mu_{PV}$ and in the conformal regime at
scales $\lym\sim m^{\rm pole}_Q < \mu < \la$ at $m_Q\ll
m^{\rm pole}_Q \ll \la$. 
\footnote{\, 
Here and below we use the perturbatively exact NSVZ $\beta$-function
\cite{NSVZ-1} corresponding to the Pauli-Villars scheme. Besides, $A\sim
B$ below means equality up to a constant factor independent of $m_Q$,
$A\ll B$ means $|A|\ll|B|$. \label{(f2)}
}
The perturbative pole  mass of quarks looks in this case as 
\bq
\hspace*{-1mm} m^{\rm pole}_Q=\frac{m_Q}{z_Q(\la,m^{\rm pole}_Q
)}\sim\la\Bigl
(\frac{m_Q}{\la}\Bigr )^{\frac{N_F}{3N_c}}\ll\la,\,\, 
z_Q(\la,\mu\ll\la)\sim\Bigl
(\frac{\mu}{\la} \Bigr )^{\gamma_Q=\frac{3N_c-N_F}{N_F}}\ll
1.\,\,\,\,\,\,\label{(2.1.3)}
\eq

The explicit dependence of the gluino condensate $\langle S \rangle$
on the current quark mass $m_Q$  and $\la$ can be found e.g. as follows.

a) One can start with $3 N_c/2 < N_F < 3N_c$ and the heavy mass
quarks, $m_Q^{\rm pole}\equiv m_Q(\mu=m_Q^ {\rm pole})\gg \la$\,, 
so that the theory is UV-free and in
the weak coupling regime at sufficiently large energies.

b) Then, to integrate out inclusively all these heavy quarks directly
in the perturbation theory at scales $\mu < \mu_H=m_Q^{\rm pole}$,
resulting in the pure Yang-Mills lower energy theory with the scale factor 
$\Lambda_{SYM}$. The value of $\Lambda_{ SYM}$ can be found from the 
matching of couplings $\alpha_{+}(\mu) $ and $\alpha_{-}(\mu)$ of the 
upper and lower theories at $\mu=\mu_H\,: \alpha_{+} (\mu_H)=\alpha_{-}
(\mu_H)$. The upper theory is always the original
one with $N_c$ colors and $N_F$ flavors, and the value of
$\alpha_{+}(\mu_H)$ can be
obtained starting with high $\mu\gg \mu_H$ and evolving down to
$\mu=\mu_H$ through the
standard perturbative RG-flow for
theory with $N_c$ colors and $N_F$ flavors of {\it massless} quarks.

But instead, the same value $\alpha_{+}(\mu)$ can be obtained
starting with $\mu\sim\la$ and going up to $\mu=\mu_H\gg \la$ with the
same perturbative RG-flow \cite{NSVZ-1} for {\it massless} quarks. I.e.\,
( ${\rm g}^2 (\mu)=4\pi\alpha(\mu)\,,$)
\bq
\frac{2\pi}{\alpha_{+}(\mu_H)}=(3N_c-N_F)\ln\,\Bigl
(\frac{\mu_H}{\la}\Bigr ) -
N_F\ln \Bigl (\frac{1}{{\it z}_{Q}(\la, \mu_H)}\Bigr )+ N_c\ln \Bigl
(\frac{{\rm g}^2(\mu=\la)}{{\rm g}^2(\mu=\mu_H)}\Bigr )+{\hat C}_{+} \,, 
\label{(2.1.4)}
\eq
where ${\hat C}_{+}$ is a constant,
${\it z}^{-1}_Q(\la, \mu_H\gg\la)={{\it z}_{Q}(\mu_H\gg\la,\la)}\ll 1$ is the 
standard perturbative renormalization factor (logarithmic in this case) of 
{\it massless} quarks in theory with $N_c$ colors and $N_F$ flavors. At the 
weak coupling $\alpha(\mu_H)\ll 1$\, (\, here $C_F=(N_c^2-1)/2N_c\,, 
\quad {\rm b}_o=(3N_c-N_F)$\, )\,:
\bbq
z_Q(\mu_H,\,\la)=C_z \Biggl(\frac{\alpha(\mu_H)}{\alpha(\la)}\Biggr
)^{2\,C_F/\bo}\Biggl ( 1+O(\alpha(\mu_H/\la)\Biggr
)\sim\Bigl(\frac{1}{\ln (\mu_H/\la)} \Bigr
)^{2\,C_F/\bo}\ll 1\,,
\eeq
where $C_z$ is a constant. 

As for the lower energy theory, in all examples considered in this
section it is the
${\cal N}=1$ SYM one with $N_c^{\,\prime}$ colors and no quarks. Its
coupling can be written in a similar way as:
\bq
\frac{2\pi}{\alpha_{-}(\mu_H)}=3N_c^{\,\prime} \ln\,\Bigl (
\frac{\mu_H}{\lym}\Bigr )+
N_c^{\,\prime}\ln\Bigl (\frac{{\rm g}^2(\mu=\lym)}{{\rm g}^2
(\mu=\mu_H)}\Bigr )+{\hat C}_{-}\,. \label{(2.1.5)} 
\eq

The purpose here and everywhere below is to trace explicitly the
dependence on
the parameters like $\mu_H/\la$ which will be finally expressed
through the
universal parameter $m_Q/\la\,,\,\,m_Q\equiv m_Q(\mu=\la)$, which can
be large $m_Q/\la\gg 1$, or small $m_Q/\la\ll 1$. So, from now on and
everywhere below the constant terms like ${\hat C}_{+}=2\pi/\alpha(\mu
=\la),\,\, {\hat C}_{-}=2\pi/\alpha(\mu=\lym)$ will be omitted if
they are ${\cal O}(1)$, as their effect is equivalent
to a redefinition of $\la$ by a constant factor. (As can be seen in
section 2.2, in the vicinity of special values $N_F/N_c$ these
constants can be parametrically different from
${\cal O}(1)$ ).

In the case considered here\,:
$N_c^{\prime}=N_c\,,\,\mu_H=m_Q^{pole}\gg \la$\,,
and one obtains then from \eqref{(2.1.3)}\,,
\eqref{(2.1.4)}\,,\eqref{(2.1.5)} within the conformal window\,:
\bq
\Lambda_{SYM}={\rm C_{SYM}(N_F,\,N_c)} (\la^{\bo}\det m_Q)^{1/3N_c},
\quad {\rm C_{SYM}(N_F,\,N_c)}=(\rm const)\,,\label{(2.1.6)}
\eq
\bbq
m_{Q}\equiv z_{Q}(\la,\,m_{Q}^{\rm pole}\gg\la) m_{Q}^{\rm pole},\quad 
m_{Q}^{\rm pole} \gg \la\,.  
\eeq

c) Lowering the scale $\mu$ down to $\mu < \Lambda_{SYM}$ and
integrating out all  gauge degrees of freedom, except for the one 
whole field $S$ itself,  one can write the
effective Lagrangian in the Veneziano-Yankielowicz (VY) form\,
\cite{VY}, from which one obtains the gluino condensate:-
\bq
\langle S \rangle=\Lambda^3_{SYM}\sim (\la^{\bo}\det m_Q)^{1/N_c}\,,
\quad
m_Q=m_{Q}(\mu=\la)\,\,. \label{(2.1.7)}
\eq

Now, the expressions \eqref{(2.1.6)}\,,\eqref{(2.1.7)} can be
continued in $m_Q$ from large $m_Q\gg \la$ to small values, $m_Q\ll\la$. 
When $m_Q$ for $m_Q\gg \la$ is some formally defined parameter\, see
\eqref{(2.1.6)}, the physical quark mass is in this case $m_Q^{\rm
pole}\gg \la$ and it does not run any more at $\mu< m_Q^{\rm pole}$,
while it has a simple and direct meaning at $m_Q < \la$: $m_Q=m_Q(\mu=\la)$.

The expression \eqref{(2.1.7)} for $\langle S \rangle$ appeared in
the literature
many times before but, to our knowledge, the exact definition of the
parameter $m_Q$ entering \eqref{(2.1.7)}, i.e. its relation with
$m_Q(\mu=\la)$ entering \eqref{(2.1.1)} which defines the theory, has
not been given. Clearly, without this explicit relation the
expressions \eqref{(2.1.6)}\,, \eqref{(2.1.7)} have no much meaning,
as the quark mass parameter $m_Q(\mu)$ is running. For instance, if 
$m_Q$ is understood as $m_Q^{\rm pole}$ 
in \eqref{(2.1.7)}, the relation $\langle S\rangle=
(\la^{\bo}\det m_Q^{\rm pole})^{1/N_c}$ will be erroneous.
(For unequal mass quarks $\det m_Q =
\Pi_{i=1}^{N_F} m_{Q,i}(\mu=\la)$ in  \eqref{(2.1.7)}).

All this becomes especially important, in particular, in the
conformal window $3N_c/2 <N_F< 3N_c$ and $m_Q\ll \la$, 
when $m_Q(\mu)$  runs at $\mu\ll \la$ in a power-like fashion:
$m_Q(\mu_2)=(\mu_1/\mu_2)^{\bo/N_F}\,m_Q(\mu_1)$. \\

d) From the Konishi anomaly equations \cite{Konishi}\,:
\bq
\langle \, \sum_{\beta=1}^{N_c}\Bigl ({\ov Q}^\beta_j\,Q^i_\beta\Bigr
)_{\mu}\rangle =\langle M^i_j(\mu)\rangle=\delta^i_j M(\mu)=\Bigl
(m^{-1}_Q (\mu)\Bigr )^i_{ j}\,\langle S \rangle \,. \label{(2.1.8)}
\eq
one obtains the explicit value of the chiral condensate:
\bbq
M\delta^i_j=\frac{\langle S\rangle}{m_Q}\,\delta^i_ j\,,\quad
M=M(\mu=\la)=
\Bigl(\la^{\bo=3N_c-N_F}\,m_Q^{\nd=N_F-N_c}\Bigr )^{1/N_c},
\eeq
\bq
\langle S \rangle=\Lambda^{3}_{SYM}\sim \Bigl (\la^{\bo}
m_Q^{N_F}\Bigr )
^{1//N_c}\,,\quad m_Q\ll\la\,.\, \label{(2.1.9)}
\eq

Now, the expressions \eqref{(2.1.6)},\, \eqref{(2.1.7)} can be
continued in $N_F$
from the region $N_F<3 N_c$ to $N_F> 3N_c$ and, together with the
Konishi anomaly relation \eqref{(2.1.2)}, these three become then
the basic universal relations for any values of quark masses and any
$N_F$.\\

Another check can be performed for $N_F < N_c-1$ and small quark
masses, $m_Q\ll\la$.\, 
In this case all quarks are higgsed (at not too large values of
$N_c$, see \,\cite{ch21}), and the gauge symmetry $SU(N_c)$ is broken
down to $SU(N_c^{\prime}=N_c-N_F)$ at the high scale $\mu_H=\mu_{\rm
gl}\gg \la$\,: $\langle {\hat Q}^{i}_{\beta}\rangle_{\mu=\mu_{\rm
gl}}=\delta^{i}_{\beta} [\hat\Pi]^{1/2}\,,\quad \langle
{\hat {\ov Q}}_{ j}^{\beta}\rangle_{\mu=\mu_{\rm gl}}=\delta_{
j}^{\beta} \Pi^{1/2}\,,\quad  [\hat\Pi]^{1/2}\gg \la$\,.

And $N_F(2N_c -  N_F)$ gluons become massive, with the mass 
$\mu^2_{\rm gl}=\Bigl ({\rm g}^2_{+}z_{+}\langle{\hat\Pi}\rangle
\Bigr )_{\mu=\mu_{\rm gl}} \gg\la^2\,, 
\quad\ {\rm g}^2_{+}=4\pi\alpha_{+}(\mu=\mu_{\rm gl})\ll 1$.

The same number of quark degrees of freedom acquire the same masses
and become the  superpartners of massive gluons (in a sense, they can be 
considered  as the heavy "constituent quarks"), and there remain 
$N_F^2$ light  complex pion fields  ${\hat \pi}^i_ j$\,  \, :
$$\,({\ov Q}_ j Q^i)_{\mu=\mu_{\rm gl}}=\hat
\Pi^i_j= z_{+}(\mu_{\rm gl},\,\la)\Pi(\mu=\la)=
(\delta^i_j\, \langle M\rangle_{\mu=\mu_{\rm gl}}+{\hat \pi}^i_{\j}\langle
M\rangle^{1/2}_{\mu=\mu_{\rm gl}}), \,\,\langle {\hat \Pi}^i_{j}
\rangle=\delta^i_ j {\hat\Pi}$$\, .

All heavy particles can be integrated out inclusively at scales
$\mu<\mu_{\rm
gl}$\,. The numerical matching of couplings at $\mu_H=\mu_{\rm
gl}$\,:\, $\alpha_{+}(\mu=\mu_{\rm gl},\la)$ in \eqref{(2.1.4)} and
 $\alpha_{-}(\mu=\mu_{\rm gl}, \langle\Lambda_L\rangle)$
in \eqref{(2.1.5)} of the lower energy pure 
${\cal N}=1$  SYM theory can be performed similarly to the 
previous   examples with
heavy quarks. But in this case we consider it will be more useful to
write the explicit form of the $\hat\Pi$-dependence of the lower
energy coupling $\alpha_{-}(\mu<\mu_{\rm gl},\Lambda_L)$ multiplying
the field strength squared of massless gluons, to see how the
multi-loop $\beta$ - function reconciles with the holomorphic
dependence of $\Lambda_L$ on the chiral superfields $\hat \Pi$\,.
This looks now as :
\bbq
\frac{2\pi}{\alpha_{-}\Bigl (\mu\leq\mu_{\rm gl},\Lambda_L\Bigr )}=
\Biggl \{3(N_c-N_F)\ln\,\Bigl (\frac{\mu}{\la}\Bigr )+(N_c-N_F)\ln
\Biggl  (\frac{ {\rm g}^2_{-}(\mu=\langle\Lambda_L\rangle)}
{ {\rm g}^2_{-} (\mu,\langle\Lambda_L\rangle)}\Biggr )
\Biggr \}+
\eeq
\bbq
+\Biggl \{\frac{3}{2}\ln \Biggl (\frac{{\rm g}_{+}^{2N_F}(\mu=\mu_{\rm
gl},\la)\det {\hat \Pi}}{\la^{2N_F}}\Biggr )+N_F\ln \Biggl (\frac{{\rm g}_{+}^2
(\mu=\la)}{{\rm g}^2_{+}(\mu=\mu_{\rm gl},\la)}\Biggr)\Biggr\}-
\eeq
\bq
-\Biggl\{\frac{1}{2}\ln \Biggl (\frac{{\rm g}^{2N_F}_{+}(\mu=\mu_{\rm
gl},\la)\det {\hat \Pi}}{\la^{2N_F}}\Biggr )+N_F\ln \Biggl
(\frac{1}{z_{+}(\mu=\mu_{\rm
gl},\,\la)}\Biggr )\Biggr\}\,,  \label{(2.1.10)}
\eq
where three terms in curly brackets in \eqref{(2.1.10)} are the
contributions of,
respectively, massless gluons, massive gluons and higgsed quarks, and
$z_{+}(\mu_{\rm gl}\gg \la,\,\la)\ll 1$ is the standard perturbative
logarithmic renormalization factor of massless quarks (see above).

It is worth noting that the dependence of the coupling
$2\pi/\alpha_{-}$ on the
quantum pion fields ${\hat\pi}^{i}_{j}$ entering ${\hat \Pi}^i_{j}$
originates only from the ${\hat \pi}$ -dependence of heavy particle
masses entering the "normal" one-loop contributions to the gluon
vacuum polarization, while the "anomalous" higher loop contributions
\cite{NSVZ-1} originating from the quark and gluon renormalization
factors $z_Q$ and $z^{\pm}_{\rm g}\sim {\rm g^2_{\pm}}$
do not contain the quantum pion fields $\hat\pi$ and enter
\eqref{(2.1.10)} as pure neutral c-numbers. This is clear from the
R-charge conservation { (see the footnote \ref{(f2)} ) and the
holomorphic dependence of F-terms on chiral quantum superfields }
(the chiral superfields here are $\Pi (\mu_1)=z_Q(\mu_1,\,\mu_2)\,\Pi
(\mu_2))$.

So, the coupling $\alpha_{-}(\mu,\,\Lambda_L )$ of the lower energy
pure Yang-Mills theory at $\mu<\mu_{\rm gl}$ and its scale factor
$\Lambda_L$ look as:
\bbq
\hspace{-5mm}\frac{2\pi}{\alpha_{-}^{W}\Bigl (\mu\leq\mu_{\rm
gl},\Lambda_L\Bigr )}=\frac{2\pi}{\alpha_
{-}\Bigl(\mu\leq\mu_{\rm gl},\Lambda_L\Bigr )}-(N_c-N_F)\ln
\frac{ {\rm g}^2_{-}(\mu=\langle\Lambda_L\rangle)}{{\rm
g}^2_{-}(\mu<\mu_{\rm gl}, \langle\Lambda_L\rangle)}=
3(N_c-N_F) \ln\Bigl ( \frac{\mu}{\Lambda_L}\Bigr )=
\eeq
\bbq
=3(N_c-N_F) \ln\Bigl ( \frac{\mu}{\la}\Bigr )+\ln \Biggl (\frac
{z^{N_F}_{+}(\mu_{\rm gl},\,\la)\det {\hat \Pi}}{\la^{2N_F}}\Biggr )\,,
\eeq
\bq
\Lambda_L^{3(N_c-N_F)}=\frac{\la^{\bo}}{z^{N_F}_{+}(\mu_{\rm
gl},\,\la)\det {\hat
\Pi}}\equiv \frac{\la^{\bo}}{\det {\Pi}}=\lym^{3(N_c-N_F)}\Biggl
(\det\frac{\langle \Pi\rangle}{\Pi}\Biggr )\,, \label{(2.1.11)}
\eq
\bq
\Pi\equiv z_{+}(\mu_{\rm gl},\,\la)\hat \Pi\,,\quad \langle
\Pi\rangle=\frac{\lym^3}{m_Q}\gg\la^2\,, \quad\langle
\Lambda_L\rangle=\lym=\Bigl (\la^{\bo}\det m_Q\Bigr )
^{1/3N_c}\,\label{(2.1.12)}\,,
\eq
and the Lagrangian at $\mu<\mu_{\rm gl}$ takes the form :
\footnote{\,
Because the gluon fields are not yet integrated completely, there are
the gluon regulator fields (implicit) whose contributions ensure the 
R-charge conservation in \eqref{(2.1.13)}.
}
\bq
L=\int \te\ote \Biggl \{2\,\rm {Tr}\sqrt {{\hat\Pi}^\dagger
{\hat\Pi}}\Biggr
\}+\Biggl [\int \te \Biggl \{ \frac{2\pi}{\alpha_{-}(\mu,\Lambda_L)}{\hat
S}+{\hat m_Q}\rm {Tr}{\hat\Pi} \Biggr \}+h.c.\Biggr ]\,, \label{(2.1.13)}
\eq
where $\hat S={\hat W}^2_{\alpha}/32\pi^2$\,,\, and ${\hat
W}_{\alpha}$ are the gauge field strengths 
of $(N_c-N_F)^2-1$ remaining massless gluon fields.

Lowering the scale $\mu$ down to $\mu < \Lambda_{SYM}$ and
integrating \, out all degrees of freedom 
of gluons, except for the one whole field $\hat S$
itself (this leaves behind a large number of gluonia with masses
$\mu_{\rm gl}\sim \Lambda_{SYM}$), one obtains the VY - form
:\bbq
L=\int \te \,\ote \, \Biggl \{ 2\,\rm {Tr}\,\sqrt {{\hat\Pi}^\dagger
{\hat\Pi}}
\,\,+(\rm {D \,\,terms\,\, of\,\, the\,\, field \,\,{\hat S}
\,})\Biggr )+
\eeq
\bq
+\Biggl [\int \te \Biggl \{ -(N_c-N_F)\,{\hat S} \Biggl ( \ln \Bigl
(\frac{\hat S}{\Lambda_{L}^3}
\Bigr )-1\Biggr )+{\hat m_Q}\rm {Tr}\, {\hat\Pi} \Biggr \}+c.c.\Biggr ]\,,
\,\,\,\mu <\Lambda_{SYM}\,.\,. \label{(2.1.14)}
\eq
It is worth noting that it is the first place where the
non-perturbative effects
were incorporated to obtain the VY\,-\,form of the superpotential
(the non-perturbative effects introduce the infrared cutoff $\sim
\lym$\,, so that the explicit dependence on $\mu$ disappears at $\mu<
\lym$),  while all previous calculations with this example were purely
perturbative. One obtains from \eqref{(2.1.14)} the gluino vacuum
condensate: $\langle \hat S\rangle=\langle
\Lambda_L^3\rangle=\Lambda_{SYM}^3=\langle S\rangle=(\la^{\bo}\det
m_Q)^{1/N_c}$\,.

Now, integrating out the last gluonium field $\hat S$ (with its mass
scale
$\sim\Lambda_{SYM}$\,) at lower energies, one obtains the Lagrangian
of pions:
\bq
L=\Biggl [2\,\rm {Tr}\,\sqrt {{\hat\Pi}^\dagger {\hat\Pi}}\Biggr
]_D+\Biggl \{\Biggl [ (N_c-N_F)\Biggl
(\frac{\la^{\bo}}{{\it z}^{N_F}_Q(\mu_{\rm gl},\,\la)\det\,\hat\Pi}
\Biggr )^{1/(N_c-N_F)}+ {\hat m}_Q\rm {Tr}\, {\hat\Pi} \Biggr ]_F\,+h.c.\Biggr
\}=\,\,\,\,\label{(2.1.15)}
\eq
\bbq
=\Biggl [2\,{\it z}_Q(\la, \mu_{\rm gl})\rm {Tr}\,\sqrt{{\Pi}^\dagger {\Pi}}
\Biggr ]_D+\Biggl \{\Biggl [ (N_c-N_F)\Biggl (\frac{\la^{\bo}}{\det\,{\Pi}}
\Biggr )^{1/(N_c-N_F)}+ {m}_Q\rm {Tr}\, {\Pi} \Biggr ]_F+h.c.\,\Biggr \}\,,
\quad\mu\ll\Lambda_{SYM}\,.
\eeq

From \eqref{(2.1.15)}, $\langle \Pi\rangle$ and pion masses look as
\bq
\langle \Pi\rangle=\la^2\Biggl (\frac{\la}{m_Q}\Bigr )^{\frac{N_c-N_F}{N_c}}
\gg\la^2\,,\quad  \mu^{\rm pole}(\Pi)=\frac{2 m_Q}{{\it z}_Q(\la, \mu_{\rm gl})}\,,
\quad m_Q\ll \mu^{\rm pole}(\Pi)\ll\lym\,.  \label{(2.1.16)}
\eq

The superpotential of the form
$(N_c-N_F)({\la^{\bo}}/\det\,{\Pi})^{1/(N_c-N_F)}$
appeared many times in the literature because, up to an absolute
normalization of the field $\Pi(\mu)={\ov Q}Q(\mu)$ (which is not
RG-invariant by itself), this is the only possible form of the
superpotential, if one is able to show that the lowest energy
Lagrangian depends on $N_F^2$ pion superfields only. But
the absolute normalization of all terms entering \eqref{(2.1.15)} has
never been carefully specified (clearly, the absolute normalization
makes sense only when both the superpotential
and the Kahler terms are absolutely normalized simultaneously). The
Lagrangian {\eqref{(2.1.15)} describes weakly interacting pions with small
masses $M_{\pi}=2 {\hat m}_Q=2 z_Q(\mu_{\rm gl},\,\la) m_Q\ll m_Q\ll
\Lambda_{SYM}\ll \la$\,.\\

On the whole, the mass spectrum contains in this case: $N_F(2N_c-N_F)$ 
very heavy gluons and "constituent quarks" with the mass 
scale  $\mu_{\rm gl}\sim \langle M\rangle^{1/2}\gg\la$, see \eqref{(2.1.9)}, 
\eqref{(2.1.12)}, \, then  $SU(N_c-N_F)$ gluonia with typical masses $\sim
\Lambda_{SYM}\ll \la$ and $N^2_F$ pions with masses $\sim m_Q
\ll\lym.$.

The form \eqref{(2.1.15)} can be continued in $N_F$ to the point
$N_F=N_c-1$ and
it predicts then the form of the pion Lagrangian for this case. Now,
the whole gauge group is higgsed at the high scale $\mu_H=\mu_{\rm
gl}\gg \la$, and the direct way to obtain \eqref{(2.1.15)} is not
through the VY\,-\,procedure, but through the calculation of the
one-instanton contribution \cite{ADS},\,\cite{SV-r}. The changes in
the mass spectrum are evident and, most important,\,-\, there is no
confinement and there are no particles
with masses $\sim\Lambda_{SYM}$ in the spectrum in this case. \\

Finally for this section, to check the universal dependence of 
$\langle S \rangle$   on $m_Q$ \eqref{(2.1.7)}, let  us consider briefly
the case $N_F > 3N_c$ and $m_Q\ll \la$. 
In this case  $\bo=(3N_c-N_F)<0$, so  that the direct  theory is 
IR-free in   the interval  $\mu_H<\mu<\la$, where  $\mu_H$ is the
highest physical mass ($\lym\ll\mu_H=m_Q^{\rm pole}\ll \la$ in this
example). I.e., its coupling which is $O(1)$ at $\mu= \la$ becomes
logarithmically small at $\mu\ll \la$. Besides, the parameter $m_Q$
has now a direct physical  meaning as the value of the running quark 
mass at $\mu=\la$,\,  $m_Q\equiv  m_Q(\mu=\la)\ll \la$.

The current quark mass $m_Q=m_Q(\mu=\la)\ll \la$ is much larger now
than the scale of its chiral condensate $\langle M \rangle^{1/2}\ll m_Q$\,,
see \eqref{(2.1.9)}, and this power hierarchy persists at lower energies as
the RG - evolution here is only logarithmic at $ \lym \ll\mu<\la$\,.
Therefore, the direct theory is at $N_F>3 N_c$ in the HQ (heavy
quark) phase, so that there is a standard weak coupling slow
logarithmic
evolution in the region $m_Q^{\rm pole}\ll \mu \ll \la ,\,\, m_Q^{\rm
pole}\equiv m_Q (\mu=m_Q^{\rm pole})=z_Q^{-1}(\la,\,\mu=m_Q^{\rm
pole})\,m_Q \gg m_Q $, where $z_Q(\la,\, \mu=m^{\rm pole}_Q)\ll 1$ is
the standard perturbative logarithmic renormalization factor of
massless quarks, and the highest physical scale is now
$m_Q^{\rm pole}$. All heavy  and weakly confined quarks 
(the string tension is $\sqrt \sigma\sim\lym\ll  m^{\rm pole}_Q $) 
can be integrated out inclusively at $\mu < m_Q^{\rm pole}$,  
their vacuum condensate $\langle M
\rangle=\langle{\ov Q} Q (\mu=\la)\rangle=\langle S\rangle/m_Q$ 
is due to a quantum one-loop contribution of these
heavy quarks  (this is the Konishi anomaly). 
This leaves behind a large number of  mesons and baryons
made of these non-relativistic quarks with masses
: ${\rm M_{meson}\sim m^{\rm pole}_Q,\,\, M_{baryon}
\sim N_c \,m^{\rm  pole}_Q}$\,. 
Evidently, there are no additional lighter pions now.

Using \eqref{(2.1.4)},\,\eqref{(2.1.5)} to match couplings at $\mu=
m^{\rm pole}_Q$\,, one obtains at lower energies $\mu < m^{\rm
pole}_Q$ the Yang-Mills Lagrangian with the scale factor of its gauge
coupling $\lym\sim (\la^{\bo}\,m_Q^{N_F})^{1/3N_c}\ll m_Q$\,, so that
this SYM theory is in the weak coupling regime at $\lym\ll
\mu<m_Q^{\rm pole}$\,. Finally, it describes strongly coupled gluonia
with masses $\mu_{\rm gl}\sim \lym \ll m_Q^{\rm pole}$, and these are
the lightest particles in this case. So, this is the end of this
short story in the direct theory.\\

\subsection{ Direct theory. Equal quark masses. 
$0<(3N_c-N_F)/N_F\ll 1$}

\numberwithin{equation}{subsection}

Let us consider a specific case of direct theory with
light equal mass quarks and $0 < \bo=(3N_c-N_F)/N_F\ll 1$.

This theory is UV free and is in the conformal regime at scales
$\mu_H<\mu<\la$,
\footnote{\,
By definition, within the conformal window, $\la$ is a scale such
that \, the coupling $a(\mu=\la)=N_c\alpha(\mu=\la)/2\pi$ is sufficiently 
close to its fixed point value $a_{*}$, i.e. $a_{*}-a(\mu=\la)=\delta a_{*},
\,\, 0 < \delta  a_{*}\ll 1$ (and similarly for the dual theory, ${\ov a}_{*}
-{\ov a}(\mu=\Lambda_q)=\delta {\ov a}_{*}$\,). The coupling $a^{*}$ 
is weak at  $\bo/N_F\ll 1$, $\gamma_Q\simeq a^{*}=(\bo/N_F)\ll 1$\,.
Here and in what follows, we trace only the leading exponential
dependence on the small parameter $\bo/N_F\ll 1$ (or $\bd/N_F\ll 1$)\,, 
i.e. factors of the order of $\exp\{-c_o(N_F/\bo)\},\, c_o\sim 1$, while the 
nonleading terms of  the order of $\exp\{-c_ {\delta} (N_F/\bo),\, c_{\delta}
\sim \delta\}$ or   pre-exponential factors
$\sim (N_c/\bo) ^{\sigma},\, \sigma=O(1)$\,, are neglected, because
this simplifies greatly all expressions. Besides, it is always
implied that even when $\bo/N_F$ or $\bd/N_F$ are $\ll 1$, these small
numbers \, do not compete {\it in any way} with the main small parameter
$m_Q/\la\ll 1$. See Appendix in \cite{ch3} for more details.  \label{(f4)}
}
where $\mu_H$ is the highest physical mass, and in this case it is
the quark pole mass $\mpQ$\,:
\bbq
\frac{\mpQ}{\la}=\frac{m_Q(\mu=\mpQ)}{\la}=\frac{m_Q}{z_Q(\la,\mpQ)
\la}\sim\frac{m_Q}{\la}\Biggl (\frac
{\la}{\mpQ}\Biggr )^{\gamma_Q}\sim\Biggl (\frac{m_Q}{\la}\Biggr
)^{\frac{1}{1+\gamma_Q}=\frac{N_F}{3N_c}}\,,
\eeq
\bq
m_Q\equiv m_Q(\mu=\la)\,, \quad\gamma_Q=\frac{\bo}{N_F},\quad
\frac{\bo=3N_c-N_F}{N_F}\ll 1, \quad z_Q(\la,\mpQ)\sim
\Bigl (\frac{\mpQ}{\la}\Bigr )^{\gamma_Q}\,, \label{(2.2.1)}
\eq
where $z_Q(\la,\mpQ\ll\la)\ll 1$ is the renormalization factor of the quark
Kahler term and $\gamma_Q$ is the quark anomalous
dimension \cite{NSVZ-1}. After integrating out inclusively
all quarks as heavy ones, a pure SYM theory with $N_c$ colors remains
at lower energies. Its scale parameter $\lym$ can be determined from
matching the couplings of the higher- and lower-energy theories at 
$\mu=\mpQ$. Proceeding as in \cite{ch1}, we obtains
\bq
3 N_c\ln\Biggl (\frac{\mpQ}{\lym} \Biggr )\approx
\frac{2\pi}{\alpha_*}\approx
N_c\frac{N_F}{\bo}\quad \ra \quad\lym\sim\exp \Bigl
\{\frac{-N_F}{3\bo}\Bigr \}\,\mpQ \ll m_Q^{\rm pole}\,.
\label{(2.2.2)}
\eq

Therefore, there are two parametrically different scales in the mass
spectrum of
the direct theory in this case. There is a large number of colorless
flavored hadrons made of weakly confined (the string tension $\sqrt
\sigma\sim \lym\ll m_Q^{\rm pole}$) and
weakly interacting {\it nonrelativistic} heavy quarks $Q\,,\,\ov Q$
with masses $m_Q^{\rm
pole}\gg \lym$\,, and a large number of gluonia with the mass scale
$\lym=\exp \{-N_F/3\bo\}m_Q^{\rm pole}\ll m_Q^{\rm pole}$.

To check the self-consistency, we estimate the scale of the gluon
masses due to a
possible quark higgsing. The quark chiral condensate at
$\mu=\la\,:\,\delta^i_j\Pi=\sum_{\beta=1}^{N_c}\langle {\ov
Q}^{\beta}_j Q^i_{\beta} (\mu=\la)
\rangle$ is given by \cite{Konishi}, see \eqref{(2.2.2)}
 \bq
\frac{\Pi}{\la^2}= \frac{\sum_{\beta=1}^{N_c}\langle {\ov
Q}^{\beta}_1Q^1_{\beta} (\mu=\la) \rangle}{\la^2} =\frac{\langle
S\rangle=\lym^3}{m_Q\la^2}\sim \exp \Bigl \{\frac{-N_F}{\bo} \Bigr \}
\Biggl (\frac{m_Q}{\la}\Biggr
)^{\frac{N_F-N_c}{N_c}}\,.\label{(2.2.3)}
\eq

If the gluons were acquired masses $\mg > m_Q^{\rm pole}$ due to
higgsing of  quarks, then the conformal renormalization group (RG) 
evolution would
stops at $\mu=\mg$. Hence, $\mg$ can be then estimated from
\bbq
\mg^2\sim {\rm g}^2_{*}(\mu=\mg)\langle 0|\Pi={\ov Q}Q (\mu=\mg)|0\rangle\,,
\quad \mg^2\sim {\rm g}^2_{*}(\mu=\mg)\Pi\Biggl (\frac{\mg}{\la} \Biggr
)^{\gamma_Q}\,,\quad
 g^2(\mu)/4\pi=\alpha(\mu)\,,
\eeq
\bq
\frac{\mg}{\la}\sim \Bigl (\frac{\Pi}{\la^2}
\Biggr)^{\frac{1}{2-\gamma_Q}}
\sim \Bigl (\frac{\Pi}{\la^2} \Biggr)^{N_F/3\nd} \sim\exp
\Bigl \{\frac{ - N_c}{2\bo} \Bigr \}\frac{\lym}{\la}\,,\quad
\mg\ll\lym\ll \mpQ\ll\la . \label{(2.2.4)}
\eq
Therefore, the scale of possible gluon masses, $\mu=\mg$, is
parametrically
smaller than the quark pole mass and the picture of $Q\,,\ov Q$ being
in the HQ (heavy
quark) phase and not higgsed but confined  is self-consistent.

This example shows that the constant $C_{SYM}$ in \eqref{(2.1.6)}
is parametrically smaller than unity in the vicinity of especial value
of $N_c/N_F$, see  \eqref{(2.2.2)}. Then {\it there are no reasons for it 
to be exactly  unity at generic values of} $N_c/N_F$.\\

\section{ Dual theory. Equal quark masses. {\boldmath $0<(3N_c-N_F)/N_F\ll 1$} }

The Lagrangian of the dual theory at the scale $\mu=|\Lambda_q|=\la$ is
taken in the form \cite{S1}
\bq
{\ov K}= {\rm Tr}\Bigl ( q^\dagger q + {\ov q}^\dagger {\ov q}\Bigr )
+\frac{1}{\mu_2^2}{\rm Tr} \Bigl (M^{\dagger}
M\Bigr)\,,\label{(3.0.1)}
\eq
\bbq
{\ov \w}=\Biggl [\,\, \frac{2\pi}{\ov \alpha(\mu)}\, {\ov
S}-\frac{1}{\mu_1}\,
{\rm Tr} \Bigl ({\ov q}\, M\, q \Bigr ) + {\rm Tr}\Bigl ( m_Q
M\Bigr )\,,
\quad {\ov S}={\ov W}_{\alpha}^2/32\pi^2\,.
\eeq
Here\,:\, the number of dual colors is ${\ov N}_c=(N_F-N_c),\, {\rm \ov
b}_o=3\nd-N_F$, and $M^i_j$ are the $N_F^2$ elementary mion fields,
${\ov a}(\mu)=\nd{\ov \alpha}(\mu)/2\pi=\nd{\ov g}^2(\mu)/8\pi^2$ is
the dual running gauge coupling,\,\, ${\rm \ov W}^b_{\alpha}$ is the
dual gluon field strength. The gluino condensates of the direct and
dual theories are observable quantities and are matched,
$\langle{-\,\ov S}\rangle=\langle S\rangle=\lym^3$.

By definition, $\mu=|\Lambda_q|$ is such a scale that the dual theory
already entered sufficiently deep into the conformal regime, i.e. both the
gauge and Yukawa couplings, ${\ov a}(\mu=|\Lambda_q|)$ and
$a_f(\mu=|\Lambda_q|)=\nd
f^2(\mu=|\Lambda_q|)/8\pi^2,\,\,f(\mu=|\Lambda_q|)=\mu_2/\mu_1$, are
already close to their fixed point values, $\ov\delta=({\ov a}_*-{\ov
a}(\mu=|\Lambda_q|)/{\ov a}_*\ll 1$ (and similarly for
$a_f(\mu=|\Lambda|_q)$\,. At $0<\bo/N_c\ll 1$ the fixed point dual
couplings are ${\ov a}_*\sim a^*_f\sim 1$, while at $0<\bd/\nd\ll 1$
they are small, ${\ov a}_*\sim a^*_f\sim (\bd)/N_F\ll 1$\,.

We take $|\Lambda_q|=\la$ for simplicity (because this does not
matter  finally but simplifies greatly all formulas, see the Appendix 
in \cite{ch3} for more details).

The mean vacuum values of $\langle M^i_{j}(\mu=\la)\rangle$ in \eqref{(3.0.1)}
and $\langle{\ov Q}_j Q^i(\mu=\la)\rangle$ 
can always be matched at $\mu=|\Lambda_q|=\la$,
at the appropriate choice of $\mu_1$ in \eqref{(3.0.1)}
\bq
M\equiv\langle M(\mu=\la)\rangle=\sum_{\beta=1}^{N_c}\langle {\ov
Q}_1^\beta Q^1_{\beta} (\mu=\la) \rangle=\frac{\langle
S\rangle}{m_Q}\,. \label{(3.0.2)}
\eq

The Konishi anomalies at $\mu=\la= - \Lambda_q$ look in this case as
\bq
\sum_{\alpha=1}^{\nd=N_F-N_c}\langle {\ov q}^{1}_\alpha
q_{1}^{\alpha}\rangle=m_Q\mu_1, \,\,
\sum_{\alpha=1}^{\nd=N_F-N_c}\langle {\ov
q}^{1}_\alpha q_{1}^{\alpha}\rangle\langle \frac{M}{\mu_1} \rangle=
- \langle {\ov S}\rangle=m_Q\langle M \rangle=\langle S\rangle.\,\,
\label{(3.0.3)}
\eq
\bq
\langle S\rangle=\lym^3= - \langle {\ov S}\rangle= - {\ov
\Lambda}_{SYM}^3\,\,\ra\,\,( {\ov \Lambda}_{SYM}= - \lym) \,\ra\,\,\,
( \la= - \Lambda_q )\,. \label{(3.0.4)} 
\eq

At $3/2<N_F/N_c<3$ this dual theory can be taken as UV free at
$\mu\gg\la$ and this requires that its Yukawa coupling at $\mu=\la,\,
f(\mu=\la)=\mu_2/\mu_1$, cannot be larger than its gauge coupling
${\ov g}(\mu=\la)$, i.e. $\mu_2/\mu_1\lesssim 1$. The same requirement 
to $\mu_2/\mu_1$ follows from the conformal
behavior of this theory at $3/2<N_F/N_c<3$ and $\lym < \mu<\la$, i.e.
$a_f(\mu=\la)\simeq a^*_f\sim {\ov a}_*=O(\bd/N_F)$. Hence, we take
$\mu_2=\mu_1$ in what follows. Therefore, only one free parameter 
$\mu_1\equiv Z_q \la$ remains in the dual
theory. It will be determined below from the explicit matching of the
gluino condensates in the direct and dual theories. We consider below
this dual theory at $\mu\leq \la$
 only where it claims to be equivalent to the direct one.\\

The dual theory is also in the conformal regime at $0<\bo/N_F\ll 1$
and ${\ov
\mu}_H<\mu<\la$\,, where ${\ov \mu}_H$ is the corresponding largest
physical mass.

Assuming that the dual quarks ${\ov q},\,q$ are in the HQ-phase and
hence ${\ov
\mu}_H=\mu_q^{\rm pole}$\,, we find their pole mass
($\nd=N_F-N_c\,,\,\,{\ov b}_0
=(3\nd-N_F)\,,\,\, \gamma_q=\bd/N_F=(2N_F-3N_c )/N_F$\,)\,:
\bq
\frac{\mu_q^{\rm pole}}{\la}\sim\frac{\mu_q}{\la} \Biggl
(\frac{\la}{\mu_q^{\rm
pole}}\Biggr )^{\gamma_q}\sim\Biggl (\frac{M}{Z_q\la^2}\Biggr
)^{N_F/3\nd}\,,\quad
\mu_q\equiv\mu_q(\mu=\la)=\frac{M}{\mu_1= Z_q\la}\,.\label{(3.0.5)}
\eq

We now integrate out inclusively all dual quarks as heavy ones at
$\mu<\mu_q^{\rm
pole}$ and determine the scale factor ${\ov\Lambda}_{SYM}$ of the
remained dual SYM theory (the dual coupling is ${\ov
a}_*=\nd\,{\ov\alpha_*}/2\pi=O(1)\,)$\,:
\bq
3\nd\ln\Biggl (\frac{\mu_q^{\rm pole}}{{-\ov \Lambda}_{SYM}} \Biggr
)\sim\frac{\nd}{\ov a_*} \sim \nd \quad \ra \quad \mu_q^{\rm
pole}\sim{\Lambda}_{SYM}\ll \mpQ\,.
\label{(3.0.6)}
\eq

One obtains from \eqref{(3.0.5)},\,\eqref{(3.0.6)},\, \eqref{(2.2.2)}
\bq
\mu_q^{\rm pole}\sim\lym \quad \ra \quad Z_q \sim
\exp \Bigl\{\frac{-N_F}{3\bo}\Bigr \} \ll 1 \,. \label{(3.0.7)}
\eq

We now find the mass $\mu_M$ of the dual mesons $M$ (mions). It can
be found from their effective Lagrangian, obtained by integrating out 
inclusivelt all   dual quarks and gluons.
Proceeding as in \cite{ch1} and integrating out inclusively all dual
quarks as heavy ones and all dual gluons via the
Veneziano-Yankielowicz (VY) procedure \cite{VY}, we obtain the
Lagrangian of mions\, (the fields $M$ are normalized e.g. in
\eqref{(3.0.2)} at $\mu=\la$ and $\bd=3\nd-N_F$)
\bq
K_M=\frac{z_M(\la,\,\mu_q^{\rm pole})}{Z_q^2\la^2}\,{\rm
Tr\,}(M^{\dagger}M)\,,\quad \w_M= -\nd \Biggl ( \frac{\det
M}{Z_q^{N_F}\la^{\bo}}\Biggr )^{1/\nd}+ {\rm Tr\,} (m_Q
M)\,,\label{(3.0.8)}
\eq
\bbq
z_M=z_M(\la,\,\mu_q^{\rm pole})\sim\Biggl (\frac{\la}{\mu_q^{\rm
pole}}
\Biggr)^{2\gamma_q}\sim\Biggl (\frac{\la}{\lym} \Biggr )^{2{\ov
b}_o/N_F}\gg 1\,,\quad
m_Q\langle M\rangle=\langle S\rangle=\lym^3\,.
\eeq
In a number of cases, it is also convenient to write the
superpotential of mions
in the form
\bq
\w_M= -\nd\, \lym^3\Biggl (\det \frac{ M}{\langle M \rangle}\Biggr
)^{1/\nd}+ {\rm
Tr\,} (m_Q M)\,.\label{(3.0.9)}
\eq

It follows from \eqref{(3.0.8)} that
\bq
\mu_M\sim\frac{Z_q^2\la^2}{z_M}\,\frac{\langle S\rangle}{\langle M\rangle^2}\sim
\frac{Z_q^2 m_Q\la^2}{z_M \langle M\rangle }\sim\lym\,.
\label{(3.0.10)}
\eq

To check that the above assumption that the dual quarks $q,\,\ov q$
are in the HQ
phase (i.e. are not higgsed) is not self-contradictory, we estimate
the mass ${\ov \mu}_{gl}$ of dual gluons due to possible higgsing of
dual quarks. The condensate of dual quarks at $\mu=\la$ is
\cite{Konishi}\,:
\bq
\sum_{\alpha_1}^{\nd}\langle {\ov q}^1_\alpha q^\alpha_1
(\mu=\la)\rangle =\mu_1 m_Q=Z_q m_Q\la \label{(3.0.11)}\,.
\eq
Therefore, the mass of dual gluons due to possible higgsing of dual
quarks can be estimated as
\bq
{\ov \mu}^2_{gl}\sim \bigl ({\ov a}_*\sim 1 \bigr )\Bigl (Z_q m_Q \la
\Bigr )
\Biggl (\frac {{\ov \mu}_{gl}} {\la}\Biggr )^{\bd/N_F}\, \quad \ra
\quad {\ov \mu}_{gl} \sim
\lym \,.\label{(3.0.12)}
\eq
Comparing with \eqref{(3.0.6)}, It is seen that ${\ov \mu}_{gl} \sim
\mu_q^{\rm pole}\sim \lym$. 
\footnote{\,
Due to the rank restriction at $N_F>N_c$, the opposite case in which
quarks were actually higgsed would result in a spontaneous 
breaking of the global flavor symmetry $SU(N_F)$ and in 
appearance of a large number of
parametrically light (pseudo) Nambu-Goldstone particles. {\it This
would be in contradiction with the Konishi anomaly
\eqref{(3.0.11)} for the dual group} which shows that, for equal mass
quarks, the mean vacuum values of quark bilinears are equal for all
flavors, i.e., as in the direct theory \eqref{(2.2.3)}, the global
flavor symmetry is not broken spontaneously.  \label{(f5)}
}

On the whole, there is only one scale $\sim \lym$ in the mass
spectrum \, of the dual  theory in this case. The masses of dual quarks 
$q,\,\ov q$, mions $M$  and dual gluonia are all $\sim \lym$.\\

Comparing the mass spectra of the direct and dual theories it is seen
that they are parametrically different. The direct quarks have large pole
masses$m_Q^{\rm
pole}/\lym\sim \exp\{N_F/3\bo\}\gg 1$ and are parametrically weakly
coupled and nonrelativistic inside hadrons (and weakly confined, the
string tension ${\sqrt \sigma}
\sim \lym\ll m_Q^{\rm pole}$\,), and therefore the mass spectrum of
low-lying mesons is Coulomb-like, with small mass differences $\delta
\mu_H/\mu_H=O(\bo^2/N_F^2)\ll 1$ between nearest hadrons. Gluonia
have masses $\sim \lym$, these are parametrically smaller than the
masses of direct flavored hadrons. At the same time, all dual hadron masses 
in the dual theory are of the same scale $\sim \lym$, and all couplings are
$O(1)$, and hence there is no reason for parametrically small mass
differences between dual hadrons.
We conclude that the direct and dual theories are not equivalent.

\section{ Direct and dual theories. Equal quark masses,\\ 
{\boldmath  $0<(2N_F-3N_c)/N_F\ll 1$}}

This case is considered analogously to those in Sections 2 and 3, and
we therefore
highlight only on differences from Sections 2 and 3. The main
difference is that at $\mu\sim \la$, the direct coupling is not small
now, $a_*=O(1)$, while both the gauge and
Yukawa couplings of the dual theory are parametrically small,\, ${\ov
a}_*\sim a_f^{*}\sim\bd/N_F\ll 1$.

The pole mass of direct quarks, $m_Q^{\rm pole}$, is the same as in
\eqref{(2.2.1)}. But $\lym$ is now parametrically the same\,: $\lym\sim 
m_Q^{\rm pole}$(as is the scale of the gluon mass due to possible quark
higgsing, $\mg\sim m_Q^{\rm pole}\sim \lym$). According to the Konishi 
anomaly for equal mass quarks \eqref{(2.1.8)}, \eqref{(2.1.9)}
and the rank restriction at $N_F > N_c$, quarks are not higgsed but
are in the HQ-phase (see footnote \ref{(f5)}\, in section 3), one
obtains that there is only one mass scale $\mu_H\sim \lym \sim
\la(m_Q/\la)^{N_F/3N_c}$ in the mass spectrum of hadrons in the
direct theory in this case.\\

We now consider the weakly coupled dual theory and recall that, by
definition, $\mu=|\Lambda_q|$ is such a scale that,
at $0<\bd/N_F\ll 1$, the dual theory at $\mu=|\Lambda_q|$ already
entered sufficiently deep into the conformal regime. I.e., both
the gauge and Yukawa couplings, ${\ov a}(\mu=|\Lambda_q|)$ and
$a_f(\mu=|\Lambda_q|)$, are already close to their small
fixed point values, $\ov\delta=({\ov a}_*-{\ov
a}(\mu=|\Lambda_q|)/{\ov a}_*\ll
1$, and similarly for $a_f(\mu=|\Lambda_q|)$.
We take $|\Lambda_q|=\la$ everywhere below for simplicity (because
this does not matter finally but simplifies greatly all formulas,
see the Appendix in \cite{ch3} for more details)

Hence, the pole mass of dual quarks looks now as
\bq
\frac{\mu_q^{\rm pole}}{\la}\sim\Biggl (\frac{\mu_q}{\la} \Biggr
)^{N_F/3\nd}\,,
\quad \frac{\mu_q= \mu_q(\mu=\la)}{\la}=\frac{M}{Z_q \la^2}\,,\quad
\frac{M}{\la^2} =\frac{\lym^3}{m_Q\la^2}=\Biggl
(\frac{m_Q}{\la}\Biggr)^{\nd/N_c}\,.\label{(4.0.1)}
\eq

After integrating out all dual quarks as heavy ones at $\mu=\mu_q^{\rm
pole}$\,, the dual SYM theory remains (together with the mions $M$).
The scale parameter ${\ov \Lambda}=\langle {\ov
\Lambda}_L(M)\rangle$ of its gauge coupling can be found from
\cite{KSV} \,:
\bq
3\nd \ln\Biggl ( \frac{\mu_q^{\rm pole}}{{-\ov \Lambda}}\Biggr
)\approx \frac{2\pi}{\ov
\alpha_*}\approx \frac{N_F}{\bd}\,\frac{\nd^2-1}{2N_F+\nd}\approx
\frac{3\nd^2}{7\,\bd}\,.\label{(4.0.2)}
\eq

Now, from matching the gluino condensates in the direct and dual
theories, we
obtain
\bq
|\ov \Lambda|=\lym \quad \ra \quad 
Z_q\sim\exp \Bigl\{\frac{-\nd}{7\,\bd} \Bigr \}\ll 1\,,
\quad \mu_q^{\rm pole}\sim\exp \Bigl \{\frac{\nd}{7\,\bd}
\Bigr \}\lym \gg
\lym\,.\label{(4.0.3)}
\eq
\vspace{2mm}

On the whole, the expressions for $\lym$ and $Z_q$ can be written in
the general case where $\bo>0,\,\bd>0$ as
\bq
\frac{\lym}{\la} \sim \exp\Bigl \{
- \frac{N_c}{\bo} \Bigr \}
\Biggl ( \frac{\det m_Q}{\la^{N_F}} \Biggr )^{1/3N_c}\,,
\quad Z_q\sim \exp \Biggl \{-\Biggl
(\frac{N_c}{\bo}+\frac{\nd}{7\,\bd}\Biggr ) \Biggr \}\,.
\label{(4.0.4)}
\eq
The symbol $\sim$ in \eqref{(4.0.4)} denotes the exponential accuracy
in dependence on the large parameters $N_c/\bo\gg 1$
or $\nd/\bd\gg 1$. Hence, if $N_c/\bo$ or $\nd/\bd$ are $O(1)$, then
the dependence on these has to be omitted from
\eqref{(4.0.4)}. For our purposes, this exponential accuracy in
\eqref{(4.0.4)} and below will be sufficient. \\

Therefore, similarly to the case of the weakly coupled direct theory
in section 2, the dual quark pole mass $\mu_q^{\rm pole}$ is 
parametrically larger here than $\lym$.

Then, proceeding as in \cite{ch1} and integrating out the dual gluons
via the VY-procedure yields the Lagrangian of mions $M$\,, which can 
be  written in the form \eqref{(3.0.8)}. The mion masses are therefore
given by
\bq
\mu_M\sim m_Q\Biggl (\frac{\mu_2^2}{z_M M} \Biggr )\sim m_Q\Biggl
(\frac{ Z_q^2\la^2}{z_M M} \Biggr ),\,\, z_M=z_M(\la,\,\mu_q^{\rm pole})
\sim\Biggl (\frac{\la}{\mu_q^{\rm pole}}\Biggr )^{2\bd/N_F}\gg 1\,.
\label{(4.0.5)}
\eq
It follows from \eqref{(4.0.5)} that
\bq
\mu_M\sim { Z_q}^2 \lym=\exp \Bigl \{ \frac{-2\nd}{7\,\bd}\Bigr
\}\lym\ll \lym\,.
\label{(4.0.6)}
\eq
Therefore, the mion masses are parametrically smaller than $\lym$.

To check that there are no self-contradictions and the dual quarks
are \, not higgsed, it remains to estimate the gluon masses due
to  possible higgsing of these dual quarks. We have
\bq
\frac{{\ov \mu}^{2}_{gl}}{\la^2}\sim { Z_q}\frac{m_Q}{\la}\Biggl (
\frac{{\ov
\mu}_{gl}}{\la}\Biggr )^{\bd/N_F}\, \quad \ra
\quad {\ov \mu}_{gl}\sim \exp \Bigl \{\frac{-\nd}{14\,\bd} \Bigr
\}\lym\ll \lym
\ll\mu_q^{\rm pole}\,. \label{(4.0.7)}
\eq
\vspace{2mm}

Therefore, there are three parametrically different mass scales in
the dual theory in this case.\\\,\, a)\, A large number of flavored 
hadrons made of weakly coupled
{\it nonrelativistic} (and weakly confined, the string tension $\sqrt
\sigma\sim \lym \ll \mu_q^{\rm pole}$) dual quarks with the pole
masses $\mu_q^{\rm pole}/\lym=\exp\{\,\nd/7\,\bd\}\gg 1$. The mass
spectrum of low lying flavored mesons is Coulomb-like, with
parametrically small mass differences $\delta \mu_H/
\mu_H=O(\bd^2/N^2_F)\ll 1$.\\
\,\, b)\, A large number of gluonia with the mass scale $\sim
\lym$.\\\,\,c)\, $N_F^2$ lightest mions $M$ with parametrically
smaller masses, $\mu_M/\lym\sim \exp \{-2\nd/7\,\bd \}\ll 1$.

At the same time, there is only one mass scale $\sim \lym$ of all
hadron masses in the direct theory which is strongly coupled here,
$a_*=O(1)$.
Clearly, the mass spectra of the direct and dual theories are
parametrically different.\\

On the whole, it follows that when an appropriate small parameter is
available ($0<\bo/ N_F\ll 1$ when the direct theory is weakly coupled, or 
$0<\bd/N_F\ll 1$ when weakly coupled is the dual theory), the mass spectra 
of the direct and dual theories are parametrically different. Therefore, there 
are no reasons for these mass spectra to become exactly the same when 
$\bo/N_F$ and $\bd/N_F$ become $O(1)$.

\section{Direct and dual theories. Equal quark masses, \\
{\hspace*{2cm} \boldmath $N_c<N_F<3N_c/2$}}

\hspace {6 mm} There are two possible ways to interpret the meaning
of the Seiberg dual theories with light quarks $m_{Q,i}\ll\la$ at 
$N_c < N_F < 3N_c/2$.\,\,\,-

{\bf a)}\, The first variant ''a''\, is similar to those which is the
only possibility in the conformal window $3N_c/2 < N_F < 3N_c$\,.
I.e., the description of all light degrees of
freedom of the direct theory in terms of effectively massless quarks
$Q,\,{\ov Q}$ and gluons remains adequate in the interval of scales
$\mu_H\ll \mu \leq \Lambda_Q$, where $\mu_H\ll \la$ is the highest
physical mass scale due to $m_Q\neq 0$, and there are no massive
particles with masses $\sim \Lambda_Q $ in the spectrum at $m_Q \ll
\la $. In comparison with the conformal behavior, the difference is
not qualitative but only quantitative: the strong coupling does not
approach the fixed value $\alpha_{*}$ at $\mu \ll \Lambda_Q$ but
continues to grow. Nevertheless, the non-perturbative contributions
are power suppressed until $\mu \gg \mu_H\,$, and one obtains
the right answers for all Green functions by resummation of standard
perturbative series with massless quarks and gluons. The dual theory is
interpreted then as a possible
alternative but equivalent (weak coupling) description. This variant
can be thought of as some formal 'algebraic duality', i.e. something
like 'the generalized change of variables'\,.

{\bf b)}\,\, The second variant \,\,''b''\, is qualitatively
different (it is usually referred to as 'confinement without chiral
symmetry breaking' at $m_{Q,i}\ra 0$, i.e. due
only to $\la\neq 0$ at $m_Q\ra 0$). It implies that, unlike the
variant 'a', the non-perturbative contributions become essential
already at $\mu\sim \la$, resulting in a high scale confinement with
the string tension $\sqrt \sigma\sim \la$ which binds direct quarks
and gluons into colorless hadron states with masses $\sim \la$. This
can be thought of, for instance, as follows. At $N_F$ close to $3N_c$
the value of $a_{*}=N_c\alpha_{*}/2\pi$ is small. As $N_F$ decreases,
$a_{*}$ increases and becomes
$\simeq 1$ at $N_F$ close to $3N_c/2$. When $N_F<3N_c/2$, the
coupling$ a(\mu\sim\la)$ exceeds some critical value $a^{(\rm crit)}=O(1)$
already at $\mu\sim \la$ and it is {\it assumed} that, for this reason, the
theory is now in another phase. The strong non-perturbative confining
 gauge interactions begin to operate at the scale $\sim \la$, resulting in
appearance of large number of colorless hadrons with masses $\sim
\la$. So, the use of old effectively massless quark and gluon fields
for description of light degrees of freedom at $\mu \ll \la$ becomes 
completely inadequate. { (This is especially visible at
$N_F=N_c+1$ where, for instance, the gauge degrees of freedom are 
not present at all amongst light ones in the dual theory)}.

Instead, the new (special solitonic ?) light degrees of freedom are
formed at the scale $\sim \la$ as a result of these strong non-perturbative
effects.These are the dual quars and gluons and dual mesons $\rm M$ 
(mions), with their sizes $\sim 1/\la$ and the internal
hardness scale $\sim \la$ (i.e. they appear as point-like at $\mu <
\la$). These new light particles are described by fields of the dual theory.  
So,this variant 'b' can be thought of as 'the physical duality', in a sense
that  \,the  dual theory is really the low
energy description of the original theory at $\mu<\la$\,.\\

{\it We consider this variant ``{\bf b}''\, as unrealistic}. Because
(at least in (S)QCD) the strong coupling $a(\mu\sim \la)\sim 1$ does not 
mean really that the scale of confining forces is $\sim \la$ . The
underlying reason is that {\it the confinement originates from the
(S)YM theory only}. There is no confinement in Yukawa-type theories
without non-Abelian gauge interactions. But pure SYM theory contains
only one dimensional parameter $\lym$. And $\lym\ra 0$ when the mass of 
even  one quark $m_{Q,k}$ tends to zero, see \eqref{(2.1.6)}. For this
reason {\it it can't give the strint tension
$\sigma^{1/2}\sim\la$, but only $\sigma^{1/2}\sim\lym\ll\la$}.
\footnote{\,
See section 7 in \cite{ch1} for additional arguments.
} 
In other words, the role of the order parameter for the confinement
plays not $\la$ by itself, but rather {\it the scale of the gluino
condensate, i.e. $\sqrt \sigma\sim\lym\sim \langle
\lambda\lambda\rangle^{1/3}$. But $\langle \lambda\lambda\rangle\sim
(\la^{\bo} m^{N_F}_Q)^{1/N_c} \ra 0$ at $m_Q\ra 0$\,. So, there
will be no confinement at all in the chiral limit $m_Q\ra 0$\,, and
the regimes at  $m_Q\ra0$ and $N_c<N_F<3N_c$ can be called more 
adequately as 'the
pure perturbative massless regimes with neither confinement, nor
chiral symmetry breaking', down to $\mu\ra 0$. They are : conformal
at $3N_c/2<N_F<3N_c$ and (very) strong coupling $a(\mu \ll \la))\gg 1$ 
at  $N_c<N_F<3N_c/2$\, (see \eqref{(5.0.4)} below)}. \\

Therefore, we will consider below the variant 'a' only in which the
nonzero particle masses arise only due to breaking of the $SU(N_F)_L\times
SU(N_F)_R\,$ and R-charge symmetries due to $m_{Q,i}\neq 0$, and
these masses are all much smaller than $\la$ at 
$m_Q\ll \la$ and $N_F> N_c$. Because in this variant the spectrum of
light (i.e. with masses $\ll \la$) particles can be calculated in
both direct and dual theories within used dynamical scenario, it
becomes possible, in addition to the 't Hooft triangles, to compare also 
the mass spectra of direcr and dual theories and values of some special
correlators in the perturbative range of energies where all particles
can still be considered as being effectively massless
($\mu_H\ll\mu\ll\la$, where $\mu_H$ is the highest physical scale due
to $m_Q\neq 0$).

These special correlators are the two-point correlators of external conserved 
currents, say, the baryon and, in the chiral limit $m_Q\ra 0$,  the 
$SU(N_F)_{left}\times SU(N_F)_{right}$  chiral flavor currents, as these can be
computed in the perturbation theory even in the strong coupling
region. Really, it is more convenient to couple these conserved
currents with the external vector fields and to consider the
corresponding external $\beta_{\rm ext}$\,-functions. Such $\beta_{\rm
ext}$-functions have the form (see e.g. \cite{KSV}) :
\bq
\frac{d}{d\,\ln \mu}\,\frac{2\pi}{\alpha_{ext}}= \sum_i T_i\,\bigl
(1+\gamma_i \bigr )\,, \label{(5.0.1)}
\eq
where the sum runs over all fields which can be considered as being
effectively massless at a given scale $\mu$, the unity in the brackets is 
due to one-loop contributions while the anomalous dimensions 
$\gamma_i$ of fields represent all higher-loop effects.

So, let us equate the values of such $\beta_{\rm ext}$-functions in
the direct and dual theories at scales $\mu_H\ll\mu \ll \la$, see section 7
in \cite{ch1}. The light particles in the direct theory are the original
quarks \, $Q,\,{\ov Q}$ and gluons, while in the dual theory these are the 
dual  quarks $q,\, {\ov q}$ and dual gluons, and the mions ${\rm M}$. For
the baryon currents one obtains:
\bq
N_F N_c\,\Bigl ( B_Q=1 \Bigr )^2\,(1+\gamma_Q)=N_F \nd \,\Bigl (
B_q=\frac{N_c}{\nd}
\Bigr )^2\,(1+\gd)\,,  \label{(5.0.2)}
\eq
while for the $SU(N_F)_L$ (or $SU(N_F)_R\,)$ flavor chiral currents
one obtains:
\bq
N_c\,(1+\gamma_Q)=\nd \,(1+\gamma_q)+N_F\,(1+\gamma_M)\,.
\label{(5.0.3)}
\eq
Here, the left-hand sides are from the direct theory while the
right-hand sides
are from the dual one, $\gamma_Q$ is the anomalous dimension of the
quark $Q$\,, while $\gamma_q$ and $\gamma_M$ are the anomalous
dimensions of the dual quark $q$ and the mion ${\rm M}$.

Now, at $\mu_{H}\ll\mu\ll \la$ the dual theory is IR-free and both
its couplings are small in this range of energies, ${\ov a}(\mu)\ll 1\,,\,
a_f(\mu)\ll 1$\,. So, $\gamma_q(\mu)\ll 1$ and $\gamma _M(\mu)\ll 1$
are both also logarithmically small at $\mu\ll\la$\,. It is seen then that 
\eqref{(5.0.2)},\, and \eqref{(5.0.3)} are incompatible with each other as 
they predict different  values for the infrared limit of $\gamma_Q$. 
Moreover, \eqref{(5.0.3)} gives $\gamma_Q$ which is
incompatible with the NSVZ $\beta$ funcion \eqref{(5.0.4)} \cite{NSVZ-1}. 

We conclude that {\it both
correlators can not be equal simultaneously in the direct and dual
theories, and so these two theories are not  equivalent}. 

We emphasize that {\it to reach this conclusion it is sufficient to account for 
one property only\,:\, the nonperturbative confining effects originate  from 
the SYM theory only and so the tension of the confining strings is  $\sigma^{1/2}\sim
\lym\ll \la$ at $m_Q\ll\la$.  Then equations \eqref{(5.0.2)}, \eqref{(5.0.3)} follow}.
(It is also worth noting that both \eqref{(5.0.2)} and  \eqref{(5.0.3)}  are fulfilled
within the conformal window).

Using the lower energy value $\gamma_Q\ra
(N_c/\nd-1)=(2N_c-N_F)/(N_F-N_c)$ from
(5.0.2) and the perturbative NSVZ $\beta$-function \cite{NSVZ-1}, one
obtains the perturbative IR-behavior of the strong coupling
$\alpha(\mu)$\,:\\\bq
\frac{d a(\mu)}{d\ln \mu}\equiv \beta(a)=-\, \frac{a^2}{1-a}\,
\frac{\bo-N_F\gamma_Q}{N_c}\,, \label{(5.0.4)}
\eq
\bbq
a(\mu)\equiv \frac{N_c\alpha(\mu)}{2\pi}, \quad \bo=(3N_c-N_F), \quad
\gamma_Q\equiv \frac{d\ln z_Q}{d \ln \mu}\,,\quad
z_Q(\la,\mu)\sim\Bigl
(\frac{\mu}{\la}\Bigr )^{\gamma_Q}\ll 1\,,
\eeq
\bbq
N_c < N_F < 3N_c/2,\,\,\, \gamma_Q=\frac{(2N_c-N_F)}{(N_F-N_c)}\,,\quad
a(\mu)\approx \Bigl
(\frac{\Lambda_Q}{\mu}\Bigr )^{\nu}\gg 1\,,\quad
\nu=\frac{3N_c-2N_F}{N_F-N_c}\,,\quad (\mu/\Lambda_Q)\ll 1\,.
\eeq

In this case, the behavior of $a(\mu/\la)$ looks as follows. As
$z=\mu/ \Lambda_Q$ decreases from large values, $a(z)$ increases first in a
standard way $\sim (1/\ln z)$. At $z=z_o\sim 1\,\,\, a(z)$ crosses unity. At
this point $\gamma_Q$ crosses the value $\bo/N_F=(3N_c-N_f)/N_F$. As a result,
{\it the $\beta$-function is smooth, it has neither pole nor zero at this
point and remains negative all the way from the UV region $z\gg 1$ to
the IR region $z\ll 1$, while $a(z)$ grows in the infrared region in a
power-like fashion, see \eqref{(5.0.4)}\,}. On the other
hand, it is not difficult to see that the IR-value of $\gamma_Q$
obtained from \eqref{(5.0.3)} with $\gamma_q\ra 0,\,\gamma_M
\ra 0$ is incompatible with the NSVZ\, $\beta$-function \eqref{(5.0.4)}.

As regards the mass spectrum of the direct SUSY $SU(N_c$) theory with 
$N_c < N_F < 3N_c/2$\,, this case is obtained from the one in the preceding 
section by a simple change of notations. The quark pole masses are \cite{ch1}
\bq
\frac{\mpQ}{\la}\sim\frac{m_Q}{\la}\Biggl
(\frac{\la}{m_Q=m_Q(\mu=\la)}\Biggr
)^{\gamma_Q}\sim\Biggl (\frac{m_Q}{\la}\Biggr
)^{\frac{1}{1+\gamma_Q}=\frac{\nd}{N_c}},\label{(5.0.5)}
\eq
while the gluon masses due to possible higgsing of quarks look as
\eqref{(5.0.4)}
\bbq
\mu_{\rm gl}^2\sim a(\mu=\mu_{\rm gl})\langle {\ov Q} Q
\rangle_{\mu=\mu_{\rm
gl}}\,,\quad a(\mu=\mu_{\rm gl})\sim\Biggl (\frac{\la}{\mu_{\rm
gl}}\Biggr )^{\nu}\,,\quad
\nu=\frac{N_F\gamma_Q-{\rm b_o}} {N_c}=\frac{3N_c-2N_F}{N_F-N_c} > 0\,,
\eeq
\bq
\langle {\ov Q} Q \rangle_{\mu=\mu_{\rm gl}}= \langle {\ov Q}
Q\rangle_{\mu=\la}\,\Biggl(\frac{ \mu_{\rm gl}}{\la} \Biggr
)^{\gamma_Q}. \label{(5.0.6)}
\eq
It follows from \eqref{(5.0.4)},\eqref{(5.0.6)} that
\bq
\frac{\mu_{\rm gl}}{\la}\sim \Biggl (\frac{m_Q}{\la}\Biggr)^{\frac{1}
{1+\gamma_Q}}\sim \frac{\mpQ}{\la}\gg\frac{\lym}{\la},\,\,\frac{1}
{1+\gamma_Q}=\frac{N_F-N_c}{N_c}\,\,{\rm for}\,\,\gamma_Q=
\frac{2N_c-N_F}{N_F-N_c}=\nu+1,\,\,\,\label{(5.0.7)}
\eq
and, as previously, the Konish anomaly \eqref{(2.1.2)} \cite{Konishi}
shows that, due to the rank restriction at $N_F > N_c$, these quarks 
are not  higgsed but confined, (see footnote \ref{(f5)}). After all quarks
are integrated out as heavy ones at
$\mu<\mpQ$, we are left with the $SU(N_c)$ SYM theory in the
strong coupling branch and with the scale factor $\lym\ll \mpQ$ of
its gauge coupling, and so on.\\

As regards the IR-free dual $SU(\nd=N_F-N_c$ ) theory, its mass
spectrum has been
described in \cite{ch1} (see section 7 therein ) and we only recall
it here briefly. The Lagrangian at $\mu=\la$ is taken in the form
\bbq
{\ov K}={\rm Tr}\Biggl ( q^\dagger e^{\ov V} q + {\ov q}^\dagger{\ov
q}\Biggr
)+\frac{{\rm Tr}\Bigl ( M^{\dagger} M\Bigr )} {\la^2}\,,\quad{\ov
\w}=\,\frac{2\pi}{{\ov \alpha}(\mu=\la)}\,{\ov S}-\frac{{\rm Tr}\Bigl
({\ov q}\, M\, q \Bigr )}{\la} + {\rm Tr} \Bigl ( m_Q M \Bigr )\,,
\eeq
\bq
\langle M(\mu=\la)\rangle =M\,,\quad {\ov a}(\mu=\la)={\nd}\,{\ov
\alpha}(\mu=\la)/2\pi=O(1)\,.\label{(5.0.8)}
\eq

The dual quarks are in the HQ phase and are therefore confined, and
their pole masses are (from now on and in this section we neglect the
logarithmic renormalization factors $z_q$ and $z_M$ for simplicity)
\bq
\frac{\mu^{\rm pole}_q}{\la}=\frac{\langle M(\mu=\la)\rangle
=\langle {\ov Q}Q(\mu=\la)\rangle}{z_q(\la, \,\mu^{\rm pole}_q)\la^2}\sim 
\Bigl (\frac{m_Q}{\la}\Bigr )^{\frac{N_F-N_c}{N_c}}.\label{(5.0.9)}
\eq

After integrating out the dual quarks as heavy ones, we are left with
the dual
gauge theory with $\nd$ colors and with the scale factor
$\langle\Lambda_L(M)\rangle=\lym$ of its gauge coupling, and with the
frozen mion fields $M$. Finally, after
integrating out the dual gluons by means of the VY- procedure
\cite{VY}, we obtain
the Lagrangian of mions
\bq
{\ov K}=\frac{1}{\la^2}\,{\rm Tr\,\Bigl (M^{\dagger}M\Bigr )}\,,\quad
{\ov \w}= -\nd \, \Biggl ( \frac {\rm \det\, M}{\la^{\bo}} \Biggr
)^{1/\nd}
+m_Q{\rm Tr  \,M}\,,\quad \mu \ll \Lambda_{SYM}\,\,.\label{(5.0.10)}
\eq
It describes the mions ${\rm M}$ with the masses
\bq
\mu_M\sim m_Q\Biggl (\frac{\la^2}{M}\Biggr )\sim m_Q \Biggl
(\frac{\la}{m_Q}\Biggr )^
{\frac{N_F-N_c}{N_c}}\,,\quad m_Q\ll \mu_M \ll
\Lambda_{SYM}\,.\label{(5.0.11)}
\eq
{\it Clearly, there is no analog of these parametrically light
particles in the  direct theory}.

\section{Direct and dual theories. Unequal quark masses, \,\, \\
$\mathbf{0<(2N_F-3N_c)/N_F\ll 1, \, N_c<\nl<3N_c/2}$} 

The standard consideration that allows "to verify" that the duality
works properly
for quarks of unequal masses is as follows \cite{S2, IS}. For
instance, we take $\nh$ quarks of the direct theory to be heavier
than the other $N_c<\nl=N_F-\nh$
quarks. Then, after integrating out these $\rm h$-flavored quarks as
heavy ones, the direct theory with $N_c$ colors and $\nl$ of $\it
l$-flavors remains at lower energies. By duality, it is equivalent to
the dual theory with $\nl-N_c$ colors and $\nl$ flavors.
On the other side, the original theory is equivalent to the dual
theory with $\nd=N_F-N_c$ colors and $N_F$ flavors. In this dual
theory, the $\rm h$-flavored dual quarks are {\it assumed} to be
higgsed, and hence the dual theory with $(N_F-N_c-\nh)=\nl-N_c$
colors and $N_F-\nh=\nl$ of $\it l$-flavors remains at lower
energies.Therefore, all looks self-consistent.\\

But we consider now this variant in more details, starting from the
left end of the conformal window, $0<{\ov {\rm b}}_0 /N_F\ll 1$. 
At $\mu < \la$, the dual theory, by definition, has already entered 
sufficiently deep the conformal regime and therefore
both its gauge and Yukawa couplings are close to their parametrically small 
frozen values, ${\ov a}_*\sim a_f^*\sim (\bd)/N_F\ll 1$. We take
$N_c<\nl<3N_c/2$ for the direct quarks $\ql\,,\oql$ to have smaller
masses $\ml$ at $\mu=\la$\,, and  other $N_h=N_F-N_l$ quarks
$\qh\,,\, \oqh$ to have larger masses $\mh\,, \,\,r\equiv \ml/\mh \ll 1$.

\hspace*{1cm}\subsection {Direct theory} 

We start from the direct theory, because, in a sense, its mass
spectrum is easier to calculate. It is in the strong coupling 
$a_{*}={\cal O}(1)$ conformal regime at $\mhp < \mu<\la$\,, 
and the largest physical mass scale $\mu_H$ is
here the pole mass of the heavier $\qh\,,\,\oqh$ quarks\,:
\bq
\frac{\mhp}{\la}\sim\frac{m_{h}}{z_{+}\la}\sim\frac{\mh}{\la}\,
\Biggl (\frac{\la}{\mhp}\Biggr)^{\gamma_Q=\bo/N_F\approx 1}
\sim\Biggl (\frac {\mh}{\la}\Biggr )^{(N_F/3N_c)\approx 1/2}\,.
\label{(6.1.1)}
\eq

The competing gluon mass $\mu_{{\rm gl},\, l}$ due to higgsing of 
$Q^l$ quarks looks as follows (at not too large $N_c$, see \cite{ch21})\,:\,
\bbq
\mu^2_{{\rm gl},\, l}\sim a(\mu_{{\rm gl},\, l}) z^{\prime}_{+}\langle
M_l\rangle,\,\,
z^{\prime}_{+}\sim
\Bigl (\frac{\mu_{{\rm gl},\, l}}{\la}\Bigr )^{\gamma_Q}\,\ra\, \frac{\mu_{{\rm
gl},\, l}}
{\la}\sim\frac{\langle M_l\rangle}{\la^2}\sim r^{\frac{N_l-N_c}{N_c}}\Bigl
(\frac{m_H}
{\la}\Bigr )^{1/2},\,\, \frac{\mu_{{\rm gl},\, l}}{\mhp}\sim
r^{\frac{N_l-N_c}{N_c}}\ll 1\,.
\eeq

After integrating inclusively these quarks out, the lower energy
 direct theory  with $N_c$ colors and
$N_c<\nl<3N_c/2$ flavors of lighter $\ql\,,\,\oql$ quarks remains.
\footnote{\,
To simplify all formulas, the value $(3N_c-2\nl)/\nl$ is considered
as $O(1)$ quantity in this section,
 }
From matching with the coupling $a_{*}=O(1)$ of the higher energy
theory, its  gauge coupling is also $O(1)$ at $\mu=\mhp$ and 
therefore the scale  parameter of this gauge coupling is 
$\Lambda^{\prime}\sim\mhp$. The
current masses of $\ql\,,\,\oql$  quarks at $\mu=\mhp$ are $\hat 
\ml\equiv\ml(\mu=\mhp)=r\,\mhp\ll  \la^{\prime}=\mhp,\,\, r\ll 1$.

Now, how to deal with this theory at lower energies ?\, As was
described in \cite{ch1} (see section 7 therein), there are two variants, "a"
and  "b".  The heaviest masses $\mu_{H}$ of the lower energy theory 
at $\mu <\Lambda^{\prime}$  are  much smaller than $\Lambda
^{\prime}$ in  the variant "a", $\mu_H\ll \Lambda^{\prime}$,   
see section 5 above. At  $\mu<\Lambda^{\prime}$ this theory enters 
{ \it smoothly}  the  strong coupling  regime with  $a(\lym\ll\mu\ll\Lambda^
{\prime})\gg 1$.  Its mass spectrum  is, in essence, the  same 
as those described above in section 5 and only the values of $N_F$ 
and the scale factor of the gauge coupling are different\,: $N_f\ra N_l,\,
\la\ra
\Lambda^{\prime}$. I.e., all $N_l$ quarks and all gluons remain effectively 
massless at  scales $m^{\rm pole}_l <\mu \ll \Lambda^{\rm \prime}$. And
quarks are weakly confined,  i.e. the  tensions of confining stringis are
much   smaller than their masses, $\sigma^{1/2}\sim \lym\ll 
m^{\rm pole}_l$.  And only the gauge coupling is very large at 
$\lym < \mu < \Lambda^{\rm \prime},\,\, a(\mu)\gg 1$.
We therefore do not consider this variant "a" in this section,
because our purpose here is to check the duality in the variant 
most favorable for its validity. 

This  is the variant "b"\, :\, "confinement without chiral 
symmetry breaking" at $m_{Q,l}\ra 0$
\,(although some general arguments have been
presented in section 7 of \cite{ch1} that this variant cannot be
realized). This amounts {\it to assuming} (because all original
$Q_l,\,{\ov Q}_l$ quarks and all gluons cannot "dissolve in the 
pure air") that due  to strong non-perturbative confining effects 
$\sim \Lambda^{\rm \prime}$  the direct  quarks  $\ql\,,\oql$ 
and gluons form a large   number of heavy   hadrons with  masses 
$\mu_H\sim \Lambda^{\rm \prime}$.   Instead,  new light 
composite particles (special solitons) appear  whose masses 
are parametrically smaller than $ \Lambda^{\rm \prime}$.
These are the dual quarks ${\hat q}_i,\, {\hat{\ov q}}^{\,j},\,\,
 i, j=1...N_l$, the dual gluons with $\nd^{\,\prime}=
\nl- N_c$  of dual colors, and the mions ${\hat
M}_l\ra {{\hat M}_j^i },\,\, i, j=1...N_l$. Their interactions at
scales $\mu<\la^{\prime}$ are described by the Seiberg dual
Lagrangian. Their dual gauge coupling ${\ov a}(\mu)$ at $\mu\sim
\Lambda^\prime\sim \mhp$ is ${\ov a}(\mu=\Lambda^\prime/(\rm 
several))\sim{\it a}_*=O(1)$, 
and therefore the scale parameter of this dual gauge
coupling is $\sim \Lambda^{\prime}\sim \mhp$. \\

The dual Lagrangian (at the scale $\mu\sim \Lambda^{\prime}\,,\, 
\langle{\ov Q}^{\it l} Q_{\it l}(\mu=\la)\rangle=\langle S\rangle/\ml =
\lym^3/\ml $\,) is
\bq
\ov K={\rm Tr\,}_{\it l} \bigl( {\hat q}^{\dagger} {\hat q}
+ {\hat {\ov q}}^{\dagger} {\hat {\ov q}}\bigr
)+\frac{1}{(\Lambda^{\prime})^2}
{\rm Tr}_l \bigl ({\hat M}^{\dagger}_l{\hat M}_l\bigr )\,,
\label{(6.1.2)}
\eq
\bbq
{\ov\w}= \frac{2\pi}{{\ov \alpha}(\Lambda^{\prime})}\, {\ov
S}-\frac{1}{\Lambda^{\prime}}\,{\rm Tr}_{\it l} \Bigl ({\hat {\ov
q}}\,{\hat M}_{\it l}\, {\hat q} \Bigr ) +{\hat \ml}\,{\rm Tr}
\, {\hat M}_ {\it l}\,;\quad {\ov S}={\ov
W}_{\beta} ^2/32\pi^2\,,\quad \langle
{\hat{\ov q}}^{\it j}\, {\hat q}_{\it i}\rangle=\delta_{\it i}^{\it
j}\,{\hat m}_{\it l} \,{\Lambda^{\prime}}\,,
\eeq
\bq
\hml\equiv\ml(\mu=\Lambda^{\,\prime})=z_{+}^{-1} \ml\,,\,\, 
\langle {\hat M}_{\it l}\rangle=\,z_{+}\langle M_{\it l} \rangle,\,\,\, 
z_{+}\sim\Biggl (\frac{\mhp}{\la}\Biggr )^{\gamma_Q=
\frac{\bo}{N_F}\approx 1} \sim\Biggl(\frac{\Lambda^{\prime}}
{\la}\Biggr )\ll 1\,. \label{(6.1.3)}
\eq

This dual theory is in the IR free weakly coupled HQ-phase at
$\mu<\Lambda^{\prime}$. At lower scales $\mulp<\mu<
\Lambda^ {\,\prime}$   both its gauge and Yukawa couplings 
decrease logarithmically (\,$\bd^{\,\prime}=
3\nd^{\,\prime}- \nl <0$) and become much less than unity at 
$\mu=\mulp\ll \Lambda^{\,\prime}$ and $r\ll 1$. The perturbative 
pole mass $\mulp$ of quarks ${\hat q}^{\it l},\, {\hat {\ov q}}_{\it \ov
l}$ is (we ignore below logarithmic RG-evolution factors):
\bq
\mulp\sim {\hat\mu_{q,\,\it l}}\sim z_{+}\Bigl ( \frac{\langle M_l
\rangle}{\Lambda^{\prime}} \Bigr ). \label{(6.1.4)}
\eq

At scales $\mu<\mulp$ all quarks ${\hat q}^{\it l},\, {\hat {\ov
q}}_{\it \ov l}$ can be integrated out inclusively as heavy ones, 
and there remains  the dual SYM theory with $\nd^{\,\prime}=
\nl-N_c$ colors and mions  ${\hat M}_{\it l}$\,. The scale factor 
${\tilde{ \Lambda}_{SYM}}=\langle {\ov \Lambda}_L ({\hat M}_
{\it l})\rangle$ (with mions ${\hat M}_{\it l}$ sitting down on ${\ov
\Lambda}_L({\hat M}_{\it l}$)\,) of its gauge coupling is determined
from the matching , see  \cite{ch1} and \eqref{(3.0.4)}
\bbq
3\nd^{\,\prime} \ln \Biggl (\frac{\mulp}{-{\tilde\Lambda}_{SYM}} \Biggr )=
\bd^{\,\prime} \ln \Biggl (\frac {\mulp}{\Lambda^{\,\prime}} \Biggr )
\quad\ra
\eeq
\bq
 - {\tilde\Lambda}_{SYM} = \langle {\ov \Lambda}_{L}({\hat M}_{\it l})\rangle =
\lym=\Biggl ( \la^{\bo}\ml^{\nl}\mh^{\nh}\Biggr
)^{1/3N_c}\,.\label{(6.1.5)}
\eq
Applying now the VY procedure to integrate out dual gluons, we obtain
the lowest-energy Lagrangian of mions ${\hat M}_{\it l}$ at $\mu < \lym$\,:
\bq
K_M\sim\frac{1}{(\la^{\,\prime})^2}{\rm Tr} \Bigl ({\hat
M}_{l}^{\dagger}{\hat M}_{l}\Bigr )\,\quad \w=\Biggl
[\,\,-(\nl-N_c)\,\Biggl (\,\frac{\det {\hat M}_{\it
l}}{(\Lambda^{\,\prime})^{(3N_c-\nl)}}
\,\Biggr )^{1/(\nl-N_c)} +\hat \ml\,{\rm Tr}\hat M_{\it l}\,\,\Biggr
]\,, \label{(6.1.6)}
\eq
From \eqref{(6.1.6)}, the masses of mions ${\hat M}_{\it l}$ are
\bq
\mu({\hat M}_{\it l})\sim  \ml\,\frac{\la^2}{\langle
{ M}_{\it l}\rangle} \sim \, \Biggl (r=\frac{\ml}{\mh}\Biggr
)^{\frac{2N_c-\nl}{N_c}}\,\,\mhp\,,\quad r\ll 1\,,\label{(6.1.7)}
\eq
\bbq
\frac{\lym}{\mulp}\sim \,\,r^{\Delta}\ll 1\,,\quad \frac{\mu({\hat
M}_{\it l})}{\lym}\sim \,r^{2\Delta}\,\ll
1\,,\quad0<\Delta=\frac{3N_c-2\nl}{3N_c}<\frac{1}{3}\,.
\eeq

Therefore, on the whole for Section 6.1, when we started from the
direct theory in the conformal region, then integrated out the 
heaviest $Q^h\,, {\ov Q}_{\it h}$ quarks
at  $\mu=\mhp$ \eqref{(6.1.1)} (weakly confined 
by strings with  the tension $\sigma^{1/2}\sim \lym\ll \mhp)$ and, 
in the variant "b"  (i.e. "confinement without chiral symmetry
breaking" in the chiral limit, see section 7 in \cite{ch1}), dualized
the remained theory of   $Q^{\it l},\, {\ov Q}_{\it  l}$ - flavors, 
the mass spectrum is as follows.\\

{\bf 1)}\, There is a large number of heavy direct  ${\it h}$\,-mesons  
$M^{\rm dir}_{\it hh}$, made of  original $Q^h\,, {\ov Q}_{\it h}$ quarks   
connected  by strings with the small tension  $\sigma^{1/2}\sim\lym$,
with masses $\sim \mhp\sim\Lambda^{\rm \prime}$, see \eqref{(6.1.1)}.\\

$\mathbf 1^{\prime}$) \, In the variant "b" considered  here, the   dualization \, 
of  $Q^{\it l}\,,{\ov Q}_{\it \ov l}$  quarks and direct  gluons leaves behind a \, 
number   of $M^{\rm dir}_{\it l l}$ mesons made of $Q^{\it l}\,, {\ov Q}_{\it l}$
quarks connected by strings with the tension $\sigma\sim\Lambda^{\rm \prime}$,
with masses $\sim\Lambda^{\rm \prime}\sim \mlp$. Besides, there is also a number
of baryons $B_{\it l},\,B_{\it lh}$ and gluonia, all also with masses $\sim
\Lambda^
{\rm \prime}$.

All other pure dual particles have smaller masses  and originate from 
dual quarks ${\hat q} ^{\it l}\,,{\hat {\ov q}}_{\it \ov l}$\,,  dual 
gluons and mions ${\hat M}_{\it l}$.\\

{\bf 2)} There is a large number of {\it l}\,-flavored dual mesons
and $b_{\it l}$-baryons made of nonrelativistic and weakly confined
(the dual string tension is $\sqrt \sigma\sim \lym\ll \mulp\ll\mhp$)\, 
${\hat q}^{\it l}\,,{\hat {\ov q}}_{\it \ov l}$ \,\, quarks, with their pole
masses
$$\mulp\sim (r)^{(\nl-N_c)/N_c}\, \mhp\sim(1/r)^{(3N_c-2\nl)/3N_c}\lym$$\, .

{\bf 3)} There is a large number of gluonia with masses $\sim \lym\ll
\mulp$.

{\bf 4)} The lightest are $\nl^2$ scalar mions ${\hat M}_{\it l}$
with masses $\mu({\hat M}_{\it l})\sim (r )^{2\Delta}\,\lym\ll\lym$. 
\vspace{2mm}

\hspace*{1cm} \subsection{Dual theory} 

We return to the beginning of this section and start directly from
the dual theory
with $SU(\nd)$ dual colors, $N_F$ dual quarks $q$ and $\ov q$, and
$N_F^2$ mions $M_{i}^j$. At the scale $\mu< \la$ the theory is
already entered the weak coupling conformal regime, i.e. its coupling
${\ov a}(\mu)$ is close to ${\ov a}_*=\nd\,
{\ov \alpha}_{*}/2\pi\approx 7\bd/3\nd \ll 1$, see footnote \ref{(f5)} 
in section 3. We first consider the case most favorable for the dual
theory, where the parameter $r=\ml/\mh$ is already taken to be
sufficiently small (see below). Then, the highest physical mass scale
$\mu_H$ in the dual theory is determined by masses of dual gluons due
to higgsing of ${\ov q}^{\it h},\,  q_{\it h}$ quarks\,: $$\mu_H^2=
\ogh^2\sim {\ov a}_{*}\langle{\ov q}^{\it h} q_{\it
h}(\mu=\ogh)\rangle.$$. The mass spectrum of the dual theory in this
phase can be obtained in a relatively standard way, similarly to
\cite{ch1, ch3}. For this reason, we will skip from now on some 
intermediate relations in similar situations in what follows. The 
emphasis will be made on new elements that have not appeared before.\\

1)\,\,\, The masses of $\nh(2\nd-\nh)$ massive dual gluons and
their scalar superpartners are
\bq
\Bigl ({\ogh}\Bigr )^2\sim \langle {\ov q}^{\it h} q_{\it h} \rangle
\Biggl(\frac{\ogh} {\la}\Biggr )^{\gamma_q}\sim \mu_1\mh
\Biggl (\frac{\ogh}{\la} \Biggr )^{\gamma_q} \sim Z_q\la\mh 
\Biggl (\frac{\ogh}{\la} \Biggr )^{\bd/N_F}\,,\label{(6.2.1)}
\eq
\bq
\frac{\ogh}{\la}\sim \exp\{-\nd/14\bd \}\, \Biggl (\,\frac{\mh}{\la}
\Biggr )^{N_F/3N_c}\ll \Biggl (\,\frac{\mh}{\la}\Biggr )^{(N_F/3N_c)\approx 1/2}
\,,
\quad{\it Z_q} \sim \exp\{-\nd/7\bd \}\ll 1\,.\label{(6.2.2)}
\eq

2)\,\,\, The $\nl\nh$ hybrid mions $M_{\it hl}$ and $\nl\nh$ nions
$N_{\it lh}$ (these are those dual ${\it l}$\,-quarks that have higgsed
colors, their partners $M_ {\it lh}$ and $ N_{\it hl}$ are implied and are 
not shown explicitly) can be treated independently of other degrees of 
freedom, and their masses are determined mainly by their
common mass term in the superpotential,
\bq
K_{\rm hybr}\simeq{\hat z}_M\,{\rm Tr}\,\Biggl (\frac{M^{\dagger}_{\it hl}
M_{\it hl} }{Z^2_q\la^2} \Biggr )+{\hat z}_q\,{\rm Tr}\,\Biggl (N^{\dagger}_
{\it lh} N_{\it lh}\Biggr )\,,\quad {\ov\w}=
\Bigl ( Z_q\mh\la\Bigr )^{1/2}{\rm Tr}\,\Biggl (\frac {M_{\it
hl}N_{\it lh}}{Z_q\la}\Biggr )\,, \label{(6.2.3)}
\eq
where ${\hat z}_q$ and ${\hat z}_M$ are the perturbative
renormalization factors of dual quarks and mions,
\bbq
{\hat z}_q={\hat z}_q(\la,\,\ogh)\sim\Biggl (\frac{\ogh}{\la} \Biggr
)^{\bd/N_F}\sim \Biggl (\frac{\mh}{\la} \Biggr )^{\bd/3N_c}\ll 1\,,\quad
{\hat z}_M={\hat z}_M(\la,\,\ogh)=1/{{\hat z}_q}^2\gg 1\,.
\eeq
Therefore,
\bq
\frac{\mu(M_{\it hl})}{\la}\sim \frac{\mu(N_{\it lh})}{\la}\sim 
\exp\{-\nd/14\bd\}
\,\Biggl (\,\frac{\mh}{\la}\Biggr )^{N_F/3N_c}\sim\frac{\it
\ogh}{\la}\,.\label{(6.2.4)}
\eq

3)\,\,\,Because the ${\ov q}^{\it h}$ and $q_{\it h}$ quarks are
higgsed, $N_{\it h}^2$ pseudo-Goldstone
mesons $N_{\it hh}$ (nions) appear. After integrating out the 
heavy  gluons and their superpartners, the
Lagrangian of remained degrees of freedom takes the form
\bbq
K\simeq {\hat z}_M {\rm Tr}\Biggl (\frac{M^{\dagger}_{\it hh}M_{\it
hh}+ M^{\dagger}_{\it ll}M_{\it ll}}{Z^2_q\la^2}\Biggr )+\,
{\hat z}_q\,2\,{\rm Tr}\sqrt {N^{\dagger}_{\it hh} N_{\it hh}}\,+{\hat
z}_q {\rm Tr\,}_{\it l}\,\Biggl ((\hat q)^{\dagger} {\hat q} + {\hat q}\ra
{\hat{\ov q}} \Biggr )\,,
\eeq
\bq
\w= \frac{2\pi}{\ov \alpha^{\,\prime}(\mu)}{\ov S}^{\,\prime}+{\rm
Tr}\,\Biggl (\frac{M_{\it hh} N_{\it hh}}{Z_q\la}\Biggr )-{\rm Tr\,}
_{\it l} \Biggl (\frac{{{\hat{\ov q}}}\, M_{\it ll}\, {\hat q}}
{Z_q\la}\Biggr )+\,{\rm Tr}
\,\Biggl ( \ml M_{\it ll}+\mh M_{\it hh}\Biggr )\,, \label{(6.2.5)}
\eq
where ${\ov S}^{\,\prime}$ includes the field strengths of the
remaining $SU( \nd^{\,\prime})$ dual gluons with $\nd^{\,\prime}=
\nd-\nh=\nl-N_c$ dual colors, and ${\hat q}^l,\, {\hat{\ov q}}_{\it \ov l}$ 
are  still active \, $l$ - flavored dual quarks
with not higgsed colors and, finally, the nions $N_{\it hh}$ are
"sitting down" inside ${\ov \alpha}^{\,\prime}(\mu)$.

At lower scales $\mu<\ogh$, the mions $M_{\it hh}$ and nions $N_{\it
hh}$ are frozen and don't evolve, while the gauge coupling
decreases logarithmically in the interval $\olp <\mu<\ogh$. The
numerical value of the pole mass of ${\hat {\ov q}}_{\it l}\,,\, {\hat q}
^{\it l}$ - quarks is
\bq
\frac{\olp}{\la}=\frac{\langle M_{\it
ll}\rangle}{Z_q\la^2}\,\frac{1}{{\hat z}}_{q}
\sim\exp \Bigl \{\, \frac{\nd}{7\bd}\,\Bigr \}\,
\Biggl [\, \Bigl ( r \Bigr )^{\frac{\nl-N_c}{N_c}}\,
\Biggl (\frac{\mh}{\la} \Biggr )^{N_F/3N_c}\,\Biggr ]
\,, \label{(6.2.6)}
\eq
\bbq
z^{\,\prime\prime}_{q}=z^{\,\prime}_{q}(\ogh\,,{\olp})\simeq
z^{\,\prime}_{q}(\mhp\,,\mulp)=z^{\,\prime}_{q}\,,
\eeq
ignoring the \, additional  logarithmic renormalization factor 
$z^{\,\prime\prime}_{q}\ll 1$ of active l-quarks.

After integrating out the quarks $\hat{\ov q}_{\it l}\,,\,
\hat{q}^{\it l}$ as heavy
ones at $\mu<\olp$, the dual $SU(\nl-N_c)$ Yang-Mills
theory remains with the scale factor of its gauge coupling $\langle
-{\ov \Lambda}_L\rangle=\lym$ (and with nions $N_{\it hh}$
and mions $M_{\it ll}$ "sitting down" on ${\ov \Lambda}_L$). Finally,
integrating out the dual gluons via the VY procedure, we obtain
(\,recalling that all fields in \eqref{(6.2.5)} and \eqref{(6.2.7)}
are normalized at $\mu=\la$, and $\lym/\la=(r)^{\nl/3N_c}\,(\mh/\la)^
{N_F/3N_c}\,,\,\,\langle N_{\it hh} \rangle=\langle{\ov q} q\rangle =
Z_q\mh\la $\,)
\bq
K={\hat z}_M {\rm Tr}\Biggl (\frac{M^{\dagger}_{\it hh}M_{\it hh}+
M^{\dagger}_{\it ll}M_{\it ll}}{Z^2_q\la^2}\Biggr )+\,{\hat
z}_q\,2\,{\rm Tr}\sqrt {N^{\dagger}_{\it hh} N_{\it
hh}}\,,\label{(6.2.7)}
\eq
\bbq
\w={\rm Tr}\,\Biggl (\frac {M_{\it hh}N_{\it hh}}{Z_q\la}\Biggr )
-(\nl-N_c)\,\lym^3\Biggl (\det \frac{\langle N_{\it hh}\rangle}{N_{\it
hh}} \det \frac{M_{\it ll}}{\langle M_{\it ll}\rangle} \Biggr
)^{1/(\nl-N_c)}+\,{\rm Tr} \,\Biggl ( \ml M_{\it ll}+\mh M_{\it hh}\Biggr )\,,
\eeq
where
$z^{\,\prime\prime}_{M}=z^{\,\prime\prime}_{M}(\ogh\,,{\olp})\simeq
z^{\,\prime}_{M}( \mgh\,,{\mulp})=z^{\,\prime}_M\gg 1$ is the
logarithmic renormalization factor of $M_{\it ll}$ mions was ignored
\footnote{\,
The second term of the superpotential in \eqref{(6.2.7)} can
equivalently be written as\,:
\bbq
-(\nl-N_c)\,\Biggl ( \frac{\det {M_{\it ll}}}{\la^{\bo}\,{\det \Bigl
(N_{\it hh}/Z_q\la\Bigr )}}\Biggr )^{1/(\nl -N_c)}
\eeq
}

The masses obtained from \eqref{(6.2.7)} look then as follows\,:
\bq
\frac{\mu(M_{\it hh})}{\la}\sim \frac{\mu(N_{\it hh})}{\la}\sim
\Biggl({\hat
z}_q\frac{|\langle N_{\it hh}\rangle|}{\la^2} \Biggr )^{1/2}
\sim \exp \Bigl \{-\nd/14\bd \Bigr \}\,\Biggl
(\,\frac{\mh}{\la}\Biggr)^{N_F/3N_c}\sim\frac{\it\ogh}{\la}\,,\label{(6.2.8)}
\eq
\bbq
\frac{\mu(M_{\it ll})}{\la}\sim Z_q^2\frac{\ml\,\la}{{\hat
z}_{M}\langle M_{\it
ll}\rangle}\sim\exp \Bigl \{-\frac{2\nd}{7\bd} \Bigr \}\Bigl (r \Bigr
)^{\frac{2N_c-\nl}{N_c}}\Biggl (\frac{\mh}{\la} \Biggr)^{N_F/3N_c}\,.
\eeq
\vspace{0.5cm}

On the whole for this Section 6.2, when we started directly from the
dual theory with
$\nd=(N_F-N_c)$ colors, $N_F=(\nl+\nh)$ quarks $\ov q\,,q$ and
$N_F^2$ mions $M^i_j$, its mass spectrum is as follows.\\

{\bf 1)\,} The sector of heavy masses includes\,:\, a)
$\nh(2\nd-\nh)$ of  massive dual 
gluons and their scalar superpartners;\, b)\, $2\nl\nh$
hybrid scalar mions $M_{\it hl}+M_{\it lh}$,\, c)\, $2\nl\nh$ hybrid
scalar nions $N_{\it lh}+ N_{\it hl}$ (these are $q^{\it l}\,,\ov
q_{\it l}$\,\, quarks with higgsed colors),\, d)\, $\nh^2$ scalar
mions $M_{\it hh}$ and $\nh^2$ of scalar nions $N_{\it hh}$. All
these particles, with {\it specific numbers of each type, definite
spins, and other quantum numbers}, have definite masses of 
the same order
\bbq
\sim \ogh\sim \exp\{-\nd/14\bd\}(\mh/\la)^{N_F/3N_c}\la\sim
\eeq
\bbq
\exp \{-\nd/14\bd\}\,\mhp\ll\mhp\sim(1/r)^{\nl/3N_c}\lym\,.
\eeq

All these particles are {\it interacting only weakly}\,, with both
their gauge and
Yukawa couplings ${\ov a}_*\sim a_{f}^{*}\sim \bd/N_F\ll 1$.\\

{\bf 1')}\,\,Because the dual quarks $q^{\it h}\,,{\ov q}_{\it h}$
are higgsed, one can imagine that solitonic excitations also appear in
the form of monopoles of the dual gauge group (its broken part), see
e.g. section 3 in \cite{ch3} and footnote 6 therein.
These dual monopoles are then confined and can, in principle, form a
number of additional hadrons $H_{\it h}^{\,\prime}$. Because the dual 
theory is weakly coupled at the scale of higgsing, $\mu\sim \ogh \sim 
\exp\{-\nd/14\bd \}\, \mhp$, the masses of these monopoles, as well as
th  tension $\sqrt {\,\ov \sigma}$ of strings confining them (with our
exponential accuracy in parametrical dependence on $\nd/\bd\gg 1$),
are also $\sim \ogh$\,. Therefore, the mass scale of these hadrons
$H_{\it h}^{\,\prime}$ is also $\sim \ogh\ll \mhp$. We even assume
(in favor of the duality) that, with respect to their quantum
numbers, \, these hadrons $H_{\it h}^ {\,\prime}$ can be identified in
some way with the direct hadrons made of the quarks ${\ov Q}^{\it h}\,, 
\qh$. But even then, the masses of $H_{\it h}^{\,\prime}$ are {\it
parametrically smaller} than those of various direct hadrons made of
the ${\ov Q}^{\it h}\,,\,\qh$ quarks\,, $\mu(H^{\,\prime}_{\it
h})\sim\exp \{-\nd/14\bd\}\,\mhp\ll \mhp$.

Besides, the chiral symmetry of $\,{\it \ov l}\,,\,\it l$\,  flavors 
and the $R$-charges of the lower energy theory at $\olp<\mu <
\ogh$ {\it remain unbroken} and the $\it l$ - flavored dual quarks are 
not higgsed and remain effectively massless  in this 
interval of scales. Therefore, no possibility is seen in this dual
theory in Section 6.2 for the appearance of $\it l$\,-flavored chiral 
hadrons  $H^{\,\prime}_{\it l}$ (mesons and
baryons) with the masses $\sim \ogh$ that could be identified
with a large number of various direct ${\it  l}$\, - flavored  chiral 
hadrons (mesons and baryons) made of ${\ov Q}^{\it l}\,, Q_{\it l}$ 
quarks (and $W_{\alpha}$) that appear when the direct theory is
dualized in variant "b" (not even speaking about their parametrically
different mass scales, $\ogh\ll \mhp$\,, see point {\bf 1'} in Section 6.1).
This shows the self-contradictory character of duality in variant "b" =
"confinement without chiral symmetry breaking".\\

All other particles in the mass spectrum of the dual theory in this
section 6.2 constitute the sector of lighter particles, with their masses 
being parametrically smaller than $\ogh$.\\

{\bf 2)} The next mass scale is formed by a large number of ${\it
l}$\,-flavored dual mesons and ${\rm b}_{\it l}\,,{\rm \ov b}_{\it l}$  
baryons made of non-relativistic (and weakly confined, the string
tension $\sqrt \sigma\sim \lym\ll \olp$)\, dual $\hat{q}^{\it
l}\,,\hat{\ov q}_{\it l}$ quarks with $\nd^{\,\prime}=(\nl-N_c)$
not higgsed colors. The pole masses of these l-quarks are $$\olp \sim
\exp\{3\nd/14\bd\}\,(r)^{(\nl-N_c)/N_c}\,{\it\ogh}/z^{\,\prime}_{q}\ll 
{\it\ogh}$$ (at $r\ll r_l$, see \eqref{(6.2.9)} below).\\

{\bf 3)} Next, there is a large number of gluonia with the mass scale
$\sim \lym\ll\olp$.\\

{\bf 4)} Finally, the lightest are $\nl^2$ scalar mions $M_{\it ll}$
with masses $$\mu(M_{\it ll})\sim \exp \{-2\nd/7\bd \} (r
)^{2\Delta}\,\lym/z^{\,\prime}_M\ll \lym\,,\quad 0 <\Delta=
\frac{3N_c-2\nl}{3N_c}<\frac{1}{3}\,.$$ 
\vspace{3mm}

Comparing the mass spectra of two supposedly equivalent descriptions
in Sections 6.1 and 6.2 above, it is seen that the masses are clearly different
parametrically, in powers
of the parameter $Z_q\sim\exp\{-\nd/7\bd\}\ll 1$. Besides, the theory
described in Section 6.2 contains very specific definite numbers of
fields with fixed quantum numbers and spins and with definite masses
$\sim\ogh\sim Z_q^{1/2} \mhp\ll\mhp$, all {\it parametrically weakly
interacting} (see point {\bf 1} in Section 6.2). No analog of
these distinguished particles is seen in Section 6.1. Instead, there
is a large number of ${\it h}$ - flavored hadrons with the masses 
$\sim\mhp$ and with various spins, all strongly interacting with 
the coupling $a(\mu\sim \mhp)\sim 1$.

Finally, and we consider this to be of special importance, {\it there
are no ${\it l}\,, l$ -flavored hadrons} $H^{\,\prime}_{\it l}$
in the theory described in Section 6.2, which have the appropriate
conserved (in the interval of scales $\olp<\mu < \ogh$)
chiral flavors ${\it \ov l}\,,{\it l}$ and $R$ - charges, such that
they can be associated with a large number of various 
flavored chiral hadrons made of
${\ov Q}^{\it \ov l}\,, Q_{\it l}$ -quarks (an $W_{\alpha}$), which
are present in Section 5.1 dualized in variant "b". This shows that
the duality in the variant "b" = "confinement without chiral symmetry 
breaking" is not self-consistent. This agrees with some general arguments
presented in \cite{ch1} (see Section 7 therein), that the duality in variant
"b"cannot be realized because, in the theory {\it with unbroken chiral flavor
symmetries and R-charges} and effectively massless quarks, the masses
of flavored and R - charged chiral hadron superfields with various
spins cannot be made "of nothing".\\

This is not the whole story, however. For the mass $\ogh$ of gluons
in the dual theory to be the largest physical mass
$\mu_H$ as was used in Section 6.2 above, the parameter $r=\ml/\mh$
has to be taken sufficiently small (see \eqref{(6.2.2)},
\eqref{(6.2.6)}; from now on, the non-leading effects due to
logarithmic factors
like $z^{\,\prime}_q\simeq z^{\,\prime\prime}_q$ are ignored)\,:
\bq
\frac{\olp}{\ogh}\ll 1 \, \ra \, r \ll r_{\it l}=\Biggl
(\,z^{\,\prime}_q\,
\exp \Bigl \{-\frac{3\nd}{14\bd}\Bigr \}\, \Biggr
)^{\frac{N_c}{\nl-N_c}}\sim \exp \Biggl 
\{-\frac{3\nd}{14\bd}\,\frac{N_c}{\nl-N_c}\,\Biggr \}\ll 1\,.
\label{(6.2.9)}
\eq
We trace below the behavior of the direct and dual theories in the
whole interval $r_{\it l}<r<1$.
\footnote{\,
For this, it is convenient to keep $\mh$ intact while $\ml$ will be
decreased, starting from $\ml=\mh$.
}

As regards the direct theory (see Section 6.1 above), its regime and
all hierarchies in the mass spectrum remain the same for any value of
$r<1$\,, i.e. the pole mass $\mhp$ of the $\qh\,,\oqh$ quarks becomes
the largest physical mass $\mu_H$ already at $r<1/(\rm several)$, and
so on.

But this is not the case for the dual theory (see Section 6.2 above).
At $r_{\it l}<r<1/ (\rm several)$\,, the pole mass $\hlp$ of $\dl\,,\odl$ 
quarks remains the largest physical mass
\bq
\frac{{\hlp}}{\la}\sim\frac{\langle M_{\it
ll}\rangle}{Z_q\la^2}\,\Biggl
(\frac{\la}{\hlp}\Biggr )^{\bd/N_F}\sim\exp \Biggl
\{\,\frac{\nd}{7\bd}\,\Biggr\}\,
\Bigl( r \Bigr )^{\Bigl (\frac{\nl-N_c}{N_c}\frac{N_F}{3\nd}\Bigr
)}\,\Biggl (\frac{\mh}{\la} \Biggr )^{N_F/3N_c}\sim
\frac{\olp}{\la}\,.\label{(6.2.10)}
\eq

Already this is sufficient to see a qualitative difference between
the direct and dual theories. The {\it hh}\,-flavored hadrons in the 
direct theory have the largest masses, while the {\it ll}\,- flavored 
hadrons are the heaviest ones in the dual theory.

We make now some rough additional estimates in Section 6.2 at
$r>r_{\it l}$. After integrating out the heaviest quarks $\dl\,,\odl$ 
at the scale $\mu\sim \hlp$, the dual theory remains
with $\nd$ colors and $\nh<\nd$ dual quarks ${\ov q}_{\it \ov h}\,,\,
q^{\it h}$ (and mions $M$), and with $\bd^{\prime\prime}=
(3\nd-\nh)>0$.  It is in the weak-coupling logarithmic regime
at $\mu^{\,\prime}_H<\mu<\hlp$\,,\,\,\, where $\mu^{\,\prime}_H$ is
the highest mass scale in the remaining theory. The new scale factor 
$\Lambda^{\,\prime}_q$ of its gauge coupling can be found from
\bq
\bd^{\,\prime\prime}\ln\Biggl (\frac{\hlp}{\Lambda^{\,\prime}_q}
\Biggr )\simeq\frac{2\pi}{\ov\alpha_*}=\frac{3\nd^2}{7\bd}\quad \ra \quad
\frac{\Lambda^{\,\prime}_q}{\hlp}\sim\exp\Biggl \{-\frac{3\nd^2}{7\bd
\bd^{\,\prime\prime}}\Biggr \}\ll 1\,. \label{(6.2.11)}
\eq

If (see below) $r_{\it l}\ll r_{\it h}<r<1/(\rm several)$, then
$\mu^{\,\prime}_H=\hhp> \Lambda^{\,\prime}_q$, where $\hhp$ 
is the pole mass of $\hd\,,\ohd$ quarks which are in the HQ-phase 
(i.e. not yet higgsed ). Roughly,  $\hhp\sim r\,\hlp$, hence
\bq
\frac{\hhp}{\Lambda^{\,\prime}_q}>1 \quad \ra \quad r>r_{\it h}\sim
\exp \Biggl\{-\frac {3\nd^2}{7\bd\bd^{\,\prime\prime}} \Biggr \}\gg r_{\it
l}\,.\label{(6.2.12)}
\eq

In the interval $r_{\it h}<r<1/(\rm several)$, the mass spectrum of
the dual theory is qualitatively
not much different from the case $r=1$ (see Section 4). The heaviest
are the {\it ll} - hadrons,
then the {\it hl} - hadrons, then the {\it hh} - hadrons, then
gluonia and, in addition, there are the
mions $M_{\it hh}\,,\,M_{\it hl}\,,\, M_{\it ll}$ with their masses
$$\mu(M_{\it hh})\sim\mh^2/\mu_{o}\,,\,\,\,\mu(M_{\it hl})\sim \mh
\ml/\mu_{o}\,,\,\,\,\mu(M_{\it ll})
\sim \ml^2/\mu_{o}\,,$$
where $$\mu_{o}= z_{M}(\la,\hlp)\lym^{3}/Z_q^2\la^2\,,\quad
z_{M}(\la,\hlp)\sim (\la^2/\mh\ml)^{\bd/3N_c}\gg 1.$$

As $r$ decreases further, a phase transition occurs from the $\rm
HQ_{\it h}$ phase to the $\rm Higgs_{\it h}$ phase at 
$r\sim r_{\it h}$\,, i.e. after $\hhp >\Lambda^{\prime}_q$
becomes $\hhp<\Lambda^{\,\prime}_q$ and the quarks $\hd\,,\ohd$ are
higgsed. But even then the $\dl\,,\odl$ quarks remain the heaviest
ones. And only when $r$ becomes $r <r_{\it l}\ll r_{\it h}\ll 1$, the
gluon mass $\ogh$ becomes the largest and the
mass spectrum of the dual theory becomes that described above in this
section 6.2.\\

\section{Direct theory. Unequal quark masses, \,\,  $\mathbf {3N_c/2<N_F<3N_c}, \\
\mathbf {(3N_c-N_F)/N_F={\cal O}(1)\,,\,\, N_c < \nl <3N_c/2}$}

Let us check first that largest observable masses have heavy quarks
$\qh\,,\oqh$. The pole mass of $\qh$ will be in this case
\bq
\mhp=\frac{m_h}{z_Q(\la,\mhp)},\,\, z_Q(\la,\mhp)\sim \bigl (
\frac{\mhp}{\la}\bigr )^
{\gamma_Q} \ra \frac{\mhp}{\la}\sim \bigl ( \frac{\mh}{\la}\bigr
)^{\frac{N_F}{3N_c}},
\,\, \gamma_Q=\frac{3N_c-N_F}{N_F},\,\,\quad \label{(7.0.1)}
\eq
The competing mass of the gluon due to possible higgsing of $\ql$
quarks (if it were the largest one) would be as follows.
\bbq
\mu_{gl,l}^2\sim z_Q(\la,\mu_{gl,l})\langle M_l \rangle,\quad z_Q\sim
(\mu_{gl,l}/\la)^{\bo/N_F}, \,\, \frac{\langle M_l\rangle}{\la^2}\sim
\bigl ( r \bigr )^{\frac{N_l-N_c}{N_c}}
\Bigl (\frac{\qh}{\la} \Bigr )^{\frac{N_F-N_c}{N_c}},\,\,
r=\frac{m_l}{m_h}\ll 1\,,
\eeq
\bq
\frac{\mu_{gl,l}}{\la}\sim \Bigl (\frac{\langle
M_l\rangle}{\la^2}\Bigr )^{N_F/3\nd} \ra
\frac{\mu_{gl,l}}{\mhp}\sim \bigl ( r \bigr
)^{\frac{N_l-N_c}{N_c}}\ll 1\,. \label{(7.0.2)}
\eq

Therefore, $\mhp$ is the largest mass. After $\qh, {\ov \qh}$ have
been integrated out in the conformal regime at $\mhp < \mu<\la$, 
all  particle masses in the lower energy theory with $N_c$ colors and
$N_c<\nl<3N_c/2$ lighter $\ql\,, \oql$ quarks are parametrically
smaller than the scale $\la^{\,\prime}\sim \mhp$ (see \eqref{(6.1.1)}
above), and at $\mu<\la^{\,\prime}$ this lower energy theory enters
the strong coupling regime with $a(\mu\ll\la^{\,\prime}) \gg 1$.
\footnote{\,
Hereafter  we use the anomalous quark
dimension $1+\gamma_Q=N_c/(N_F-N_c)$ and the strong coupling 
$a(\mu)\gg 1$, see section 5 for explanations.
}
This lower-energy theory is in the HQ-phase (see footnote
\ref{(f5)}\,) with $N_l > N_c$ flavors this time, and the pole mass
of $\ql\,,\oql$ quarks is
\bbq
\frac{\mlp}{\la^{\,\prime}}\sim\Biggl
(\frac{\ml(\mu=\la^{\,\prime})}{\la^{\,\prime}}=r\,\Biggr ) \Biggl
(\frac{\la^{\,\prime}}{\mlp}\Biggr )^{\gamma^{\,\prime}_Q}\quad \ra
\quad
\frac{\mlp} {\la^{\,\prime}}\sim\Bigl (r \Bigr )^{(\nl-N_c)/N_c}\ll
1\,,\quad r\ll 1\,,
\eeq
\bq
{\rm b}^{\,\prime}_o=3N_c-\nl,\quad
1+\gamma^{\,\prime}_Q=\frac{N_c}{\nl-N_c}\,,
\quad \nu=\frac{N_F \gamma_Q^{\,\prime}-{\rm b}^{\,\prime}_{\rm
o}}{N_c} =\frac{3N_c-2\nl}{\nl-N_c}\,. \label{(7.0.3)}
\eq

The coupling $a_{+}(\mu=\mlp)$ is large 
\bq
a_{+}(\mu=\mlp)\sim\Biggl (\frac{\la^{\prime}}{\mu=\mlp} \Biggr
)^{\nu}\sim\Bigl
(\,\frac{1}{r}\, \Bigr )^{(3N_c-2\nl)/N_c}\gg 1\,.\label{(7.0.4)}
\eq

Hence, after integrating out all $\ql\,,\oql$ quarks as heavy ones at
$\mu<\mlp$, we are left with the pure $SU(N_c)$ SYM theory, but it 
is now in the {\it strong coupling
branch}. This is somewhat unusual, but there is no contradiction
because this perturbative strong coupling regime with
$a_{SYM}(\mu)\gg1$ is realized in a restricted interval of scales
only, \,$\lym\ll \mu<\mlp\ll \la^{\prime}$. It follows from the NSVZ $\beta$  
function \cite{NSVZ-1} that the coupling $a_{SYM}(\mu\gg \lym)\gg 1$ is 
then given  by
\bbq
a_{SYM}(\mu=\mlp)=\Biggl (\frac{\mu=\mlp}{\lambda_{SYM}} \Biggr )^3=
a_{+}(\mu=\mlp)\,\ra  
\eeq
\bq
\frac{\lambda_{SYM}}{\la}=\frac{\lym}{\la}=\Biggl [ \Bigl
(\frac{\ml}{\la}\Bigr )^{\nl}\Bigl (\frac{\mh}{\la}\Bigr
)^{\nh}\Biggr]^{\frac{1}{3N_c}}, \label{(7.0.5)}
\eq
\bq
\frac{\lym}{\mlp}=r^{\frac{3N_c-2\nl}{3N_c}}\ll 1,\quad
a_{SYM}(\lym\ll\mu<\mlp)=\Bigl (\frac{\mu}{\lym} \Bigr )^3\,,
\label{(7.0.6)}
\eq
and it now {\it decreases} from $a_{SYM}(\mu=\mlp)\gg 1$ to
$a_{SYM}(\mu\sim\lym)\sim 1$ as $\mu$ decreases, after which the
non-perturbative effects of pure SYM stop the perturbative
RG-evolution at $\mu\sim\lym$.

Therefore, decreasing the scale $\mu$ from $\mu=\mlp$ to $\mu <\lym$,
integrating
out all gauge degrees of freedom except for the one whole field $S\sim
W_{\alpha}^2$\,, and using the VY-form of the superpotential of the
field $S$ \cite{VY}, we obtain the  universal value of $\lym$ and the 
standard gluino condensate, $\langle   S\rangle=\lym^3$.

To verify the self-consistency, we have to estimate the scale
$\mu_{\rm gl}$ of
the possible higgsing of $\ql\,,\oql$ quarks. This estimate looks as,
see \eqref{(7.0.3)}
\bq
\mu_{\rm gl}^2\sim a_{+}(\mu=\mu_{\rm gl})\langle \oql \ql
\rangle_{\mu=\mu_{\rm
gl}}\sim \Bigl (\mlp \Bigr )^2\,,\quad a_{+}(\mu=\mu_{\rm
gl})\sim\Biggl ( \frac{\la^{\,\prime}\sim\mhp}{\mu_{\rm gl}} \Biggr
)^{\nu}\,, \label{7.0.7)}
\eq
\bbq
\langle \oql \ql \rangle_{\mu=\mu_{\rm gl}}\sim\langle \oql \ql
\rangle_{\mu=\la}\,\Biggl(\frac{ \la^{\,\prime}}{\la} \Biggr
)^{\gamma_Q=(\bo/N_F)}\,\Biggl (\frac{\mu_{\rm
gl}}{\la^{\,\prime}} \Biggr )^ {\gamma^{\,\prime}_Q}\,,\quad \langle
\oql \ql
\rangle_{\mu=\la}=\frac{\lym^3}{\ml}\,.
\eeq

As before, the Konishi anomaly shows that $\ql\,,\oql$ quarks are not
higgsed at $N_l > N_c$ and the $HQ_{\it l}$ phase is self-consistent
\ref{(f5)}.\\

On the whole, all quarks of the direct theory are in the HQ phase
(and are therefore confined, but the string tension is small in
comparison with quark masses, $ \sqrt
\sigma\sim \lym\ll\mlp\ll\mhp$. This is ``the weak confinement''\,.
The mass spectrum then includes \,:\, 1)\, a large number of heavy
$\it hh$\,-flavored mesons with the mass scale
$\sim \mhp\ll \la$\,;\,\, 2)\, a large number of hybrid $\it
hl$\,-mesons and baryons $B_{\it
hl}\,, {\ov B}_{\it hl}$ with the same mass scale $\sim \mhp$\,;\,\,
3)\, a large number
of $\it ll$\,-flavored mesons and baryons $B_{\it l}\,,{\ov B}_{\it
l}$ with the mass
scale $\sim\mlp\ll\mhp$\,, and finally,\,\, 4)\, gluonia, which are
the lightest, with their mass scale $\sim \lym \ll \mlp$.

\section  {Direct and dual theories. Unequal quark masses. \\
\boldmath $N_c<N_F<3N_c/2,\,\,\nl>N_c$}

As regards the direct theory, the mass spectrum in this case with
$r=\ml/\mh\ll 1$ is not much different from the one in the preceding 
section. All quarks are in the (very) strongly coupled HQ phase and 
are not higgsed but confined with  $a(\mu\ll\la)\gg 1$, and the 
highest physical mass is the pole mass of {\it h} - quarks
\bq
\mhp=\mh\Biggl (\frac{\la}{\mhp} \Biggr )^{\gamma_{+}}\quad \ra \quad
\frac{\mhp}{\la}=
\Biggl (\frac{\mh}{\la}\Biggr )^{(N_F-N_c)/N_c}\,,\quad
\gamma_{+}=\frac{(2N_c-N_F)}{(N_F-N_c)}\,.\label{(8.01)}
\eq

After integrating out $\qh\,,\oqh$ quarks as heavy ones at
$\mu<\mhp$,  the lower energy theory remains with $N_c$ colors 
and $N_c<\nl<3N_c/2$\, {\it l} - quarks,
in the strong coupling regime with $a(\mlp<\mu<\mhp)\gg 1$, see
sections 5, 6, 7. The next independent physical scale is the pole mass 
of {\it l} - quarks, $\gamma_{-}=(2N_c-\nl)/(\nl-N_c)$,
\bq
\mlp\sim\Biggl (\ml(\mu=\mhp)=r\,\mhp \Biggr )\,\Biggl
(\frac{\mhp}{\mlp} \Biggr
)^{\gamma_{-}} \quad \ra \quad\Bigl (\frac{\mlp}{\mhp}\Bigr
)\sim\Bigl( r\Bigr )^{\frac{\nl-N_c}{N_c}}\ll 1\,. \label{(8.0.2)}
\eq

After integrating out the {\it l} - quarks as heavy ones at
$\mu<\mlp$, leaves us with the $SU(N_c)$ SYM theory in the 
strong-coupling branch with the large gauge
coupling. The scale factor $\lym^{\,\prime}$ of its gauge 
coupling is determined from the matching (see sections 6, 7)
\bq
\Biggl (\frac{\mlp}{\lym^{\,\prime}}\Biggr )^3=\Biggl
(\frac{\la}{\mhp} \Biggr)^{\nu_{+}}
\Biggl (\frac{\mhp}{\mlp} \Biggr )^{\nu_{-}} \quad \ra \quad
\lym^{\,\prime}=\lym=\Biggl (\la^ {\bo}\ml^{\nl}\mh^{\nh}\Biggr
)^{1/3N_c}\,.\label{(8.0.3)}
\eq
\bbq
\nu_{+}=\frac{3N_c-2N_F}{N_F-N_c}\,,\quad
\nu_{-}=\frac{3N_c-2\nl}{\nl-N_c}>\nu_{+}\,\,.
\eeq

To verify the self-consistency, we also estimate the gluon masses due
to possible higgsing of the $\qh$ and/or $\ql$ - quarks. As for 
$\qh$- quarks\,,
\bq
\hspace*{-3mm}\frac{\mgh^2}{\la^2}\sim \Biggl
[\,a_{+}(\mu=\mgh)=\Biggl (\frac{\la}{\mgh} \Biggr )^{\nu_{+}}\,
\Biggr ]\,\frac{\langle{\ov Q}^{\it h}\qh \rangle}{\la^2}\,\Biggl
(\frac{\mgh}{\la}\Biggr ) ^{\gamma_{+}}\, \ra \,\,
\frac{\mgh}{\mhp}\sim \Bigl (\, r \,\Bigr )^{\nl/N_c}\ll
1\,,\label{(8.0.4)}
\eq
and hence there is no problem, but for $\ql$ - quarks, we now
have
\bq
\frac{\mgl^2}{\la^2}\sim \Biggl [\,a_{-}(\mu=\mgl)=\Biggl
(\frac{\la}{\mhp} \Biggr
)^{\nu_{+}}\Biggl (\frac{\mhp}{\mgl} \Biggr )^{\nu_{-}}\,\Biggr ]\,
\frac{\langle{\ov Q}^{\it l}\ql \rangle}{\la^2}\,\Biggl
(\frac{\mhp}{\la}\Biggr )
^{\gamma_{+}}\Biggl ( \frac{\mgl}{\mhp}\Biggr
)^{\gamma_{-}}\,\label{(8.0.5)}
\eq
\bq
\ra \quad \frac{\mgl}{\la} \sim \frac{\langle{\ov Q}^{\it l}\ql
\rangle}{\la^2}\,\sim\,\frac{\mlp}{\la}\sim\Bigl (
r\Bigr)^{\frac{\nl-N_c}{N_c}}\,\,\frac{\mhp}{\la}\,.\label{(8.0.6)}
\eq
As before, because $N_l > N_c$, it is seen from the Konishi anomaly
\eqref{(2.1.2)} that l-qurks are not higgsed due to the rank restrictions,
see footnote \ref{(f5)}.

On the whole in the direct theory, all quarks are in the HQ phase and
the mass spectrum consists of\,:\, a)\, a large number of {\it hh} and hybrid 
{\it hl} - mesons and baryons with the mass scale $\sim \mhp\sim \la\,(\mh/\la)
^{(N_F-N_c)/N_c}\ll \la$\,,\,\, b)\, a
large number of {\it ll} - mesons and baryons with the mass scale
$\sim \mlp\sim (r)^{(\nl-N_c)/N_c}\,\mhp\ll \mhp$\,;\, all quarks are
weakly confined, i.e. the string tension $\sqrt\sigma\sim\lym$ is
much  smaller than their masses\,,\,\, c)\, gluonia, which are the
lightest,  with the masses $\sim \lym \ll \mlp$.\\

In the IR-free dual theory there are
several regimes depending on the value of $r=\ml/\mh< 1$.\\

i) At $r_1=(\mh/\la)^{(3N_c-2N_F)/2\nl}<r<1$,\, the hierarchy of
masses looks
as\,: $\mu_{q,\,\it l}>\mu_{q,\,\it h}>\mu_{\rm gl}^{\it h}$\,, where
$\mu_{q,\,\it l}$ and
$\mu_{q,\,\it h}$ are the pole masses of dual quarks and $\mu_{\rm
gl}^{\it h}$
is the gluon mass due to possible higgsing of $q_{\it h},\, {\ov
q}_{\it h}$ quarks (all logarithmic renormalization effects are
ignored here and below in this section for simplicity):
\bbq
\frac{\muhp}{\la}\sim \Bigl ( r \Bigr )^{\frac{N_l}{N_c}}
\Bigl ( \frac{m_h}{\la} \Bigr )^{\frac{N_F-N_c}{N_c}}\,.
\eeq
\bq
\frac{\mulp}{\la}\sim \Biggl (\frac{\langle{\ov Q}_{\it l}Q_{\it
l}(\mu=\la) \rangle}{\la^2}\Biggr )^{\frac{1}{1+\gamma_{+}}}
\sim\Bigl(r \Bigr )^{\frac{\nl-N_c}{N_c}}\,
\Bigl (\frac{\mhp}{\la}\Bigr )^{\frac{N_F-N_c}{N_c}},
\label{(8.0.7)}\eq
\bq
\frac{\muhp}{\la}\sim \frac{\mu_{q,\,\it h}}{\la}
\sim\frac{\langle{\ov Q}_{\it h}Q_{\it
h}(\mu=\la)\rangle}{\la^2}\sim\Bigl (r \Bigr
)^{\frac{\nl}{N_c}}\,\Bigl(\frac{\mh}{\la}\Bigr
)^{\frac{N_F-N_c}{N_c}}\,<\, \frac{\mulp}{\la}, \label{(8.0.8)}
\eq
\bbq
\mu_{\rm gl}^{\it h}\sim |\langle{\ov q}_{\it h}q_{\it h}
\rangle|^{1/2}\sim (\mh\la)^{1/2}\,<\,\muhp\,.  
\eeq

This hierarchy shows that all dual quarks are in the HQ phase and not
higgsed but confined by the dual $SU(\nd > 1)$ SYM..\\

ii) At $r_2=(\mh/\la)^{(3N_c-2N_F)/2(\nl-N_c)}<r<r_1=(\mh/\la)^
{(3N_c-2N_F)/2\nl}$,\,
the hierarchy of masses looks as\,: $\mu_{q,\,\it l}>\mu_{\rm gl}^{\it
h}>\mu_{q,\,\it h}$. This shows that ${\ov q}_{\it l},\, q_{\it l}$
quarks are in the HQ phase and are the heaviest ones, while ${\ov q}_
{\it h},\, q_{\it h}$ quarks are higgsed. The phase transition of ${\ov q}_{\it
h},\, q_{\it h}$ quarks from the ${\rm HQ}_{\it h}$ phase to the ${\rm
Higgs}_{\it h}$ phase occurs in the region $r\sim r_1$.\\

iii) At $r<r_2=(\mh/\la)^{(3N_c-2N_F)/2(\nl-N_c)}$,\, the hierarchy
of  masses looks as\,: $\mu_{\rm gl}^{\it h}>\mu_{q,\,\it l}>\mu_{q,\,\it h}$.
Here, the ${\ov q}_{\it h},\, q_{\it h}$ quarks are higgsed and
$\mu_{\rm gl}^{\it h}$ is the highest mass, while the ${\ov q}_{\it l},\, 
q_{\it l}$ quarks are lighter and are in the HQ phase. We now give some
details.\\

i) After the heaviest dual $\it l$ - quarks are integrated out at
$\mu<\mu_{q,\,\it l}$\,, a theory remains with $\nd$ colors and
$\nh<\nd$ quarks ${\ov q}_{\it h},\, q_{\it h}$ (and mions $M$), and
with the scale factor of its gauge coupling
\bq
\Bigl (\Lambda^{\,\prime}_q \Bigr )^{{\ov
b}^{\,\prime}_o}=\la^{\bd}\,\mu^{\nl}_{q,\,\it l}\,, \quad {\ov
b}^{\,\prime}_o=(3\nd-\nh)>0\,,
\quad \Biggl (\frac{\Lambda^{\,\prime}_q}{\mu_{q,\,\it h}}\Biggr
)^{{\ov  b}^{\,\prime}_o/\nd}=\Bigl (\frac{r_1}{r}\Bigr
)^{2\nl/N_c}<1\,.\label{(8.0.9)}
\eq
Hence, after integrating out the quarks ${\ov q}_{\it h},\, q_{\it
h}  $ \, as heavy ones with masses $\mu_{q, \,\it h}>\Lambda^
{\,\prime}_q$ and then gluons via the
VY - procedure, we obtain the Lagrangian of mions
\bq
\ov K= \frac{1}{\la^2}\,{\rm Tr}\,\Bigl (M^{\dagger} M\Bigr )\,,
\quad {\ov \w}=-\nd \,
\Biggl ( \frac {\rm \det\, M}{\la^{\bo}} \Biggr )^{1/\nd}+{\rm Tr}\,
(m_Q M)\,\,.\label{(8.0.10)}
\eq
It describes mions ${\rm M}$ with the masses
\bq
\mu(M^{i}_{j})\sim \frac{m_i m_j\la^2}{\lym^3}< \lym\,\,,\quad
i,\,j={\it l\,,\,h}\,.\label{(8.0.11)}
\eq
\vspace{3mm}

ii) The first step of integrating out the heaviest $\it l$ - quarks
with masses
$\mu_{q,\,\it l}$ is the same. But now, at $r_2<r<r_1$, the next
physical mass is $\ogh>\Lambda^{\,\prime}_q$
due to higgsing of dual $\it h$ - quarks, with $\nd\ra (\nd-\nh)$ and
formation of  $N_{\it h}^2$ nions $N_{\it hh}$. After integrating out
the heavy higgsed gluons
and their superpartners with masses $\ogh\sim (\mh\la)^{1/2}$\,, and
not higgsed gluons via the VY - procedure,
the Lagrangian of the remaining degrees of freedom takes the form
\bbq
{\ov K}=\frac{{\rm Tr}\,\Bigl (M^{\dagger}M\Bigr )}{\la^2}+2\,{\rm
Tr}\sqrt{N^{\dagger}_{\it hh}N_{\it hh}}\,,\quad {\ov\w}=
-(\nl-N_c)\,\Biggl (\frac{\det {M_{\it ll}}}{\la^{\bo}\,{\det 
\Bigl (-N_{\it hh}/\la\Bigr )}}\Biggr )^{1/(\nl-N_c)}+
\eeq
\bq
+\, \frac{1}{\la}{\rm Tr}\, N_{\it hh}\Biggl ( M_{\it hh}-M_{\it
hl}M^{-1}_{\it ll}M_{\it lh}\Biggr )+{\rm Tr}\Bigl (\ml M_{\it ll}+\mh 
M_{\it hh}\Bigr )\,.\label{(8.0.12)}
\eq
From this, the masses of mions $M_{\it hh},\,M_{\it ll}$,\, hybrids
$M_{\it lh},\, M_{\it hl}$, and nions $N_{\it hh}$ are
\bq
\mu(M_{\it hh})\sim \mu(N_{\it hh})\sim\Bigl (\mh \la \Bigr
)^{1/2}\sim \ogh\,,\label{(8.0.13)}
\eq
\bbq
\mu(M_{\it lh})\sim \mu(M_{\it hl})\sim
\frac{\ml\mh\la^2}{\lym^3}\,,\quad
\mu(M_{\it ll})\sim \frac{\ml^2\la^2}{\lym^3}\,.
\eeq
\vspace{2mm}

iii) The heaviest particles in this region $r<r_2$ are higgsed gluons
and their
superpartners with masses $\ogh\sim (\mh \la )^{1/2}$. After these
have been integrated out, there remain
$\nd^{\,\prime}=(\nl-N_c)$ dual colors and $\nl$ flavors with active
not higgsed colors, and the regime at $\mu<\ogh$ is IR - free logarithmic,
$\bd^{\,\prime}=(3\nd^ {\,\prime}-\nl)=(2\nl-3N_c)<0$. In this theory, 
the next independent physical scale is given by the mass $\mu_{q,\,\it l}$ 
of the $\nl$ active $\it l$ - quarks with unhiggsed colors.

The masses of $2\nl\nh$ hybrid mions $M_{\it hl}+M_{\it lh}$ and
$2\nl\nh$ nions
$N_{\it lh}+N_{\it hl}$ (these are the dual ${\it l}$\,-quarks that
have higgsed colors) are determined mainly
by their common mass term in the superpotential\,:
\bq
K_{\rm hybr}\simeq{\rm Tr}\,\Biggl (\frac{M^{\dagger}_{\it hl}M_{\it
hl} } {\la^2}\Biggr )+{\rm Tr}\,\Bigl (N^{\dagger}_{\it lh} N_{\it lh}
\Bigr)\,,\quad \w=\Bigl (\mh\la\Bigr )^{1/2}{\rm Tr}\,\Biggl (\frac 
{M_{\it hl}N_{\it lh}}{\la}\Biggr )\,,\label{(8.0.14)}
\eq
and hence their masses are
\bq
\mu(M_{\it hl})\sim \mu(M_{\it lh})\sim \mu (N_{\it hl})\sim \mu
(N_{\it lh})\sim \ogh\,.\label{(8.0.15)}
\eq

Passing to lower scales and integrating out first active $\it l$
-quarks as heavy ones with masses $\mu_{q,\,\it l}$ and then  
not higgsed gluons  via the VY-procedure, we obtain
\bq
K=\frac{{\rm Tr}\,\Bigl (M^{\dagger}_{\it ll} M_{\it
ll}+M^{\dagger}_{\it hh} M_{\it hh}\Bigr )}{\la^2}+2\,{\rm Tr}\sqrt
{N^{\dagger}_{\it hh}N_{\it hh}}\,, \label{(8.0.16)}
\eq
\bbq
{\ov\w}=-(\nl-N_c)\,\Biggl ( \frac{\det {M_{\it
ll}}}{\la^{\bo}\,{\det\Bigl (-N_{\it hh}/\la\Bigr )}}\Biggr )^
{1/(\nl-N_c)}+\, \frac{{\rm
Tr}\,( N_{\it hh}M_{\it hh})}{\la}+{\rm Tr}\Bigl(\ml
M_{\it ll}+\mh M_{\it hh}\Bigr )\,.
\eeq
From \eqref{(8.0.16)}, finally, the masses are given by
\bq
\mu (N_{\it hh})\sim \mu (M_{\it hh})\sim \ogh \sim \Bigl (\mh \la
\Bigr
)^{1/2}\,,\quad \mu(M_{\it ll}) \sim
\frac{\ml^2\la^2}{\lym^3}\,.\label{(8.0..17)}
\eq

\section{ Direct theory. Unequal quark masses. \\ 
{\hspace*{1cm} \boldmath $N_c<N_F<3N_c/2\,,\,\,\nl< N_c-1$}}

In this regime, at $r=\ml/\mh\ll 1$\,, the highest physical scale
$\mu_{H}$ is determined by
the gluon masses $\mgl$ that arise due to higgsing of the
$\ql\,,\oql$ - quarks\,:\,
\footnote{\, 
Here and below, the value of\, 
$r$\, is taken to be not too small,\, such that\,
$\mgl\ll \la$, i.e. $r_H\ll r\ll 1,\, r_H=(\mh/\la)^{\rho},\, \rho=\nd/
(N_c-\nl)$. At $r$ so small that $r\ll r_H,\,
\mgl\gg\la$, the quarks $\ql\,,\oql$ are higgsed in the logarithmic
weak coupling region and the form of the RG flow is different, but 
the regime is qualitatively the same for $\mgl\ll \la$ or
$\mgl\gg \la$, and nothing happens as $\mgl$ overshoots $\la$ in the
scenario  considered here.

Besides, l-quarks are higgsed only at not too large values of $N_c$,
see \cite{ch21}.
}
\bbq
\frac{\mgl^2}{\la^2}\sim \Biggl [\,a_{+}(\mu=\mgl)=\Biggl
(\frac{\la}{\mgl} \Biggr
)^{\nu_{+}}\,\Biggr ]\,
\frac{\langle{\ov Q}_{\it l}\ql \rangle}{\la^2}\,\Biggl
(\frac{\mgl}{\la}\Biggr
)^{\gamma_{+}}\,\quad \ra
\eeq
\bq
\frac{\mgl}{\la} \sim \frac{\langle{\ov Q}^{\it l} Q_{\it
l}\rangle}{\la^2}\sim\frac{\lym^3}{m_l\la^2}\sim\,\Bigl (\,
\frac{1}{r}\,\Bigr
)^{(N_c-\nl)/N_c}\,\Biggl (\frac{\mh}{\la} \Biggr
)^{(N_F-N_c)/N_c}\ll 1\,. \quad \gamma_{+}-\nu_{+}=1\,.\ \
\label{(9.0.1)}
\eq

The lower energy theory includes\,: unbroken color $SU(\hat
N_c=N_c-\nl)$ SYM,\, $2\nl\nh$ hybrids $\Pi_{\it hl}+{\Pi}_{\it lh}$,\, 
$\nh$ flavors of active $\it
h$-quarks with unbroken colors and, finally,\, $\nl^2$ pion fields
$\Pi_{\it ll},\, \langle \Pi_{\it ll}\rangle= \langle {\ov Q}_{\it l}
Q^l rangle$. Their Lagrangian at $\mu\lesssim \mgl$ we write in the
form (all fields are normalized at $\mu=\la$)\,:
\bbq
K=z^{+}_{Q}(\la,\mgl)\Bigl [K_{\it ll}+K_{\rm hybr}+\dots\Bigr
]+z^{+}_{Q}(\la,\mgl)K_{\it h},\,\,\, K_{\it h}={\rm Tr}\Bigl
({\sq}^{\dagger} {\sq}
+({\sq}\ra {\oq} )\Bigr )\,,
\eeq
\bbq
K_{\rm hybr}={\rm Tr}\Biggl (\Pi^{\dagger}_{{\it l}{\it
h}}\,\frac{1}{\sqrt{\Pi_{\it ll}\Pi^{\dagger}_{\it ll}}}\,\Pi_{{\it
l}{\it  h}}+\Pi_{{\it h}{\it l}}\,\frac{1}{\sqrt{\Pi^{\dagger}_{\it
ll}\Pi_{\it ll}}}\,\Pi^{\dagger}_{{\it h}{\it l}}\Biggr ),\quad K_{\it
ll}=
2\,{\rm Tr}\sqrt{\Pi^{\dagger}_{\it ll}\Pi_{\it ll}}\,\,,
\eeq
\bq
W=\frac{2\pi}{{\hat\alpha(\mu)}}\, {\hat S}+W_{\Pi}+\mh {\rm
Tr}\,\Bigl ({\oq}{\sq}\,\Bigr )\,,\quad W_{\Pi}=\ml {\rm Tr}\,\Pi_{\it ll}+\mh
{\rm Tr}\,\Bigl (\Pi_{\it hl}\frac{1}{\Pi_{\it
ll}}\Pi_{\it lh}\Bigr )\,,\label{(9.0.2)}
\eq
where ${\hat\alpha}(\mu)$ is the gauge coupling of unbroken
$SU(N_c-N_l)$ (with the pions $\Pi_{\it ll}$ sitting down inside), 
$\hat S$ is the kinetic term of not higgsed
gluons, $\Pi_{\it ll}$ is the $\nl\times\nl$ matrix of pions
originated due to higgsing of ${\ov Q}_{\it l}, Q_{\it l}$ quarks,
$\Pi_{\it lh}$ and $\Pi_{\it hl}$ are the $\nl\times\nh$
matrices of the hybrid pions (in essence, these are the quarks ${\ov
Q}_{\it h}, Q_{\it h}$ with higgsed colors), $\oq$ and $\sq$ are
still  active $\it h$-quarks with unbroken colors, and dots denote
residual D-term interactions which play no role in what follows and
will be ignored. $z^+_Q=z_Q(\la\,,\mgl)$ in \eqref{(9.0.2)} is the
numerical value of the quark renormalization factor due to a perturbative 
evolution in  the range of scales $\mgl<\mu<\la$,
\bbq
z^{+}_Q=z^{+}_Q(\la\,,\mgl)=z^{+}_Q(\langle\Pi_{\it
ll}\rangle^{\dagger},\langle\Pi_{\it ll}\rangle)\sim\Biggl
(\frac{\mgl}{\la}\Biggr )
^{\gamma_{+}}\ll 1\,,\quad \gamma_{+}=(2N_c-N_F)/(N_F-N_c)\,.
\eeq

The numbers of colors and flavors have already changed in the
threshold region $\mgl/(\rm several)<\mu<(\rm several)\mgl\,,\,N_F
\ra {\hat N}_F=N_F-\nl=\nh
\,,\,N_c\ra {\hat N}_c=N_c-\nl$\,, while the coupling $\hat
\alpha(\mu)$ did not change essentially and remains $\sim
\alpha(\mu=\mgl)\gg 1$. within this  threshold region. Therefore, the
new quark anomalous dimension $\gamma_{-}(\hat N_c\,,\hat N_F
= \nh\,,\hat a\gg 1)$ and the new $\hat\beta$ -function have the form
\bq
\frac{d {\hat a}(\mu)}{d\ln \mu}=-\,\, \nu_{-}\,\hat a(\mu)\,,\quad
\nu_{-}=\frac{\hat N_F\, \gamma_{-}-{\hat b}_{o}}{\hat
N_c}=\frac{3\hat N_c-2\hat
N_F}{\hat N_F-\hat N_c}=\nu_{+}-\frac{\nl}{N_F-N_c}\,,
\label{(`9.0.3)}
\eq
\bbq
{\hat b}_{o}=(3\hat N_c-\hat N_F)=(\bo-2\nl) > 0,,\quad
\gamma_{-}=\frac{2\hat
N_c-\hat N_F} {\hat N_F-\hat N_c}=\gamma_{+}-\frac{\nl}{N_F-N_c}\,\,.
\eeq

Depending on the value of $\hat N_F/\hat N_c$\,, the lower energy
theory is in different regimes. We consider here the case  with the
strong coupling regime: { $1<\hat N_F/\hat N_c<3/2$\,.}\quad In this case,\,
$\nl<(3N_c-2N_F),\,\nu_{-}>0$, and hence the coupling $\hat a(\mu)$
continues to increase with decreasing $\mu$\,, but more slowly than before.

The next physical scale $\mu^{\prime}_H$ is given by the pole mass of
active $\oq\,,\sq$ quarks
\bbq
\frac{\mu^{\prime}_H}{\la}=\frac{\mhp}{\la}\sim\Biggl
[\,\frac{\mh(\mu=\mgl)}{\la}\sim\frac{\mh}{\la}\Biggl
(\frac{\la}{\mgl}
\Biggr )^{\gamma_{+}}\,\Biggr ]\Biggl (\frac{\mgl}{\mhp} \Biggr
)^{\gamma_{-}}\quad \ra
\eeq
\bq
\frac{\mhp}{\mgl}\sim r\ll 1\,,\quad\quad \frac{\mhp}{\la}\sim\Bigl
(\, r\,\Bigr
)^{\nl/N_c}\Biggl (\frac{\mh}{\la}\Biggr )^{(N_F-N_c)/N_c}\gg
\frac{\lym}{\la}\,.\label{(9.0.4)}
\eq

After integrating out these $\oq\,,\sq$ quarks as heavy ones at
$\mu<\mhp$, we are
left with the $SU(\hat N_c)$ SYM theory in the strong-coupling branch
${\hat a}_{SYM}(\mu=\mhp)\gg 1$ (and pions). The scale factor of its
gauge coupling ${\hat\Lambda}_{SYM}=\langle \Lambda_{L}(\Pi_{\it
ll})\rangle$ is determined from
\bq
{\hat a}_{SYM}(\mu=\mhp)=\Biggl
(\frac{\mhp}{{\hat\Lambda}_{SYM}}\Biggr )^3={\hat
a}(\mu=\mhp)=\Biggl (\frac{\la}{\mgl}\Biggr )^{\nu_{+}}\Biggl
(\frac{\mgl}{\mhp}\Biggr )^{\nu_{-}}\ra
{\hat\Lambda}_{SYM}=\lym. \label{(9.0.5)}
\eq

Finally, after integrating out $SU(N_c-\nl)$ gluons at $\mu < \lym$,
the Lagrangian of pions looks then at $\mu < \lym$ as
\bq
\hspace*{-2mm} {\hat K}=z^+_Q\Bigl (K_{\it ll}+K_{\rm hybr}\Bigr
),\,\, {\hat W}=(N_c-\nl)\Biggl ( \frac{\la^{\bo}\mh^{\nh}}{\det
\Pi_{\it ll}}\Biggr )^{\frac{1}{(N_c-\nl)}}+\ml
{\rm Tr}\,\Pi_{\it ll}+\mh{\rm Tr}\Bigl (\Pi_{\it hl}\frac{1}{\Pi_{\it
ll}}\Pi_{\it lh}\Bigr ).
\,\,\,\label{(9.0.6)}
\eq

From this, the masses of $\Pi_{\it hl}\,,{\Pi}_{\it lh}$ and $\Pi_{\it
ll}$ are
\bq
\mu(\Pi_{\it hl})=\mu({\Pi}_{\it lh})\sim \Bigl (r
\Bigr)^{\gamma_{-}}\,\mhp\,,\quad
\mu(\Pi_{\it ll})\sim \frac{\ml}{z^+_Q}\sim r\, \mu(\Pi_{\it
hl})\,.\label{(10.7)}
\eq

To check the self-consistency, i.e. that the active $\oq\,,\sq$
quarks  are indeed in the HQ  phase and are not higgsed, we 
estimate the possible value of the gluon mass $\mgh$
\bq
\mgh^2\sim \Biggl [\,{\hat a}(\mu=\mgh)=\Biggl (\frac{\la}{\mgl}
\Biggr)^{\nu_{+}}\Biggl
(\frac{\mgl}{\mgh}\Biggr
)^{\nu_{-}}\,\Biggr]\langle{\oq}\sq\rangle_{\mu=\mgh}\,\,\,,
\label{(9.0.8)}
\eq
\bq
\frac{\langle {\oq}\sq\rangle_{\mu=\mgh}}{\la^2}\sim\,
r\,\frac{\mgl}{\la}\,\Biggl(\frac{\mgl}{\la}\Biggr
)^{\gamma_{+}}\Biggl ( \frac{\mgh}{\mgl}\Biggr )^{\gamma_{-}}\quad\ra
\quad \mgh\sim \mhp\,\,\,, \label{(9.0.9)}
\eq
as could be expected. As before, because $N_h > {\hat N}_c$, the
Konishi anomaly and the rank restriction show that $\sq$ -quarks are 
{\it not higgsed}, but confined, see  footnote \ref{(f5)}.

As was shown above in the text, in those regions of the parameter
space where an additional small parameter is available (this is
$0<\bo/N_F=(3N_c-N_F)/N_F\ll 1$ at the right
end of the conformal window, or its dual analog
$0<\bd/N_F=(2N_F-3N_c)/N_F\ll 1$
at the left end), there are {\it parametrical differences} in the
mass  spectra of direct and dual theories, and therefore they are
clearly not equivalent. In fact, this implies that even
when both $\bo/N_F\sim \bd/N_F\sim 1$, there are no reasons for 
these two theories to become exactly the same.
\footnote{\,
But to see the possible differences more clearly, it is insufficient
in this case to make rough estimates of particle masses up to
non-parametric factors $O(1)$ as has been done in this paper. One has
either to resolve the mass spectra in more detail or to calculate
some  Green's functions in both theories and to compare them.
}

Besides, as was shown in section 5, one can trace unavoidable
internal  inconsistencies of the Seiberg duality in the variant
"confinement without chiral symmetry breaking". This variant  
implies   that at $N_c<N_F<3N_c/2$ and in the chiral limit 
$m_Q\ra 0$ the direct  quarks and gluons 
form  a  large number of massive hadrons with masses
$\sim \la$\,, while {\it new light composite solitons with 
masses  $\mu_i\ll\la$ appear, described by the dual theory}.

As regards the mass spectra of the direct theory in the dynamical
scenario of this paper, their main features are as follows (for 
$\bo/N_F=O(1)$\,, see section 2 for $\bo/N_F\ll 1$).\\ 
1) In all cases considered, there is a large number of gluonia with
masses $\sim \lym=(\la^{\bo}\det m_Q)^{1/3N_c}$.\\
2) When all quark masses are equal, they are in the HQ phase (i.e.
not  higgsed but confined, the string tension $\sqrt\sigma\sim\lym
\lesssim m_Q^{\rm pole}$\,), for the whole 
interval $N_c<N_F<3N_c$\,, and hence form a large number of
various hadrons with the mass scales\,: a) $\sim m_Q^{\rm pole}\sim
\lym$ at $3N_c/2<N_F<3N_c$\,, and \, b)\,
$\sim m_Q^{\rm pole}\sim \la(m_Q/\la)^ {(N_F-N_c)/N_c}\gg \lym$ at
$N_c<N_F<3N_c/2$. {\it There are no additional lighter pions $\pi^{j}_{i}$ 
with masses $\mu_{\pi}\ll m_Q^{\rm pole}$\,, for all $N_c<N_F<3N_c$}\,.

3) The case with $\nl$ flavors of smaller masses $\ml$ and
$\nh=N_F-\nl$ flavors
with larger masses $\mh\,,\,0<\ml<\mh\ll \la$ was also considered.
When $\nl>N_c$\,, all quarks are also in the HQ phase for all
$N_c<N_F<3N_c$\,, and form a large number of hadrons whose masses
depend on their flavor content , but there are no any additional
lighter pions also.

4) Only when $\nl<N_c$\,, the {\it l} - flavored quarks $Q_{\it
l}\,,{\ov Q}^{\it
\ov l}$ are higgsed, $SU(N_c)\ra SU(N_c-\nl)$, and there $\nl^2$
lighter pions $\pi^{\it
l}_{\it l}$ appear, while the heavier {\it h} - flavored quarks
$Q_{\it h}\,,{\ov Q}^{\it h}$ always
remain in the HQ phase . In this case, the mass spectra and some new
regimes with unusual properties of the RG flow were presented in
sections 7-9.\\

We have described above not all possible regimes, but only those that
reveal some  qualitatively new features. Some additional details can be 
found in  \cite{ch3}.  Besides, we hope that if needed, a reader can deal 
with other regimes  using the methods  from this review. \\

On the whole, the mass spectra of the direct ${\cal N}=1$ SQCD with
$SU(N_c)$ colors and $N_c<N_F<3N_c$ flavors of massless or light 
quarks  and its   Seiberg's dual variant  with $\nd=N_F-N_c$ dual colors 
have been calculated and compared in the framework of simple 
dynamical  scenario \cite{ch3} described in Introduction.

{\bf I).}\,\,In the region $N_c < N_F < 3N_c/2$, as was shown in section 7 of 
\cite{ch1},  {\it the two-point correlators of conserved currents are different \, in the 
direct and  dual theories  in the range of energies $\mu_H < \mu < \la \, $where all  
light  quarks in both theories are  effectively massless}.

{\bf II.)}\,\, {\it The parametrical differences in the mass spectra of the
direct and  dual theories were traced in the whole range $N_c < N_F < 3N_c$ 
for quarks with nonzero  masses}. \\

Let us emphasize that, {\it in all cases considered, no internal
inconsistences or contradictions with previously proven results were 
found  in calculations of mass spectra both of the
direct and dual theories within the framework of used scenario from 
\cite{ch3} }. Let us recall also that {\it this dynamical scenario satisfies all
those tests which were used as checks of the Seiberg hypothesis about 
the equivalence of the direct and dual theories}. This shows, in particular, 
that {\it all these tests, although necessary, may well be insufficient}.\\

\numberwithin{equation}{section}
\numberwithin{equation}{subsection}
\numberwithin{equation}{subsubsection}

\section{\bf \large Phase transitions between confinement and higgs phases in ${\cal N}=1$ \\
$\mathbf{{\cal N}=1\,\, SU(N_c)}$ SQCD with $\mathbf{1\leq N_F\leq N_c-1}$   quark flavors}.

\vspace*{1mm}

Considered in this section is the standard ${\cal N}=1 \, SU(N_c)$ SQCD 
with $1\leq N_F\leq N_c-1$ equal mass quark flavors in the fundamental
representation. {\it The gauge invariant order parameter} $\rho$ was 
introduced in \cite{ch21} distinguishing confinement (with $\rho=0$) and 
higgs (with  $\rho\neq 0$) phases.

Using a number of independent arguments for different variants of
transition between the confinement and higgs regimes, it was shown in
\cite{ch21}  that transitions between these regimes are not crossovers but 
the phase transitions.  Besides, it was argued that these phase transitions 
are of the first  order.

This is opposite to the conclusion of the E. Fradkin and S.H. Shenker
paper \cite{FS} that the transition between the confinement and higgs
regimes is the crossover, not the phase transition. And although the 
theories considered in \cite{ch21}  and in \cite{FS} are different, an 
experience shows that there is a  widely spread opinion
that the conclusion of \cite{FS} is applicable to all QCD-type
theories: both lattice and continuum, both not supersymmetric and
supersymmetric. This opinion is in contradiction with the results of
the paper \cite{ch21}.

\numberwithin{equation}{subsection}

\subsection{Introduction}

\hspace*{4mm} Considered in \cite{ch21} was the standard ${\cal N}=1$
SQCD with $SU(N_c)$ colors and $1\leq N_F\leq N_c-1$ flavors of equal mass
quarks with the mass parameter $m_Q=m(\mu=\la)$ in the Lagrangian, where 
$\la$ is the scale factor of the gauge coupling in the UV region, see e.g. 
section 2.1.

The purpose was to show that, in this theory with light quarks with
fixed $m_Q\ll\la$, there is {\it the phase transition} from the
region of not too large $N_c$ where all quarks are higgsed with
$\mu_{\rm gl}\gg\la$ \eqref{(10.2.1)}, with the gauge invariant order
parameter $\rho_{\rm higgs}\neq 0$, to the region of sufficiently
large $N_c$ \eqref{(10.3.1)} where $\rho_{HQ}=0$ and all quarks are in 
the HQ (heavy quark) phase and not higgsed but confined.

Besides, it was shown that, at fixed $N_c$, there is {\it the phase
transition} from the region $m_Q\gg\la$ where all heavy quarks are 
not higgsed but  confined and
$\rho_{HQ}=0$, to the region of sufficiently small $m_Q\ll\la$ where
they all are higgsed with $\rho_{\rm higgs}\gg\la$.

And finally, using {\it independent arguments based on realization of
the global flavor symmetry $SU(N_F)$}, it was shown that, at fixed
$N_c$ and $N_F=N_c-1$, there is
{\it the phase transition} from the region $m_Q\gg\la$ where all
quarks are not higgsed but confined to the region of sufficiently
small $m_Q\ll\la$ where they all are higgsed and not confined.\\

For all this, let us recall first in short some properties of the
standard ${\cal N}=1$ SQCD with $SU(N_c)$ colors and $1\leq N_F 
<3 N_c$ flavors of light equal mass quarks, see e.g. section 2. It is 
convenient to start e.g. with $3 N_c/2 < N_F < 3N_c$
and the scale $\mu=\la$. The Lagrangian looks as
\bq
K={\rm Tr}\,\Bigl (Q^\dagger e^{V} Q+ (Q\ra {\ov Q}) \Bigr )\,, \quad {\cal
W}=\frac{2\pi}{\alpha(\mu=\la)} S+m_Q {\rm Tr}\,({\ov Q} Q)\,.
\label{(10.1.1)}
\eq
\hspace*{-2 mm} Here: $m_Q=m_Q(\mu=\la)$ is the mass parameter 
(taken as real positive), $S=\sum_{A,\beta}
W^{A,\,\beta}W^{A}_{\beta}/32\pi^2$, where $W^A_{\beta}$ is the
gauge field strength, $A=1...N_c^2-1,\, \beta=1,2$,\, $a(\mu)=N_c
g^2(\mu)/8\pi^2=N_c\alpha(\mu)/2\pi$ is the gauge coupling with its
scale factor $\la$. Let us take now $m_Q\ra 0$ and evolve to the UV
Pauli-Villars (PV) scale $\mu_{PV}$ to define the parent UV theory.
The only change in comparison with \eqref{(10.1.1)} will be the
appearance of the corresponding logarithmic renormalization factor
$z(\la, \mu_{PV})\gg 1$  for {\it massless} quarks  in the Kahler term 
and the logarithmic evolution of the gauge coupling: $\alpha(\mu=\la)
\ra\alpha(\mu=\mu_{PV})\ll \alpha(\mu=\la)$, while the scale factor
$\la$ of the gauge coupling remains the same. Now, the parameter
$m_Q$  is continued from zero to some nonzero value, e.g. $0 <
m_Q\ll\la$. And this will be {\it a definition} of the parent UV theory.

The Konishi anomaly \cite{Konishi} for this theory looks as
(everywhere below the repeated indices are summed)
\bbq
\quad m_Q(\mu)=z^{-1}_Q(\la, \mu)m_Q\,,\quad m_Q\equiv
m_Q(\mu=\la)\,,\quad
M^i_j(\mu)= z_Q(\la, \mu) M^i_j\,,\quad M^i_j\equiv M^i_j
(\mu=\la)\,.\eeq
\bq
m_Q(\mu)\langle M^i_j(\mu)\rangle=\delta^i_j \langle S\rangle\,,\quad
i,j=1...N_F\,,\quad \langle M^i_j\rangle=\langle{\ov Q}^\beta_j
Q^i_\beta\rangle=\delta^i_j\langle M\rangle\,,\quad \alpha=1...N_c\,.
\label{(10.1.2)}
\eq
Evolving now to lower energies, the regime is conformal at $m^{\rm
pole}_Q < \mu<\la$ and the perturbative pole mass of quarks looks as
~\footnote{
\, Here and below the perturbatively exact NSVZ $\beta$-function
\cite{NSVZ-1} is used. 

In \eqref{(10.1.3)} and below $A\sim B$ means equality up to a
constant factor independent of $m_Q$ and $A\ll B$ means $|A|\ll |B|$.
}
\bq
m^{\rm pole}_Q=\frac{m_Q}{z_Q(\la,m^{\rm pole}_Q )}\sim \la\Bigl
(\frac{m_Q}{\la}\Bigr )^{\frac{N_F}{3N_c}}\ll\la\,, \label{(10.1.3)}
\eq
\bbq
z_Q(\la, \mu\ll\la)\sim\Bigl (\frac{\mu}{\la} \Bigr )^{\gamma_Q^{\rm
conf}=\frac{3N_c-N_F}{N_F}}\ll 1.
\eeq
Integrating then inclusively all quarks as heavy at $\mu < m^{\rm
pole}_Q$ \eqref{(10.1.3)},
\footnote{
It is seen from the Konishi anomaly \eqref{(10.1.2)}  that 
{\it the global flavor symmetry $SU(N_F)$ is not broken spontaneously 
in  ${\cal N}=1$\, $SU(N_c)$ SQCD
for equal mass quarks}. Therefore, due to the rank restriction at
$N_F> N_c$, all quarks are not higgsed but confined.
}
there remains $SU(N_c)$ SYM with the scale factor $\lym$
\eqref{(10.1.4)} of its
coupling. Integrating then all gluons via the Veneziano-Yankielowicz
(VY) procedure \cite{VY}, one obtains the gluino condensate, see
\eqref{(2.1.6)}, \eqref{(2.1.7)}
\bq
\lym=\Bigl (\la^{3N_c-N_F}m^{N_F}_Q \Bigr )_{,}^{\frac{1}{3N_c}}\quad
\langle S\rangle=\lym^3=\Bigl (\la^{3N_c-N_F}m^{N_F}_Q \Bigr
)^{\frac{1}{N_c}}=m_Q\langle M \rangle\, \label{(10.1.4)} 
\eq
\bbq
\langle M^i_j \rangle=\delta^i_j M\,,\quad  M=\la^2\Bigl
(\frac{m_Q}{\la} \Bigr )^{\frac{N_F-N_c}{N_c}}\,.
\eeq

Now \eqref{(10.1.4)} can be continued to $1 \leq N_F < N_c$
considered  in this section.\\

\subsection{The Higgs phase}

In this range $1 \leq N_F < N_c$, the weak coupling Higgs phase at
$\mu_{\rm gl}\gg\la$ for light quarks with $0 < m_Q\ll\la$ looks as
follows, see e.g. section 2 in \cite{ch1}. All quarks are higgsed
i.e.form a constant coherent condensate in a vacuum state, at the high
scale
$\mu\sim\mu_{\rm gl}\gg\la$ in the logarithmic weak coupling regime.
~\footnote{
\,We ignore from now on for simplicity all logarithmic factors and
trace only the power dependence on $m_Q/\la$ and $N_c$.
}
And the perturbative pole masses of $N_F(2N_c-N_F)$ massive gluons
look as
\bbq
\Bigl (\frac{\mu_{\rm gl}}{\la}\Bigr )^2\sim g^2(\mu=\mu_{\rm
gl})\,z_Q(\la,\mu=\mu_{\rm gl})\, \frac{\rho_{\rm higgs}^2}
{\la^2}\sim\frac{1}{N_c}\frac{\langle M\rangle}{\la^2}\sim
\eeq
\bq
\sim \frac{1}{N_c}\frac{\langle S\rangle}{m_Q\la^2}\sim \frac{1}{N_c}\Bigl
(\frac{\la}{m_Q} \Bigr )^{\frac{N_c-N_F}{N_c}}\gg 1\,.
\label{(10.2.1)}
\eq
\bbq
\quad g^2(\mu=\mu_{\rm gl})\approx \frac{8\pi^2}{(3N_c-N_F)\ln\Bigl
(\mu_{\rm gl}/\la\Bigr )}\sim \frac{1}{N_c}\,,\quad z_Q(\la,\mu=\mu_{\rm
gl})\sim\Bigl (\ln\frac{\mu_{\rm gl}}{\la}\Bigr )^{\frac{N_c}{3N_c-N_F}}\sim
1\,.
\eeq

Higgsing of all $N_F$ quarks with $1\leq N_F\leq N_c-2$ flavors at
$\rho_{\rm higgs}=\la\Bigl (\frac{\la}{m_Q} \Bigr
)^{\frac{N_c-N_F}{2N_c}}\gg\la$\eqref{(10.2.1)} breaks spontaneously
separately the global $SU(N_F)$
and global $SU(N_c)$ down to $SU(N_c-N_f)$, but there remains
unbroken  diagonal $SU(N_F)_{C+F}$ global symmetry. Besides, the gluons
from remained $SU(N_c-N_F)$ SYM do not receive large masses
$\sim\rho_{higgs}$ and remain (effectively) massless at scales
$\mu>\lym$.\\

Dealing with higgsed quarks, to obtain \eqref{(10.2.1)}, we first
separate out the  Goldstone fields from quark fields $Q^i_{\alpha}$ 
normalized at the  scale $\la$ (a  part of these Goldstone fields or all 
of them will be eaten by gluons  when quarks are higgsed)
\bq
Q^i_{\alpha}(x)=\Bigl (V_{\rm Goldst}^{SU(N_c)}(x)\Bigr
)^{\beta}_{\alpha}\,{\hat Q} ^i_{\beta}(x)\,,\quad {\hat
Q}^i_{\beta}(x)=\Bigl (V_{\rm Goldst}^{SU(N_c)}(x)^\dagger \Bigr
)^\gamma_\beta Q^i_{\gamma}(x)\,, \label{(10.2.2)}
\eq
\bbq
{\hat Q}^i_{\beta}(x)=\Bigl ( U_{\rm global}^{SU(N_c)}\Bigr
)^{\delta}_{\beta}
\Bigl ( U_{\rm global}^{SU(N_F)}\Bigr )^i_j {\tilde Q}^j_{\delta}(x),
\quad \alpha, \beta, \gamma, \delta=1...N_c,\,\, i, j=1...N_F,
\eeq
where $V_{\rm Goldst}^{SU(N_c)}(x)$ is the $N_c\times N_c$ unitary
$SU(N_c)$  matrix of Goldstone fields.

But the physical degrees of freedom of massive scalar superpartners
of \, massive gluons and light pion fields $\Pi^i_j$ \eqref{(2.1.13)},
originating \, from combined physical degrees of freedom of $Q$
 and  ${\ov Q}$,
remain in ${\hat Q}^i_{\beta}(x)$ and ${\hat {\ov Q}_i^{\,\beta}(x)}$.

This can be checked by direct counting. The quark fields $Q$ and ${\ov
Q}$ have  $4 N_F N_c$ real physical degrees of freedom on the whole. From
these,  $N_F(2N_c-N_F)$ real Goldstone degrees of freedom are eaten by
massive  gluons. (The extra $(N_c-N_F)^2-1$ real Goldstone modes not
eaten by remaining perturbatively massless $SU(N_c-N_F)$ gluons remain 
not physical due to the gauge invariance of the Lagrangian). The same
number $N_F(2N_c-N_F)$ of combined real physical degrees of freedom
of $Q$ and ${\ov Q}$ form scalar superpartners of massive gluons. And remaining
$2 N_F^2$ combined real degrees of freedom of $Q$ and ${\ov Q}$ form
$N_F^2$ complex physical degrees of freedom of light colorless pions $\Pi^i_j$.

And then, with the standard choice of vacuum of spontaneously broken
global symmetry, we replace ${\hat Q}^i_{\beta}(x)$ in \eqref{(10.2.2)},
containing remained degrees of freedom, by its mean vacuum value 
(at $\mu=\la$)
\bq
\langle {\hat Q}^i_{\beta}(x)\rangle =\langle {\hat
Q}^i_{\beta}(0)\rangle=\delta^i_{\beta}\,\rho_{\rm higgs}, \,\,
\frac{\rho_{\rm higgs}}{\la}=\Bigl (\frac{\la}{m_Q}\,\Bigr
)^{\frac{N_c-N_F}{2 N_c}}\gg 1,\,\, i=1...N_F,\,\, \beta=1...N_c\,.
\label{(10.2.3)}
\eq
And similarly $\langle {\hat {\ov
Q}_i^{\,\beta}(x)}\rangle=\delta_i^{\,\beta}\,\rho_{\rm higgs}$.\\

Under pure gauge transformations, see \eqref{(10.2.2)}\,:
\bq
Q^i_\alpha (x)\ra \Bigl ( V_{\rm pure\, gauge}^{SU(N_c)}(x)\Bigr
)^\beta_\alpha
Q^i_\beta (x),\quad V_{\rm Goldst}^{SU(N_c)}(x)\ra V_{\rm pure\,
gauge}^{SU(N_c)}(x) V_{\rm Goldst}^{SU(N_c)}(x). \label{(10.2.4)}
\eq
That is, these are $Q^i_\alpha (x)$ and Goldstone fields which are
transformed in \eqref{(10.2.2)},\eqref{(10.2.4)}, while ${\hat
Q}^i_{\beta}(x)$ stays intact under pure gauge transformations and is
the gauge invariant quark field. And $\rho_{\rm higgs}\neq 0$ in
\eqref{(10.2.3)} is the {\it gauge invariant order parameter} for
higgsed scalar quarks, while
$\rho_{\rm HQ}=0$ if quarks are in the HQ (heavy quark) phase and not
higgsed, see \eqref{(11.1.1)},\eqref{(11.1.2)}. ( And, in particular, $V_{\rm
pure\, gauge}^{SU(N_c)}(x) V_{\rm Goldst}^{SU(N_c)}(x)=I_{SU(N_c)}$
in  the so called "unitary gauge"\, where $I_{SU(N_c)}$ is the unity matrix).\\

Under the replacement $Q\ra V_{\rm Goldst}^{SU(N_c)}(x){\hat Q}$
\eqref{(10.2.2)}, the covariant derivative $i
D_{\nu}(A)Q=(i\partial_{\nu}+A_{\nu})Q$
is replaced by $V_{\rm Goldst}^{SU(N_c)}(x) i D_{\nu}(B){\hat Q}$, with
$$B_{\nu}(x)=\Bigl [ \Bigl ( V_{\rm Goldst}^{SU(N_c)}(x)\Bigr )^{\dagger}
A_{\nu}(x)
V_{\rm Goldst}^{SU(N_c)}(x)+i\Bigl ( V_{\rm Goldst}^{SU(N_c)}(x)\Bigr
)^{\dagger} \\
\partial_{\nu} V_{\rm Goldst}^{SU(N_c)}(x)\Bigr ]\,.$$ \\
Now, the fields ${\hat Q}$ and $B_{\nu}$ \, are invariant under $SU(N_c)$ pure
gauge \, transformations \eqref{(10.2.4)}.

When all $N_F$ quarks $Q$ and $\ov Q$ are higgsed, $N_F(2N_c-N_F)$
Goldstone modes in $V_{\rm Goldst}^{SU(N_c)}(x)$ \eqref{(10.2.2)} are eaten 
by gluons \, .The remaining light $SU(N_c-N_f)$ SYM gluon fields $B_{\nu}$
are gauge invariant with respect to the original $SU(N_c)$ gauge
transformations.  But there appears the emergent standard $SU(N_c-N_F)$ 
gauge invariance of the  lower energy $SU(N_c-N_F)$ SYM Lagrangian 
$L_{SYM}(B_{\nu})$ written in  terms of fields $B_{\nu}(x)$.

The gauge invariant pole masses of massive gluons are as in
\eqref{(10.2.1)}.\\

{\it The gauge invariant order parameter $\rho_{\rm higgs}\neq 0$ in
\eqref{(10.2.3)} is the counter-example to a widely spread opinion
that the gauge invariant order parameter for higgsed scalar quarks in
the fundamental representation does not exist}. Besides, as
pointed out in section (6.2) in \cite{ch17}, the attempt to use as
the  gauge invariant order parameter the mean vacuum value of the {\it
colorless} composite operator $\langle M\rangle^{1/2}$
\eqref{(2.1.2)}, instead of the gauge invariant but {\it colorful}
$\langle {\hat Q}^i_{\beta}\rangle=\delta^i_{\beta}\rho$, is erroneous.
The reason is that $\langle M\rangle^{1/2} > 0$ \eqref{(2.1.2)} is small but 
nonzero due to quantum loop and nonperturbative effects even for heavy 
quarks with
$m_Q\gg\la,\,\,N_F<N_c$, see e.g. \eqref{(2.1.8)},\eqref{(2.1.9)}. Such quarks 
are in the HQ(=heavy quark)-phase and {\it they are not higgsed really, i.e.
$\rho_{\rm HQ}=0,\,\,\rho_{\rm HQ}\neq \langle M\rangle^{1/2}$, see
\eqref{(11.1.1)},\eqref{(11.1.2)}}. Or e.g., using in lattice
calculations for QCD-type not supersymmetric theories with {\it heavy
scalar not higgsed quarks} the mean vacuum value of the {\it
colorless} gauge invariant composite operator $"V"=\langle\,
\sum_{\alpha=1}^{N_c}\sum_{i=1}^{N_F} (\phi^\dagger)^\alpha_i
\phi^i_\alpha\,\rangle^{1/2} > 0$ as the order parameter (instead of
$\langle {\hat \phi}^{\,i}_\alpha\rangle=0$ for such heavy confined
quarks). This is also misleading because $"V"\neq 0$ in all regimes
due to various quantum effects.

Using $\langle M\rangle^{1/2}$ \eqref{(2.1.2)} or $"V"$ as order
parameters creates an illusion that the transition between the confinement 
and higgs regimes is the analytic crossover, while it is really the
non-analytic \, phase transition .

Unlike the {\it analytical} dependence of mean vacuum values of
lowest  components of {\it colorless} chiral superfields, e.g.
$\sum_{\beta=1}^{N_c}\langle{\ov Q}^\beta_j
Q^i_\beta\rangle=\delta^i_j M(N_c, N_F, m_{Q,i})$, on chiral
parameters of the superpotential, the mean vacuum values of lowest 
components of gauge invariant but
{\it colorful} chiral superfields, e.g. $\langle{\hat
Q}^i_\beta\rangle=\delta^i_{\beta}\,\rho(N_c, N_F, m_{Q, i})$, depend
{\it non-analytically} on these parameters. E.g., $\rho$ is nonzero
at  fixed $N_c$ and sufficiently light quarks but zero for either
sufficiently heavy quarks, or for light quarks and sufficiently large
$N_c$. In general, the gauge invariant order parameter $\langle{\hat
Q}^i_\beta\rangle=\delta^i_{\beta}\,\rho$ is nonzero if quarks are
higgsed and is zero if they are not higgsed, independently of whether
they are confined or not.

That the order parameter is $\langle {\hat Q}^i_{\beta}\rangle$ and
not $\langle M\rangle^{1/2}$ is especially clearly seen in D-terms of fermions 
of the Lagrangian \eqref{(2.1.1)}\,: $\Bigl\{
(Q^\dagger)_i^\beta\,\lambda_\beta^\gamma\,\chi^i_\gamma\,\, + {\rm
h.c.}\Bigr\}+(Q\ra {\ov Q})$. The nonzero mass term of fermions
(superpartners of massive bosons due to higgsed quarks) expressed in
terms gauge invariant fields with hats, see
\eqref{(10.2.2)},\eqref{(10.2.3)}, looks then as\,: $\sim\Bigl\{ [\,\langle\,
({\hat Q}^\dagger)_i^\sigma\,\rangle=\delta_i^{\sigma}\,\rho_{\rm
higgs}\neq0\,]\,{\hat\lambda}_\sigma^\tau\,{\hat\chi}^{\,i}_\tau\,\,
+{\rm
h.c.}\Bigr \}+({\hat Q}\ra {\hat{\ov Q}}),\,\, {\hat\lambda}=\Bigl (
V^{SU(N_c)}_{\rm Goldst}(x)\Bigr )^{\dagger}\lambda \Bigl (
V^{SU(N_c)}_{\rm Goldst}(x)\Bigr )$, where $\chi$ is the fermionic
superpartner of $Q$.\\

At $1\leq N_F\leq N_c-2$, due to higgsed quarks, $N_F(2N_c-N_F)$
gluons and the same number of their ${\cal N}=1$ superpartners acquire 
masses $\mu_{\rm
gl}\gg\la$ and decouple at $\mu<\mu_{\rm gl}$. There remain at lower
energies local ${\cal N}=1$\,\, $SU(N_c-N_F)$ SYM and $N^2_F$ light
complex pion fields
$\Pi^i_j(x)\,:\,\,M^i_j(x)=\delta^i_j\langle M\rangle
+\Pi^i_j(x),\,\, \langle\Pi^i_j(x)\rangle=0,\,\, i,j=1...N_F$. After
integrating out all heavy particles with masses $\sim \mu_{\rm gl}\gg\la$, 
the scale factor of $SU(N_c-N_F)$ SYM looks as, see section 2 in \cite{ch1} 
and \eqref{(2.1.6)}
\bq
\Lambda^3_{SYM}=\Bigl (\frac{\la^{3N_c-N_F}}{\det M} \Bigr
)^{\frac{1}{N_c-N_F}},\quad M^i_j=\langle M^i_j\rangle(\mu=\la)+\Pi^i_j\,.
\label{(10.2.5)}
\eq

Lowering energy to $\mu\sim \Lambda_{SYM}$ and integrating all
$SU(N_c-N_F)$ gluons via the VY procedure \cite{VY}, the Lagrangian
of $N^2_F$ light  pions $\Pi^i_j$ looks as
\footnote{
\,The whole $SU(N_c)$ group is higgsed at $N_F=N_c-1$ and all
$N_c^2-1$ gluons are heavy. {\it There is no confinement}. The last 
term in the superpotential
\eqref{(10.2.6)} is then due to the instanton contribution
\cite{ADS}.  For $1\leq N_F\leq N_c-2$ the instanton contribution to
superpotential  from the broken part of $SU(N_c)$ is zero due to extra
gluino zero modes. The nonperturbative term in the superpotential
\eqref{(10.2.6)} originates from nonperturbative effects in the ${\cal
N}=1\,\, SU(N_c-N_F)$ SYM, see \cite{VY}, section 2 in \cite{ch1} and 
section 2 above.
}
\bq
K_M=2\,z_Q(\la,\mu=\mu_{\rm gl}) {\rm Tr}\, \sqrt{M^\dagger
M}\,,\quad{\cal
W}_{\Pi}= m_Q{\rm Tr}\,M +(N_c-N_F)\Bigl (\frac{\la^{3N_c-N_F}}{\det
M} \Bigr
)^{\frac{1}{N_c-N_F}},  \label{(10.2.6)}
\eq
where $z_Q(\la,\mu=\mu_{\rm gl}\gg\la)\gg 1$ is the quark logarithmic
renormalization factor.

From this, $\langle M^i_j\rangle$ and the pion masses are
\bq
\langle M^i_j \rangle=\delta^i_j \la^2\Bigl (\frac{\la}{m_Q} \Bigr
)^{\frac{N_c-N_F}{N_c}}\,,\quad\mu^{\rm pole}(\Pi)=\frac{
2 m_Q}{z_Q(\la,\mu=\mu_{\rm gl})}\ll\lym\ll\la\,. \label{(10.2.7)}
\eq

On the whole. All quarks are higgsed and the mass spectrum at $1\leq
N_F\leq N_c-2$ looks as follows.\, a) $SU(N_F)_{\rm adj}\,\,{\cal N}=1$
multiplet of heavy
not confined gluons with the mass \eqref{(10.2.1)};\, b) one heavy
${\cal N}=1$ multiplet of $SU(N_F)_{\rm singl}$ not confined gluon
with the mass of the same order \eqref{(10.2.1)};\, c)
$2N_F(N_c-N_F)$\, ${\cal N}=1$ multiplets of heavy $SU(N_F)\times
SU(N_c-N_F)$ bifundamental gluons (hybrids) with masses
$\sim\eqref{(10.2.1)}$ , which behave as quarks with $N_F$ flavors
with respect to confining them not higgsed by quarks ${\cal
N}=1$ $SU(N_c-N_F)$ SYM and are weakly coupled and weakly 
confined, \, d) a number of ${\cal N}=1$ $SU(N_c-N_F)$ SYM 
gluonia with the typical mass scale ${\cal O}(\lym)\ll\la$ \eqref{(2.1.6)}
(except for the case $N_f=N_c-1$);\, e) $N^2_F$ light colorless complex pions
$\Pi^i_j$ with masses $\sim m_Q\ll\lym$ \eqref{(10.2.7)}.

\subsection{The heavy quark (HQ) phase}

It is seen from \eqref{(10.2.1)}  that at $\mu\gg\la$
the value of the running gluon mass $\mu_{\rm gl}(\mu)\gg\la$ 
decreases with increasing $N_c$ and fixed
$(m_Q/\la)\ll 1$.  And at sufficiently large number of colors,
\bq
\frac{N_c}{N_c-N_F}\ln (N_c) \gg \ln(\frac{\la}{m_Q})\gg 1\,,
\label{(10.3.1)}
\eq
$\mu_{\rm gl}(\mu\sim\la)$ will be much smaller than $\la$. This
means  that even quarks with large $(\rho_{\rm higgs}/\la)=\Bigl 
(\la/m_Q\,\Bigr )^{(N_c-N_F)/2 N_c}\gg 1$ are not higgsed then 
in the weak coupling regime at
$\mu\gg\la$. And now, at such $N_c$ \eqref{(10.3.1)}, {\it all quarks
and gluons will remain effectively massless in
some interval of scales} $\mu_H < \mu < \la$. Recall also that
considered ${\cal N}=1$ SQCD is outside the conformal window at $N_F
<3N_c/2$ \cite{S1}. Therefore, to see whether quarks are really able
to \, give by higgsing such  mass to gluons which will stop the 
perturbative  massless RG-evolution,  we have to
consider the region $\mu\ll\la$ where the theory entered smoothly 
into a {\it  perturbative strong coupling regime} with
$a(\mu\ll\la)=N_c\alpha(\mu)/2\pi\gg 1$.

Let us recall a similar situation at $N_c < N_F < 3N_c/2$ considered in 
section 7 of \cite{ch1} (only pages 18 - 21 including the footnote 18 in
arXiv:0712.3167\, [hep-th]). As pointed out therein, when decreasing scale 
$\mu$ crosses \, $\mu\sim\la$ from above, the increasing perturbative coupling
$a(\mu)$  crosses unity from below. But for (effectively) massless quarks and 
gluons the perturbatively exact NSVZ$\beta$-function \cite{NSVZ-1}
\bq
\frac{d a(\mu)}{d\ln \mu} = \beta(a)=-\, \frac{a^2}{1-a}\,
\frac{(3N_c-N_F)-N_F\gamma_Q(a)} {N_c},\quad a(\mu)=\frac{N_c
g^2(\mu)}{8\pi^2}=\frac{N_c\alpha(\mu)}{2\pi}\, \label{(10.3.2)}
\eq
{\it cannot change its sign by itself (and cannot become frozen at zero
outside the conformal window) and behaves smoothly}. I.e., when 
increased $a(\mu)$ \, crosses
unity from below and denominator in \eqref{(10.3.2)} crosses zero,
the  increased quark anomalous dimension $\gamma_Q(\mu)$ crosses
$(3N_c-N_F)/N_F$ from below, so that the $\beta$-function behaves 
smoothly and remains negative at $\mu < \la$.  The coupling 
$a(\mu\ll\la)$   continues to increase with decreasing $\mu$
\vspace*{-2mm}
\bbq
\frac{d a(\mu)}{d\ln \mu} = \beta(a)\ra - \nu\, a< 0,\,\, \nu=\Bigl
[\frac{N_F}{N_c}(1+\gamma^{\rm str}_Q)-3\Bigr ]={\rm const} > 0,
\eeq
 \bq 
a(m^{\rm pole}_Q\ll\mu\ll\la)\sim\Bigl (\frac{\la}{\mu} \Bigr )^{\nu\, >\, 0}\gg 1.
\,\,\label{(10.3.3)}
\eq
In section 7 of \cite{ch1} (see also \cite{ch3},\cite{Session}) the
values
$\gamma^{\rm str}_Q=(2N_c-N_F)/(N_F-N_c) >
1,\,\,\nu=(3N_c-2N_F)/(N_F-N_c) > 0$ at
$\mu\ll\la$ and $N_c < N_F < 3N_c/2$ have been found from matching of
definite two point correlators in the direct $SU(N_c)$ theory and in
$SU(N_F-N_c)$ Seiberg's dual \cite{S2}. In our case here with $1 \leq
N_F < N_c$ the dual theory does not exist.  So that, unfortunately, we 
cannot find  the concrete value $\gamma^{\rm str}_Q$.  (Nevertheless, 
the behavior $\gamma^{\rm str}_Q\ra N_c/|N_c-N_F|$ at
$N_F\ra N_c$ looks appropriate).  But, as will be  
shown below, for our purposes it will be 
sufficient to have the only  condition $\nu > 0$ in  \eqref{(10.3.3)}. 

Let us look now whether at large $N_c$ \eqref{(10.3.1)} a
potentially possible higgsing of quarks, even with large 
$(\rho_{\rm   higgs}/\la)=\Bigl (\la/m_Q\,\Bigr
)^{(N_c-N_F)/2 N_c}\gg 1$, can give gluons such mass which will
stop  the perturbative massless RG-evolution. At large $N_c$
\eqref{(10.3.1)}, such running gluon mass would look
at $\mu\ll\la$ as, see  \eqref{(10.2.1)},\eqref{(10.3.3)},
\eqref{(10.3.6)},\eqref{(10.3.7)},
\bq
\frac{\mu^2_{\rm gl}(\mu\ll\la, N_c)}{\mu^2}\sim
\frac{a(\mu\ll\la)}{N_c}\,
z_Q(\la,\mu\ll\la)\,\frac{\rho_{\rm higgs}^2}{\mu^2}\sim
\,\,\label{(10.3.4)}
\eq
\bbq
\Bigl (\frac{\mu}{\la} \ll 1\Bigr )^{\Delta >\, 0 }\Bigl
[\frac{1}{N_c}\Bigl
(\frac{\la}{m_Q}\Bigr )^{\frac{N_c-N_F}{N_c}}\ll 1 \Bigr ] \ll 1,
\eeq
\bbq
\Delta=\frac{N_c-N_F}{N_c}(1+\gamma^{\rm str}_Q) > 0\,,\quad
z_Q(\la,\mu\ll\la)\sim \Bigl (\frac{\mu}{\la}\Bigr )^{\gamma^{\rm
str}_Q\,>\,2}\ll
1\,,\quad {\rm at}\quad m^{\rm pole}_Q < \mu \ll
\la\,,
\eeq
\bbq
\hspace*{-1.5mm}\frac{\mu^2_{\rm gl}(\mu < m^{\rm pole}_Q,
N_c)}{\mu^2}\sim
\frac{a^{(str, pert)}_{SYM}(\mu < m^{\rm pole}_Q)}{N_c} z_Q(\la,
m^{\rm pole}_Q)\,\frac{\rho_{\rm higgs}^2}{\mu^2}\sim
\eeq
\bq
\sim\frac{1}{N_c} \frac{\mu}{m^{\rm pole}_Q}\ll 1,\, {\rm at}\,\,
\lym< \mu < m^{\rm pole}_Q. \,\,\label{(10.3.5)}
\eq

It is seen from \eqref{(10.3.4)},\eqref{(10.3.5)} that, at fixed
$(m_Q/\la)\ll 1$ and large $N_c$ \eqref{(10.3.1)}, even with large 
$(\rho_{\rm higgs}/\la)=\Bigl
(\la/m_Q\,\Bigr )^{(N_c-N_F)/2 N_c}\gg 1$ \eqref{(10.2.3)} the
potentially possible gluon masses become too small. I.e., with
increasing $N_c$, in some vicinity $\mu\sim\la$ the gluon masses
become $\mu_{\rm gl}(\mu\sim\la) < \la$. And this inequality is only
strengthened with decreasing $\mu:\,\,\mu_{\rm gl}(\mu\ll\la)\ll 
\mu$.  And so,  {\it the potentially possible gluon mass terms in the
Lagrangian from still higgsed quarks become too small and dynamically
irrelevant. The gluons become effectively massless}. I.e.,
potentially  higgsed quarks become unable to give such masses to gluons
which will stop the perturbative massless RG-evolution (and there is no pole
 in the gluon propagator).

Moreover, with fixed $(m_Q/\la)\ll 1$ and increasing $N_c$, the
numerical values
of $\langle {\hat{\ov Q}}\rangle$ and $\langle {\hat Q}\rangle$
(together with the
gluon mass term in the Lagrangian from higgsed quarks) {\it drop then
to zero} at some value of increasing $N_c$, somewhere in the region
$\mu_{\rm gl}(\mu\sim\la) \sim \la$. And
remain zero at this (or larger) value of $N_c$ at smaller
$\mu\,:\,\lym < \mu\ll\la$. {\it The drop of the order parameter
$\rho$ from $\rho\neq 0$ to $\rho=0$ is the phase transition}. The
physical reason for this is that {\it when gluons become
perturbatively massless (and effectively massless because their
nonperturbative masses from SYM are small, $\sim\lym$), the physical,
i.e. path dependent, phases of colored quark fields $\hat Q$ and
${\hat{\ov Q}}$ become freely fluctuating due to interactions with
such gluons}.

And although the mean value $\langle
M\rangle=\sum_{\alpha=1}^{N_c}\langle{\ov
Q}^\alpha_1 Q^1_\alpha\rangle (\mu=\la)=\langle S\rangle/m_Q\gg\la^2$
\eqref{(10.1.4)} remains the same, {\it it becomes nonfactorizable}
because gluons become (effectively) massless and quarks become
not higgsed. And all this shows that the assumption about higgsed
quarks  with $\langle{\hat Q}\rangle=\langle{\hat{\ov Q}}\rangle\neq 0$
becomes not self-consistent at fixed $m_Q/\la\ll 1$ and sufficiently
large $N_c$ \eqref{(10.3.1)}.

This regime (i.e. the HQ-phase) with light quarks at fixed
$m_Q/\la\ll 1$ and large $N_c$ \eqref{(10.3.1)} is qualitatively the same 
as those for heavy quarks with $m_Q/\la\gg 1$, see section 11.4.1 below. 
They are also not higgsed, i.e. $\langle{\hat{\ov Q}}\rangle=\langle {\hat
Q}\rangle=0$,  but confined and decouple as heavy at $\mu < m_Q^{\rm
pole}$, in the weak coupling region where gluons are (effectively) massless. 
And  nonzero value of $\langle M\rangle=\sum_{\alpha=1}^{N_c}\langle{\ov
Q}^\alpha_1 Q^1_\alpha\rangle (\mu=\la)=\langle S\rangle/m_Q$
\eqref{(10.1.4)} is also not due to higgsed quarks but arises from the
one quark loop Konishi anomaly for quarks in the HQ (heavy quark) phase.

The meaning and properties of the operator $M^i_j$ are very different
for higgsed
or not higgsed at large $N_c$ \eqref{(10.3.1)} quarks. While
$M^i_j=[\,\delta^i_j\rho_{\rm higgs}^2=\delta^i_j
\la^2 (\la/m_Q)^{(N_c-N_F)/N_c}]+\Pi^i_j,\,\, \langle
\Pi^i_j\rangle=0$, where
$\Pi^i_j$ is the one-particle operator of the light pion for higgsed
quarks, for not higgsed quarks with
$\rho_{HQ}=0\,\, M^i_j$ is the two-particle quark operator, its mean
value $\langle M^i_j\rangle$ becomes {\it nonfactorizable} and originates
from the one quark loop Konishi anomaly, see \eqref{(10.1.2)}, 
\eqref{(10.1.4)} and section 4.1.

Therefore, let us look in this case of not higgsed quarks on the
increasing with  decreasing $\mu < \la$ running quark mass 
$m_Q(\mu < \la)$ and on  possible value of the quark perturbative pole 
mass. It looks as, see  \eqref{(10.3.3)}
\bbq 
m_Q(\mu\ll\la)=\frac{m_Q}{z_Q(\la,\mu\ll\la)}\,,\quad z_Q(\la,\mu\ll\la)
\sim \Bigl (\frac{\mu}{\la} \Bigr )^{\gamma^{\rm str}_Q\, >\, 2}\ll 1\,,
\eeq
\bq
m^{\rm pole}_Q=\frac{m_Q}{z_Q(\la,m^{\rm pole}_Q)}\quad\ra \quad
m^{\rm pole}_Q\sim\la \Bigl (\frac{m_Q}{\la} \Bigr )^{0 \, <
\frac{1}{1+\gamma^{\rm str}_Q}\, < \,\frac{1}{3}\,} \ll \la\,.
\label{(10.3.6)}
\eq
As a result, {\it all quarks are not higgsed and decouple as heavy}
at  $\mu < m^{\rm pole}_Q$. There remains at lower energies the 
${\cal N}=1\,\, SU(N_c)$ SYM {\it in the perturbative strong coupling 
branch}.   From the NSVZ $\beta$-function \cite{NSVZ-1}
\bbq
\frac{d a^{(str,\, pert)}_{SYM}(\mu\gg\lym)}{d\ln\mu}= -
\frac{3\,\Bigl (a^{(str,\, pert)}_{SYM}(\mu\gg\lym)\Bigr )
^2}{1-a^{(str,\, pert)}_{SYM}(\mu\gg\lym)}\ra 3\, a^{(str,\,
pert)}_{SYM}(\mu)\,,
\eeq
\bq
a^{(str,\, pert)}_{SYM}(\mu\gg\lym)\sim\Bigl (\frac{\mu}{\lym}\Bigr
)^3 \gg 1\,,
\quad a^{(str,\, pert)}_{SYM}(\mu\sim\lym)={\cal O}(1)\,.
\label{(10.3.7)}
\eq

The scale factor of $\lym$ of the gauge coupling is determined from
matching, see \eqref{(10.3.3)},\eqref{(10.3.6)},\\\eqref{(10.3.7)}
\bbq
a_{+}(\mu=m^{\rm pole}_Q)=\Bigl (\frac{\la}{m^{\rm pole}_Q} \Bigr
)^{\nu}=a^{(str,\, pert)}_{SYM}(\mu=m^{\rm pole}_Q)=\Bigl
(\frac{m^{\rm pole}_Q)}{\lym} \Bigr )^3\,\ra
\eeq
\bq
\ra\, \lym=\Bigl (\la^{3N_c-N_F}m^{N_F}_Q \Bigr )^{1/3N_c}\,,
\label{(10.3.8)}
\eq
as it should be, see \eqref{(10.1.4)}. Besides, as a check of
self-consistency, see \eqref{(10.3.3)},\eqref{(10.3.6)}, \eqref{(10.3.8)}
\bq
\Bigl (\frac{\lym}{m^{\rm pole}_Q}\Bigr )^3\sim\Bigl (\frac{m_Q}{\la}
\Bigr )^{\omega\, >\, 0}\ll 1\,,\quad \omega= \frac{\nu > 0}{(1+
\gamma^{\rm str}_Q)} > 0 \,. \label{(10.3.9)}
\eq
as it should be. At $\mu < \lym$ the perturbative RG evolution stops
due to  non-perturbative effects $\sim\lym$ in the pure ${\cal
N}=1\,\,\,SU(N_c)$ SYM.

On the whole, the mass spectrum at $1\leq N_F\leq N_c-1$ and large
$N_c$ \eqref{(10.3.1)} looks as follows.\, a)\, All quarks are not
higgsed (i.e. $\rho_{HQ}=0$ and their color charges are not screened
due to $\rho_{HQ}\neq 0$) but decouple as heavy at $\mu<m^{\rm
pole}_Q$ and are weakly confined. There is a number of quarkonia with
the typical mass scale ${\cal O} (m^{\rm pole}_Q) \ll\la$
\eqref{(10.3.6)}\eqref{(10.3.9)}, with different spins and other
quantum numbers. {\it This mass scale is checked by
eqs.\eqref{(10.3.8)},\eqref{(10.3.9)}. Integrating inclusively all
these hadrons (i.e. equivalently, integrating inclusively all quarks
as heavy at $\mu= m^{\rm pole}_Q$), we obtain the well known
beforehand right value of $\lym$} \eqref{(2.1.4)}.

The confinement originates only from the ${\cal N}=1\,\, SU(N_c)$ SYM
and so the typical string tension is $\sigma^{1/2}_{SYM}\sim\lym\ll m^{\rm
pole}_Q\ll\la$.
\footnote{
\, There is no confinement in Yukawa-type theories without gauge
interactions. Confinement originates {\it only} from (S)YM sector. 
The ${\cal N}=1$ SYM is the theory with only one dimensional parameter 
$\lym$. Therefore, it cannot  give a string tension $\sigma^{1/2}\sim\la$ 
but only $\sigma^{1/2}\sim\lym\ll\la$ at $m_Q\ll\la$. \label{(f17)}
}
\\

b)\,\, There is a number of $SU(N_c)$ gluonia with the typical mass
scale $\sim \lym$ \eqref{(10.3.8)}. It is seen from the above that the mass
spectra at $\mu^{\rm pert}_{\rm gl}\gg\la$ \eqref{(10.2.1)} or
$\mu^{non-pert}_{\rm gl}\sim\lym\ll\la$ at large $N_c$
\eqref{(10.3.1)} are qualitatively different.\\

Now, about a qualitative difference between the analytic crossover
and  not analytic phase transition. The gauge invariant order parameter
for  quark higgsing is $\rho_{\rm higgs}\neq 0$ \eqref{(10.2.3)}. As
pointed out below \eqref{(10.3.5)}, the perturbative mass terms of
gluons in the Lagrangian originating from higgsed quarks drop to zero
because $\rho_{HQ}=0$ drops to zero at large $N_c$ \eqref{(10.3.1)}.
This is due to freely fluctuating physical quark fields phases from
interaction with effectively massless gluons (with small
nonperturbative masses $\sim\lym$), see also \eqref{(11.1.2)}. I.e.,
{\it quarks become unhiggsed at so large} $N_c$. While $\rho_{\rm
higgs}\neq 0$ and large at not too large $N_c$ from \eqref{(10.2.1)}
because the corresponding gluons are heavy, $\mu_{\rm gl}\gg\la$.
Therefore, with fixed $(m_Q/\la)\ll 1$ and increasing $N_c$, there is
the phase transition somewhere in the region $\mu_{\rm
gl}(\mu\sim\la)\sim\la$.\\

The additional arguments for a phase transition between these two
regions of $N_c$ (as opposite to an analytical crossover) look as follows.

Let us suppose now that, for the analytical crossover instead of the
phase transition, the quarks would remain higgsed at large $N_c$
\eqref{(10.3.1)} and even with still sufficiently large ${\tilde
\rho}^{\,2}_{\rm higgs}(\mu\sim\la)\sim\la^2
(\la/m_Q)^{(N_c-N_F)/N_c}$
(i.e. ignoring all given above arguments for $\rho=0$). As can be
seen  from \eqref{(10.3.4)}, \, \eqref{(10.3.5)},\eqref{(10.3.7)},
($\mu_{0}^{2}=z_{Q}(\la, m^{\rm pole}_Q)\rho_{higgs}^2\ll\lym^2$, and
even $N_c\mu_{\rm gl}^{2}(\mu\sim\lym)/\lym^2\sim \lym/m_Q^{\rm
pole}\ll 1$), the {\it additional} effects from supposedly still
higgsed quarks will be then \, parametrically small and dynamically 
irrelevant   for the RG-evolution from
$\mu=\la/(\rm several)$ down to $\mu\sim\lym$. So, the RG-evolution
in\eqref{(10.3.4)},\eqref{(10.3.6)} will remain valid in the range
$m^{\rm pole}_Q < \mu<\la$ where all quarks and gluons are
(effectively) massless. And the RG-evolution in
\eqref{(10.3.5)},\eqref{(10.3.7)} (after quarks decoupled as heavy)
will also remain valid in the range $\lym<\mu < m^{\rm pole}_Q$ where
all gluons remain (effectively) massless. At $\mu\sim\lym$ the larger
nonperturbative effects $\sim\lym$ from ${\cal N}=1 \, \, SU(N_c)$
SYM  come into a game and stop the perturbative RG-evolution with
(effectively) massless gluons.

For the case of decoupled as heavy not higgsed quarks with
$\rho_{HQ}=0$ (as described above), a widely spread opinion
(supported  by lattice calculations) is that the confinement of massive
quarks {\it originates from higgsing (i.e. condensation) of magnetically
charged solitons in $SU(N_c)$ (S)YM}.

But then, for sufficiently heavy \eqref{(10.3.9)} but still 'slightly
higgsed' quarks giving the supposed electric mass 
$\mu_{\rm gl}^{2}(\mu\sim\lym)/\lym^2 \ll 1$ to
gluons, the regime would be self-contradictory. There would be then
in the whole ${\cal N}=1\, \,SU(N_c)$ SYM {\it simultaneously} such
'slightly  higgsed' quarks and higgsed magnetically charged solitons
with the much larger condensate $\rho_{\rm magn}\sim\lym$. But these
magnetically charged solitons and quarks are mutually nonlocal. For
this reason, such solitons will keep quarks confined and will prevent
them from condensing in the vacuum state.\\

In \cite{FS} the special (not supersymmetric) QCD-type lattice
$SU(N_c)$ gauge theory with $N_F=N_c$ flavors of scalar quarks
$\Phi^i_{\beta}$ in the fundamental representation was considered. In
the unitary gauge, all $N_c^2+1$ remained degrees of freedom of these
quarks were replaced by one constant parameter $|v|\,:\,
\Phi^i_{\beta}\, \ra\, \delta^i_{\beta}|v|,\, \beta=1...N_c,\,\,
i=1...N_F=N_c$.\,  I.e., all quarks  were higgsed {\it by hands}
even at small $|v|\neq 0$ and all $N^2_c-1$ gluons received electric
masses $g |v|$. The matter potential is zero. All other   $N^2_c+1$
quark physical dynamical degrees of freedom were deleted by hands.
The  region with the large values of $|v|$ was considered as the higgs
regime, while those with small $|v|$ as the confinement one. The
conclusion of \cite{FS} was that the transition between the higgs and
confinement regimes is the analytic crossover, not the non-analytic
phase transition.

Let us note that this model of E. Fradkin and S.H. Shenker \cite{FS}
with such {\it permanently higgsed at all $\,0 < g |v| < \infty$ non-dynamical
scalar "quarks" $\Phi^i_{\beta} = \delta ^i_{\beta}|v|$ looks unphysical and
is incompatible with {\it normal models with dynamical} electrically charged 
scalar quarks $\phi^i_{\beta}$ with all $2N_c^2$ their real physical degrees 
of freedom}. This model \cite{FS} is really the pure Stueckelberg
$SU(N_c)$ YM-theory {\it with no dynamical electric quarks and with
{\it massive all electric gluons with fixed by hands masses} $g|v| >
0$ in the bare perturbative Lagrangian}, see eq.(4.1) in \cite{FS}.
\footnote{\,
Unlike ${\cal N}=1$ SQCD, in non-supersymmetric $N_F=N_c$ QCD with
normal dynamical scalar quarks with all their degrees of freedom, these
scalar quarks are not massless even in the limit $m_Q\ra 0$. They acquire
non-perturbative dynamical masses
$\sim\Lambda_{YM}\sim\Lambda_{QCD}$. And, connected with this, 
{\it  there is confinement with the string tension}  $\sigma^{1/2}
\sim\Lambda_{YM}\sim\Lambda_{QCD}$ in this limit}.

For this reason, in any case, the electric flux emanating from the
test (anti)quark becomes {\it exponentially suppressed} at distances
$L > l_0=(g|v|)^{-1}$ from the source. And so, {\it the potentially 
possible confining string tension will be also exponentially
suppressed at distances $L > l_0$ from sources}. And external test
quark-antiquark
pair will be not connected then by one common {\it really confining
string} at large distance between them. These quark and antiquark can
move then practically independently of each other and can be
registered alone in two different detectors at large distance between
one another. I.e., in any case,  in this Stueckelberg theory
\cite{FS},  at fixed $|v| > 0$, {\it there is no genuine confinement}
which prevents appearance of one (anti)quark in the far detector.

As pointed out above, higgsed magnetically charged solitons ensuring
confinement in YM, are mutually nonlocal with electrically charged quarks. 
For this reason, such solitons with the much larger vacuum condensate
$\rho_{\rm magn}\sim\Lambda_{YM} \sim\Lambda_{QCD}\gg g|v|$ {\it will
keep normal dynamical scalar quarks with all their physical degrees
off  reedom confined and will prevent them from condensing with $|v|\neq
0$ in the vacuum state}. See also section 5 in \cite{ch23} for more
details.\\

On the whole, a number of arguments were presented in \cite{ch21}
that, for quarks with $ 1\leq N_F\leq N_c-1$ and fixed $0 < m_Q/\la\ll 1$, 
there is not the analytical crossover but the non-analytic phase transition 
at sufficiently large $N_c$ \eqref{(10.3.1)} between the phases with higgsed 
or not higgsed but confined quarks. As was argued
above, the perturbative mass term of gluons in the Lagrangian,
proportional to the order parameter $\rho_{\rm higgs}$ of higgsed
quarks \eqref{(10.2.3)}, is nonzero and large at values of $N_c$ in
\eqref{(10.2.1)} and becomes not simply small but zero at
sufficiently  large $N_c$ \eqref{(10.3.1)}, because the order parameter
$\rho$ drops to zero and quarks become unhiggsed (see also
\eqref{(11.1.1)},\eqref{(11.1.2)}). The nonzero gluon masses $\sim\lym$ 
are then of nonperturbative origin from SYM, not due to higgsed quarks.\\

\section{\bf \large Additional independent arguments for  the phase  
transitions at \,\, \\ $\mathbf{1\leq N_F \leq N_c-1}$}

\subsection{The phase transition vs crossover}  

{with increasing $N_c$ at fixed  $m_Q\ll\la$}.

\numberwithin{equation}{subsection}

As emphasized in the text below \eqref{(10.3.5)}, for all $1\leq
N_F\leq N_c-1$, the HQ (heavy quark)-phase of not higgsed but 
confined quarks with $m_Q\ll\la$, large $N_c$ \eqref{(10.3.1)} and 
$m_Q^{\rm pole}\gg\lym$ \eqref{(10.3.6)},\eqref{(10.3.9)},
is qualitatively the same as the HQ-phase of heavy not higgsed 
but confined quarks with $m_Q\gg\la$.

For heavy quarks with $m^{\rm pole}_Q\gg\la$ and for light quarks
with \, $m_Q\ll\la$ and large $N_c$ \eqref{(10.3.1)} in this HQ-phase, 
we  can add also the  following. Let
us write, see \eqref{(2.1.2)},\eqref{(10.3.6)},\eqref{(10.3.9)}, the
normalization point is
$\mu=\mu_{p}=m_Q^{\rm pole}\gg\lym$ :
\bbq
\hspace*{-4mm} \Bigl [\, Q^i_\alpha(x)\Bigr ]_{\mu_{p}}=\Bigl ( V_{\rm
Goldst}(x)\Bigr)^{\beta}_{\alpha}\Bigl [\,\langle{\hat
Q}^i_{\beta}(x)\rangle +\delta {\hat Q}^i_\beta(x)\Bigr
]_{\mu_{p}},\,\, \Bigl [\,{\ov Q}_i^{\alpha}(x)\Bigr
]_{\mu_{p}}=\Bigl[\,\langle{\hat{\ov Q}}_i^{\beta}(x)\rangle +\delta
{\hat{\ov
Q}}_i^\beta(x)\Bigr ]_{\mu_{p}}\Bigl ( V_{\rm
Goldst}^{\dagger}(x)\Bigr )_{\beta}^{\alpha},
\eeq
\bq
\langle \delta {\hat Q}^i_\alpha(x)\rangle_{\mu_{p}}=\langle \delta
{\hat{\ov
Q}}_i^\alpha(x)\rangle_{\mu_{p}}\equiv 0\,. \label{(11.1.1)}
\eq
\bbq
\langle M^i_j(x)\rangle_{\mu_{p}}=\sum_{\alpha=1}^{N_c}\langle {\ov
Q}^\alpha_j(x)
Q^i_\alpha(x)\rangle_{\mu_{p}}=\Bigl [\,\sum_{\alpha=1}^{N_c}\langle
{\hat {\ov
Q}}_j^\alpha\rangle_{\mu_{p}} \langle {\hat
Q}^i_\alpha\rangle_{\mu_{p}}=\delta^i_j\, \rho_{\rm HQ}^2\,\Bigr ]
+\sum_{\alpha=1}^{N_c}\langle \delta {\hat{\ov Q}}_j^\alpha(x) \delta
{\hat Q}^i_\alpha(x)\rangle_{\mu_{p}}\,,
\eeq
\bbq
\langle {\hat Q}^i_\alpha\rangle_{\mu_{p}}=\delta^i_\alpha\rho_{\rm
HQ}\,,\quad
\langle {\hat {\ov
Q}}^\alpha_j\rangle_{\mu_{p}}=\delta^\alpha_j\rho_{\rm HQ}.
\eeq

Now, the one quark loop contribution of such quarks with $m_Q^{\rm
pole}\gg\lym$ \eqref{(10.3.9)} looks as
\bq
\sum_{\alpha=1}^{N_c}\langle \delta {\hat{\ov Q}}_j^\alpha(x) \delta
{\hat Q}^i_\alpha(x)\rangle_{\mu_{p}}=\delta^i_j\,\frac{\langle
S\rangle}{m^{\rm
pole}_Q}\quad \ra \quad  \rho_{\rm HQ}=0\,,  \label{(11.1.2)}
\eq
i.e. {\it it saturates the Konishi anomaly} \eqref{(2.1.8)}.
Therefore, the
equations \eqref{(11.1.1)},\eqref{(11.1.2)} show that not only 
quarks  with $m_Q\gg\la$, but {\it in the whole region of
the HQ-phase \eqref{(10.3.1)} even light quarks with $m_Q\ll\la$ are
not higgsed}: $\langle {\hat Q}^i_{\alpha}\rangle_{\mu_{p}} =
\delta^i_\alpha \rho_{\rm HQ}=0,\,\,\langle {\hat {\ov
Q}}_j^{\alpha}\rangle_{\mu_{p}}=\delta_i^{\alpha}
\rho_{\rm HQ}=0$. This is independent confirmation of presented in
section 3 arguments that {\it quarks in the HQ-phase are not higgsed}, 
i.e.  $\rho_{\rm HQ}=0$. Because, at fixed $N_c$ and sufficiently small 
$m_Q$, in the  whole region  $m_Q\ll\la,\,\,\mu_{\rm gl}\gg\la$ the 
value of the gauge invariant order parameter $\rho_{\rm higgs}$ is 
nonzero and  large, see 
\eqref{(10.2.1)},\eqref{(10.2.3)}, this shows that {\it  the  order
parameter $\rho$ behaves non-analytically: it is nonzero in the
region of the higgs phase and zero either at large $m_Q\gg\la$ or at
small  $m_Q\ll\la$ but large $N_c$ \eqref{(10.3.1)} in the region of the
HQ-phase}. This independently confirms that there is not the analytic 
crossover but  non-analytic phase transition between regimes of 
confined or higgsed  quarks.

\subsection{The phase transition vs crossover with decreasing $m_Q$}

{from  $m_Q\gg\la$ to $m_Q\ll\la$ at fixed $N_c\,,\,N_F=N_c-1$}.

\numberwithin{equation}{subsection}

\hspace*{-4mm} Heavy quarks with $m_Q\gg\la$ have large masses and
small mean vacuum value $\langle
M\rangle=\la^2(\la/m_Q)^{\frac{1}{N_c}}\\
\ll\la^2\ll\lym^2\ll m^2_Q$, see \eqref{(10.1.2)},\eqref{(10.1.4)}. For this 
reason, they  are in the HQ-regime and are not higgsed, i.e. 
$\langle{\hat{\ov  Q}}^{\,\alpha _i}\rangle=
\langle{\hat Q}^i_{\alpha}\rangle=0$, see also \eqref{(11.1.1)},
\eqref{(11.1.2)}. {\it They are weakly confined, see footnote
\ref{(f5)}}, and decouple as heavy in the weak coupling regime at
$\mu< m^{\rm pole}_Q=m_Q/z_Q(\la, m^{\rm pole}_Q)\gg\la$, where
$z_Q(\la,  m^{\rm pole}_Q)\gg 1$ is the logarithmic renormalization factor. 
The  scale factor $\lym$  \eqref{(2.1.4)} of remained ${\cal N}=1\,\, 
SU(N_c)$ SYM is  determined  from matching of logarithmically small 
couplings  $a_{+}(\mu=m^{\rm  pole}_Q)=a_{SYM}(\mu=m^{\rm
pole}_Q)$.
\footnote{
The nonzero gluon masses originate only in ${\cal N}=1$ SYM due to
nonperturbative effects, so that their typical values are 
$\mu_{\rm gl}^{\rm non-pert}\sim\lym$.
}
The small nonzero non-factorizable value of $\langle M\rangle=\langle {\ov Q}
_1^{\beta} Q^1_{\beta}\rangle$ \, originates from the one
quark loop Konishi anomaly \eqref{(2.1.8)} for quarks in the HQ-phase, 
not due to \, "slightly higgsed"\, heavy quarks.\\

{\it The global $SU(N_F)$ is unbroken}. There is in the spectrum a
number of heavy flavored quarkonia with typical masses ${\cal O}(m_Q)\gg\lym$ 
and  different
quantum numbers. For instance, the quark-antiquark bound states with
different spins and other quantum numbers are in the adjoint or
singlet representations of unbroken global $SU(N_F)$. It is important
that, due to a confinement, {\it there are no particles in the
spectrum in the $SU(N_F)$} (anti)fundamental representation of
dimensionality $N_F$. Besides, there are in the spectrum a number of
$SU(N_F)$ singlet gluonia with typical masses $\sim\lym=\la
(m_Q/\la)^{N_F/3N_c},\, \la\ll\lym\ll m_Q$. \\

The light quarks with $N_F=N_c-1$ flavors with their $4N_F N_c$ real
degrees of freedom have small current masses $m_Q\ll\la$ and large 
mean vacuum  values
$\langle M^i_j\rangle=\sum_{\alpha=1}^{N_c}\langle{\ov
Q}^{\,\alpha}_jQ^i_\alpha\rangle
=\delta^i_j\la^2 (\la/m_Q)^{\frac{1}{N_c}}\gg
\la^2\gg\lym^2,\,\,\mu_{\rm gl}\gg\la$, see
\eqref{(2.1.8)}\eqref{(2.1.9)}. They all are higgsed in this case in
the weak coupling region and the whole global color group $SU(N_c)$
is  broken. The quark mean vacuum values of
$\langle{\hat Q}^i_{\alpha}\rangle$ and $\langle{\hat{\ov
Q}}^{\,{\alpha}}_i\rangle$ look as, see section 2.1\,:
\bq
\langle{\hat Q}^i_\alpha\rangle=\delta^i_\alpha\,\omega,\,\,
\langle{\hat{\ov Q}}^{\,\alpha}_i\rangle=\delta^\alpha_j\,\omega,\,
\,\omega=\la(\la/m_Q)^{1/2N_c}\gg
\la, \,\, i=1...N_F,\,\, \alpha=1...N_c\,.  \label{(11.2.1)}
\eq
\vspace*{1mm}

Let us present now the additional independent arguments that, at
fixed  $N_c,\, N_F=N_c-1$, it is 
not the crossover but phase transition between
regions of $m_Q\gg\la$ and $m_Q\ll\la,\,\mu_{\rm gl} \gg \la$.

From \eqref{(11.2.1)}, the unbroken global symmetry looks now as:\,
$SU(N_F)\times SU(N_c)\times U(1)_B\ra SU(N_F)_{F+C}\times U(1)_{\tilde B}$\,, 
i.e. the color-flavor locking. {\it There is no confinement.} All
$N_c^2-1=N_F^2+2 N_F$ heavy gluons (which "ate" $N_F^2+2 N_F$
massless  Goldstone degrees of freedom from quarks) and the same number
of their  ${\cal N}=1$ scalar superpartners acquired large masses
\eqref{(10.2.1)}, \eqref{(11.2.1)}. They form 2 adjoint
representations of $SU(N_F)$ plus two $SU(N_F)$ singlets. Plus, and
this is most important, else $2 N_F$ heavy gluons
$(A_{\mu})^i_{\alpha=N_c},\, (A_{\mu})^{\alpha=N_c}_{i},\, i=1...N_F$
and $2 N_F$ their ${\cal N}=1$ scalar superpartners. These $4 N_f$
form {\it two fundamental and two antifundamental representations of
$SU(N_F)$ with dimensionality $N_F$} each. And finally, there are
$N^2_F$ light complex pions $\Pi^i_j,\, i,j=1...N_F$ with small
masses  $\sim m_Q$ \eqref{(10.2.7)} which form the adjoint and singlet
representations of $SU(N_F)$. {\it Therefore, there are only fixed
numbers of particles with fixed quantum numbers in the spectrum}.

The mass terms of $N_c^2-1$ heavy massive gluons in the Lagrangian
look as:
\bbq
M^2_{\rm gl} =2 g^2(\mu_{\rm gl})\,z_Q(\la,\mu=\mu_{\rm
gl})\,\sum_{i=1}^{N_F}\sum_{\alpha,\gamma=1}^{N_c}\langle{\,\Bigl
({\hat Q}^{\,\dagger}\Bigr
)}^{\,\alpha}_i\rangle\Biggl\{\sum_{\beta=1}^{N_c}(A_{\mu})_\alpha^\beta\
(A_{\mu})^\gamma_\beta
\Biggr\}\langle\,{\hat Q}^{\,i}_\gamma\,\rangle=
\eeq
\bq
K\,\sum_{i=1}^{N_F}\sum_{\beta=1}^{N_c}(A_{\mu})^i_\beta\,(A_{\mu})_i^\beta\,,
\,\, K= 2 g^2(\mu_{\rm gl})\,z_Q(\la,\mu=\mu_{\rm gl})\Biggl (
\omega^2=\la^2(\la/m_Q)^{1/N_c}\Biggr )\gg\la^2\,. \label{(11.2.2)}
\eq

From \eqref{(11.2.2)}, the masses of gluons in different
representations of unbroken global $SU(N_F)$ are {\it different}\,:
\bq \mu^2_{\rm gl}\bigl(SU(N_F)_{\rm adj}\bigr )=K,\,\, \mu^2_{\rm
gl}\bigl(SU(N_F)_{\rm singl}\bigr )=\frac{1}{ N_c} K,\,\, \label{(11.2.3)}
\eq
\bbq
\mu^2_{\rm gl}\bigl(SU(N_F)_{\rm fund}\bigr )=\mu^2_{\rm
gl}\bigl(SU(N_F)_{\rm anti-fund}\bigr )=\frac{1}{2} K,\,
\eeq
(and the same for their scalar superpartners). It is seen from
\eqref{(11.2.3)} that $N_c^2-1$ heavy gluons have different 
masses and do not form one adjoint
representation of global $SU(N_c)$ at $N_c > 2$.\\

From comparison of mass spectra properties in regions $m_Q\gg\la$ and
$m_Q\ll\la,\mu_{\rm gl} \gg \la$ \eqref{(10.2.1)} it is seen that, although the
unbroken global symmetry $SU(N_F)$ is the same, but {\it realized are its 
different representations}. In the case of heavy confined quarks with 
$m_Q\gg \la$ in the  HQ-phase  there are no particles in the spectrum in 
the  (anti)fundamental representation of $SU(N_F)$, while in the case of
light higgsed quarks such representations are present. E.g., for
fixed  $N_c$, we can start with the case of heavy quarks with
$m_Q\gg\la$ and  to diminish continuously $m_Q$ until $m_Q\ll\la$. And
when reaching the appropriately small value of $m_Q$, such that $\mu_{\rm gl}
\gtrsim \la$ \eqref{(11.2.2)} \eqref{(11.2.3)}, all quarks become higgsed and
the
behavior of the mass spectrum under unbroken global $SU(N_F)$
transformations changes discontinuously (because the dimensions of
representations can not change continuously). {\it This
jump is impossible in the case of crossover (which is analytic), this
means the phase transition between the confinement and higgs
phases}.\\

In other words. The fraction $R_{\rm fund}$ of particles in the
(anti)fundamental representation in the mass spectrum can serve 
in the case considered as the order
parameter. This fraction is zero in the confinement region where
quarks with $\rho_{HQ}=0$ \eqref{(11.1.2)} are not higgsed. While
this  fraction is the {\it nonzero constant} in the region with higgsed
quarks with $\rho_{higgs}=\omega \gg\la$ \eqref{(11.2.1)}. I.e.,
$R_{\rm fund}$ behaves non-analytically. This is a clear sign of the
phase transition, because this fraction would behave analytically in
the case of crossover.

At the same time, the dependence of bilinear mean vacuum value
$\langle M^i_j\rangle= \delta^i_j \langle M\rangle$ \eqref{(2.1.2)}, 
\eqref{(10.2.7)} on $m_Q$ is analytic, but this does not mean that there 
can not be the phase transition. The qualitative difference is that
$\langle M^i_j\rangle=\sum_{\alpha=1}^{N_c}\langle {\hat {\ov
Q}^\alpha_j}\rangle \langle {\hat Q}^i_\alpha\rangle=\delta^i_j\,
\omega^2\neq 0$, see \eqref{(11.2.1)}, \eqref{(11.2.3)}, {\it i.e. 
factorizes} for higgsed quarks with $m_Q\ll\la,\,\mu_{\rm
gl} > \la$ (the order parameter is $\omega=\la (\la/m_Q)^{\frac{1}{2
N_c}}\gg\la$ \eqref{(11.2.1)}\,). While for non-higgsed but weakly
confined (see footnote \ref{(f5)}) quarks in the HQ-phase with
$m_Q\gg\la$ or with $\lym\ll m^{\rm pole}_Q\ll\la$ and large $N_c$
\eqref{(10.3.1)},\eqref{(10.3.9)}, this bilinear mean value $\langle
M^i_j\rangle$ {\it becomes non-factorizable. It originates then from
the one quark loop Konishi anomaly \eqref{(2.1.2)}and all $\langle
{\hat{\ov Q}}^a_i\rangle=\langle {\hat Q}^i_a\rangle=0$},
see the text under \eqref{(10.3.5)} and
\eqref{(11.1.1)},\eqref{(11.1.2)}.\\

Let us present now the additional arguments that the above described
phase transitions are of the first order. Suppose that, vice versa, they
are  e.g. of the second order. Then, at fixed $0 < m_Q/\la\ll 1$ and
increasing $N_c$, there is a finite width region of $N_c$ around
$\mu_{\rm gl}(\mu\sim\la)\sim\la$ where the order parameter
$\rho_0(N_c,\,m_Q/\la)=\rho(N_c,\,m_Q/\la)/\la$ changes continuously
with $N_c$, e.g. from its large value $\approx\rho_{\rm higgs}$ in 
\eqref{(11.2.1)} in the higgs phase down to zero in the HQ-phase.
\footnote{
The order parameter $\rho_0$ is defined in \eqref{(10.2.3)},\eqref{(11.1.2)} 
at the scale $\mu=\la$ and {\it is independent of the scale factor} $\mu$.
The whole dependence of $\mu_{\rm gl}(\mu)$ on the scale $\mu$ at fixed
$\rho_0=\rho(N_c,\,m_Q/\la)/\la$ originates from the running quark
renormalization factor $z_{Q}(\mu)$ and from the
running coupling $a(\mu)$,
see\eqref{(10.2.1)},\eqref{(10.3.3)}-\eqref{(10.3.7)}.
}
Consider now some vicinity of the point where $\rho_0(N_c,\,m_Q/\la)$
reaches zero. This vicinity is such that $0 < \rho_0(N_c,\,m_Q/\la) < 1 \ll
\rho_{\rm higgs}$, with $\rho_{\rm higgs}\gg 1$ from \eqref{(10.2.3)}. I.e.
$\rho_0(N_c,\,m_Q/\la)$ is not large but nonzero within it. And, at
fixed value of $N_c$ within this interval, the gluon mass terms in
the  Lagrangian from higgsed quarks are nonzero and became such that 
$0 <\mu_{\rm gl}(\mu\sim\la)\ll\la$ (and remain $\mu_{\rm gl}(\mu)\ll\mu$
at smaller $\mu$) and are {\it dynamically irrelevant}.

Then quarks are already confined but nevertheless remain "slightly
higgsed"\, at fixed $0 <\rho_0(N_c,\,m_Q/\la) < 1$. Then, because the 
gluon mass terms in the
Lagrangian from higgsed quarks are too small and dynamically
irrelevant, the perturbative RG-evolution to smaller scales down to
$\mu\sim\lym$ is still described by \eqref{(10.3.4)}-\eqref{(10.3.7)}, the
only difference is that fixed $\rho_0(N_c,\,m_Q/\la)$ instead of its
large value \eqref{(11.2.1)} is now much smaller, $0 <
\rho_0(N_c,\,m_Q/\la) < 1$.

As was argued in section 10.3, this variant with confined and simultaneously 
"slightly higgsed"\, quarks with fixed $\rho_0(N_c,\,m_Q/\la)\neq
0$ is self-contradictory for the vacuum state. I.e. the second order
phase transition can't be realized. The only self-consistent variant is the
first order phase transition. I.e., with fixed $m_Q/\la$ and increased $N_c$,
there is {\it the point} within the region $\mu_{\rm gl}(\mu\sim\la)\sim\la$ 
where the order parameter $\rho_{\rm higgs}$ drops from its value
\eqref{(10.2.3)}
{\it to zero}. And these reasonings are applicable to both types of
phase transitions described in sections 10 and 11. The only
differenceis that either $\rho_{0}(N_c,\,m_Q/\la)$ changes with
varying $N_c$ at  fixed $m_Q/\la$, or it changes with varying $m_Q/\la$
at fixed $N_c$ and $N_F$.\\
\vspace*{2mm}

The conclusion of the paper \cite{ch21} about the phase transition
between the confinement and higgs regimes is opposite to the conclusion 
of the paper of E. Fradkin and S.H. Shenker \cite{FS} that the transition
between
these regimes is the crossover, not the phase transition.
And although the theories considered in \cite{ch21} and in \cite{FS}
are different, an experience shows that there is a widely spread opinion that
the
conclusion of \cite{FS} is applicable to all QCD-type theories: both
lattice and continuum, and both not supersymmetric and
supersymmetric.This opinion is in contradiction with the results of
the paper \cite{ch21}.\\

Another types of phase transitions in ${\cal N}=2$ SQCD are described
in sections 6.1, 6.2, 7, 8 of \cite{ch17}.\\

\section{\bf \large Mass spectra in $\mathbf{ {\cal N}=1\,  SU(N_c)}$ SQCD 
with $\boldmath{N_F=N_c}$ \,\, \\  and   problems  with S-confinement}

\numberwithin{equation}{subsection}

\subsection{Introduction} 

\hspace*{4mm} N. Seiberg proposed in \cite{S1,IS} the low energy form
of the direct $SU(N_c)$\, ${\cal N}=1$ SUSY QCD theory (SQCD) with 
$N_F=N_c$ light  quark flavors. While the very high energy direct theory 
contained $2 N_F N_c$  colored complex quark fields $Q^i_{\beta},\,
{\ov Q}^{\alpha}_i\,\,\,\beta=1...N_c,\,\, i=1...N_F=N_c$ and
$N^2_c-1$ gluons, see \eqref{(12.2.2)}, the proposed in \cite{S1} low
energy one at the scale $\mu < \la$ ($\la={\rm const}$ is the scale
factor of the direct theory gauge coupling in the
UV region ) contained instead only $N_F^2$ colorless
complex mesons  $M^i_j=\sum_{\beta=1}^{N_c}({\ov Q}^{\beta}_j
Q^i_{\beta}),\,\,i,j=1...N_F$ and
two colorless complex baryons $B\sim Q^{N_c},\, {\ov B}\sim {\ov
Q}^{\,N_c}$, with one constraint.

This regime with e.g. equal quark masses $m_{Q,i}=m_Q\ra 0$ and
$\langle M^i_j\rangle=\delta^{i}_{j} \la^2,\,\, i,j=1...N_F$, see
\eqref{(12.2.4)} was called  {\it "confinement with 
spontaneous chiral symmetry breaking"}, see
e.g. section 4.2 in \cite{IS}. The implied physical interpretation
was  the following. At the scale $\mu \sim \la$ the gauge coupling is
sufficiently large: $a(\mu=\la)=N_c g^2(\mu=\la)/8\pi^2={\cal O}(1)$.
And {\it all massless or light quarks and all gluons are confined,
with the assumed large string tension} $\sigma^{1/2}\sim\la$. They
form then hadrons most of which have masses $\sim\la$ due to such
string tension. And, at $m_Q\ll\la$, only special composites:
$N_c^2-1$ independent mesons from $M^i_j=\sum_{\beta=1}^{N_c} ({\ov
Q}^{\beta}_j Q^i_{\beta}),\,\,i,j=1...N_F=N_c$ and two baryons $B\sim
Q^{N_c}$ and
${\ov B}\sim {\ov Q}^{N_c}$ are light, with masses $\sim m_Q\ll\la$.
For these reasons, only these enter the low energy Lagrangian of the direct
theory at the scale $\mu < \la$. As a test, it was checked that the
'tHooft triangles are matched, see \cite{S1}. This Seiberg's proposal
is  usually called "S-confinement"\, in the literature.\\

The above line of reasonings was criticized in \cite{ch1}, see
section  7 therein.
\footnote{\,
It was pointed out in section 7 of\cite{ch1} that {\it the confinement 
of colored
particles originates only from pure SYM}. There is no confinement in
Yukawa-type theories without gauge interactions. But the lower energy 
pure SYM theory contains only one universal dimensional parameter 
$\langle\lym\rangle=\Bigl(\la^{3N_c-N_F}\det \mi\Bigr )^{1/3
N_c}\ll\la$ at $\mi\ll\la$,\, see \eqref{(12.2.3)}. So that, {\it its string
tension at $\mi\ll\la$ can not be as large as $\sigma^{1/2}\sim\la$, but is
much smaller:\, $\sigma^{1/2}\sim\lym\ll\la$. And $\lym\ra 0$ if even
one $m_{Q,i}\ra 0$, in which limit there is no confinement at all}.
\label{(f21)}
}

And this argument from footnote \ref{(f21)} is applicable to all
values $N_F > 0$ at all $\mi\ll\la$. Therefore, {\it the interpretation 
e.g.  of the dual theory  with $\mi=m_Q$ as the
low energy form of the direct theory \eqref{(12.2.2)} at $m_Q\ll\la$
due to strong confinement with $\sigma^{1/2}\sim\la$ is not correct.
It should be understood as  independent theory. And mass spectra of
these two theories have to be calculated and compared to see whether
they are equivalent or not}.

And similarly for the case $N_F=N_c+1$. The theory with $N_F^2$
colorless mesons  $M^i_j=\sum_{\beta=1}^{N_c}({\ov Q}^{\beta}_j 
Q^i_{\beta})$ and $2 N_F$ colorless  baryons $B^i\sim Q^{N_c+1},\,\, 
{\ov B}\sim {\ov Q}^{N_c+1}$ was  proposed by N. Seiberg
as the low energy form of this direct theory and similarly
interpreted \, in \cite{S1},\,\cite{IS} as the regime with "confinement
without spontaneous chiral symmetry breaking"\, at e.g.
$m_{Q, i}=m_Q\ra 0$.\\ \vspace*{2mm}

The low energy mass spectra of the direct standard $SU(N_c),\,
\,N_F=N_c$ $\,\,{\cal N}=1$ SQCD theory and its N. Seiberg's dual variant
\cite{S1} considered as {\it two independent theories}, were calculated in
\cite{ch23}. Within the dynamical scenario \cite{ch3} used in
\cite{ch23}, it was shown that these two mass spectra are
parametrically different, both for equal or unequal small quark
masses  $m_{Q,i}\neq 0$. Therefore, it was concluded in \cite{ch23}
 that the proposal by N. Seiberg \cite{S1} of his $N_F=N_c$ dual theory 
 of $N_F^2-1$ mesons $M^i_j$ and two baryons $B,\,{\ov B}$ as the low
energy form of the direct theory is erroneous (and similarly for the
case $N_F=N_c+1$, see Appendix {\bf B} in \cite{Session}). \\

\vspace*{2mm}

In \cite{FS} the very special non-supersymmetric lattice $SU(N_c)$
QCD  theory with $N_F=N_c$ scalar defective "quarks" in the unitary
gauge: \, $\Phi^i_{\beta}= \delta^i_{\beta}(|v|={\rm const} >
0),\,\,i,{\beta}=1,...,N_F=N_c$, was considered by E. Fradkin and
S.H.Shenker. The conclusion of \cite{FS} was that the transition
between the confinement at $0 < |v|\ll\Lambda_{QCD}$ and higgs at
$|v|\gg\Lambda_{QCD}$ regimes in this theory is the analytic
crossover, not the non-analytic phase transition. And although the
theory considered in \cite{FS} was very specific, the experience
shows  that up to now there is a widely spread opinion that this
conclusion has general applicability.

This model used in \cite{FS} was criticized in \cite{ch23} as
incompatible with and qualitatively different from the standard
non-SUSY $SU(N_c)$ $\,\,N_F=N_c$ QCD theory with standard scalar
quarks $\phi^i_{\beta}$ with all $2 N_c^2$ their physical real
degrees  of freedom. It was emphasized in \cite{ch23} that this model 
\cite{FS} is  really the Stueckelberg $SU(N_c)$ pure YM-theory with no 
dynamical electric quarks and with
massive all $N_c^2-1$ electric gluons with fixed by hands nonzero
masses. {\it There is no genuine confinement in this theory, it stays
permanently in the completely higgsed phase only}. And this is a
reason for a crossover in this theory. While in the theory with
standard scalar quarks there is the phase transition between the
confinement and higgs phases.

Besides, the arguments presented in \cite{IS} by K. Intriligator and
N. Seiberg for the standard direct $SU(N_c)$, $\,\,N_F=N_c\,\,\, {\cal
N}=1$ SQCD in support of the crossover from \cite{FS} were criticized
in \cite{ch23} as erroneous, see section 12.5 below.\\

\vspace*{2mm}

In \cite{S2} N.Seiberg introduced the dual "magnetic"\, $SU({\ov
N}_c=N_F-N_c)$
gauge theory with e.g. $N_c+2\leq N_F < 3N_c$ dual quarks $q^a_i,\,\,
{\ov q}^i_a,\,\,i=1...N_F, \,\,a=1...{\ov N}_c$. And proposed the
following. - a) {\it "The quarks and gluons of one theory can be 
interpreted as  solitons
(non-Abelian magnetic monopoles) of the elementary fields of the
other  theory}\, \cite{S2}"\,.\, b) These two theories, although clearly
different at scales $\mu > \la$, {\it become
equivalent} at $\mu < \la$. I.e., {\it the dual theory is the
equivalent alternative description of the direct one} at $\mu < \la$.
{\it "When one of the theories is Higgsed by an expectation value of
a  squark, the other theory is confined. Massless glueballs, baryons and
Abelian magnetic monopoles in the confining description are the
weakly  coupled elementary quarks (i.e. solitons of the confined quarks)
in the dual Higgs description}\, \cite{S2}"\,.

The Seiberg duality \cite{S1}, \cite{S2} passed a number non-trivial
checks, mainly  in the effectively massless regime: the matching of 't Hooft
triangles  and behavior in conformal regimes, and correspondences under
appropriate mass deformations. But up to now no direct proof
has been given that the direct and dual theories are really
equivalent. The reason
is that such a proof needs real understanding of and a control over
the dynamics of both theories in the strong coupling regimes. E.g.,
at  $N_c+2 \leq N_F < 3N_c$,  {\it no way was shown up to now how to 
obtain  solitons with quantum numbers of dual quarks in the direct theory
(and  vice versa)}.  Hence, the Seiberg proposal of equivalence of
direct  and dual theories at $\mu < \la$ remains a hypothesis up to now, 
see  e.g. Introduction in arXiv:0811.4283 [hep-th] \cite{ch3} and
references \cite{AR,G,A,K,V}.\\

For the reasons described above, the approach in \cite{ch23} was to
consider the direct and Seiberg's dual theories {\it as independent
theories, to calculate their mass spectra and to compare}. This is
then the direct {\it additional check} of possible equivalence of
these two theories at $\mu < \la$.

For this purpose, the dynamical scenario was introduced in \cite{ch3}
which allows to calculate the mass spectra in ${\cal N}=1$ SQCD-type 
theories even in strong
coupling regimes. The only dynamical assumption of this scenario was
that quarks with $\mi\neq 0$ in such theories can be in two
different {\it standard} phases only\,: this is either the HQ (heavy
quark) phase where they are not higgsed but confined at $\lym\neq 0$,
or the higgs phase where they form nonzero coherent condensate in the
vacuum breaking the color symmetry, at the appropriate values of the
Lagrangian parameters. The word "standard" implies also that, in
addition to the ordinary mass spectrum described e.g. in
\cite{ch3},\cite{ch19} and below in this paper, in such ${\cal N}=1$
theories without elementary colored adjoint scalars (unlike the very
special ${\cal N}=2$ SQCD with its additional elementary colored
scalar fields $X^{adj}$ and enhanced supersymmetry)  {\it no
additional parametrically lighter solitons (e.g. magnetic monopoles
or dyons) are formed at those scales where the {effectively} massless
regime is broken explicitly by nonzero particle masses}.
(It is worth  noting \, that the appearance of {\it additional} light 
solitons will  influence the 't Hooft triangles of this ${\cal N}=1$
theory). Let us emphasize also that  {\it this dynamical scenario
satisfies all those tests which were used as checks of the Seiberg
hypothesis about the equivalence of the direct and dual theories at
scales $\mu < \la$}. This shows, in particular, that all these tests,
although necessary, may well be insufficient.

Within the framework of this dynamical scenario, in was shown e.g. in
\cite{ch3}, \cite{ch17}, \cite{ch13}, \cite{ch16} and in Appendix {\bf B} of \cite{Session} 
that  mass spectra of the direct theories and their Seiberg' dual variants
are parametrically different in many cases under control.
Therefore, {\it the Seiberg dual $SU({\ov N}_c=N_F-N_c)$ theory is
not\,  the genuine  or equivalent low energy form of the direct theory at 
$\mu < \la$. It
is the independent theory}. And, in each case, the mass spectra of
the  direct and dual theories should be calculated and compared to see
whether they are really equivalent or not. This was a purpose of
\cite{ch23} for the case $N_F=N_c$, see this section below.

\subsection{Direct theory with $N_F=N_c$}.

Let us start with the direct theory at $\mu=\la$ with $N_c$ colors
and  $N_F=N_c+1$ quark flavors with e.g. {\it the real positive masses}: 
$0 < \mi(\mu=\la)=m_{Q,i}\ll\la,\,\,i=1...N_c,\,\,
m_{Q,N_c+1}(\mu=\la)=\la$. The Lagrangian at $\mu=\la$ looks as
\bbq
K_{N_F=N_c+1}={\rm Tr}_{N_F=N_c+1}\Bigl (Q^\dagger e^{V}Q+Q\ra {\ov Q}
\Bigr ),\quad {\cal W}_{N_F=N_c+1}=\frac{2\pi}{\alpha(\mu=\la)} S+
\eeq
\bq
+\sum_{i=1}^{N_F=N_c}\mi M^i_i +\la M^{N_c+1}_{N_c+1}\,,\quad \mi\ll\la\,,
\label{(12.2.1)}
\eq
\bbq
M^i_j=\sum_{\beta=1}^{N_c}({\ov Q}^{\beta}_j Q^i_{\beta})(\mu=\la)\,,
\quad
i,j=1...N_F=N_c\,, \quad M^{N_c+1}_{N_c+1}=\sum_{\beta=1}^{N_c}({\ov
Q}^{\beta}_{i=N_c+1} Q^{i=N_c+1}_{\beta})(\mu=\la)\,,
\eeq
where $S=\sum_{A,\gamma} W^{A,\,\gamma} W^{A}_{\gamma}/32\pi^2$,\,\,
$W^A_{\gamma}$ is the gauge field strength, $A=1...N_c^2-1,\,
\gamma=1,2$,\,
$a(\mu)=N_c g^2(\mu)/8\pi^2=N_c\alpha(\mu)/2\pi$ is the gauge
coupling with its scale factor $\la={\rm const}$.

Integrating out the last heaviest quark, we obtain at $\mu=\la$ the
Lagrangian of
the direct theory with $N_F=N_c$ flavors of light quarks (and with
the same scale  factor $\la$ of the gauge coupling)
\bq
K_{N_F=N_c}={\rm Tr}_{N_F=N_c}\Bigl (Q^\dagger Q+Q\ra {\ov Q} \Bigr
),\quad {\cal W}_{N_F=N_c}=\frac{2\pi}{\alpha(\mu=\la)} S+\sum_{i=1}
^{N_F=N_c}\mi M^i_i\,.\label{(12.2.2)}
\eq

To find the mean vacuum value of the {\it colorless} gluino
condensate  $\langle S\rangle_{N_c}$ in this theory, we use 
that it depends analytically on the numerical values $\mi$ of quark 
masses. For this reason, we can start  now with all very heavy quarks with 
$\mi\sim m_{Q,j}\gg \la$ in the  weakly coupled HQ(heavy quark) phase, 
i.e. not higgsed but confined, and integrate
them  out. There remains the pure $SU(N_c)$ SYM lower energy theory 
with  the  scale factor of its gauge coupling $\lym$ and the gluino mean
vacuum value, see \cite{VY} and section 2 in \cite{ch1},
\bq
\lym^3=\langle
S\rangle_{N_c}=\frac{\sum_{\gamma=1}^{2}\sum_{A=1}^{N_c^2-1}\langle
\lambda^{A,\gamma}\lambda^A_{\gamma}\rangle}{32\pi^2}=
\Bigl (\la^{3N_c-N_F=2N_c} \Pi_{i=1}^{N_F=N_c}\mi\Bigr )^{1/N_c}.
\label{(12.2.3)}
\eq
The values $\mi$ in \eqref{(12.2.3)} can be continued now
analytically to desired values, e.g. $0 < \mi\ll\la$.

From the Konishi anomalies \cite{Konishi}, see \eqref{(12.2.2)}-
\eqref{(12.2.4)}
\bq
\langle M^i_j\rangle=\sum_{\beta=1}^{N_c}\langle {\ov Q}^{\beta}_j
Q^i_{\beta} \rangle=\delta^i_j\frac{\langle
S\rangle_{N_c}}{\mi}=\delta^i_j \la^2\frac{\det^{1/N_c} \mi\equiv
\Bigl (\Pi_{i=1}^{N_F=N_c}\mi\Bigr )^{1/N_c}}{\mi}\,.\label{(12.2.4)}
\eq

Besides, from \eqref{(12.2.4)}
\bq
\langle \det M^i_j\rangle=\Pi_{i=1}^{N_c}\langle
M^i_i\rangle=\frac{\Bigl (\langle
S\rangle_{N_c}\Bigr )^{N_c}}{\det m_{Q,i}} = \la^{2N_c}\,,
\label{(12.2.5)}
\eq
{\it for all $\mi \neq 0$, and even in the chiral limit
$\mi\,\ra\,0$}.  The theory \eqref{(12.2.2)} approaches the 
moduli space In this  limit. \\

Let us consider now the colorless baryon operator
\bq
B=\frac{1}{N_c !\,\la^{N_c-1}}\det Q^i_\beta=\frac{1}{N_c
!\,\la^{N_c-1}}\sum_{i=1}^{N_F=N_c}\sum_{\beta=1}^{N_c}
\epsilon_{i_1...i_{N_c}}\epsilon^{\beta_1...\beta_{N_c}}
Q^{i_1}_{\beta_1}...Q^{i_{N_c}}_{\beta_{N_c}}, \label{(12.2.6)}
\eq
and similarly ${\ov B}$ with $Q\ra {\ov Q}$. Let us start first with
heavy equal
mass quarks with $\mi=m_Q\gg\la$. If all such quarks were higgsed
then, by definition, $\langle B\rangle=\langle {\hat
Q}^1_1\rangle...\langle{\hat
Q}^{N_c}_{N_c}\rangle\neq 0$. Here $\langle{\hat Q}^i_{\beta}
\rangle=\delta^i_{\beta}\rho_{\rm higgs}$, where $\rho_{\rm
higgs}\neq0$ is the gauge invariant order parameter introduced in 
section 2 of  \cite{ch21}. But,
clearly, all such heavy quarks are in the HQ (heavy quark) phase,
i.e.{\it not higgsed but confined} by $SU(N_c)$ SYM\,:\, $\langle
{\hat{\ov Q}}_i^{\beta}\rangle=\delta_i^{\beta}\rho_{HQ}=0,\,\,
\langle {\hat Q}^i_{\beta}\rangle=\delta^i_{\beta}\rho_{HQ}=0,\,\,\,
i,\beta=1...N_c$, see section 4.1 in \cite{ch21}. Therefore, $\langle
B\rangle=\langle{\ov B}\rangle=0$ for such equal mass heavy quarks.
While $\langle M^1_1\rangle=\langle M^i_i\rangle=\la^2\neq \langle
{\hat Q}^1_1\rangle\langle{Q}^{1}_{1}\rangle=0$ for such heavy equal
mass quarks, see \eqref{(12.2.4)}. The reason is that $\langle
M^i_i\rangle$ is non-factorizable in this case and is nonzero due to
the Konishi anomaly.
\footnote{\,
There is a qualitative difference between $\langle
B\rangle=\langle{\ov B}\rangle$
and $\langle M^i_i\rangle$. This last is non-factorizable in the
HQ-phase with
$\rho_{HQ}=0$, but is nonzero even for heavy non-higgsed quarks in
this HQ-phase due to the one quark loop Konishi anomaly,
\eqref{(12.2.4)}, see section 4.1 in \cite{ch21}. But there is no
analog of the Konishi anomaly for $\langle B\rangle=\langle{\ov
B}\rangle$. \label{(f22)}
}

But, as mean vacuum values of lowest components of any {\it
colorless}  chiral superfields, e.g. colorless $\langle{\rm Tr}\,
\lambda\lambda\rangle$ or $\langle M^i_i\rangle$, the mean vacuum
values $\langle B\rangle=\langle{\ov B}\rangle$ of these {\it
colorless} chiral
superfields also {\it depend analytically on values of the chiral
superpotential parameters $\mi$}. Therefore, for these equal mass
quarks, the values of $\langle B\rangle =\langle{\ov
B}\rangle$ can be continued analytically from the regions of heavy
quark mass values where definitely $\langle B\rangle=\langle {\ov
B}\rangle=0$ to {\it small nonzero mass values}, and both $\langle
B\rangle$ and $\langle{\ov B}\rangle$ will remain equal zero.

In particular, it follows from the above that {\it even all light
equal mass  quarks are in the HQ-phase and not higgsed}
\bbq
\langle {\hat Q}^i_{\beta}\rangle=\langle{\hat {\ov
Q}}^{\beta}_i\rangle=0,\,\,\,
i,\beta=1... N_F=N_c,\,\,\, {\rm for}\,\,\, 0 < \mi=m_Q\ll\la,.
\eeq
\bq
\,\, {\rm and\,\,\,even\,\,\,in\,\,\,the\,\,\,limit}\,\, m_Q\ra 0\,.
\,\label{(12.2.7)}
\eq
As it is seen from  \eqref{(12.2.4)},\eqref{(12.2.5)},  it is
impossible to have all $\langle M^i_i\rangle > \la^2$ for all
$N_F=N_c$ unequal
mass quarks. Heavier H-quarks will have $\langle M^H_H\rangle <
\la^2$. And because these quarks were not higgsed even for equal 
mass  quarks with $\langle M^H_H\rangle = \la^2$
\eqref{(12.2.4)},\eqref{(12.2.5)},\eqref{(12.2.7)}, they definitely
will remain non-higgsed at $\langle M^H_H\rangle < \la^2$.
But when even one H-quark is not higgsed, then
$\langle B\rangle=\Pi_{i=1}^{N_F=N_c}\langle Q^i_{i}\rangle=0$ also
for unequal mass quarks with $m_{Q,i} > 0$, and even in the chiral
limit $m_{L,H}\ra 0$ with fixed ratio $0 < r=m_L/m_H\ll 1$, see
section 12.2.2, footnote \ref{(f22)} and \eqref{(12.4.1)}. And the
same for $\langle{\ov B}\rangle=\langle B\rangle=0$.

\numberwithin{equation}{subsubsection}

\subsubsection{Light equal mass  quarks}

Let us consider first just this case with all light equal mass
quarks,  $0 < \mi=m_Q\ll\la\,,\,\,\,i=1...N_F=N_c$. As shown above, see
\eqref{(12.2.7)}, {\it they all are in the HQ (heavy quark) phase,
i.e. not higgsed but confined} by $SU(N_c)$ SYM. Lowering the scale
$\mu$ from $\mu=\la$ down to $\mu=m^{\rm pole}_{Q}\ll\la$, the
Lagrangian at $\mu=m^{\rm pole}_{Q}$ has the form
\bq
K_{N_F=N_c}=z_Q(\la,\mu=m^{\rm pole}_{Q}){\rm Tr}_{N_F=N_c}\Bigl
(Q^\dagger e^{V} Q+Q\ra
{\ov Q} \Bigr )\,, \label{(12.2.1.1)}
\eq
\bbq
{\cal W}_{N_F=N_c}=\frac{2\pi}{\alpha(\mu=m^{\rm pole}_{Q})} S+ m_Q
\sum_{i=1}^{N_c} M^i_i\,, \quad z_Q(\la,\mu=m^{\rm pole}_{Q} \ll\la)\sim
\Bigl(\frac{\mu=m^{\rm pole}_{Q}}{\la} \Bigr
)^{\gamma^{\rm str}_Q \,>\, 2}\ll 1,
\eeq
\bbq
\quad m^{\rm pole}_{Q}=\frac{m_Q}{z_Q(\la,\mu=m^{\rm pole}_{Q})}\sim
\la \Bigl
(\frac{m_{Q}}{\la} \Bigr )^{0\, <\, \frac{1}{1+\gamma^{\rm str}_Q}\,
<\,\frac{1}{3}}\ll\la\,,
\eeq
where $\gamma^{\rm str}_Q$ is the quark anomalous dimension, see
\eqref{(12.2.1.2)}.

As explained in section 7 of \cite{ch1} or in section 3 of
\cite{ch21}, in the considered UV-free ${\cal N}=1$ SQCD with
$N_F=N_c$, if light quarks are not higgsed at scales $\mu > \la$, see
\eqref{(12.2.7)}, then this theory {\it enters smoothly} at $\mu < \la$
into  the perturbative strong coupling regime with the gauge coupling
$a(\mu\ll\la)=N_c
g^2(\mu\ll\la)/8\pi^2\gg 1$ and with effectively massless all quarks
and gluons at $m^{\rm pole}_{Q} < \mu < \la$. The perturbatively
exact  NSVZ $\beta$ function \cite{NSVZ-1} looks then as
\bq
\frac{d a(\mu\ll\la)}{d\ln \mu} = \beta(a)=-\, \frac{a^2}{1-a}\,
\frac{(3N_c-N_F)-N_F\gamma^{\rm str}_Q(a)} {N_c}\,\,\, \ra \,\,\, -\,
\nu\, a\,<\, 0 \label{(12.2.1.2)}
\eq
\bbq
\nu=\Bigl [\frac{N_F}{N_c}(1+\gamma^{\rm str}_Q) - 3\Bigr ]=\Bigl
[\gamma^{\rm str}_Q-2\Bigr ]={\rm const} > 0\,, \quad
a(\mu\ll\la)\sim\Bigl (\frac{\la}{\mu} \Bigr )^{\nu\, >\, 0}\gg 1\,.
\eeq
In section 7 of \cite{ch1} the values $\gamma^{\rm
str}_Q=(2N_c-N_F)/(N_F-N_c) > 1,\,\,\nu=(3N_c-2N_F)/(N_F-N_c) > 0$ 
at $\mu\ll\la$ and $N_c < N_F < 3N_c/2$ have  been found 
from matching of definite two point correlators in the
direct $SU(N_c)$ theory and in Seiberg's dual \cite{S2}. 
But there is no guarantee that  this expresion is applicable
 in our case here with $N_F = N_c$. So that,  unfortunately, 
we cannot give the concrete value $\gamma^{\rm str}_Q$.
(Nevertheless, the behavior $\gamma^{\rm str}_Q\ra N_c/|N_F-N_c|$ at
$N_F\ra N_c$ looks very appropriate).  But, as will be shown below, for our 
purposes it will be sufficient  to  have the only condition $\nu > 0$ in
\eqref{(12.2.1.2)}.

At $\mu <  m^{\rm pole}_Q$ all quarks decouple as heavy and there
remains at lower energies $SU(N_c)$ SYM in the perturbative strong
coupling branch. From the NSVZ  $\beta$ function \cite{NSVZ-1}
\bbq
\frac{d a^{(str,\, pert)}_{SYM}(\mu\gg\lym)}{d\ln\mu}= -\,\,
\frac{3\,\Bigl (a^{(str,\, pert)}_{SYM}(\mu\gg\lym)\Bigr )
^2}{1-a^{(str,\, pert)}_{SYM}(\mu\gg\lym)}\ra 3\, a^{(str,\,
pert)}_{SYM}(\mu)\,,
\eeq
\bq
a^{(str,\, pert)}_{SYM}(\mu\gg\lym)\sim\Bigl (\frac{\mu}{\lym}\Bigr
)^3 \gg 1\,,
\quad a^{(str,\, pert)}_{SYM}(\mu\sim\lym)={\cal O}(1)\,.
\label{(12.2.1.3)}
\eq
The scale factor $\lym$ of the gauge coupling is determined from
matching, see
\eqref{(12.2.1.1)},\eqref{(12.2.1.3)}
\bq
a_{+}(\mu=m^{\rm pole}_Q)=\Bigl (\frac{\la}{m^{\rm pole}_Q} \Bigr
)^{\nu}=a^{(str,\, pert)}_{SYM}(\mu=m^{\rm pole}_Q)=\Bigl
(\frac{m^{\rm
pole}_Q)}{\lym} \Bigr )^3\, \ra \label{(12.2.1.4)}
\eq
\bbq
\ra \,\,\lym^3=\Bigl (\la^{2N_c} m^{N_c}_Q \Bigr
)^{1/N_c}=m_Q\la^2,\,\,\,
\eeq
takes its {\it universal value}  as it should be, see \eqref{(12.2.3)}. 
Besides, as a check of  self-consistency, see
\eqref{(12.2.1.1)},\eqref{(12.2.1.2)},\eqref{(12.2.1.4)}
\bq
\Bigl (\frac{\lym}{m^{\rm pole}_Q}\Bigr )^3\sim\Bigl (\frac{m_Q}{\la}
\Bigr )^{\omega\, >\, 0}\ll 1\,,\quad \omega= \frac{\nu >
0}{(1+\gamma^{\rm str}_Q)} > 0 \,, \label{(12.2.1.5)}
\eq
as it should be. At $\mu < \lym$ the perturbative RG evolution stops
due to nonperturbative effects $\sim\lym$ in the pure ${\cal
N}=1\,\,\,SU(N_c)$ SYM.\\

After integrating {\it inclusively} all quarks as heavy at $\mu <
m^{\rm pole}_{Q}$ and then all gluons at $\mu < \lym$ via the VY
procedure \cite{VY}, the low energy superpotential at
$\mu <\lym$ 
\bq
{\cal W}=N_c\Bigl (\la^{2N_c} m^{N_c}_Q \Bigr )^{1/N_c}=N_c
m_Q\la^2=N_c \lym^3 \label{(12.2.1.6)}
\eq
takes its {\it universal value}  as it should be.

Therefore, the mass spectrum of the direct theory for this case looks
as follows.\\
a)\,All quarks are in the HQ (heavy quark) phase. i.e. {\it not
higgsed but confined}, and decouple as heavy at $\mu < m^
{\rm pole}_Q$, see \eqref{(12.2.1.1)}.\\
b)\, There is a number of quarkonia with the typical mass scale ${\cal
O}(m^{\rm pole}_Q)\gg \lym$. {\it This mass scale is checked by
\eqref{(12.2.1.4)},\eqref{(12.2.1.5)}. Integrating inclusively all
these quarkonia (i.e. equivalently, integrating inclusively all
quarks as heavy at $\mu= m^{\rm pole}_Q$), we obtain the well known
beforehand right value of $\lym$, see \eqref{(12.2.1.4)}}.\\ 
c)\, There is a number of $SU(N_c)$ gluonia with the typical mass scale 
${\cal O}(\lym)$. In the chiral limit $m_Q\ra 0$, the masses of {\it all}
quarks, gluons and hadrons {\it coalesce to zero}. Because the tension 
of the  $SU(N_c)$ confining string is $\sigma^{1/2}\sim\lym\ra 0$, 
{\it there  is no confinement} in this chiral limit, see Introduction in
section 12.1 and  footnote \ref{(f21)}. For more details see section 12.5\,.

\numberwithin{equation}{subsubsection}

\subsubsection{Light unequal mass  quarks}

Consider now unequal quark masses \,:
$\mi=\ml,\,\,i=1,...,N_L,\,\,\,m_{Q,i}=\mh,\,\, i=N_L+1,...,N_F
,\,\,\,N_L+N_H=N_F=N_c$, with $0 < m_L\ll m_H\ll\la$. In this case,
see \eqref{(12.2.4)}-\eqref{(12.2.5)},
\bq
\langle M_L\rangle=\langle M^L_L\rangle=\frac{\langle
S\rangle_{N_c}}{\ml}=\la^2\Bigl ( \frac{1}{r}\Bigr
)^{\frac{N_H}{N_c}},\,\,\langle M_H\rangle=\langle M^H_H\rangle=
\frac{\langle S\rangle_{N_c}}{\mh}=\la^2\,r^{\frac{N_L}{N_c}},\,\,
\label{(12.2.2.1)}
\eq
\bbq
\langle M_L\rangle\gg\la^2,\,\,\,\,\, \langle
M_H\rangle\ll\la^2.\,\,\, \langle M_L\rangle^{N_L}\langle
M_H\rangle^{N_H}=\la^{2 N_c},\,\, 0 < r=\ml/\mh\ll 1,\,\, \langle
B\rangle= \langle{\ov B}\rangle=0\,.
\eeq
\bbq
(M_{L,H})^i_j=(M_{L,H})^i_j(\rm adj)+ \delta^i_j\, M_{L,H}({\rm
singl}),\,\, M_{L,H}({\rm
singl})=\frac{1}{N_{N_L,N_H}}\sum_{i=1}^{N_L,N_H} M^i_i\,\,,
\eeq
\bbq
\langle (M_{L,H})^i_j(adj)\rangle=0, \,\, \langle M_{L,H}({\rm
singl})\rangle=\langle M_{L,H}\rangle\,.
\eeq

In this case, at not too large $N_c$, see \cite{ch21}, \,
$Q^L,\,{\ov Q}_L$ quarks are higgsed at the large scale 
$\mu=\mu_{\rm gl,L}\gg\la$
in the weak coupling region. There are $N_L(2N_c-N_L)$ heavy gluons
(and their superpartners) with masses (ignoring logarithmic
renormalization factors),
\bq
\frac{\mu^2_{\rm gl,\, L}}{\la^2}\sim \frac{1}{N_c}\frac{\rho^2_{\rm
higgs}}{\la^2}\sim \frac{1}{N_c}\frac{\langle M_L({\rm
singl})\rangle}{\la^2}\sim \frac{1}{N_c}\frac{\langle
S\rangle_{N_c}}{m_L\la^2}\sim\frac{1}{N_c}\Bigl ( \frac{1}{r}\Bigr
)^{\frac{N_H}{N_c}}\gg 1\,,\label{(12.2.2.2)}
\eq
\bbq
\langle {\hat Q}^i_{\beta}(x)\rangle =\langle {\hat
Q}^i_{\beta}(0)\rangle=\delta^i_{\beta}\,\rho_{\rm higgs}\,, \quad
\frac{\rho^2_{\rm higgs}= \langle M_L({\rm
singl})\rangle}{\la^2}=\Bigl ( \frac{1}{r}\Bigr )^{\frac{N_H}{N_c}}
\gg 1\,,\quad i=1...N_L\,,\,\,\,\beta=1...N_c,
\eeq
where ${\hat Q}^i_{\beta}(x)$ is the {\it colored} gauge invariant
quark field and  $\rho_{\rm higgs}$ is the gauge invariant order 
parameter, see  section 2 in \cite{ch21}.

In the chiral limit $m_{L,H}\ra 0$ with fixed ratio $0 < r=m_L/m_H
\ll1$, the Lagrangian \eqref{(12.2.2)} is invariant under global
$SU(N_{F})_{left}\times SU(N_{F})_{right}$ flavor
transformations under which quarks transform in the fundamental 
and  anti-fundamental representations:\, $Q^{left}_{\beta},\, {\ov
Q}^{\beta}_{right}$. The generators of global V- or A- flavor
transformations are, respectively, $(right+left)$ and $(right-left)$.
The upper and lower flavor indices denote fundamental and
anti-fundamental representations. Besides, in what follows, e.g.
$Q^{L,H}_{\beta}$ and ${\ov Q}^{\beta}_{L,H}$ will denote left $Q$
quarks with flavor indices $L=(i=1,...,N_L)$ or
$H=(i=N_L+1,...,N_F)$,  and similarly for right $\ov Q$ quarks. And
similarly for the color  indices:\, $L=(\beta=1,...,N_L)$ or $H=(\beta=
N_L+1,...,N_c)$, e.g.  $Q^L_L$.

At $1\leq N_L\leq N_c-2$ and small nonzero $m_{L,H}$ there remains at
lower
energies $\mu < \mu_{\rm gl.L}$\,:\, $SU(N^\prime_c=N_c-N_L=N_H\geq
2)$ gauge group with the scale factor $\Lambda_H$ of its gauge
coupling, $N^2_L$ colorless genuine complex pseudo-Goldstone pions
$\pi^L_L(adj)$ (sitting on $\Lambda_H$) remained from
higgsed $Q^L_L,\, {\ov Q}^L_L$ quarks, $2N_L N_H$ equal mass hybrid
genuine complex pseudo-Goldstone pions $\pi^H_L,\,\pi^L_H$ (in
essence, these are non-active $Q^H_L,\, {\ov Q}^L_H$ quarks with
higgsed L-colors), and $2 N^2_H$ of still active non-higgsed but
confined  ${\mathbf{Q}^H_H},\, {\mathbf{\ov Q}^H_H}$ quarks with $N_H$ 
flavors and $N_H$ non-higgsed colors.

The fields of hybrid pions canonically normalized at $\mu=\la$ are
defined as
\bq
\pi^H_L=\frac{1}{\langle
M_L\rangle^{1/2}}\sum_{\beta=1}^{N_L}\langle{\ov
Q}_L^{\beta}\rangle Q^H_{\beta},\quad \pi^L_H=\frac{1}{\langle
M_L\rangle^{1/2}}
\sum_{\beta=1}^{N_L}{\ov Q}^{\beta}_H\langle Q^L_{\beta}\rangle, \,\,
\label{(12.2.2.3)}
\eq
\bbq
H=(N_L+1,...,N_F),\quad \beta=(1,...,N_L).\,\,
\eeq
After all heavy particles with masses $\sim \mu_{\rm gl,L}$ decoupled
at $\mu <  \mu_{\rm gl,L}$, the scale factor $\Lambda_H$ of the remained
$SU(N_H)$ gauge theory is
\bq
\langle\Lambda_H^{3 N^\prime_c-N^\prime_F=2 N_H}\rangle=\frac{\la^{2
N_c}}{\langle
M_L\rangle^{N_L}},\,\, \langle\Lambda_H^2\rangle= \la^2
r^{N_L/N_c}=\langle M_H\rangle \ll \la^2\,,\,\,
\Lambda^{2N_H}_H=\frac{\la^{2 N_c}}{\det{M^L_L}}\,.
\label{(12.2.2.4)}
\eq

The heavy hybrid gluons $(A_{\mu})^{L}_{H},\,\,(A_{\mu})_{L}^{H}$
with  masses $\mu_{\rm gl,L}\gg\la$ and lighter active H-quarks 
${\mathbf{Q}^H_H},\,{\mathbf{\ov Q}^H_H}$ with $N_H\geq 2$
non-higgsed colors are {\it weakly confined} by $SU(N_H)$ SYM 
(the  string tension is
$\sigma^{1/2}\sim\langle\lym\rangle$, much smaller than their
perturbative pole masses). These hybrid gluons decouple as heavy at
$\mu < \mu_{\rm gl,L}$. Quarks ${\mathbf{Q}^H_H},\,{\mathbf{\ov Q}^
H_H}$   are in the HQ (heavy quark) phase, i.e. not higgsed but confined 
and decouple at $\mu < m^{\rm pole}_H\ll\la\ll \mu_{\rm gl,L}$.

The Lagrangian of lighter particles looks at $\mu= \mu_{\rm gl,L}$
as
\bq
K=2 z_Q(\la,\mu_{\rm gl,L}) {\rm Tr}\sqrt{(M^L_L)^\dagger
M^L_L}+z_Q(\la,\mu_{\rm
gl,L}){\rm Tr}\Bigl ( (\pi^H_L)^\dagger \pi^H_L+(\pi_H^L)^\dagger
\pi_H^L\Bigr ) + \label{(12.2.2.5)}
\eq
\bbq
+ z_Q(\la,\mu_{\rm gl,L}){\rm Tr}\Bigl ( (\hh)^\dagger {\mathbf{Q}^H_H} +
({\mathbf{\ov Q}^H_H})^\dagger {\mathbf{\ov Q}^H_H} \Bigr )\,,
\eeq
\bbq
{\cal W}={\cal W}_{SU(N_H)}^{SYM} +m_L{\rm Tr} M_L + m_H {\rm Tr}
\Bigl ( \pi^H_L \pi^L_H+{\mathbf{\ov Q}^H_H}{\mathbf{Q}^H_H}   \Bigr)\,,
\eeq
where $z_Q(\la,\mu_{\rm gl,L})\sim \log^{1/2} (\mu_{\rm gl,L}/\la)\gg
1$ is the logarithmic renormalization factor of quarks.

{\bf A)}\, If $\la\gg m_H \gg \langle\Lambda_H\rangle=\langle
M_H\rangle^{1/2}$,  see \eqref{(12.2.2.4)}\,: active HH-quarks 
${\mathbf{Q}^H_H},\, {\mathbf{\ov Q}^H_H}$ 
are in the HQ(heavy quark) phase and decouple as heavy at $\mu < 
m^{\rm pole}_H=m_H/z_Q(\la, m^{\rm pole}_H)\gg \langle\Lambda_H
\rangle$ {\it  in the weak coupling regime}, where $z_Q(\la, m^{\rm pole}_H)
\ll 1$  is the logarithmic renormalization factor of active HH-quarks.

There remain at lower energies: $SU(N_H)$ SYM theory, $N_L^2$ light
genuine colorless complex $\pi^L_L$ pions \eqref{(12.2.2.5)},
\eqref{(12.2.2.8)} and light genuine hybrid complex pions $\pi^H_L,
\pi^L_H$ \eqref{(12.2.2.3)}.

As a check, the gluon mass due to potentially possible higgsing of
active HH-quarks looks as ( with logarithmic accuracy), see
\eqref{(12.2.2.4)},
\bq
m^{\rm pole}_H=\frac{m_H}{z_Q(\la, m^{\rm pole}_H)}\sim m_H\gg
\langle\Lambda_H\rangle\gg\langle\lym\rangle\,, 
\label{(12.2.2.6)}
\eq
\bbq
z_Q(\la, m^{\rm pole}_H)\sim\log^{-1/2} (\la/m^{\rm pole}_H)\ll 1\,, \,\,\,
\frac{\mu^2_{\rm  gl,\,H}}{m^{2,\,\rm pole}_H}\sim\frac{\langle
M_H\rangle=\langle\Lambda_H^2\rangle}{m^{2,\,\rm pole}_H
\sim m^2_H}\ll 1\,,
\eeq
where $z_Q(\la, m^{\rm pole}_H)\ll 1$ is the quark perturbative
logarithmic renormalization factor.

This shows that active HH-quarks ${\mathbf{Q}^H_H},\, {\mathbf{\ov Q}^H_H}$ 
are in the HQ (heavy quark) phase, i.e. not higgsed but confined by $SU(N_H)$
SYM.\\

After integrating inclusively decoupled at $\mu < m^{\rm pole}_H$
active HH-quarks, the scale factor ${\hat\Lambda}_{N_H}$ of the gauge
coupling of remained $SU(N_H)$ SYM is, see \eqref{(12.2.2.4)},
\bq
\langle{\hat\Lambda}_{N_H}^{3}\rangle=\langle\Lambda_H^{2}\rangle
m_H=\langle
S\rangle=\la^2 \Bigl ( m_L^{N_L} m_H^{N_H} \Bigr
)^{1/N_c}=\langle\lym^{3}\rangle\,,\quad
{\hat\Lambda}_{N_H}^{3}=\Biggl (\frac{\la^{2N_F} m_H^{N_H}}{\det
M_L}\Biggr ) ^{1/N_H}\,, \label{(12.2.2.7)}
\eq
\bbq
\Bigl (\frac{m^{\rm pole}_H\sim m_H}{\langle\Lambda_H^{2}\rangle}
\Bigr ) \gg  1\,\ra\, r^{N_L/N_c}\ll \frac{m^2_H}{\la^2}\ll 1 \, \ra\, 
\Bigl (\frac{\langle\lym\rangle}{m_H} \Bigr )^3=
\Bigl (\frac{\la}{m_H}\Bigr )^2 \Biggl [ r^{N_L/N_c} \ll \frac{m^2_H}
{\la^2}\Biggr ] \ll 1\,.
\eeq

Lowering the scale down to $\sim\langle\lym\rangle$ and integrating
inclusively all $SU(N_H)$ gluons at $\mu=\langle \lym\rangle$ via the
VY-procedure \cite{VY},
the lower energy Lagrangian at $\mu=\langle \lym\rangle$ looks as
\bq
K=2 z_Q(\la,\mu_{\rm gl,L}) {\rm Tr}\sqrt{(M^L_L)^\dagger
M^L_L}+z_Q(\la,\mu_{\rm
gl,L}){\rm Tr}\Bigl ( (\pi^H_L)^\dagger \pi^H_L+(\pi_H^L)^\dagger
\pi_H^L\Bigr )\,, \label{(12.2.2.8)}
\eq
\bbq
{\cal W}=\Biggl [ N_H ({\hat\Lambda}_{N_H})^3=N_H\Biggl
(\frac{\la^{2N_F} m_H^{N_H}}{\det M_L}\Biggr )^{1/N_H}\Biggr ]+
m_L{\rm Tr} M_L + m_H {\rm Tr}\Bigl ( \pi^H_L \pi^L_H \Bigr)\,, \quad
\langle {\cal W}\rangle=N_c\langle S\rangle_{N_c}\,.
\eeq
\bbq
M^L_L=\langle M_L\rangle+\langle M_L\rangle^{1/2}\pi^L_L\,,\quad
2\,{\rm Tr}\sqrt{(M^L_L)^{\dagger} M^L_L}\ra {\rm Tr}\Bigl
(\,(\pi^L_L)^ \dagger \pi^L_L \Bigr )\,,
\eeq
where $\pi^L_L$ and $\pi^L_H,\, \pi^H_L$ fields are canonically
normalized at
$\mu=\la$, see \eqref{(12.2.2.3)}. At $\mu < \mu_{\rm gl,L}$ the
RG-evolution of
$N^2_L$ pions $M^L_L$ and $2 N_LN_H$ hybrid pions $\pi^H_L,\,\pi^L_H$
is frozen, see \eqref{(12.2.2.5)}.

From \eqref{(12.2.2.8)}, the particle masses are
\bbq
\mu^{\rm pole}\Bigl (\pi^L_L ({\rm adj}) \Bigr )=\frac{2 m_L}{
z_Q(\la,\mu_{\rm gl,L})} \ll \langle\lym\rangle\,,\quad\mu^
{\rm pole}\Bigl (\pi^L_L ({\rm singl})
\Bigr )= \Bigl ( \frac{N_L}{N_H}+
1\Bigr )\frac{2 m_L}{ z_Q(\la,\mu_{\rm \,gl,L})}\,,
\eeq
\bbq
M^L_L ({\rm singl})=\frac{1}{N_L}\sum_{i=1}^{N_L} M^i_i,\,\,\,\,
\langle M^L_L
({\rm singl})\rangle=\la^2\Bigl ( \frac{1}{r}\Bigr
)^{\frac{N_H}{N_c}}\,,\quad
\eeq
\bq
\mu^{\rm pole} (\pi^L_H)=\mu^{\rm pole} (\pi_L^H)=\frac{m_H}{
z_Q(\la,\mu_{\rm
gl,L})}=m^{\rm pole}_{H,hybr}\,.\,\,\label{(12.2.2.9)}
\eq

The mass spectrum looks in this case as follows. - \\
a)\, There are $N_L(2N_c - N_L)$\, ${\cal N}=1$ multiplets of real
heavy gluons with masses $\mu_{\rm gl,L}\gg\la$, see
\eqref{(12.2.2.2)}. $N_L^2$ of $(A_\mu)^L_L$ gluons are weakly
coupled  and not confined, while $2 N_L N_H$ heavy hybrids
$(A_{\mu})^{L}_{H},\,\ (A_{\mu})_{L}^{H}$ are weakly coupled and
weakly confined by the low energy $SU(N_H)$ SYM.\\
b)\,There are $2 N_L N_H$ $\,{\cal N}=1$ multiplets of complex
genuine  hybrid pseudo-Goldstone pions $\pi^L_H, \pi^H_L$
with masses $\mu^{\rm pole}(\pi^L_H)=\mu^{\rm pole}(\pi_L^H)=
m^{\rm pole}_{H,hybr}$ which are not confined, see \eqref{(12.2.2.9)}. \\
c)\, There is a number of hadrons made from weakly coupled and weakly
confined by  $SU(N_H)$ SYM non-relativistic active quarks ${\mathbf{Q}
^H_H},\, {\mathbf{\ov Q }^H_H}$ 
with non-higgsed colors, with the typical mass scale ${\cal O} (m^{\rm
pole}_H)\gg\langle\Lambda_H\rangle=\langle M\rangle^{1/2}_H$, see
\eqref{(12.2.2.6)}. Their genuine quark-antiquark bound states have
masses $\approx 2 m^{\rm pole}_H$, and $N^2_H-1$\, ${\cal N}=1$
multiplets of genuine complex $\pi^H_H(adj)$ pseudo-Goldstone 
pions are among them.\\
d)\,There is a number of gluonia of $SU(N_H)$ SYM with the typical
mass scale ${\cal O} (\langle\lym\rangle)$, see \eqref{(12.2.2.7)}.\\
e)\, The lightest are $N_L^2\,\, \,{\cal N}=1$ multiplets of genuine
complex pseudo-Goldstone pions $\pi^L_L(adj)$ with masses
\eqref{(12.2.2.9)}.

Because the tension of the $SU(N_H)$ confining string is
$\sigma^{1/2}\sim\langle \lym
\rangle\ra 0$ \eqref{(12.2.2.7)}, {\it there is no confinement in the
chiral limit $m_{L,H}\ra 0$ with fixed $r=m_L/m_H \ll 1$}.\\

\vspace*{1mm}

{\bf B)}\, If $m_H \ll \langle\Lambda_H\rangle\ll\la$, then active
HH-quarks ${\mathbf{Q}^H_H},\, {\mathbf{\ov Q}^H_H}$ with 
non-higgsed colors are also in the  HQ-phase, i.e. not higgsed but 
confined, but theory enters now at  $\langle\lym\rangle < \mu < 
\langle\Lambda_H\rangle$ the perturbative  strong
coupling regime with the large gauge coupling $a(\langle\lym
\rangle\ll \mu\ll\langle  \Lambda_H\rangle)\gg 1$. These active 
HH-quarks decouple then as
heavy \, at $\mu < m^{\rm pole}_H$, see section 3 in \cite{ch21}.

{\it The only difference with previous case $m_H\gg \Lambda_H$ is in
the value of perturbative pole mass} $m^{\rm pole}_H$,
\bq
m_H(m^{\rm pole}_H < \mu < \langle\Lambda_H\rangle)\sim
\frac{m_H}{\Biggl [
z_Q (\langle\Lambda_H\rangle, m^{\rm pole}_H < \mu <
\langle\Lambda_H\rangle)
\sim \Bigl (\mu/\langle\Lambda_H\rangle\Bigr ) ^{\gamma^{\rm str}_Q\,
>\, 2} < 1 \Biggr ]}\,,
\label{(12.2.2.10)}
\eq
\bbq
m^{\rm pole}_H\sim\frac{m_H}{z_Q(\langle\Lambda_H\rangle, m^{\rm
pole}_H)}\,\ra \,\langle\lym\rangle \ll m^{\rm pole}_H\sim
\langle\Lambda_H\rangle\Bigl
(\frac{m_H}{\langle\Lambda_H\rangle} \Bigr )^{0\, <
\frac{1}{1+\gamma^{\rm
str}_Q}\, < \frac{1}{3}\,}\ll \langle\Lambda_H\rangle\,.
\eeq

After these active HH-quarks decoupled at $\mu=m^{\rm pole}_H$\,
\eqref{(12.2.2.10)},
the scale factor ${\langle\hat\Lambda_{N_H}}\rangle$ of the $SU(N_H)$
SYM gauge
coupling is determined from matching, see \eqref{(12.2.1.2)},
\eqref{(12.2.1.3)},
\bq
a_{+}(\mu=m^{\rm pole}_H)=\Bigl
(\frac{\langle\Lambda_{H}\rangle}{m^{\rm pole}_H}
\Bigr )^{\nu=(\gamma^{\rm str}_Q-2)\,>\,0}=a^{(str,\,
pert)}_{SYM}(\mu=m^{\rm
pole}_H)=\Bigl (\frac{m^{\rm
pole}_H)}{\langle\hat\Lambda_{N_H}\rangle} \Bigr )^3\, \ra
\label{(12.2.2.11)}
\eq
\bbq
\ra \,\,\langle{\hat\Lambda_{N_H}}^3\rangle\sim \la^2 r^{N_L/N_c} 
m_H =\la^2 \Bigl
( m_L^{N_L} m_H^{N_H} \Bigr )^{1/N_c}=\langle\lym^3\rangle\,,
\eeq
as it should be, see \eqref{(12.2.2.7)}.

Besides, as for a potentially possible higgsing of these HH-quarks, see
\eqref{(12.2.1.1)}-\eqref{(12.2.1.4)},\eqref{(12.2.2.4)},\\\eqref{(12.2.2.10)}
\bq
\hspace*{-5mm} K_{N_H}(\mu=\langle\Lambda_H\rangle)=Z_H {\rm
Tr}_{N_H}\Bigl (
({\mathbf{Q}^H_H})^\dagger {\mathbf{Q}^H_H} + ({\mathbf{\ov Q}^H_H})
^\dagger {\mathbf{\ov Q}^H_H} \Bigr ), \,\,Z_H\sim \log^{-1/2}
(\la/\langle\Lambda_H\rangle) \ll 1, \quad \label{(12.2.2.12)}
\eq
\bbq
\frac{\mu^2_{{\rm gl},\,H}(m^{\rm pole}_H < \mu <
\langle\Lambda_H\rangle)}{\mu^2}\sim \frac{Z_H}{\mu^2}\Biggl [
a(\langle\Lambda_H\rangle, \mu)
z_Q(\langle\Lambda_H\rangle,\mu)\sim\frac{\mu^2}{\langle\Lambda^2_H
\rangle}\Biggr ]\langle M_H\rangle \sim Z_H \ll 1\,.
\eeq

It is seen from \eqref{(12.2.2.12)} that, even if these HH-quarks
were  higgsed, they  would be unable to give such mass to $SU(N_H)$ 
gluons which would  stop  the
perturbative RG-evolution and there would be no pole in the gluon
propagator at $m^{\rm pole}_H < \mu < \langle\Lambda_H\rangle$. And
$\mu^2_{{\rm gl},H}(\mu=m^{\rm  pole}_H)/m^{\rm 2,\, pole}_H \ll
1$. This shows that these HH-quarks are really in the HQ (heavy
quark)  phase, i.e. not higgsed but confined by $SU(N_H)$ SYM. 
At $\mu < m^{\rm pole}_H$ these active HH-quarks decouple.

And  as a check of self-consistency, see \eqref{(12.2.2.10)},
\bbq
\Bigl (\frac{m_H}{\langle\Lambda_H\rangle} \Bigr ) \ll 1\,\ra\,
\Biggl[ \Bigl (\frac{1}{r}\Bigr )^{N_L/2N_c}\ll\frac{\la}{m_H}\,\Biggr
] \ra\Bigl
(\frac{{\langle\hat\Lambda_{N_H}\rangle}=\langle\lym\rangle}
{m^{\rm pole}_H} \Bigr )^{3}\sim
\eeq
\bq 
\Biggl [\frac{m_H}{\la} \Bigl (\frac{1}{r}\Bigr )^{N_L/2N_c}\ll 1
\Biggr ]^{\frac{(\gamma^{\rm str}_Q-2) =\, \nu}{(1+\gamma^{\rm
str}_Q)}\, > \,\, 0}\ll 1, \label{(12.2.2.13)}
\eq
as it should be.

The mass spectrum in this case '{\bf B}' differs from '{\bf A}' only
in the HH-sector which is strongly coupled now. Confined by $SU(N_H)$
SYM active quarks ${\mathbf{Q}^H_H},\, {\mathbf{\ov Q}^H_H}$ are strongly 
coupled and weakly confined and form a number of HH-hadrons with 
the typical mass scale $O (m^{\rm pole}_H)$ \eqref{(12.2.2.10)}.  
And $N^2_H-1\,  {\cal {N}}=1$  multiplets of genuine complex $\pi^H_H(adj)$ 
pseudo-Goldstone pions are among these HH-hadrons.

{\it The typical mass scale $O(m^{\rm pole}_H)$\,\eqref{(12.2.2.10)} 
of HH-hadrons is checked by \eqref{(12.2.2.11)}.
Integrating inclusively all these HH-hadrons (i.e. equivalently,
integrating inclusively all ${\mathbf{Q}^H_H},\, {\mathbf{\ov Q}^H_H}$ 
quarks as heavy at $\mu=
m^{\rm pole}_H$)\, \eqref{(12.2.2.10)}, we obtain the well known
beforehand right value of $\langle\lym\rangle$}, see \eqref{(12.2.2.11)}.\\

In the chiral limit $m_{L,H}\ra 0$ with fixed $0 < r=m_L/m_H\ll
1$,\,\, $\langle {\hat Q}^i_
{\beta} (0)\rangle=\delta^i_{\beta}\,\rho_{\rm higgs}$ in
\eqref{(12.2.2.2)} breaks {\it spontaneously} global $SU(N_F)_A\ra
SU(N_H)_A$, so that there will be $N_L^2$ $\,\,{\cal N}=1$ multiplets
of massless genuine complex Goldstone pions $\pi^L_L$ (see section12.4)
and $2N_L N_H$ $\,\,{\cal N}=1$ multiplets of massless genuine
complex  Goldstone hybrid pions $\pi^L_H+\pi^H_L$. (The supersymmetry
doubles  the number of real Goldstone particles).

All particle masses in the HH-sector also coalesce to zero, in spite
of the strong gauge coupling. Because the tension of the $SU(N_H)$
confining string is $\sigma^{1/2}\sim\langle\lym\rangle\ra 0$
\eqref{(12.2.2.11)}, {\it there is no confinement} in this chiral
limit. For more details see section 12.4.

\numberwithin{equation}{subsection}

\subsection{Dual Seiberg's theory with $N_F=N_c$}

Let us recall first once more that in his paper \cite{S1} N. Seiberg
proposed the
$N_F=N_c$\, ${\cal N}=1$ SQCD theory with $N^2_F$ mesons $M^i_j$ and
two $B,\,{\ov
B}$ baryons (with one constraint), and $N_F=N_c+1$ SQCD theory with
$N_F^2=(N_c+1)^2$ mesons $M^i_j$ and $2 N_F=2 (N_c+1)$ baryons
$B_i,\,\,{\ov B}^j$ as {\it the low energy forms at the scale $\mu <
\la$ of, respectively, direct SQCD theories with $N_F=N_c$ and
$N_F=N_c+1$ light quark flavors with $m_{Q,i}\ll\la$}. In \cite{IS}
the regime at $N_F=N_c$ and $\mu < \la$ was called ''confinement with
the (spontaneous at $m_{Q,i}\ra 0$) chiral symmetry breaking'' and
those at $N_F=N_c+1$ as ''confinement without the (spontaneous) 
chiral symmetry  breaking''\,.

This implied that in these two strongly coupled at $\mu\sim\la$
direct theories the light direct quarks and gluons are confined by
strings with the tension $\sigma^{1/2}\sim\la$ and form hadrons 
with masses $\sim\la$, and there remain at lower energies only 
corresponding light  mesons and baryons. (These regimes were 
called 
latter in the literature as 'S-confinement'\,).\\

This proposal and interpretation were criticized in \cite{ch1} (see
pages 18 and 19 in arXiv:0712.3167\, [hep-th]) and in \cite{ch3} 
(see Introduction  and sections
7-9 in arXiv:0811.4283 [hep-th]). The reason was the following. The
confinement originates {\it only} from SYM sector, see footnote
\ref{(f17)}. And the $SU(N_c)$ SYM theory has only
one dimensional parameter $\lym=(\la^{3 N_c-N_F}
\Pi_{i=1}^{N_F}{m_{Q,i}})^{1/3 N_{c}}\ll\la$ at $m_{Q,i}\ll\la$.
Therefore, it can not give the string with the tension
$\sigma^{1/2}\sim\la$, but only with $\sigma^{1/2}\sim\lym\ll\la$.
And  e.g. direct light equal mass quarks and direct gluons {remain 
light and  confined by strings with $\sigma^{1/2}\sim\lym\ll\la$
and can't form hadrons with masses} $\sim\la$, see section 12.2\,.

For this reason, in papers \cite{ch1,ch3} and in \cite{ch23} we
considered  at $\mu < \la$ the direct theories with quarks and gluons
and  proposed by N. Seiberg \cite{S1} dual theories with mesons and
baryons  as {\it independent theories}. Then, the mass spectra of the
direct and dual theories have to be calculated and compared, to see
whether they  are equivalent or not.

It was shown explicitly in Appendix 'B' in \cite{Session} how the
proposed by N.  Seiberg low energy dual theory with $N_F^2=
(N_c+1)^2$ mesons $M^i_j$  and $2 N_F=2
(N_c+1)$ baryons $B_i,\,\,{\ov B}^j$ originates not as the low energy
form of the direct $SU(N_c)$ theory with $N_F=N_c+1$, but from his
dual $SU({\ov N}_c=N_F-N_c=2)$ SQCD theory \cite{S2} with 
$N_F=N_c+2$  dual quarks at the large mass of last direct quark
$m_{Q,i=N_c+2}=\la$. Similarly, it was shown in \cite{ch23}  
how the proposed by
N. Seiberg in \cite{S1} dual $N_F=N_c$ SQCD theory with $N_c^2-1$
mesons $M^i_j$ and $B,\,{\ov B}$ baryons (but with different Kahler
terms) originates not as the low energy form of the direct $SU(N_c)$
theory with $N_F=N_c$, but from the dual $N_F=N_c+1$ theory in
\cite{Session} at the large mass of last direct quark
$m_{Q,i=N_c+1}=\la$.

\numberwithin{equation}{subsubsection}

\subsubsection{Light equal mass quarks}

Let us remind now the properties of the Seiberg dual $\,{\cal N}=1$
SQCD theory with $N^2_F=(N_c+1)^2$ mesons $M^i_j$,\,\, $2 (N_c+1)$
baryons $B_i,\,\, {\ov B}^i$ and $\mi=m_Q \ll\la,\,\,
i=1,...,N_c,\,\,m_{Q,N_c+1}=\la$, which was obtained from Seiberg's
dual $SU({\ov N}_c=N_F-N_c=2)$ $\,\,N_F=N_c+2$ theory \cite{S2}
increasing mass of one quark in the direct $SU(N_c)\,\, N_F=N_c+2$
theory, see appendix 'B' in \cite{Session},
and \eqref{(12.2.1)} above.

The dual Kahler terms at $\mu=\la$ look as
\bq
{\wt K}_{N_c+1}= {\rm Tr}_{N_c+1}\frac{M^\dagger
M}{\la^2}+\sum_{i=1}^{N_F=N_c+1}
\Bigl ( (B^\dagger)^i B_i +({\ov B}^\dagger)_i\,{\ov B}^{\,i}\Bigr ),
\label{(12.3.1.1)}
\eq
and the dual superpotential is
\bq
{\wt {\cal W}}_{N_c+1}=m_Q\sum_{i=1}^{N_c} M^i_i +\la
M^{N_c+1}_{N_c+1}+{\rm Tr}_{N_c+1}\,({\ov B}\frac{M}{\la}
B)-\frac{\det_{N_c+1} M}{\la^{2N_c-1}}\,,\label{(12.3.1.2)}
\eq
where the baryons $B_i$ and ${\ov B}^i$ with $N_F=N_c+1$ flavors are
really the remained light dual quarks, see appendix 'B' in
\cite{Session}\,:
$B_i=\sum_{\alpha,\beta=1}^{2}\epsilon_{\alpha\beta}[\,\langle {\hat
q}^{\,\alpha}_{j=N_c+2}\rangle/\la=\delta^{\alpha,1}]\,{\hat
q}^{\,\beta}_i= {\hat q}^{\,\beta=2}_i,\,\,i=1,...,N_c+1$ (and
similarly ${\ov B}$).

After most dual particles with masses $\sim\la$ decoupled at $\mu <
\la$, see \eqref{(12.3.1.2)}, there remained only $N_c^2+1$ mesons
$M^i_j,\,\, i,j=1,...,N_F=N_c$ and $M^{N_c+1}_{N_c+1}$, and two
baryons $B=B_{i=N_c+1}$ and ${\ov B}={\ov B}^{\,i=N_c+1}$. The lower
energy Lagrangian looks as
\bq
{\wt K}_{N_c}(\mu=\la)= {\rm Tr}_{N_c}\frac{M^\dagger M}{\la^2}+
\frac{\Bigl
({M^{N_c+1}_{N_c+1}}\Bigr )^{\dagger}\,M^{N_c+1}_{N_c+1}}{\la^2}
+ (B^\dagger B+B\ra {\ov B})\,, \label{(12.3.1.3)}
\eq
\bq
{\wt {\cal W}}_{N_c}=M^{N_c+1}_{N_c+1}\Bigl (\la-\frac{\det_{N_c}
M}{\la^{2N_c-1}}
+\frac{{\ov B} B}{\la}\Bigr )+m_Q\sum_{i=1}^{N_F=N_c} M^i_i+...\,\,
\label{(12.3.1.4)}
\eq
Dots in \eqref{(12.3.1.4)} denote terms which are irrelevant in what
follows for our purposes.

The mean vacuum values from \eqref{(12.3.1.3)}.\eqref{(12.3.1.4)} look  
as, compare with \eqref{(12.2.1)}\,,\eqref{(12.2.4)}\,,\eqref{(12.2.1.6)},
\bq
\frac{\langle\det_{N_c} M\rangle=\Pi_{i=1}^{N_F=N_c}\langle
M^i_i\rangle}{\la^{2N_c-1}}=\la, \,\,\,\langle M^i_i\rangle=\la^2,
\,\,\, i=1,...,N_F, \,\,\,\langle B\rangle=\langle{\ov B}\rangle=0,
\label{(12.3.1.5)}
\eq
\bq
M^i_j=M^i_j(adj)+ \delta^i_j M_s ,\,\,\,
M_s=\frac{1}{N_c}\sum_{i=1}^{N_F=N_c} M^i_i,\,\,\,
\langle M^i_j(adj)\rangle=0,\,\,\, \langle M_s
\rangle=\la^2,\,\,\label{(12.3.1.6)}
\eq
\bq
\langle \frac{\partial {\wt {\cal W}}_{N_c}}{\partial M^i_i}\rangle=0
\ra \langle
M^{N_c+1}_{N_c+1}\rangle=m_Q\la=\frac{\langle S\rangle_{N_c}}{\la},
\label{(12.3.1.7)}
\eq
\bbq
\langle {\wt {\cal W}}_{N_c}\rangle=m_Q\sum_{i=1}^{N_c}\langle
M^i_i\rangle=N_c
m_Q\la^2=N_c\langle S\rangle_{N_c},\,\, 
\eeq
Because colorless $\langle B\rangle$ and $\langle{\ov B}\rangle$
depend analytically on $m_Q\neq 0$, they remain zero also in the
limit $m_Q\ra 0$.

From \eqref{(12.3.1.3)},\eqref{(12.3.1.4)} the masses of heavier
particles look as
\bq
\mu^{\rm pole}\Bigl ( M_s\Bigr ) = \mu^{\rm pole}\Bigl (
M^{N_c+1}_{N_c+1}\Bigr
)\sim \la\,, \label{(12.3.1.8)}
\eq
while the masses of lighter particles look as (up to logarithmic
renormalization factors)
\bq
\mu^{\rm pole}\Bigl (M^i_j(\rm adj)\Bigr )\sim m_Q\ll \la, \quad
\mu^{\rm
pole}(B)=\mu^{\rm pole}({\ov B})\sim m_Q\ll\la\,.\label{(12.3.1.9)}
\eq

Let us put attention that in \cite{S1} the Kahler terms of two
physical fields $\delta M^{N_c+1}_{N_c+1}=(M^{N_c+1}_{N_c+1}- \langle
M^{N_c+1}_{N_c+1}\rangle\,)$ and $\delta M_s=(M_s-\langle
M_s\rangle\,)$ are excluded by hands from \eqref{(12.3.1.3)} (in
order  to match the 't Hooft triangles). Then $\delta M^{N_c+1}_{N_c+1}$
becomes the auxiliary field and the physical field $\delta M_s$ is
excluded by the constraint $\det_
{N_c} M-\la^{2N_c-2}\,{\ov B}B =\la^{2N_c}$. While here these two
fields are
normal physical fields and this looks at least much more natural. \\

In the chiral limit $m_Q\ra 0$ the dual Lagrangian
\eqref{(12.3.1.3)},\eqref{(12.3.1.4)}
is $SU(N_F=N_c)_{left}\times SU(N_F=N_c)_{right}\times
U(1)_{B,V}\times U(1)_R$
invariant. (The R-charges of scalars $M$ and $B,\,{\ov B}$ in
\eqref{(12.3.1.4)} are zero, R-charge of scalar $M^{N_c+1}_{N_c+1}$
is  2). But $\langle M_s\rangle=\la^2$ \eqref{(12.3.1.6)} breaks
spontaneously $SU(N_F=N_c)_A$. And $N_F^2-1$ ${\cal N}=1$ multiplets
of Goldstone particles $M^i_j(\rm adj)$ become massless in the chiral
limit.

The symmetry $U(1)_{N_F=N_c,A}$ is broken explicitly by the term
$\sim\det_{N_F=N_c}M$ in \eqref{(12.3.1.4)}. {\it The coupling of
$M_s$ with $M^{N_c+1}_{N_c+1}$ gives them both large masses} $\sim\la$.

In this limit $m_Q\ra 0$, the whole moduli 'space' is exhausted by
one  point:\, $\langle M_s\rangle=\la^2,\,\, \langle M^i_j(\rm
adj)\rangle=0, \,\, \langle B\rangle=\langle{\ov B}\rangle=0$.

And finally, about the anomalous 't Hooft triangles. In the chiral
limit $m_Q\ra 0$ all particle masses of the direct theory coalesce to
zero. In the dual theory masses of $\mu^{\rm pole}\Bigl (M^i_j(\rm
adj)\Bigr )$ and $\mu^{\rm pole}(B)=\mu^{\rm pole}({\ov B})$ tend to
zero \eqref{(12.3.1.9)}, while $\mu^{\rm pole}\Bigl ( M_s\Bigr ) =
\mu^{\rm pole}\Bigl ( M^{N_c+1}_{N_c+1})\Bigr )\sim \la$
\eqref{(12.3.1.8)}. It is not difficult to check that at scales
$\mu\ll\la$ all 't Hooft triangles of direct and dual theories are
matched, see \cite{AKM},\cite{S1}.

At $0 < m_Q\ll\la $ the 't Hooft triangles are matched only in the
range of scales
$m^{\rm pole}_Q\ll\mu\ll\la$ \eqref{(12.2.1.1)}, where all particles
of the direct and dual theories are effectively massless. At scales
$\mu < m^{\rm pole}_Q$ triangles are not matched because the mass
spectra of direct and dual theories are qualitatively different.

\numberwithin{equation}{subsubsection}

\subsubsection{Light unequal mass quarks}

Consider now the case with unequal masses\,:
$\mi=\ml,\,\,i=1,...,N_L,\,\,\,m_{Q,k}=\mh,\,\, k=N_L+1,...,N_c,
N_L+N_H=N_F=N_c$, $0 < m_L\ll m_H\ll\la$, with fixed $0 <
r=m_L/m_H\ll1$.

The Lagrangian at $\mu=\la$ is taken as
\bq
{\wt K}_{N_c+1}= {\rm Tr}_{N_c+1}\frac{M^\dagger
M}{\la^2}+\sum_{i=1}^{N_F=N_c+1}\Bigl ( (B^\dagger)^i B_i +({\ov
B}^\dagger)_i\,{\ov B}^{\,i}\Bigr ),\label{(12.3.2.1)}
\eq
\bbq
{\wt {\cal W}}_{N_c+1}=m_L\sum_{i=1}^{N_L} (M_L)^i_i +
m_H\sum_{i=1}^{N_H}
(M_H)^i_i + \la M^{N_c+1}_{N_c+1}+{\rm Tr}_{N_c+1}\,({\ov
B}\frac{M}{\la}
B)-\frac{\det_{N_c+1} M}{\la^{2N_c-1}}\,.
\eeq

From \eqref{(12.3.2.1)}\,:
\bq
\langle M_L\rangle=\la^2\Bigl (\frac{1}{r} \Bigr
)^{N_H/N_c}\gg\la^2\,,\quad
\langle M_H\rangle=\la^2\Bigl (r \Bigr )^{N_L/N_c}\ll \la^2\,,
\label{(12.3.2.2)}
\eq
\bbq
\langle (\det M)_{N_c} \rangle=\langle M_L\rangle^{N_L}\langle
M_H\rangle^{N_H}=\la^{2N_c}\,,\quad \langle
M^{N_c+1}_{N_c+1}\rangle=\frac{m_L\langle
M_L\rangle}{\la}=\frac{m_H\langle M_H\rangle}{\la}=\frac{\langle
S\rangle_{N_c}}{\la}\,,
\eeq
\bbq
\langle (\det M)_{N_c+1} \rangle=\la^{2N_c}\Biggl [\langle
M^{N_c+1}_{N_c+1}\rangle=\frac{\langle S\rangle_{N_c}}{\la}\Biggr
]\,,\quad 0 < r=m_L/m_H \ll 1\,,
\eeq
compare with \eqref{(12.2.2.1)},
\bq
\langle B_L\rangle=\langle{\ov B}_L\rangle=\langle
B_H\rangle=\langle{\ov
B}_H\rangle=\langle B\rangle=\langle{\ov B}\rangle=0\,,
\label{(12.3.2.3)}
\eq
\bbq
\langle {\wt {\cal W}}\rangle=m_L\sum_{i=1}^{N_L}\langle M_L\rangle +
m_H\sum_{i=1}^{N_H}\langle M_H\rangle=N_c\langle S\rangle_{N_c}\,,
\eeq
\bq
\langle S\rangle_{N_c}=\la^2 m_L^{N_L/N_c}
m_H^{N_H/N_c}=r^{N_l/N_c}\la^2 m_H \,,
\label{(12.3.2.4)}
\eq
compare with \eqref{(12.2.1.6)}. 

The particle masses look at $\mu\sim\la$ as, 
\bq
\mu(B_L)=\mu({\ov B}_L)\sim\frac{\langle M_L\rangle}{\la}\sim
\la\Bigl(\frac{1}{r}\Bigr )^{N_H/N_c}\gg\la\,, \, \label{(12.3.2.5)}
\eq
\bbq
\mu(B_H)=\mu({\ov B}_H)\sim\frac{\langle M_H\rangle}{\la}
\sim\la\Bigl (r \Bigr )^{N_L/N_c}\ll \la,
\eeq
\bq
\quad \mu(B=B_{i=N_c+1})=\mu({\ov B}={\ov B}^{i=N_c+1})
\sim\frac{\langle
S\rangle_{N_c}}{\la^2}\sim r^{N_L/N_c} m_H \ll\la\,,
\label{(12.3.2.6)}
\eq
\bbq
(M^L_L)^i_j=(M^L_L)(\rm adj)^i_j+ \delta^i_j (M_s)^L_L,\,\, (M_s)^L_L
=\frac{1}{N_L}\sum_{i=1}^{N_L} M_{L,i}, \,\, (M^H_H)^i_j=(M^H_H(\rm
adj)^i_j+ \delta^i_j (M_s)^H_H,
\eeq
\bbq
(M_s)^H_H=\frac{1}{N_H}\sum_{i=1}^{N_H} M_{H,i}, \quad \langle
(M_s)^H_H \rangle=\langle M_H\rangle\,,\quad \langle M^i_j(\rm
adj)\rangle=0\,, \quad \langle (M_s)^L_L \rangle=\langle
M_L\rangle\,,\eeq
\bq
\mu\Bigl ( (M_s )^H_H\Bigr )\sim \mu\Bigl (M^{N_c+1}_{N_c+1} \Bigr
)\sim\frac{\la}{r^{N_L/N_c}}\gg\la\,,\quad \mu\Bigl ( (M_s )^L_L\Bigr
)\sim \,m_L \ll\la\,, \label{(12.3.2.7)}
\eq
\bq
\hspace*{-3mm} \mu\Bigl (M^L_L(\rm adj)\Bigr )\sim \frac{\langle
(M^{N_c+1}_{N_c+1} \rangle \la^3}{\langle M^L_L\rangle^2}\sim
r^{(N_c+N_H)/N_c}\,m_H\ll\la,\, \mu\Bigl (M^H_H(\rm adj)\Bigr
)\sim \frac{m_H}{ r^{ N_L/N_c}}\lessgtr \la,\label{(12.3.2.8)}
\eq
\bq
\mu\Bigl ( M^L_H(\rm adj)\Bigr )=\mu\Bigl ( M_L^H(\rm adj)\Bigr )\sim
r^{N_H/N_c}  m_H\ll\la\,. \label{(12.3.2.9)}
\eq
Therefore, the heavy particles with masses $\gg\la$ at small $\mi\neq
0$ and $r\ll  1$ are:
$B_L,\,{\ov B}_L, \, (M_s)^H_H, \\ M^{N_c+1}_{N_c+1}$ and maybe
$M^H_H(\rm adj)$, while the particle with masses $\ll\la$ are
\bq
B_H,\,\,{\ov B}_H,\,\, B,\,\, {\ov B},\,\,(M_s)^L_L,\,\, M^L_L(\rm
adj),\,\,M^L_H(\rm adj)=M^H_L(\rm adj),\, {\rm and\, may\, be}\,
M^H_H(\rm adj). \label{(12.3.2.10)}
\eq

In the chiral limit $m_{L,H}\ra 0$ with fixed $0 < r=m_L/m_H\ll 1$,
the Lagrangian
${\wt {\cal W}}_{N_c+1}$ \eqref{(12.3.1.2)} is invariant under global
$SU(N_F=N_c)_{left}\times SU(N_F=N_c)_{right}\times U(1)_{V,B}\times
U(1)_R$. The mean values $\langle (M_s)^L_L\rangle =\langle
M_L\rangle\neq \langle (M_s)^H_H\rangle=\langle M_H\rangle$
\eqref{(12.3.2.2)},\eqref{(12.3.2.6)} break spontaneously
$SU(N_F=N_c)_{left}$ and
$SU(N_F=N_c)_{right}$. The residual symmetry is $SU(N_L)_V\times
SU(N_H)_V\times
U(1)_{V,L}\times U(1)_{V,H}\times U(1)_R$. Therefore, $N_L^2-1\,\,
M^L_L(\rm adj),\,\,N_H^2-1\,\, M^H_H(\rm adj)$ and $2N_L N_H$ hybrids
are the genuine Goldstone particles and are massless in this limit,
see \eqref{(12.3.2.8)}, \eqref{(12.3.2.9)}. Because the supersymmetry
doubles the number of Goldstones, the spontaneous breaking of
$SU(N_F=N_c)_V$ does not increase the number of Goldstones. 
The term  $\sim \det_{N_c}M$
in \eqref{(12.3.2.1)} is not invariant under $U(1)_{A,L}$ and
$U(1)_{A,H}$. The mass
$\mu^{\rm pole}\Bigl ( (M_s )^L_L\Bigr )$ tends to zero in this
limit,  see \eqref{(12.3.2.7)}. But the mass of $(M_s)^H_H$ is nonzero
in this  chiral limit, see \eqref{(12.2.7)}.
This differs from the direct theory, see section 2.2, where masses of
the whole  HH-sector tend to zero at $m_H\ra 0$.

The whole moduli space is exhausted in this limit by the line of
different values of $0 < r < \infty$.

The 't Hooft triangles of the direct and dual theories are matched in
this chiral limit at scales $\mu < \mu^{\rm pole }(B_H)$. But at
$\mu^{\rm pole }(B_H) < \mu < \la$ \eqref{(12.3.2.5)} the baryons
$B_H,\,{\ov B}_H$ give additional contributions to the triangles
$R^1=R^3$
and $RB^2$ and spoil the matching.

At small $0 < m_L\ll m_H\ll\la$, the values of 't Hooft triangles
depend on ranges of the scale considered. Because the mass spectra of
particles with masses $\ll\la$ are qualitatively
different in the direct and dual theories, the triangles are not
matched.

\numberwithin{equation}{subsection}

\subsection{Direct theory:\, (pseudo)Goldstone particles}

{\bf A)} Equal quark masses. In the chiral limit $m_Q\ra 0$, the
non-anomalous global symmetry of the Lagrangian \eqref{(12.2.2)} is
$SU(N_c)\times SU(N_F=N_c)_{left}\times
SU(N_F=N_c)_{right}\times U(1)_R\times U(1)_{V,B}$. Recall that {\it
all quarks are not higgsed} in this case, see \eqref{(12.2.7)}. All
quark masses $m^{\rm pole}_Q$ \eqref{(12.2.1.1)} tend to zero.
Besides, $\langle M^i_j(adj)\rangle=0, \,\, \langle M({\rm
singl})\rangle=\la^2$, see the end of section 12.2.1.

$\langle M({\rm singl})\rangle=\la^2$ breaks {\it spontaneously} the
whole
$SU(N_F)_A$ global symmetry, while the global $SU(N_F)_V$ and
$U(1)_R\times
U(1)_{V,B}$ symmetries remain unbroken (remind that R-charge of
scalars $M^i_j$ is zero). As a result, there will be
$N^2_F-1$\,\,${\cal N}=1$ complex multiplets of genuine (i.e.
non-anomalous) massless Goldstone pions $\pi^i_j(adj)$ (the
supersymmetry doubles the minimal number of real Goldstone fields
predicted by the Goldstone theorem).

The global Abelian symmetry $U(1)_{N_F,A}$ is also broken
spontaneously by $\langle M_s \rangle=\la^2$. But this symmetry 
is anomalous, i.e. it is explicitly  broken in addition.

As it is seen from \eqref{(12.2.3)}, the tension of the confining
string $\sigma^{1/2}\sim \lym\ra 0$, i.e. {\it there is no
confinement in this chiral limit} $m_Q\ra 0$. Moreover, the 
masses of all quarks
and gluons ($\mu^{\rm non-pert}_{\rm gl}\sim\lym\ra 0$) and hadrons
made from them coalesce to zero. I.e., {\it the whole mass spectrum
consists only from massless quarks and gluons}, see footnotes
\ref{(f17)},\,\ref{(f23)}. For these reasons,
all 't Hooft triangles are matched automatically.\\

At small $0 < m_Q\ll\la$ all quarks are still not higgsed but are
weakly confined now by $SU(N_c)$ SYM, i.e. $\sigma^{1/2}\sim 
\lym\ll m^{\rm  pole}_Q\ll\la$, see \eqref{(12.2.7)},\eqref{(12.2.1.1)},
\eqref{(12.2.1.5)}. Hadrons made from them have typical
masses ${\cal O} (m^{\rm pole}_Q)\ll\la$. And $N_F^2-1$ genuine
pseudo-Goldstone pions $\pi^i_j(adj)$ are among these hadrons. 
All quarks and gluons on  the one hand, and hadrons on the other 
one form two full bases in the same space. And, in the cases of {\it
inclusive} integrations, we can use those basis which is more
convenient. At scales ${\cal O} (m^{\rm pole}_Q)\ll \mu \ll\la$ all
particles in the mass spectrum, i.e. hadrons, quarks and gluons, are
effectively massless. And in this range of scales two sets of 't
Hooft  triangles, those at scales $\mu\gg\la$ and those at ${\cal O}
(m^{\rm pole}_Q)\ll \mu \ll\la$, are matched automatically in the basis 
of quarks and gluons.

At scales $\mu \ll m^{\rm pole}_Q$ 't Hooft triangles are changed
because all quarks decouple as heavy, but this is normal. This is a
result of not spontaneous but the {\it explicit}
$SU(N_F)_A$ global symmetry breaking.\\

{\bf B)} Unequal quark masses. When all quark masses tend to zero
with  fixed ratio $0 < r=m_L/m_H \ll1$ as in section 2.2, the global
symmetry of the Lagrangian is the same as in {\bf A} above. But the
spontaneous breaking is different, see \eqref{(12.2.2.1)}. {\it
LL-quarks with e.g. $0 < N_L < N_c-1$ are now higgsed in the weak
coupling regime}, $\langle {\hat Q}^i_{\beta}\rangle=\delta^i_{\beta}
\rho_{\rm higgs},\,\,\, \langle {\hat {\ov
Q}}_i^{\,\beta}\rangle=\delta_i^{\beta} \rho_{\rm higgs},\,\,
i,\beta=1...N_L,\,\,\rho_{\rm higgs}\gg\la$, see
\eqref{(12.2.2.1)},\eqref{(12.2.2.2)}. Now, not only the whole global
$SU(N_F)_{A,F}$ is broken spontaneously by higgsed LL-quarks and
$\langle (M_{L,H})({\rm singl})\rangle=\langle M_{L,H}\rangle\neq
0,\,\langle M_{L}\rangle\neq \langle M_{H} \rangle$\,, but also global
$SU(N_c)_C\times SU(N_F)_{V,F}\times U(1)_{V,B}$. The unbroken
non-anomalous part looks as, see \eqref{(2.2.1)}\,~:\,
$SU(N_L)_{V,C+F}\times SU(N_H)_{V,F}\times U(1)_{V, B_H}\times
U(1)_{{\tilde V}, F+C}\times U(1)_R$. Both $U(1)_{A,L}$ and
(\,$U(1)_{A,H}$ are broken spontaneously but, besides, they both are
anomalous.

Now, the vector baryon charge is redefined: $B\ra B_H$, such that
$B_L$-charge is
zero for all $2 N^2_L$ higgsed \, LL-quarks $Q^i_{\beta},\, {\ov
Q}^{\beta}_i,\,\,i,{\beta}=1...N_L$ with L-flavors and L-colors. And
the nonzero generator of $B_H$-charge is normalized in the same way
as  $B$\,: ${\rm Tr}\,(B^2)={\rm Tr}\, (B^2_H)$. I.e., the generators of
vector $B,\, {\tilde V}$ and $B_H$ look as
\bq
B= diag\, (\underbrace{1}_{N_F}),\,\,{\tilde V}= diag\,
(\underbrace{\, - 1}_{N_L}; \,\underbrace{N_L/N_H}_{N_H}),\,
B_H=(N_F/N_H)^{1/2}\, diag\, (\,\underbrace{0}_{N_L};
\,\underbrace{1}_{N_H})\,. \label{(12.4.1)}
\eq

All $2 N_L N_H$ hybrid $LH+HL$ generators of $SU(N_F)_{V,F}$ are
broken spontaneously by $\langle (M_{L,H})({\rm singl})\rangle=\langle
M_{L,H}\rangle\neq 0,\,\, \langle M_L\rangle
\neq \langle M_H\rangle$ \,\eqref{(12.2.2.1)}. But the Goldstone
theorem determines only a {\it minimal} number of genuine real
Goldstone fields. While supersymmetry doubles the
number of Goldstone fields. And so, spontaneously broken vector
hybrid  generators do not increase the number of genuine Goldstone
fields.

Except for $N_L(2N_c-N_L)$\, ${\cal N}=1$ multiplets of heavy gluons,
the masses of all other lighter particles, quarks, gluons and hadrons
coalesce to zero in this chiral limit. And masses of $(N_F^2-1)$
${\cal N}=1$ multiplets of genuine complex Goldstone fields among
them. {\it Because $\langle\lym\rangle\ra 0$, there is no
confinement}.
\footnote{\,
The massless HH genuine Goldstone particles, see sections 2.2,
considered as bound state fields $\sum_{\beta=1}^{N_H}({\mathbf{\ov Q}
^H_H})^{\beta}_j  ({\mathbf{Q}^H_H})^i_{\beta})$ of not higgsed
massless HH quarks with approaching
zero binding energy, have binding radius $R_{\rm bind}\ra \infty$ and
are indistinguishable from the state of two unbound HH quarks. And
this concerns  the approaching to zero masses of all other HH hadrons 
made from  massless HH quarks. \label{(f23)}
}

There are at small $0 < m_L\ll m_H\ll\la$ the following light fields
(see section 2.2):\, a)\, $2 N_H^2$ complex ${\cal N}=1$ multiplets
of  weakly confined active ${\mathbf{Q}^H_H},\, {\mathbf{\ov Q}^H_H}$ 
quarks   with $N_H\times N_H$ color and flavor degrees of freedom. 
They form HH  hadrons with typical masses ${\cal O}(m^{\rm pole}_H)$\,\,
\eqref{(12.2.2.6)},\eqref{(12.2.2.10)},
and $N^2_H-1$\, ${\cal N}=1$ multiplets of genuine complex
pseudo-Goldstone pions $\pi^H_H(adj)$ are among these hadrons;\,\, b)
${\cal N}=1$ multiplet of $SU(N_H)$ gluons with non-perturbative
masses $\sim\langle\lym\rangle$.\, Besides, there are not
confined:\,\, c)\, $N_L^2$\, ${\cal N}=1$ multiplets of genuine
complex $\pi^L_L(adj)$ pseudo-Goldstone fields with masses $\sim
m_L\ll\la$ (with logarithmic accuracy, see \eqref{(12.2.2.9)}\,);\,
d)\, $2 N_L N_H$ \, ${\cal N}=1$ multiplets of genuine complex
pseudo-Goldstone hybrids $\pi^H_L,\,\pi^L_H$ with masses $\sim
m_H\ll\la$ (with logarithmic accuracy, see \eqref{(12.2.2.9)}\,).

As for the 't Hooft triangles, see the end of section 12.2.

\numberwithin{equation}{subsection}

\subsection{Regimes on the moduli space}

Consider now the direct theory with $N_{L,H},\,\,N_L+N_H=N_F=N_c$
quarks with masses ${m_{L,H}\ra 0}$ at fixed ratio $0 < r=m_L/m_H =
{\rm const} < \infty$, see section 12.2,
\bbq
{\cal W}_{\rm matter}=\sum_{i=1}^{N_c}\mi M^i_i \ra 0,\,\, \langle
S\rangle=\langle\lym^3\rangle \ra 0,\,\, \langle \det
M^i_j\rangle=\la^{2
N_c},\,\,
\eeq
\bq
\langle B\rangle=\langle {\ov B}\rangle=0,\,\, m_{L,H}\ra 0\,,.
\label{(12.5.1)}
\eq
The moduli $\langle M_{L,H}\rangle$ are fixed in this limit, see
\eqref{(12.2.2.1)}.

The theory \eqref{(12.2.2)} approaches now, at least, to the subspace
with $\langle  B\rangle=\langle {\ov B}\rangle=0$ of the whole moduli 
space. And  this will be sufficient for our purposes.

{\bf{A)}}\, The region $0 < r=m_L/m_H\ll 1$. The direct theory
approaches in the limit $m_{L,H}\ra 0$ to those region of the moduli
space where LL-quarks $Q^L_L,
{\ov Q}^L_L$ are higgsed, see \eqref{(12.2.2.1)},\eqref{(12.2.2.2)},
\footnote{\,
Here and below: at not too large values of $N_c$, see \cite{ch21}.
If, at fixed $0 < r\ll 1$, $N_c$ is so large that $\mu_{\rm {gl,L}}/\la^2\ll
1$,see \eqref{(2.2.2)}, then even LL-quarks are in the HQ (heavy quark)
phase, i.e. not higgsed  but confined  at $\mi\neq 0$. Then
all quarks are in the HQ phase with $m^{\rm pole}_H\gg m^{\rm
pole}_L\gg\lym$,\, $\lym$ is as in \eqref{(12.2.2.11)}, and with $\langle {\hat
Q}^i_a\rangle=\langle {\hat{\ov Q}}_i^{\,a}\rangle=0$. And similarly for 
H-quarks at $r\gg 1$, see \eqref{(2.2.1)}. \label{(f24)}
}
while HH-quarks $Q^H_H,\, {\ov Q}^H_H$ with H-flavors and H-colors
are not. And $m^{\rm pole}_H\ra 0$, see \eqref{(12.2.2.6)},\\
\eqref{(12.2.2.10)}. Moreover, the scale
factor of the unbroken by higgsed LL-quarks low energy non-Abelian
$SU(N_c-N_L=N_H\geq 2)$ SYM gauge coupling
$\langle{\hat\Lambda}_{N_H}\rangle=\langle\lym\rangle\, \ra\,0$, see
\eqref{(12.2.2.7)},\eqref{(12.2.3)}. Because the confining
$SU(N_H\geq2)$ SYM theory has only one dimensional parameter
$\langle\lym\rangle$, the tension of its confining string is
$\sigma^{1/2}\sim \langle\lym\rangle \ra 0$. I.e., {\it there is no
confinement in the direct theory in this region of the moduli space}.

{\bf{B)}}\, Let us now consider another region of the moduli space
where  $r=m_L/m_H \gg 1$. There, vice versa, see \eqref{(12.2.2.1)},
quarks $Q^H_H,\, {\ov Q}^H_H$ are higgsed while quarks $Q^L_L,\, {\ov
Q}^L_L$ with L-flavors and L-colors are not, with $N_L \leftrightarrow
N_H,\,\,\mu^2_{\rm gl,\, L}\leftrightarrow \mu^2_{\rm gl,\,\, H},\,\,
m^{\rm pole}_H\leftrightarrow m^{\rm pole}_L$. And also the scale
factor of the non-Abelian $SU(N_L\geq 2)$ SYM gauge coupling is
$\langle\lym\rangle\, \ra\,0$
\eqref{(12.2.3)}. I.e., {\it there is no confinement in the direct
theory also in this another region of the moduli space}.\\

In the paper \cite{FS} of E. Fradkin and S.H. Shenker, the special
(not supersymmetric) QCD-type lattice $SU(N_c)$ gauge theory with
$N_F=N_c$ flavors of scalar ``quarks''\, $\Phi^i_{\beta}$ in the
bi-fundamental representation was considered. In the unitary gauge,
all remained
$N_c^2+1$ physical real degrees of freedom of these quarks were {\it
deleted by hands} and replaced by one constant parameter $|v| >
0\,:\,\Phi^i_{\beta} = \delta ^i_{\beta}|v|,\,\,
\beta=1,...,N_c,\,\, i=1,...,N_F=N_c$.\, I.e., all such "quarks" are
massless, with no self-interactions and {\it permanently higgsed by
hands} even at small $g|v|\ll\Lambda_{QCD}$, see page 3694 and
eq.(4.1) for the bare perturbative Lagrangian in \cite{FS}. (Here
$\Lambda_{QCD}={\rm const}$ is the analog of $\la={\rm const}$ in
${\cal N}=1$ SQCD).
And all $N^2_c-1$ electric gluons received {\it fixed masses} $\sim g
|v|$.The region with the large values of $g|v|\gg\Lambda_{QCD}$ was
considered in \cite{FS} as the higgs regime, while those with small
$0< g|v|\ll\Lambda_{QCD}$ as the confinement one. The conclusion of
\cite{FS} was that the transition between the higgs and confinement
regimes is the analytic crossover, not the non-analytic phase transition. 
And although the theory considered in \cite{FS} was very specific, the 
experience shows  that up to now there is a widely spread opinion that 
this  conclusion has general applicability both to lattice and continuum 
theories, and to non-supersymmetric and supersymmetric ones. And not 
only for the QCD with "defective" quarks from \cite{FS}, {\it but also for 
normal scalar quarks with all their degrees of freedom}.\\

Let us note that this model of E. Fradkin and S.H. Shenker \cite{FS}
with such {\it permanently higgsed by hands at all $\,0 < g |v| <
\infty$ non-dynamical scalar "quarks" $\Phi^i_{\beta} =
\delta ^i_{\beta}|v|$ looks unphysical and is incompatible with {\it
normal models with dynamical} electrically charged scalar quarks
$\phi^i_{\beta}$ with all $2 N_F N_c$ their real physical degrees of
freedom}. This model \cite{FS} is really the Stueckelberg pure
$SU(N_c)$  YM-theory {\it with no dynamical electric quarks and with {\it
massive  all $N_c^2-1$ electric gluons with fixed by hands masses} $g|v|
> 0$  in the bare perturbative Lagrangian}, see eq.(4.1) in \cite{FS}.

For this reason, {\it in any case}, the electric flux emanating from
the test (anti)quark becomes {\it exponentially suppressed} at
distances $L > l_0=(g|v|)^{-1}$ from the source. And so, {\it the
potentially possible confining string tension will be also
exponentially  suppressed at distances $L > l_0$ from sources}. 
And e.g. external  heavy test quark-antiquark pair will be not 
connected then by one  common {\it really confining string} at 
large distance  between them. These quark and antiquark can move 
then practically  independently of each other and can be registered 
alone in two  different detectors at large distance between one another. 
I.e., {\it in any case}, in this Stueckelberg theory \cite{FS}, at all fixed
$|v| > 0$, {\it there is no genuine confinement} which prevents appearance 
of one (anti)quark in the far detector. See also page 9 in v.6 of
\cite{ch21}.\\

Now, about non-perturbative effects in the {\it standard
non-supersymmetric} $SU(N_c)$  pure YM-theory. 
The common opinion (supported by lattice calculations)
is that there are magnetically  charged solitons {\it
condensing in the vacuum state with the density $\rho^2_
{\rm  magn}\sim\Lambda^2_{YM}$}. This condensation
leads to real confinement of normal electrically charged
particles by the confining strings with the typical tension
$\sigma^{1/2}\sim\Lambda_{\rm YM}$.

But, as was emphasized in \cite{ch21}, {\it responsible for confinement
condensed magnetically  charged solitons  and mutually non-local 
with  them electric scalar  quarks can not
condense simultaneously in the vacuum state}. For this reason,
because  the condensate of all defective electric "quarks"
$\Phi^i_{\beta}=\delta ^i_{\beta}|v|$ is {\it strictly fixed by
hands} at $\rho_{\rm electr}=|v| > 0$ in \cite{FS}, {\it this
prevents  mutually nonlocal with them magnetically  charged solitons  
to  condense  simultaneously with such "quarks"\, in the vacuum state. 
I.e.,  definitely, there is then no confinement of electric charges at all
$0< |v| < \infty$ in the Stueckelberg $SU(N_c)$ YM-theory used in
\cite{FS}. All electric charges are really not confined but
screened}.  Therefore, the conclusion of \cite{FS} about an analytical
crossover  between the "regime with confinement" at $0 < g|v|\ll
\Lambda_{YM}$  (according \cite{FS}) and the higgs regime at $g|v|
\gg\Lambda_{YM}$  is  not surprising. {\it This Stueckelberg theory 
used in \cite{FS} with  such defective non-dynamical electric "quarks" 
$\Phi^i_{\beta} =  \delta ^i_{\beta}|v|,\,\, |v| > 0$ stays permanently in the
completely  higgsed by hands phase with massive all electric gluons,
with no condensation of magnetically charged solitons and no confinement 
of  electrically  charged particles}.\\

To describe all this in more details, we need more detailed
picture of the confinement mechanist of electric charges in the
standard pure $SU(N_c)$ YM theory. We describe now the proposed 
model   for this confinement mechanism.
~\footnote{\,
Nevertheless, we expect that this model well may be realistic.
\label{(f25)}
}

This mechanism is qualitatively similar to those which is realized in
${\cal N}=2$ SYM theory
softly broken to ${\cal N}=1$ SYM by the small but nonzero mass term
$\mx {\rm Tr}\, (X^{\rm adj})^2$ of the $SU(N_c)$ adjoint scalar
superfield $(X^{\rm adj})^i_j$ \cite{SW1, SW2, DS}. But, unlike the
${\cal N}=2$ SYM, where there is this {\it elementary} adjoint field
$({\hat X}^{\rm adj})^i_j$ which condenses in the vacuum resulting in
$SU(N_c)\ra U(1)^{N_c-1}$,  in the ordinary pure $SU(N_c > 2)$ YM 
this role is played by the {\it  composite} adjoint field  $(H^{\rm adj})^
i_j(x)$
\bq
(H^{\rm adj})^i_j(x)=\frac{\Lambda_{YM}^{-3}}{16\pi}\sum_{A,B,C=1}^{N_c^2-1}
(T^{A})^i_j d^{ABC} G^B_{\mu\nu}(x) G^{C,\,\mu\nu}(x)=
\Biggl (V_{\rm Goldst}(x){\hat H}^{\rm adj}(x)V^{\dagger}(x)_{\rm Goldst}(x)
\Biggr )^i_j\,, \quad \label{(12.5.2)}
\eq
\bbq
\langle ( {\hat H}^{\rm adj} )^i_j(x)\rangle=\langle\Biggl ( V^{\dagger}
_{\rm Goldst} (x){H}^{\rm adj}(x) V(x)_{\rm Goldst}(x)
\Biggr )^i_j\rangle = {\rm diag} (\rho_1,...,\rho_{N_c}), \,\,
\sum_{k=1}^{N_c} \rho_k=0, \,\, \rho_k={\cal O}(\Lambda_{YM})\,,
\eeq
where $({\hat H}^{\rm adj})^i_j(x)$ is  {\it the
colored but gauge invariant field and  constants $\rho_k$ are gauge 
invariant  order parameters}, see section 2 in \cite{ch21} and section 5 in
\cite{ch23}.  As a result, 
$SU(N_c > 2)\ra  U(1)^{N_c-1}$, all charged electric gluons acquire masses
$\mu_{\rm electr}\sim g\rho_{\rm electr}\sim g\Lambda_{YM}$ and there
appear $N_c-1$ independent scalar magnetic monopoles $M_n$. {\it
These  monopoles also condense in the vacuum} with the density
$\rho^2_{\rm  magn}\sim \Lambda^2_{YM}$ and give masses
$\mu_{\rm magn}\sim {\tilde g}\Lambda_{YM}$ to all $N_c-1$ dual
magnetic photons. This leads to the genuine confinement of all
charged  electric massive $SU(N_c)$ gluons by strings with the tension
$\sigma^{1/2}\sim \Lambda_{YM}$.
\footnote{\,
Instead of $(H^{\rm adj})^i_{j}(x)$ in \eqref{(12.5.2)}, in ${\cal N}=1$
$SU(N_c  > 2)$ SYM this will be its analog - the chiral superfield 

$$\hspace*{8mm} (h^{\rm adj})^i_{j}(x)=\Lambda_{SYM}^{-2}
\sum_{A,B,C=1}^{N_c^2-1} \sum_{\gamma=1}^{2}(T^{A})^i_j d^{ABC}
W^{B,\,\gamma} W^{C}_{\gamma}/32\pi^2$$ 

and $\Lambda_{YM}\ra \Lambda_{SYM}$.\label{(f26)}
}
\\

Let us introduce now into this Abelian $U(1)^{N_c-1}$ dual gauge
theory the large size $L$ Wilson loop with the quark-antiquark pair
of  external heavy test electric quarks. Each quark
$Q_{\beta},\,\,\beta=1,...,N_c$ is characterized then by its electric
charge vector ${\vec E}_{\beta}(Q)$ with $N_c-1$ components in the
base of $N_c-1$ independent standard Cartan charges. And similarly,
each of $N_c-1$ {\it condensed} magnetic monopoles $M_n,\,\,
n=1,...,N_c-1$ is characterized by its magnetic charge vector ${\vec
M}_n({\rm Mon})$ with $N_c-1$ components. All quarks are mutually
local between themselves. And $N_c-1$ magnetic monopoles are mutually
local between themselves. But electrically charged quarks and
magnetic  monopoles are, in general, mutually non-local. Their proper
charge  vectors satisfy the standard Dirac charge quantization conditions
\bq
{\vec E}_{\beta}(Q){\vec M}_n({\rm Mon})=Z_{\beta,n}/2\,,\quad
\beta=1,...,N_c\,,\,\, n=1,...,N_c-1\,, \label{(12.5.3)}
\eq
where $Z_{\beta,n}$ is the integer number (including zero), depending
on the choice of
$\beta$ and $n$. And the vector ${\vec F}_{\beta}(Q)$ of the total
electric flux of $N_c-1$ Abelian electric photons emanating from
electrically charged quark $Q_{\beta}$ is ${\vec F}_{\beta}(Q)= {\vec
E}_{\beta}(Q)$.\\

In the standard pure $SU(N_c)$ YM, when these $N_c-1$ magnetic
monopoles are condensed in the vacuum state, then $N_c-1$
corresponding dual magnetic photons acquire masses $\mu_{\rm
magn}\sim{\tilde g} \rho_{\rm magn}\sim {\tilde g}\Lambda_{YM}$. But
condensation of magnetic monopoles can not screen the fluxes ${\vec
F}_{\beta}(Q)$ of electric photons emanating from heavy test quarks
$Q_{\beta}$. As a result, these constant electric fluxes ${\vec
F}_{\beta}(Q)={\vec E}_{\beta}(Q)$ are concentrated then within the
confining tube of radius $R^{\perp}_{\rm magn}(Q)\sim \mu^{-1}_{\rm
magn}$ connecting quark and antiquark at the distance $L\gg
R^{\perp}_{\rm magn}$ between them. {\it The quantization conditions
\eqref{(12.5.3)} are necessary for self-consistency and stability of
these confining tubes}.\\

The properties of the Stueckelberg theory \cite{FS} are qualitatively
different. All $N_c^2-1$
electric gluons have now {\it non-dynamical but fixed by hands
masses}  $g|v| > 0$ in the bare perturbative Lagrangian, even at
$g|v|\ll\Lambda_{\rm YM}$. For this reason, if
there were confinement, all total electric flux vectors ${\vec
F}_{\beta}(Q,l)$ emanating along the tube from these test
(anti)quarks, and so the (anti)quark corresponding electric charge
vectors ${\vec E}_{\beta}(Q,l)= {\vec F}_{\beta}(Q,l)$, would become  
screened.  And their numerical values 
would  become dependent on distances
along the tube from the sources.  And they all 
would   become exponentially
small at distances $l > (g|v|)^{-1}$ along the tube from the
sources. {\it This dependence on $l$ at $g|v|\neq 0$ of all $N_c$
charge vectors ${\vec E}_{\beta}(Q,l)$ would violate the charge 
quantization conditions \eqref{(12.5.3)} which were necessary for the
self-consistency and stability} of the confining electric tube
connecting quark and antiquark.\\

In other words, the normal self-consistent stable tubes with the
constant electric charge vectors ${\vec E}_{\beta}(Q)$ \eqref{(12.5.3)}
along  the tube, originating from $\mu_{\rm magn}\neq 0$ due to
condensation of magnetic monopoles, {\it can not be formed  at}
$g|v| > 0$. But varying total electric quark flux ${\vec F}_{\beta}(Q,l)=
{\vec E}_{\beta}(Q,l)$ along the tube still can not  be  screened
by these condensed magnetic monopoles. As a result, $\mu_{\rm magn}$ 
{\it has to be zero} at fixed by hands $g|v|\neq 0$ for
self-consistency, i.e. these normal dynamical magnetic monopoles {\it
can not condense} then in vacuum and can not produce 
nonzero masses $\mu_{\rm
magn}\neq 0$ of corresponding dual magnetic photons, fixed $g|v|\neq
0$ prevents this condensation. This agrees with the statement in
\cite{ch21} that {\it mutually non-local normal dynamical magnetic
monopoles and normal dynamical electric quarks can not condense
simultaneously in the vacuum state}. As a result, {\it at all $0 <
g|v| <\infty$, there is not confinement but screening of all electric
charges in the theory} considered in \cite{FS}. And, as a result,
there is not the phase transition but crossover between regions with
$g|v|\ll\Lambda_{\rm YM}$ and $g|v|\gg\Lambda_{\rm YM}$ in the theory
considered by E. Fradkin and S.H. Shenker in \cite{FS}.\\

A widely spread opinion (supported by lattice calculations) is that,
in non-supersymmetric $SU(N_c)$ QCD (or in ${\cal N}=1$ SQCD) with
{\it normal dynamical scalar quarks}  $\phi^i_{\beta},\, i=1,...,N_F,\, \beta=
1,...,N_c$, with all $2 N_F  N_c$ their physical real degrees
of freedom and with sufficiently large bare and/or dynamical masses,
the confinement of such non-higgsed quarks originates from higgsing
(i.e. condensation) of $N_c-1$ magnetically charged solitons of $SU(N_c)$ QCD,
with the typical magnetic condensate $\rho_{\rm magn}\sim\Lambda_{YM}
\sim\Lambda_{QCD}$. But such higgsed   magnetically charged solitons,
ensuring confinement in YM, are mutually nonlocal with these
electrically charged scalar quarks. For this reason, such  magnetically charged 
solitons  with the much larger vacuum condensate $\rho_{\rm
magn}\sim\Lambda_{YM}\sim\Lambda_{QCD}\gg g|v|$ {\it will keep normal
dynamical scalar electric quarks} $\phi^i_{\beta}$ {\it confined and
will prevent them from condensing with} $|v|\neq 0$ {\it in the
vacuum \, state}. This allows then to form the normal self-consistent and
stable  confining tubes with constant electric flux vectors 
${\vec F}_{\beta}(Q)={\vec E}_{\beta}(Q)$ \eqref{(12.5.3)}.\\

The normal dynamical particles with larger vacuum condensate
overwhelm. E.g., the normal scalar magnetic monopoles with the 
vacuum condensate $\rho_{\rm magn} >
\rho_{\rm electr}$ will be really condensed in vacuum with $\rho_{\rm
magn}\neq 0$, keep normal electrically charged scalar quarks mutually
non-local with them confined and prevent these last from condensing
in  vacuum, i.e. $\rho_{\rm electr}=0$ then. And vice versa at
$\rho_{\rm magn} < \rho_{\rm electr}$.
\footnote{\,
In the standard non-supersymmetric $SU(N_c),\,\,N_F=N_c$ QCD with
normal dynamical
light scalar quarks $\phi^i_{\beta}$ with all $2 N_c^2$ their
physical  real degrees of freedom, these quarks are not massless even 
at  $m_{\phi}= 0$ and zero self-interaction potential in the bare
perturbative Lagrangian. Even at small $|v|\ll\Lambda_{QCD}$ 
(where  $|v|$ is e.g. a {\it potentially possible} dynamical quark
condensate,  $\langle{\hat\phi}^i_{\beta}\rangle
=\delta^i_{\beta}|v|\,$), they acquire in this case non-perturbative
dynamical masses $\mu^2_{\phi}\sim V_{HQ}\sim
\Lambda^2_{QCD}\sim\Lambda^2_{YM},\quad
V_{HQ}\delta^i_j=\frac{1}{N_c}\sum_{\beta=1}^{N_c}
\langle(\phi^\dagger)_j^{\beta}\phi^i_{\beta}
\rangle_{HQ}$.  And, connected with this, at $|v|
\ll\rho_{\rm magn}\sim\Lambda_{YM}\sim\Lambda_{QCD}$ {\it there is
confinement of all electrically charged particles with the string
tension} $\sigma^{1/2}\sim\rho_{\rm
magn}\sim\Lambda_{YM}\sim\Lambda_{QCD}$. So that,
$\sigma^{1/2}\sim\rho_{\rm magn}\sim\Lambda_{QCD}\sim\Lambda_{YM}\gg
|v|$ will keep really such standard quarks confined and will prevent
them from condensing with $|v|\neq 0$ in vacuum. Such {\it standard
quarks will be really in the HQ(heavy quark)-phase at
$|v|\ll\Lambda_{QCD}$, i.e. not higgsed but confined} in this case,
with $\langle{\hat\phi}^i_{\beta}\rangle_{HQ}=\delta^i_{\beta}
|v|_{HQ}=0$ really.  And $V_{HQ}\sim\Lambda^2_{YM}$ is {\it non-factorizable} 
for such  quarks in the  HQ-phase with $\mu_{\phi}\sim [\,V_{HQ}\,]^{1/2}\sim
\Lambda_{QCD}\neq |v|_{HQ}=0$. This is qualitatively similar to section 4.1 in
\,\cite{ch21} with standard massive confined quarks in the HQ (heavy
quark)-phase.\, Standard scalar quarks $\phi^i_{\beta}$ with $m_{\phi}= 0$ and
zero  self-interaction quark potential in the bare perturbative Lagrangian
will be really higgsed in the weak coupling regime at e.g. $\langle
{\hat\phi}^i_{\beta}\rangle_{higgs}=\delta^i_ {\beta}|v|_{\rm
higgs},\,\,\,i,\beta=1,...,N_F=N_c$, with $g|v|_{\rm
higgs}\gg\Lambda_{QCD}$. All $N_c^2-1$
electric gluons will acquire then large masses $\sim g|v|_{\rm
higgs}\gg\Lambda_{QCD}$ in this case, because $V_{higgs}\sim |v|_{\rm
higgs}^2\gg\Lambda^2_{QCD}$ will be {\it
factorized} now. $N_c^2$ quark real degrees of freedom will also
acquire dynamical masses $\sim g|v|_{\rm higgs}\gg\Lambda_{QCD}$ and
form one adjoint and one singlet representations of $SU(N_F=N_c)$.
And  there will be {\it the phase transition} between the regions with
values of the gauge invariant order parameter $\rho=|v|_{\rm
higgs}\gg\Lambda_{QCD}$ in the higgs phase and $\rho=|v|_{HQ}=0$ in
the confinement phase. This is similar to the phase transitions in
\cite{ch21}. \, While in ${\cal N}=1$ SQCD, non-higgsed quarks have $m^{\rm
pole}_Q\ra  0$ at $m_Q\ra 0$ \eqref{(12.2.1.1)}, (or   $m^{\rm pole}_{H}\ra 0$ 
at $m_H\ra 0$ \eqref{(12.2.2.6)},\eqref{(12.2.2.10)}, and the tension of
confining  string is $\sigma^{1/2}\sim\lym\ra 0$ in the limit 
$\mi\ra 0$ with fixed  $r=m_L/m_H$. And so, there is no confinement, 
even if the chiral symmetry of 
the SUSY Lagrangian is broken spontaneously, see
\eqref{(12.2.3)}-\eqref{(12.2.5)},\eqref{(12.2.2.1)}. \label{(f27)}
}
\\

And model of the confinement  described above illustrates explicitly the 
mechanism  underlying the statement in \cite{ch21} that mutually non-local 
normal   magnetic monopoles and
normal scalar quarks with all their physical degrees of freedom {\it
can not be condensed simultaneously in the vacuum state}.\\

In support of the conclusion of E. Fradkin and S.H. Shenker \cite{FS}
about the crossover between the confinement and higgs regimes, it was
written by K. Intriligator and N. Seiberg in \cite{IS} for ${\cal
N}=1\,SU(N_c)$ SQCD with $N_F=N_c$ the following (for the moduli
space, with $m_{Q,i}\ra 0)$.\, -

''For large expectation values of the fields, \, 
\footnote{\,
the values of $\langle M^i_i ={\ov Q}_i Q^i \rangle$ are implied \label{(f28)}
}
a Higgs description is most natural while, for small expectation
values, it is more natural to interpret the theory as 'confining'...
\footnote{\,
the quotes are set here because the string between heavy test quark
and antiquark
breaks with increased distance between them due to the birth of a
light dynamical quark-antiquark pair from the vacuum
}
Because these theories (i.e. ${\cal N}=1$ SUSY QCD) have matter
fields  in the fundamental representation of the gauge group, as
mentioned in  the introduction, there is no invariant distinction 
between the Higgs  and the confining phases \cite{FS}. It is possible 
to smoothly  interpolate from one interpretation to the other''.

In other words. {\it Because one can move completely smoothly (i.e.
analytically)} on the moduli space from e.g. the region {\bf A} with
$\langle M_L\rangle\gg\la^2,\, \langle
M_H\rangle\ll\la^2$, see \eqref{(12.2.2.1)}, where LL-quarks $Q^L_L,
{\ov Q}^L_L$ are
higgsed while (according to \cite{IS}) HH-quarks $Q^H_H,\, {\ov
Q}^H_H$ are confined, to another region {\bf B} with $\langle
M_L\rangle\ll\la^2,\, \langle M_H\rangle\gg\la^2$ where, vice versa,
quarks $Q^H_H, {\ov Q}^H_H$ are higgsed while quarks $Q^L_L,\, {\ov
Q}^L_L$ are confined, this is the independent confirmation of the
conclusions of paper \cite{FS} that there is the analytic crossover
between the confinement and higgs regimes, not the non-analytic phase
transition.\\

There are two loopholes in these arguments. It is right that quarks
$Q^L_L, {\ov Q}^L_L$
or $Q^H_H, {\ov Q}^H_H$ are higgsed on the moduli space in regions
{\bf A} or {\bf B} with $\langle M_L\rangle\gg\la^2$ or $\langle
M_H\rangle\gg\la^2$ respectively (at not too large
$N_c$, see section 10). But first, as emphasized above in this
section, the quarks with $\langle M_{L,H}\rangle\ll\la^2$ are not
confined. {\it There are regions on the moduli
space where some quarks are higgsed, but there are no regions where
some quarks
are confined. There is no confinement on the moduli space of the
direct theory because the string tension $\sigma^{1/2}\sim\lym\ra 0$}
at $\mi\ra 0$, see \eqref{(12.2.3)},\eqref{(12.5.1)} and footnotes
\ref{(f17)}, \ref{(f21)}. Therefore, unlike the statements in
\cite{IS}, the transitions between the regions {\bf A} and {\bf B} of
the moduli space are not the transitions between the higgs and
confinement regimes. These are transitions between regimes of higgsed
or not higgsed some quarks, but in all regions all quarks are not
confined.\\

Second, let us start on the moduli space, i.e. at $m_{L,H}\ra 0$ with
fixed $0 < r=m_L/m_H < \infty$ in the region {\bf A} and move to the
region {\bf B}. In the region {\bf A} at $\mu_{\rm
gl, L}\gg\la$, LL-quarks are higgsed, see
\eqref{(2.2.1)},\eqref{(2.2.2)}, while HH-quarks are {\it not higgsed
and not confined} (because $\sigma^{1/2}\sim\lym\ra 0$ at $\mi\ra
0$).  In the  region {\bf B} at $\mu_{\rm gl, H}\gg\la$, vice versa, 
HH-quarks   are  higgsed while
{\it LL-quarks are not higgsed and not confined}. On the way between
regions {\bf A} and {\bf B}\,, there is a point $r=1$ with equal mass
quarks and $\langle M^i_i\rangle=\la^2$. As shown in \eqref{(12.2.7)}, 
all quarks are {\it not higgsed} at this point at
$m_Q > 0$, even at not large $N_c$. And this remains so even in the
chiral limit $m_Q\ra 0$. In this limit, the whole mass spectrum
consists of massless quarks and gluons, see above in this section.
All particles are {\it not higgsed and not confined} (because
$\lym\ra0$). Although $\langle M^i_i\rangle=\la^2$, as it follows from
the Konishi anomaly \eqref{(2.1.8)}, but it is {\it non-factorizable},
i.e  $\langle M^i_i\rangle=\la^2\neq \sum_{\beta=1}^{N_c}\langle{\ov
Q}^{\beta}_i\rangle\langle Q^i_{\beta}\rangle=0$ at $r=1$.\\

In section 2 of \cite{ch21} the {\it colored} gauge invariant order
parameter was introduced. It looks as:\, $\langle {\hat Q}^i_{\beta}
\rangle=\delta^i_{\beta} \rho,\,\, \langle{\hat {\ov
Q}}^{\,{\beta}}_i\rangle=\delta^{\beta}_i
\rho,\,\,i=1,...,N_F,\,\beta=1,...,N_c$,
where ${\hat Q}^i_{\beta}$ is the {\it colored but gauge invariant
quark field}. The gauge invariant order parameter $\rho$ behaves {{\it
non-analytically}. It is nonzero for higgsed quarks. While it is zero
if quarks are not higgsed, independently of whether they are confned
or not}, see \cite{ch21} for all details.

At fixed $N_c$ and $0 < r=m_L/m_H\ll 1$, such that $\langle
M_L\rangle\gg\la^2,\,\mu^2_{\rm gl,L}\gg \la^2$
\eqref{(12.2.2.1)},\eqref{(12.2.2.2)}, LL-quarks are definitely
higgsed, while HH-quarks with $\langle M_H\rangle\ll\la^2$ are
definitely not higgsed (because they are not higgsed even at $\langle
M_H\rangle=\la^2$). And similarly, at $r\gg 1$ HH-quarks are higgsed,
while LL-quarks are not.

As it is seen from \eqref{(12.2.2.1)},\eqref{(12.2.2.2)}, at some
fixed $N_c$, to have
definitely higgsed LL-quarks, $r$ has to be sufficiently small: \, $r
< r_{L}(N_c)$, where $r_{L}(N_c)$ is sufficiently smaller than unity
and  is such that $\mu_{\rm gl,L}$ is sufficiently
larger than $\la$. And similarly for HH-quarks. To have definitely
higgsed HH-quarks, $r$ has to be sufficiently large:\, $r >
r_H(N_c)$,  where $r_{H}(N_c)$ is sufficiently larger
than unity and is such that $\mu_{\rm gl,H}$ is sufficiently larger
than $\la$. Therefore, at fixed $N_c$, there are {\it two phase
transitions at the points $r=r_L(N_c)$ and $r=r_H(N_c)$ on the way
from the region {\bf A} to {\bf B}} where, respectively, the
LL-quarks \, become  unhiggsed and HH-quarks become higgsed. 
And  there is the finite width
region $r_L(N_c) < r < r_H(N_c)$ where all quarks are not higgsed.
The gauge invariant order parameter $\rho$
\cite{ch21} is nonzero outside the region $r_L(N_c) < r <
r_H(N_c),\,\,\rho_{\rm higgs}\neq 0$, while it is zero inside it,
$\rho_{HQ}=0$.

So that, the arguments of K. Intriligator and N. Seiberg in
\cite{IS} ,  put forward in support of the crossover \cite{FS}, about an
analytical  movement along the moduli space and an absence of 
phase  transitions are erroneous.

\subsection {\bf \large Conclusions to Part I} 

\hspace*{5mm} {\bf 1)}\,\, As shown above, the mass spectra of particles with
masses $<\la$ of the direct $SU(N_c),\,\,\, N_F=N_c$ $\,\,{\cal N}=1$
SQCD theory and of Seiberg's dual one \cite{S1} are parametrically
different, both for equal or unequal quark masses $m_{Q,i}\neq 0$. 
Therefore,  these two theories are not equivalent.

So that, the proposal by N. Seiberg \cite{S1} of his non-gauge
$N_F=N_c$ dual ${\cal N}=1$ SQCD theory with only mesons $M^i_j$ 
and  baryons $B,\,{\ov B}$ (with
one constraint) as the genuine low energy form of the direct
$SU(N_c)\,\, N_F=N_c$ $\,\,{\cal N}=1$ SQCD theory due to the strong
confinement with the string tension $\sigma^{1/2}\sim\la$ (the so
called "S-confinement") is erroneous. And similarly for the case
$N_F=N_c+1$, see Appendix "B" in \cite{Session}. 

As for differences  in mass spectra of the direct and Seiberg's dual 
${\cal N}=1$ SQCD  at other values $N_c+1 < N_F < 3N_c $, 
see  e.g. \cite{ch1,ch3}.  \\
{\bf a)}\, It was shown in section 7 of \cite{ch1} that two-point correlators
of conserved  in the chiral limit $m_Q\ra 0$ baryon and $SU(N_F)_{left}
\times SU(N_F)_{right}$ currents are diferent in the direct and Seiberg's
dual theories at $N_c <  N_F < 3N_c/2$. And arguments were given  that 
it is impossible to write for this case in the direct theory the nonsingular
effective Lagrangian of hadrons with masses $\sim\la$ which preserves 
the chiral symmetry and R-charge conservation.\\
{\bf b)}\, It was shown in \cite{ch3} that at the left or right ends of conformal 
window, i.e. at $0 < (2N_F-3N_c)/N_F\ll 1$ or at $0 < (3N_c-N_F)/N_F\ll 1$, the
mass spectra of the direct and Seiberg's dual theories are parametrically different.
Then, there are no reasons that they become exactly the same at generic 
values  of  $N_F/N_c$ within the conformal window.\\

{\bf 2)}\,\, E. Fradkin and S.H. Shenker considered in \cite{FS} the very
special  non-supersymmetric lattice $SU(N_c)$ QCD theory with 
$N_F=N_c$  scalar  bifundamental quarks $\Phi^i_{\beta},\,\,i=1...N_F
=N_c,\,\,\beta=1...N_c$. In the  unitary gauge, from $2 N_c^2$ physical 
real degrees of freedom of  these quarks
$\Phi^i_{\beta}$, were retained only $N_c^2-1$ real Goldstone modes
eaten by gluons. All other $N_c^2+1$ real physical degrees of freedom
were deleted by hands: $\Phi^i_{\beta}=
\delta^i_{\beta}|v|,\,\,|v|={\rm const > 0}$. The conclusion of
\cite{FS} was that the transition between the confinement regime at
small $0 < |v|\ll\Lambda_{QCD}$ (according to \cite{FS}) and the
higgs  regime at large $|v|\gg\Lambda_{QCD}$ is the analytic crossover,
not  the non-analytic phase transition.

This model used in \cite{FS} was criticized in \cite{ch23}
(see also  page 9 in v.6 of \cite{ch21}\,) as {\it incompatible with and
qualitatively different from} the standard $SU(N_c),\,\,N_F=N_c$ QCD
theory with standard scalar quarks $\phi^i_{\beta}$ with all $2 N_c^2$  
their physical real degrees of freedom, see footnote \ref{(f27)}.
Really, this theory \cite{FS} is {\it the Stueckelberg  $SU(N_c)$ YM
theory with  no dynamical quarks and with massive all $N_c^2-1$
electric gluons with fixed by hands masses $g|v| > 0$}. In this
theory,  in any case,  the total electric flux  emanating   e.g. from 
external  heavy test (anti)quark is exponentially   
screened  at large  distances  due to fixed by hands nonzero 
masses of all  electric gluons and {\it there 
is  no genuin confinement at all $|v|\neq 0$}. In other words, this
theory\cite{FS} with such defective "quarks"\, {\it stays really
permanently in  the completely higgsed phase at all $|v|\neq 0$, 
and this is a reason  for the crossover between regions   $0 < |v|
\ll\Lambda_{QCD}$ and $|v|\gg\Lambda_{QCD}$ in \cite{FS}}.\\

Besides, the arguments presented in \cite{IS} by K. Intriligator and
N. Seiberg for the standard direct $SU(N_c),\,\, N_F=N_c$ $\,\,\,{\cal N}=1$
SQCD theory  in support of conclusions \cite{FS} about crossover between
the confinement and higgs regimes,  are
criticized in \cite{ch23} as erroneous.\\

\addcontentsline{toc}{section}
{\bf \Large{Part II. \,\, Mass spectra in $\mathbf{{\cal N}=1}$\,  SQCD \,  with \\ \,
additional colorless but flavored fields}}

\addcontentsline{toc}{section} 
{\bf \large Part IIa. $\mathbf{1\leq N_F< N_c}$ and
$\mathbf{3N_c/2 < N_F < 2N_c}$} 

\begin{center} 
\bf \large Part II.\,\, Mass spectra in $\mathbf{\cal N}=1$ 
SQCD \\ \,\, with additional colorless but flavored fields 
\end{center}

\begin{center} 
\bf \large Part IIa. $\mathbf {1\leq N_F< N_c}$ and
$\mathbf {3N_c/2 < N_F < 2N_c}$ 
\end{center}

\numberwithin{equation}{section}

\section  {\bf  Introduction to Part II}

Considered is the ${\cal N}=1$ SQCD-type theory with $SU(N_c)$
colors and $1\leq  N_F < 3N_c$ flavors of equal mass $0< m_Q\ll\la$
quarks $\,Q^i_\beta,{\ov Q}^{\,\beta}_j,\,\beta=1...N_c,\, i,j=1...N_F$. 
Besides, it includes $N^2_F$  additional 
colorless but flavored fields $\Phi_{i}^{j},\,\, i, j=1...N_F$, with
the large mass parameter $\mph\gg\la$, interacting with quarks
through the Yukawa coupling in the superpotential. The mass
spectra of this direct $\Phi$-theory are first calculated at
$1\leq N_F<N_c$ where the quarks are weakly coupled, in all
different vacua with the unbroken or spontaneously broken 
flavor  symmetry $U(N_F)\ra U(n_1)\times U(n_2)$.

Further, the mass spectra of this $\Phi$-theory and its
Seiberg's dual variant, the $d\Phi$-theory, are calculated at
$N_c < N_F<3N_c$ and at various  values of $\mph/\la\gg 1$   
using the dynamical scenario from  \cite{ch3}.  

This scenario  assumes that, for such ${\cal N}=1$ SQCD-type theories, 
the  quarks  can be in two different {\it standard} phases only\,:\, either this 
is  the HQ (heavy quark) phase with $\langle Q\rangle=
\langle {\ov Q}\rangle=0$ where quarks are not higgsed but
confined, or they are higgsed with some components 
$\langle Q^i_\beta\rangle\neq 0$, at appropriate values of the 
Lagrangian parameters. 

At $1 < N_F < N_c$ and $m_Q\ll\la$  all quarks $Q,\, {\ov Q}$
are higgsed in the perturbative weak coupling regime (at not
too large values of $N_c$, see \cite{ch21}),
so that there is no need  for additional 
dynamical assumptions.  But really, although  the 
gauge couplings  $a_{*}$  and ${\ov a}_{*}$, in  
general, are not small  at $\mu\lesssim\la$ within  
the conformal region $3N_c/2 < N_F < 3 N_c$
and are $\sim 1$, the dynamics of the  conformal  regime 
is sufficiently simple and well understood,  so that {\it there 
is also  no  real need for any additional  dynamical assumptions}. 

Within the conformal   window, clearly, the  direct SQCD 
theory with light quarks and gluons {\it enters smoothly} from the 
weak   coupling regime at $\mu\gg\la$ to conformal regime at 
$\mu < \la$.     Nothing happens with its mass spectrum  and 
{\it all its light quarks and gluons remain effectively massless} (as it was \,
at  $\mu > \la$). And so, {\it the dual theory at $\mu < \la$ is not the low 
energy form  of  the direct one in the interval $3N_c/2 < N_F <3N_c$.  
It has  to be considered as the independent theory}.  Its mass spectrum 
has to  be  calculated and compared with the direct one  to see whether 
these   two  theories are equivalent or not.

It was shown then \cite{ch19} that {\it the mass spectra of the direct 
and dual  theories are {\it parametrically different} in the vicinity of the 
left  end $0 < (2N_F-3N_c)/N_f\ll 1$ of the conformal window.
Then, there are no reasons that they will become exactly the same  
in the whole region}  $ (2N_F-3N_c)/N_f\sim 1$.\\

Besides it was shown in the direct $\Phi$-theory that a
qualitatively new phenomenon takes place: under appropriate
conditions, the seemingly heavy and dynamically irrelevant
fields $\Phi$ `return back' and  {\it  there appear two additional
generations of light $\Phi$-particles with small masses}
$\mu(\Phi)\ll\la$, \cite{ch19}.

\hspace*{3mm} The main purpose of papers \cite{ch19},
\cite{ch13} and \cite{ch16} was to calculate the mass spectra 
in two  ${\cal N}=1$ SQCD-type theories with additional 
scalar fields :  the direct $\Phi$-theory and Seiberg's dual variant
\cite{S1,S2}, the ${d\Phi}$-theory. In the next section  the
definitions of these direct $\Phi$- and dual $d\Phi$-theories and 
their most general properties are presented in some details.

In section 15 considered are the mass spectra of the direct
$\Phi$-theory at $1 \leq N_F< N_c$. It is shown for these values of
$N_F$ that all quarks interactions have logarithmically small
couplings and so all calculations do not require any additional
dynamical assumptions and are, in a sense, standard and
straightforward.

Starting from section 16 considered are both the direct and dual
theories with $3N_c/2 < N_F < 3N_c$. In section 16 exact results are
given for multiplicities of dufferent vacua and the nontrivial parametric
behavior of quark and gluino condensates in different vacua and
at different values of $\mph/\la\gg 1$, where $\mph$ is the
large mass parameter of the fields $\Phi$ and $\la$ is the scale
factor of the gauge coupling. These results for the quark and
gluino condensates constitute a base for further calculations
of mass spectra.

In section 17 discussed is the new nontrivial phenomenon of 
the  appearance (at the appropriate conditions) of additional
generations of {\it light} colorless $\Phi$-particles in the
direct theory. It is shown that, due to their Yukawa
interactions with light quarks, the seemingly heavy and
dynamically irrelevant fields $\Phi$ (fions) with the large
original mass parameter $\mph(\mu\sim\la)\gg\la$ can 'return
back', and there appear two additional generations of light
$\Phi$-particles with small masses $\mu^{\rm pole}(\Phi)\ll\la$.

Sections 18-23 deal with calculations of mass spectra in the
direct and dual theories at $3N_c/2<N_F<2N_c$ where both
theories are in the strong coupling conformal regimes 
with both fixed couplings $a_{*}=(N_c g^2/8\pi^2)\sim 
{\ov a}_{*}\sim  1$,  in general, at the scale $\mu=\la$. 
 
It was shown that, as in the standard SQCD \cite{ch3}, the mass
spectra of the direct theory and its Seiberg's dual variant  {\it
are parametrically  different at the left end of the conformal 
window},  i.e.  at $0 < (2N_F-3N_c)/N_F\ll  1$.

In sections 24-29 these direct and dual theories are considered at 
$2N_c < N_F <3 N_c$. It is shown that their mass spectra are
also parametrically different

And finally for this Part II, the direct and dual theories are considered 
in sections 30-37 at  $N_c < N_F < 3N_c/2$, where the dual theory  
at $\mu \ll\la$  is IR-free  and  logarithmically 
weakly coupled  while direct theory  is in the very strong \, 
coupling regime with $a(\mu\ll\la)\gg 1$. And  {\it only in this case} 
the dynamical  scenario from \cite{ch3} was really  needed to calculate 
the mass spectra of the  direct theory.\\
 
At present, unfortunately, no way is known to obtain 
{\it  direct solutions} (i.e. without any additional assumptions) of
${\cal N}=1$ SQCD-type theories in strong coupling regimes.
Therefore, to calculate mass spectra of ${\cal N}=1$ theories in
such cases one has to introduce and use some {\it assumptions}
about the dynamics of these theories in the strong coupling
regions. In other words, one has to rely on a definite dynamical
scenario.

Calculations of mass spectra of the very strongly coupled 
direct theory  in \cite{ch16}  were based on the dynamical scenario 
introduced in  \cite{ch3} for calculations of mass spectra in such 
regime  in the standard direct  ${\cal N}=1$ SQCD 
(i.e. without  additional scalar fields).

We recall that this dynamical scenario introduced in \cite{ch3}
assumes that in considered ${\cal N}=1$ SQCD-type theories,
the quarks can be in two {\it standard} phases  only. These are: 
a) the HQ (heavy quark)  phase where they are not higgsed but 
confined, with  $\langle Q^i_{\beta}\rangle=\langle {\ov Q}_i^
{\beta} \rangle=0$\,\, b) the Higgs phase where 
they are  higgsed, with some components 
$\langle Q^i_{\beta}\rangle\neq 0$. 
Moreover, the  word `standard'  implies that these two phases 
are realized in a standard way, even in the strong coupling region 
$a\gtrsim 1$. This means that, unlike e.g. ${\cal N}=2$ SQCD with its 
very special properties, in these ${\cal N}=1$ SQCD-type theories 
without elementary adjoint colored scalar superfields, {\it there appear 
no  additional non-standard parametrically lighter solitons (e.g.
parametrically lighter magnetic monopoles or dyons) in the
spectrum in the strong coupling region where quarks either 
decouple as heavy or are higgsed}. \\

We would like to emphasize that {\it this dynamical scenario from
\cite{ch3} satisfies all those tests which were used as checks of Seiberg's
hypothesis about equivalence of the direct and dual theories}. 
This shows that {\it all these checks, although necessary,  may well be
insufficient}. 

Moreover,  this scenario  looks self-consistent, without internal contradictions 
and  is not  in contradiction with  known at present proven results. \\

The direct $\Phi$ and dual $d\Phi$ - theories considered in
papers \cite{ch19,ch13,ch16}  have much in common with the 
standard ${\cal N}=1$
SQCD (and its dual variant) considered in \cite{ch3}. It is
implied that the reader is familiar with the paper \cite{ch3}
and with the calculation methods used therein. These methods 
(by  the way, sufficiently standard, the non-standard is only the
dynamical scenario itself) are heavily used in these papers
\cite{ch19},\cite{ch13},\cite{ch16}. For this reason, some
technical details are omitted in the text below, and we refer to
\cite{ch3} where all additional technical details of similar
calculations can be found.

\section  { Definitions and some generalities}

\subsection{\quad Direct $\mathbf \Phi$ - theory}

\numberwithin{equation}{subsection}

\hspace*{4mm} The field content of this direct ${\cal
N}=1\,\,\,\Phi$ - theory includes $SU(N_c)$ gluons and $0<
N_F<3N_c$ flavors of quarks ${\ov Q}_j, Q^i$. Besides, there are
$N^2_F$ colorless but flavored fields $\Phi^{j}_{i}$ (fions)
with the large mass parameter $\mph\gg\la$.

The Lagrangian at the scale $\mu=\la$ looks as
 (the exponents with gluons in the Kahler term K
are implied here and everywhere below) 
\bq
K={\rm Tr}\,\Bigl (\Phi^\dagger \Phi\Bigr
)+{\rm Tr}\Bigl (\,Q^\dagger Q+(Q\ra {\ov Q})\,\Bigr),\,
W=\frac{2\pi}{\alpha(\mu=\la)} S+W_{\Phi}+W_{Q}.  \quad
\label{(14.1.1)}
\eq
\bbq
W_{\Phi}=\frac{\mph}{2}\Biggl [{\rm Tr}
\,(\Phi^2)-\frac{1}{\nd}\Bigl ({\rm Tr}\,\Phi\Bigr )^2\Biggr],
\quad W_Q={\rm Tr}\,{\ov Q}(m_Q-\Phi) Q.
\eeq
Here\,: $\mph=\mph(\mu=\la)$ and $m_Q=m_Q(\mu=\la)$ are the mass
parameters,  $S=W^{A}_{\beta}W^{A,\,
\beta}/32\pi^2$ where $W^A_{\beta}$ is
the gauge field strength, $A=1...N_c^2-1,\, \beta=1,2$,\,
$a(\mu=\la)=N_c g^2(\mu=\la)/8\pi^2\sim 1$ at $\bo/N_F\sim 1$ is the
gauge coupling with its scale factor $\la, \,\, f(\mu=\la)=1$ 
is the Yukawa coupling,
$a_f(\mu)= N_F f(\mu)^2/8\pi^2$,\, $\bo=3N_c-N_F,\,\, \nd=(N_F-N_c)$. 
The field $\Phi^j_i$ is in the adjoint representation of $SU(N_F)$.
This normalization of fields is used everywhere below in the main
text. Besides, the perturbatively exact NSVZ $\beta$-function for
massless SUSY theories \cite{NSVZ-1,NSVZ-2} is used in this paper.

Therefore, finally, the $\Phi$-theory we deal with has the
parameters\,: $N_c\,,\,0<N_F<3N_c\,,\,\mph$,\,
$\la,\, m_Q$, with the {\it strong hierarchies}
$\mph\gg\la\gg m_Q$. Everywhere below in the text the mass
parameter $\mph$ will be varied while $m_Q$ and $\la$ will stay
intact.\\

The Konishi anomalies \cite{Konishi} from \eqref{(14.1.1)} for the
$i$-th flavor look as (${\it i}=1\, ...\, N_F$)
\bbq
\langle\Phi_{i}\rangle\langle\frac{\partial \cw_{\Phi}}{\partial
\Phi_{i}}\rangle=0\,,\quad
\langle m_{Q,i}^{\rm tot}\rangle\langle {\ov Q}_i
Q^i\rangle=\langle S\rangle\,,\quad \langle m_{Q,\,i}^{\rm
tot}\rangle=m_Q-\langle\Phi^i_{i}\rangle,\quad ({\rm no\,\,summation
\,\,over\,\, i}).
\eeq
\bq
\langle\Phi^{i}_{j}\rangle=\frac{1}{\mph}\Biggl ( \langle{\ov
Q}_j Q^i\rangle
-\delta_{j}^{i}\frac{1}{N_c}{\rm Tr}\,\langle\QQ\rangle\Biggr
)\,,\quad \langle{\ov Q}_j Q^i\rangle
=\delta_{j}^{i}\langle{\ov Q}_i Q^i\rangle\,, \label{(14.1.2)}
\eq
and, in cases with $\mu_H<\la$,\, $\langle m_{Q,i}^{\rm
tot}\rangle$ is the value of the quark total running mass at
$\mu=\la$.

At all scales $\mu$ until the field $\Phi$ remains too heavy and
non-dynamical, i.e. until its perturbative running mass
$\mu_{\Phi}^{\rm pert}(\mu)>\mu$, it can be integrated out and
the Lagrangian takes the form

\bbq
K=z_Q(\la, \mu){\rm Tr}\Bigl (Q^\dagger Q+Q\ra {\ov Q}\Bigr ),
\quad \cw =\frac{2\pi}{\alpha(\mu,\la)}S+\cw_Q\,,
\eeq
\bq
W_Q=m_Q{\rm Tr}({\ov Q} Q)-\frac{1}{2\mph}\Biggl ({\rm
Tr}\,({\ov Q}Q)^2-\frac{1}{N_c}\Bigl({\rm Tr}\,{\ov Q} Q
\Bigr)^2 \Biggr ).\label{(14.1.3)}
\eq

The Konishi anomalies from \eqref{(14.1.3)} for the i-th flavor
look as
\bbq
\langle S\rangle=\langle\frac{\lambda\lambda}{32\pi^2}\rangle=
\langle {\ov Q}_i\frac{\partial \cw_Q}{\partial {\ov
Q}_i}\rangle=m_Q\langle {\ov Q}_i Q^i
\rangle-\frac{1}{\mph}\Biggl (\sum_j\langle{\ov Q}_i
Q^j\rangle\langle{\ov Q}_j Q^i\rangle-\frac{1}{N_c}\langle{\ov
Q}_i Q^i\rangle\langle {\rm Tr}\,{\ov Q} Q \rangle\Biggr )=
\eeq
\bq
=\langle {\ov Q}_i Q^i \rangle\Biggl [\,m_Q-\frac{1}{\mph}\Biggl
( \langle {\ov Q}_i Q^i \rangle-\frac{1}{N_c}\langle {\rm Tr}
\,{\ov Q} Q \rangle\Biggr ) \Biggr ]= \langle m_{Q,i}^{\rm tot}\rangle
\,\langle {\ov Q}_i Q^i\rangle\,,\quad i=1\,...\, N_F\,,  \label{(14.1.4)}
\eq
\bbq
\langle {\ov Q}_i\frac{\partial W_Q}{\partial {\ov Q}_i}-{\ov
Q}_j\frac{\partial W_Q}{\partial
{\ov Q}_j}\rangle=\langle{\ov Q}_i Q^i -{\ov Q}_j Q^j\rangle
\Biggl [\,m_Q-\frac{1}{\mph}\Biggl ( \langle {\ov Q}_i Q^i+{\ov
Q}_j Q^j\rangle-\frac{1}{N_c}\langle {\rm Tr}\,{\ov Q} Q
\rangle\Biggr ) \Biggr ]=0\,.
\eeq

It is most easily seen from \eqref{(14.1.4)} that there are only
two types of vacua\,: a) the vacua with the unbroken flavor
symmetry, $\langle{\ov Q}_j Q^i\rangle=\delta^i_j\langle{\ov
Q}Q\rangle$,\, b) the vacua with the spontaneously broken flavor
symmetry, and the breaking is of the type $U(N_F)\ra
U(n_1)\times U(n_2)$ only: $\langle{\ov Q}_j
Q^i\rangle=\delta^i_j\langle {\ov Q}_1
Q^1\rangle\equiv\delta^i_j\langle\Qo\rangle,\, i,j=1,...n_1,\,\,
\langle{\ov Q}_j Q^i\rangle=\delta^i_j\langle {\ov Q}_2
Q^2\rangle\equiv\delta^i_j\langle\Qt\rangle,\,
i,j=n_1+1,...N_F$. In these vacua one obtains from \eqref{(14.1.4)}
\bbq
\langle {\Qo+\Qt-\frac{1}{N_c}{\rm Tr}\, {\ov Q} Q}\rangle_{\rm
br}=m_Q\mph,\quad
\langle S\rangle_{\rm br}=\frac{1}{\mph}\langle\Qo\rangle_{\rm
br}\langle\Qt\rangle_{\rm br},\quad
\langle\Qo\rangle_{\rm br}\neq\langle\Qt\rangle_{\rm br}\,,
\eeq
\bq
\langle m^{\rm tot}_{Q,1}\rangle_{\rm
br}=m_Q-\langle\Phi_1\rangle_{\rm
br}=\frac{\langle\Qt\rangle_{\rm br}}{\mph},\quad \langle m^{\rm
tot}_{Q,2}\rangle_{\rm br}=m_Q-\langle\Phi_2\rangle_{\rm
br}=\frac{\langle\Qo\rangle_{\rm br}}{\mph}\,.\label{(14.1.5)}
\eq

We now recall details of the dynamical scenario introduced in
\cite{ch3} and used in this paper for calculations of mass spectra, 
both in the  direct and dual theories.\\

1) Recall first that perturbatively exact NSVZ $\beta$-function
\cite{NSVZ-1} predicts values  of quark
anomalous dimensions, $\gamma_Q$ of the direct quark or
$\gamma_q$ of the dual one, in the conformal regime at
$3N_c/2<N_F<2N_c\,$. In the direct $\Phi$-theory
\bq
\frac{d a^{-1}(\mu)}{d\ln\mu}=\hat\beta(a)=\frac{\bo-N_F\gamma_Q}
{1-a}\ra 0\,\,\ra\,\,\gamma_Q \ra\frac{\bo}
{N_F}\,,\,\, \bo=3N_c-N_F>0\,,\,\, a(\mu)=\frac{N_c
g^2(\mu)}{8\pi^2}\,.\label{(14.1.6)}
\eq

Therefore, the renormalization factor of the quark Kahler term
is also known exactly in the conformal regime:
$z_Q(\la, \mu)\ra (\mu/\la)^{\gamma_Q}\ll 1$ at $\mu\ll\la$, while
$a(\mu\ll\la)\ra a_{*}={\rm const},\,\,a_{*}=O(1)$ in general.
In the direct theory, when the fion field $\Phi$ is effectively
massless and participates actively in the conformal regime, its
anomalous dimension and renormalization factor are also 
known exactly from the conformal symmetry:
$\gamma_{\Phi}= - 2\gamma_Q,\,\,
z_{\Phi}(\la, \mu)=(\la/\mu)^{2\gamma_Q}\gg 1$ at
$\mu\ll\la$. In the dual theory, correspondingly:
$\gamma_q=\bd/N_F\,,\,\,\gamma_M=-2\gamma_q\,,\,\,
\bd=3\nd-N_F=2N_F-3N_c>0$, and the dual gauge coupling ${\ov
a}(\mu\ll\la)\ra {\ov a}_{*}={\rm const}$. But at the left end
of the conformal window there appears {\it additional small
parameter}\,: $0<\bd/N_F=(2N_F-3N_c)/N_F\ll 1,\,\,
\gamma_q=\bd/N_F\approx {\ov a}_{*}\ll 1$. The {\it explicit
parametric dependence of various particle masses on this small
parameter} is widely used in the text. It allows to trace {\it
the parametric differences in mass spectra of direct and dual
theories}. \\

2)  The  effectively massless  conformal regimes are  broken 
explicitly  at some lower scales $\mu_i\ll\la$ 
by nonzero particles masses. These may be e.g.
the quark pole masses $m^{\rm pole}_{Q,i},\,\,i=1\, {\rm or}\,\,
2,$ or gluon masses $\mu^{\rm pole}_{{\rm gl},\,i}$ due to
higgsed quarks. And {\it this is a first place where we need to  use the  
simple assumption of the dynamical scenario from}  \cite{ch3}.  It 
states that,   unlike the very specific  ${\cal N}=2$ SQCD, in considered 
${\cal N}=1$ SQCD-type theories
without elementary colored adjoint scalar fields, the dynamics is 
really  standard, i.e. {\it no additional  parametrically lighter
solitons (e.g. magnetic monopoles or dyons) are formed at those
scales where the (effectively) massless regime is broken explicitly by 
the quark masses or gluon masses originating from higgsed quarks}.
\footnote{\,
Note that appearance of such {\it additional light solitons}
will change the 't Hooft triangles at lower energies.
\label{(f30)}
}
I.e., in this respect, the dynamics is {\it qualitatively
similar} to those in the weak coupling regime.\\

3) Finally, to deal with the ${\cal N}=1$ SYM theory,
originating after decoupling of heavy particles at lower
 energies,  we use the effective superpotential proposed 
 by  Veneziano-Yankielowicz \cite{VY}.\\

The use of quark condensates values  $\langle\Qo\rangle$
and $\langle\Qt\rangle$ in various vacua calculated in section
16, the known RG evolution in the weak coupling or 
the superconformal regime and described above simple 
assumption of the used dynamical  scenario are  sufficient to 
calculate the typical mass spectra of  direct and dual theories  
at $1 < N_F < 3N_c$. 
4) Moreover, we explicitly calculate below the 
{\it  parametric dependencies} of particle masses on the {\it
additional small parameter}, $0<\bd/N_F=(2N_F-3N_c)/N_F\ll 1$,
appearing at the left end of the conformal window, see 
below and section 4 in \cite{ch3} for more details. This
allows to trace explicitly the {\it parametric differences in
mass spectra of the direct and Seiberg's dual theories.} \\

\numberwithin{equation}{subsection}

\subsection{\quad Dual $\mathbf {d\Phi}$ - theory}

In parallel with the direct $\Phi$ - theory with
$N_c<N_F<3N_c$\,,\, we consider also the Seiberg dual variant
\cite{S1,S2} (the $d\Phi$ - theory), with the dual Lagrangian at
$\mu=\la$
\bq
K={\rm Tr}\,\Phi^\dagger \Phi+ {\rm Tr}\Bigl (
q^\dagger q + (q\ra\ov q)\, \Bigr )+{\rm Tr}\,
\frac{M^{\dagger} M}{\mu_2^2}\,,\quad
W=\, \,\frac{2\pi}{\ov \alpha(\mu=\la)}\, {\ov S}+
{\wt W}_M+ W_q\,,\label{(14.2.1)}
\eq
\bbq
{\wt W}_M=\frac{\mph}{2}\Biggl [{\rm
Tr}\,(\Phi^2)-\frac{1}{\nd}\Bigl ({\rm Tr}\,\Phi\Bigr )^2\Biggr
]+ {\rm Tr}\, M(m_Q-\Phi),\quad W_q= -\,\frac{1}{\mu_1}\,
{\rm Tr} \Bigl ({\ov q}\,M\, q \Bigr )\,.
\eeq
Here\,:\, the number of dual colors is $\nd=N_F-N_c,\,
\bd=3\nd-N_F$, and $M^i_j\ra ({\ov Q}_j Q^i)$ are the $N_F^2$
elementary mion fields, ${\ov a}(\mu)=\nd{\ov
\alpha}(\mu)/2\pi=\nd{\ov g}^2(\mu)/8\pi^2$ is the dual running
gauge coupling (with its scale parameter $\Lambda_q$),\,\,${\ov
S}={\rm \ov w}^{b}_{\beta}{\rm \ov w}^{b,\,\beta}/32\pi^2$,\,\,
${\rm \ov w}^b_{\beta}$ is the dual gluon field strength. The
gluino condensates of the direct and dual theories are matched
in all vacua, $\langle{-\,\ov S}\rangle=\langle S\rangle=\lym^3$, 
as well as $\langle M_j^i(\mu=\la)\rangle=\langle{\ov Q}_j Q_i (\mu=
\la)\rangle$,  and the scale parameter $\Lambda_q$ of the dual gauge 
coupling  is taken as $\Lambda_q= - \la$, see \eqref{(3.0.4)} and 
Appendix in \cite{ch3}  for more details. At $3/2<N_F/N_c<2$ this 
dual theory can be  taken as UV free at $\mu\gg\la$, and this requires 
that its  dual Yukawa coupling at $\mu=\la,\, {\ov f} (\mu=\la)=\mu_2/\mu_1$, 
cannot  be larger than its gauge coupling ${\ov g}(\mu=\la)$, i.e.
$\mu_2/\mu_1\lesssim 1$. The same requirement to the value of
the Yukawa coupling follows from the conformal behavior of this
theory at $3/2<N_F/N_c<2$ and $\mu<\la$, i.e.
${\ov f}(\mu=\la)=1=\mu_2/\mu_1\approx f_{*}$ at $\bd/N_F=O(1)$. 
We  consider below this dual theory at $\mu\leq\la$ only, where it
claims to be equivalent to the direct $\Phi$ - theory. As was
explained in \cite{ch3}, one has to take $\mu_1\sim\la$ at
$\bd/N_F=(3\nd-N_F)/N_F=O(1)$ in \eqref{(14.2.1)} to match the
gluino condensates in the direct and dual theories. Therefore,
$\mu_2\sim\mu_1\sim\la$ in this case also. But to match the
gluino condensates in the direct and dual theories at the left
end of the conformal window, i.e. at $0<\bd/N_F\ll 1$, one has
to take $(\mu_2/\mu_1)^2\approx f^2_*=O(\bd/N_F)\ll 1$ and
$\mu_1\sim Z_q\la\ll\la,\, Z_q\sim\exp\{-\nd/7\bd\}\ll 1$ (with
the exponential accuracy, i.e. powers of the small parameter
$0<\bd/N_F\ll 1$ are not traced here and only the powers of
$Z_q$ are traced, this is sufficient for our purposes, so that
at $\bd/N_F=O(1)$ one has to put $Z_q\ra 1$, see \cite{ch3} 
 for  more details.\\

Really, the fields $\Phi$ remain always too heavy and
dynamically irrelevant in this $d\Phi$ - theory at $3
N_c/2<N_F<2 N_c$, so that they can be integrated out once and
forever and, finally, we write the Lagrangian of the dual theory
at $\mu=\la$ in the form
\bbq
K= {\rm Tr}\Bigl ( q^\dagger q +(q\ra\ov q) \Bigr )+{\rm
Tr}\,\frac{M^{\dagger}M}{Z^2_q\la^2}\,,\quad
\cw=\, \,\frac{2\pi}{\ov \alpha(\mu=\la)}\, {\ov
s}+\cw_M+\cw_q\,,
\eeq
\bq
\cw_M=m_Q{\rm Tr}\,M -\frac{1}{2\mph}\Biggl [{\rm Tr}\, (M^2)-
\frac{1}{N_c}({\rm Tr}\, M)^2 \Biggr ]\,,\quad
\cw_q= -\,\frac{1}{Z_q\la}\,\rm {Tr} \Bigl ({\ov q}\,M\, q \Bigr
)\,.  \label{(14.2.2)}
\eq

The Konishi anomalies for the $i$-th flavor look here as (${\it
i}=1\, ...\, N_F$)
\bq
\langle M_i\rangle\langle N_i\rangle=Z_q\la\langle
S\rangle\,,\quad \frac{\langle
N_i\rangle}{Z_q\la}=m_Q-\frac{1}{\mph}\Bigl (\langle
M_i-\frac{1}{N_c}{\rm Tr}\,M \rangle\Bigr )=\langle m_{Q,i}^{\rm
tot}\rangle\,,\label{(14.9)}
\eq
\bbq
\langle N_i\rangle\equiv\langle{\ov q}^i
q_i(\mu=\la)\rangle\,,\quad {\rm no\,\, summation\,\, over\,\, i}\,.
\eeq

In vacua with the broken flavor symmetry these can be rewritten
as
\bbq
\langle M_1+M_2-\frac{1}{N_c}{\rm Tr}\, M\rangle_{\rm
br}=m_Q\mph,\quad
\langle S\rangle_{\rm br}=\frac{1}{\mph}\langle M_1\rangle_{\rm
br}\langle M_2\rangle_{\rm br},\quad
\langle M_1\rangle_{\rm br}\neq\langle M_2\rangle_{\rm br}\,,
\eeq
\bq
\frac{\langle N_1\rangle_{\rm br}}{Z_q\la}=\frac{\langle
S\rangle_{\rm br}}{\langle M_{1}\rangle_{\rm br}}=\frac{\langle
M_{2}\rangle_{\rm br}}{\mph}=m_Q-\frac{1}{\mph}\Bigl (\langle
M_{1}-\frac{1}{N_c}{\rm Tr}\, M\rangle_{\rm br} \Bigr )=\langle
m^{\rm tot}_{Q,1}\rangle_{\rm br}\,,\label{(14.2.3)}
\eq
\bbq
\frac{\langle N_2\rangle_{\rm br}}{Z_q\la}=\frac{\langle
S\rangle_{\rm br}}{\langle M_{2}\rangle_{\rm br}}=\frac{\langle
M_{1}\rangle_{\rm br}}{\mph}=m_Q-\frac{1}{\mph}\Bigl (\langle
M_{2}-\frac{1}{N_c}{\rm Tr}\, M\rangle_{\rm br} \Bigr )=\langle
m^{\rm tot}_{Q,1}\rangle_{\rm br}\,.
\eeq

\section {Vacua, condensates and mass spectra at $\mathbf 
{0< N_F< N_c}$}

\hspace*{4mm} Clearly, there is no dual theory for this range of
$N_F$ values. Moreover (see below in this section), in vacua of
the direct theory with the unbroken flavor symmetry all quarks
are higgsed in the logarithmically small coupling regime
with the large masses of higgsed gluons, $\mu_{\rm gl}\gg \la$
(at not too large values of $N_c$, see \cite{ch21} ).
In vacua with the broken flavor symmetry all quarks: a) are
either also higgsed in the weak (logarithmically small) coupling
regime with $\mu_{\rm gl,\,\it i}\gg \la,\, i=1,2$\,, at
$\la\ll\mph\ll {\tilde\mu}_{\Phi}$,\,\, ${\tilde\mu}_{\Phi}\sim
\la (\la/m_Q)^{(\bo-n_1)/n_1}\gg\la$\,;\,\, b) or, in $\rm
br1$-vacua with $1\leq n_1\leq [N_F/2]$, the quarks ${\ov
Q}_1,\, Q^1$ with flavors $n_1$ are higgsed in the weak
(logarithmically small) coupling regime, while the quarks ${\ov
Q}_2,\, Q^2$ with flavors $n_2=N_F-n_1$ are in the HQ-phase at
$\mph\gg {\tilde\mu}_{\Phi}$, they are weakly confined (i.e. the
tension of confining string originated from unbroken
$SU(N_c-n_1)$ color group is much smaller than quark masses,
$\sqrt\sigma\ll m_{Q,\, perturb}^{\rm pole}$) and also
perturbatively logarithmically weakly coupled and
non-relativistic inside hadrons (in $\rm br2$-vacua
$n_1\leftrightarrow n_2$). Therefore, finally, in all vacua and
at all values $\la\ll\mph$, the quarks are parametrically weakly
coupled and their dynamics is simple and qualitatively evident.

For this reason, {\it we need no any additional assumptions
about the quark dynamics to calculate the mass spectra at}\, $0<
N_F< N_c$. In other words, because the HQ- and Higgs-phases of
quarks are at logarithmically weak couplings, there is no need
to mention about any assumed dynamical scenario at all (it is
really needed to calculate the mass spectra in the strong
coupling region only).

The calculations methods used below in this section have much in
common with those in the standard SQCD with $m_Q/\la\ll 1$ and
$0< N_F< N_c$ in section 2 of \cite{ch1}. It is implied that the
reader is familiar with \cite{ch1}, so that some technical ins
and outs are omitted below (see section 2 in \cite{ch1} for much
more details). But really, as mentioned above, because all
quarks are parametrically weakly coupled, all calculations in
this section 15 are highly standard and, we hope, self-evident.

\subsection{ Unbroken flavor symmetry}

There is $N_{\rm unbrok}=(2N_c-N_F)$ such vacua and all quarks
are higgsed in all of them, but the hierarchies in the mass
spectrum are parametrically different depending on the value of
$\mph$ (see below). In any case, all $N_F^2$ fions are very
heavy and dynamically irrelevant in these vacua at scales
$\mu<\mu^{\rm pole}_{1}(\Phi)=f^2\mph$ and can be
integrated out from the beginning.

All quarks are higgsed at the high scale $\mu=\mu_{\rm gl},\,
\la\ll\mu_{\rm gl}\ll\mu^{\rm pole}_1(\Phi)=\mu_{\Phi}/z_{\Phi}
\simeq \mu_{\Phi}$,
\bq
\mu^2_{\rm gl}\sim  g^2(\mu=\mu_{\rm gl})z_Q(\la, \mu_{\rm gl})
\langle\Pi\rangle,\quad \langle\Pi\rangle=\langle{\ov Q}_1
Q^1(\mu=\la)\rangle\equiv\langle{\ov Q} Q\rangle, \quad
g^2(\mu)=4\pi\alpha(\mu),\label{(15.1.1)}
\eq
where (in the approximation of leading logs and
$C_F=(N_c^2-1)/2N_c\approx N_c/2$)
\bq
\frac{2\pi}{\alpha(\mu_{\rm gl})}\approx \bo\ln{\frac{\mu_{\rm
gl}}{\la}}\,\,,\,\, z_Q(\la, \mu_{\rm gl})\sim\Bigl
(\frac{\alpha(\la)}{\alpha(\mu_{\rm gl})}\Bigr)^{2C_F/\bo}\sim
\Bigl (\ln\frac{\mu_{\rm gl}}{\la}\Bigr )^{N_c/\bo}\gg 1\,,\quad
\bo=3N_c-N_F\,.\label{(15.1.2)}
\eq

Hence, after integrating out all heavy higgsed gluons and their
superpartners at $\mu<\mu_{\rm gl}$ one remains with the
$SU(N_c-N_F)$ pure Yang-Mills theory with the scale factor
$\lym$ of its gauge coupling. Finally, after integrating out
remained gluons at $\mu<\lym$ via the Veneziano-Yankielowicz
(VY) procedure \cite{VY} (see section 2 in \cite{ch1} for
more details), one obtains the Lagrangian of $N_F^2$ pions
\bq
K=2 z_Q(\la, \mu_{\rm gl})\,{\rm Tr}\,\sqrt
{\Pi^{\dagger}\Pi}\,,\quad \cw=\nd S+\cw_{\Pi}\,,\label{(15.1.3)}
\eq
\bbq
S=\Biggl (\,\frac{\la^{\bo}}{\det \Pi}\,\Biggr
)^{\frac{1}{N_c-N_F}}\,,\quad
\cw_{\Pi}=m_Q{\rm Tr}\,\Pi -\frac{1}{2\mph}\Biggl [{\rm Tr}\,
({\Pi}^2)- \frac{1}{N_c}({\rm Tr}\, \Pi)^2 \Biggr ],
\eeq
\bbq
\langle\Pi^i_j\rangle=\delta^i_j\,\langle\Pi\rangle=\delta^i_j\,\langle
{\ov Q}_1 Q^1(\mu=\la)\rangle,\quad i,j=1\,...\,N_F\,.
\eeq

It follows from \eqref{(15.1.3)} that depending on the value of
$\mph/\la \gg 1$\, there are two different regimes.\\

{\bf i)}\,\, At $\la\ll\mph \ll
\mo=\la(\la/m_Q)^{(2N_c-N_F)/N_c},\,\,\mo\gg\la$, the term
$m_Q{\rm Tr}({\ov Q}Q)$ in the superpotential \eqref{(15.1.3)}
gives only  small correction and one obtains
\bq
\langle \Pi \rangle_{\rm o}\sim \la^2\Bigl
(\frac{\mph}{\la}\Biggr )^{\frac{N_c-N_F}{2N_c-N_F}}\gg \la^2\,.
\label{(15.1.4)}
\eq

There are $(2N_c-N_F)$ such vacua, this agrees with \cite{CKM}.
\footnote{\,
To see that there are just $2N_c-N_F$ vacua and not less, one
has to separate slightly all quark masses, $m_Q\ra
m_Q^{(i)},\,i=1...N_F,\, 0<(\delta
m_Q)^{ij}=(m_Q^{(i)}-m_Q^{(j)})\ll {\ov m}_Q$. All quark mass
terms give only small power corrections to \eqref{(15.1.4)}, but just these
corrections show the $Z_{2N_c-N_F}$ multiplicity of vacua.
}
The masses of heavy gluons and their superpartners are given in
\eqref{(15.1.1)}, while from \eqref{(15.1.3)} the pion masses are
\bq
\mu_{\rm o}(\Pi)\sim \frac{\langle \Pi \rangle_{\rm
o}}{z_Q(\la, \mu_{\rm gl})\mph}\sim \frac{\la}{z_Q(\la, 
\mu_{\rm gl})}\Bigl (\frac{\la}{\mph}\Biggr
)^{\frac{N_c}{2N_c-N_F}}\ll m_Q\,.\label{(15.1.5)}
\eq
Besides, the scale of the gluino condensate of unbroken
$SU(N_c-N_F)$ is
\bq
\lym=\langle S\rangle^{1/3}\sim \Biggl (\frac{\la^{\bo}}{\det
\langle\Pi\rangle_{\rm o}}\Biggr )^{\frac{1}{3(N_c-N_F)}}\sim
\la\Bigl(\frac{\la}{\mph}\Biggr
)^{\frac{N_F}{3(2N_c-N_F)}},\,\,  \mu_{\rm o}(\Pi)\ll \lym\ll
\la\ll \mu_{\rm gl},\,\,\,\,\label{(15.1.6)}
\eq
and there is a large number of gluonia with the mass scale $\sim
\lym$ (except for the case $N_F=N_c-1$ when the whole gauge
group is higgsed, there is no confinement and no gluonia with 
masses $\sim \lym$
and the non-perturbative superpotential in \eqref{(15.1.3)}
originates not from the unbroken $SU(N_c-N_F)$ but directly from
the instanton contribution \cite{ADS}).\\

{\bf ii)}\,\,\, $(2N_c-N_F)$ vacua split into two groups of
vacua with parametrically different mass spectra at $\mph\gg
\mo$. There are $N_c$\,\, SQCD vacua with $\langle \Pi
\rangle_{\rm SQCD}\sim\la^2(\la/m_Q)^{(N_c-N_F)/N_c}$ differing
by $Z_{N_c}$ phases of the quark and gluino condensates (in
these, the last term $\sim \Pi^2/\mph$ in the superpotential
\eqref{(15.1.3)} can be neglected).  And $(N_c-N_F)$ of nearly
degenerate classical vacua with parametrically larger
condensates $\langle \Pi \rangle_{\rm cl}\sim m_Q\mph$. In
these, the first non-perturbative quantum term $\sim S$ in the
superpotential \eqref{(15.1.3)} gives only small corrections with
$Z_{N_c-N_F}$ phases, but the multiplicity of vacua originates
just from these small corrections. The properties of SQCD vacua
have been described in detail in chapter 2 of \cite{ch1}, the
pion masses are $\mu_{\rm SQCD}(\Pi)\sim
m_Q/z_Q(\la, \mu^{SQCD}_{\rm gl})\ll m_Q$ therein, where
$z_Q(\la, \mu^{SQCD}_{\rm gl})\gg 1$ is the logarithically large
perturbative renormalization factor. In $(N_c-N_F)$ classical
vacua the gluon and pion masses are given  in
\eqref{(15.1.5)} but now
\bq
\langle \Pi \rangle_{\rm cl}\sim  m_Q\mph \gg \la^2\,,\quad
\mu_{\rm cl}(\Pi)\sim \frac{m_Q}{z_Q(\la, \mu^{\rm cl}_{\rm
gl})}\ll m_Q\,,\label{(15.1.7)}
\eq
and in all vacua (except for the case $N_F=N_c-1$ ) there is a
large number of gluonia with the mass scale
\bq
\sim \lym^{\rm SQCD}=\langle S \rangle^{1/3}\sim\Biggl
(\frac{\la^{\bo}}{\det \langle\Pi\rangle_{\rm SQCD}}\Biggr
)^{\frac{1}{3(N_c-N_F)}}\sim \la\Biggl
(\frac{m_Q}{\la}\Biggr)^{N_F/3N_c}\quad {\rm in\,\, N_c
\,\,\,SQCD \,\, vacua}\,,\label{(15.1.8)}
\eq
\bq
\sim \lym^{\rm class}\sim\Biggl (\frac{\la^{\bo}}{\det
\langle\Pi\rangle_{\rm cl}}\Biggr )^{\frac{1}{3(N_c-N_F)}}\sim
\la\Biggl (\frac{\la^2}{m_Q\mph}\Biggr
)^{\frac{N_F}{3(N_c-N_F)}}\,\, {\rm in\,\, (N_c-N_F)
\,\,classical\,\, vacua}.\label{(15.1.9)}
\eq

Finally, the change of regimes ${\bf i}\leftrightarrow {\bf ii}$
occurs at
\bq
\Biggl (\frac{\mo}{\la}\Biggr )^{\frac{N_c-N_F}{2N_c-N_F}}\sim
\frac{m_Q \mo}{\la^2}\gg 1\quad \ra \quad
\mo \sim \la\Biggl (\frac{\la}{m_Q}\Biggr
)^{\frac{2N_c-N_F}{N_c}}\gg\la\,.\label{(15.1.10)}
\eq
\vspace*{2mm}

\subsection{ Spontaneously broken flavor symmetry\,: $U(N_F)\ra
U(n_1)\times U(n_2)$}

\hspace*{4mm} The quark condensates $\langle{\ov Q}_j
Q^i\rangle\sim \delta^i_j C_i$ split into two groups in these
vacua with the spontaneously broken flavor symmetry\,: there are
$1\leq n_1\leq [N_F/2]$ equal values
$\langle\Pi_1\rangle=\langle{\ov Q}_1
Q^1\rangle\equiv\langle\Qo\rangle$ and $n_2=(N_F-n_1)\geq n_1$
equal values $\langle\Pi_2\rangle=\langle{\ov Q}_2
Q^2\rangle\equiv\langle\Qt\rangle\neq \langle\Qo\rangle$ (unless
stated explicitly, here and everywhere below in the text it is
implied that $1-(n_1/N_c),\,\, 1-(n_2/N_c)$ and $(2N_c-N_F)/N_c$
are all $O(1)$\,). And there will be two different phases,
depending on the value of $\mph/\la \gg 1$\, (see below).\\

 At $\la\ll\mph\ll\mo$ all qualitative
properties are similar to those for unbroken symmetry. All
quarks are higgsed at high scales $\mu_{\rm gl, 1}\sim \mu_{\rm
gl, 2}\gg \la$ and the low energy Lagrangian has the form
\eqref{(15.1.3)}. The term $m_Q{\rm Tr} ({\ov Q}Q)$ in the
superpotential in \eqref{(15.1.3)} gives only small corrections,
while \eqref{(14.1.4)} can be rewritten here in the form
\bbq
\langle\Pi_1+\Pi_2\rangle_{\rm br}=\frac{1}{N_c}{\rm
Tr}\,\langle \Pi\rangle_{\rm br}+m_Q\mph\approx
\frac{1}{N_c}\langle n_1\Pi_1+n_2\Pi_2\rangle_{\rm br}\,\, \ra
\,\, \Bigl (1-\frac{n_1}{N_c}\Bigr )\langle\Pi_1\rangle_{\rm
br}\approx -\Bigl (1-\frac{n_2}{N_c}\Bigr
)\langle\Pi_2\rangle_{\rm br},
\eeq
\bq
\langle S\rangle_{\rm br}=\Biggl
(\frac{\la^{\bo}}{\langle\Pi_1\rangle^{n_1}_{\rm
br}\langle\Pi_2\rangle^{n_2}_{\rm br}}\Biggr )^
{\frac{1}{N_c-N_F}}=\frac{\langle\Pi_1\rangle_{\rm
br}\langle\Pi_2\rangle_{\rm br}}{\mph}\,,\label{(15.2.1)}
\eq
\bq
\mu^2_{\rm gl, 1}\sim\mu^2_{\rm gl, 2}\sim g^2(\mu=\mu_{\rm
gl})z_Q(\la,\mu_{\rm gl})
\langle\Pi_{1,2}\rangle_{\rm br}\gg\la^2,\,\,\,\, \langle \Pi_1 \rangle_{\rm
br}\sim\langle\Pi_2\rangle_{\rm br}\sim \la^2\Biggl
(\frac{\mph}{\la}\Biggr )^{\frac{N_c-N_F}{2N_c-N_F}}\,.
\label{(15.2.2)}
\eq

The pion masses in this regime look as follows, see
\eqref{(15.2.7)}\,:\, a) due to the spontaneous breaking of the
flavor symmetry, $U(N_F)\ra U(n_1)\times U(n_2)$, there always
will be $2n_1 n_2$ exactly massless Nambu-Goldstone particles
and in this case these are the hybrids $\Pi_{12}$ and
$\Pi_{21}$; \,b) other  $n_1^2+n_2^2$\,\, `normal' pions have
masses as in \eqref{(15.2.7)}.

There are
\bq
N^{\rm tot}_{\rm brok}=\sum_{n_1=1}^{n_1=[N_F/2]}N_{\rm
brok}(n_1)=\sum_{n_1=1}^{n_1=[N_F/2]}
(2N_c-N_F){\ov C}^{\, n_1}_{N_F}\,,\quad C^{\,
n_1}_{N_F}=\frac{N_F!}{n_1!\, n_2!} \label{(15.2.3)}
\eq
such vacua (the factor $2N_c-N_F$ originates from 
$Z_{2N_c-N_F}$\,, for even $N_F$ the last term with
$n_1=N_F/2$ enters \eqref{(15.2.3)} with the additional factor
$1/2$, i.e. ${\ov C}^{\, n_1}_{N_F}$ differ from the standard
$C^{\,n_1}_{N_F}$  only by ${\ov C}^{\,n_1={\rm
k}}_{N_F=2{\rm k}}=C^{\,n_1={\rm k}}_{N_F=2{\rm k}}/2$ ), so
that the total number of vacua is
\footnote {\,
By convention, we ignore the continuous multiplicity of vacua
due to the spontaneous flavor symmetry breaking. Another way,
one can separate slightly all quark masses,
so that all Nambu-Goldstone bosons will acquire small masses
$O(\delta m_Q)\ll {\ov m}_Q$.
}
\bq
N_{\rm tot}=\Bigl ( N_{\rm unbrok}=2N_c-N_F\Bigr )+N^{\rm
tot}_{\rm brok}\,,\label{(15.2.4)}
\eq
this agrees with \cite{CKM}.

 The change of the regime in these vacua
with the broken symmetry occurs at
$\mo\ll\mph\ll{\tilde\mu}_{\Phi}$\,, see
\eqref{(15.2.10)}, when all quarks are still higgsed
but there appears a large hierarchy between the values of quark
condensates at $\mph\gg\mo$\,, see \eqref{(15.2.5)}. Instead of
$\langle \Pi_1 \rangle\sim \langle \Pi_2 \rangle $, they look
now as: \\

{\bf a)}\,\, $\rm{br}1$\, ($\rm{br}1=\rm{breaking}-1$) - vacua
\bq
\hspace*{-4mm}\langle \Pi_1 \rangle_{\rm br1}\approx \Biggl
(\rho_1=\frac{N_c}{N_c-n_1}\Biggr ) m_Q\mph\gg \la^2,\,\,
\langle \Pi_2 \rangle_{\rm br1}\approx\la^2\Bigl
(\frac{\la}{m_Q\rho_1} \Bigr )^{\frac{N_c-n_2}{N_c-n_1}}\Bigl
(\frac{\la}{\mph} \Bigr )^{\frac{n_1}{N_c-n_1}}\ll \langle \Pi_1
\rangle_{\rm br1}.\,\,\,\,\label{(15.2.5)}
\eq

Unlike the mainly quantum $\langle\Pi\rangle_{\rm o}$ or mainly
classical $\langle\Pi\rangle_{\rm cl}$ vacua with unbroken
symmetry, these vacua are pseudo-classical\,: the largest value
of the condensate $\langle \Pi_1 \rangle_{\rm br1}\sim m_Q\mph$
is classical while the smaller value of $\langle
\Pi_2\rangle_{\rm br1}\sim\langle S\rangle_{\rm br1}/m_Q$ is of
quantum origin. There are $N_{\rm
br1}(n_1)=(N_c-n_1){\ov C}_{N_F}^{\, n_1}$ such vacua at given
values of $n_1$ and $n_2$.\\

{\bf b)}\,\, $\rm{br2}$ - vacua. These are obtained from
\eqref{(15.2.5)} by $n_1\leftrightarrow n_2$ and there are $N_{\rm
br2}(n_2)=(N_c-n_2){\ov C}_{N_F}^{\, n_2}$ such vacua. Of
course, the total number of vacua, $N_{\rm brok}=N_{\rm
br1}(n_1)+N_{\rm br2}(n_2)=(2N_c-N_F){\ov C}_{N_F}^{\, n_1}$
remains the same at $\mph\lessgtr\mo$.\\

We consider $\rm br1$ vacua (all results in $\rm br2$ vacua can
be obtained by $n_1\leftrightarrow n_2$). In the range
$\mo\ll\mph\ll {\tilde\mu}_{\Phi}$ (see below) where all quarks
are higgsed finally, the masses of higgsed gluons look now as
\bq
\mu^2_{\rm gl,1}\sim g^2(\mu=\mu_{\rm gl,1})z_Q(\la, \mu_{\rm
gl,1})\langle\Pi_1\rangle\gg\mu^2_{\rm gl,2 }\,.\label{(15.2.6)}
\eq
The superpotential in the low energy Lagrangian of pions looks
as in \eqref{(15.1.3)}, but the Kahler term of pions is different.
We write it in the form \,:\, $K\sim z_Q(\la,\mu_{\rm gl,1}){\rm
Tr}\sqrt{\Pi^{\dagger}_z\Pi_z}$\,. The $N_F\times N_F$ matrix 
$\Pi_z$ of pions looks
as follows. Its $n_2\times n_2$ part consists of fields
$z^{\prime}_Q(\mu_{\rm gl,1},\mu_{\rm gl,2})\Pi_{22}$, where
$z^{\prime}_Q\ll 1$ is the perturbative logarithmic
renormalization factor of ${\oq}_2,\, {\sq}^2$ quarks with
not higgsed colors which appears due to their additional RG
evolution in the range of scales $\mu_{\rm gl,2}<\mu<\mu_{\rm
gl,1}$, while at $\mu=\mu_{\rm gl,2}$ they are also higgsed. All
other  pion fields $\Pi_{11}, \Pi_{12}$ and $\Pi_{21}$ are
normal. As a result, the pion masses look as follows. $2n_1n_2$
hybrid pions $\Pi_{12}$ and $\Pi_{21}$ are massless, while the
masses of $n_1^2$\, $\Pi_{11}$ and $n_2^2$\, $\Pi_{22}$ are
\bq
\mu(\Pi_{11})\sim \frac{m_Q}{z_Q(\la,\mu_{\rm gl,1})}\,,\quad
\mu(\Pi_{22})\sim \frac{m_Q}{z_Q(\la,\mu_{\rm \,gl,1})z^{\prime}
_Q(\mu_{\rm gl,1},\mu_{\rm gl,2})}\,,\label{(15.2.7)}
\eq
where $z_Q$ and $z^{\rm \prime}_Q$ are both logarithmic.
Finally, the mass scale of gluonia from the not higgsed
$SU(N_c-N_F)$ group is $\sim \lym^{\rm (br1)}$\,, where
\bq
(\lym^{\rm (br1)})^3=\langle S\rangle_{\rm br1}=\frac{\langle
\Pi_1\rangle_{\rm br1}\langle \Pi_2\rangle_{\rm br1}}{\mph}\sim
m_Q\langle\Pi_2\rangle_{\rm br1} \sim\la^3\Biggl
(\frac{\la}{\mph}\Biggr)^{\frac{n_1} {N_c-n_1}}\Biggl
(\frac{m_Q}{\la}\Biggr)^{\frac{n_2-n_1}{N_c-n_1}}\,.\label{(15.2.8)}
\eq

At scales $\mu_{\rm gl,2}\ll\mu<\mu_{\rm gl,1}\sim
\langle \Pi_1 \rangle^{1/2}\sim (m_Q\mph)^{1/2}$ (ignoring
logarithmic factors) the light degrees of freedom include\,:\, a)
$SU(N_c-n_1)$ gluons,\,\,b) active quarks ${\oq}_2,\, {\sq}^2$ with
not higgsed colors and $n_2<(N_c-n_1)$ flavors,\,\, c) $n_1^2$ pions
$\Pi_{11}$,\,\, d) $2n_1 n_2$ hybrid pions $\Pi_{12}$ and $\Pi_{21}$
(in essence, these are the quarks ${\ov Q}_2,\, Q^2$ with
higgsed colors in this case). The scale factor ${\Lambda_1}$ of
the gauge coupling in this lower energy theory is
\bq
\Lambda^{{\rm b}^{\prime}_{\rm o}}_1\sim\la^{\bo}/\det
\Pi_{11}\,,\quad {\rm b}^{\prime}_{\rm o}=3(N_c-n_1)-n_2\,,\quad
\bo=3N_c-N_F\,.\label{(15.2.9)}
\eq
The scale of the perturbative pole mass of ${\oq}_2,\, {\sq}^2$
quarks is $m_Q^{\rm pole}\sim m_Q$\,, while the scale of
$\mu_{\rm gl,2}$ is $\mu_{\rm
gl,2}\sim\langle{\oq}_2{\sq}^2\rangle^{1/2}=
\langle\Pi_2\rangle^{1/2}$\,, with $\langle\Pi_2\rangle\ll
\langle\Pi_1\rangle$ given in \eqref{(15.2.5)}. Hence, the
hierarchy at $\mo\ll\mph\ll{\tilde\mu}_{\Phi}$ looks as\,
$m_Q\ll {\Lambda_1}\ll\mu_{\rm gl, 2}
\sim\langle\Pi_2\rangle^{1/2}$ and active ${\oq}_2,\, {\sq}^2$
quarks are also higgsed, while at $\mph\gg {\tilde\mu}_{\Phi}$
the hierarchy looks as  $\langle\Qt\rangle^{1/2}\ll
{\Lambda_1}\ll m_Q$ and the active quarks ${\oq}_2,\, {\sq}^2$
become too heavy, they are not higgsed but are in the $\rm HQ_2$
(heavy quark) phase. The phase changes~ at
\bq
\langle \Pi_2 \rangle^{1/2} \sim m_Q\sim
{\langle\Lambda_1\rangle}\sim\lym^{(\rm br1)} \,\,\ra\,\,
{\tilde\mu}_{\Phi}\sim \la \Biggl (\frac{\la}{m_Q}\Biggr
)^{\frac{ \bo-n_1}{n_1}}\gg \mo\,.\label{(15.2.10)}
\eq

Hence, we consider now this $Higgs_1-HQ_2$ phase realized at
$\mph>{\tilde\mu}_{\Phi}$. For this it is convenient to retain
all fields $\Phi$ although, in essence, they are too heavy and
dynamically irrelevant. After integrating out all heavy higgsed
gluons and ${\ov Q}_1, Q^1$ quarks, we write the Lagrangian at
$\mu^2=\mu^2_{\rm gl,1}\sim N_c g^2(\mu=\mu_{\rm
gl,1})z_Q(\la,\mu_{\rm gl,1})\langle\Pi_1\rangle$ in the form
\bq
K=\Bigl [\,\frac{z_{\Phi}}{f^2}{\rm
Tr}(\Phi^\dagger\Phi)+z_Q(\la,\mu^2_{\rm gl,1})\Bigl
(K_{\Pi}+K_{{\sq}_2}\Bigr )\,\Bigr ], \label{(15.2.11)}
\eq
\bbq
K_{{\sq}_2}={\rm Tr}\Bigl ({\sq}^{\dagger}_2 {\sq}^2
+({\sq}^2\ra
{\oq}_2 )\Bigr )\,, \quad K_{\Pi}= 2{\rm
Tr}\sqrt{\Pi^{\dagger}_{11}\Pi_{11}}+K_{\rm hybr},
\eeq
\bbq
K_{\rm hybr}={\rm Tr}\Biggl
(\Pi^{\dagger}_{12}\frac{1}{\sqrt{\Pi_{11}\Pi^{\dagger}_{11}}}\Pi_{12}+
\Pi_{21}\frac{1}{\sqrt{\Pi^{\dagger}_{11}\Pi_{11}}}\Pi^\dagger_{21}\Biggr
),
\eeq
\bbq
\cw=\Bigl [\frac{2\pi}{\alpha(\mu_{\rm gl,1})}{\textsf S}\Bigr
]+\frac{\mph}{2}\Biggl [{\rm Tr}\, (\Phi^2) -\frac{1}{\nd}\Bigl
({\rm Tr}\,\Phi\Bigr)^2\Biggr ]+{\rm Tr}\Bigl ({\oq}_2m^{\rm
tot}_{{\sq}_2}{\sq}^2\Bigr )+\cw_{\Pi},
\eeq
\bbq
\cw_{\Pi}= {\rm Tr}\Bigl (m_Q\Pi_{11}+m^{\rm
tot}_{{\sq}_2}\,\Pi_{21}\frac{1}{\Pi_{11}}\Pi_{12}\Bigr )-
{\rm Tr}\Bigl
(\Phi_{11}\Pi_{11}+\Phi_{12}\Pi_{21}+\Phi_{21}\Pi_{12} \Bigr ),
\quad m^{\rm tot}_{{\sq}_2}=(m_Q-\Phi_{22}).
\eeq

In \eqref{(15.2.11)}: ${\oq}_2,\, \sq^2$  are the active ${\ov Q}_2, 
Q^2$ guarks  with not higgsed colors ($\textsf S$ 
is th field strength squared of not higgsed gluons), $\Pi_{12},
\Pi_{21}$ are the hybrid pions (in essence, these are the ${\ov
Q}_2, Q^2$ guarks with higgsed colors), $z_Q(\la,\mu^2_{\rm
gl,1})\gg 1$ is the corresponding perturbative logarithmic
renormalization factor of massless quarks,   see \eqref{(15.1.2)}.
Evolving now down in the scale and integrating ${\oq}_2,\, \sq^2$
quarks as heavy ones at $\mu<m^{\rm pole}_{{\sq}_2}$ and then
not higgsed gluons at $\mu<\lym^{(\rm br1)}$ one obtains the
Lagrangian of pions and fions
\bq
K=\Bigl [\frac{z^{\rm \prime}_{\Phi}}{f^2}{\rm
Tr}(\Phi^\dagger\Phi)+z_Q(\la,\mu^2_{\rm gl,1})K_{\Pi}\Bigr ],\,
\label{(15.2.12)}
\eq
\bbq
W=(N_c-n_1){\textsf S}+\frac{\mph}{2}\Biggl [{\rm Tr} (\Phi^2)
-\frac{1}{\nd}\Bigl ({\rm Tr}\,\Phi\Bigr)^2\Biggr ]+W_{\Pi}\,,
\quad\quad S=\Biggl [\frac{\la^{\bo}\det m^{\rm
tot}_{{\sq}_2}}{\det \Pi_{11}}\Biggr ]^{\frac{1}{N_c-n_1}}\,,
\eeq

We start with determining the masses of hybrids $\Pi_{12},
\Pi_{21}$ and $\Phi_{12}, \Phi_{21}$. They are mixed and their
kinetic and mass terms look as
\bq
K_{\rm hybr}={\rm Tr}\Bigl
[\phi^{\dagger}_{12}\phi_{12}+\phi^{\dagger}_{21}\phi_{21}+
\pi^{\dagger}_{12}\pi_{12}+\pi^{\dagger}_{21}\pi_{21} \Bigr ], 
\label{(15.2.13)}
\eq
\bbq
W_{\rm hybr}={\rm Tr}\Bigl
(m_{\phi}\phi_{12}\phi_{21}+m_{\pi}\pi_{12}\pi_{21}-m_{\rm
\phi\pi}(\phi_{12}\pi_{21}+\phi_{21}\pi_{12})\Bigr )\,,
\eeq
\bbq
m_{\phi}=\frac{f^2\mph}{z^{\prime}_{\Phi}},\quad
m_{\pi}=\frac{m_Q-\langle\Phi_{2}\rangle}{z_Q}=\frac{\langle\Pi_1\rangle}
{\mph  z_Q}\sim\frac{m_Q}{z_Q}\ll m_{\phi}\,,\quad z_Q=z_Q(\la,\mu_{\rm
gl,1})\,,
\eeq
\bq
m_{\rm \phi\pi}=\Bigl (\frac{f^2\langle\Pi_1\rangle}{z_Q z^{\prime}_{\Phi}}
\Bigr)^{1/2}, \quad m_{\rm \phi\pi}^2=m_{\phi}m_{\pi}\,.  \label{(15.2.14)}
\eq

Hence, the scalar potential looks as
\bq
V_S=|m|^2\cdot |\Psi^{(-)}_{12}|^2+0\cdot |\Psi^{(+)}_{12}|^2
+(12\rightarrow 21),\quad |m|=(|m_{\phi}|+|m_{\pi}|)\,,
\label{(15.2.15)}
\eq
\bbq
\Psi^{(-)}_{12}=\Bigl (c\,\phi_{12}-s\,\pi_{12} \Bigr ),\quad
\Psi^{(+)}_{12}=\Bigl (c\,\pi_{12}+s\,\phi_{12} \Bigr ),\quad
c=\Bigl (\frac{|m_{\phi}|}{|m|}\Bigr )^{1/2},\quad s=\Bigl
(\frac{|m_{\pi}|}{|m|}\Bigr)^{1/2}\,.
\eeq
Therefore, the fields $\Psi^{(-)}_{12}$ and $\Psi^{(-)}_{21}$
are heavy, with the masses $|m|\approx |m_{\phi}|\gg\la$, while
the fields $\Psi^{(+)}_{12}$ and $\Psi^{(+)}_{21}$ are massless.
But the mixing is really parametrically small, so that the heavy
fields are mainly $\phi_{12}, \phi_{21}$ while the massless ones
are mainly $\pi_{12}, \pi_{21}$.
\footnote{\,
Everywhere below in the text we neglect mixing when it is small.}

And finally from \eqref{(15.2.12)}, the pole mass of pions
$\Pi_{11}$ is
\bq
\mu(\Pi_{11})\sim \frac{\langle\Pi_1\rangle}{z_Q(\la,\mu_{\rm
gl,1})\mph}\sim\frac{m_Q}{z_Q(\la,\mu_{\rm gl,1})}\,.
\label{(15.2.16)}
\eq

On the whole for this $Higgs_1-HQ_2$ phase the mass spectrum
looks as follows at $\mph\gg{\tilde\mu}_{\Phi}$\,. a) The
heaviest are $n_1(2N_c-n_1)$ massive gluons and the same number
of their scalar superpartners with the masses $\mu_{\rm gl,1}$,
see \eqref{(15.2.6)}, these masses originate from the higgsing of
${\ov Q}_1, Q^1$ quarks. b) There is a large number of
22-flavored hadrons made of weakly interacting and weakly
confined non-relativistic ${\oq}_2, {\sq}^2$ quarks with not higgsed
colors (the tension of the confining string originated from the
unbroken $SU(N_c-n_1)$ color group is $\sqrt\sigma\sim\lym^{(\rm
br1)}\ll m^{\rm pole}_{\sq,2}$, see \eqref{(15.2.8)}, the scale of
their masses is $m^{\rm pole}_{\sq,2}\sim m_Q/[z_Q(\la,\mu_{\rm
gl,1})z^{\prime}_Q(\mu_{\rm gl,1}, m^{\rm pole}_{{\sq}_2}]$,
where $z_Q\gg 1$ and $z^{\rm \prime}_Q\ll 1$ are the
corresponding massless perturbative logarithmic renormalization
factors. c) There are $n_1^2$ pions $\Pi_{11}$ with the masses
\eqref{(15.2.16)}, $\mu(\Pi_{11})\ll m^{\rm pole}_{\sq,2}$. d)
There is a large number of gluonia made of gluons with not higgsed
colors, the scale of their masses is $\sim\lym^{(\rm br1)}$, see
\eqref{(15.2.8)}. e) The hybrids $\Pi_{12}, \Pi_{21}$ are
massless.

All $N^2_F$ fions $\Phi_{ij}$ remain too heavy and dynamically
irrelevant, their pole masses are $\mu^{\rm
pole}_1(\Phi)\sim \mph\gg\mu_{\rm gl,1}$.

\section{Quark and gluino condensates and multiplicities of
vacua at $\mathbf{N_c<N_F<2N_c}$}

\hspace{3mm} To obtain the numerical values of the quark
condensates $\langle{\ov Q}_j Q^i\rangle=\delta
^i_j\langle ({\ov Q}Q)_i\rangle$ at $N_c<N_F<2N_c$ (but only for
this purpose), the simplest way is to use the known {\it exact
form} of the non-perturbative contribution to the superpotential
in the standard SQCD with the quark superpotential $m_Q{\rm
Tr}({\ov Q}Q)$ and without the fions $\Phi$. It seems clear that
at sufficiently large values of $\mph$ and fixed $\la$ among the vacua 
of the $\Phi$-theory there should be $N_c$ vacua of ordinary SQCD 
(i.e. without fields $\Phi$) in which, definitely, all fions $\Phi$ are too 
heavy and dynamically irrelevant.  Therefore, they all can be integrated 
out and {\it the exact superpotential can be written as}
\bq
{\mathcal W}=
-\nd\Bigl (\frac{  \det {\QQ} }{ \la^{\bo} }\Bigr )^{1/\nd}
+m_Q{\rm Tr}\,{\QQ} 
-\frac{1}{2\mph} \Biggl [ {\rm Tr}\,\Bigl ( \QQ^2 \Bigr ) - \frac{1}{N_c}\Bigl
({\rm Tr}\,\QQ \Bigr )^2
\Biggr ], \label{(16.0.1)}
\eq
where $m_Q=m_Q(\mu=\la),\, \mph=\mph(\mu=\la)$, see sections 14 and 15  above
and sections 3 and 7 in~ \cite{ch1}).

Indeed, at sufficiently large $\mph$ and fixed $\la$,  there are $N_c$ vacuum
solutions for \eqref{(16.0.1)} with the unbroken $SU(N_F)$ flavor
symmetry. In these, the last term in \eqref{(16.0.1)} gives a small
correction only and can be neglected and one obtains
\bq
\langle{\ov Q}_j Q^i\rangle_{SQCD}\approx\delta^i_j\frac{1}{m_Q}\Bigl
(\lym^{(\rm
SQCD)}\Bigr )^3=\delta^i_j\frac{1}{m_Q}\Bigl
(\la^{\bo}m_Q^{N_F}\Bigr)^{1/N_c}\,,\label{(16.0.2)}
\eq
as it should be.

Now, using the holomorphic dependence of the exact
superpotential on the chiral superfields $({\ov Q}_j Q^i)$ and
on chiral parameters $m_Q$ and $\mph$, the exact form
\eqref{(16.0.1)} can be used to find the values of the quark
condensates $\langle{\ov Q}_j Q^i\rangle$ in all other  vacua of
the $\Phi$ - theory and at all other  values of $\mph>\la$. It is
worth recalling only that, in general, as in the standard SQCD
\cite{ch1,ch3}: \eqref{(16.0.1)} {\it is not the superpotential of
the genuine low energy Lagrangian describing lightest particles,
it determines only the values of the vacuum condensates}
$\langle{\ov Q}_j Q^i\rangle$. The genuine low energy
Lagrangians in different vacua will be obtained below in
sections 18-23, both in the direct and dual theories. \\

\subsection{Vacua with the unbroken flavor symmetry}

One obtains from \eqref{(16.0.1)} that at $\la\ll\mph\ll \mo$ there
are two groups of such vacua with parametrically different
values of condensates, $\langle{\ov Q}_j
Q^i\rangle_L=\delta^i_j\langle{\ov Q}
Q\rangle_L$ and $\langle{\ov Q}_j
Q^i\rangle_S=\delta^i_j\langle{\ov Q} Q\rangle_S$.

{\bf a}) There are $(2N_c-N_F)$ L - vacua (L=large),  with
\bq
\langle{\ov Q}Q(\mu=\la)\rangle_L\sim \la^2\Biggl
(\frac{\la}{\mph}\Biggr )^{\frac{\nd}{2N_c-N_F}}\ll \la^2\,.
\label{(16.1.1)}
\eq
In these quantum L -vacua the second term in the superpotential
\eqref{(16.0.1)} gives numerically only  small correction.

{\bf b}) There are $(N_F-N_c)$ classical S - vacua (S=small)
with
\bq
\langle{\ov Q}Q(\mu=\la)\rangle_S\approx -\frac{N_c}{\nd}\,
m_Q\mph\,.\label{(16.1.2)}
\eq
In these S - vacua, the first non-perturbative term in the
superpotential \eqref{(16.0.1)} gives only small corrections with
$Z_{N_F-N_c}$ phases, but just these corrections determine the
multiplicity of these $(N_F-N_c)$ nearly degenerate vacua. On
the whole, there are
\bq
N_{\rm unbrok}=(2N_c-N_F)+(N_F-N_c)=N_c \label{(16.1.3)}
\eq
vacua with the unbroken flavor symmetry at $N_c<N_F<2N_c$.\\

One obtains from \eqref{(16.0.1)} that at $\mph\gg \mo$ the above
$(2N_c-N_F)$ L - vacua and $(N_F-N_c)$ S - vacua degenerate into
$N_c$ SQCD vacua \eqref{(16.0.2)}.

The value of $\mo$ is determined from the matching
\bbq
\Biggl [\langle\QQ\rangle_L\sim \la^2\Biggl
(\frac{\la}{\mo}\Biggr )^{\frac{\nd}{2N_c-N_F}}\Biggr ]\sim
\Biggl [\langle\QQ\rangle_S\sim m_Q\mo\Biggl ]\sim \Biggl
[\langle\QQ\rangle_{\rm SQCD}\sim \la^2\Bigl
(\frac{m_Q}{\la}\Bigr )^{\frac{\nd}{N_c}}\Biggl ]\quad\ra
\eeq
\bq
\ra \mo\sim \la\Bigl (\frac{\la}{m_Q}\Bigr
)^{\frac{2N_c-N_F}{N_c}}\gg \la\,.\label{(16.1.4)}
\eq

\subsection{Vacua with the spontaneously broken flavor symmetry}

In these, there are $n_1$ equal condensates $\langle{\ov
Q}_1Q^1(\mu=\la)\rangle\equiv\langle\Qo
\rangle$ and $n_2\geq n_1$ equal condensates $\langle{\ov Q}_2
Q^2(\mu=\la)\rangle\equiv\langle
\Qt\rangle\neq\langle\Qo\rangle$. The simplest way to find the
values of quark condensates in these vacua is to use
\eqref{(14.1.5)}. We rewrite it here for convenience
\bbq
\langle\Qo+\Qt-\frac{1}{N_c}{\rm Tr}\,\QQ\rangle_{\rm
br}=m_Q\mph\,,
\eeq
\bq
\langle S\rangle_{\rm br}=\Bigl
(\frac{\det \langle\QQ\rangle_{\rm br}=\langle\Qo\rangle^{\rm
{n}_1}_{\rm br}\langle\Qt\rangle^{\rm {n}_2}_{\rm
br}}{\la^{\bo}}\Bigr )^{1/\nd}=\frac{\langle\Qo\rangle_{\rm
br}\langle\Qt\rangle_{\rm br}}{\mph}\,.\label{(16.2.1)}
\eq
Besides, the multiplicity of vacua will be shown below at given
values of $n_1$ and $n_2\geq n_1$.\\

{\bf 16.2.1} The region $\la\ll\mph\ll\mo$.\\

{\bf a)} At $n_2\lessgtr N_c$, including $n_1=n_2=N_F/2$ for
even $N_F$ but excluding $n_2=N_c$\,, there are $(2N_c-N_F){\ov
C}^{\,n_1}_{N_F}$ Lt - vacua (Lt=L -type) with the parametric
behavior of condensates 
\bq
(1-\frac{n_1}{N_c})\langle\Qo\rangle_{\rm Lt}\approx
-(1-\frac{n_2}{N_c})\langle\Qt\rangle_{\rm Lt}\sim \la^2\Biggl
(\frac{\la}{\mph}\Biggr )^{\frac{\nd}{2N_c-N_F}},\label{(16.2.1)}
\eq
i.e. as in the L - vacua above but $\langle\Qo\rangle_{\rm
Lt}\neq\langle\Qt\rangle_{\rm Lt}$ here.

{\bf b)} At $n_2>N_c$ there are $(n_2-N_c)C^{n_1}_{N_F}$ $\rm
br2$ - vacua (br2=breaking-2) with, see \eqref{(16.2.1)},
\bq
\langle\Qt\rangle_{\rm br2}\sim m_Q\mph,\,
\langle\Qo\rangle_{\rm br2}\sim \la^2\Bigl
(\frac{\mph}{\la}\Bigr )^{\frac{n_2}{n_2-N_c}}\Bigl
(\frac{m_Q}{\la}\Bigr )^{\frac{N_c-n_1}{n_2-N_c}},\,
\frac{\langle\Qo\rangle_{\rm br2}}{\langle\Qt\rangle_{\rm
br2}}\sim \Bigl (\frac{\mph}{\mo}\Bigr
)^{\frac{N_c}{n_2-N_c}}\ll 1.\quad\,\,\,\label{(16.2.3)}
\eq

{\bf c)} At $n_1=\nd,\, n_2=N_c$ there are $(2N_c-N_F)
C^{n_1=\nd}_{N_F}$ 'special' vacua with, see \eqref{(16.2.1)},
\bq
\langle\Qo\rangle_{\rm
spec}=\frac{N_c}{2N_c-N_F}(m_Q\mph)\,,\quad
\langle\Qt\rangle_{\rm spec}\sim \la^2\Bigl
(\frac{\la}{\mph}\Bigr
)^{\frac{\nd}{2N_c-N_F}},\,\,\label{(16.2.4)}
\eq
\bbq
\frac{\langle\Qo\rangle_{\rm spec}}{\langle\Qt\rangle_{\rm
spec}}\sim\Bigl (\frac{\mph}{\mo}\Bigr
)^{\frac{N_c}{2N_c-N_F}}\ll 1\,.
\eeq

On the whole, there are (\,$\theta(z)$ is the step function\,)
\bq
N_{\rm brok}(n_1)=\Bigl
[(2N_c-N_F)+\theta(n_2-N_c)(n_2-N_c)\Bigr ]{\ov
C}^{\,n_1}_{N_F}=\label{(16.2.5)}
\eq
\bbq
=\Bigl [(N_c-\nd)+\theta(\nd-n_1)(\nd-n_1)\Bigr ]{\ov
C}^{\,n_1}_{N_F}\,,
\eeq
( ${\ov C}^{\,n_1}_{N_F}$ differ from the standard
$C^{\,n_1}_{N_F}$ only by ${\ov C}^{\,n_1={\rm k}}_{N_F=2{\rm
k}}=C^{\,n_1={\rm k}}_{N_F=2{\rm k}}/2$, see \eqref{(15.2.3)}\, )
vacua with the broken flavor symmetry $U(N_F)\ra U(n_1)\times
U(n_2)$, this agrees with \cite{CKM}. \\ 

{\bf 16.2.2} The region $\mph\gg\mo$.\\

{\bf a)} At all values of $n_2\lessgtr N_c$, including
$n_1=n_2=N_F/2$ at even $N_F$ and the `special' vacua with
$n_1=\nd,\, n_2=N_c$, there are $(N_c-n_1){\ov C}^{\,n_1}_{N_F}$
$\rm br1$ - vacua (br1=breaking-1) with, see \eqref{(16.2.1)},
\bq
\langle\Qo\rangle_{\rm br1}\sim m_Q\mph\,,\quad
\langle\Qt\rangle_{\rm br1}\sim \la^2\Bigl
(\frac{\la}{\mph}\Bigr )^{\frac{n_1}{N_c-n_1}}\Bigl
(\frac{\la}{m_Q}\Bigr
)^{\frac{N_c-n_2}{N_c-n_1}}\,,\label{(16.2.6)}
\eq
\bbq
\frac{\langle\Qt\rangle_{\rm br1}}{\langle\Qo\rangle_{\rm
br1}}\sim \Bigl (\frac{\mo}{\mph}\Bigr
)^{\frac{N_c}{N_c-n_1}}\ll 1\,.
\eeq

{\bf b)} At $n_2<N_c$, including $n_1=n_2=N_F/2$, there are also
$(N_c-n_2){\ov C}^{\,n_2}_{N_F}=(N_c-n_2){\ov
C}^{\,n_1}_{N_F}$\,\, $\rm br2$ - vacua with, see \eqref{(16.2.1)},
\bq
\langle\Qt\rangle_{\rm br2}\sim m_Q\mph\,,\quad
\langle\Qo\rangle_{\rm br2}\sim \la^2\Bigl
(\frac{\la}{\mph}\Bigr )^{\frac{n_2}{N_c-n_2}}\Bigl
(\frac{\la}{m_Q}\Bigr
)^{\frac{N_c-n_1}{N_c-n_2}}\,,\label{(16.2.7)}
\eq
\bbq
\frac{\langle\Qo\rangle_{\rm br2}}{\langle\Qt\rangle_{\rm
br2}}\sim \Bigl (\frac{\mo}{\mph}\Bigr
)^{\frac{N_c}{N_c-n_2}}\ll 1\,.
\eeq

On the whole, there are
\bq
N_{\rm brok}(n_1)=\Bigl [(N_c-n_1)+\theta
(N_c-n_2)(N_c-n_2)\Bigr ]{\ov C}^{\,n_1}_{N_F}= \label{(16.2.8)}
\eq
\bbq
=\Bigl [(N_c-\nd)+\theta (\nd-n_1)(\nd-n_1)\Bigr ]{\ov
C}^{\,n_1}_{N_F}
\eeq
vacua. As it should be, the number of vacua at $\mph\lessgtr
\mo$ is the same.\\

As one can see from the above, all quark condensates become
parametrically the same at $\mph\sim\mo$. Clearly, this region
$\mph\sim\mo$ is very special and most of the quark condensates
change their parametric behavior and hierarchies at
$\mph\lessgtr\mo$. For example, the br2 - vacua with
$n_2<N_c\,,\,\,\langle\Qt \rangle\sim
m_Q\mph\gg\langle\Qo\rangle$ at $\mph\gg\mo$ evolve into the L -
type vacua with $\langle
\Qt\rangle\sim\langle\Qo\rangle\sim \la^2
(\la/\mph)^{\nd/(2N_c-N_F)}$ at $\mph\ll\mo$, while the br2 -
vacua with $n_2>N_c\,,\,\,\langle\Qt\rangle\sim
m_Q\mph\gg\langle\Qo\rangle$ at $\mph\ll\mo$ evolve into the br1
- vacua with $\langle\Qo\rangle\sim m_Q\mph\gg\langle\Qt\rangle$
at $\mph\gg\mo$, etc. 
The exception is the special vacua with
$n_1=\nd,\, n_2=N_c$\,. In these, the parametric behavior
$\langle\Qo\rangle\sim m_Q\mph, \,\langle\Qt\rangle\sim
\la^2(\la/\mph)^{\nd/(2N_c-N_F)}$ remains the same but the
hierarchy is reversed at $\mph\lessgtr\mo\, :\,
\langle\Qo\rangle/\langle\Qt\rangle\sim
(\mph/\mo)^{N_c/(2N_c-N_F)}$.\\

The total number of all vacua at $N_c<N_F<2N_c$ is
\bq
N_{\rm tot}=\Bigl ( N_{\rm unbrok}=N_c \Bigr )+\Bigl ( N_{\rm
brok}^{\rm tot}=\sum_{n_1=1}^{[N_F/2]}N_{\rm brok}(n_1)
\Bigr )=\sum_{k=0}^{N_c}(N_c-k)C^{\,k}_{N_F}\,, \label{(16.2.9)}
\eq
this agrees with \cite{CKM}\,.
\footnote{\,
But we disagree with their `derivation' in their section 4.3. There is
no their ${\cal N}_2$ vacua with $\langle
M^i_i\rangle\langle{\ov q}^i q_i\rangle/\la=\langle
S\rangle=0,\,\, i=1,...N_F$ (no summation over $i$) in the
$SU(N_c)$ dual theory at $m_Q\neq 0$. In all $N_{\rm tot}$ vacua
in both direct and dual theories\,:\, $\langle\det
M/\la^{\bo}\rangle^{1/\nd}=\langle\det {\ov
Q}Q/\la^{\bo}\rangle^{1/\nd}=\langle S\rangle\neq 0$ at $m_Q\neq
0$ (see sections 18-23 below ). Really, the
superpotential (4.48) in \cite{CKM} contains all $N_{\rm
tot}={\cal N}_1+{\cal N}_2$ vacua.
}

Comparing this with the number of vacua
\eqref{(15.2.3)},\eqref{(15.2.4)} at $N_F<N_c$ it is seen that, for
both $N_{\rm unbrok}$ and $N_{\rm brok}^{\rm tot}$ separately,
the multiplicities of vacua at $N_F<N_c$ and $N_F>N_c$ are not
analytic continuations of each other.\\

The analog of \eqref{(16.0.1)} in the dual theory with
$\la= - \Lambda_q$ is obtained by the
replacement ${\ov Q} Q(\mu=\la)\ra M(\mu=\la)$, so that $\langle
M(\mu=\la)\rangle=\langle {\ov Q} Q(\mu=\la)\rangle$ in all
vacua and multiplicities of vacua are the same.

\section{Fions $\mathbf{\Phi}$ in the direct theory\,: one or
three generations}

\hspace{3mm} At $N_c<N_F<2N_c$ and in the interval of scales
$\mu_H<\mu<\la$ ( $\mu_H$ is the largest physical mass in the
quark-gluon sector), the quark and gluon fields are effectively
massless. Because the quark renormalization factor $z_Q(\la,\mu
\ll\la)= (\mu/\la)^{\gamma_Q >\, 0}\ll 1$ in the  Kahler term 
decreases in
this case in a {\it power fashion} with lowering energy due to
the perturbative RG evolution, it is seen from \eqref{(14.1.3)}
that the role of the 4-quark term $({\ov Q}Q)^2/\mph$ increases
with lowering energy. Hence, while it is irrelevant at the scale
$\mu\sim\la$ because $\mph\gg \la$, the question is whether it
becomes dynamically relevant in the range of energies
$\mu_H\ll\mu\ll \la$. For this, we estimate the scale $\mu_o$
where it becomes relevant in the massless theory (see section 7
in \cite{ch1} for the perturbative strong coupling regime with
$a(\mu\sim\la)\sim 1,\, a(\mu\ll\la)\sim (\la/\mu)^{\nu\,>\, 0}\gg
1$ at $N_c<N_F<3N_c/2$\,)
\bq
\frac{\mu_o}{\mph}\frac{1}{z^2_Q(\la,\mu_o)}=\frac{\mu_o}{\mph}\Bigl
(\frac{\la}{\mu_o}\Bigr )^{2\gamma_Q}\sim 1\quad\ra \quad
\frac{\mu_o}{\la}\sim \Bigl (\frac{\la}{\mph}\Bigr
)^{\frac{1}{(2\gamma_Q-1)}}\,\,, \label{(17.0.1)}
\eq
\bbq
\gamma^{\rm conf}_Q=\frac{\bo}{N_F}\,\ra\,\frac{\mu^{\rm
conf}_o}{\la}\sim \Bigl (\frac{\la}{\mph}\Bigr
)^{\frac{N_F}{3(2N_c-N_F)}}\,\,, \quad \gamma^{\rm
strong}_Q=\frac{2N_c-N_F}{\nd}\,\ra\,
\frac{\mu^{\rm strong}_o}{\la}\sim \Bigl (\frac{\la}{\mph}\Bigr
)^{\frac{\nd}{(5N_c-3N_F)}}\,\,.
\eeq

Hence, if $\mu_H\ll\mu_o$, then at scales $\mu<\mu_o$ the
four-quark terms in the superpotential \eqref{(14.1.3)} cannot be
neglected any more and we have to account for them. For this, we
have to reinstate the fion fields $\Phi$ and to use the
Lagrangian \eqref{(14.1.1)} in which the Kahler term at
$\mu_H<\mu\ll\la$ looks as
\bq
K=\Bigl [z_{\Phi}(\la,\mu){\rm Tr}\,(\Phi^\dagger
\Phi)+z_Q(\la,\mu){\rm Tr}\Bigl (Q^\dagger Q+(Q\ra {\ov Q})\Bigr
)\Bigr ],\,\, z_Q(\la,\mu)=\Bigl (\frac{\mu}{\la}\Bigr
)^{\gamma_Q}\ll 1. \label{(17.0.2)}
\eq

We recall that even at those scales $\mu$ that the running
perturbative mass of fions $\mu_{\Phi}(\mu)\equiv\mph/
z_{\Phi}(\la,\mu)\gg \mu$ and so they are too heavy and
dynamically irrelevant, the quarks and gluons remain effectively
massless and active. Therefore, due to the Yukawa interactions
of fions with quarks, the loops of still active light quarks
(and gluons interacting with quarks) still induce the running
renormalization factor $z_{\Phi}(\la,\mu)$ of fions at all those
scales until quarks are effectively massless, $\mu>\mu_H$. But,
in contrast with a very slow logarithmic RG evolution at
$N_F<N_c$ in section 15, the perturbative running mass of fions
decreases now at $N_c<N_F<2N_c$ and $\mu<\la$ monotonically 
in a power fashin  with diminishing scale (see below),
$\mph(\mu\ll \la)=\mph/
z_{\Phi}(\la,\mu)\sim\mph(\mu/\la)^{|\gamma_{\Phi}|>1}\ll \mph$.
Nevertheless, until $\mph(\mu)\gg \mu$, the fields $\Phi$ remain
heavy and do not influence the RG evolution. But, when
$\mu_H\ll\mu_o$ and $\mph(\mu)\sim\mph/z_{\Phi}(\la,\mu)$ is the
main contribution to the fion mass
\footnote{\,
the cases when the additional contributions to the masses of
fions from other perturbative or non-perturbative terms in the
superpotential are not small in comparison with
$\sim\mph/z_{\Phi}(\la,\mu)$ have to be considered 
separately  \label{(f35)}
}
,
the quickly decreasing mass $\mph(\mu)$ becomes $\mu^{\rm
pole}_2(\Phi)=\mph(\mu=\mu^{\rm pole}_2(\Phi))$ and
$\mph(\mu<\mu^{\rm pole}_2(\Phi))< \mu$, so that\,: 1) there is
a pole in the fion propagator at $p=\mu^{\rm pole}_2(\Phi)$
(ignoring here and below a nonzero fion width, in any case the
nonzero width can have only massive particle), this is a second
generation of fions (the first one is at $\mu^{\rm
pole}_1(\Phi)\gg\la$\,; 2) the fields $\Phi$
{\it become effectively massless at $\mu<\mu^{\rm pole}_2(\Phi)$
and begins to influence the perturbative RG evolution}. In other
words, the seemingly `heavy' fields $\Phi$ {\it return back},
they become effectively massless and dynamically {\it relevant}.
Here and below the terms `relevant' and `irrelevant' (at a given
scale $\mu$\,) will be used in the sense of whether the running
mass $\sim\mph/z_{\Phi}(\la,\mu\ll\la)$ of fions at a given
scale $\mu$ is $<\mu$, so that they are effectively massless and
participate actively in interactions at this scale\,, or they
remain too heavy with the running mass $>\mu$ whose interactions
at this scale give only small corrections.

It seems clear that, at $N_c < N_F < 2N_c$,  {\it the physical reason why 
the $4$-quark
terms in the superpotential \eqref{(14.1.3)} become relevant at
scales $\mu<\mu_o$ is that the fion field $\Phi$ which was too
heavy and so dynamically irrelevant at $\mu>\mu_o,\,
\mph(\mu>\mu_o)>\mu$\,, becomes effectively massless at
$\mu<\mu_o,\, \mph(\mu<\mu_o)<\mu$\,, and begins to participate
actively in the RG evolution, i.e. it becomes relevant}. In
other words, the four quark term in \eqref{(14.1.3)} 'remembers'
about fions and signals about the scale below which the fions
become effectively massless, $\mu_o=\mu^{\rm pole}_2(\Phi)$.
This allows us to find the value of $z_{\Phi}(\la,\mu_o)$,
see \eqref{(17.0.1)},
\bbq
\frac{\mph}{z_{\Phi}(\la,\mu_o)}\approx\mu_o\,,\quad
z_{\Phi}(\la,\mu_o<\mu\ll\la)=\Bigl (\frac{\mu}{\la}\Bigr )^
{\gamma_{\Phi}}\,,
\eeq
\bq
\gamma_{\Phi}=-2\gamma_Q\,<\,0\,,\quad
f^2=f^2(\mu=\la)= 1\,.\label{(17.0.3)}
\eq

The perturbative running mass
$\mph(\mu)=\mph/z_{\Phi}(\la,\mu\ll\la)\ll\mph$ of fions
continues to decrease strongly with diminishing $\mu$ at all
scales $\mu_H<\mu<\la$ until quarks remain effectively massless,
and becomes frozen only at scales below the quark physical mass,
when the heavy quarks decouple.

Hence, if $\mu_H\gg\mu_o$\,, there is no pole in the fion
propagator at momenta $p<\la$ because the running fion mass is
too large in this range of scales, $\mph(p>\mu_o)>p$. The fions
remain dynamically irrelevant in this case at all momenta
$p<\la$.

But when $\mu_H\ll\mu_o$, {\it there will be not only the second
generation of fions at $p=\mu^{\rm pole}_2(\Phi)=\mu_o$ but also
a third generation at $p\ll\mu_o$}. Indeed, after the heavy
quarks decouple at momenta $p<\mu_H\ll\mu_o$ and the
renormalization factor $z_{\Phi}(\la,\mu)$ of fions becomes
frozen in the region of scales where the fions already became
relevant, $z_{\Phi}(\la,\mu<\mu_H)\sim z_{\Phi}(\la,\mu\sim
\mu_H)$, the frozen value $\mph(\mu<\mu_H)$ of the running
perturbative fion mass is now $\mph(\mu\sim\mu_H)\ll p_H=\mu_H$.
Hence, {there is one more pole in the fion propagator} at
$p=\mu^{\rm pole}_3(\Phi)\sim \mph(\mu\sim\mu_H)\ll \mu_H$.

On the whole, in a few words for the direct theory (see the
footnote \ref{(f35)} for reservations).\\
{\bf a)} The fions remain dynamically irrelevant and there are
no poles in the fion propagator at momenta $p<\la$ if
$\mu_H\gg\mu_o$.\\
{\bf b)} If $\mu_H\ll\mu_o\ll\la$, there are two poles in the
fion propagator at momenta $p\ll\la$\,:\, $\mu^{\rm
pole}_2(\Phi)\sim \mu_o$ and $\mu^{\rm pole}_3(\Phi)\sim
\mph/z_{\Phi}(\la,\mu_H)\ll\mu^{\rm pole}_2(\Phi)$ (here and
everywhere below in similar cases, - up to corrections due to
nonzero decay widths of fions). In other words, the
fions appear in three generations in this case (we recall that
there is always the largest pole mass of fions $\mu^{\rm
pole}_1(\Phi)\gg\la$). Hence, the fions are
effectively massless and dynamically relevant in the range of
scales $\mu^{\rm pole}_3(\Phi)<\mu<\mu^{\rm pole}_2(\Phi)$.

Moreover, once the fions become effectively massless and
dynamically relevant with respect to internal interactions, they
begin to contribute simultaneously to the external anomalies (
the 't Hooft triangles in the external background fields).

The case $\mu_H\sim\mu_o$ requires additional information. The
reason is that at scales $\mu\lesssim\mu_H$, in addition to the
canonical kinetic term $\Phi^{\dagger}_R p^2\Phi_R$
(R=renormalized) of fions, there are also terms $\sim
\Phi^{\dagger}_R p^2(p^2/\mu_H^2)^k\Phi_R$ with higher powers of
momenta induced by loops of heavy quarks (and gluons). If
$\mu_H\ll\mu_o$, then the pole in the fion propagator at
$p=\mu^{\rm pole}_2(\Phi)=\mu_o$ is definitely there and,
because $\mph(\mu=\mu_H)\ll\mu_H$, these additional terms are
irrelevant in the region $p\sim\mph(\mu=\mu_H)\ll\mu_H$ and the
pole in the fion propagator at $p=\mu^{\rm
pole}_3(\Phi)=\mph(\mu=\mu_H)\ll\mu_H$ is also guaranteed. But
$\mph(\mu\sim\mu_H)\sim\mu_H$ if $\mu_H\sim\mu_o$, and these
additional terms become relevant. Hence, whether there is a pole
in the fion propagator in this case or not depends on all these
terms.\\

Now, if $\mu_H<\mu_o$ so that the fions become relevant at
$\mu<\mu_o$, the question is\,: what are the values of the quark
and fion anomalous dimensions, $\gamma_Q$ and $\gamma_\Phi$, in
the massless perturbative regime at $\mu_H<\mu<\mu_o$\,?

To answer this question, we use the approach used in \cite{ch1}
(see section 7 therein). For this, we introduce first the corresponding
massless Seiberg dual theory \cite{S2}. Our direct theory
includes at $\mu_H<\mu<\mu^{\rm conf}_o$ not only the original
effectively massless in this range of scales quark and gluon
fields, but also the active $N_F^2$ fion fields $\Phi^j_i$ as
they became now also effectively massless, so that the effective
superpotential becomes nonzero and includes the Yukawa term
${\rm Tr}\,({\ov Q}\Phi Q)$. Then, the massless dual theory with
the same 't Hooft triangles includes only the massless qual
quarks ${\ov q},\, q$ with $N_F$ flavors and the dual
$SU(\nd=N_F-N_c)$ gluons. Further, one equates two NSVZ
$\,{\widehat\beta}_{ext}$ - functions of the external baryon and
$SU(N_F)_{L,R}$ - flavor vector fields in the direct and dual
theories,
\bq
\frac{d}{d\,\ln
\mu}\,\frac{2\pi}{\alpha_{ext}}={\widehat\beta}_{ext}=
-\frac{2\pi}{\alpha^2_{ext}}\,\beta_{ext}= \sum_i T_i\,\bigl
(1+\gamma_i\bigr )\,,\label{(17.0.4)}
\eq
where the sum runs over all fields which are effectively
massless at scales $\mu_H<\mu<\mu_o$, the unity in the brackets
is due to one-loop contributions while the anomalous dimensions
$\gamma_i$ of fields represent all higher-loop effects, $T_i$
are the coefficients. It is worth noting that these general NSVZ
forms \eqref{(17.0.4)} of the external 'flavored'
$\widehat\beta$-functions are independent of the kind of
massless perturbative regime of the internal gauge theory, i.e.
whether it is conformal, or the strong coupling or the IR free
one.

The effectively massless particles in the direct theory here are
the original quarks $Q,\,{\ov Q}$ and gluons and, in addition,
the fions $\Phi^j_i$, while in the dual theory these are the
dual quarks $q,\, {\ov q}$ and dual gluons only.

It is clear that, in comparison with the standard SQCD without
the fion fields (see section 7 in \cite{ch1}), the addition of
the fion fields with zero baryon charge does not influence
${\widehat\beta}_{ext}$ for the baryon charge, so that in the
whole interval $\mu_H<\mu<\la$ it remains the same as in
\cite{ch1}
\bq
N_F N_c\,\Bigl ( B_Q=1 \Bigr )^2\,(1+\gamma_Q)=N_F \nd \,\Bigl (
B_q=\frac{N_c}{\nd}
\Bigr )^2\,(1+\gamma_q)\,. \label{(17.0.5)}
\eq

The form of \eqref{(17.0.4)} for the $SU(N_F)_L$ flavor charge at
scales $\mu_H<\mu<\mu_o$ where the fion fields became
effectively massless and relevant differs from those in
\cite{ch1}, now it looks as
\bq
N_c\,(1+\gamma_Q)+N_F\,(1+\gamma_\Phi)=\nd
\,(1+\gamma_q)\,.\label{(17.0.6)}
\eq
In \eqref{(17.0.5)},\eqref{(17.0.6)}: the left-hand sides are from the
direct theory while the right-hand sides are from the dual one,
$\gamma_Q$ and $\gamma_\Phi$ are the anomalous dimensions of the
quark $Q$ and fion $\Phi$\,, while $\gamma_q$ is the anomalous
dimension of the dual quark $q$.

The massless dual theory is in the conformal regime at
$3N_c/2<N_F<2N_c$\,, so that $\gamma^{\rm conf}_q=\rm{{\ov
b}_o}/N_F=(3\nd-N_F)/N_F$. Therefore, one obtains from
\eqref{(17.0.5)},\eqref{(17.0.6)} that $\gamma^{\rm
conf}_Q=\bo/N_F=(3N_c-N_F)/N_F$ and $\gamma^{\rm
conf}_\Phi=-2\gamma^{\rm conf}_Q$, i.e. while only the
quark-gluon sector of the direct theory behaves conformally at
scales $\mu^{\rm conf}_o<\mu< \la$ where the fion fields $\Phi$
remain heavy and irrelevant, the whole theory including the
fields $\Phi$ becomes conformal at scales $\mu_H<\mu< \mu^{\rm
conf}_o$ where fions become effectively massless and relevant.
\footnote{\,
This does not mean that nothing changes at all after the fion
field $\Phi$ begins to participate actively in the perturbative
RG evolution at $\mu_H<\mu<\mu^{\rm conf}_o$. In particular, the
frozen fixed point values of the gauge and Yukawa couplings
$a^*$ and $a_{f}^*$ will change.
}

In the region $N_c<N_F<3N_c/2$ the situation with
\eqref{(17.0.5)},\eqref{(17.0.6)} is somewhat different. The massless
direct theory is now in the strong gauge coupling regime
starting from $\mu<\la$,\, $a(\mu\ll\la)\sim
(\la/\mu)^{\nu\,>\,0}\gg 1$, see section 7 in \cite{ch1} and \cite{ch3}, 
while  the massless dual theory is in the IR free logarithmic regime.

Therefore, $\gamma_q$ is logarithmically small at $\mu\ll\la,\, 
\gamma_q\ra 0$, and one obtains in this case from \eqref{(17.0.5)} 
for the baryon charge the same value of $\gamma^{\rm strong}_
Q(\mu_H\ll\mu\ll\la)$ as in \cite{ch1}
\bq
\gamma^{\rm strong}_Q(\mu_H\ll\mu\ll\la)=\frac{2N_c-N_F}{\nd}\,,
\quad a(\mu_H\ll\mu\ll\la)\sim (\la/\mu)^{\nu}\gg 1\,, \label{(17.0.7)}
\eq
\bq
\quad \nu=\frac{N_F}{N_c}\gamma^{\rm strong}_Q-3=\frac{3N_c-2N_F}{\nd}>0\,. 
\label{(17.0.8)}
\eq
In other words, the value of the quark anomalous dimension $\gamma^
{\rm strong}_Q(\mu_H\ll\mu\ll\la)$ in the $\Phi$-theory is the same as in the 
standard SQCD, independently of whether the field $\Phi$ is relevant or not.

Additional details can be found in \cite{ch16}  (see also Appendix 1).
\vspace*{2mm}

\begin{center}
\bf \LARGE  Mass spectra at $\mathbf {3N_c/2 < N_F < 2 N_c}$ 
\end{center}

\section{Direct theory. Unbroken flavor symmetry}

\subsection{\quad  L - vacua}

\hspace*{4mm} The theory entered the conformal regime at the scale 
$\mu < \la$. In these $(2N_c-N_F)$ vacua with the unbroken flavor symmetry 
$U(N_F)$ the current quark mass at $\la\ll\mph\ll\mo$ looks as, see
\eqref{(16.1.1)},\eqref{(18.1.1)}
\bq
\langle m^{\rm tot}_Q\rangle_L\equiv\langle m^{\rm
tot}_Q(\mu=\la)\rangle_{L}=m_Q-\langle\Phi\rangle_L=
m_Q+\frac{\nd}{N_c}\frac{({\ov Q}Q)_L}{\mph}\,, \label{(18.1.1)}
\eq
\bbq
({\ov Q}Q)_L\sim\la^2\Bigl (\frac{\la}{\mph}\Bigr )^{\frac{\nd}{2N_c-N_F}}\gg
m_Q\mph,\quad \langle m^{\rm tot}_Q\rangle_L\sim \la\Bigl (\frac{\la}{\mph}
\Bigr)^{\frac{N_c}{2N_c-N_F}}\,,
\eeq
\bbq
\mql=\frac{\langle m^{\rm tot}_Q\rangle_{L}}{z_Q(\la,m^{\rm pole}_{Q,\,L})}\sim
\la \Bigl (\frac{\la}{\mph}\Bigr )^{\frac{N_F}{3(2N_c-N_F)}}\sim\lym^{(L)}\,,
\quad z_Q(\la,\mu\ll\la)\sim\Bigl (\frac{\mu}{\la}\Bigr )^{\bo/N_F}\ll 1\,.
\eeq
We compare $\mql$ with the gluon mass due to possible higgsing of quarks. 
This last looks as
\bq
\mgl^2\sim \Bigl (a_{*}\sim 1\Bigr ) z_Q(\la,\mgl)({\ov Q}Q)_L \quad \ra \quad
\mgl\sim
\mql\sim\lym^{(L)}=\langle S\rangle^{1/3}_L\,. \label{(18.1.2)}
\eq
Hence, qualitatively, the situation is the same as in the standard SQCD
\cite{ch3}. And one can use here the same reasonings, see the footnote 
 \ref{(f5)} or footnote 3 in 3 in \cite{ch3}. 
In the case considered, there are only $(2N_c-N_F)$ these isolated
 L-vacua with {\it unbroken flavor symmetry}. If quarks were higgsed in these 
L-vacua, then {\it the flavor symmetry will be necessary broken
spontaneously} \, due to the rank restriction because $N_F>N_c$. 
There will  appear genuine exactly massless Nambu-Goldstone fields 
$\Pi$ (pions), so that there will be a continuous family of non-isolated 
vacua.  And besides,  this will be in contradiction with   the Konishi 
anomaly which predicts equal values of $\langle \QQ\rangle_L$ 
for  all flavors in these  L-vacua. So, we conclude that quarks  are not 
higgsed but are in the HQ (heavy quark) phase and are confined. 

Therefore (see sections 3, 4 in \cite{ch3}), after integrating out all quarks
as \, heavy  ones at $\mu<\mql$ and then all $SU(N_c)$ gluons at
$\mu<\lym^{(L)}=\mql/(\rm several)$ via the Veneziano-Yankielowicz (VY)
procedure \cite{VY}, we obtain the Lagrangian of fions
\bq
K=z_{\Phi}(\la,\mql){\rm Tr}\,(\Phi^\dagger\Phi)\,, \quad
z_{\Phi}(\la,\mql)\sim\frac{1}
{z^2_Q(\la,\mql)}\sim\Bigl (\frac{\la}{\mql}\Bigr )^{2\bo/N_F}\gg 1\,,
\label{(18.1.3)}
\eq
\bbq
\cw=N_c S+\frac{\mph}{2}\Biggl [{\rm Tr}\,(\Phi^2)-\frac{1}{\nd}\Bigl ({\rm
Tr}\,\Phi\Bigr )^2\Biggr ]\,,
\quad S=\Bigl (\la^{\bo}\det m^{\rm tot}_{Q}\Bigr )^{1/N_c}\,,\quad m^{\rm
tot}_{Q}=(m_Q-\Phi)\,,
\eeq
and one has to choose the L - vacua in \eqref{(18.1.3)}.

There are two contributions to the mass of fions in \eqref{(18.1.3)}, the
perturbative one from the term $\sim \mph\Phi^2$ and the non-perturbative one
from $\sim S$, and both are parametrically the same, $\sim \lym^{(L)}\gg m_Q$.
Therefore,
\bq
\mu(\Phi)\sim\frac{\mph}{z_{\Phi}(\la,\mql)}\sim\mql\sim\lym^{(L)}\,.
\label{(18.1.4)}
\eq
Besides, see \eqref{(17.0.1)}, because
\bq
\mu^{\rm conf}_o\sim\la \Bigl (\frac{\la}{\mph}\Bigr
)^{\frac{N_F}{3(2N_c-N_F)}}\sim\mql\sim\lym^{(L)}\,, \label{(18.1.5)}
\eq
and fions are dynamically irrelevant at $\mu^{\rm conf}_o<\mu<\la$ and can
become relevant only at the scale $\mu<\mu^{\rm conf}_o$, it remains unclear in
these L - vacua whether there is a pole in the fion propagators at
$p\sim\mu^{\rm
conf}_o\sim\mql$. May be yes but maybe not.\\

On the whole for the mass spectrum in these L - vacua. The quarks ${\ov Q}, Q$
are confined and strongly coupled here, the coupling being $a_*\sim 1$.
Parametrically, there is only one scale $\sim \lym^{(L)}$ in the mass spectrum
at $\la\ll\mph\ll\mo$. And there is no parametrical guaranty that there is the
second generation of fions with the pole masses $\mu_2^{\rm
pole}(\Phi)\sim\lym^{(L)}$.

The condensate $\langle{\ov Q}Q\rangle_L$ and the quark pole mass $\mql$ become
frozen at their SQCD values at $\mph\gg\mo,\, \langle{\ov
Q}Q\rangle_{SQCD}\sim\la^2(m_Q/\la)^{\nd/N_c},\,\, m^{\rm pole}_{SQCD}\sim
\lym^{(SQCD)}\sim\la(m_Q/\la)^{N_F/3N_c}$ \cite{ch3}, while $\mph$ increases
and \, $\mu^{\rm conf}_o\ll m^{\rm pole}_{Q,SQCD}$ decreases, see
\eqref{(17.0.1)}.
Hence, the perturbative contribution $\sim\mph/z_{\Phi}(\la,\mql)\gg m^{\rm
pole}_{Q,SQCD}$ to the fion mass becomes dominant at $\mph\gg\mo$ and the fion
fields will be dynamically irrelevant at $\mu<\la$.\\

Finally, it is worth emphasizing for all what follows that, unlike the dual
theory, {\it in all vacua of the direct theory the mass spectra remain
parametrically the same at $\,\bd/N_F=O(1)$ or $\,\bd/N_F\ll 1$}.\\

\subsection{\quad  S - vacua}

In these $\nd$ vacua the quark mass at $\la\ll\mph\ll\mo$ looks as, see
\eqref{(16.1.2)},
\bbq
\frac{\langle m^{\rm tot}_Q(\mu=\la)\rangle_S}{\la}\sim \frac{\langle
S\rangle_S}{\la\langle{\ov Q}Q\rangle_S}\sim \Bigl (\frac{\langle{\ov
Q}Q\rangle_S}{\la^2}\Bigr )^{N_c/\nd}\sim \Bigl (\frac{m_Q\mph}{\la^2}\Bigr
)^{N_c/\nd}\,,
\eeq
This has to be compared with the gluon mass due to possible higgsing of quarks
\bq
\mgs^2\sim z_Q(\la,\mgs)\langle\QQ_ S\rangle\ra\mgs\sim 
m^{\rm pole}_{Q,S} \sim\lym^{(S)},\,\,  z_Q(\la,\mgs)\sim
\Bigl (\frac{\mgs}{\la}\Bigr )_{.}^{\frac{\bo}{N_F}} \,\,\,\label{(18.2.1)}
\eq

For the same reasons as in previous section, it is clear that quarks will not
be\, higgsed in these vacua at $N_F>N_c$ (as otherwise the flavor symmetry will
be ], broken spontaneously). Hence, as in \cite{ch3}, we conclude here also that
the
pole mass of quarks is the largest physical mass, i.e. $\mu_H=
m^{\rm pole}_{Q,S} =(\rm several)\mgs$.

But, in contrast with the L - vacua, the fion fields {\it become dynamically
relevant in these S - vacua} at scales $\mu<\mu^{\rm conf}_o$, see
\eqref{(18.2.1)}, if
\bq
\mu^{\rm conf}_o\sim\la\Bigl (\frac{\la}{\mph}\Bigr
)^{\frac{N_f}{3(2N_c-N_F)}}\gg  m^{\rm pole}_{Q,S}
\quad\ra \quad  {at}\quad
\la\ll\mph\ll\mo\,.\label{(18.2.2)}
\eq

Therefore, there is the second generation of $N_F^2$ fions with the pole masses
\bq
\mu_2^{\rm pole}(\Phi)\sim \mu_o^{\rm conf}\gg 
m^{\rm pole}_{Q,S}\sim\lym^{(S)}\,.\label{(18.2.3)}
\eq

Nevertheless, see section 17, the theory remains in the conformal regime and 
quark and fion anomalous dimensions remain the same in the whole range of
$m^{\rm pole}_{Q,S}<\mu<\la$ of scales, but fions  become effectively massless 
at $\mu<\mu^{\rm conf}_o$ and begin to contribute to the 't Hooft triangles.

The RG evolution of the quark and fion fields becomes frozen at scales
$\mu< m^{\rm pole}_{Q,S}$ because the heavy   quarks decouple. Proceeding as 
before, i.e. integrating out first all quarks as heavy ones at $\mu < m^{\rm
pole}
_{Q,S}=(\rm several)\lym^{(S)}$ and then all $SU(N_c)$ gluons at
$\mu<\lym^{(S)}$,
one obtains the Lagrangian of fions as in \eqref{(18.1.3)}, with a replacement
$z_Q(\la,
\mql)\ra\, z_Q(\la, m^{\rm pole}_{Q,S})$ (and the S-vacua have to be chosen
therein).

Because fions became relevant at $m^{\rm pole}_{Q,S}\ll\mu\ll\mu^{\rm conf}_o$,
one
could expect that their running mass will be much smaller than $m^{\rm
pole}_{Q,S}$.
This is right, but only for $\mu^{\rm pert}_{\Phi}\sim \mph/z_Q(\la, m^{\rm
pole}_{Q,S})\ll
m^{\rm pole}_{Q,S}$. But there\, is \, also additional {\it non-perturbative}
contribution
to the fion mass originating from \, the region of scales $\mu\sim m^{\rm
pole}_{Q,S}$
and it is dominant in these S - vacua,
\bq
\mu(\Phi)\sim\frac{1}{z_{\Phi}(\la, m^{\rm pole}_{Q,S})}\,\frac{\langle
S\rangle_S}
{\langle  m^{\rm tot}_Q\rangle^2_S} \sim m^{\rm pole}_{Q,S}\,,
\quad z_{\Phi}(\la, m^{\rm pole}_{Q,S})\sim \Bigl (\frac{\la}{m^{\rm
pole}_{Q,S}}
\Bigr)^{2\bo/N_F}\,.\label{(18.2.4)}
\eq
Therefore, despite the fact that the fions are definitely dynamically relevant
in the range of scales $m^{\rm pole}_{Q,S}\ll\mu\ll\mu^{\rm conf}_o\ll\la$ at 
$\la\ll\mph\ll\mo$, whether there is 
the third generation of fions, i.e.  whether \, there is a pole in the fion 
propagator at $p=\mu_3^{\rm  pole}(\Phi)\sim m^{\rm pole}_{Q,S}
\sim\lym^{(S)}$ remains unclear. \\

On the whole for the mass spectra in these S - vacua. The largest are 
masses\, of the second generation fions, $\mu_2^{\rm pole}(\Phi)\sim\la\Bigl
(\la/\mph\Bigr )^{N_F/3(2N_c-N_F)}\gg m^{\rm pole}_{Q,S}$. The scale of all
other  masses is  $\sim m^{\rm pole}_{Q,S}\sim\lym^{(S)}$. There is no parametrical
guaranty that there is the third generation of fions with the pole masses 
$\mu_3^{\rm pole}(\Phi)\sim\lym^{(S)}$. May be yes, but maybe not.\\

The vacuum condensates $\langle{\ov Q}Q\rangle_{S}$ and $m^{\rm pole}_{Q,S}$
evolve into
their \, independent of $\mph$ SQCD-values at $\mph\gg\mo$,
\bq
\langle{\ov Q}Q\rangle_{SQCD}\sim \la^2\Bigl (\frac{m_Q}{\la}\Bigr
)^{\nd/N_c}\,,\quad m^{\rm pole}_{Q,SQCD}\sim\la\Bigl (\frac{m_Q}{\la}\Bigr
)^{N_F/3N_c}\,,\label{(18.2.5)}
\eq
and the perturbative contribution $\sim \mph/z_Q(\la,m^{\rm pole}_{Q,SQCD})$ to
the fion mass becomes dominant. Hence, because $m^{\rm pole}_{Q, SQCD}\gg
\mu^{\rm conf}_o$,  the fions fields become dynamically irrelevant 
at all scales $\mu<\la$ when $\mph\gg\mo$.\\

\section{Dual theory. Unbroken flavor symmetry}

\subsection{\quad  L - vacua,\,\,\, $\bd/N_F\ll 1$}

Let us recall, see e.g. section 4 in \cite{ch3},
that the Lagrangian of the dual theory at $\mu=\la$ and
$0<\bd/N_F\ll 1,\,\,\bd=3\nd-N_F$, looks as
\bbq
K= {\rm Tr}\Bigl ( q^\dagger q +(q\ra\ov q) \Bigr )+{\rm
Tr}\,\frac{M^{\dagger}M}{Z_q^2\la^2}\,,\,\,
\cw=\, -\,\frac{2\pi}{\ov \alpha(\mu=\la)}\, {\ov
s}+\cw_M+\cw_q,\,
Z_q\sim\exp \Bigl\{-\frac{\nd}{7\bd}\Bigr\}\ll 1\,,
\eeq
\bq
\cw_M=m_Q{\rm Tr}\,M -\frac{1}{2\mph}\Biggl [{\rm Tr}\, (M^2)-
\frac{1}{N_c}({\rm Tr}\, M)^2 \Biggr ]\,,\quad
\cw_q= -\,\frac{1}{Z_q\la}\,\rm {Tr} \Bigl ({\ov q}\,M\, q \Bigr
)\,. \label{(19.1.1)}
\eq

Because $\la^2/\mph\ll\la$, the mions are effectively massless
and dynamically relevant at $\mu\sim\la$ (and so in some range
of scales below $\la$). By definition, $\mu\sim\la$ is such a
scale that the dual theory already entered sufficiently deep the
conformal regime, i.e. the dual gauge coupling ${\ov
a}(\mu=\la)=\nd{\ov\alpha}(\mu=\la)/2\pi$ is sufficiently close
to its small frozen value, $\ov\delta=[{\ov a}_*-{\ov
a}(\mu\sim\la)]/{\ov a}_*\ll 1$, and $\ov\delta$ is neglected
everywhere below in comparison with 1 for simplicity (and the
same for the Yukawa coupling ${\ov a}_f=\nd{\ov\alpha}_f/2\pi$),
see \cite{ch3} and Appendix therein). The fixed point value
of the dual gauge coupling is ${\ov a}_*\approx 7\bd/3\nd\ll 1$
\cite{KSV}.

We recall also that the mion condensates are matched to the
condensates of direct quarks {\it in all vacua}, $\langle
M^i_j(\mu=\la)\rangle=\langle{\ov Q}_j Q^i(\mu=\la)\rangle$\,.
Hence, in these L - vacua
\bq
\langle M\rangle_L\sim\la^2\Bigl (\frac{\la}{\mph}\Bigr
)^{\frac{\nd}{2N_c-N_F}}\,,\quad\langle
N\rangle_L\equiv\langle{\ov q}q(\mu=\la)\rangle=
\frac{Z_q\la\langle S\rangle_L}{\langle M\rangle_L}\sim Z_q\la^2
\Bigl (\frac{\la}{\mph}\Bigr
)^{\frac{N_c}{2N_c-N_F}}\,,\label{(19.1.2)}
\eq
\bbq
Z_q\sim\exp \Bigl\{-\frac{1}{3{\ov a}_{*}}\Bigr\}\sim \exp
\Bigl\{-\frac{\nd}{7\bd}\Bigr\}\ll 1,
\eeq
and here and everywhere below, as in \cite{ch3}, a parametric
dependence on the small parameter $\bd/N_F\ll 1$ is traced with
an exponential accuracy only (i.e. powers of $\bd/N_F$ are not
traced, only powers of $Z_q$).

The current mass $\mu_{q,L}\equiv \mu_{q,L}(\mu=\la)$ of dual
quarks ${\ov q},\,q$ and their pole mass in these $(2N_c-N_F)$ L
- vacua are, see \eqref{(19.1.1)},
\bbq
\frac{\mu_{q,L}}{\la}=\frac{\langle
M\rangle_L}{Z_q\la^2}\sim\frac{1}{Z_q}\Bigl
(\frac{\la}{\mph}\Bigr )^{\frac{\nd}{2N_c-N_F}}\,,\quad\mu^{\rm
pole}_{q,L}=\frac{\mu_{q,L}}{z_q(\la,\mu^{\rm
pole}_{q,L})}\,,\quad z_q(\la,\mu^{\rm pole}_{q,L})\sim\Bigl
(\frac{\mu^{\rm pole}_{q,L}}{\la}\Bigr )^{\bd/N_F}\,,
\eeq
\bq
\mu^{\rm pole}_{q,L}\sim\la\Bigl (\frac{\mu_{q,L}}{\la}\Bigr
)^{N_F/3\nd}\sim\frac{\la}{Z_q}\Bigl (\frac{\la}{\mph}\Bigr
)^{\frac{N_F}{3(2N_c-N_F)}}\sim\frac{1}{Z_q}\lym^{(L)}\gg\lym^{(L)}
\,, \quad \la\ll\mph\ll\mo\,,\label{(19.1.3)}
\eq
\bbq
\frac{\mu_{q,L}}{\la}\sim\frac{1}{Z_q}\Bigl
(\frac{m_Q}{\la}\Bigr )^{\frac{\nd}{N_c}}\,,\quad \mu^{\rm
pole}_{q,L}\sim\frac{\la}{Z_q}\Bigl (\frac{m_Q}{\la}
\Bigr )^{\frac{N_F}{3N_c}}\sim\frac{1}{Z_q}\lym^{(SQCD)}\gg
\lym^{(SQCD)}\,,\quad \mph\gg\mo\,,
\eeq
while the gluon mass due to possible higgsing of dual quarks
looks at $\la\ll\mph\ll\mo$ as
\bq
{\ov\mu}_{\rm gl, L}\sim\Bigl [{\ov a}_*\langle N\rangle_{L}\,
z_q(\la,{\ov\mu}_{\rm gl}\Bigr]^{1/2}\sim Z_q^{1/2}\la
\Bigl (\frac{\la}{\mph}\Bigr )^{N_F/3(2N_c-N_F)}\sim
Z_q^{3/2}\mu^{\rm pole}_{q,L}\ll \mu^{\rm pole}_{q,L}\,.
\label{(19.1.4)}
\eq
Therefore, the dual quarks are definitely in the HQ phase in
these L - vacua at $\bd/\nd\ll 1$.

With decreasing scale the perturbative running mass
$\mu_{M}(\mu)$ of mions
\bq
\mu_{M}(\mu)\sim\frac{Z_q^2\la^2}{\mph
z_M(\mu)}=\frac{Z_q^2\la^2}{\mph}\Bigl (\frac{\mu}{\la}\Bigr
)^{2\bd/N_F} \label{(19.1.5)}
\eq
decreases but more slowly than the scale $\mu$ itself because
$\gamma_M= - (2{\rm\bd}/N_F),\,\,|\gamma_M| <  1$ at
$3/2<N_F/N_c<2$\,, and $\mu_{M}(\mu)$ becomes frozen at
$\mu<\mu^{\rm pole}_{q,L}$,\, $\mu_M(\mu<\mu^{\rm
pole}_{q,L})=\mu_M(\mu=\mu^{\rm pole}_{q,L})$.

After integrating out all dual quarks as heavy ones at
$\mu<\mu^{\rm pole}_{q,L}$ and then all $SU(\nd)$ gluons at
$\mu<\lym^{(L)}$ via the Veneziano-Yankielowicz (VY) procedure
\cite{VY}, the Lagrangian of mions looks as
\bbq
K=\frac{z^{(L)}_M(\la,\mu^{\rm pole}_{q,L})}{Z_q^2\la^2}{\rm
Tr}\,(M^\dagger M)\,,\quad z^{(L)}_M(\la,\mu^{\rm
pole}_{q,L})\sim\Bigl (\frac{\la}{\mu^{\rm pole}_{q,L}}\Bigr
)^{2\bd/N_F}\,,\quad
S=\Biggl (\,\frac{\det M}{\la^{\bo}}\,\Biggr )^{1/\nd},
\eeq
\bq
W= -\nd S+m_Q{\rm Tr}\,M -\frac{1}{2\mph}\Biggl [{\rm Tr}\,
(M^2)- \frac{1}{N_c}({\rm Tr}\, M)^2 \Biggr ]\,. \label{(19.1.6)}
\eq
There are two contributions to the mass of mions in
\eqref{(19.1.6)}, the perturbative one from the term $\sim
M^2/\mph$ and non-perturbative one from $\sim S$.
Both are parametrically the same and the total contribution
looks as
\bq
\mu^{\rm pole}(M)\sim \frac{Z_q^2\la^2}{z_M(\la,\mu^{\rm
pole}_{q,L})\mph}\sim Z_q^2\lym^{(L)}\ll \lym^{(L)}\ll \mu^{\rm
pole}_{q,L}\,, \label{(19.1.7)}
\eq
and this parametrical hierarchy guarantees that the mass
$\mu^{\rm pole}(M)$ in \eqref{(19.1.7)} is indeed the {\it pole
mass} of mions.\\

On the whole, the mass spectrum in these dual L - vacua looks as
follows at $\la\ll\mph\ll\mo$. a) There is a large number of
heaviest flavored hadrons made of weakly interacting and weakly
confined (the tension of the confining string originating from
the unbroken $SU(\nd)$ SYM is
$\sqrt\sigma\sim\lym^{(L)}\ll\mu^{\rm pole}_{q,L}$)
non-relativistic quarks ${\ov q}, q$ with the pole masses
$\mu^{\rm pole}_{q,L}/\lym^{(L)}\sim \exp (\nd/7\bd)\gg 1$. The
mass spectrum of low-lying flavored mesons is Coulomb-like with
parametrically small mass differences
$\Delta\mu_H/\mu_H=O(\bd^2/\nd^2)\ll 1$. b) A large number of
gluonia made of $SU(\nd)$ gluons with the mass scale
$\sim\lym^{(L)}\sim\la (\la/\mph)^{N_F/3(2N_c-N_F)}$. c) $N_F^2$
lightest mions with parametrically smaller masses $\mu^{\rm
pole}(M)/\lym^{(L)}\sim \exp (-2\nd/7\bd)\ll 1$.

At $\mph\gg\mo$ these L - vacua evolve into the vacua of the
dual SQCD theory (dSQCD), see section 4 in \cite{ch3}.\\

\subsection{\quad  S - vacua,\,\,\, $\bd/N_F\ll 1$}

The current mass $\mu_{q,S}\equiv \mu_{q,S}(\mu=\la)$ of dual
quarks ${\ov q},\,q$ at the scale $\mu=\la$ in these $(N_F-N_c)$
dual S-vacua is, see \eqref{(16.1.2)},
\bq
\mu_{q,S}=\frac{\langle
M\rangle_S=\langle\QQ\rangle_S}{Z_q\la}\sim\frac{m_Q\mph}{Z_q\la}\,,\quad
Z_q\sim \exp \Bigl\{-\frac{\nd}{7\bd}\Bigr\}\ll
1\,.\label{(19.2.1)}
\eq

In comparison with the L - vacua in section 19.1, a qualitatively
new element here is that $\mu^{\rm pole}(M)$ is the largest
mass, $\mu^{\rm pole}(M)\gg\mu^{\rm pole}_{q,S}$, in the wide
region $\la\ll\mph\ll Z_q^{\,3/2}\mo$. In this region: a) the mions 
are effectively massless
and dynamically relevant at scales $\mu^{\rm
pole}(M)\ll\mu\ll\la$,\,\, b) there is a pole in the mion
propagator at the momentum $p=\mu^{\rm pole}(M)$,
\bbq
\mu^{\rm pole}(M)=\frac{Z_q^2\la^2}{z_M(\la,\mu^{\rm
pole}(M))\mph}\,,\quad z_M(\la,\mu^{\rm pole}(M))\sim\Bigl
(\frac{
\la}{\mu^{\rm pole}(M)}\Bigr )^{2\bd/N_F}\,,
\eeq
\bq
\mu^{\rm pole}(M)\sim Z_q^2\la\Bigl (\frac{\la}{\mph}\Bigr
)^{\frac{N_F}{3(2N_c-N_F)}}\,,\quad Z_q^2\sim
\exp\{-\frac{2\nd}{7\bd}\}\ll 1\,. \label{(19.2.2)}
\eq

The mions then become too heavy and dynamically irrelevant at
$\mu\ll\mu^{\rm pole}(M)$. Due to this, they decouple from the
RG evolution of dual quarks and gluons and from the 't Hooft
triangles, and (at $\mph$ not too close to $\mo$ to have enough
"time" to evolve, see \eqref{(19.2.7)}\,) the remained dual theory
of $N_F$ quarks ${\ov q}, q$ and $SU(\nd)$ gluons evolves into a
new conformal regime with a new smaller value of the frozen
gauge coupling, ${\ov a}^{\,\prime}_*\approx\bd/3\nd$.
It is worth noting that, in spite of that mions
are dynamically irrelevant at $\mu<\mu^{\rm pole}(M)$, their
renormalization factor $z_M(\mu<\mu^{\rm pole}(M))$ still runs
in the range of scales $\mu^{\rm pole}_{q,S}<\mu<\mu^{\rm
pole}(M)$ being induced by loops of still effectively massless
dual quarks and gluons.

The next physical scale is the perturbative pole mass of dual
quarks
\bbq
\mu^{\rm pole}_{q,S}=\frac{\langle
M\rangle_S}{Z_q\la}\,\frac{1}{z_q(\la,\mu^{\rm
pole}_{q,S})}\,,\quad
z_q(\la,\mu^{\rm pole}_{q,S})=\Bigl (\frac{\mu^{\rm
pole}_{q,S}}{\la}\Bigr )^{\bd/N_F}{\ov\rho}_S\,,
\eeq
\bbq
{\ov\rho}_S=\Bigl (\frac{{\ov a}_*}{{\ov a}^{\,\prime}_*}\Bigr
)^{\frac{\nd}{N_F}}\exp\Bigl\{\frac{\nd}{N_F}
\Bigl (\frac{1}{{\ov a}_*}-\frac{1}{{\ov a}^{\,\prime}_*}\Bigr
)\Bigr\}\sim \frac{{\ov Z}_q}{Z_q}\ll 1\,,\quad
{\ov Z}_q\sim \exp \Bigl\{-\frac{\nd}{\bd}\Bigr\}\sim \Bigl
(Z_q\Bigr )^{7}\ll Z_q\,,
\eeq
\bq
\mu^{\rm pole}_{q,S}\sim \frac{1}{{\ov
Z}_q}\lym^{(S)}\gg\lym^{(S)}\,,\quad \lym^{(S)}=\la\Bigl
(\frac{m_Q\mph}{\la^2}
\Bigr )^{N_F/3\nd}\,. \label{(19.2.3)}
\eq

This has to be compared with the gluon mass due to possible
higgsing of ${\ov q}, q$
\bq
{\ov\mu}^{\,2}_{\rm gl, S}\sim z_q(\la,{\ov\mu}_{\rm gl,
S})\langle{\ov q}q\rangle_S\,\,\ra\,\,{\ov\mu}_{\rm gl,
S}\sim{\ov Z}_q^{\,1/2}\lym^{(S)}\ll\lym^{(S)}\ll\mu^{\rm
pole}_{q,S}. \label{(19.2.4)}
\eq
The parametric hierarchy in \eqref{(19.2.4)} guarantees that the
dual quarks are in the HQ phase in these S - vacua.

Hence, after integrating out all quarks at $\mu<\mu^{\rm
pole}_{q,S}$ and, finally, $SU(\nd)$ gluons at $\mu<\lym^{(S)}$,
the Lagrangian looks as in \eqref{(19.1.6)} but with a replacement
$z^{(L)}_M (\la,\mu^{\rm pole}_{q,L})\ra z^{(S)}_M (\la,\mu^{\rm
pole}_{q,S})$,
\bq
z^{(S)}_M(\la,\mu^{\rm pole}_{q,S})=\frac{1}{z^2_q(\la,\mu^{\rm
pole}_{q,S})}\sim\frac{1}{z^2_q(\la,\mu^{\rm
pole}_{q,S})}\sim\frac{Z_q^{\,2}}{{\ov Z}_q^{\,2}}\Bigl
(\frac{\la}{\mu^{\rm pole}_{q,S}}\Bigr )^{2\bd/N_F}\,.
\label{(19.2.5)}
\eq
The contribution of the term $\sim M^2/\mph$ in the
superpotential \eqref{(19.1.6)} to the frozen low energy value
$\mu(M)$ of the running mion mass is dominant at $\mph/\mo\ll 1$
and is
\bq
\hspace*{-5mm}\mu(M)=\frac{Z_q^2\la^2}{z^{(S)}_M(\la,\mu^{\rm
pole}_{q,S})\mph}\sim
\frac{{\ov Z}_q^{\,2}\la^2}{\mph}\Bigl (\frac{\mu^{\rm
pole}_{q,S}}{\la}\Bigr )^{\frac{2\bd}{N_F}}\ll\mu^{\rm
pole}(M)\,. \label{(19.2.6)}
\eq
The requirement of self-consistency looks in this case as
\bq
\frac{\mu(M)}{\mu^{\rm pole}_{q,S}}\sim {\ov Z}_q^{\,3}\Bigl
(\frac{\mo}{\mph}\Bigr )^{N_c/\nd}\gg 1\quad\ra\quad
\frac{\mph}{\mo}\ll {\ov Z}^{\, 3/2}_q\sim\exp
\Bigl\{-\frac{3\nd}{2\bd}\Bigr\}\ll Z_q^{3/2}\,, \label{(19.2.7)}
\eq
the meaning of \eqref{(19.2.7)} is that only at this condition the
range of scales between $\mu^{\rm pole}(M)$ in \eqref{(19.2.2)} and
$\mu^{\rm pole}_{q,S}\ll\mu^{\rm pole}(M)$ in \eqref{(19.2.3)} is
sufficiently large that theory has enough "time" to evolve from
${\ov a}_*=7\bd/3\nd$ to ${\ov a}^{\,\prime}_*=\bd/3\nd$. There
is no pole in the mion propagator at the momentum
$p=\mu(M)\gg\mu^{\rm pole}_{q,S}$.

The opposite case with $\mu^{\rm pole}_{q,S}\gg\mu^{\rm
pole}(M)$ is realized if the ratio $\mph/\mo$ is still $\ll 1$
but is much larger than $Z_q^{\,3/2}\gg{\ov Z}_q^{\,3/2}$, see
\eqref{(19.2.8)} below. In this case the theory at $\mu^{\rm
pole}_{q,S}<\mu<\la$ remains in the conformal regime with ${\ov
a}_*=7\bd/3\nd$ and the largest mass is $\mu^{\rm pole}_{q,S}$.
One has in this case instead of
\eqref{(19.2.2)},\eqref{(19.2.3)},\eqref{(19.2.7)}
\bbq
{\ov\rho}_S\sim 1\,,\quad \mu^{\rm
pole}_{q,S}\sim\frac{1}{Z_q}\lym^{(S)}\,,\quad \frac{\mu^{\rm
pole}
(M)}{\mu^{\rm pole}_{q,S}}\sim Z^3_q\Bigl (\frac{\mo}{\mph}\Bigr
)^{N_c/\nd}\,,
\eeq
\bq
\frac{\mu^{\rm pole}(M)}{\mu^{\rm pole}_{q,S}}\ll 1\quad\ra\quad
Z_q^{\,3/2}\ll\frac{\mph}{\mo}\ll 1\,. \label{(19.2.8)}
\eq

On the whole, the mass spectrum in these $\nd$ dual S - vacua
looks as follows at $\la\ll\mph\ll{\ov Z}_q^{\,3/2}\mo$. a) The
heaviest are $N_F^2$ mions with the pole masses \eqref{(19.2.2)}.
b) There is a large number of flavored hadrons made of weakly
interacting and weakly confined (the tension of the confining
string is $\sqrt\sigma\sim\lym^{(S)}\ll\mu^{\rm
pole}_{q,S}\ll\mu^{\rm pole}(M)$) non-relativistic dual quarks
${\ov q}, q$ with the perturbative pole masses \eqref{(19.2.3)}.
The mass spectrum of low-lying flavored mesons is Coulomb-like
with parametrically small mass differences
$\Delta\mu_H/\mu_H=O(\bd^2/\nd^2)\ll 1$. b) A large number of
gluonia made of $SU(\nd)$ gluons with the mass scale
$\sim\lym^{(S)}\sim\la (m_Q\mph/\la^2)^{N_F/3\nd}$.

The mions with the pole masses \eqref{(19.2.2)} remain the heaviest
ones, $\mu^{\rm pole}(M)\gg\mu^{\rm pole}_{q,S}$, at values
$\mph$ in the range ${\ov Z}_q^{\,3/2}\mo\ll\mph\ll
Z_q^{\,3/2}\mo$, while the value $\mu^{\rm pole}_{q,S}$ varies
in a range $\lym^{(S)}/Z_q\ll\mu^{\rm
pole}_{q,S}\ll\lym^{(S)}/{\ov Z}_q$\,. Finally, in a close
vicinity of $\mo,\,\, Z_q^{\,3/2}\mo\ll\mph\ll\mo$, the
perturbative pole mass of quarks, $\mu^{\rm
pole}_{q,S}\sim\lym^{(S)}
/Z_q\gg\lym^{(S)}$, becomes the largest one, while the pole
masses of mions $\mu^{\rm pole}(M)\ll\mu^{\rm pole}_{q,S}$
become as in \eqref{(19.2.8)}.

At $\mph\gg\mo$ these S - vacua evolve into the vacua of dSQCD,
see section 4 in \cite{ch3}.

\section{Direct theory. Broken flavor symmetry,
$\mathbf{\la\ll\mph\ll\mo}$}

\subsection{\quad  L - type vacua}

\hspace*{4mm} The quark condensates are parametrically the same
as in the L - vacua with unbroken flavor symmetry in section 18,
\bq
(1-\frac{n_1}{N_c})\langle\Qo\rangle_{Lt}\approx
-(1-\frac{n_2}{N_c})\langle\Qt\rangle_{Lt},\quad \langle
S\rangle=\frac{\langle\Qo\rangle\langle\Qt\rangle}{\mph}\,,
\label{(20.1.1)}
\eq
\bbq
\langle\Qo\rangle_{Lt}\sim\langle\Qt\rangle_{Lt}\sim\la^2\Bigl
(\frac{\la}{\mph}\Bigr )^{\frac{\nd}{2N_c-N_F}}\,.
\eeq

All quarks are in the HQ phase and are confined and the
Lagrangian of fions looks as in \eqref{(18.1.3)}, but one has to
choose the L - type vacua with the broken flavor symmetry in
\eqref{(18.1.3)}. Due to this, the masses of
hybrid fions $\Phi_{12}, \Phi_{21}$ are qualitatively different,
they are the Nambu-Goldstone particles here and are massless.
The "masses" of $\Phi_{11}$ and $\Phi_{22}$ are parametrically
as in \eqref{(18.1.4)},
\bq
\mu(\Phi_{11})\sim\mu(\Phi_{22})\sim\frac{\mph}{z_{\Phi}(\la,
m^{\rm pole}_Q)}\sim m^{\rm pole}_{Q,1}\sim m^{\rm
pole}_{Q,2}\sim\lym^{(L)}\sim\la \Bigl (\frac{\la}{\mph}\Bigr
)^{\frac{N_F}{3(2N_c-N_F)}}\,, \label{(20.1.2)}
\eq
and hence there is no guaranty that these are the pole masses of
fions, see section 18. May be yes, but maybe not.

On the whole, there are only two characteristic scales in the
mass spectra in these L - type vacua. The hybrid fions
$\Phi_{12}, \Phi_{21}$ are massless
while all other masses are $\sim\lym^{(L)}$.\\

\subsection{\quad  $\rm br2$ vacua}

\hspace*{4mm} The condensates of quarks look as
\bq
\langle\Qt\rangle_{\rm br2}\approx \Bigl
(\rho_2=-\frac{n_2-N_c}{N_c}\Bigr )m_Q\mph,\,\,\,
\langle\Qo\rangle_{\rm br2}\sim \la^2\Bigl(\frac{\mph}{\la}\Bigr
)^{\frac{n_2}{n_2-N_c}}\Bigl (\frac{m_Q}{\la}\Bigr
)^{\frac{N_c-n_1}{n_2-N_c}},\label{(20.2.1)}
\eq
\bbq
\frac{\langle\Qo\rangle_{\rm br2}}{\langle\Qt\rangle_{\rm
br2}}\sim \Bigl (\frac{\mph}{\mo}\Bigr)^{\frac{N_c}{n_2-N_c}}\ll 1
\eeq
in these vacua with $n_2>N_c\,, 1\leq n_1<\nd$\,. Hence, the
largest among the masses smaller than $\la$ are the masses of
the $N_F^2$ second generation fions, see \eqref{(17.0.1)},
\bq
\mu^{\rm pole}_2(\Phi_i^j)=\mu_o^{\rm conf}\sim \la\Bigl
(\frac{\la}{\mph}\Bigr )^{\frac{N_F}{3(2N_c-N_F)}}\,,
\label{(20.2.2)}
\eq
while some other possible characteristic masses look here as
\bq
\langle m^{\rm tot}_{Q,1}\rangle_{\rm
br2}=\frac{\langle\Qt\rangle_{\rm br2}}{\mph}\sim m_Q\,,\quad
m^{\rm pole}_{Q,1}\sim\la\Bigl(\frac{m_Q}{\la}\Bigr
)^{N_F/3N_c}\gg {\tilde m}^{\rm pole}_{Q,2}\,, \label{(20.2.3)}
\eq
\bbq
\mu^2_{\rm gl, 2}\sim z_Q(\la,\mu_{\rm gl, 2})\langle\Qt\rangle_{\rm
br2},\quad
z_Q(\la, \mu_{\rm gl, 2})\sim\Bigl (\frac{\mu_{\rm gl, 2}}{\la}\Bigr
)^{\frac{\bo}{N_F}}\ll 1\,,
\eeq
\bq
\mu_{\rm gl, 2}\sim\la\Bigl(\frac{m_Q\mph}{\la^2}\Bigr
)^{N_F/3\nd}\gg\mu_{\rm gl, 1}\,,\quad \frac{\mu_{\rm gl, 2}}{m^{\rm
pole}_{Q,1}}\sim
\Bigl (\frac{\mph}{\mo}\Bigr )^{\frac{N_F}{3\nd}}\ll 1\,,
\label{(20.2.4)}
\eq
where $m^{\rm pole}_{Q,1}$ and ${\tilde m}^{\rm pole}_{Q,2}$ are
the pole masses of quarks ${\ov Q}_1, Q^1$ and ${\ov Q}_2, Q^2$
and $\mu_{\rm gl, 1},\, \mu_{\rm gl, 2}$ are the gluon masses due to
possible higgsing of these quarks. Hence, the largest mass is 
$m^{\rm pole}_{Q,1}$ and the overall phase is $HQ_1-HQ_2$.

The lower energy theory at $\mu<m^{\rm pole}_{Q,1}$ has $N_c$
colors and $N_F^\prime =n_2>N_c$ flavors of quarks ${\ov Q}_2,
Q^2$. In the range of scales $m^{\rm pole}_{Q,2}<\mu<m^{\rm
pole}_{Q,1}$, it will remain in the conformal regime at
$n_1<\bd=(2N_F-3N_c)/2$, while it will be in the strong coupling
regime at $n_1>\bd/2$, with the gauge coupling $a(\mu\ll m^{\rm
pole}_{Q,1})\gg 1$. We do not consider the strong coupling
regime here and for this reason we take $\bd/\nd=O(1)$
in this subsection and consider $n_1<\bd/2$ only.

After the heaviest quarks ${\ov Q}_1, Q^1$ decouple at
$\mu<m^{\rm pole}_{Q,1}$, the pole mass of quarks ${\ov Q}_2,
Q^2$ in the lower energy theory looks as
\bq
m^{\rm pole}_{Q,2}=\frac{1}{z^{\,\prime}_Q(m^{\rm
pole}_{Q,1},m^{\rm
pole}_{Q,2})}\Biggl(\,\frac{\langle\Qo\rangle_{\rm
br2}}{\langle\Qt\rangle_{\rm br2}}\, m^{\rm pole}_{Q,1}\,\Biggr
)\sim\lym^{(\rm br2)}\,, \label{(20.2.5)}
\eq
\bbq
z^{\,\prime}_Q(m^{\rm pole}_{Q,1},m^{\rm pole}_{Q,2})\sim\Bigl
(\frac{m^{\rm pole}_{Q,2}}{m^{\rm pole}_{Q,1}}\Bigr
)^{\frac{3N_c-n_2}{n_2}}\ll 1\,.
\eeq

Hence, after integrating out quarks ${\ov Q}_1, Q^1$ at
$\mu<m^{\rm pole}_{Q,1}$ and then quarks ${\ov Q}_2, Q^2$ and
$SU(N_c)$ gluons at $\mu<\lym^{(\rm br2)}$, the Lagrangian of
fions looks as
\bq
K=z_{\Phi}(\la,m^{\rm pole}_{Q,1})\,{\rm Tr}\,\Bigl
[\,\Phi_{11}^\dagger \Phi_{11}+\Phi_{12}^\dagger
\Phi_{12}+\Phi_{21}^\dagger \Phi_{21}+z^{\,\prime}_{\Phi}(m^{\rm
pole}_{Q,1},m^{\rm pole}_{Q,2})\Phi_{22}^\dagger
\Phi_{22}\,\Bigr ]\,, \label{(20.2.6)}
\eq
\bbq
z_{\Phi}(\la,m^{\rm pole}_{Q,1})\sim\Bigl (\frac{\la}{m^{\rm
pole}_{Q,1}}\Bigr )^{\frac{2(3N_c-N_F)}
{N_F}}\gg 1\,,\quad z^{\,\prime}_{\Phi}(m^{\rm
pole}_{Q,1},m^{\rm pole}_{Q,2})\sim\Bigl (\frac{m^{\rm
pole}_{Q,1}}{m^{\rm pole}_{Q,2}}\Bigr
)^{\frac{2(3N_c-n_2)}{n_2}}\gg 1\,,
\eeq
\bq
\cw=N_c S+\cw_{\Phi}\,,\quad m^{\rm tot}_Q= (m_Q-\Phi)\,,\quad
\label{(20.2.7)}
\eq
\bbq
S=\Bigl (\la^{\bo}\det m^{\rm tot}_Q\Bigr )^{1/N_c}\,,\quad
\cw_{\Phi}=\frac{\mph}{2}\Bigl ({\rm Tr}\,(\Phi^2)-
\frac{1}{\nd}({\rm Tr}\,\Phi)^2 \,\Bigr ).
\eeq
From \eqref{(20.2.6)},\eqref{(20.2.7)}, the main contribution to the
mass of the third generation fions $\Phi_{11}$ gives the term
$\sim\mph\Phi^2_{11}$,
\bq
\mu^{\rm pole}_3(\Phi_{11})\sim\frac{\mph}{z_{\Phi}(\la,m^{\rm
pole}_{Q,1})}\sim\Bigl (\frac{\mph}{\mo}\Bigr )m^{\rm
pole}_{Q,1}\,, \label{(20.2.8)}
\eq
while the third generation hybrid fions $\Phi_{12}, \Phi_{21}$
are massless, $\mu^{\rm pole}_3(\Phi_{12})=\mu^{\rm
pole}_3(\Phi_{21})=0$. As for the third generation fions
$\Phi_{22}$, the main contribution to their masses comes from
the non-perturbative term $\sim S$ in the superpotential
\eqref{(20.2.7)}
\bq
\mu_3(\Phi_{22})\sim\frac{\langle S\rangle}{\langle m^{\rm
tot}_{Q,2}\rangle^2}\frac{1}{z_{\Phi}(\la,m^{\rm pole}_{Q,1})
z^{\,\prime}_{\Phi}(m^{\rm pole}_{Q,1},m^{\rm pole}_{Q,2})}\sim
m^{\rm pole}_{Q,2}\sim\lym^{(\rm br2)}. \label{(20.2.9)}
\eq
In such a situation there is no guaranty that there is a pole in
the propagator of $\Phi_{22}$ at the momentum $p\sim m^{\rm
pole}_{Q,2}$. May be yes but maybe not, see section 17.\\

\section{Dual theory. Broken flavor symmetry,
$\mathbf{\la\ll\mph\ll\mo}$}

\subsection{\quad L - type vacua, $\,\,\bd/N_F\ll 1$}

\hspace*{4mm} The condensates of mions and dual quarks look here
as
\bbq
\langle M_1+M_2-\frac{1}{N_c}{\rm
Tr}\,M\rangle_{Lt}=m_Q\mph\quad\ra\quad \frac{\langle
M_1\rangle_{Lt}}{\langle M_2\rangle_{Lt}}\approx
-\,\frac{N_c-n_1}{N_c-n_2}\,,
\eeq
\bbq
\langle M_1\rangle_{Lt}\langle\qo\rangle_{Lt}=\langle
M_2\rangle_{Lt}\langle\qt\rangle_{Lt}=Z_q\la\langle
S\rangle_{Lt},\quad\langle S\rangle_{Lt}=\frac{\langle
M_1\rangle_{Lt}\langle M_2\rangle_{Lt}}{\mph}\,.
\eeq

I.e., all condensates are parametrically the same as in the L -
vacua with unbroken flavor symmetry in section 19.1 and the
overall phase is also $HQ_1-HQ_2$. The pole masses of dual
quarks are as in \eqref{(19.1.3)}, the Lagrangian of mions is as in
\eqref{(19.1.6)} and the pole masses of mions $M_{11}$ and $M_{22}$
are as in \eqref{(19.1.7)}. But the masses of hybrid mions $M_{12}$
and $M_{21}$ are qualitatively different here. They are the
Nambu-Goldstone particles now and are exactly massless,
$\mu(M_{12})=\mu (M_{21})=0$.\\

\subsection{\quad  $\rm br2$ vacua, $\,\,\bd/N_F=O(1)$}

\hspace*{4mm} In these vacua with $n_2>N_c\,, 1\leq n_1<\nd$ the
condensates of mions and dual quarks look as
\bbq
\langle M_1\rangle_{\rm br2}=\langle\Qo\rangle_{\rm br2}\sim
\la^2\Bigl(\frac{\mph}{\la}\Bigr )^{\frac{n_2}{n_2-N_c}}\Bigl
(\frac{m_Q}{\la}\Bigr )^{\frac{N_c-n_1}{n_2-N_c}},
\eeq
\bq
\langle M_2\rangle_{\rm br2}=\langle\Qt\rangle_{\rm br2}\approx
-\,\frac{n_2-N_c}{N_c}\, m_Q\mph\,,\quad
\frac{\langle M_1\rangle_{\rm br2}}{\langle M_2\rangle_{\rm
br2}}\sim \Bigl (\frac{\mph}{\mo}\Bigr
)^{\frac{N_c}{n_2-N_c}}\ll 1\,, \label{(21.2.1)}
\eq
\bbq
\langle\qo\rangle_{\rm br2}=\langle{\ov q}^1
q_1(\mu=\la)\rangle_{\rm br2}=\frac{\la\langle S\rangle_{\rm
br2}}{\langle M_1\rangle_{\rm br2}}=\frac{\la\langle
M_2\rangle_{\rm br2}}{\mph}\sim m_Q\la\gg\langle \qt\rangle_{\rm
br2}\,.
\eeq

From these, the heaviest are $N_F^2$ mions $M^i_j$ with the pole
masses
\bq
\hspace*{-4mm}\mu^{\rm
pole}(M)=\frac{\la^2/\mph}{z_M(\la,\mu^{\rm pole}(M))}\sim
\la\Bigl (\frac{\la}{\mph}\Bigr
)^{\frac{N_F}{3(2N_c-N_F)}}\,,\label{(21.2.2)}
\eq
\bbq
z_M(\la,\mu^{\rm pole}(M))\sim\Bigl (\frac{\la}{\mu^{\rm
pole}(M)}\Bigr )^{\frac{2\bd}{N_F}}\gg 1,
\quad \bd=3\nd-N_F\,,
\eeq
while some other possible characteristic masses look as
\bq
\mu_{q,2}=\frac{\langle
M_2\rangle}{\la}\sim\frac{m_Q\mph}{\la},\quad {\tilde\mu}^{\rm
pole}_{q,2}\sim\la
\Bigl (\frac{m_Q\mph}{\la^2}\Bigr )^{N_F/3\nd}\gg\mu^{\rm
pole}_{q,1}\,, \label{(21.2.3)}
\eq
\bbq
{\ov\mu}_{\rm gl,1}\sim\la\Bigl
(\frac{\langle\qo\rangle}{\la^2}\Bigr )^{N_F/3N_c}\sim
\la\Bigl(\frac{m_Q}{\la}\Bigr )^{N_F/3N_c}\gg{\ov\mu}_{\rm
gl,2}\,,\quad\frac{{\ov\mu}_{\rm gl,1}}{{\tilde\mu}^{\rm
pole}_{q,2}}
\sim \Bigl(\frac{\mo}{\mph}\Bigr )^{N_F/3\nd}\gg 1\,,
\eeq
where $\mu^{\rm pole}_{q,1}$ and ${\tilde\mu}^{\rm pole}_{q,2}$
are the perturbative pole masses of quarks ${\ov q}^1, q_1$ and
${\ov q}^2, q_2$ and ${\ov\mu}_{\rm gl,1},\, {\ov\mu}_{\rm
gl,2}$ are the gluon masses due to possible higgsing of these
quarks. Hence, the largest mass is ${\ov\mu}_{\rm gl,1}$ and the
overall phase is $Higgs_1-HQ_2$.

After integrating out all higgsed quarks ${\ov q}^1, q_1$ and heavy 
gluons,  we write the dual Lagrangian at 
$\mu={\ov\mu}_{{\rm gl},\,1}$ as
\bbq
K= z_M(\la,{\ov\mu}_{{\rm gl},\,1}){\rm Tr}\,\frac{M^\dagger
M}{\la^2}+  z_q(\la,{\ov\mu}_{{\rm gl},\,1}){\rm Tr}\,\Bigl
[\,2\sqrt{N_{11}^\dagger  N_{11}}+K_{\rm hybr}+
\Bigl ({\textsf{q}}^{\dagger}_2 {\textsf{q}}_2
+({\textsf{q}}_2\ra {\ov{\textsf{q}}}^2 )\Bigr )\,\Bigr ]\,,
\eeq
\bq
K_{\rm hybr}=\Biggl (N^{\dagger}_{12}\frac{1}{\sqrt{N_{11}
N^{\dagger}_{11}}} N_{12}+
N_{21}\frac{1}{\sqrt{N^{\dagger}_{11} N_{11}}}
N^\dagger_{21}\Biggr ),\quad z_q(\la,\muo)=\Bigl
(\frac{\muo}{\la}\Bigr )^{\bd/N_F}\,,\label{(21.2.4)}
\eq
\bbq
z_M(\la,\muo)=1/z^2_q(\la,\muo),\quad \cw=\Bigl
[\frac{2\pi}{{\ov\alpha}(\mu)}{\ov{\textsf s}}
\Bigr ]-\frac{1}{\la}{\rm Tr}\,\Bigl (\ov{\textsf{q}}_2
M_{22}\textsf{q}_2\Bigr) - \cw_{MN}+\cw_{M},
\eeq
\bbq
\cw_{MN}=\frac{1}{\la}{\rm Tr}\,\Bigl (M_{11}N_{11}+M_{21}
N_{12}+N_{21}
M_{12}+M_{22} N_{21}\frac{1}{N_{11}} N_{12}\Bigr )\,,
\eeq
where the nions (dual pions) $N_{11}$ originate from higgsing of
${\ov q}^1, q_1$ dual quarks while $\ov{\textsf{q}}_2, \textsf{q}_2$
are the active quarks ${\ov q}^2, q_2$ with not higgsed colors,
$\ov{\textsf s}$ is the field strength of not higgsed dual gluons
and the hybrid nions $N_{12}$ and $N_{21}$ are, in essence, the
quarks ${\ov q}^2, q_2$ with higgsed colors, $\cw_M$ is given in
\eqref{(14.2.2)}. The lower energy theory at $\mu<\muo$ has
$\nd^{\,\prime}=\nd-n_1$ colors and $n_2>N_c$ flavors,
$\bd^{\,\prime}=\bd-2n_1<\bd$. We consider here only the case
$\bd^{\,\prime}>0$ when it remains in the conformal window. In
this case the value of the pole mass $\mu^{\rm pole}_{q,2}$ in this
lower energy theory is
\bq
\mu^{\rm pole}_{q,2}\sim \frac{\langle
M_2\rangle}{\la}\frac{1}{z_q(\la,{\ov\mu}_{{\rm gl},\,1})
z^{\,\prime}_q(\muo,\mu^{\rm pole}_{q,2})} \sim\lym^{(\rm br2)}\,,
\quad z^{\,\prime}_q(\ov{\mu}_{{\rm gl},\,1},\mu^{\rm pole}_{q,2})\sim
\Bigl (\frac{\mu^{\rm pole}_{q,2}} { {\ov\mu}_{{\rm gl},\,1} } \Bigr)
^{\bd^{\,\prime}/n_2}\ll 1\,. \label{(21.2.5)}
\eq

The fields $N_{11}, N_{12}, N_{21}$ and $M_{11}, M_{12}, M_{21}$
are frozen and do not evolve at $\mu<\muo$. After integrating
out remained not higgsed quarks $\ov{\textsf q}_2, {\textsf q}_2$ 
as heavy ones and  not higgsed gluons at $\mu<\lym^{(\rm br2)}$ 
the Lagrangian of mions and nions looks as, see \eqref{(21.2.4)},
\bbq
K=z_M(\la,\muo){\rm Tr}\,K_M+z_q(\la,\muo)\Bigl
[\,2\sqrt{N_{11}^\dagger N_{11}}+K_{\rm hybr}\,\Bigr ],\,\,
z^{\,\prime}_M (\muo,\qtp)\sim\Bigl (\frac{\muo}{\qtp}\Bigr
)^{\frac{2\bd^{\,\prime}}{n_2}}\gg 1,
\eeq
\bq
K_M=\frac{1}{\la^2}\Bigl (M_{11}^\dagger M_{11}+M_{12}^\dagger
M_{12}+M_{21}^\dagger M_{21}+
z^{\,\prime}_M (\muo,\qtp)M_{22}^\dagger M_{22}\Bigr )\,,
\label{(21.2.6)}
\eq
\bbq
\cw=\nd^{\,\prime}S-\cw_{MN}+\cw_M\,,\quad S=\Bigl (\lym^{(\rm
br2)}\Bigr )^3 \Biggl (\det\frac{\langle
N_1\rangle}{N_{11}}\det\frac{M_{22}}{\langle M_2\rangle}\Biggr
)^{1/\nd^{\,\prime}},\quad \lym^{(\rm br2)}\sim\Bigl (m_Q\langle
M_1\rangle\Bigr )^{1/3}.
\eeq

From \eqref{(21.2.6)}, the "masses" of mions look as
\bq
\mu(M_{11})\sim\mu(M_{12})\sim\mu(M_{21})\sim\frac{\la^2}
{z_M(\la,\muo)\mph}\sim\Bigl
(\frac{\mo}{\mph}\Bigr )\muo\gg\muo\,,\label{(21.2.7)}
\eq
\bq
\mu(M_{22})\sim\frac{\la^2}{z_M(\la,\muo)z^\prime_M
(\muo,\qtp)\mph}\sim \Bigl (\frac{\mo}{\mph}\Bigr
)^{\frac{3N_c-n_2}{3(n_2-N_c)}}\,\muo\gg \muo\,,\label{(21.2.8)}
\eq
while the pole masses of nions $N_{11}$ are
\bq
\mu^{\rm pole}(N_{11})\sim \frac{\mph\langle N_1\rangle_{\rm
br2}}{z_q(\la,\muo)\la^2}\sim\Bigl (\frac{\mph}{\mo}\Bigr
)\muo\,,\label{(21.2.9)}
\eq
and the hybrid nions $N_{12}, N_{21}$ are massless,
$\mu(N_{12})=\mu(N_{21})=0$. The mion "masses"
\eqref{(21.2.7)},\eqref{(21.2.8)} are not the pole masses but simply
the low energy values of mass terms in their propagators, the
only pole masses are given in \eqref{(21.2.2)}.\\

\subsection{\quad   $\rm br2$ vacua, $\bd/N_F\ll 1$}

Instead of \eqref{(21.2.2)}, the pole mass of mions is
parametrically smaller now,
\bq
\mu^{\rm pole}(M)=\frac{Z_q^{\, 2}\la^2/\mph}{z_M(\la,\mu^{\rm
pole}(M))}\sim Z_q^{\, 2}\la\Bigl (\frac{\la}{\mph}\Bigr
)^{\frac{N_F}{3(2N_c-N_F)}},\quad \frac{\mu^{\rm
pole}(M)}{\mu^{\rm pole}_2(\Phi)}\sim Z_q^{\, 2}\ll 1\,,
\label{(21.3.1)}
\eq
while instead of \eqref{(21.2.3)} we have now
\bq
\mu_{q,2}=\frac{\langle
M_2\rangle}{Z_q\la}\sim\frac{m_Q\mph}{Z_q\la},\quad
{\tilde\mu}^{\rm pole}_{q,2}\sim\frac{\la}{Z_q}\Bigl
(\frac{m_Q\mph}{\la^2}\Bigr )^{N_F/3\nd}\gg\mu^{\rm
pole}_{q,1}\,, \quad Z_q\sim\exp \Bigl\{\frac{-\nd}{7\,\bd} \Bigr \}\ll 1\,.
\label{(21.3.2)}
\eq
\bq
{\ov\mu}_{\rm gl,1}\sim \la\Bigl (\frac{\langle
N_1\rangle}{\la^2}\Bigr )^{N_F/3N_c}\sim
Z_q^{1/2}\la\Bigl(\frac{m_Q}{\la}\Bigr
)^{N_F/3N_c}\gg{\ov\mu}_{\rm gl,2}\,,\quad \frac{{\ov\mu}_{\rm
gl,1}}{m^{\rm pole}_{Q,1}}\sim Z_q^{1/2}\ll 1\,,\label{(21.3.3)}
\eq
\bq
\quad\frac{{\ov\mu}_{\rm gl,1}}{{\tilde\mu}^{\rm
pole}_{q,2}}\sim Z_q^{3/2}\Bigl(\frac{\mo}{\mph}\Bigr
)^{N_F/3\nd}\gg 1\,,\quad \la\ll {\mph}\ll
Z_q^{\,3/2}\mo\,,\quad Z_q\sim\exp\{-\frac{\nd}{7\bd}\}\ll
1\,.\label{(21.3.4)}
\eq

Hence, at the condition \eqref{(21.3.4)}, the largest mass is
${\ov\mu}_{\rm gl,1}$ and the overall phase is also
$Higgs_1-HQ_2$. But now, at $\bd/\nd\ll 1$, it looks unnatural
to require $\bd^{\, \prime}=(\bd-2n_1)>0$. Therefore, with
$n_1/\nd=O(1)$, the lower energy theory at $\mu<{\ov\mu}_{\rm
gl,1}$ has $\bd^{\, \prime}<0$ and is in the logarithmic IR free
regime in the range of scales $\qtp<\mu<{\ov\mu}_{\rm gl,1}$.
Then instead of \eqref{(21.2.5)} (ignoring all logarithmic
renormalization factors),
\bq
\lym^{(\rm br2)}\ll\qtp\sim \frac{\langle M_2\rangle_{\rm
br2}}{Z_q\la}\frac{1}{z_q(\la,\muo)}\sim\frac
{\mph}{Z_q^{3/2}\mo}\,{\ov\mu}_{\rm gl,1}\ll {\ov\mu}_{\rm
gl,1}\,.\label{(21.3.5)}
\eq

The Lagrangian of mions and nions has now the form \eqref{(21.2.6)}
with accounting additionally for $Z_q$ factors, and with a
replacement $z^{\,\prime}_M (\muo,\qtp)\sim 1$, and so
$\mu(M_{22})\sim\mu(M_{11})\sim\mu(M_{12})\sim\mu(M_{21})$ now,
see \eqref{(21.2.7)},\eqref{(21.2.8)},\eqref{(21.3.4)},
\bq
\mu(M^i_j)\sim\frac{Z_q^2\la^2}{z_M(\la,\muo)\mph}\sim
Z_q^{3/2}\Bigl (\frac{\mo}{\mph}\Bigr
)\muo\gg\muo\,,\label{(21.3.6)}
\eq
while, instead of \eqref{(21.2.9)}, the mass of nions looks now as
\bq
\mu^{\rm pole}(N_{11})\sim \frac{\mph\langle N_1\rangle_{\rm
br2}}{z_q(\la,\muo)\la^2}\sim Z_q^{1/2}\Bigl
(\frac{\mph}{\mo}\Bigr )\,\muo\,.\label{(21.3.7)}
\eq

On the whole for the mass spectra in this case. a) The heaviest
are $N_F^2$ mions with the pole masses \eqref{(21.3.1)} (the
'masses' \eqref{(21.3.6)} are not the pole masses but simply the
low energy values of mass terms in the mion propagators). \, b)
The next are the masses \eqref{(21.3.3)} of $n_1(2\nd-n_1)$
higgsed gluons and their superpartners. \, c) There is a large
number of flavored hadrons, mesons and baryons, made of
non-relativistic and weakly confined (the string tension is
$\sqrt{\sigma}\sim\lym^{(\rm br2)}\ll\qtp$\,) quarks $\odq^2,
\dq_2$ with not higgsed colors. The mass spectrum of low-lying
flavored mesons is Coulomb-like with parametrically small mass
differences, $\Delta\mu_H/\mu_H=O(\bd^{\,2}/N^2_F)\ll 1$.\, d) A
large number of gluonia made of $SU(\nd-n_1)$ gluons with the
mass scale $\sim\lym^{(\rm br2)}$.\, e) $n_1^2$ nions $N_{11}$
with the masses \eqref{(21.3.7)}.\, f) The hybrid nions $N_{12},
N_{21}$ are the Nambu-Goldstone particles here and are
massless.\\

\section{\hspace*{-2mm} Direct theory. Broken flavor symmetry,
$\mathbf{\mo\ll\mph\ll\frac{\la^2}{m_Q}}$}

\subsection{\quad  $\rm br1$ vacua}

\hspace*{4mm} The values of quark condensates are here
\bq
\langle\Qo\rangle_{\rm br1}\approx\frac{N_c}{N_c-n_1}\,
m_Q\mph\,,\quad \langle\Qt\rangle_{\rm br1}\sim \la^2\Bigl
(\frac{\la}{\mph}\Bigr )^{\frac{n_1}{N_c-n_1}}\Bigl
(\frac{m_Q}{\la}\Bigr )^{\frac{n_2-N_c}{N_c-n_1}}\,,
\label{(22.1.1)}
\eq
\bbq
\frac{\langle\Qt\rangle_{\rm br1}}{\langle\Qo\rangle_{\rm
br1}}\sim\Bigl (\frac{\mo}{\mph}\Bigr )^{\frac{N_c}{N_c-n_1}}\ll
1\,.
\eeq
From these, the values of some potentially relevant masses 
look as
\bbq
\mgo^2\sim \Bigl (a_*\sim 1\Bigr
)z_Q(\la,\mgo)\langle\Qo\rangle_{\rm br1} \,,\quad
z_Q(\la,\mgo)\sim
\Bigl (\frac{\mgo}{\la}\Bigr )^{\bo/N_F}\,,
\eeq
\bq
\mgo\sim\la \Bigl (\frac{m_Q\mph}{\la^2}\Bigr
)^{N_F/3\nd}\gg\mgt\,, \label{(22.1.2)}
\eq
\bbq
\langle m^{\rm tot}_{Q,2}\rangle=\frac{\langle\Qo\rangle_{\rm
br1}}{\mph}\sim m_Q\,,\quad {\tilde m}^{\rm
pole}_{Q,2}=\frac{\langle m^{\rm tot}_{Q,2}\rangle_{\rm
br1}}{z_Q(\la,{\tilde m}^{\rm pole}_{Q,2})}\,,
\eeq
\bq
{\tilde m}^{\rm pole}_{Q,2}\sim\la\Bigl (\frac{m_Q}{\la}\Bigr
)^{N_F/3N_c}\gg m^{\rm pole}_{Q,1},\quad\,\,\, \frac{{\tilde
m}^{\rm pole}_{Q,2}}{\mgo}\sim\Bigl (\frac{\mo}{\mph}\Bigr
)^{N_F/3\nd}\ll 1\,. \label{(22.1.3)}
\eq
Hence, the largest mass is $\mgo$ due to higgsing of ${\ov Q}_1,
Q^1$ quarks and the overall phase is $Higgs_1-HQ_2$.

The lower energy theory at $\mu<\mgo$ has $N_c^\prime=N_c-n_1$
colors and $n_2\geq N_f/2$ flavors. At $2n_1<\bo$ it remains in
the conformal window with ${\rm b}_o^\prime>0$, while at
$2n_1>\bo,\,\,{\rm b}_o^\prime<0$ it enters the logarithmic IR
free perturbative regime.

We start with $\bo^\prime>0$. Then the value of the pole mass of
quarks $\oqt,\,\sqt$ with not higgsed colors looks as
\bbq
\ma=\frac{\langle m^{\rm tot}_{Q,2}\rangle_{\rm
br1}}{z_Q(\la,\mgo)z_Q^{\,\prime}(\mgo,
\ma)}\,,\quad z_Q^{\,\prime}(\mgo,\ma)\sim\Bigl
(\frac{\ma}{\mgo}\Bigr )^{{\rm b}_o^\prime/n_2}\,,
\eeq
\bq
\ma\sim\la\Bigl (\frac{\la}{\mph}\Bigr
)^{\frac{n_1}{3(N_c-n_1)}}\Bigl (\frac{m_Q}{\la}\Bigr
)^{\frac{n_2-n_1}{3(N_c-n_1)}}\sim\lym^{(\rm
br1)}\,.\label{(22.1.4)}
\eq
It is technically convenient to retain all fion fields $\Phi$
although, in essence, they are too heavy and dynamically
irrelevant at $\mph\gg\mo$. After integrating out all heavy
higgsed gluons and quarks ${\ov Q}_1, Q^1$, we write the
Lagrangian at $\mu=\mu_{\rm gl,1}$ in the form
\bbq
K=\Bigl [\,z_{\Phi}(\la,\mgo){\rm
Tr}(\Phi^\dagger\Phi)+z_Q(\la,\mu^2_{\rm gl,1})\Bigl
(K_{{\sq}_2}+K_{\Pi}\Bigr )\,\Bigr ],\quad
z_{\Phi}(\la,\mgo)=1/z^2_Q(\la,\mgo)\,,
\eeq
\bq
K_{{\sq}_2}={\rm Tr}\Bigl ({\sq}^{\dagger}_2 {\sq}^2
+({\sq}^2\ra
{\oq}_2 )\Bigr )\,, \quad K_{\Pi}= 2{\rm
Tr}\sqrt{\Pi^{\dagger}_{11}\Pi_{11}}+K_{\rm hybr},
\label{(22.1.5)}
\eq
\bbq
K_{\rm hybr}={\rm Tr}\Biggl
(\Pi^{\dagger}_{12}\frac{1}{\sqrt{\Pi_{11}\Pi^{\dagger}_{11}}}\Pi_{12}+
\Pi_{21}\frac{1}{\sqrt{\Pi^{\dagger}_{11}\Pi_{11}}}\Pi^\dagger_{21}\Biggr
),
\eeq
\bbq
\cw=\Bigl [\frac{2\pi}{\alpha(\mu_{\rm gl,1})}{\textsf S}\Bigr
]+\frac{\mph}{2}\Biggl [{\rm Tr}\, (\Phi^2) -\frac{1}{\nd}\Bigl
({\rm Tr}\,\Phi\Bigr)^2\Biggr ]+{\rm Tr}\Bigl ({\oq_2}m^{\rm
tot}_{{\sq}_2}{\sq}^2\Bigr )+\cw_{\Pi},
\eeq
\bbq
\cw_{\Pi}= {\rm Tr}\Bigl (m_Q\Pi_{11}+m^{\rm
tot}_{{\sq}_2}\,\Pi_{21}\frac{1}{\Pi_{11}}\Pi_{12}\Bigr )-
{\rm Tr}\Bigl
(\Phi_{11}\Pi_{11}+\Phi_{12}\Pi_{21}+\Phi_{21}\Pi_{12} \Bigr ),
\quad m^{\rm tot}_{{\sq}_2}=(m_Q-\Phi_{22}),
\eeq
In \eqref{(22.1.5)}: $\oqt,\, \sqt$ and $\textsf V$ are the active
${\ov Q}_2, Q^2$ quarks and gluons with not higgsed colors
($\textsf S$ is their field strength squared), $\Pi_{12},
\Pi_{21}$ are the hybrid pions (in essence, these are the quarks
${\ov Q}_2, Q^2$ with higgsed colors), $z_Q(\la,\mu^2_{\rm
gl,1})$ is the corresponding perturbative renormalization factor
of massless quarks, see \eqref{(22.1.2)}, while
$z_{\Phi}(\la,\mgo)$ is that of fions. Evolving now down in the
scale and integrating out at $\mu<\lym^{(\rm br1)}$ quarks
$\oq_2,\, \sq^2$ as heavy ones and not higgsed gluons via the
VY-procedure, we obtain the Lagrangian of pions and fions
\bbq
K=\Bigl [z_{\Phi}(\la,\mgo){\rm
Tr}\Bigl(\Phi^\dagger_{11}\Phi_{11}+\Phi^\dagger_{12}\Phi_{12}+
\Phi^\dagger_{21}\Phi_{21}+z^{\,\prime}_{\Phi}(\mgo,\ma)\Phi^
\dagger_{22}\Phi_{22}\Bigr
)+z_Q(\la,\mu^2_{\rm gl,1})K_{\Pi}\Bigr ],\,
\eeq
\bq
\cw=(N_c-n_1)S+W_{\Phi}+W_{\Pi}\,,\quad S=\Biggl
[\frac{\la^{\bo}\det m^{\rm tot}_{{\sq}_2}}{\det \Pi_{11}}\Biggr
]^{\frac{1}{N_c-n_1}}\,, \label{(22.1.6)}
\eq
\bbq
\cw_{\Phi}=\frac{\mph}{2}\Biggl [{\rm Tr} (\Phi^2)
-\frac{1}{\nd}\Bigl ({\rm Tr}\,\Phi\Bigr)^2\Biggr ],\quad
z^{\,\prime}_{\Phi}(\mgo,\ma)\sim\Bigl (\frac{\mgo}{\ma}\Bigr
)^{2{\rm b}_o^\prime/n_2}\,.
\eeq
We obtain from \eqref{(22.1.6)} that all fions are heavy with the
"masses"
\bq
\mu(\Phi_{11})\sim\mu(\Phi_{12})\sim\mu(\Phi_{21})\sim\frac{\mph}{z_{\Phi}
(\la,\mgo)}\sim\Bigl
(\frac{\mph}{\mo}\Bigr )^{N_c/\nd}\mgo\gg\mgo\,, \label{(22.1.7)}
\eq
\bq
\mu(\Phi_{22})\sim\frac{\mph}{z_{\Phi}(\la,\mgo)z^{\,\prime}_{\Phi}
(\mgo,\ma)}\sim
\Bigl (\frac{\mph}{\mo}\Bigr )^{\frac{N_c}{N_c-n_1}}\, \ma\gg
\ma\,. \label{(22.1.8)}
\eq
These are not the pole masses but simply the low energy values
of mass terms in their propagators. All fions are dynamically
irrelevant at all scales $\mu<\la$. The mixings of
$\Phi_{12}\leftrightarrow\Pi_{12},\,
\Phi_{21}\leftrightarrow\Pi_{21}$ and
$\Phi_{11}\leftrightarrow\Pi_{11}$ are parametrically small and
neglected. We obtain then for the masses of pions $\Pi_{11}$
\bq
\mu(\Pi_{11})\sim\Bigl (\frac{\mo}{\mph}\Bigr
)^{\frac{N_c(\bo-2n_1)}{3\nd(N_c-n_1)}}\,\lym^{(\rm
br1)}\sim\Bigl (\frac{\mo}{\mph}\Bigr
)^{\frac{N_c(\bo-2n_1)}{3\nd(N_c-n_1)}}\, \ma\ll \ma\,,
\label{(22.1.9)}
\eq
and, finally, the hybrids $\Pi_{12}, \Pi_{21}$ are massless,
$\mu(\Pi_{12})=\mu(\Pi_{21})=0$.

At $2n_1>\bo$ the RG evolution at $\ma<\mu<\mgo$ is only slow
logarithmic (and is neglected). We replace then
$z^{\,\prime}_Q(\mgo,\ma)\sim 1$ in \eqref{(22.1.4)} and
$z^{\,\prime}_{\Phi}(\mgo,\ma)\sim 1$ in \eqref{(22.1.8)} and
obtain
\bq
\mu(\Phi_{22})\sim\mu(\Phi_{11})\sim\Bigl (\frac{\mph}{\mo}\Bigr
)^{N_c/\nd}\mgo\gg\mgo\,, \label{(22.1.10)}
\eq
\bbq
\mu(\Pi_{11})\sim \ma\sim\frac{m_Q}{z_Q(\la,\mu^2_{\rm
gl,1})}\sim\la\Bigl (
\frac{\la}{\mph}\Bigr )^{\bo/3\nd}\Bigl (\frac{m_Q}{\la}\Bigr
)^{2\,\bd/3\nd}\sim\Bigl (\frac{\mo}{\mph}
\Bigr )^{N_c/\nd}\mgo\ll\mgo.
\eeq
\bq
\frac{\lym^{(\rm br1)}}{\ma}\sim\Bigl (\frac{\mo}{\mph}\Bigr
)^{\Delta}\ll
1,\quad\Delta=\frac{N_c(2n_1-\bo)}{3\nd(N_c-n_1)}>0\,.
\label{(22.1.11)}
\eq
\vspace*{2mm}

\subsection{\quad  $\rm br2$ vacua}

\hspace*{4mm} At $n_2<N_c$ there are also $\rm br2$ - vacua. All
their properties can be obtained by a replacement
$n_1\leftrightarrow n_2$ in formulas of the preceding section.
The only difference is that, because $n_2\geq N_F/2$ and
so $2n_2>\bo$, there is no analog of the conformal regime at
$\mu<\mu_{\rm gl,1}$ with $2n_1<\bo$. I.e. at $\mu<\mu_{\rm
gl,2}$ the lower energy theory will be always in the
perturbative IR free logarithmic regime and the overall phase
will be $Higgs_2-HQ_1$.

\section{\hspace*{-2mm}  Dual theory. Broken flavor symmetry,
$\mathbf{\mo\ll\mph\ll\frac{\la^2}{m_Q}}$}

\subsection{\quad $\rm br1$ vacua, $\,\,\bd/N_F\ll 1$}

We recall that condensates of mions and dual quarks in these
vacua are
\bq
\langle M_1\rangle_{\rm br1}\approx\frac{N_c}{N_c-n_1}\,
m_Q\mph\,,\quad \langle M_2\rangle_{\rm br1}\sim \la^2\Bigl
(\frac{\la}{\mph}\Bigr )^{\frac{n_1}{N_c-n_1}}\Bigl
(\frac{m_Q}{\la}\Bigr )^{\frac{n_2-N_c}{N_c-n_1}}\,,
\label{(23.1.1)}
\eq
\bbq
\frac{\langle M_2\rangle_{\rm br1}}{\langle M_1\rangle_{\rm
br1}}\sim\Bigl (\frac{\mo}{\mph}\Bigr )^{\frac{N_c}{N_c-n_1}}\ll1\,,
\eeq
\bbq
\langle N_2\rangle_{\rm br1}\equiv\langle {\ov q}^2
q_2(\mu=\la)\rangle_{\rm br1}=Z_q\frac{\langle M_1\rangle_{\rm
br1}\la}{\mph}\sim Z_q m_Q\la\gg\langle N_1\rangle_{\rm br1}\,,
\eeq
and some potentially relevant masses look here as
\bq
\langle\qo\rangle=\langle{\ov q}^1
q_1(\mu=\la)\rangle=\frac{\langle M_1\rangle_{\rm
br1}}{Z_q\la}\sim\frac{m_Q\mph}{Z_q\la}\,,\quad
\frac{\langle\qt\rangle}{\langle\qo\rangle}= \frac{\langle
M_2\rangle_{\rm br1}}{\langle M_1\rangle_{\rm br1}}\ll
1\,,\label{(23.1.2)}
\eq
\bbq
Z_q\sim\exp \Bigl\{-\frac{1}{3{\ov a}_{*}}\Bigr\}\sim \exp
\Bigl\{-\frac{\nd}{7\bd}\Bigr\}\ll 1\,,
\eeq
\bq
\qop\sim \frac{\la}{Z_q}\Bigl (\frac{m_Q\mph}{\la^2}\Bigr
)^{N_F/3\nd}\gg\qtp\,,\quad
\frac{\lym^{(\rm br1)}}{\qop}\sim Z_q\Bigl
(\frac{\mo}{\mph}\Bigr )^{\frac{n_2 N_c}{3\nd(N_c-n_1)}}\ll
1\,,\label{(23.1.3)}
\eq
\bbq
\mut\sim\la\Bigl (\frac{\langle N_2\rangle}{\la^2}\Bigr
)^{N_F/3N_c}\sim Z_q^{1/2}\la
\Bigl (\frac{m_Q}{\la}\Bigr )^{N_F/3N_c}\gg\muo\,,
\eeq
\bq
\frac{\mut}{\qop}\sim Z_q^{3/2}\Bigl (\frac{\mo}{\mph}\Bigr
)^{N_F/3\nd}\ll 1\,.\label{(23.1.4)}
\eq
Hence, the largest mass is $\qop$ while the overall phase is
$HQ_1-HQ_2$. We consider below only the case $n_1<\bo/2$, so
that the lower energy theory with $\nd$ colors and
$N^\prime_F=n_2$ flavors at $\mu<\qop$ remains in the conformal
window.

After integrating out the heaviest quarks ${\ov q}^1, q_1$ at
$\mu<\qop$ and ${\ov q}^2, q_2$ quarks at $\mu
<\qtp$ and, finally, all $SU(\nd)$ dual gluons at
$\mu<\lym^{(\rm br1)}$, the Lagrangian of mions looks as
\bq
K=\frac{z_M(\la,\qop)}{Z^2_q\la^2}\,{\rm Tr}\Bigl [\,
M_{11}^\dagger M_{11}+M_{12}^\dagger M_{12}+M_{21}^\dagger
M_{21}+z^{\,\prime}_{M}(\qop,\qtp)
M_{22}^\dagger M_{22} \Bigr ]\,, \label{(23.1.5)}
\eq
\bbq
\cw=-\nd S+\cw_M\,,\quad\quad S=\Bigl (\frac{\det
M}{\la^{\bo}}\Bigr )^{1/\nd}\,,
\quad \lym^{(\rm br1)}=\langle S\rangle^{1/3}\sim\Bigl
(m_Q\langle M_2\rangle\Bigr )^{1/3}\,.
\eeq
\bbq
\cw_M=m_Q{\rm Tr} M-\frac{1}{2\mph}\Bigl [\,{\rm
Tr}(M^2)-\frac{1}{N_c}({\rm Tr} M)^2 \Bigr ]\,,
\quad z_M(\la,\qop)\sim \Bigl (\frac{\la}{\qop}\Bigr
)^{2\,\bd/N_F}\gg 1\,.
\eeq

From \eqref{(23.1.5)}\,: the hybrids $M_{12}$ and $M_{21}$ are
massless, $\mu(M_{12})=\mu(M_{21})=0$, while the pole mass of
$M_{11}$ is (compare with \eqref{(22.1.9)}\,)
\bq
\mu^{\rm
pole}(M_{11})\sim\frac{Z^2_q\la^2}{z_M(\la,\qop)\mph}\,,\quad
\frac{\mu^{\rm pole}(M_{11})}{\lym^
{(\rm br1)}}\sim Z^2_q\Bigl (\frac{\mo}{\mph}\Bigr
)^{\frac{N_c(\bo-2n_1)}{3\nd(N_c-n_1)}}\ll 1\,.\label{(23.1.6)}
\eq

The parametric behavior of $\qtp$ and
$z^{\,\prime}_{M}(\qop,\qtp)$ depends on the value
$\mph\lessgtr{\tilde\mu}_{\Phi,1}$ (see below). We consider
first the case $\mph\gg{\tilde\mu}_{\Phi,1}$ so that, by
definition, the lower energy theory with $\nd$ colors and $n_2$
flavors had enough "time" to evolve and entered already the new
conformal regime at $\qtp<\mu\ll\qop$, with ${\rm\ov
b\,}^\prime_o/\nd=(3\nd-n_2)/\nd=O(1)$ and ${\ov
a\,}^\prime_*=O(1)$. Hence, when the quarks ${\ov q}^2, q_2$
decouple as heavy ones at $\mu<\qtp$, the coupling ${\ov
a}_{YM}$ of the remained $SU(\nd)$ Yang-Mills theory is ${\ov
a}_{YM}\sim {\ov a\,}^\prime_*=O(1)$ and this means that
$\qtp\sim\lym^{(\rm br1)}$. This can be obtained also in a
direct way. The running mass of quarks ${\ov q}^2, q_2$ at
$\mu=\qop$ is, see \eqref{(23.1.1)}-\eqref{(23.1.3)},
\bq
\mu_{q,2}(\mu=\qop)=\frac{\langle M_2\rangle_{\rm br1}}{\langle
M_1\rangle_{\rm br1}}\,\qop\,,\quad \qtp=\frac{\mu_{q,2}
(\mu=\qop)}{z^{\,\prime}_q(\qop,\qtp)}\sim \lym^{(\rm
br1)}\sim\Bigl (m_Q\langle M_2\rangle\Bigr
)^{1/3}\,,\label{(23.1.7)}
\eq
\bbq
z^{\,\prime}_q(\qop,\qtp)=\Bigl (\frac{\qtp}{\qop}\Bigr
)^{\frac{{\rm\ov b\,}^\prime_o}{n_2}}\rho\,,\quad \rho=\Bigl
(\frac{{\ov a\,}_*}{{\ov a\,}^\prime_*}\Bigr
)^{\frac{\nd}{n_2}}\exp\Bigl\{\frac{\nd}{n_2}\Bigl
(\frac{1}{{\ov a\,}_*}-\frac{1}{{\ov a\,}^\prime_*}\Bigr )\Bigr
\}
\sim \exp\Bigl\{\frac{\nd}{n_2}\frac{1}{{\ov a\,}_*}\Bigr\}\gg
1\,.
\eeq
We obtain from \eqref{(23.1.5)} that the main contribution to the
mass of mions $M_{22}$ originates from the non-perturbative term
$\sim S$ in the superpotential and, using
\eqref{(23.1.5)},\eqref{(23.1.7)},
\bq
z^{\,\prime}_{M}(\qop,\qtp)=\frac{{\ov a}_f(\mu=\qop)}{{\ov
a}_f(\mu=\qtp)}\Bigl (\frac{1}{z^{\,\prime}_q
(\qop,\qtp)}\Bigr )^2\sim\Bigl
(\frac{1}{z^{\,\prime}_q(\qop,\qtp)}\Bigr )^2\,,\label{(23.1.8)}
\eq
\bq
\mu^{\rm pole}(M_{22})\sim\frac{Z_q^2\la^2}{z_M(\la,\qop)
z^{\,\prime}_{M}(\qop,\qtp)}\Biggl (\frac{\langle
S\rangle}{\langle M_2\rangle^2}=\frac{\langle
M_1\rangle}{\langle M_2\rangle}\frac{1}{\mph}\Biggr )_{\rm
br1}\,\,\sim \lym^{(\rm br1)}\sim\qtp\,.\label{(23.1.9)}
\eq

We consider now the region $\mo\ll\mph\ll{\tilde\mu}_{\Phi,1},\,
2n_1\lessgtr\bo$ where, by definition, $\qtp$ is too close to
$\qop$, so that the range of scales $\qtp<\mu<\qop$ is too small
and the lower energy theory at $\mu<\qop$ has no enough "time"
to enter a new regime (conformal at $2n_1<\bo$ or strong
coupling one at $2n_1>\bo$) and remains in the weak coupling
logarithmic regime. Then, ignoring logarithmic effects in
renormalization factors, $z^{\,\prime}_q(\qop,\qtp)\sim
z^{\,\prime}_{M}(\qop,\qtp)\sim 1$, and keeping as always only
the exponential dependence on $\nd/\bd$\,:
\bbq
\qtp\sim\frac{\langle M_2\rangle_{\rm br1}}{\langle
M_1\rangle_{\rm br1}}\,\qop\,,\quad\quad \frac{\lym^{(\rm
br1)}}{\qtp}\ll 1\quad\ra\quad
\mo\ll\mph\ll{\tilde\mu}_{\Phi,1}\,,
\eeq
\bq
{\tilde\mu}_{\Phi,1}\sim\exp\Bigl\{\frac{(N_c-n_1)}{2n_1}\frac{1}{{\ov
a\,}_*}\Bigr\}\mo \gg \mo\,.\label{(23.1.10)}
\eq

The pole mass of mions $M_{22}$ looks in this case as
\bq
\frac{\mu^{\rm pole}(M_{22})}{\mu^{\rm
pole}(M_{11})}\sim\frac{\langle M_1\rangle_{\rm br1}}{\langle
M_2\rangle_{\rm br1}}\gg 1,\quad\frac{\mu^{\rm
pole}(M_{22})}{\lym^{(\rm br1)}}\sim Z^2_q\Bigl
(\frac{\mph}{\mo}\Bigr )^{\frac{2n_1 N_c}{3\nd(N_c-n_1)}}\ll
1\,.\label{(23.1.11)}
\eq

On the whole, see \eqref{(23.1.10)}, the mass spectrum at
$\mo\ll\mph\ll{\tilde\mu}_{\Phi,1}$ and $2n_1\lessgtr\bo$ looks
as follows. a) There is a large number of heaviest hadrons made
of weakly coupled (and weakly confined, the tension of the
confining string is $\sqrt{\sigma}\sim\lym^{(\rm br1)}\ll\qop)$
non-relativistic quarks ${\ov q}^1, q_1$, the scale of their
masses is $\qop$, see \eqref{(23.1.3)}.\, b) The next physical
mass scale is due to $\qtp\,:\,\, \lym^{(\rm
br1)}\ll\qtp\ll\qop$. Hence, there is also a large number of
hadrons made of weakly coupled and weakly confined
non-relativistic quarks ${\ov q}^2, q_2$, the scale of their
masses is $\qtp$, see \eqref{(23.1.10)}, and a large number of
heavy hybrid hadrons with the masses $\sim (\qop+\qtp)$. Because
all quarks are weakly coupled and non-relativistic in all three
flavor sectors, $"11",\, "22"$ and $"12+21"$, the mass spectrum
of low-lying flavored mesons is Coulomb-like with parametrically
small mass differences $\Delta\mu_H/\mu_H=O(\bd^2/\nd^2)\ll
1$.\, c) A large number of gluonia made of $SU(\nd)$ gluons,
with the mass scale $\sim\lym^{(\rm br1)}\sim\Bigl (m_Q\langle
M_2\rangle\Bigr )^{1/3}$, see \eqref{(23.1.1)}.\,
d) $n^2_2$ mions $M_{22}$ with the pole masses $\mu^{\rm
pole}(M_{22})\ll\lym^{(\rm br1)}$, see \eqref{(23.1.11)}.\, 
 e) $n^2_1$ mions $M_{11}$ with the pole masses $\mu^{\rm
pole}(M_{11})\ll\mu^{\rm pole}(M_{22})$, see
\eqref{(23.1.6)},\eqref{(23.1.9)},\eqref{(23.1.11)}.\, f) $2n_1n_2$
 hybrids $M_{12}, M_{21}$ are massless, $\mu(M_{12})=\mu(M_{21})=0$.

The pole mass of quarks ${\ov q}^2, q_2$ is smaller at
${\tilde\mu}_{\Phi,1}\ll\mph\ll \la^2/m_Q$ and $2n_1<\bo$, and
stays at $\qtp\sim\lym^{(\rm br1)}$, while the mass of mions
$M_{22}$ is larger and also stays at $\mu(M_{22})\sim\lym^{(\rm
br1)}$.\\

\subsection{\quad $\rm br2$ vacua,\,\,\, $\,\bd/N_F\ll 1$}

\hspace*{4mm} The condensates of mions look in these br2 - vacua
as in \eqref{(23.1.1)} with the exchange $1\leftrightarrow 2$. The
largest mass is $\qtp$,
\bq
\qtp\sim \frac{\la}{Z_q}\Bigl (\frac{m_Q\mph}{\la^2}\Bigr
)^{N_F/3\nd}\gg\qop\,,\quad
\frac{\lym^{(\rm br2)}}{\qtp}\sim Z_q\Bigl
(\frac{\mo}{\mph}\Bigr )^{\frac{n_1 N_c}{3\nd(N_c-n_2)}}\ll
1\,,\label{(23.2.1)}
\eq
and the overall phase is $HQ_1-HQ_2$. After decoupling the
heaviest quarks ${\ov q}^2, q_2$ at $\mu<\qtp$ the lower energy
theory remains in the weak coupling logarithmic regime at, see
\eqref{(23.1.10)},
\bq
\frac{\lym^{(\rm br2)}}{\qop}\ll 1 \quad\ra\quad \mo\ll\mph\ll
{\tilde\mu}_{\Phi,2}\,,\quad
\frac{{\tilde\mu}_{\Phi,2}}{\mo}\sim\exp\Bigl\{\frac{(N_c-n_2)}{2n_2}\frac{1}{{\ov
a\,}_*}\Bigr\}\gg 1\,.\label{(23.2.2)}
\eq
Hence, the mass spectra in this range $\mo<\mph\ll
{\tilde\mu}_{\Phi,2}$ can be obtained from corresponding
formulas by the replacements $n_1\leftrightarrow
n_2$.

But because $n_2\geq N_F/2$, the lower energy theory with
$1<n_1/\nd<3/2$ is in the strong coupling regime at
$\mph\gg{\tilde\mu}_{\Phi,2}$, with ${\ov a}(\mu)\gg 1$ at
$\lym^{(\rm br2)}\ll\mu\ll\qtp$. We do not consider the strong
coupling regime here. \\

\numberwithin{equation}{section}
\numberwithin{equation}{subsection}

\addcontentsline{toc}{section}
{\bf \large Part IIb. \,\,   $\mathbf{2 N_c <N_F < 3 N_c}.$}  

\begin{center}  
\bf \large Part IIb. \,\,   $\mathbf{2 N_c <N_F < 3 N_c}$.  
\end{center}

\section{Introduction  }   

\hspace*{3mm} Considered here in sections 24-29 is the ${\cal N}=1$ SQCD-type 
theory with $SU(N_c)$ colors (and their Seiberg's dual with $SU(N_F-N_c)$
dual \, colors) and $N_F$ flavors of light quarks, and with $N_F^2$ additional
colorless flavored fields $\Phi^j_i$ (fions). But now with $N_F$ in the
different range $2N_c<N_F<3N_c$.  The multiplicities of various vacua,  
quark and gluino  condensates and mass spectra    are found.

\hspace*{3mm} This region $2N_c < N_F < 3N_c$ was considered in \cite{ch13}. 
As shown in \cite{ch13}, the multiplicities of various vacua and mass spectra
therein \,differ even qualitatively from those in the region $3 N_c/2 < N_F < 2
N_c$ \cite{ch19}.
 
Recall that the Lagrangian of the direct $\Phi$-theory at the scale $\mu=\la$
looks as
\footnote{\,
The gluon exponents are always implied in the Kahler terms. Besides, here and
everywhere below in the text we neglect for simplicity all RG-evolution effects
if they are logarithmic only. \label{(f37)}
}
\bbq
K={\rm Tr}\,\Bigl (\Phi^\dagger \Phi\Bigr )+{\rm Tr}\Bigl (\,Q^\dagger Q+(Q\ra
{\ov Q})\,\Bigr )\,,\quad {\cal W}=\frac{2\pi}{\alpha(\mu=\la)} S+{\cal W}_{\rm \ matter}\,,
\quad {\cal W}_{\rm matter}={\cal W}_{\Phi}+{\cal W}_Q\,,
\eeq
\bq
{\cal W}_{\Phi}=\frac{\mph}{2}\Biggl [{\rm Tr}\,(\Phi^2)-\frac{1}{\nd}\Bigl
({\rm Tr}\,\Phi\Bigr )^2\Biggr ],\,\, {\cal W}_Q={\rm Tr}\,({\ov Q} m_Q^{\rm
tot} Q),\,\, m_Q^{\rm tot}= (m_Q-\Phi),\,\, \nd= N_F-N_c.
\quad\label{(24.0.1)}
\eq
Here\,: the gauge group is $SU(N_c)$,\, $\mph$ and $m_Q$ are the mass
parameters, $S=W^{A}_{\beta}W^{A,\,\beta}/32\pi^2$, where $W^A_{\beta}$ 
is the gauge field strength, $A=1...N_c^2-1,\, \beta=1,2$,\, $a(\mu)=N_c\alpha(
\mu)/2\pi=N_c g^2(\mu)/8\pi^2$ is the gauge coupling with its scale factor
$\la$.

Besides, in addition to the direct $\Phi$-theory in \eqref{(24.0.1)}, we
calculate all this also \, in Seiberg's dual variant \cite{S2,IS}. The
Lagrangian of the dual $d\Phi$-theory at the scale $\mu=\la$ looks as
\bbq
K={\rm Tr}\,(\Phi^\dagger \Phi)+ {\rm Tr}\Bigl ( q^\dagger q + (q\ra\ov q)\,
\Bigr )+{\rm Tr}\, \Bigl (\frac{M^{\dagger}M}{\la^2}\Bigr )\,,\quad {\cal W}=\,
\,\frac{2\pi}{\tilde \alpha(\mu=\la)}\, {\wt S}+{\cal W}_{\Phi}+{\cal W}_q\,,
\eeq
\bq
{\cal W}_q={\rm Tr}\, \Bigl ( M (m_Q-\Phi)\Bigr) -\frac{1}{\la}\,\rm {Tr} \Bigl
({\ov q}\,M\, q \Bigr )\,, \label{(24.0.2)}
\eq
Here\,:\, the number of dual colors is ${\ov N}_c=(N_F-N_c)$ and $M^i_j\ra 
({\ov Q}_j Q^i)$ are the $N_F^2$ elementary mion fields, ${\tilde
a}(\mu)=\nd{\tilde
\alpha}(\mu)/2\pi=\nd{\tilde g}^2(\mu)/8\pi^2$ is the dual gauge coupling (with
its scale parameter $\Lambda_q= - \la$),\,\,${\wt S}= {\wt W}^{B}_{\beta}\,
{\wt
W}^{B,\,\beta}/32\pi^2$,\,\, ${\wt W}^B_{\beta}$ is the dual gluon field
strength, $B=1...\nd^2-1$. The gluino condensates of the direct and dual
theories are matched as well as $\langle M^i_j\rangle$ and $\langle{\ov Q}_j
Q^i\rangle\equiv\sum_{a=1}^{N_c}\langle{\ov Q}^{\,a}_j Q^i_{a} \rangle$,
\bq
\langle{-\,\wt S}\rangle=\langle S\rangle,\,\, \langle M^i_j\rangle
\equiv\langle M^i_j(\mu=\la)\rangle=\delta^i_j \langle M_i\rangle=
\langle{\ov Q}_j Q^i\rangle\equiv\langle{\ov Q}_j Q^i (\mu=\la)\rangle
=\delta^i_j\langle(\QQ)_i \rangle. \,\,\,\label{(24.0.3)}
\eq
In sections 25-29 the hierarchies of parameters entering
\eqref{(24.0.1)},\eqref{(24.0.2)} are $m_Q\ll\la\ll\mph$,\, 
$\mph$ is varied while $m_Q$ and $\la$ stay intact.

In those cases when the fields $\Phi$ are too heavy and 
dynamically irrelevant, they can be integrated out and the 
matter Lagrangians of the direct and dual theories take the form
\bq
K={\rm Tr}\Bigl (\,Q^\dagger Q+(Q\ra {\ov Q})\,\Bigr ),\,\, W_Q=m_Q{\rm
Tr}({\ov Q} Q)-\frac{1}{2\mph}\Biggl (\sum_{i,j}\,({\ov Q}_j Q^i)({\ov Q}_i
Q^j)-\frac{1}{N_c}\Bigl({\rm Tr}\,{\ov Q} Q \Bigr)^2 \Biggr ), \label{(24.0.4)}
\eq
\bq
K={\rm Tr}\Bigl ( q^\dagger q + (q\ra\ov q)\,\Bigr )+{\rm Tr}\,
\frac{M^{\dagger}M}{\la^2}\,,\label{(24.0.5)}
\eq
\bbq
{\cal W}_{\rm matter}={\cal W}_M-\frac{1}{\la}\,\rm {Tr} \Bigl ({\ov q}\,M\, q
\Bigr )\,,\quad {\cal W}_M=m_Q \rm {Tr}\,M-\frac{1}{2\mph}\Biggl [{\rm Tr}\,
(M^2)- \frac{1}{N_c}({\rm Tr}\, M)^2 \Biggr ].
\eeq

It was shown in \cite{ch13} that in all considered in  sections 25-29 vacua of
both the direct and dual theories quarks are either in the weak coupling or
conformal regimes with the coupling constants $\lesssim 1$. The dynamics of 
these regimes is sufficiently simple and well understood, so that there was 
{\it no real need to use for calculations of mass spectra the dynamical
scenario,
introduced  in} \cite{ch3}.~
~\footnote{\,
In those cases when the coupling is small and in the case of conformal regime
this scenario is literally standard. The use of this scenario is really needed
only for calculations of mass spectra in the regions of
parametrically large coupling $a\gg 1$. Besides, it is worth noting that the
appearance of additional parametrically light composite solitons will influence
the 't Hooft triangles. \label{(f38)}
}

As shown in sections 25-29, many properties of both direct 
$\Phi$-theories and dual $d\Phi$-theories at $2N_c<N_F<3N_c$, 
i.e. multiplicities of vacua and mass spectra differ from those 
calculated in \cite{ch19} for $3N_c/2<N_F<2N_c$.

\section {Quark condensates and multiplicities of vacua  at $\mathbf {N_F>2N_c}$}

\hspace*{5mm} For the reader convenience, we reproduce here first some useful
formulas for the theories \eqref{(24.0.1)},\eqref{(24.0.2)}.

The Konishi anomalies \cite{Konishi} for the $i$-th flavor look in the direct
$\Phi$-theory  as
\bq
\langle\Phi^j_{i}\rangle\langle\frac{\partial W_{\Phi}}{\partial
\Phi^j_{i}}\rangle=0\,,\quad
\langle m_{Q,i}^{\rm tot}\rangle\langle \QQ_i\rangle=\langle S\rangle\,, \quad
\langle\Phi^j_{i}\rangle=\delta^j_i \langle\Phi_{i}\rangle, \quad \langle
m_{Q,\,i}^{\rm tot}\rangle=m_Q-\langle\Phi_{i}\rangle\,, \label{(25.0.1)}
\eq
\bbq
\langle\Phi_j^i\rangle=\frac{1}{\mph}\Biggl ( \langle{\ov Q}_j Q^i
\rangle-\delta^i_j\frac{1}{N_c}{\rm Tr}\,\langle\QQ\rangle\Biggr ),\,\,
\langle{\ov Q}_j Q^i \rangle\equiv\sum_{a=1}^{N_c}\langle{\ov Q}^{\,a}_j
Q^i_{a}\rangle =\delta^i_j\langle \QQ_i\rangle,\,\, {\it i,j}=1\, ...\,
N_F\,,
\eeq
and $\langle m_{Q,i}^{\rm tot}\rangle$ is the value of the quark running mass
at \, the scale $\mu=\la$.

The values of the quark condensates for the $i$-th flavor, $\langle{\ov Q}_i
Q^i\rangle$, in various vacua can be obtained from the standard effective
superpotential $W^{\rm eff}_{\rm tot}(\Pi)$ depending only on quark bilinears
$\Pi^i_j=({\ov Q}_j Q^i)$. ( It is worth recalling that this is {\it not} a
genuine low energy superpotential, \eqref{(25.0.3)} can be used {\it only} for
finding the values of mean vacuum values $\langle{\ov Q}_j Q^i\rangle$. The
genuine low energy superpotentials in each vacuum are given below in the text
).\bq
W^{\rm eff}_{\rm tot}(\Pi)=W_Q - \nd S, \,\, S=\Bigl (\frac{\det{\ov Q}
Q}{\la^{\bo}}\Bigr )^{1/\nd} \equiv \lym^3, \,\, \bo=3N_c-N_F\,,\quad
\nd=N_F-N_c\,. \label{(25.0.2)}
\eq
Because $\langle{\ov Q}_j Q^i\rangle$ is holomorphic in $\mph$, the values of
$\langle{\ov Q}_j Q^i \rangle$ obtained in a standard way from \eqref{(25.0.2)}
can be  used both at large and small values of $\mph$.

For the vacua with the spontaneously broken flavor symmetry, $U(N_F)\ra
U(n_1)\times U(n_2),\,\,1\leq n_1\leq N_F/2,\,\, n_2\geq N_F/2$, the most
convenient way to find the quark condensates is to use the relations following
from \eqref{(25.0.2)}:
\bbq
\langle \Qo+\Qt-\frac{1}{N_c}{\rm Tr}\, ({\ov Q} Q)\,\rangle_{\rm
br}=m_Q\mph,\quad
\langle S\rangle_{\rm br}=\Bigl (\frac{\det \langle{\ov Q} Q\rangle_{\rm
br}}{\la^{\bo}}\Bigr )^{1/\nd}=\frac{\langle\Qo \rangle_{\rm
br}\langle\Qt\rangle_{\rm br}}{\mph},
\eeq
\bq
\det \langle{\ov Q} Q\rangle_{\rm br}=\langle\Qo\rangle^{{\rm n}_1}_{\rm
br}\,\langle\Qt\rangle^{{\rm n}_2}_{\rm br}\,,\quad
\Qo\equiv\sum_{a=1}^{N_c}{\ov Q}^{\,a}_1 Q^1_{a}\,,\quad
\Qt\equiv\sum_{a=1}^{N_c}{\ov Q}^{\,a}_2 Q^2_{a}\,,\label{(25.0.3)}
\eq
\bbq
\langle m^{\rm tot}_{Q,1}\rangle_{\rm br}=m_Q-\langle\Phi_1\rangle_{\rm
br}=\frac{\langle\Qt\rangle_{\rm br}}{\mph},\quad \langle m^{\rm
tot}_{Q,2}\rangle_{\rm br}=m_Q-\langle\Phi_2\rangle_{\rm
br}=\frac{\langle\Qo\rangle_{\rm br}}{\mph}\,.
\eeq

The Konishi anomalies for the $i$-th flavor look in the dual $d\Phi$-theory
\eqref{(24.0.2)} as
\bq
\langle M_i\rangle \langle {\ov q}^{\,i} q_i\rangle=\la\langle S\rangle\,,\quad
\frac{\langle{\ov q}^{\,i} q_i\rangle}{\la}=\langle m_{Q,i}^{\rm
tot}\rangle=m_Q-\frac{1}{\mph}\,\langle M_i-\frac{1}{N_c}{\rm Tr}\,M
\rangle\,,\quad {\it i}=1\, ...\, N_F\,.\label{(25.0.4)}
\eq

In vacua with the broken flavor symmetry from  \eqref{(24.0.5)} 
\ (remind that $\langle M^i_j\rangle=\delta^i_j\langle \, M_i\rangle,\,\,
\langle M_1\rangle=\langle\Qo\rangle,\, \langle  M_2\rangle=\langle\Qt\rangle$)
\bbq
\langle M_1+M_2-\frac{1}{N_c}{\rm Tr}\, M\rangle_{\rm br}=m_Q\mph,\quad\langle
S\rangle_{\rm br}
=\Bigl (\frac{\det \langle M\rangle_{\rm br}}{\la^{\bo}}\Bigr)^{1/\nd}=
\frac{1}{\mph}\langle M_1\rangle_{\rm br}\langle M_2\rangle_{\rm br}\,,
\,\,
\eeq
\bq
\frac{\langle\qo\rangle_{\rm br}}{\la}=\frac{\langle S\rangle_{\rm br}}{\langle
M_{1}\rangle_{\rm br}}=\frac{\langle M_{2}\rangle_{\rm br}}{\mph}=\langle 
m^{\rm  tot}_{Q,1}\rangle_{\rm br},\quad  \frac{\langle \qt\rangle_{\rm br}}
{\la}=\frac{\langle S\rangle_{\rm br}}{\langle  M_{2}\rangle_{\rm br}}=
\frac{\langle M_{1}\rangle_{\rm br}}{\mph}=\langle m^{\rm  tot}_{Q,2}\rangle
_{\rm br}\,,\label{(25.0.5)}
\eq
\bbq
\frac{\langle\qt\rangle_{\rm br}}{\langle\qo\rangle_{\rm
br}}=\frac{\langle\Qo\rangle_{\rm br}}{\langle\Qt \rangle_{\rm br}}\,,\quad
\qo\equiv\sum_{b=1}^{\nd}{\ov q}_{b}^{\,1} q^{b}_1\,,\quad
\qt\equiv\sum_{b=1}^{\nd} {\ov q}_{b}^{\,2} q^b_2\,,\quad \nd\equiv N_F-N_c\,.
\eeq

\subsection{Vacua with the unbroken flavor symmetry}

One obtains from \eqref{(25.0.2)} at $m_Q\ll\la,\,\,\mph\lessgtr \mo$ and with
$\langle{\ov Q}_j Q^i\rangle=\delta^i_j
\langle{\ov Q} Q\rangle\,,\,\, \langle{\ov Q} Q\rangle=\sum_{a=1}^{N_c}{\ov
Q}^{\,a}_1 Q^1_{a}\,$. -

{\bf a)} There are only $\nd=(N_F-N_c)$ nearly classical S - vacua at
$\mph\ll\mo$ with
\footnote{\,
Here and everywhere below\,: $A\approx B$ has to be understood as an equality
neglecting smaller power corrections, and $A\ll B$ has to be understood as
$|A|\ll |B|$. \label{(f39)}
}
\bq
\langle\QQ\rangle_S\equiv\langle\QQ(\mu=\la)\rangle_S\approx -\frac{N_c}{\nd}\,
m_Q\mph\,,\quad \mo=\la\Bigl (\frac{m_Q}{\la}\Bigr
)^{\frac{N_F-2N_c}{N_c}}\ll\la\,,\label{(25.1.1)}
\eq
\bbq
\langle\lym^{(S)}\rangle^3\equiv\langle S\rangle_S=\Bigl
(\frac{\det\langle\QQ\rangle_S}{\la^{\rm \bo}}\Bigr )^{1/\nd}\sim\la^3\Bigl
(\frac{m\mph}{\la^2}\Bigr )^{N_F/\nd}\,.
\eeq

{\bf b)} There are $(N_F-2N_c)$ quantum L - vacua at $\mph\gg\mo$ with
\bq
\langle\QQ\rangle_L\equiv\langle\QQ(\mu=\la)\rangle_L\sim \la^2\Biggl
(\frac{\mph}{\la}\Biggr )^{\frac{\nd}{N_F-2N_c}}\,,\label{(25.1.2)}
\eq
\bbq
\langle\lym^{(L)}\rangle^3\equiv\langle S\rangle_L=\Bigl
(\frac{\det\langle\QQ\rangle_L}{\la^{\rm \bo}}\Bigr )^{1/\nd}\sim\la^3\Bigl
(\frac{\mph}{\la}\Bigr )^{\frac{N_F}{N_F-2N_c}}\,.
\eeq

{\bf c)} There are $N_c$ quantum QCD  vacua at $\mph\gg\mo$ with
\bq
\langle\QQ\rangle_{QCD}=\langle\QQ(\mu=\la)\rangle_{QCD}\approx \,
\frac{\langle S\rangle_{\rm QCD}\equiv \langle
\lym^{(\rm QCD)}\rangle^3}{m_Q}\approx\frac{1}{m_Q}\Bigl
(\la^{\bo}m_Q^{N_F}\Bigr)^{1/N_c}\,.\label{(25.1..3)}
\eq

The total number of vacua with the unbroken flavor symmetry is
\bq
N^{\rm tot}_{\rm unbr}=(N_F-2N_c)+N_c=\nd\,.\label{(25.1.4)}
\eq

\subsection{Vacua with the broken flavor symmetry, $U(N_F)\ra U(n_1)
\times U(n_2)$}

One obtains from \eqref{(25.0.3)},\eqref{(25.0.4)} for such vacua. -

{\bf a)} There are $(n_1-N_c){\ov C}^{\,\rm n_1}_{N_F}$ br1 -vacua
\footnote {\,
${\ov C}^{\, n_1}_{N_F}$ differ from the standard $C^{\,
n_1}_{N_F}=N_F!/[n_1!\,n_2!]$ only for ${\ov C}^{\,n_1={\rm k}}_{N_F=2{\rm
k}}=C^{\,n_1={\rm k}}_{N_F=2{\rm k}}/2$.

Besides, by convention, we ignore the continuous multiplicity of vacua due to
the spontaneous flavor symmetry breaking. Another way, one can separate
slightlyall quark masses, so that all Nambu-Goldstone particles will acquire
small
masses $O(\delta m_Q)\ll {\ov m}_Q$.
}
at $N_c<n_1\leq [N_F/2]$ and $\mph\ll\mo$ with
\bq
\langle\Qo\rangle_{\rm br1}\approx \frac{N_c}{N_c-n_1} m_Q\mph,\quad \langle\Qt
\rangle_{\rm br1}\sim \la^2\Bigl (\frac{\mph}{\la}\Bigr
)^{\frac{n_1}{n_1-N_c}}\Bigl (\frac{\la}{m_Q}\Bigr
)^{\frac{n_2-N_c}{n_1-N_c}}\,,\label{(25.2.1)}
\eq
\bbq
\langle S\rangle_{\rm br1}=\frac{\langle\Qo\rangle_{\rm
br1}\langle\Qt\rangle_{\rm br1}}{\mph}\sim
\la^3\Bigl (\frac{\mph}{\la}\Bigr )^{\frac{n_1}{n_1-N_c}}\Bigl
(\frac{\la}{m_Q}\Bigr
)^{\frac{n_2-n_1}{n_1-N_c}}\,,\quad\frac{\langle\Qt\rangle_{\rm
br1}}{\langle\Qo\rangle
_{\rm br1}}\sim \Bigl (\frac{\mph}{\mo}\Bigr )^{\frac{N_c}{n_1-N_c}}\ll 1\,.
\eeq

{\bf b)} There are at $\mph\ll\mo$ $\,(n_2-N_c){\ov C}^{\,\rm n_1}_{N_F}$
br2-vacua at all values $1\leq n_1\leq [N_F/2]$ with
\bq
\langle\Qt\rangle_{\rm br2}\approx \frac{N_c}{N_c-n_2} m_Q\mph\,,\quad
\langle\Qo\rangle_{\rm br2}\sim \la^2\Bigl (\frac{\mph}{\la}\Bigr
)^{\frac{n_2}{n_2-N_c}}\Bigl (\frac{m_Q}{\la}\Bigr
)^{\frac{N_c-n_1}{n_2-N_c}}\,,\label{(25.2.2)}
\eq
\bbq
\langle\lym^{(\rm br2)}\rangle^3\equiv\langle S\rangle_{\rm br2}\sim\la^3\Bigl
(\frac{\mph}{\la}\Bigr )^{\frac{n_2}{n_2-N_c}}\Bigl (\frac{m_Q}{\la}\Bigr
)^{\frac{n_2-n_1}{n_2-N_c}}\,,\quad
\frac{\langle\Qo\rangle_{\rm br2}}{\langle\Qt\rangle_{\rm br2}}\sim \Bigl
(\frac{\mph}{\mo}\Bigr )^{\frac{N_c}{n_2-N_c}}\ll 1\,.
\eeq

On the whole, there are ($\,\theta(z)$ is the step function)
\bq
N^{\rm tot}_{\rm br}(n_1)=\Bigl [ (n_2-N_c)+\theta(n_1-N_c)(n_1-N_c)\Bigr ]{\ov
C}^{\,\rm n_1}_{N_F},\,\,
N^{\rm tot}_{\rm br}=\sum_{n_1=1}^{[N_F/2]}N^{\rm tot}_{\rm br}(n_1)\,,\,\,
n_1+n_2=N_F, \label{(25.2.3)}
\eq
vacua with the broken flavor symmetry at $\mph\ll\mo$.

\vspace{1mm}

{\bf c)} There are $(N_c-n_1)C^{\rm\,n_1}_{N_F}$ br1 -vacua at $1\le n_1<N_c$
and $\mph\gg\mo$ with
\bq
\langle\Qo\rangle_{\rm br1}\approx \frac{N_c}{N_c-n_1} m_Q\mph,\quad
\langle\Qt\rangle_{\rm br1}\sim \la^2\Bigl (\frac{\la}{\mph}\Bigr
)^{\frac{n_1}{N_c-n_1}}\Bigl (\frac{m_Q}{\la}\Bigr
)^{\frac{n_2-N_c}{N_c-n_1}}\,,\label{(25.2.4)}
\eq
\bbq
\langle\lym^{(\rm br1)}\rangle^3\equiv\langle S\rangle_{\rm br1}\sim\la^3\Bigl
(\frac{\la}{\mph}\Bigr )^{\frac{n_1}{N_c-n_1}}\Bigl (\frac{m_Q}{\la}\Bigr
)^{\frac{n_2-n_1}{N_c-n_1}}\,,\quad
\quad \frac{\langle\Qt\rangle_{\rm br1}}{\langle\Qo\rangle_{\rm br1}}\sim \Bigl
(\frac{\mo}{\mph}\Bigr )^{\frac{N_c}{N_c-n_1}}\ll 1\,.
\eeq

{\bf d)} There are $(N_F-2N_c){\ov C}^{\rm\,n_1}_{N_F}$ Lt (i.e. L - type)
vacuaat $n_1\neq N_c$ and $\mph\gg\mo$ with
\bq
(1-\frac{n_1}{N_c})\langle\Qo\rangle_{\rm Lt}\approx
-(1-\frac{n_2}{N_c})\langle\Qt\rangle_{\rm Lt}\sim \la^2\Biggl
(\frac{\mph}{\la}\Biggr )^{\frac{\nd}{N_F-2N_c}},\label{(25.2.5)}
\eq
i.e. as in the L - vacua in \eqref{(25.1.2)} but $\langle\Qo\rangle_{\rm
Lt}\neq\langle\Qt\rangle_{\rm Lt}$ here.

{\bf e)} There are $(N_F-2N_c) C^{\rm\,n_1}_{N_F}$ special vacua at $n_1=N_c$
and $\mph\gg\mo$ with
\bq
\langle\Qt\rangle_{\rm spec}=\frac{N_c}{2N_c-N_F} m_Q\mph\,,\quad
\langle\Qo\rangle_{\rm spec}= \la^2
\Bigl (\frac{\mph}{\la}\Bigr )^{\frac{\nd}{N_F-2N_c}}\,,\label{(25.2.6)}
\eq
\bbq
\langle\lym^{(\rm spec)}\rangle^3\equiv\langle S\rangle_{\rm spec}= m\la^2\Bigl
(\frac{\mph}{\la}\Bigr )^{\frac{\nd}{N_F-2N_c}}\,,\quad
\frac{\langle\Qt\rangle_{\rm spec}}{\langle\Qo\rangle_{\rm spec}}= \Bigl
(\frac{\mo}{\mph}\Bigr )^{\frac{N_c}{N_F-2N_c}}\ll 1\,.
\eeq

As one can see from the above, similarly to \cite{ch19} with $N_c<N_F<2N_c$,
all \, quark condensates become parametrically the same at $\mph\sim\mo$, see
\eqref{(25.1.1)}. Clearly, just this region $\mph\sim\mo$ and not $\mph\sim\la$
\, is  very special and most of the quark condensates change their parametric 
behavior and hierarchies at $\mph\lessgtr\mo$. For example, the br1 - vacua with
\bbq
1\le n_1<N_c\,,\quad \langle\Qo \rangle\sim m_Q\mph\gg\langle\Qt\rangle
\quad{\rm at}\quad\mph\gg\mo
\eeq
evolve into br2 - vacua with
\bbq
\langle\Qt\rangle\sim m_Q\mph\gg\langle\Qo\rangle \quad{\rm at}
\quad\mph\ll\mo,\eeq
while the br1 - vacua with
\bbq
n_1>N_c\,,\quad\langle\Qo\rangle\sim m_Q\mph\gg\langle\Qt\rangle\quad {\rm
at}\quad \mph\ll\mo
\eeq
evolve into the L-type  vacua with
\bbq
\langle\Qo\rangle\sim\langle\Qt\rangle\sim\la^2\,
(\mph/\la)^{\nd/(N_F-2N_c)}\quad {\rm at}\quad \mph\gg\mo\,,
\eeq
etc. The exceptions are the special vacua with $n_1=N_c\,,\, n_2=\nd$\,. In
these, the parametric behavior
\bbq
\langle\Qt\rangle\sim m_Q\mph\,, \quad\langle\Qo\rangle\sim
\la^2\,(\mph/\la)^{\nd/(N_F-2N_c)}
\eeq
remains the same but the hierarchy is reversed at $\mph\lessgtr\mo$\, :
\bbq
\langle\Qo\rangle/\langle\Qt\rangle\sim (\mph/\mo)^{N_c/(N_F-2N_c)}\,.
\eeq

On the whole, there are
\bq
N^{\rm tot}_{\rm br}(n_1)=\Bigl [ (N_F-2N_c)+\theta (N_c-n_1)(N_c-n_1)\Bigr ]
{\ov  C}^{\,\rm n_1}_{N_F},\quad N^{\rm tot}_{\rm br}=\sum_{n_1=1}^{[N_F/2]}
N^{\rm tot}_{\rm br}(n_1)  \label{(25.2.7)}
\eq
vacua with the broken flavor symmetry at $\mph\gg\mo$. The total number of
 vacua \, is the same at $\mph\lessgtr\mo$\,, as it should be.

We point out finally that the multiplicities of vacua at $N_F>2N_c$ are {\it not\, 
the analytic  continuations of those at} $N_c<N_F<2N_c$\,, see e.g.
\cite{ch19}.

\section{Direct theory. Unbroken flavor symmetry.\\ 
$\mathbf{\hspace*{1cm}  2N_c<N_F<3N_c}$}

\hspace*{5mm} It is worth noting first that in all calculations below in the
text we use the NSVZ $\beta$-functions \cite{NSVZ-1} ( remind also
the \, footnote \ref{(f37)}).

Because $\mo=\la(m_Q/\la)^{(N_F-2N_c)/N_c}\ll\la$ at $m_Q\ll\la$ while
$\mph\gg\la$, we are here in the region $\mph\gg\mo$. 

There are  $(N_F-2N_c)$ L - vacua  \eqref{(25.1.2)}  with the quark condensates
\bq
\frac{\langle\QQ\rangle_L}{\la^2}\sim\Biggl (\frac{\mph}{\la}\Biggr
)^{\frac{\nd}{N_F-2N_c}}\gg 1\,. \label{(26.0.1)}
\eq
So, the gluon masses due to possible higgsing of some quark flavors are large
(neglecting here and below in this section all logarithmic factors due to the
RG-evolution), 
\bq
\mu^{\rm pole}_{\rm gl, L}\sim \langle\QQ\rangle_L^{1/2}\gg\la.
\label{(26.0.2)}
\eq

But the quark masses are even larger,
\bq
m_{Q,\,L}^{\rm pole}\sim\mtq_L\sim\langle\Phi\rangle_L\sim\frac{\langle\QQ\rangle_L}
{\mph}\sim \la\Bigl (\frac{\mph}{\la} \Bigr )^{\frac{N_c}{N_F-2N_c}}\gg\la,\,\,
\frac{\mu^{\rm pole}_{\rm gl,\,L}}{m_{Q,\,L}^{\rm\, pole}}\sim \Bigl (\frac{\la}{\mph}
\Bigr )^{\frac{3N_c-N_F}{2(N_F-2N_c)}}\ll 1. \quad\,\,\label{(26.0..3)}
\eq
Therefore, all quarks are in the HQ (heavy quark) phase and are not higgsed but
confined. This is self-consistent in these vacua with the unbroken global
$SU(N_F>N_c)$ symmetry, as otherwise higgsed quarks $\langle Q^i_a\rangle\neq 0$
will break spontaneously $SU(N_F)$ due to the rank restriction.

After integrating out all quarks as heavy ones at scales $\mu<m^{\rm
pole}_Q,\,\,m^{\rm pole}_Q\gg\la$ {\it in the weak coupling regime}, there
remain $N_F^2$ fions $\Phi$ and the $SU(N_c)$ SYM with the scale factor of its
gauge coupling
\bq
\langle\lym^{(L)}\rangle=\Bigl (\la^{\bo}\det \langle m^{\rm tot}_Q\rangle \Bigr )
^{1/3N_c}\,,\quad\frac{\langle\lym^{(L)}\rangle}{\la}\sim\Bigl (\frac{\mph}{\la} 
\Bigr )^{\frac{N_F}{3(N_F-2N_c)}} \ll \frac{m_{Q,\,L}^{\rm pole}}{\la}. \label{(26.0.4)}
\eq

After integrating out all gluons at the scale $\mu<\langle\lym^{(L)}\rangle$
through the Veneziano-Yankielowics (VY) procedure \cite{VY}, the Lagrangian
of $N_F^2$ fions looks as, see  \eqref{(24.0.1)} for ${\cal W}_{\Phi}$, 
for ${\cal W}_{\Phi}$,
\bq
K_{\Phi}\sim{\rm Tr}\,(\Phi^\dagger\Phi),\,\, {\cal W}= N_c\Bigl (\la^{\bo}\det
m^{\rm tot}_{Q,\,L} \Bigr )^{1/N_c}+{\cal W}_{\Phi}\,,\quad m^{\rm
tot}_Q=m_Q-\Phi,\,\, \langle m^{\rm tot}_Q\rangle_L\approx
-\langle\Phi\rangle_L\,.\,\label{(26.0.5)} 
\eq
From \eqref{(26.0.5)} the fion masses are $\mu^{\rm
pole}_{1}(\Phi)\sim\mph\gg\la$.

On the whole, the mass spectrum looks in these $(N_F-2N_c)$ L - vacua as
follows:\\
a) there is a large number of hadrons made of weakly coupled and weakly confined
non-relativistic quarks with masses $m^{\rm pole}_{Q,\,L}\sim\la(\mph/\la)^{N_c/(N_F-2N_c)}
\gg\la$\, (the  tension of the confining string originating from the $SU(N_c)$ SYM is
$\sqrt\sigma\sim\langle\lym^{(L)}\rangle\ll m^{\rm pole}_{Q,\,L})$\,;\\
b) a large number of gluonia with the mass scale $\sim
\langle\lym^{(L)}\rangle\sim\la (\mph/\la)^{N_F/3(N_F-2N_c)}\ll m^{\rm
pole}_{Q,\,L}$\,;\\
c) the lightest are $N^2_F$ fions with the masses $\mu^{\rm
pole}(\Phi)\sim\mph\ll\langle\lym^{(L)}\rangle$.

The overall hierarchies look as
\bq
\la\ll\mu_{1}^{\rm pole}(\Phi)\sim\mph\ll\langle\lym^{(L)}\rangle
\ll m^{\rm\, pole}_{Q,\,L}\sim  \la\Bigl (\frac{\mph}{\la} \Bigr )^{\frac{N_c}{N_F-2N_c}}.  
\label{(26.0.6)}
\eq

\section{Dual theory. Unbroken flavor symmetry.\\ $\hspace*{1 cm}
\mathbf{2N_c<N_F<3N_c\,,\,\mph\gg\la}$}

\hspace*{5mm} The dual theory in the UV region $\mu>\la$ and at $2N_c<N_F<3N_c$
is taken as UV free. The largest mass in the case considered is that of dual
quarks (neglecting here and below in this section all logarithmic factors due
to ], the RG-evolution)
\bq
\mu^{\rm pole}_{q,\, L}\sim \frac{\langle
M\rangle_L=\langle\QQ\rangle_L}{\la}\sim\la\Biggl (\frac{\mph}{\la}\Biggr
)^{\frac{\nd}{N_F-2N_c}}\gg\la,\label{(27.0.1)}
\eq
while the gluon masses due to possible higgsing of dual quarks are smaller
\bbq
{\ov\mu}^{\,2}_{\rm {gl,\, L}}\sim \langle{\ov q}q \rangle_L=\mtq\la\sim
\langle\Phi\rangle\la\sim\frac{\la\langle\QQ\rangle_L}{\mph}\sim\la^2\Biggl
(\frac{\mph}{\la}\Biggr )^{\frac{N_c}{N_F-2N_c}}\,,
\eeq
\bq
\frac{{\ov\mu}_{\rm gl}}{\mu^{\rm pole}_{q,\, L}}\sim \Bigl (\frac{\la}{\mph}\Biggr
)^{\frac{2N_F-3N_c}{2(N_F-2N_c)}}\ll 1\,.\label{(27.0.2)}
\eq

Therefore, the dual quarks are in the Hq (heavy quark) phase and are not
higgsed \, but confined. This is self-consistent, because otherwise the global
flavor symmetry $SU(N_F>\nd)$ will be broken spontaneously due to the rank
restriction.  After they are integrated out at scales $\mu<\mu^{\rm pole}_q$ {\it
in the weak
coupling regime}, there remain $N_F^2$ fions $\Phi$, $N_F^2$ mions $M$ 
and the  $SU(\nd)$ SYM with the scale factor of its gauge coupling
\bq
\lym^{(L)}=\Bigl (\la^{\bd}\det\frac{M}{\la} \Bigr )^{1/3\nd},\,\,
\frac{\langle\lym^{(L)}\rangle}{\la} \sim\Bigl (\frac{\mph}{\la} \Bigr
)^{\frac{N_F}{3(N_F-2N_c)}}\gg \frac{\mph}{\la}\gg 1,\,\,
\bd=3\nd-N_F>0\,.\label{(27.0.3)}
\eq

After integrating out all dual gluons at scales $\mu<\langle\lym^{(L)}\rangle$
through the Veneziano-Yankielowics (VY) procedure \cite{VY}, the Lagrangian
looks as, see \eqref{(24.0.1)} for ${\cal W}_{\Phi}$,
\bq
K\sim {\rm Tr}\,\Bigl (\frac{M^\dagger M}{\la^2}+ \Phi^\dagger\Phi\,\Bigr
)\,,\quad{\cal W}=-\nd
\Bigl (\la^{\bd}\det\frac{M}{\la}\Bigr )^{1/\nd}+{\rm Tr}\,(m_Q-\Phi)M+{\cal
W}_{\Phi}\,.\label{(27.0.4)}
\eq

All fions $\Phi$ have masses $\mu^{\rm pole}(\Phi)\sim \mph$ and can be
integrated out at scales $\mu<\mph$, resulting in the lower energy Lagrangian
of \, mions
\bq
K_M\sim {\rm Tr}\,\frac{M^\dagger M}{\la^2}\,,\,\,\quad {\cal W}= -\nd\Bigl
(\frac{\det M}{\la^{\bo}}\Bigr )^{1/\nd}+{\cal W}_{M}\,,\label{(27.0.5)}
\eq
\bbq
{\cal W}_M=m_Q{\rm Tr}\,M -\frac{1}{2\mph}\Biggl [{\rm Tr}\, (M^2)-
\frac{1}{N_c}({\rm Tr}\, M)^2 \Biggr ].
\eeq
From \eqref{(27.0.5)} the mion masses are
\bq
\mu^{\rm pole}(M)\sim\frac{\la^2}{\mph}\ll\la \,.\label{(27.0.6)}
\eq

On the whole, the mass spectrum looks in these $(N_F-2N_c)$ dual L - vacua as
follows:\\
a) there is a large number of hadrons made of the weakly coupled and weakly
confined non-relativistic dual quarks with masses $\mu^{\rm
pole}_q\sim\la(\mph/\la)^{\nd/(N_F-2N_c)}$ (the tension of the confining string
originating from the $SU(\nd)$ dual SYM is $\sqrt\sigma\sim
\langle\lym^{(L)}\rangle\ll \mu^{\rm pole}_q)$\,;\\
b) a large number of gluonia with the mass scale $\sim
\langle\lym^{(L)}\rangle\sim\la (\mph/\la)
^{N_F/3(N_F-2N_c)}$\,;\\
c) there are $N_F^2$ fions with masses $\mu^{\rm pole}(\Phi)\sim \mph$\,;\\
d) the lightest are $N^2_F$ mions with masses $\mu^{\rm pole}(M)\sim
(\la^2/\mph)\ll\la$.

The overall hierarchies look here as:
\bbq
\mu^{\rm pole}(M)\ll\la\ll\mu^{\rm
pole}(\Phi)\sim\mph\ll\langle\lym^{(L)}\rangle\ll
\mu^{\rm pole}_{q,\, L}\sim \la\Biggl (\frac{\mph}{\la}\Biggr
)^{\frac{\nd}{N_F-2N_c}}\,.
\eeq

Comparing the mass spectra in these L - vacua of the direct theory in section
26 \, and the dual one in this section it is seen that they are parametrically
different.

We would like to emphasize also that quarks in both direct and dual theories
are \, simultaneously weakly coupled in these L - vacua, so that their dynamics is
clear and {\it we need not any additional dynamical assumptions to calculate
the \, mass spectra}.

\section{Direct theory. Broken flavor symmetry. \,\,\\ $\hspace*{1cm}
\mathbf{2N_c<N_F<3N_c\,,\,\mph\gg\la}$}

\subsection{Lt - vacua}

\hspace*{5mm} The only qualitative difference with the L - vacua in section 26
is  that the flavor symmetry is broken spontaneously in these Lt - vacua,
$\langle\Qo\rangle\neq\langle\Qt\rangle$, see \eqref{(25.2.5)}, and
fions $\Phi_1^2$ and $\Phi_2^1$ are the Nambu-Goldstone
particles here and are exactly massless.

\subsection{br1 - vacua}

\hspace*{5mm} In these vacua with $n_1<N_c$ and $\mo\ll\la\ll\mph\ll
\la^2/m_Q$, \, the regime is conformal at scales $m^{\rm pole}_{Q,2}\ll\mu\ll\la$
(see below)
and potentially most important masses look here as follows
(\,$z_Q(\la,\mu<\la)\ll 1$ is the quark renormalization factor in the conformal
regime).

The gluon masses due to possible higgsing of quarks are, see \eqref{(25.2.4)}
\bq
\hspace*{-3mm}\Bigl (\mu^{\rm pole}_{\rm gl,1}\Bigr )^2\sim z_Q(\la,\mu^{\rm
pole}_{\rm gl,1})\langle\Qo\rangle_{\rm br1}\sim\la^2\Bigl
(\frac{m_Q\mph}{\la^2}\Bigr )^{\frac{N_F}{3\nd}}\gg\mu^2_{\rm gl,2},\,\,
z_Q(\la,\mu^{\rm pole}_{\rm gl,1})\sim\Bigl (\frac{\mu^{\rm pole}_{\rm
gl,1}}{\la}\Bigr )^{\frac{3N_c-N_F}{N_F}}\ll 1,\,\,\,\label{(28.2.1)}
\eq
while the quark masses are
\bq
\quad \langle m^{\rm tot}_{Q,2}\rangle=\frac{\langle\Qo\rangle_{\rm
br1}}{\mph}\sim m_Q\,,\quad {\tilde m}^{\rm pole}_{Q,2}\sim\frac{\langle m^{\rm
tot}_{Q,2}\rangle}{z_Q(\la,{\tilde m}^{\rm pole}_{Q,2})}\sim\la\Bigl
(\frac{m_Q}{\la}\Bigr )^{N_F/3N_c}\gg m^{\rm pole}_{Q,1}\,,\label{(28.2.2)}
\eq
\bq
\frac{{\tilde m}^{\rm pole}_{Q,2}}{\mu^{\rm pole}_{\rm gl,1}}\sim \Bigl
(\frac{\mo}{\mph} \Bigr )^{N_F/3\nd}\ll 1\,.\label{(28.2.3)}
\eq

Therefore, the quarks $Q^1, {\ov Q}_1$ are higgsed and the overall phase is
$Higgs_1-HQ_2$. If we take $2n_1<\bo=(3N_c-N_F)$, then ${\rm b}^
\prime_{\rm o}=(\bo-2n_1)>0$ and the lower
energy theory with $SU(N_c-n_1)$ colors and $n_2$ flavors will be in the
conformal regime at scales $m^{\rm pole}_{Q,2}\sim\langle\lym^{(\rm
br1)}\rangle\ll\mu\ll\mu^{\rm pole}_{\rm gl,1}$. Then all results for the mass
spectra will be the same as in section 10.1 of \cite{ch19} (and only the ranges
of $N_F$ are different in \cite{ch19} and here).

\section{Dual theory. Broken flavor symmetry. \,\,\\ $\hspace*{1cm}
\mathbf{2N_c<N_F<3N_c\,,\,\mph\gg\la}$}

\subsection{Lt - vacua}

\hspace*{5mm} The main difference with the L - vacua in section 27 is that the
flavor symmetry is broken spontaneously in these dual Lt - vacua, $\langle
M_1\rangle\neq\langle M_2\rangle$. The fions have masses $\mu^{\rm
pole}(\Phi)\sim\mph\gg\la$ and are dynamically irrelevant at scales $\mu<\mph$.
The low energy Lagrangian of mions is \eqref{(27.0.5)}, but the mion masses
look ], now as
\bq
\mu^{\rm pole}(M_1^1)\sim \mu^{\rm
pole}(M_2^2)\sim\frac{\la^2}{\mph}\ll\la,\quad \mu^{\rm pole}(M_1^2)=\mu^{\rm
pole}(M_2^1)=0\,.\label{(29.1.1)}
\eq

Clearly, the parametric differences remain in mass spectra of the direct and
dual theories in these Lt-vacua.

\subsection{br1 - vacua}

\hspace*{5mm} Not going into any details we give here the results only. The
overall phase is $Hq_1-Hq_2$ (i.e. heavy confined quarks). The regime at
$\mu<\la$ is conformal, and the massless Nambu-Goldstone particles here are the
mions $M_1^2$ and $M_2^1$. The mass spectra are as in section 11.1 of
\cite{ch19}, the only difference is that $Z_q\ra 1$ here because
$\bd=(3\nd-N_F)/N_F=O(1)$ now (and only the ranges of $N_F$ are different in
\cite{ch19} and here).

More details can be found in the original paper \cite{ch13}.

\addcontentsline{toc}{section} 
{ \bf \large Part IIc. $\mathbf{N_c < N_F <3N_c/2}$ } .

\begin{center}
\bf \large Part IIc. $\mathbf{N_c < N_F <3N_c/2}$  
\end{center}


\section{\bf  Introduction}

This part continues studies of ${\cal N}=1$ SQCD-type theories (see
above). Aso considered here are the $SU(N_c)$ SQCD-type (direct)
theory (and its Seiberg's dual variant with $SU(\nd=N_F-N_c)$ dual 
colors),  with $N_F$ flavors of quarks ${\ov Q}_j, Q^i$ with the 
small mass parameter $m_Q\ll\la$ in the Lagrandian.
Besides, there are $N_F^2$ additional colorless
but flavored fields $\Phi^j_i$ with the large mass parameter
$\mph\gg\la$. But now considered in details is {\it the region
$N_c<N_F<3N_c/2$} where the UV free direct $SU(N_c)$ theory
is very strongly coupled at scales $\mu<\la$, with the parametrically
large gauge coupling $a(\mu\ll\la)\gg 1$.
The mass spectra of this direct theory in various vacua and at
different values of $\mph$ are calculated in this case within the
dynamical scenario introduced by the author in \cite{ch3}. This
scenario assumes that massive quarks with $m_{Q}\neq 0$ in such 
${\cal N}=1$ SQCD-type   theories without elementary colored adjoint 
scalars can be in two {\it  standard} phases only. These are either the 
HQ (heavy quark) phase  where they are confined or the Higgs phase 
where they are condenced in the vaxuum state.

Similarly to previous studies of this theory within the conformal
window $3N_c/2<N_F<3N_c$, it is shown here that, in the direct
$SU(N_c)$ theory, due to the powerlike RG evolution, the seemingly
heavy and dynamically irrelevant at scales $\mu<\la$ fields $\Phi^j_i$
can become light and relevant at lower energies, and there appear then
two additional generation of light $\Phi$-particles with masses
$\mu_{2,3}^{\rm pole}(\Phi)\ll\la$.

The calculated mass spectra of this strongly coupled at $\mu<\la$
direct $SU(N_c)$ theory were compared to those of its weakly coupled
at $\mu<\la$ Seiberg's dual $SU(N_F-N_c)$ variant and appeared to be
{ \it  parametrically different}.

All results obtained within the used dynamical scenario from
\cite{ch3} look self-consistent. In other words, no internal
contradictions were encountered in all cases considered. It is worth
also to remind that {\it this dynamical scenario  satisfies all those tests 
which  were used as checks of
the Seiberg hypothesis about the equivalence of the direct and dual
theories. The parametrical differences of mass spectra of the direct
and dual theories show, in particular, that all these tests, although
necessary, may well be insufficient}. See also  \cite{Session}).\\

Recall that the Lagrangian of this (direct) $\Phi$-theory at the scale
$\mu=\la$ has the form
\footnote{\,
The gluon exponents are always implied in the Kahler terms. Besides,
here and everywhere below in the text we neglect for simplicity all
RG-evolution effects if they are logarithmic only.
}
\bbq
K={\rm Tr}\, (\Phi^\dagger \Phi )+{\rm Tr}\Bigl (\,Q^\dagger
Q+Q\ra{\ov Q}\,\Bigr )\,,\quad {\cal
W}=\frac{2\pi}{\alpha(\mu=\la)}S+{\cal W}_{\rm matter}\,,
\eeq
\bq
{\cal W}_{\rm matter}={\cal W}_{\Phi}+{\cal W}_Q,\, {\cal
W}_{\Phi}=\frac{\mph}{2}\Biggl [{\rm
Tr}\,(\Phi^2)-\frac{1}{\nd}\Bigl ({\rm Tr}\,\Phi\Bigr )^2
\Bigg ],\, {\cal W}_Q={\rm Tr}\,\Bigl ({\ov Q}(m_Q-\Phi) Q 
\Bigr).\label{(30.0.1)}
\eq

Here\,: $\mph\gg\la$ and $m_Q\ll\la$ are  mass parameters, the
traces in \eqref{(30.0.1)} are over color and/or flavor indices,
$S=\sum_{A,\gamma} W^{A}_{\gamma}W^{A,\,\gamma}/32\pi^2$, where
$W^A_{\gamma}$ is the gauge field strength, $A=1...N_c^2-1,\,
\gamma=1,2$,\, $a(\mu)=N_c \alpha(\mu)/2\pi=N_c g^2(\mu)/8\pi^2$ is the
running gauge coupling with its scale factor $\la$, $\,Q^i_{\beta}, {\ov
Q}_j^{\,\beta},\,\, \beta=1...N_c,\,\, i, j=1...N_F$ are the quark fields. This 
normalization of fields  is used everywhere below in the  text. Besides, 
the perturbatively exact   NSVZ $\beta_{NSVZ}$  functions for (effectively) 
massless SUSY theories \cite{NSVZ-1,NSVZ-2} are used in this paper.

All dynamical properties of theory \eqref{(30.0.1)}\,: the RG evolution,
phase states, mass spectra etc., depends essentially on the value of
$N_F/N_c$. For instance, it enters at $\mu<\la$\,: a) the weakly
coupled IR free logarithmic regime with the gauge coupling
$a(\mu\ll\la)\sim 1/\log(\la/\mu)\ll 1$ at $N_F>3N_c$;\, b) the
strongly coupled conformal regime with $a(\mu\ll\la)=a_{*}={\rm co
nst}=O(1)$ (in general) at $3N_c/2<N_F<3N_c$;\, c) the (very) strongly
coupled regime with $a(\mu\ll\la)\sim (\la/\mu)^{\nu_Q\,>\,0}\gg 1,\,
\nu=(3N_c-2 N_F)/(N_F-N_c)$ at $N_c<N_F<3N_c/2$. Besides, the mass
spectra at given $N_F/N_c$ depend essentially on the considered
vacuum.

In parallel with the direct $\Phi$-theory \eqref{(30.0.1)}, we study also
proposed by Seiberg's \cite{S1,S2,IS} its dual variant, the
$d\Phi$-theory with $SU(\nd=N_F-N_c)$ dual colors, $N_F$ flavors of
dual quarks $q_i^{\beta}, {\ov q}^{\,j}_{\beta},\,\,\beta=1...\nd$, 
and with $N_F^2$  additional colorless but flavored elementary fields 
$M^i_j \ra ({\ov  Q}_j Q^i)$. Its Lagrangian at $\mu=\la$ looks as, 
see \eqref{(30.0.1)}  for ${\cal W}_{\Phi}$,
\bbq
{\ov K}={\rm Tr}\,(\Phi^\dagger\Phi)+ {\rm Tr}\,\Bigl
(\frac{M^{\dagger}M}{\la^2}\Bigr )+{\rm Tr}\,\Bigl ( q^\dagger q +
(q\ra{\ov q})\Bigr )\,,\quad
\ov{\cal W}=\, \,\frac{2\pi}{{\ov\alpha}(\mu=\la)}\,{\ov S}+{\ov{\cal
W}}_{\rm matter}\,,
\eeq
\bq
{\ov{\cal W}}_{\rm matter}={\cal W}_{\Phi}+{\cal W}_{\Phi M}+{\cal
W}_q,\, {\cal W}_{\Phi M}={\rm Tr}\,(m^{\rm tot}_Q M)= {\rm Tr} \Bigl (
(m_Q-\Phi)M\Bigr ),\, {\cal W}_q= - \rm {Tr} \Bigl ({\ov
q} \frac{M}{\la}  q \Bigr ). \label{(30.0.2)}
\eq
Here\,:\, the number of dual colors is ${\ov N}_c=(N_F-N_c)$,\,
$M^i_j$ are $N_F^2$ Seiberg's elementary mion fields, $M^i_j\ra ({\ov
Q}_j Q^i)$,\,\, ${\ov a}(\mu)=\nd{\ov \alpha}(\mu)/2\pi=\nd{\ov
g}^2(\mu)/8\pi^2$ is the dual running gauge coupling (with its scale
parameter $\Lambda_q= - \la$),\,\,${\ov S}= \sum_{B,\beta}{\rm\ov
W}^{\,B}_{\beta}\,{\rm \ov W}^{\,B,\,\beta}/32\pi^2$,\,\, ${\rm \ov
W}^{\,B}_{\beta}$ is the dual gluon field strength, $B=1...\nd^2-1,
\,\,\beta=1,\,2$.

The gluino condensates of the direct and dual theories are matched,
$\langle{-\,\ov S}\rangle=\langle S\rangle=\lym^3$, as well as
$\langle M^i_j\rangle\equiv\langle M^i_j(\mu=\la)\rangle=\langle{\ov
Q}_j Q^i (\mu=\la)\rangle\equiv\langle{\ov Q}_j Q^i\rangle$, and the
scale parameter $\Lambda_q$ of the dual gauge coupling is taken as
$ ( - \Lambda_q)=\la$. At $N_c<N_F<3N_c/2$ this dual theory is IR free and
logarithmically weakly coupled at $\mu<\la,\,\, {\ov a}(\mu\ll\la)\sim
1/{\log(\la/\mu)\ll 1}$.

We studied these theories \eqref{(30.0.1)} and \eqref{(30.0.2)} in 
\cite{ch19,ch13} at values of $N_F$ in the range  $3N_c/2<N_F<3N_c$, 
i.e. within the conformal window. The purpose of  article \cite{ch16} 
was to consider in details the range $N_c<N_F<3N_c/2$.\\

According to Seiberg's view of the standard direct (i.e. without
fields $\Phi^i_j$)\, ${\cal N}=1$ SQCD at
$N_c+1<N_F<3N_c/2$, with the scale factor $\la$ of $SU(N_c)$ gauge
coupling and direct quarks with $m_Q=0$ (or with $0 < m_Q\ll\Lambda$),
the regime of the direct theory at $\mu<\la$ is in this case: {\it
`confinement without chiral symmetry breaking'} (as far as small
$m_Q\neq 0$ can be neglected). And {\it the dual theory is considered
as  the  lower energy form of the direct theory}. This
means that all
direct quarks remained massless (or light), but hadrons made from
these quarks and direct gluons {\it acquired large
masses $\sim\la$ due to mysterious confinement with the string
tension}  $\sigma^{1/2}\sim\la$, and decoupled at $\mu<\la$. Instead of 
them,  there mysteriously appeared massless (or light) composite solitons.
These last are particles of the dual theory.\\

This picture was questioned in \cite{ch1} (see section 7 therein). It
was argued that, with the unbroken at $m_Q\ra 0$  chiral flavor symmetry
$SU(N_F)_L\times SU(N_F)_R$ and unbroken R-charge, it is impossible to
write at $\mu\sim\la$ the nonsingular superpotential of the effective
Lagrangian of massive flavored hadrons with masses $\sim\la$ made from
direct massless (or light) quarks.~
\footnote{\,
This is similar to our ordinary QCD with the scale factor $\Lambda$ of
the $SU(3)$ gauge coupling. Then, with confinement and with three
massless (or light) quarks, but without breaking of chiral symmetry
$SU(3)_L\times SU(3)_R$, it is impossible e.g. to have in the
effective hadron Lagrangian at $\mu\sim\Lambda$ the massive nucleons
with the mass $\sim\Lambda$, as the term $\sim\Lambda {\ov N} N$ in
the potential is incompatible with the unbroken chiral symmetry. And
the situation in ${\cal N}=1$ SQCD is even more restrictive because
the superpotential is holomorphic and due to additional R-charge
conservation.
}

We also recall here the following. There is no confinement in
Yukawa-type theories without gauge interactions. The confinement
originates {\it only} from the (part of) YM or ${\cal N}=1$ SYM in
${\cal N}=1$ SQCD-type theories unbroken by (possibly) higgsed quarks.
And because ${\cal N}=1$ SYM has only one dimensional parameter
$\langle\lym\rangle=\langle S\rangle^{1/3}$, the string tension is
$\sigma^{1/2}\sim\langle\lym\rangle$. But in the standard ${\cal N}=1$
SQCD the value of $\lym$ is well known: $\lym=(\Lambda_{Q}^{\bo}
\det {m_Q})^{1/3N_c}\ll\Lambda_Q$ at $m_Q\ll\la$ Therefore, such 
SYM cannot produce  confinement with 
the string tension $\sim\Lambda_Q$, but only with 
$\sigma^{1/2}\sim\lym\ll \la$  (and there is no
confinement at all at $m_Q\ra 0$).
\footnote{\,
And the same for the direct SQCD-type $\Phi$-theory considered here:\,
$\langle\lym\rangle=(\la^{\bo}\det\langle m^{\rm
tot}_Q\rangle)^{1/3N_c}\ll\la$. Therefore, such SYM cannot produce
confinement with $\sigma^{1/2}\sim\la$, only with
$\sigma^{1/2}\sim\langle\lym\rangle \ll \la$.
}

For these reasons, as was argued in detail in section 7 of \cite{ch1}
and in Conclusions of \cite{Session}, the UV free direct theory
\eqref{(30.0.1)} with $N_c<N_F<3N_c/2$ flavors of light quarks enters
{\it smoothly} at $\mu<\la$ into the effectively massless perturbative
(very) strongly coupled regime with $a(\mu\ll\la)\sim
(\la/\mu)^{\nu_Q\,>\,0}\gg 1$ (and NSVZ $\beta$-function allows this).
This is qualitatively similar to situation in the conformal window at
$3N_c/2<N_F<3N_c$, where the direct theory enters smoothly at
$\mu<\la$ into the conformal regime with the coupling
$a(\mu\ll\la)={\rm const}\sim 1$, and with {\it all its light quarks and
gluons remaining effectively massless}.

For a description of the RG evolution and calculations of mass spectra
in various vacua in this strong coupling regime we use the dynamical
scenario introduced in \cite{ch3}. This scenario assumes that quarks
in such ${\cal N}=1$ SQCD-type theories can be in two {\it standard}
phases only. These are either the HQ (heavy quark) phase where
$\langle{\ov Q}\rangle=\langle Q \rangle=0$ and they are confined, or
the Higgs phase where they form nonzero coherent condensate
$\langle{\ov Q}\rangle=\langle Q \rangle\neq 0$ breaking the color
symmetry. The word {\it standard} implies here also that, unlike e.g.
very special ${\cal N}=2$ SQCD with enhanced supersymmetry, in such
${\cal N}=1$ theories without elementary colored adjoint scalars, no
{\it additional}
\footnote{\,
i.e. in addition to the massless particles due to
spontaneously broken global flavor symmetry
}
parametrically light solitons (e.g. magnetic monopoles or dyons) are
formed at those scales where quarks decouple as heavy or are higgsed.

Within this scenario, were calculated the mass spectra of the strongly
coupled direct theory \eqref{(30.0.1)} in different vacua. It was shown in
\cite{ch16} (as well as in \cite{ch19,ch13}\,) that the use of
this scenario leads to the results for the mass spectra which do not
contradict to any proven results and look self-consistent, i.e. no
internal inconsistences are incountered in all cases considered.

Similarly to our previous studies of this theory \eqref{(30.0.1)}  within 
the conformal window at $3N_c/2<N_F<3N_c$ in \cite{ch13,ch19}, it was
shown in \cite{ch16} that, due to a strong power-like RG evolution at scales
$\mu<\la$ in the direct theory, the seemingly heavy and dynamically
irrelevant fields $\Phi^j_i$ with $\mph\gg\la$ can become light and
there appear then two additional generations of light $\Phi$-particles
with $\mu_{2,3}^{\rm pole}(\Phi)\ll\la$.

In parallel, we calculate the mass spectra in the Seiberg's dual
theory \eqref{(30.0.2)}. But, as explained above, this "dual"\, theory
has to be considered {\it not} as the literal or equivalent low energy 
form of the direct
theory at scales $\mu<\la$, but simply as {\it a definite independent
theory}. This IR free theory with $N_c<N_F<3N_c/2$ is logarithmically
weakly coupled at $\mu<\la$ and so needs no additional dynamical
assumptions for calculations of its mass spectra. The mass spectra of 
both  direct and dual theories are then calculated and compared.

As described below in the text, the comparison of mass spectra of
direct and dual theories \eqref{(30.0.1)} and \eqref{(30.0.2)} shows,
similarly to the standard direct and dual ${\cal N}=1$ SQCD theories
(i.e. those in \eqref{(30.0.1)} and \eqref{(30.0.2)} but without fields
$\Phi^j_i$, see \cite{ch3}), that these two mass spectra are {\it
parametrically different}. It is worth to recall that the dynamical
scenario from \cite{ch3} used in this article satisfies all those
tests which were used as checks of the Seiberg hypothesis about the
equivalence of the direct and dual theories. This shows, in
particular, that all these tests, although necessary, may well be
insufficient.

The paper is organized as follows. Because the global flavor symmetry
$U(N_F)$ is unbroken or broken spontaneously as $U(N_F)\ra
U(\no)\times U(\nt)$ in different vacua of
\eqref{(30.0.1)},\eqref{(30.0.2)}, we consider these cases separately.
Besides, because the parametric behavior of quark condensates 
and the  whole dynamics are quite different in two regions
$\mph\gtrless\mo=\la(\la/m_Q)^{(2N_c-N_F)/N_c}$,   
we also  consider these regions separately.  See contents.
\vspace*{1mm}

\begin{center} 
\bf \Large The region $\mathbf{\la\ll\mph\ll\mo=\la (\la/m_Q)
^{(2N_c-N_F)/N_c}}$
\end{center}

\section{Unbroken flavor symmetry, \,  L-vacua}

\subsection{\quad Direct theory}

The quark condensates (here and everywhere below always at the scale
$\mu=\la$) in these L - vacua with the multiplicity $(2N_c-N_F)$
look as, see section 16 
(\,$\, \langle S\rangle$ is the gluino condensate 
summed over all its colors, $\bo=3N_c-N_F$\,)
\bq
\langle {\ov Q} Q (\mu=\la)\rangle_{L}\sim\la^2\Bigl
(\frac{\la}{\mph}\Bigr )^{\frac{N_F-N_c}{2N_c-N_F}}\ll\la^2,\,\,
\langle S\rangle_L=\Bigl (\frac{ \langle\det{\ov
Q}Q\rangle_L}{\la^{\rm \bo}}\Bigr )^{\frac{1}{N_F-N_c}}\sim\la^3\Bigl
(\frac{\la}{\mph}\Bigr )_{,}^{\frac{N_F}{2N_c-N_F}}\,\,\label{(31.1.1)}
\eq
while from the Konishi anomaly \cite{Konishi}
\bq
\mtl\equiv\langle m_Q-\Phi\rangle_L=\frac{\langle
S\rangle_L}{\QQ_L}\sim\la\Bigl (\frac{\la}{\mph}\Bigr
)^{\frac{N_c}{2N_c-N_F}}\ll\la\,.\label{(31.1.2)}
\eq

As was argued in detail in section 7 of \cite{ch1} and in Conclusions
of \cite{Session}, this UV free direct $\Phi$-theory with the gauge
coupling $a(\mu\gg\la)\equiv N_c g^2(\mu\gg\la)/8\pi^2\ll 1$ enters
smoothly the strong coupling regime at the scale $\mu<\la$, with the
gauge coupling $a(\mu\ll\la)\gg 1$. The values of anomalous
dimensions $\gamma_Q,\, \gamma_{\Phi}$ in the direct theory and
$\gamma_q\,,\,\gamma_M$ in the dual one   at $N_c<N_F<3N_c/2$ 
and $\mu\ll\la$ are,   see  \cite{ch1,ch16} and  Appendix 1,
\bbq
 \gamma_Q=\frac{2N_c-N_F}{N_F-N_c}\,,\quad \gamma_{\Phi}= - 2 \gamma_Q\,, 
\quad \gamma_q\ra 0\,, \quad \gamma_M\ra 0\,.
\eeq

The potentially important masses look then as follows\,: \\ 
a) the quark pole mass
\bq
m^{\rm pole}_{Q,L}=\frac{\mtl}{z_Q(\la,m^{\rm
pole}_{Q,L})}\sim\la\Bigl (\frac{\la}{\mph}\Bigr
)^{\frac{N_F-N_c}{2N_c-N_F}},\, z_Q(\la,m^{\rm
pole}_{Q,L})=\Bigl (\frac{m^{\rm pole}_{Q,L}}{\la}\Bigr
)^{\gamma_Q},\,\gamma_Q=\frac{2N_c-N_F}{N_F-N_c}>1;
\,\,\,\,\,\,\label{(31.1.3)}
\eq
\\
b) the gluon mass due to possible higgsing of quarks
\bq
\Bigl (\mu^{\rm pole}_{{\rm gl},L}\Bigr )^2\sim a(\mu^{\rm pole}_{{\rm
gl},L})\,z_Q(\la,\mu^{\rm pole}_{{\rm gl},L})\QQ_L\,,\quad
z_Q(\la,\mu^{\rm pole}_{{\rm gl},L})=\Bigl (\frac{\mu^{\rm pole}_{{\rm
gl},L}}{\la}\Bigr )^{\gamma_Q}\ll 1,\,\, \label{(31.1.4)}
\eq
\bbq
\frac{d\,a(\mu\ll\la)}{d\log\mu}=\beta_{NSVZ}(a)=\frac{a^2(\mu)}{a(\mu)-1}\,
\frac{\bo-N_F\gamma_Q}{N_c}\quad
\xrightarrow{a(\mu)\gg 1}\quad -\,\nu_Q\,a(\mu),\quad \bo=3N_c-N_F\,,
\eeq
\bbq
\nu_Q=\frac{N_F\gamma_Q-\bo}{N_c}=\frac{3N_c-2N_F}{N_F-N_c}=
\gamma_Q-1>0\,, \quad \, a(\mu=\mu^{\rm \,pole}_{{\rm gl},L}\ll\la)
\sim\Bigl (\frac{\la}{\mu^ {\rm pole}_{{\rm  \,gl},L}}\Bigr)^{\nu_Q}\gg 1,
\eeq
\bbq
a(\mu^{\rm pole}_{{\rm gl},L})\,z_Q(\la,\mu^{\rm pole}_{{\rm
gl},L})\sim\frac{\mu^{\rm pole}_{{\rm gl},L}}{\la},\quad \mu^{\rm
pole}_{{\rm gl},L}\sim \frac{\QQ_L}{\la}\sim\la\Bigl
(\frac{\la}{\mph}\Bigr )^{\frac{N_F-N_c}{2N_c-N_F}}\sim m^{\rm
pole}_{Q,L}\,.
\eeq

Because the global non-Abelian flavor symmetry $SU(N_F)$ is unbroken
in these L-vacua, this means that {\it the quarks in such a case are
not higgsed due to the rank restriction} $N_c<N_F$, as otherwise the
global flavor symmetry $SU(N_F)$ will be broken spontaneously.
Therefore, {\it the overall phase is in this case HQ} (heavy confined
quarks).
\footnote{\,
The same reasonings are used also everywhere below in similar cases.
\label{f4}
}

The Lagrangian at scales $\mu$ such that $\mos<\mu<\la$ looks as
\bq
K=z_{\Phi}(\la,\mu)\,{\rm Tr}
(\Phi^\dagger\Phi\,)+z_Q(\la,\mu){\rm Tr}\,\Bigl (Q^\dagger Q+Q\ra
\bar Q\Bigr ),\, z_{\Phi}(\la,\mu)=1/z^2_Q(\la,\mu),\,\,\label{(31.1.5)}
\eq
\bbq
z_Q(\la,\mu\ll\la)=\Bigl (\frac{\mu}{\la}\Bigr )^{\gamma_Q}=\Bigl
(\frac{\mu}{\la}\Bigr )^{\frac{\bb}{N_F-N_c}}\ll 1,\,\,
z_{\Phi}(\la,\mu)=\Bigl (\frac{\mu}{\la}\Bigr
)^{\gamma_{\Phi}=-2\gamma_Q}=\Bigl (\frac{\la}{\mu}\Bigr
)^{\frac{2(2N_c-N_F)}{N_F-N_c}}\gg 1.
\eeq
\bq
\w={\cal W}_{\Phi}+{\rm Tr}\,(\,{\ov Q}\,m_Q^{\rm tot} Q\,)\,,\quad
{\cal W}_{\Phi}=\frac{\mph}{2}\Bigl ({\rm
Tr}\,(\Phi^2)-\frac{1}{\nd}({\rm Tr}\,\Phi)^2\Bigr ),\quad m_Q^{\rm
tot}=(m_Q-\Phi)\,.\label{(31.1.6)}
\eq
Therefore, the running perturbative mass of $\Phi$ is
$\mu_{\Phi}(\mu\ll\la)=\mph/z_{\Phi}(\la,\mu)\ll\mph$ and, if nothing
prevents, the field $\Phi$ becomes dynamically relevant at scales
$\mu<\mu_o^{\rm str},\,\mu_{\Phi}(\mu=\mu_o^{\rm str})=\mu_o^{\rm
str}$,
\bbq
\mu_o^{\rm str}=\la\Bigl (\frac{\la}{\mph}\Bigr
)^{\frac{1}{2\gamma_Q-1}}=\la\Bigl (\frac{\la}{\mph}\Bigr
)^{\frac{\nd}{5N_c-3N_F}}\,,\quad
\gamma_Q=\frac{2N_c-N_F}{\nd}\,,\quad \nd=N_F-N_c\,,
\eeq
\bq
\frac{\mu_o^{\rm str}}{m^{\rm pole}_{Q,L}}\sim\Bigl
(\frac{\mph}{\la}\Bigr )^{\Delta}\gg 1\,,\quad \Delta=
\frac{\nd(3N_c-2N_F)}{(2N_c-N_F)(5N_c-3N_F)}>0\,,\label{(31.1.7)}
\eq
and there is the second generation of all $N_F^2$ fions $\Phi^j_i$
with $\mu_{2,L}^{\rm pole}(\Phi)\sim\mu_o^{\rm str}\gg m_{Q,L}^{\rm
pole}$. Besides, even at lower scales
$m^{\rm pole}_{Q,L}\ll\mu\ll\mu_o^{\rm str}$ where the fields
$\Phi^j_i$ became already effectively massless (in the sense
$\mu_{\Phi}(\mu)\ll\mu$ ) and so dynamically relevant, the quark
anomalous dimension $\gamma_Q=(\bb)/\nd$ remains the same, as well as
$\gamma_{\Phi}=-2\gamma_Q$, see Appendix 1\,.

At $\mu<m^{\rm pole}_{Q,L}$ all quarks decouple as heavy ones and the
RG evolution of all fields $\Phi$ becomes frozen, but this happens in
the region where they are already relevant, i.e. the running mass of
fions $\mu_{\Phi}(\mu)$ is $\mu_{\Phi}(\mu=m^{\rm pole}_{Q,L})\ll
m^{\rm pole}_{Q,L}$. This means that there is the third generation of
all $N_F^2$ fions with $\mu^{\rm pole}_{3,L}(\Phi)\ll m^{\rm
pole}_{Q,L}$,  see below.

The lower energy theory at $\mu<m^{\rm pole}_{Q,L}$ is ${\cal N}=1\,\,
SU(N_c)$ SYM in the strong coupling regime (plus $N_F^2$ colorless
fions $\Phi$). The scale factor $\langle\lym\rangle_L$ of its gauge
coupling is determined from the matching, see
\eqref{(31.1.3)},\eqref{(31.1.8)} and section 7 in \cite{ch1}\,:
\bbq
\frac{d\, a^{\rm str}_{YM}(\mu)}{d\log\mu}=\beta^{YM}_{NSVZ}(a^{\rm
str}_{YM})=\frac{3( a^{\rm str}_{YM})^2}{a^{\rm
str}_{YM}-1}\,\,\,\,\xrightarrow{a^{\rm str}_{YM}\gg 1}\,\,\, 3\,
a^{\rm str}_{YM}(\mu)\,, \quad \,a^{\rm str}_{YM}(\mu\gg\lym)\sim\Bigl
( \frac{\mu}{\lym}\Bigr )^3\gg 1,
\eeq
\bq
a_{+}(\mu=m^{\rm pole}_{Q,L}\ll\la)\sim\Bigl (\frac{\la}{m^{\rm
pole}_{Q,L}}\Bigr )^{\nu_Q}=a^{\rm str}_{YM}
(\mu=m^{\rm pole}_{Q,L}\gg\lym)\sim\Bigl (\frac{m^{\rm
pole}_{Q,L}}{\langle\lym^{(L)}\rangle}\Bigr )^3\gg 1\,.\label{(31.1.8)}
\eq
From \eqref{(31.1.3)},\eqref{(31.1.8)}
\bq
\langle\lym^{(L)}\rangle^{3}\sim\la^3\Bigl (\frac{\la}{\mph}\Bigr
)^{\frac{N_F}{2N_c-N_F}}\,,\quad
\frac{\langle\lym^{(L)}\rangle}{m^{\rm pole}_{Q,L}}\sim\Bigl
(\frac{\la}{\mph}\Bigr )^{\frac{3N_c-2N_F}{3(2N_c-N_F)}}\ll
1\,,\label{(31.1.9)}\eq
as it should be because, see \eqref{(31.1.1)},
\bbq
\langle\lym^{(L)}\rangle^{3}\equiv\langle S\rangle_L=\Biggl (\frac{\det
\langle{\ov Q}Q\rangle_L}{\la^{\rm \bo}}\Biggr )^{1/\nd}\sim \la^3\Bigl
(\frac{\la}{\mph}\Bigr )^{\frac{N_F}{2N_c-N_F}}.
\eeq

After lowering the scale down to $\mu<\langle\lym^{(L)}\rangle$ and integrating
out all gauge degrees of freedom via the VY (Veneziano-Yankielowicz)-procedure
\cite{VY}, the low energy Lagrangian looks as, see
\eqref{(31.1.5)},\eqref{(31.1.6)},
\bq
K=z_{\Phi}(\la,m^{\rm pole}_{Q,L})\,{\rm Tr}(\Phi^\dagger\Phi\,),\,
z_{\Phi}(\la,m^{\rm pole}_{Q,L})=\frac{1}{z^2_Q(\la,m^{\rm pole}_{Q,L})}=\Bigl
(\frac{\la}{m^{\rm pole}_{Q,L}}\Bigr )^{\frac{2(2N_c-N_F)}{N_F-N_c}}\gg
1,\,\,\,\,\label{(31.1.10)}
\eq
\bbq
{\cal W}={\cal W}_{\Phi}+{\cal W}_{\rm non-pert}\,,\quad {\cal W}_{\rm
non-pert}=N_c\Bigl (\lym^{(L)}\Bigr )^3=N_c\Bigl (\la^{\rm \bo}\det m^{\rm
tot}_Q\Bigr )^{1/N_c}\,.
\eeq
From \eqref{(31.1.10)}, the pole masses of $N_F^2$ third generation fions
$\Phi^j_i$ are (the contributions to $\mu^{\rm pole}_{3,L}(\Phi)$ from ${\cal
W}_{\Phi}$ and ${\cal W}_{\rm non-pert}$ are parametrically the same)
\bq
\mu^{\rm pole}_{3,L}(\Phi)\sim\frac{\mph}{z_{\Phi}(\la,m^{\rm
pole}_{Q,L})}\sim\frac{\la^2}{\mph}\,,\quad
\quad\frac{\mu^{\rm pole}_{3,L}(\Phi)}{\langle\lym^{(L)}\rangle}\sim\Biggl
(\frac{\la}{\mph}\Biggr )^{\frac{2(3N_c-2N_F)}{3(2N_F-N_c)}}\ll
1\,.\label{(31.1.11)}
\eq

On the whole for the case considered.\\
1)\,\, All quarks ${\ov Q}_j, Q^i$ are in the HQ (heavy quark) phase and weakly
confined (i.e. the tension of the confining string originating from ${\cal
N}=1\,\, SU(N_c)$ SYM is much smaller than quark masses,
$\sqrt\sigma\sim\langle\lym^{(L)}\rangle\ll m^{\rm pole}_{Q,L}$, see
\eqref{(31.1.3)},\eqref{(31.1.9)}.\\
2)\,\, There is a large number of $SU(N_c)$ gluonia with the mass scale
$\sim\langle\lym^{(L)}\rangle=\langle S\rangle^{1/3}_{L}$, see
\eqref{(31.1.9)}.\\
3) There are two generations of $N_F^2$ fions $\Phi^j_i$ with masses
\bq
\mu^{\rm pole}_{2,L}(\Phi)\sim\mu_o^{\rm
str}\sim\la\Bigl(\frac{\la}{\mph}\Bigr)^{\frac{\nd}{5N_c-3N_F}},\quad \mu^{\rm
pole}_{3,L}(\Phi)\sim\frac{\la^2}{\mph}\,.\label{(31.1.12)}
\eq
The overall mass hierarchies look as
\bq
\mu^{\rm pole}_{3,L}(\Phi)\ll\langle\lym^{(L)}\rangle\ll m^{\rm
pole}_{Q,L}\ll\mu^{\rm pole}_{2,L}(\Phi)\ll\la\ll\mu^{\rm
pole}_{1}(\Phi)\sim\mph\,.\label{(31.1.13)}
\eq

\subsection{\quad Dual theory}

The mass spectra in the Seiberg dual IR free and logarithmically weakly coupled
at $\mu<\la$ $SU(\nd)$ theory \eqref{(30.0.2)} were described for this case in
section 7.1 of \cite{ch16}. For the reader convenience and completeness we
reproduce here the results. \\
1) All $N_F^2$ fions $\Phi^j_i$ have large masses $\mu_1^{\rm
pole}(\Phi)\sim\mph\gg\la$ (with logarithmic accuracy) and are dynamically
irrelevant at all lower scales.\\
2) All dual quarks ${\ov q}^j, q_i$ are in the overall Hq (heavy quark)
phase. There is a large number of hadrons made of weakly interacting,
non-relativistic and weakly confined dual quarks. The scale of their masses,
neglecting logarithmic RG-evolution factors, is $\mu^{\rm pole}_{q,L}\sim
\langle M\rangle_{\rm L}/\la=\QQ_{\rm L}/\la\sim m^{\rm pole}_{Q,L}$, see
\eqref{(31.1.3)} (the tension of the confining string originating from ${\cal
N}=1\,\, SU(\nd)$ SYM is much smaller, $\sqrt
\sigma\sim\langle\lym^{(L)}\rangle\ll\mu^{\rm pole}_{q,L}$, see
\eqref{(31.1.9)}\,). \\
3) A large number of gluonia made of $SU(\nd)$ gluons with their mass scale
$\sim\langle\lym^{(L)}\rangle=\langle S\rangle_L^{1/3}\sim \la
(\la/\mph)^{N_F/3(2N_c-N_F)}$.\\
4) $N_F^2$ Seiberg's mions $M^i_j$ with masses $\mu^{\rm
pole}_{L}(M)\sim\la^2/\mph$.\\

The mass hierarchies look here as $\mu^{\rm
pole}_{L}(M)\ll\langle\lym^{(L)}\rangle\ll\mu^{\rm pole}_{q}\ll\la\ll\mu^{\rm
pole}_{1}(\Phi)\sim\mph$.\\

Comparing the mass spectra of the direct and dual theories we note the
following.\\
1) The weakly confined dual quarks ${\ov q}^j, q_i$ inside dual hadrons are
non-relativistic and parametrically weakly coupled (the dual coupling ${\ov a}$
at the scale of the Bohr momentum is logarithmically small). Therefore, the
Coulomb mass splittings of the low lying hadrons are also parametrically small,
i.e. $\delta M_{H}/M_{H}\sim {\ov a}^{\,2}\ll 1$. There is nothing similar in
the direct theory with the strongly coupled quarks ${\ov Q}_j, Q^i$.\\
2) In the range of scales $\mu_o^{\rm str}\ll\mu\ll\la$ the effectively
massless  flavored particles in the direct theory are only quarks ${\ov Q}_j,
Q^i$, while all $N_F^2$ fions $\Phi^j_i$ have large running masses 
$\mu_{\Phi}(\mu)>\mu$ and are dynamically irrelevant. In the dual theory the 
effectively massless flavored particles are the dual quarks ${\ov q}^j, q_i$
and \, $N_F^2$ mions $M^i_j$. Therefore, the anomalous 't Hooft triangles
$SU^3(N_F)_L$    are the same in the  direct and dual theories \cite{S2}.\\
3) In the range of scales $\mu^{\rm pole}_{q,L}\sim m^{\rm
pole}_{Q,L}\ll\mu\ll\mu_o^{\rm str}$ the effectively massless flavored
particles  in the dual theory remain the same, while $N_F^2$ fions become
effectively  massless (i.e. the running mass $\mu_{\Phi}(\mu)$ of all fions is
$\mu_{\Phi}(\mu)\ll\mu$\,) and give now additional contributions e.g. to
$SU^{\,3}(N_F)_L$ triangles in the direct theory. Therefore, the
$SU^{\,3}(N_F)_L$ triangles do not match now in the direct and dual theories.\\
4) At scales $\mu\ll m^{\rm pole}_{Q,L}\sim\mu^{\rm pole}_{q,L}$ all direct and
dual quarks decouple. In the range of scales $\mu^{\rm pole}_{3,L}(\Phi)\sim
\mu^{\rm pole}_{L}(M)\ll\mu\ll m^{\rm pole}_{Q,L}$ the effectively massless
flavored particles in the direct theory are $N_F^2$ third generation fions
$\Phi^j_i$, while in the dual theory these are $N_F^2$ mions $M_j^i$.
Therefore,  the values of $SU^{\,3}(N_F)_L$ triangles differ in sign in the
direct  and dual theories.

\section{Unbroken flavor symmetry,\,  S-vacua}

\subsection{\quad Direct theory}

The quark condensates in these S (small)-vacua look as, see section 16,
\bq
\QQ_S\simeq -\frac{N_c}{\nd}\, m_Q\mph\,,\quad \langle S\rangle_S=\Bigl
(\frac{\det \QQ_S}{\la^{\rm \bo}}\Bigr )^{1/\nd}\sim\la^3\Bigl
(\frac{m_Q\mph}{\la^2}\Bigr )^{N_F/\nd}\,.\label{(32.1.1)}
\eq

The direct theory is strongly coupled at $\mu<\la$. Proceeding as in 
section 31.1  we obtain for this S-vacuum, see
\eqref{(31.1.3)},\eqref{(31.1.4)},\eqref{(31.1.8)},\eqref{(32.1.1)},
\bq
\langle m_Q^{\rm tot}\rangle_S=\frac{\langle S\rangle_S}{\QQ_S}\sim \la\Bigl
(\frac{\QQ_S}{\la^2}\Bigr )^{N_c/\nd},\,\, m_{Q,S}^{\rm pole}=\frac{\langle
m_Q^{\rm tot}\rangle_S}{z_Q(\la,m_{Q,S}^{\rm pole})}
\sim \frac{\QQ_S}{\la}\sim\frac{m_Q\mph}{\la}\ll\la\,,\label{(32.1.2)}
\eq
\bbq
\Bigl (\mu^{\rm pole}_{{\rm gl},S}\Bigr )^2\sim a(\mu^{\rm pole}_
{{\rm gl},S})z_Q(\la,\mu^{\rm pole}_{{\rm gl},S})\,\QQ_S\sim\frac{\mu^{\rm
pole}_{{\rm  gl},S}}{\la}\,\QQ_S\quad\ra\quad\mu^{\rm pole}_{{\rm gl},S}\sim 
\frac{\QQ_S}{\la}\sim\frac{m_Q\mph}{\la}\sim m_{Q,S}^{\rm pole}\,.
\eeq
Therefore, for the same reasons of the rank restrictions and unbroken global
flavor symmetry as in L-vacua in section 31.1, the overall phase is HQ.
Besides,\bq
\frac{m_{Q,S}^{\rm pole}}{\mu_o^{\rm str}}\sim\frac{m_Q}{\la}\Bigl
(\frac{\mph}{\la}\Bigr )^{\frac{2(2N_c-N_F)}{5N_c-3N_F}}\ll\frac{m_Q}{\la}\Bigl
(\frac{\mo}{\la}\Bigr )^{\frac{2(2N_c-N_F)}{5N_c-3N_F}}\sim\Bigl
(\frac{m_Q}{\la}\Bigr )^{\frac{\nd(3N_c-2N_F)}{N_c(5N_c-3N_F)}}\ll 1,\,\,
\la\ll\mph\ll\mo,\,\,\,\,\,\,\,\,\label{(32.1..3)}
\eq
so that the running mass of all fions $\Phi^j_i$ is $\mu_{\Phi}(\mu<\mu_o^{\rm
str})<\mu$, all $N_F^2$ fions become dynamically relevant at $\mu<\mu_o^{\rm
str}$ and there is the second generation of fions with $\mu_{2,S}^{\rm
pole}\sim\mu_o^{\rm str}=\la(\la/\mph)^{\nd/(5N_c-3N_F)}\gg m_{Q,S}^{\rm
pole}$. At $\mu<m_{Q,S}^{\rm pole}$ all quarks decouple as heavy and, 
proceeding \, as in \eqref{(31.1.8)}, we obtain the scale factor of remained 
${\cal N}=1\,\,  SU(N_c)$  SYM
\bq
a_{+}(\mu=m^{\rm pole}_{Q,S})=\Bigl (\frac{\la}{m^{\rm pole}_{Q,S}}\Bigr
)^{\nu_Q=(3N_c-2N_F)/\nd}=a^{\rm str}_{YM}(\mu=m^{\rm pole}_{Q,S})=\Bigl
(\frac{\mu=m^{\rm pole}_{Q,S}}{\langle\lym^{(S)}\rangle}\Bigr )^3\gg
1\,.\label{(32.1.4)}
\eq
From \eqref{(32.1.2)},\eqref{(32.1.4)}
\bbq
\langle\lym^{(S)}\rangle^{3}\sim\la^3\Bigl (\frac{\langle m_Q^{\rm
pole}\rangle_S}
{\la} \Bigr )^{N_F/\nd}\sim\la^3\Bigl (\frac{m_Q\mph}{\la^2}\Bigr)^{N_F/\nd},
\eeq
\bq
\frac{\langle\lym^{(S)}\rangle}{m^{\rm pole}_{Q,S}}\sim\Bigl(\frac
{m_Q\mph}{\la^2}\Bigr )^{(3N_c-2N_F)/3\nd}\ll\Bigl (\frac{m_Q}{\la}\Bigr
)^{(3N_c-2N_F)/3N_c}\ll 1,\,\,\,\label{(32.1.5)}
\eq
as it should be because, see \eqref{(32.1.1)},
\bq
\langle\lym^{(S)}\rangle^{3}\equiv\langle S\rangle_S=\Biggl (\frac{\det
\langle{\bar Q}Q\rangle_S}{\la^{\rm \bo}}\Biggr )^{1/\nd}\sim\la^3\Bigl
(\frac{m_Q\mph}{\la^2}\Bigr )^{N_F/\nd}\,.\label{(32.1.6)}
\eq

The low energy Lagrangian at $\mu<\langle\lym^{(S)}\rangle$ has
the same form  as in \eqref{(31.1.10)}, but now in S - vacua. In this 
case the main  contribution to  masses of third generation fions 
originates from the nonperturbative term in\eqref{(31.1.10)} and is
\bq
\mu^{\rm pole}_{3,S}(\Phi)\sim\frac{1}{z_{\Phi}(\la,m_{Q,S}^{\rm
pole})}\,\frac{\langle S\rangle_S}{\langle m_{Q,S}^{\rm
tot}\rangle^2}\sim\la\Bigl (\frac{m_Q\mph}{\la^2}\Bigr
)^{(2N_c-N_F)/\nd}\,,\label{(32.1.7)}
\eq
\bq
\frac{\mu^{\rm pole}_{3,S}(\Phi)}{\langle\lym^{(S)}\rangle}\sim
\Bigl(\frac{m_Q\mph}{\la^2}\Bigr )^{2(3N_c-2N_F)/3\nd}<
\Bigl(\frac{m_Q\mo}{\la^2}\Bigr )^{2(3N_c-2N_F)3\nd}\sim
\Bigl (\frac{m_Q}{\la} \Bigr)^{2(3N_c-2N_F)/3N_c}\ll 1. 
\,\,\,\label{(32.1.8)}
\eq

The overall mass hierarchies look as
\bq
\mu^{\rm pole}_{3,S}(\Phi)\ll\langle\lym^{(S)}\rangle\ll m^{\rm
pole}_{Q,S}\ll\mu^{\rm pole}_{2,S}(\Phi)\sim\mu_o^{\rm str}
\ll\la\ll\mu_1^{\rm pole}(\Phi)\sim\mph\,.\label{(32.1.9)}
\eq

\subsection{\quad Dual theory}

The mass spectra in the Seiberg dual $SU(\nd)$ theory for this case were also
described in section 7.2 of \cite{ch16}. The results look as follows. \\
1) All $N_F^2$ fions $\Phi^j_i$ have large masses $\mu_1^{\rm
pole}(\Phi)\sim\mph\gg\la$ (with logarithmic accuracy) and are dynamically
irrelevant at all lower scales.\\
2) All dual quarks ${\ov q}^j, q_i$ are in  the overall Hq (heavy quark)
phase. There is a large number of hadrons made of weakly interacting
non-relativistic and weakly confined dual quarks, the scale of their masses
(with logarithmic accuracy) is $\mu^{\rm pole}_{q,S}\sim \langle M\rangle_{\rm
S}/\la=\QQ_{\rm S}/\la\sim (m_Q\mph)/\la\sim m^{\rm pole}_{Q,S}$, see
\eqref{(31.1.3)},\eqref{(32.1.1)}\, (the tension of the confining string
originating
from ${\cal N}=1\,\, SU(\nd)$ SYM is much smaller, $\sqrt
\sigma\sim\langle\lym^{(S)}\rangle\ll\mu^{\rm pole}_{q,S}\sim m^{\rm
pole}_{Q,S}$\,).\\
3) A large number of gluonia made of $SU(\nd)$ gluons with their mass scale
$\sim\langle\lym^{(S)}\rangle=\langle S\rangle_S^{1/3}\sim \la
(m_Q\mph/\la^2)^{N_F/3\nd}$.\\
4) $N_F^2$ Seiberg's mions $M^i_j$ with masses $\mu^{\rm
pole}_{S}(M)\sim\la^2/\mph$.\\

The hierarchies of masses (except for $\mu^{\rm pole}_{1}(\Phi)\sim\mph\gg\la$)
look here as:\\
a) $\la\gg\mu^{\rm pole}_S(M)\gg\mu^{\rm pole}_q\gg\lym^{(\rm S)}$\quad at
\quad$\la\ll\mph\ll{\mu_{\Phi}^\prime}=\la (\la/m_Q)^{1/2}\,$;\\
b) $\la\gg\mu^{\rm pole}_q\gg\mu^{\rm pole}_S(M)\gg\lym^{(\rm S)}$\quad at
\quad${\mu_{\Phi}^\prime}\ll\mph\ll {\tilde\mu}_{\Phi}=\la
(\la/m_Q)^{N_F/(4N_F-3N_c)}\,$;\\
c) $\la\gg\mu^{\rm pole}_q\gg\lym^{(\rm S)}\gg\mu^{\rm pole}_S(M)$\quad at
\quad${\tilde\mu}_{\Phi}\ll\mph\ll\mo=\la (\la/m_Q)^{(2N_c-N_F)/N_c}\,$\,.

It is seen that, at least, $\mu^{\rm pole}_{3,S}(\Phi)\sim\la\Bigl
(m_Q\mph/\la^2\Bigr )^{(2N_c-N_F)/\nd}$ \eqref{(32.1.7)} in the direct theory
and
$\mu^{\rm pole}_S(M)\sim\la^2/\mph$ in the dual one are parametrically
different  (the 't Hooft triangles $SU^3(N_F)_L$ are also different).

\section{Broken flavor symmetry, \, L-type vacua}

The main qualitative difference compared with the L-vacua in section 2 is the
spontaneous breaking of the flavor symmetry, $U(N_F)\ra U(n_1)\times U(n_2)$,
see section 16,
\bbq
\langle ({\ov Q}Q)_1\rangle_{Lt}\equiv\langle {\ov Q}_1 Q^1\rangle_{Lt}=\langle
M_1\rangle_{Lt}\neq \langle ({\ov Q}Q)_2\rangle_{Lt}=\langle {\ov Q}_2
Q^2\rangle_{Lt}=\langle M_2\rangle_{Lt}\,,
\eeq
\bq
(1-\frac{\no}{N_c})\langle ({\ov Q}Q)_1\rangle_{Lt}\simeq -\,
(1-\frac{\nt}{N_c})\langle ({\ov Q}Q)_2\rangle_{Lt}\sim\QQ_L\sim \la^2\Bigl
(\frac{\la}{\mph}\Bigr )^{\nd/(\bb)}\ll\la^2\,.\label{(33.0.1)}
\eq

For this reason, unlike the L-vacua, the fions $\Phi^1_2,\, \Phi^2_1$ in the
direct theory and mions $M^1_2,\, M^2_1$ in the dual one are the
Nambu-Goldstone \, particles and are massless. Except for this, all other masses in
these Lt-vacua of the direct and dual theories are parametrically the same as in 
L-vacua.  Therefore, all differences between the direct and dual theories 
described above  for the L-vacua remain in L-type vacua also.

\section{Broken flavor symmetry, \, br2 vacua}

\subsection{\quad Direct theory}

The quark condensates of the $SU(N_c)$ theory look in these vacua with
$\nt>N_c,\, \no<\nd$ as, 
\bq
\langle\Qt\rangle_{\rm br2}\simeq\frac{N_c}{N_c-\nt} m_Q\mph\,,\quad
\langle\Qo\rangle_{\rm br2}\sim\la^2
\Bigl (\frac{m_Q}{\la}\Bigr )^{\frac{N_c-n_1}{n_2-N_c}}\Bigl
(\frac{\mph}{\la}\Bigr )^{\frac{n_2}{n_2-N_c}}\,,\label{(34.1.1)}
\eq
\bbq
\frac{\langle\Qo\rangle_{\rm br2}}{\langle\Qt\rangle_{\rm br2}}\sim\Bigl
(\frac{\mph}{\mo}\Bigr )^{\frac{N_c}{n_2-N_c}}\ll 1\,,\quad
\mo=\la\Bigl(\frac{\la}{m_Q}\Bigr )^{(2N_c-N_F)/N_c}\,,
\eeq
\bbq
\langle\lym^{(\rm br2)}\rangle^3\equiv\langle S\rangle_{\rm br2}=\Bigl
(\frac{\langle\Qo\rangle_{\rm br2}^{\no}\langle\Qt\rangle_{\rm
br2}^{\nt}}{\la^{3N_c-N_F}}\Bigr )^{1/\nd}
\sim\la^3\Bigl (\frac{m_Q}{\la}\Bigr )^{\frac{n_2-n_1}{n_2-N_c}}\Bigl
(\frac{\mph}{\la}\Bigr )^{\frac{n_2}{n_2-N_c}}\,.
\eeq
From \eqref{(34.1.1)} and the Konishi anomalies,
\bbq
\qma_{\rm br2}=m_Q-\langle\Phi_1\rangle_{\rm br2}=\frac{\langle\Qt\rangle_{\rm
br2}}{\mph}\sim m_Q\gg \qmb_{\rm br2}=\frac{\langle\Qo\rangle_{\rm
br2}}{\mph}\sim\Bigl(\frac{m_Q}{\la}\Bigr )^{\frac{N_c-n_1}{n_2-N_c}}\Bigl
(\frac{\mph}{\la}\Bigr )^{\frac{N_c}{n_2-N_c}},
\eeq
\bq
m_{Q,1}^{\rm pole}=\frac{\qma_{\rm br2}}{z_Q^{+}(\la,m_{Q,1}^{\rm
pole})}\sim\la\Bigl (\frac{m_Q}{\la}\Bigr )^{\nd/N_c},\,\,
z_Q^{+}(\la,m_{Q,1}^{\rm pole})=\Bigl (\frac{m_{Q,1}^{\rm pole}}{\la}\Bigr
)^{\gamma_Q^{+}}\,\, \gamma_Q^{+}=\frac{2N_c-N_F}{N_F-N_c}>
1,\,\,\,\label{(34.1.2)}
\eq
while the gluon mass from the possible higgsing of ${\ov Q}_2, Q^2$ quarks
looks here as, see \eqref{(31.1.4)},
\bbq
\Bigl ({\mu}^{\rm pole}_{gl,2}\Bigr )^2\sim\Biggl
[a_{+}(\mu={\mu}_{gl,2})=\Bigl(\frac{\la}{{\mu}^{\rm pole}_{gl,2}}\Bigr
)^{\nu^{+}_Q}\Biggr ]
z_Q^{+}(\la,{\mu}^{\rm pole}_{gl,2})\langle\Qt\rangle_{\rm br2}\,, \quad
\nu^{+}_Q=\gamma^{+}_Q-1=\frac{3N_c-2N_F}{N_F-N_c}>\,0\,,
\eeq
\bq
a_{+}({\mu}_{gl,2})\,z_Q^{+}(\la,{\mu}^{\rm pole}_{gl,2})\sim\frac{{\mu}^{\rm
pole}_{gl,2}}{\la}, \,\,
{\mu}^{\rm pole}_{gl,2}\sim\frac{\langle\Qt\rangle_{\rm
br2}}{\la}\sim\frac{m_Q\mph}{\la}
\gg\mu_{gl,1}\,,\quad\frac{{\mu}^{\rm pole}_{gl,2}}{m_{Q,1}^{\rm
pole}}\sim\frac{\mph}{\mo}\ll 1.\,\,\,\,\label{(34.1.3)}
\eq

Besides, see \eqref{(31.1.7)}, because the largest mass in the quark-gluon
sector
is $m_{Q,1}^{\rm pole}$ and
\bq
\frac{m_{Q,1}^{\rm pole}}{\mos}\sim\Bigl (\frac{m_Q}{\la}\Bigr
)^{\frac{\nd}{N_c}} \Bigl (\frac{\mph}{\la}\Bigr
)^{\frac{\nd}{5N_c-3N_F}}\ll\Bigl (\frac{m_Q}{\la}\Bigr )^{\frac{\nd}{N_c}}
\Bigl (\frac{\mo}{\la}\Bigr )^{\frac{\nd}{5N_c-3N_F}}\sim\Bigl
(\frac{m_Q}{\la}\Bigr )^{\frac{\nd(3N_c-2N_F)}{N_c(5N_c-3N_F)}\,>0}\ll
1,\,\,\,\,\,\label{(34.1.4)}
\eq
\bbq
\mos=\la\Bigl (\frac{\la}{\mph}\Bigr )^{\frac{1}{2\gamma^{+}_Q-1}}=\la\Bigl
(\frac{\la}{\mph}\Bigr )^{\frac{\nd}{5N_c-3N_F}}\ll\la\,,\quad \nd=N_F-N_c\,,
\eeq
there are $N_F^2$ of 2-nd generation fions $\Phi^j_i$ with masses $\mu_2^{\rm
pole}(\Phi)\sim\mos=\la(\la/\mph)^{\nd/(5N_c-3N_F)}\gg m_{Q,1}^{\rm pole}$.

After the heaviest quarks ${\ov Q}_1, Q^1$ decouple at $\mu<m_{Q,1}^{\rm
pole}$,  the RG evolution of $\Phi^1_1, \Phi^2_1, \Phi^1_2$ fions is frozen,
while the new quantum numbers are
\bq
N_F^\prime=N_F-n_1=n_2,\,\, N_c^\prime=N_c,\,\,
\gamma_Q^{-}=\frac{2N_c-N_F+n_1}{n_2-N_c}>\gamma_Q^{+},\,\,
\nu^{-}_Q=\frac{3N_c-2N_F+2n_1}{n_2-N_c}>\nu^{+}_Q. \,\,\,\, \label{(34.1.5)}
\eq
From \eqref{(34.1.5)}, the pole mass of ${\ov Q}_2, Q^2$ quarks is, see
\eqref{(34.1.2)},
\bq
m_{Q,2}^{\rm pole}=\frac{\qmb_{\rm br2}}{z_Q^{+}(\la,m_{Q,1}^{\rm pole})
z_Q^{-}(m_{Q,1}^{\rm pole},m_{Q,2}^{\rm
pole})}\sim\frac{m_Q\mph}{\la}\,,\label{(34.1.6)}
\eq
\bbq
z_Q^{-}(m_{Q,1}^{\rm pole},m_{Q,2}^{\rm pole})=
\Bigl (\frac{m_{Q,2}^{\rm pole}}{m_{Q,1}^{\rm pole}}\Bigr
)^{\gamma_Q^{-}}\,,\quad
\frac{m_{Q,2}^{\rm pole}}{m_{Q,1}^{\rm pole}}\sim\frac{\mph}{\mo}\ll 1\,.
\eeq
Besides, because $n_2>N_c$ in these br2-vacua, the quarks ${\ov Q}_2, Q^2$ are
not higgsed  due to the rank restriction, as otherwise the flavor
symmetry $U(n_2)$ will be further broken spontaneously. Therefore,  the
overall phase is $HQ_1-HQ_2$. All quarks are not higgsed but weakly
confined, the confinement originates from ${\cal N}=1\,\, SU(N_c)$ \,SYM, so
that the scale of the confining string is $\sqrt\sigma\sim\langle\lym^{(\rm
br2)}\rangle\ll m_{Q,2}^{\rm pole}$, see \eqref{(34.1.1)},\eqref{(34.1.6)}.

After the quarks ${\ov Q}_2, Q^2$ decouple at $\mu<m_{Q,2}^{\rm pole}$, there
remain ${\cal N}=1 \,\, SU(N_c)$ SYM and $N_F^2$ fions $\Phi^j_i$. The value of
the scale factor $\langle\lym^{\rm (br2)}\rangle$ of SYM is determined from the
matching
\bbq
a_{-}(\mu=m_{Q,2}^{\rm pole})=\Bigl (\frac{\la}{m_{Q,1}^{\rm pole}}\Bigr
)^{\nu^{+}_Q}\Bigl (\frac{m_{Q,1}^{\rm pole}}{m_{Q,2}^{\rm pole}}\Bigr
)^{\nu^{-}_Q}=a^{\rm str}_{YM}(\mu=m_{Q,2}^{\rm pole})=
\Bigl (\frac{m_{Q,2}^{\rm pole}}{\langle\lym^{\rm (br2)}\rangle}\Bigr )^3\quad
\ra
\eeq
\bq
\quad \ra\quad \Bigl (\langle\lym^{\rm (br2)}\rangle\Bigr
)^3\sim\la^3\Bigl(\frac{m_Q}{\la}\Bigr )^
{\frac{n_2-n_1}{n_2-N_c}}\Bigl (\frac{\mph}{\la}\Bigr
)^{\frac{n_2}{n_2-N_c}}\,,\label{(34.1.7)}
\eq
as it should be, see \eqref{(34.1.1)},
\bbq
\frac{\langle\lym^{\rm (br2)}\rangle}{m_{Q,2}^{\rm pole}}\sim\Bigl
(\frac{m_Q}{\la}\Bigr )^{\frac{3N_c-2N_F+n_1}{3(n_2-N_c)}}\Bigl
(\frac{\mph}{\la}\Bigr )^{\frac{3N_c-2N_F+2n_1}{3(n_2-N_c)}}\ll\Bigl
(\frac{m_Q}{\la}\Bigr )^{\frac{3N_c-2N_F+n_1}{3(n_2-N_c)}}\Bigl
(\frac{\mo}{\la}\Bigr )^{\frac{3N_c-2N_F+2n_1}{3(n_2-N_c)}}\sim 
\eeq
\bbq
\sim\Bigl (\frac{m_Q}{\la}\Bigr
)^{\frac{(3N_c-2N_F)(\nd-n_1)}{3N_c(n_2-N_c)}\,>\,0}\ll 1\,,\quad
n_1<\nd\,,\quad n_2>N_c\,. 
\eeq

After integrating out all gluons at $\mu<\langle\lym^{\rm (br2)}\rangle$ via
the \, VY procedure \cite{VY}, the lower energy Lagrangian of $N_F^2$ fions
$\Phi^j_i$ looks as, see \eqref{(30.0.1)} for ${\cal W}_{\Phi}$ and
\eqref{(34.1.3)},\eqref{(34.1.6)},
\bq
K_{\Phi}=z_{\Phi}^{+}(\la,m_{Q,1}^{\rm pole})\, {\rm Tr}\,\Bigl [
(\Phi^1_1)^\dagger \Phi^1_1+\Bigl ((\Phi^1_2)^\dagger
\Phi^1_2+(\Phi^2_1)^\dagger \Phi^2_1)+z_{\Phi}^{-}(m_{Q,1}^{\rm
pole},m_{Q,2}^{\rm pole})\, (\Phi^2_2)^\dagger \Phi^2_2 \Bigr
],\label{(34.1.8)}
\eq
\bbq
{\cal W}={\cal W}_{\Phi}+{\cal W}_{\rm{non-pert}},\quad {\cal
W}_{\rm{non-pert}}=N_c\Bigl (\la^{\rm \bo}\det m_Q^{\rm tot}\Bigr
)^{1/N_c},\quad m_Q^{\rm tot}=(m_Q-\Phi)\,.
\eeq

From \eqref{(34.1.8)}, the masses of the third generation fions look as, see
\eqref{(34.1.1)},\eqref{(34.1.2)},\eqref{(34.1.6)},
\bq
\mu^{\rm pole}_{3}(\Phi_1^1)\sim\frac{\mph}{z_{\Phi}^{+}(\la,m_{Q,1}^{\rm
pole})}=\mph\Bigl (z_Q^{+}(\la,m_
{Q,1}^{\rm pole})\Bigr )^2\sim \mph\Bigl (\frac{m_Q}{\la}\Bigr
)^{\frac{2(2N_c-N_F)}{N_c}}\,,\label{(34.1.9)}
\eq
\bbq
z_{\Phi}^{\pm}(\mu_1,\mu_2)=(\frac{\mu_2}{\mu_1})^{\gamma_{\Phi}^{\pm}\,=
\,-2\gamma_Q^{\pm}}\,,\quad\frac{\mu^{\rm
pole}_{3}(\Phi_1^1)}{m_{Q,1}^{\rm pole}}\ll\frac{\mu^{\rm
pole}_{3}(\Phi_1^1)}{m_{Q,2}^{\rm pole}}\sim\Bigl (\frac{m_Q}{\la}\Bigr
)^{\frac{3N_c-2N_F}{N_c}}\ll 1\,,
\eeq
\bq
\mu^{\rm pole}_{3}(\Phi_2^2)\sim\frac{\langle S\rangle_{\rm br2}}{\qmb_{\rm
br2}^2}\,\frac{1}{z_{\Phi}^{+}(\la,m_{Q,1}^{\rm pole})z_{\Phi}^{-}(m_{Q,1}^{\rm
pole},m_{Q,2}^{\rm pole})}\sim\frac{\langle\Qt\rangle_{\rm
br2}}{\langle\Qo\rangle_{\rm br2}}\,
\frac{\mu^{\rm pole}_{3}(\Phi_1^1)}{z_{\Phi}^{-}(m_{Q,1}^{\rm
pole},m_{Q,2}^{\rm pole})}\,,\label{(34.1.10)}
\eq
\bbq
\mu^{\rm pole}_{3}(\Phi_2^2)\sim\la\Bigl (\frac{m_Q}{\la}\Bigr
)^{\frac{2N_c-N_F}{n_2-N_c}}\Bigl
(\frac{\mph}{\la}\Bigr)^{\frac{2N_c-N_F+n_1}{n_2-N_c}},\quad\quad 
\frac{\mu^{\rm pole}_{3}(\Phi_2^2)}{\langle\lym^{\rm (br2)}\rangle}\ll 1\,,
\eeq
(the main contribution to $\mu^{\rm pole}_{3}(\Phi_1^1)$ originates from the
term ${\cal W}_{\Phi}\sim \mph{\rm Tr}\,(\Phi^2)$ in \eqref{(34.1.8)}, while
the \, main contribution to $\mu^{\rm pole}_{3}(\Phi_2^2)$ originates from ${\cal
W}_{\rm non-pert}$ in \eqref{(34.1.8)}\,).

The fions $\Phi_1^2$ and $\Phi^1_2$ are the Nambu-Goldstone particles and are
massless.\\

On the whole for this case the mass spectrum looks as follows.\\
1) Among the masses smaller than $\la$, the largest are the masses of $N_F^2$
2-nd generation fions,
$\mu^{\rm pole}_2(\Phi)\sim\mos\sim\la(\la/\mph)^{\nd/(5N_c-3N_F)}$.\\
2) Next are masses of ${\ov Q}_1, Q^1$ quarks, $m_{Q,1}^{\rm pole}\sim\la
(m_Q/\la)^{\nd/N_c}$, see \eqref{(34.1.2)}.\\
3) The masses of ${\ov Q}_2, Q^2$ quarks are $m_{Q,2}^{\rm pole}\sim
m_Q\mph/\la\ll m_{Q,1}^{\rm pole}$, see \eqref{(34.1.6)}.\\
4) There is a large number of $SU(N_c)$ SYM gluonia with the mass scale
$\sim\langle\lym^{\rm (br2)}\rangle\ll m_{Q,2}^{\rm pole}$, see
\eqref{(34.1.7)}\,.\\
5) The 3-rd generation fions $\Phi_1^1$ have masses $\mu^{\rm
pole}_{3}(\Phi_1^1)\sim\mph (m_Q/\la)^{2(2N_c-N_F)/N_c}\ll m_{Q,1}^
{\rm pole}$, see  \eqref{(34.1.9)}.\\
6) The 3-rd generation fions $\Phi_2^2$ have masses \eqref{(34.1.10)}, 
$\mu^{\rm  pole}_{3}(\Phi_2^2)\ll\langle\lym^{\rm (br2)}\rangle$.\\
7) The 3-rd generation fions $\Phi_1^2, \Phi_2^1$ are the massless
Nambu-Goldstone particles.

The overall hierarchy of nonzero masses look as
\bbq
\mu^{\rm pole}_{3}(\Phi_2^2)\ll\langle\lym^{\rm (br2)}\rangle\ll m_{Q,2}^{\rm
pole}\ll m_{Q,1}^{\rm pole}\ll\mu^{\rm pole}_{2}(\Phi)\ll\la\ll\mu_1^{\rm
pole}(\Phi)\sim\mph,
\eeq
\bbq
\mu^{\rm pole}_{3}(\Phi_2^2)\ll\mu^{\rm pole}_{3}(\Phi_1^1)\ll m_{Q,2}^{\rm
pole}\ll m_{Q,1}^{\rm pole}\,,\quad \la\ll\mph\ll\mo=\la(\la/
m_Q)^{(2N_c-N_F)/N_c}.
\eeq
\vspace*{1mm}

Let us point out also the following, see section 8.1 in \cite{ch13}. The dual
$SU(N_c)$ theory in br1-vacua considered therein looks as (see eq.(8.1.10) in
\cite{ch13}\,)
\bbq
K={\rm Tr}\,(\Phi^\dagger\Phi)+\frac{1}{\lt^2}{\rm Tr}\,(\tp^\dagger \tp)+{\rm
Tr}\,({\tq}^\dagger \tq+
{\otq}^\dagger {\otq}),\quad \tm=\frac{N_c}{\nd}\,m\,,\quad \wmu=-\mx\,,\quad
m\ll\mx\ll\lm\,,
\eeq
\bbq
W_{\rm matter}={\cal W}_{\Phi}+ {\rm Tr}\,\,(\tm-\Phi)\tp\,-\frac{1}{\lt}{\rm
Tr}\,(\,{\otq}\, \tp \tq \,)\,,
\quad {\cal W}_{\Phi}=\frac{\wmu}{2}\Bigl (\,{\rm
Tr}\,(\Phi^2)-\frac{1}{N_c}({\rm Tr}\,\Phi)^2\Bigr )\,.
\eeq

Let us compare now the mass spectra of this theory with those of
\eqref{(30.0.1)},
with the substitution (see eq.(8.1.24) in \cite{ch13}, both these $SU(N_c)$
theories with $N_c<N_F<3N_c/2$ quark flavors are strongly coupled at 
scales \,$\mu<\la$):
\bq
\la= - \lt=(\lm)\Biggl (\frac{\lm}{\mx}\Biggr )^{\frac{\nd}{3N_c-2N_F}},\quad
m_Q=m\,\frac{\lt}{\mph}\,, \quad \mph=\frac{\lt^2}{\mx}\,,
\quad m\ll\mx\ll\lm\,. \label{(34.1.11)}
\eq
The particle masses smaller than $\mph$ look then as follows\,:\\
a) the masses $m^{\rm pole}_{Q,1}$ of ${\ov Q}_1, Q^1$ quarks, see
\eqref{(34.1.2)}, compare with eq.(8.1.11) in \cite{ch13},
\bbq
m^{\rm pole}_{Q,1}\sim\la \Bigl (\frac{m_Q}{\la}\Bigr
)^{\frac{\nd}{N_c}}\sim\lm(\frac{m}{\lm})^{{\frac{\nd}{N_c}}}\sim
\mu_{\tq,1}^{\rm pole}\,, \hspace*{9cm} \label{(34.1.12.a)}
\eeq
b) the masses $m^{\rm pole}_{Q,2}$ of ${\ov Q}_2, Q^2$ quarks, see
\eqref{(34.1.6)}, compare with eq.(8.1.11) in \cite{ch13},
\bbq
m_{Q,2}^{\rm pole}\sim \frac{m_Q\mph}{\la}\sim m\sim \mu_{\tq,2}^{\rm
pole}\,,\hspace*{11.5cm} \label{(34.1.12.b)}
\eeq
c) the mass scale of gluonia, see  \eqref{(34.1.7)}, compare with
eq.(8.1.8), eq.(8.1.9)  in \cite{ch13},
\bbq
\langle\lym^{\rm (br2)}\rangle^3\sim\la^3\Bigl(\frac{m_Q}{\la}\Bigr
)^{\frac{n_2-n_1}{n_2-N_c}}\Bigl (\frac{\mph}{\la}\Bigr
)^{\frac{n_2}{n_2-N_c}}\sim \mx m^2\Bigl
(\frac{m}{\lm}\Bigr)^{\frac{2N_c-N_F}{\nt-N_c}}\sim\langle\lym^
{\rm(br1)}\rangle^3\,,\,\hspace*{3.5cm} \label{(34.1.12.c)}
\eeq
d) The masses of 3-rd generation fions $\Phi_1^1$, see \eqref{(34.1.9)},
compare \,with eq.8.1.16) in \cite{ch13},
\bbq
\mu^{\rm pole}_{3}(\Phi_1^1)\sim\mph \Bigl (\frac{m_Q}{\la}\Bigr
)^{\frac{2(2N_c-N_F)}{N_c}}\sim\mx\Bigl (\frac{m}{\lm} \Bigr
)^{\frac{2(2N_c-N_F)}{N_c}}\sim\mu^{\rm pole}({M}_1^1)\,,\hspace*{5cm}
\label{(34.1.12.d)}
\eeq
e) The masses of 3-rd generation fions $\Phi_2^2$, see \eqref{(34.1.10)}, 
compare \,with eq.(8.1.17) in \cite{ch13},
\bbq
\mu^{\rm pole}_{3}(\Phi_2^2)\sim\la\Bigl (\frac{m_Q}{\la}\Bigr
)^{\frac{2N_c-N_F}{n_2-N_c}}\Bigl (\frac{\mph}{\la}\Bigr
)^{\frac{2N_c-N_F+n_1}{n_2-N_c}}\sim\mx\Bigl (\frac{m}{\lm}\Bigr
)^{\frac{2N_c-N_F}{n_2-N_c}}\sim\mu^{\rm pole}({M}_2^2)\,,
\hspace*{2.8cm} \label{(34.1.12.e)}
\eeq
f) The hybrids $\Phi^1_2$ and $M^1_2$ are massless.\\

Therefore, with the choice of parameters as in \eqref{(34.1.11)} and with
correspondences $Q\leftrightarrow\textsf{q}$ and $\Phi\leftrightarrow M$, 
the  mass spectra coincide in the direct $SU(N_c)$ theory
\eqref{(30.0.1)} (this is the direct theory eq.(8.1.18) in \cite{ch13}\,) and in
the  dual $SU(N_c)$ theory eq.(8.1.10) in \cite{ch13}.  Both are strongly 
coupled  at $N_c < N_F < 3N_c/2$ $\,SU(N_c)$ theories.  Much more details 
can  be found in sections 8.1-8.3  of \cite{ch13}.

\subsection{\quad Dual theory}  

For the reader convenience and to make this article self-contained, we
reproduce below in short the results from the section 7.3 in \cite{ch16} for
the \, mass spectra of the weakly coupled dual theory \eqref{(30.0.2)} (with
\eqref{(34.1.11)},
the spectrum of masses  in the case "A"\, below is the same
as in the theories eq.(8.1.7) and eq.(8.1.19) in \cite{ch13}). \\

{\bf A)} The range $\la\ll\mph\ll \la(\la/m_Q)^{1/2}$\\

The overall phase is $Higgs_1-Hq_2$ in this range due to
higgsing of dual quarks $q_1,{\ov q}^1$, $SU(\nd)\ra SU(\nd-n_1)$. On the whole
for this case the mass spectrum looks as follows.

1) The heaviest (among the masses $<\la$) are $N_F^2$ mions $M^i_j$ with 
masses  $\mu^{\rm pole}(M)\sim \la^2/\mph\,$.

2) There are $n_1(2\nd-n_1)$ massive dual gluons (and their super-partners)
with masses \\ ${\ov\mu}^{\,\rm pole}_{\rm gl,1}\sim\langle N_1\rangle^{1/2}
\sim (m_Q\la)^{1/2}$.

3) There is a large number of hadrons made of weakly interacting
non-relativistic and weakly confined dual quarks $q^{\,\prime}_2$ and ${\ov
q}^{\,\prime,\, 2}$ with unbroken colors, the scale of their masses is
$\mu^{\rm pole}_{q,2}\sim m_Q\mph/\la\sim m_{Q,2}^{\rm pole}$, see
\eqref{(34.1.6)}, \, (the tension of the confining string is much smaller, $\sqrt
\sigma\sim\langle\lym^{(\rm br2)}\rangle\ll\mu^{\rm pole}_{q,2}$, see
\eqref{(34.1.7)}).

4) The masses of $n_1^2$ nions $N_1^1$ (dual pions) are also $\mu^{\rm
pole}(N_1^1)\sim m_Q\mph/\la$.

5) There is a large number of $SU(\nd-n_1)$ SYM gluonia, the scale of their
masses is \\ $\sim\langle\lym^{(\rm br2)}\rangle\sim (m_Q\langle
M_1\rangle_{\rm br2})^{1/3}=(m_Q\langle\Qo\rangle_{\rm br2})^{1/3}$, see
\eqref{(34.1.1)},\eqref{(34.1.7)}.

6) Finally, $2n_1 n_2$ Nambu-Goldstone hybrid nions $N_1^2, N_2^1$ are
massless. The overall hierarchy of nonzero masses looks as:
\bq
\langle\lym^{(\rm br2)}\rangle\ll\mu(N_1^1)\sim\mu^{\rm
pole}_{q,2}\ll{\ov\mu}_{\rm gl,1}\ll\mu^{\rm pole}(M)\ll\la\,.\label{(34.2.1)}
\eq

{\bf B)} The range $\la(\la/m_Q)^{1/2}\ll\mph\ll
\mo=\la(\la/m_Q)^{(2N_c-N_F)/N_c}$\\

The largest mass have in this case $q_2, {\ov q}^{\,2}$ quarks, 
\bq
\mu^{\rm pole}_{q,2}\sim\frac{\langle M_2\rangle_{\rm
br2}}{\la}=\frac{\langle\Qt\rangle_{\rm
br2}}{\la}\sim\frac{m_Q\mph}{\la}\,.\label{(34.2.2)}
\eq
After integrating them out at $\mu<\mu^{\rm pole}_{q,2}$, the new 
scale factor  of the gauge coupling is
\bq
\Bigl (\Lambda^{\prime\prime}\Bigr )^{3\nd-n_1}=\la^{\bd}\Bigl
(\frac{m_Q\mph}{\la}\Bigr )^{n_2},
\quad\Lambda^{\prime\prime}\ll \mu^{\rm pole}_{q,2}\,,\quad \bd=3\nd-N_F\,,
\quad n_1<\nd\,.\label{(34.2.3)}
\eq

In the range $(\la^3/m_Q)^{1/2}\ll\mph\ll {\mu}_{\Phi}^{\prime\prime}=\la\Bigl
(\la/m_Q\Bigr )^{(3N_c-N_F-n_1)/2{\rm n}_2}\ll\mo$ the hierarchies look as:
${\ov\mu}_{\rm gl,1}\gg\Lambda^{\prime\prime}\gg\mu^{\rm pole}_{q,1}$, and
therefore the quarks $q_1, {\ov q}^{\,1}$ are higgsed in the weak coupling
region at $\mu={\ov\mu}_{\rm gl,1}\gg\Lambda^{\prime\prime}$,  the overall
phase is also $Higgs_1-Hq_2$. The mass spectrum looks in this region
as follows (as previously, all mass values are given below up to logarithmic
factors). - \\
1) There are $2n_1n_2$ not confined dual quarks $q_2, {\ov q}^{\,2}$ with
broken \, colors, their masses (up to logarithmic factors) are $\mu^{\rm
pole}_{q,2}\sim\langle M_2\rangle_{\rm br2}/\la\sim m_Q\mph/\la$.\\
2) The quarks $q_2^{\prime}, {\ov q}^{\,\prime,\,2}$ with unbroken
$SU(\nd-n_1)$  dual colors are confined and there is a large number of dual
hadrons made from
these weakly coupled and weakly confined non-relativistic quarks, the scale of
their masses is also $\mu_{H}\sim (m_Q\mph/\la)$ (the tension of the confining
string originated from $SU(\nd-n_1)\,\,\, {\cal N}=1$ SYM is
$\sqrt\sigma\sim\langle\lym^{(\rm br2)}\rangle\ll\mu^{\rm pole}_{q,2} $).\\
3) $n_1(2\nd-n_1)$ massive gluons (and their super-partners) due to higgsing
$SU(\nd)\ra SU(\nd-n_1)$ by $q_1, {\ov q}^{\,1}$ quarks, $\,\,{\ov\mu}^{\,\rm
pole}_{\rm gl,1}\sim (m_Q\la)^{1/2}\ll\mu^{\rm pole}_{q,2} $.\\
3) $n_1^2$ nions $N_1^1$ and $n_1^2$ mions $M_1^1$ also have masses 
$\mu^{\rm  pole}(N_1^1)\sim\mu^{\rm pole}(M_1^1)\sim (m_Q\la)^{1/2}$.\\
4) There is a large number of gluonia from $SU(\nd-n_1)\,\,\, {\cal N}=1$ SYM,
the scale of their masses is
$\sim \langle\lym^{(\rm br2)}\rangle\sim (m_Q\langle M_1\rangle_{\rm
br2})^{1/3}\sim\la (m_Q/\la)^{({\rm n_2-n_1})/3(\nd-{\rm n}_1)} \,
(\mph/\la)^{{\rm  n}_2/3(\nd-{\rm\, n}_1)}$.\\
5) $n_2^2$ mions $M_2^2$ have masses $\mu^{\rm pole}(M_2^2)\sim\la^2/\mph$.\\
6) $2n_1 n_2$ mions $M_1^2$ and $M_2^1$ are Nambu-Goldstone particles and
are massless, $\mu(M_1^2)=\mu(M_2^1)=0$.

The overall hierarchy of nonzero masses looks in this range
$(\la^3/m_Q)^{1/2}<\mu<{\mu}_{\Phi}^{\prime\prime}$ as
\bq
\mu(M_2^2)\ll\langle\lym^{(\rm br2)}\rangle\ll{\ov\mu}_{\rm
gl,1}\sim\mu(N_1^1)\sim\mu(M_1^1)\ll
\mu^{\rm pole}_{q,2}\ll\la\,.\label{(34.2.4)}
\eq

But the hierarchies look as $\mu^{\rm pole}_{q,1}\sim (\langle M_1\rangle_{\rm
br2}/\la)\gg\Lambda^{\prime
\prime}\gg{\ov\mu}_{\rm gl,1}$ at $\mu_{\Phi}^{\prime\prime}\ll\mph\ll\mo$, the
quarks $q_1, {\ov q}^{\,1}$ are then too heavy and not higgsed, and  the
overall phase is $Hq_2-Hq_1$ (heavy quarks). The mass spectrum looks
in this region as follows.\\
1) All quarks are confined and there is a large number of dual hadrons made
from these weakly coupled and weakly confined non-relativistic quarks, the 
scale \, of their masses is $\mu^{\rm pole}_{q,2}\sim (m_Q\mph/\la)\gg\mu^{\rm
pole}_{q,1}\sim (\langle M_1\rangle_{\rm br2}/\la)=(\langle\Qo\rangle_{\rm
br2}/\la)$, see \eqref{(34.1.1)}, (the tension of the confining string
originatedfrom $SU(\nd)\,\,\, {\cal N}=1$ SYM is
$\sqrt\sigma\sim\langle\lym^{(\rm
br2)}\rangle\ll\mu^{\rm pole}_{q,1}\ll\mu^{\rm pole}_{q,2})$.\\
2) A large number of gluonia from $SU(\nd)\,\,\, {\cal N}=1$ SYM, the scale of
their masses is
$\sim \langle\lym^{(\rm br2)}\rangle\sim (m_Q\langle M_1\rangle_{\rm
br2})^{1/3}\sim\la (m_Q/\la)^{({\rm n_2-n_1})/3(\nd-{\rm n}_1)}(\mph/\la)^{{\rm
n}_2/3(\nd-{\rm n}_1)}$.\\
3) $n_1^2$ mions $M_1^1$ with masses $\mu^{\rm pole}(M_1^1)\sim \la^2\langle
M_2\rangle/\mph\langle M_1\rangle$.\\
4) $n_2^2$ mions $M_2^2$ with masses $\mu^{\rm pole}(M_2^2)\sim
(\la^2/\mph)\ll\mu^{\rm pole}(M_1^1)$.\\
5) $2\no\nt$ mions $M_1^2$ and $M_2^1$ are the massless 
Nambu-Goldstone  particles. \\

The overall hierarchy of nonzero masses look in this range
$\mu_{\Phi}^{\prime\prime}<\mph<\mo$ as
\bq
\mu^{\rm pole}(M_2^2)\ll\mu^{\rm pole}(M_1^1)\ll\langle\lym^{(\rm
br2)}\rangle\ll\mu^{\rm pole}_{q,1}\ll\mu^{\rm
pole}_{q,2}\ll\la\,.\label{(34.2.5)}
\eq

Comparing with the direct theory in section 34.1 it is seen that the mass
spectra are parametrically different.

\section{Unbroken flavor symmetry, \, QCD vacua}

\subsection{Direct theory.}

The direct theory is in the strong coupling regime at $\mu<\la$. The quark
condensates look here as, see section 16,
\bq
\QQ_{\rm QCD}\simeq\la^2\Bigl (\frac{m_Q}{\la}\Bigr )^{\nd/N_c},\,\,
\langle\lym^{\rm (QCD)}\rangle ^3
\equiv\langle S\rangle_{\rm QCD}\simeq\Bigl (\la^{\rm \bo}m_Q^{N_F}\Bigr
)^{1/N_c},\,\, \bo=3N_c-N_F\,.\label{(35.1.1)}
\eq
The potentially important masses look here as
\bbq
\langle m^{\rm tot}_Q\rangle_{\rm QCD}=\langle m_Q-\Phi\rangle_{\rm
QCD}=\frac{\langle S\rangle_{\rm QCD}}{\QQ_{\rm QCD}}\simeq m_Q\,,
\eeq
\bq
m^{\rm pole}_{Q,\rm QCD}=\frac{\langle m^{\rm tot}_Q\rangle_{\rm
QCD}}{z_Q(\la,m^{\rm pole}_{Q,\rm QCD})}\sim \la\Bigl (\frac{m_Q}{\la}\Bigr
)^{\nd/N_c},\,\, z_Q(\la,m^{\rm pole}_{Q,\rm QCD})=\Bigl (\frac{m^{\rm
pole}_{Q,\rm QCD}}{\la}\Bigr )^{\gamma_Q=(2N_c-N_F)/\nd}\,,\label{(35.1.2)}
\eq
\bq
\mu^2_{\rm gl,QCD}\sim a_{+}(\mu=\mu_{\rm gl,QCD}) 
z_Q(\la,\mu_{\rm gl,QCD})\QQ_{\rm QCD}\ra\mu_{\rm gl,QCD}
\sim\frac{\QQ_{\rm QCD}}{\la}\sim m^{\rm pole}_{Q,\rm QCD}\,,
\label{(35.1.3)}
\eq

\bbq
a_{+}(\mu=\mu_{\rm gl,QCD})\sim \Bigl (\frac{\la}{\mu_{\rm gl,QCD}}\Bigr
)^{\nu_Q=(3N_c-2N_F)/\nd},\quad
a_{+}(\mu=\mu_{\rm gl,QCD})\,z_Q(\la,\mu_{\rm gl,QCD})\sim\frac{\mu_{\rm
gl,QCD}}{\la}\,.
\eeq
As before for vacua with the unbroken flavor symmetry, this implies from the
rank restriction at $N_F>N_c$ that quarks are not higgsed but confined and 
the overall phase is HQ (heavy quark).

Besides, see \eqref{(34.1.4)} for $\mos$ and \eqref{(35.1.2)},
\bq
\frac{m^{\rm pole}_{Q,\rm QCD}}{\mos}\sim \Bigl (\frac{\mph}{{\hat
\mu}_{\Phi,\rm QCD}}
\Bigr )^{\frac{\nd}{5N_c-3N_F}},\,\, {\hat \mu}_{\Phi,\rm QCD}=\la\Bigl
(\frac{\la}{m_Q}\Bigr )^{\frac{5N_c-3N_F}{N_c}}\gg\mo,\,\,
\frac{1}{2}<\frac{5N_c-3N_F}{N_c}<2.\,\,\, \label{(35.1.4)}
\eq

It follows from \eqref{(35.1.4)} that\,:\, a) $\,{m^{\rm pole}_{Q,\rm
QCD}}<\mos$
in the range $\mo<\mph<{\hat \mu}_{\Phi,\rm QCD}$ and so there will be two
additional generations of $\Phi$-particles, $\mu_{2}^{\rm pole}(\Phi)\sim\mos$
and $\mu_{3}^{\rm pole}(\Phi)\ll m^{\rm pole}_{Q,\rm QCD}$, and all $N_F^2$
fions $\Phi^j_i$ will be dynamically relevant at scales $\mu_{3}^{\rm
pole}(\Phi)<\mu<\mu_{2}^{\rm pole}(\Phi)$;\, b)\, while ${m^{\rm pole}_{Q,\rm
QCD}}>\mos$ at $\mph>{\hat \mu}_{\Phi,\rm QCD}$ and all fions will be too heavy
and dynamically irrelevant at all scales $\mu<\mu_{1}^{\rm
pole}(\Phi)\sim\mph$.\, Proceeding further in a "standard" way, i.e.
integrating \, out first all quarks \, as heavy ones at $\mu<m^{\rm pole}_]
{Q,\rm QCD}$,  the scale  factor of the lower  energy 
$SU(N_c)$ SYM in the strong coupling regime is determined from the
matching
\bbq
a_{+}(\mu=m^{\rm pole}_{Q,\rm QCD})=\Bigl (\frac{\la}{m^{\rm pole}_{Q,\rm
QCD}}\Bigr )^{(3N_c-2N_F)/\nd}=a^{\rm str}_{\rm YM}(\mu=m^{\rm pole}_{Q,\rm
QCD})=\Bigl (\frac{m^{\rm pole}_{Q,\rm QCD}}{\langle\lym^{(\rm
QCD)}\rangle}\Bigr )^3\quad\ra
\eeq
\bq
\ra\quad\langle\lym^{(\rm QCD)}\rangle=\Bigl (\la^{3N_c-N_F}m_Q^{N_F}\Bigr
)^{1/3N_c} \label{(35.1.5)}
\eq
as it should be, see \eqref{(35.1.1)}.

Integrating out now all $SU(N_c)$ gluons at $\mu<\lym^{(\rm QCD)}$ via the
VY-procedure \cite{VY},
we obtain the low energy Lagrangian of $N_F^2$ fions $\Phi^j_i$
\bq
K=z_{\Phi}(\la,m^{\rm pole}_{Q,\rm QCD})\,{\rm Tr}\,(\Phi^\dagger\Phi),  
z_{\Phi}(\la,m^{\rm pole}_{Q,\rm QCD})=\Bigl(\frac{m^{\rm pole}_Q}{\la}\Bigr
)^{\gamma_{\Phi}}=\Bigl(\frac{\la}{m^{\rm pole}_Q}\Bigr )^{2(\bb)/\nd}\gg
1,\,\,\,\,\label{(35.1.6)}
\eq
\bbq
{\cal W}={\cal W}_{\Phi}+N_c\Bigl(\la^{\rm \bo}\det m^{\rm tot}_Q\Bigr
)^{1/N_c}\,,\quad m^{\rm tot}_Q=m_Q-\Phi\,,\quad{\cal
W}_{\Phi}=\frac{\mph}{2}\Biggl [{\rm Tr}\,(\Phi^2)-\frac{1}{\nd}\Bigl ({\rm
Tr}\,\Phi\Bigr )^2\Biggr ].
\eeq
The main contribution to the masses of $N_F^2$ 3-rd generation fions at
$\mo<\mph<{\hat\mu}_{\Phi,\rm QCD}$ originates from the term ${\cal W}_{\Phi}$
in \eqref{(35.1.6)}
\bq
\mu_{3}^{\rm pole}(\Phi)\sim\frac{\mph}{z_{\Phi}(\la,m^{\rm pole}_{Q,\rm QCD})}
\sim\mph\Bigl(\frac{m_Q}{\la}\Bigr )^{2(\bb)/N_c}\,,\label{(7.7)}
\label{(35.1.7)}
\eq
\bbq
\quad \frac{\mu_{3}^{\rm pole}(\Phi)}{m^{\rm pole}_{Q,\rm QCD}}
\sim\frac{\mph}{\la}
\Bigl(\frac{m_Q}{\la}\Bigr
)^{\frac{5N_c-3N_F}{3N_c}}=\frac{\mph}{{\hat\mu}_{\Phi,\rm QCD}}<1\,.
\eeq

On the whole for the mass spectra in these $N_c$ QCD-vacua at
$\mph>\mo=\la(\la/m_Q)^{(\bb)/N_c}$. The overall phase is HQ, all quarks
${\ov Q}_j, Q^i$ are not higgsed but confined.

{\bf A}) The range $\mo=\la(\la/m_Q)^{(\bb)/N_c}<\mph<{\hat\mu}_{\Phi,\rm
QCD}=\la (\la/m_Q)^{(5N_c-3N_F)/N_c}$.\\
1) The largest among masses $<\la$ are masses of $N_F^2$ second 
generation  fions \, $\Phi^j_i$ with $\mu_2^{\rm
pole}(\Phi)\sim\mos=\la(\la/\mph)^{\nd/(5N_c-3N_F)}$.\\
2) The next are masses $m^{\rm pole}_{Q,\rm QCD}\sim \la (m_Q/\la)^{\nd/N_c}$
of strongly interacting but weakly confined quarks (the tension of the 
confiningstring originating from ${\cal N}=1\,\, SU(N_c)$ SYM is
$\sqrt\sigma\sim\lym^{(\rm QCD)}\ll m^{\rm pole}_{Q,\rm QCD}$).\\
3) There are $N_F^2$ third generation fions $\Phi^j_i$ with $\mu_3^{\rm
pole}(\Phi)\sim\mph (m_Q/\la)^{2(\bb)/N_c}$.
\\
4) A large number of gluonia from ${\cal N}=1\,\, SU(N_c)$ SYM, the scale of
their masses is $\lym^{(\rm QCD)}\sim (\la^{\rm \bo}m_Q^{N_F})^{1/3N_c}$.\\

The mass hierarchies look as
\bq
\mu_3^{\rm pole}(\Phi)\ll m^{\rm pole}_{Q,\rm QCD}\ll\mu_2^{\rm
pole}(\Phi)=\mos\ll\la\ll\mu_1^{\rm pole}(\Phi)\sim\mph\,,\quad \lym^{(\rm
QCD)}\ll m^{\rm pole}_{Q,\rm QCD}\, .\label{(35.1.8)}
\eq

{\bf B}) The range $\mph>{\hat\mu}_{\Phi,\rm QCD}=\la
(\la/m_Q)^{(5N_c-3N_F)/N_c}\gg{\tilde\mu}_{\Phi,\rm QCD}\gg\mo$\,.\\
The difference is that there are no additional generations of all fions in this
case because $m^{\rm pole}_{Q,\rm QCD}>\mos$, and all fions remain too heavy
and  dynamically irrelevant at all scales $\mu<\mu_1^{\rm pole}(\Phi)\sim\mph$.

\subsection{Dual theory.}

The dual theory is in the IR free logarithmic regime at scales
$\mu_{q,QCD}^{\rm pole}<\mu<\la$ (as before, all logarithmic factors are
ignored \, for simplicity).  The potentially important masses look
here as follows. The masses of dual quarks
\bq
\mu_{q,QCD}^{\rm pole}\sim\frac{\langle M\rangle_{\rm QCD}}{\la}=\frac{\langle
{\ov Q}Q\rangle_{\rm QCD}}{\la}\sim\la\Bigl(\frac{m_Q}{\la}\Bigr
)^{\frac{\nd}{N_c}}\sim m^{\rm pole}_{Q,\rm QCD}.\label{(35.2.1)}
\eq
The masses of dual gluons due to possible higgsing of dual quarks
\bq
{\ov\mu}^{\,\rm pole}_{\rm gl,QCD}\sim \langle ({\ov q}q)\rangle^{1/2}_{\rm
QCD}\sim (m_Q\la)^{1/2},\quad
\frac{{\ov\mu}^{\,\rm pole}_{\rm gl,QCD}}{\mu_{q,QCD}^{\rm pole}}\sim\Bigl
(\frac{m_Q}{\la}\Bigr )^{\frac{3N_c-2N_F}{2N_c}}\ll 1\,.\label{(35.2.2)}
\eq
Therefore, the overall phase is Hq (heavy quark), all quarks ${\ov q}^j,
q_i$ are not higgsed but confined. All $N_F^2$ fions $\Phi^j_i$ have large
masses $\mu^{\rm pole}_{1}(\Phi)\sim\mph\gg\la$ and are irrelevant at scales
$\mu<\mph$, they can be integrated out from the beginning. After integrating
out  all dual quarks as heavy ones at $\mu<\mu_{q,QCD}^{\rm pole}$ and then all
dual gluons at $\mu<\langle\lym^{(QCD)}\rangle$ via the VY procedure
\cite{VY}, \, the low energy Lagrangian of $N_F^2$ mions $M^i_j$ looks as
\bq
K\sim \,{\rm Tr}\,\frac{M^\dagger M}{\la^2},\,\, {\cal W}={\cal W}_M-\nd\Bigl
(\frac{\det M}{\la^{\rm \bo}}\Bigr )^{1/\nd},\,\, {\cal W}_M=m_Q {\rm
Tr}\,M-\frac{1}{2\mph}\Bigl [{\rm Tr}\,(M^2)-\frac{1}{N_c}\Bigl ({\rm Tr}\,
M\Bigr )^2\Bigr ].\,\,\,\, \label{(35.2.3)}
\eq
The main contribution to the masses of $N_F^2$ mions originates from the
nonperturbative term in \eqref{(35.2.3)}
\bq
\mu^{\rm pole}(M)\sim\frac{\langle S\rangle_{\rm QCD}\la^2}{\langle
M\rangle^2_{\rm QCD}}\sim\la\Bigl(\frac{m_Q}{\la}\Bigr )^{(\bb)/N_c}\,,\quad
\frac{\mu^{\rm pole}(M)}{\lym^{(QCD)}}\sim\Bigl (\frac{m_Q}{\la}\Bigr
)^{\frac{2(3N_c-2N_F)}{3N_c}}\ll 1\,.\label{(35.2.4)}
\eq

Comparing with the direct theory in section 35.1 it is seen that at
$\mo<\mph<{\hat\mu}_{\Phi,\rm QCD}$ the masses $\mu_3^{\rm pole}(\Phi)$
\eqref{(35.1.7)} and $\mu^{\rm pole}(M)$ \eqref{(35.2.4)} are parametrically
different (the triangles $SU^3(N_F)_L$ are also different). Besides, there are
no particles with masses smaller than $\langle\lym^{(QCD)}\rangle$ at
$\mph>{\hat\mu}_{\Phi,\rm QCD}$ in the direct theory while $\mu^{\rm
pole}(M)\ll\langle\lym^{(QCD)}\rangle$ in the dual one, see \eqref{(35.2.4)}.

\section{Broken flavor symmetry.\, Direct theory.}

\subsection{\quad br1 and special vacua with $1\leq n_1<{\rm
Min}\,(N_F/2,\,3N_c-2N_F)$}

The condensates look in this case as, see section 16,
\bq
\langle\Qo\rangle_{\rm br1}\simeq\frac{N_c}{N_c-\no}
m_Q\mph\,,\quad\langle\Qt\rangle_{\rm br1}\sim\Bigl(\frac{m_Q}{\la}\Bigr
)^{\frac{\nt-N_c}{N_c-\no}}\Bigl(\frac{\la}{\mph}\Bigr
)^{\frac{\no}{N_c-\no}}\,,\label{(36.1.1)}
\eq
\bbq
\langle\lym^{\rm (br1)}\rangle^3\equiv\langle S\rangle_{\rm br1}=\frac{
\langle\Qo\rangle_{\rm br1} \langle\Qt\rangle_{\rm br1}}{\mph}\sim
\Bigl(\frac{m_Q}{\la}\Bigr )^{\frac{\nt-\no}{N_c-\no}}
\Bigl(\frac{\la}{\mph}\Bigr )_{,}^{\frac{\no}{N_c-\no}}\quad
\frac{\langle\Qt\rangle_{\rm br2}}{\langle\Qo\rangle_{\rm br2}}\sim\Bigl
(\frac{\mo}{\mph}\Bigr )^{\frac{N_c}{N_c-\no}}\ll 1\,.
\eeq

The potentially important masses look as\,:

a) the quark masses,
\bq
\qma\ll\qmb=\langle m_Q-\Phi_2\rangle=\frac{\langle\Qo\rangle}{\mph}\sim
m_Q,\quad {\tilde m}^{\rm pole}_{Q,2}=\frac{\qmb}{z_Q^{+}(\la,\qmb)}\sim
\la\Bigl(\frac{m_Q}{\la}\Bigr )^{\nd/N_c},\label{(36.1.2)}
\eq

b) the gluon masses due to possible higgsing of quarks,
\bbq
(\mgo)^2\sim a_{+}(\mgo)\, z_Q^{+}(\la,\mgo)\,\langle\Qo\rangle,\quad
a_{+}(\mgo)=\Bigl (\frac{\la}{\mgo}
\Bigr )^{\nu_{+}},\,\, z_Q^{+}(\la,\mgo)=\Bigl (\frac{\mgo}{\la}\Bigr
)^{\gamma^{+}_Q},
\eeq
\bq
\mgo\sim\frac{\langle\Qo\rangle}{\la}\sim\frac{m_Q\mph}{\la},\quad\,
\frac{{\tilde m}^{\rm pole}_{Q,2}}{\mgo}\sim\frac{\mo}{\mph}\ll
1,\quad\quad\gamma^{+}_Q=\frac{2N_c-N_F}{N_F-N_c},\quad
\nu^{+}=\frac{3N_c-2N_F}{N_F-N_c}\,.\label{(36.1.3)}
\eq

Therefore, the quarks $Q^1, {\ov Q}_1$ are higgsed and  the overall phase
is $Higgs_1-HQ_2$ in this case. Besides,
\bq
\frac{\mgo}{\mos}\sim\Bigl (\frac{\mph}{\mu_{\Phi,1}}\Bigr
)^{\frac{2(2N_c-N_F)}{5N_c-3N_F}}>1\,\,\,
{\rm only \, at}\,\,\, \mph>\mu_{\Phi,1}=\la\Bigl (\frac{\la}{m_Q}\Bigr
)^{\frac{5N_c-3N_F}{2(2N_c-N_F)}},\label{(36.1.4)}
\eq
\bbq
\mos=\la\Bigl (\frac{\la}{\mph}\Bigr )^{\frac{1}{2\gamma_Q^{+}-1}}=\la\Bigl
(\frac{\la}{\mph}\Bigl
)^{\frac{\nd}{5N_c-3N_F}}\,,\quad\frac{\mo}{\mu_{\Phi,1}}\sim\Bigl
(\frac{m_Q}{\la}\Bigr )^{\frac{\nd(3N_c-2N_F)}{2N_c(2N_c-N_F)}}\ll
1,\quad\mo=\la\Bigl (\frac{\la}{m_Q}\Bigr )^{\frac{2N_c-N_F}{N_c}}\,,
\eeq
this shows that the fions $\Phi^1_1,\Phi_1^2$ and $\Phi_2^1$ become too heavy
and dynamically irrelevant at all scales $\mu<\mu_1^{\rm pole}(\Phi)\sim
\mph\gg\la$ when $\mph>\mu_{\Phi,1}$ only (the situation with $\Phi^2_2$ is
different, see below), while at $\mo<\mph<\mu_{\Phi,1}$ all $N_F^2$ fions
become relevant at the scale $\mu<\mu^{\rm
pole}_2(\Phi)=\mos=\la(\la/\mph)^{\nd/(5N_c-3N_F)}$ and there appears
additional second generation of all $N_F^2\,\,\Phi$-particles.\\

{\bf A)} \,\, The region $\mo\ll\mph\ll\mu_{\Phi,1}=\la
(\la/m_Q)^{\frac{5N_c-3N_F}{2(2N_c-N_F)}}$.

Because quarks ${\ov Q}_1, Q^1$ are higgsed at $\mu\sim\mgo$, \,
the lower energy theory at $\mu<\mgo$ has the unbroken gauge
group  $SU(N_c-\no)$,\, $n_2$ flavors of quarks ${\ov Q}^{\,\prime}_2,
Q^{\,\prime,\,2}$ with unbroken colors, $N_F^2$ fions $\Phi^j_i$, $\,\no^2$
pions $\Pi^1_1$ (these originated from higgsing of ${\ov Q}_1, Q^1$\,), and
finally $2\no\nt$ hybrid pions $\Pi^1_2,\, \Pi^2_1$ (these in essence are
quarks  ${\ov Q}_2, Q^2$ with broken colors).

In case $1\leq \no<{\rm Min}(N_F/2,\,3N_c-2N_F)$, i.e. either any $1\leq
\no<N_F/2$ at $N_c<N_F<6N_c/5$, or $\no<(3N_c-2N_F)$ at $6N_c/5<N_F<3N_c/2$,
the lower energy theory at $\mu<\mgo$ remains in the strong coupling region
with:$N_c^\prime=(N_c-\no),\,\,N_F^\prime=\nt,\,\, 1<
N_F^\prime/N_c^\prime<3/2\,,\,\, {\rm
b}^\prime_o=(3N_c^\prime-N_F^\prime)=(3N_c-N_F-2\no)>0$.

The anomalous dimensions and the gauge coupling look in this lower energy
 theory as, see Appendix A for the values of anomalous dimensions in
the strong coupling region,
\bq
\gamma_Q^{-}=\frac{2N_c^\prime-N_F^\prime}{N_F^\prime-N_c^\prime}=\frac{2N_c-N_F-
\no}{(N_F-N_c)=\nd}= \gamma_Q^{+}-\frac{\no}{\nd}\,>\,1\,,\quad
\gamma_{\Phi}^{\pm}=-2\gamma_Q^{\pm}\,,\label{(36.1.5)}
\eq
\bbq
a_{-}(\mu<\mgo)=\Bigl [a_{+}(\mu=\mgo)=\Bigl (\frac{\la}{\mgo}\Bigr
)^{\nu_{+}}\Bigr ]\Bigl (\frac{\mgo}{\mu}\Bigr )^{\nu_{-}},\quad
\nu_{+}=\frac{3N_c-2N_F}{\nd},\quad \nu_{-}=\nu_{+}-\frac{\no}{\nd}>0\,.
\eeq
The pole mass of ${\ov Q}^{\,\prime}_2, Q^{\,\prime,\,2}$ quarks and $\mgt$ due
to possible higgsing of these quarks look then as, see
\eqref{(36.1.1)},\eqref{(36.1.3)},
\bbq
m_{Q,2}^{\rm pole}=\frac{\qmb}{z^{+}_Q(\la,\mgo)\,z^{-}_Q(\mgo,m_{Q,2}^{\rm
pole})}\,,\,\, z^{+}_Q(\la,\mgo)=\Bigl (\frac{\mgo}{\la}\Bigr
)^{\gamma^{+}_Q},\,\, z^{-}_Q(\mgo,m_{Q,2}^{\rm pole})=\Bigl
(\frac{m_{Q,2}^{\rm pole}}{\mgo}\Bigr )^{\gamma^{-}_Q}\,,
\eeq
\bq
m_{Q,2}^{\rm pole}\sim\Bigl (\frac{\la}{\mph}\Bigr
)^{\frac{\no}{N_c-\no}}\,\Bigl (\frac{m_Q}{\la}\Bigr
)^{\frac{\nd-\no}{N_c-\no}}\sim\frac{\langle\Qt\rangle}{\la}\,,\quad\quad
\frac{m_{Q,2}^{\rm pole}}
{\mgo}\sim\Bigl (\frac{\mo}{\mph}\Bigr )^{\frac{N_c}{N_c-\no}}\ll
1\,,\label{(36.1.6)}
\eq
\bbq
(\mgt)^2\sim \rho_{+}\rho_{-}\langle\Qt\rangle,\,\, \rho_{+} =a_{+}(\mgo)\,
z_Q^{+}(\la,\mgo)
\sim\frac{\mgo}{\la},\,\, \rho_{-} \sim\Bigl (\frac{\mgo}{\mgt}\Bigr )^{\nu^-}
z_Q^{-}(\mgo,\mgt)=\frac{\mgt}{\mgo}\,,
\eeq
\bbq
\rho_{+}\rho_{-}\sim\frac{\mgt}{\la}\,,\quad \mgt\sim m_{Q,2}^{\rm pole}\sim
\frac{\langle\Qt\rangle} {\la}\sim \Bigl(\frac{m_Q}{\la}\Bigr
)^{\frac{\nt-N_c}{N_c-\no}}\Bigl(\frac{\la}{\mph}\Bigr
)^{\frac{\no}{N_c-\no}}\,.
\eeq
Because $\nt-(N_c^\prime=N_c-\no)=N_F-N_c>0$, the rank restriction shows that
the qurks ${\ov Q}^{\,\prime}_2, Q^{\,\prime,\,2}$ are not higgsed because
otherwise the global $U(\nt)$ flavor symmetry will be {\it additionally} broken
spontaneously (and this will be wrong).

Therefore, after integrating out remained active quarks ${\ov Q}^{\,\prime}_2,
Q^{\,\prime,\,2}$ with unbroken $N_c-\no$ colors as heavy ones at
$\mu<m_{Q,2}^{\rm pole}$, the scale factor of the lower energy $SU(N_c)$ SYM in
the strong coupling regime is determined from the matching, see
\eqref{(36.1.3)},\eqref{(36.1.5)},\eqref{(36.1.6)},
\bq
a_{-}(\mu=m_{Q,2}^{\rm pole})=\Bigl (\frac{\la}{\mgo}\Bigr )^{\nu_{+}}\,\Bigl
(\frac{\mgo}{m_{Q,2}^{\rm pole}}\Bigr )^{\nu_{-}}=a_{\rm YM}^{(\rm
br1)}(\mu=m_{Q,2}^{\rm pole})=\Bigl (\frac{m_{Q,2}^{\rm
pole}}{\lambda_{YM}}\Bigr )^3\gg 1\,,\label{(36.1.7)}
\eq
and is
\bq
\lambda_{YM}^3=\Bigl (\langle\lym^{(\rm br1)}\rangle\Bigr )^3=\la^3\Bigl
(\frac{m_Q}{\la}\Bigr )^{\frac{\nt-\no}{N_c-\no}}\Bigl (\frac{\la}{\mph}\Bigr
)^{\frac{\no}{N_c-\no}}=\langle S\rangle_{\rm br1}=\frac{\langle\Qo\rangle_{\rm
br1}\langle\Qt\rangle_{\rm br1}}{\mph}\,,\label{(36.1.8)}
\eq
as it should be, while
\bq
\frac{\langle\lym^{(\rm br1)}\rangle}{m_{Q,2}^{\rm pole}}=\Bigl
(\frac{m_Q}{\la}\Bigr )^{\frac{3N_c-2N_F+\no}{3(N_c-\no)}}\Bigl
(\frac{\mph}{\la}\Bigr )^
{\frac{2\no}{3(N_c-\no)}}=\Bigl (\frac{\mph}{{\mu}_{\Phi,3}}\Bigr
)^{\frac{2\no}{3(N_c-\no)}}\ll 1 \label{(36.1.9)}
\eq
\bbq
{\rm at}\,\quad \mph\ll {\rm
Min}\,\Bigl\{\frac{\la^2}{m_Q},\,{\mu}_{\Phi,3}\Bigr\},\quad
{\mu}_{\Phi,3}=\la(\frac{\la}{m_Q}\Bigr )^{\frac{3N_c-2N_F+\no}{2\no}}\,,
\eeq
(at $\mph>\la^2/m_Q$ the quarks are higgsed in the weak coupling 
regime, see  \eqref{(36.1.1)}).

Integrating out finally all $SU(N_c-\no)$ gluons at $\mu<\langle\lym^{(\rm
br1)}\rangle\ll m_{Q,2}^{\rm pole}$ via the VY-procedure \cite{VY}, the lower
energy Lagrangian of $N_F^2$ 3-rd generation fions $\Phi_i^j$,\, $\no^2$ pions
$\Pi^i_j$ and $2\no\nt$ hybrid pions $\Pi^1_2,\, \Pi^2_1$ (in essence, these
are \, ${\ov Q}_2, Q^2$ quarks with broken colors) look as, see \cite{ch16},
\bq
K_{\rm tot}=K_{\Phi}+K_{\Pi}+K^{\rm hybr}_{\Pi}\,,\label{(36.1.10)}
\eq
\bbq
K_{\Phi}=z_{\Phi}^{+}(\la,\mgo)\,{\rm Tr}\,\Bigl
[(\Phi_1^1)^\dagger\Phi_1^1+\Bigl
(\Phi_1^2)^\dagger\Phi_1^2+(\Phi_2^1)^\dagger\Phi_2^1\Bigr )+z_{\Phi}
^{-}(\mgo,m_{Q,2}^{\rm pole})\,(\Phi_2^2)^\dagger\Phi_2^2\Bigr ]\,,
\eeq
\bbq
K_{\Pi}\simeq K^{\rm (loop)}_{\Pi} =\,{\rm
Tr}\,\frac{(\Pi_1^1)^\dagger\Pi_1^1}{\la^2}\,,\quad
K^{\rm hybr}_{\Pi}=z_{Q}^{+}(\la,\mgo)\,{\rm Tr}\,\Bigl
(\,\,\frac{(\Pi_1^2)^\dagger\Pi_1^2+(\Pi_2^1)^\dagger
\Pi_2^1}{|\langle\Pi_1\rangle|=|\langle\Qo\rangle|}\,\,\Bigr )\,,
\eeq
\bq
{\cal W}_{\rm tot}={\cal W}_{\Phi}+{\cal W}_{\Phi\Pi}+{\cal W}_{\rm
non-pert}\,,\label{(36.1.11)}
\eq
\bbq
{\cal W}_{\Phi}=\frac{\mph}{2}\Bigl [\,{\rm Tr}\, (\Phi^2)-\frac{1}{\nd}\,\Bigl
({\rm Tr}\,\Phi\Bigr )
^2\Bigr ]\,,\quad {\cal W}_{\rm non-pert}=(\nd-\no)\Bigl (\frac{\la^{\rm
\bo}\det m_{Q,2}^{\rm tot}}{\det\Pi_1^1}\Bigr )^{\frac{1}{N_c-\no}}\,,
\eeq
\bbq
{\cal W}_{\Phi\Pi}=m_Q\,{\rm Tr}\,\Bigl [
\Pi_1^1+\Pi^2_1\frac{1}{\Pi_1^1}\Pi^1_2\Bigr ]-
\,{\rm Tr}\,\Bigl [\, \Pi_1^1\Phi_1^1+\Pi_1^2\Phi_2^1
+\Pi_2^1\Phi_1^2+\Phi_2^2\,(\Pi^2_1\frac{1}{\Pi_1^1}\Pi^1_2)\,\Bigr ]\,.
\eeq

We obtain from \eqref{(36.1.10)},\eqref{(36.1.11)} for the particle masses. -

The masses of the 3-rd generation fions $\mu^{\rm
pole}_{3}(\Phi_1^1)\sim\mu^{\rm pole}_{3}(\Phi_1^2)\sim\mu^{\rm
pole}_{3}(\Phi_2^1)$ look as
\bq
\mu^{\rm pole}_{3}(\Phi_1^1)=\frac{\mph}{z_{\Phi}^{+}(\la,\mgo)}\sim\la\Bigl
(\frac{m_Q}
{\la}\Bigr )^{2(2N_c-N_F)/\nd}\Bigl (\frac{\mph}{\la}\Bigr
)^{(3N_c-N_F)/\nd}\,,\label{(36.1.12)}
\eq
\bbq
\frac{\mu^{\rm pole}_{3}(\Phi_1^1)}{\mgo}\sim\Bigl
(\frac{\mph}{\mu_{\Phi,1}}\Bigr )^{2(2N_c-N_F)/\nd}
\ll 1\,,
\eeq
while for $\mu^{\rm pole}_{3}(\Phi_2^2)$,
\bq
\mu^{\rm
pole}_{3}(\Phi_2^2)\sim\frac{\mph}{z_{\Phi}^{+}(\la,\mgo)z_{\Phi}^{-}
(\mgo,m_{Q,2}^{\rm pole})}\sim
\la\Bigl (\frac{m_Q}{\la}\Bigr )^{\frac{2(2N_c-N_F)}{N_c-\no}}\,
\Bigl (\frac{\mph}{\la}\Bigr )^{\frac{N_c+\no}{N_c-\no}}\,,\label{(36.1.13)}
\eq
\bbq
\frac{\mu^{\rm pole}_{3}(\Phi_2^2)}{\mu^{\rm
pole}_{3}(\Phi_1^1)}=\frac{1}{z_{\Phi}^{-}(
\mgo,m_{Q,2}^{\rm pole})}\sim\Bigl (\frac{m_{Q,2}^{\rm pole})}{\mgo}\Bigr
)^{2\gamma^{-}_Q}\sim\Bigl (\frac{\mo}{\mph}\Bigr
)^{\frac{2N_c}{N_c-\no}(\gamma^{-}_Q\,>\,1)}\ll 1\,,
\eeq
\bbq
\frac{\mu^{\rm pole}_{3}(\Phi_2^2)}{m_{Q,2}^{\rm pole}}\sim\Bigl
(\frac{\mph}{\mu_{\Phi,2}} \Bigr )^{\frac{N_c+2\no}{N_c-\no}}\ll 1\quad {\rm
at}\quad \mph\ll\mu_{\Phi,2},\quad\mu_{\Phi,1}\ll
\mu_{\Phi,2}=\la\Bigl (\frac{\la}{m_Q}\Bigr )^{\frac{5N_c-3N_F+\no}{N_c+2\no}}
\eeq
(the main contribution to $\mu^{\rm pole}_{3}(\Phi_2^2)$ in \eqref{(36.1.13)}
originates from the term $\sim \mph{\rm Tr}\,(\Phi_2^2)^2$ in
\eqref{(36.1.11)}, \, while the contribution from ${\cal W}_{\rm non-pert}$ is 
much  smaller\,).

The smallest nonzero masses have $\no^2$ pions $\Pi^1_1$,
\bq
\mu^{\rm pole}(\Pi_1^1)\sim\frac{\la^2}{\mph}\,,\quad\quad \frac{\mu^{\rm
pole}(\Pi_1^1)}{\mu^{\rm pole}_{3}(\Phi_2^2)}\sim\Bigl (\frac{\mo}{\mph}\Bigr
)^{\frac{2N_c}{N_c-\no}}\ll 1 \label{(36.1.14)}
\eq
(the main contribution to $\mu^{\rm pole}(\Pi_1^1)$ in \eqref{(36.1.14)}
originates from the mixing term $\sim {\rm Tr}\,(\Pi_1^1\Phi_1^1)$ in
\eqref{(36.1.11)}, after integrating out heavier $\Phi^1_1$, while the
contribution
to $\mu^{\rm pole}(\Pi_1^1)$ from ${\cal W}_{\rm non-pert}$ is much smaller\,).

Finally, $2\no\nt$ multiplets $\Pi_1^2$ and $\Pi_2^1$ are Nambu-Goldstone
particles and are massless.\\

{\bf B)}\,\, The region $\mu_{\Phi,1}=\la
(\la/m_Q)^{\frac{5N_c-3N_F}{2(2N_c-N_F)}}\ll\mph\ll\la^2/m_Q$\,.\\

Because $\mgo>\mos=\la(\la/\mph)^{\nd/(5N_c-3N_F)}$ in this case, the running
mass $\mu_{\Phi}(\mu)$ of all $N_F^2$ fions $\Phi^j_i$ remains too large at the
scale $\mu=\mgo,\,\, \mu_{\Phi}(\mu=\mgo)>\mgo$, so that they are still
irrelevant at this scale. Because all heavy higgsed quarks ${\ov Q}_1, Q^1$
decouple at $\mu<\mgo$, the RG evolution of fions $\Phi_1^1, \Phi_1^2$ and
$\Phi_2^1$ is frozen at scales $\mu<\mgo$, so that they remain irrelevant at
all \, scales $\mu<\mu_1^{\rm pole)}(\Phi)\sim\mph\gg\la$, there are no 
additional  generations of these particles.

But as for $\Phi_2^2$\,, the situation is different because the quarks ${\ov
Q}^{\,\prime}_2, Q^{\,\prime,\,2}$ with unbroken colors are still active at
$\mu<\mgo$ and the running mass of fions $\Phi_2^2$ continues to decrease with
decreasing scale (until ${\ov Q}^{\,\prime}_2, Q^{\,\prime,\,2}$ decouple at
$\mu<m_{Q,2}^{\rm pole}$). As a result, there still will be two additional
generations of $n_2^2\,\, \Phi_2^2$-particles in the range
$\mu_{\Phi,1}<\mph<\mu_{\Phi,2}<\la^2/m_Q$, see \eqref{(36.1.13)},
\bq
\mu^{\rm pole}_{2}(\Phi_2^2)={\hat\mu}_o^{\,\rm str}=\la\Bigl
(\frac{\la}{{\hat\mu}_{\Phi}}\Bigr )^{\frac{1}{2\gamma^{-}_Q-1}}=\la\Bigl
(\frac{\la}{m_Q}\Bigr )^{\frac{2n_1}{5N_c-3N_F-2\no}}\Bigl
(\frac{\la}{\mph}\Bigr )^{\frac{\nd+2\no}{5N_c-3N_F-2\no}}\,,\label{(36.1.15)}
\eq
\bbq
{\hat\mu}_{\Phi}=\mph\Bigl (\frac{\mgo}{\la}\Bigr
)^{2(\gamma_Q^{+}-\gamma_Q^{-})}=
\la\Bigl(\frac{m_Q}{\la}\Bigr )^{\frac{2\no}{\nd}}\Bigl (\frac{\mph}{\la}\Bigr
)^{\frac{\nd+2\no}{\nd}},
\quad \frac{\mu^{\rm
pole}_{2}(\Phi_2^2)}{\mgo}=\Bigl(\frac{\mu_{\Phi,1}}{\mph}\Bigr
)^{\frac{2(2N_c-N_F)}{5N_c-3N_F-2\no}\,>\,0}\ll 1\,.
\eeq
\bq
\frac{m_{Q,2}^{\rm pole}}{\mu^{\rm pole}_{2}(\Phi_2^2)}\sim\Bigl
(\frac{\mph}{\mu_{\Phi,2}}\Bigr
)^{\frac{\nd(N_c+2\no)}{(N_c-\no)(5N_c-3N_F-2\no)}>\,0},\,\,
\frac{\mu_{\Phi,1}}{\mu_{\Phi,2}}\sim\Bigl (\frac{m_Q}{\la}\Bigr
)^{\frac{(3N_c-2N_F)(5N_c-3N_F-2\no)}{2(2N_c-N_F)(N_c+2\no)}\,>\,0}\ll
1\,.\label{(36.1.16)}
\eq

As it is seen from \eqref{(36.1.16)}, all $\nt^2$ fions $\Phi_2^2$ remain
dynamically irrelevant only at $\mph>\mu_{\Phi,2}\gg\mu_{\Phi,1}\gg\mo$,
because the quarks ${\ov Q}^{\,\prime}_2, Q^{\,\prime,\,2}$ with unbroken colors\,
decouple in this case before there appears the second generation of
$\Phi_2^2$\,, while in the range $\mu_{\Phi,1}\ll\mph\ll\mu_{\Phi,2}$ there
appear two additional generations of $\nt^2\,\, \Phi_2^2$-particles, and fions
$\Phi_2^2$ become dynamically relevant in the interval of scales $\mu^{\rm
pole}_{3}(\Phi_2^2)<\mu<\mu^{\rm pole}_{2}(\Phi_2^2)$, see
\eqref{(36.1.13)},\eqref{(36.1.15)}.

But to keep $\mu_{\Phi,2}<\la^2/m_Q$ (the RG flow becomes different at
$\mph>\la^2/m_Q$ because $\mgo\sim (m_Q\mph)/\la$ becomes larger than $\la$ 
and the quarks ${\ov Q}_1, Q^1$ will be higgsed in the weak coupling regime) it
should be: $8N_c/7<N_F<3N_c/2$. Therefore, at $N_c<N_F<8N_c/7$ the $\nt^2$
fions $\Phi_2^2$ are relevant in the range of scales $\mu^{\rm
pole}_{3}(\Phi_2^2)<\mu<
\mu^{\rm pole}_{2}(\Phi_2^2)$ in the whole interval $\mo<\mph<\la^2/m_Q$ and
there
are  two additional generations, $\mu^{\rm pole}_2(\Phi_2^2)$ and $\mu^{\rm
pole}_3(\Phi_2^2)$, see e.g. \eqref{(36.1.4)}. But at
$8N_c/7<N_F<3N_c/2$ these two additional generations of particles $\Phi_2^2$
exist only in the range $\mu_{\Phi,1}<\mph<\mu_{\Phi,2}$ and disappear at
$\mu_{\Phi,2}<\mph<\la^2/m_Q$, so that there are no additional generations of
any $\Phi$-particles at all at $\mph>\mu_{\Phi,2}$\,.\\

All formulas in special vacua with $\no=\nd,\, \nt=N_c$ can be obtained simply
substituting $\no=\nd$ in all expressions in this section 36.1.

\subsection{\quad  br1 and special vacua with $3N_c-2N_F < n_1< N_F/2$}

Consider now the case $\no > 3N_c-2N_F$, this requires $6N_c/5< N_F < 3N_c/2$
since $\no < N_F/2$ (recall also that we ignore for simplicity all RG evolution
effects if they are logarithmic only). In this case $3/2 <
N_F^\prime/N_c^\prime< 3\,,\,\,\, N_F^\prime=N_F-\no=\nt\,,\,\,\,N_c^\prime=
N_c-\no$. This means that, after the quarks ${\ov Q}_1, Q^1$ are higgsed at
$\mu\sim\mgo\sim m_Q\mph/\la\ll\la$ in the strong coupling regime
$a_{+}(\mu=\mgo)\gg 1$, the theory remains with the same large coupling
$a_{-}(\mu=\mgo)=a_{+}(\mu=\mgo)$, but the numbers of colors
$N_c^\prime=N_c-\no$ and flavors $N_F^\prime=N_F-\no=\nt$ in the lower energy
theory are now such that the coupling $a_{-}(\mu)$ begins {\it to decrease}
with diminishing scale because the $\beta$-function becomes {\it positive} at
$\mu<\mgo$ (until the gauge coupling is large, $a_{-}(\mu)\gg 1$), while the
quark anomalous dimension looks now as
$\gamma^{-}_Q(N^\prime_c,N^\prime_F,a(\mu)\gg
1)=(2N^\prime_c-N^\prime_F)/(N^\prime_F-N^\prime_c)$,
\bq
\frac{d a_{-}(\mu)}{d\ln\mu}=\frac{a^2_{-}(\mu)[\,
3-\frac{N_F^\prime}{N_c^\prime}(1+\gamma^{-}_Q) ]}
{a_{-}(\mu)-1}\simeq a_{-}(\mu)\Bigl [\frac{\,\no-(3N_c-2N_F)}{\nd}\Bigr ]>0,
\label{(36.2.1)}
\eq
\bbq
0\,<\,\gamma^{-}_Q=\frac{2N_c-N_F-\no}{\nd}\,<\, 1,\,\,
\gamma^{-}_{\Phi}=-2\gamma^{-}_Q \,\,\, {\rm at}
\,\,\, (3N_c-2N_F)<\no<(2N_c-N_F)\,,\,\, \frac{6N_c}{5}< N_F <
\frac{4N_c}{3}\,,\eeq
\bbq
-\,\frac{1}{2}<\,\gamma^{-}_Q=\frac{2N_c-N_F-\no}{\nd}\,<\,0,\,\,
\gamma^{-}_{\Phi}=0 \,\,\, {\rm at}\,\,\, (2N_c-N_F)<\no<\frac{N_F}{2}\,,\quad
\frac{4N_c}{3}< N_F < \frac{3N_c}{2}\,.
\eeq

The scale factor $\Lambda^\prime_Q$ of the gauge coupling $a_{-}(\mu)$ is
determined from the matching
\bbq
a_{+}(\mu=\mgo)=\Bigl (\frac{\la}{\mgo}\Bigr
)^{\frac{3N_c-2N_F}{N_F-N_c}}=a_{-}(\mu=\mgo)=\Bigl
(\frac{\mgo}{\Lambda^\prime_Q}\Bigr )^{\frac{\no-(3N_c-2N_F)}{N_F-N_c}}\gg 1\,,
\eeq
\bq
\Lambda^\prime_Q=\la\Bigl (\frac{m_Q\mph}{\la^2}\Bigr
)^{\frac{\no}{\no-(3N_c-2N_F)}\,>\, 1}\ll\mgo\sim
\frac{m_Q\mph}{\la}\ll\la\,.\label{(36.2.2)}
\eq
Therefore (if nothing prevents, see below), the theory will finally enter
smoothly at the scale $\mu<\Lambda^\prime_Q$ into the conformal regime with the
frozen gauge coupling $a_{*}\sim 1$. But, if $m_{Q,2}^{\rm
pole}\gg\Lambda^\prime_Q$, see \eqref{(36.1.9)}, the quarks ${\ov
Q}^{\,\prime}_2,
Q^{\,\prime,\,2}$ with unbroken colors will decouple still in the strong
coupling regime $a_{-}(\mu=m_{Q,2}^{\rm pole})\gg 1$ and the lower energy gauge
theory will be ${\cal N}=1\,\, SU(N_c-\no)$ SYM in the strong coupling regime,
with the scale factor of its gauge coupling determined as previously from the
matching
\bbq
a_{-}(\mu=m_{Q,2}^{\rm pole})=\Bigl (\frac{m_{Q,2}^{\rm
pole}}{\Lambda^\prime_Q}\Bigr )^{\frac{\no-(3N_c-2N_F)}{N_F-N_c}\,>\,0}=a^{(\rm
br1)}_{YM}(\mu=m_{Q,2}^{\rm pole})=
\Bigl (\frac{m_{Q,2}^{\rm pole}}{\lambda_{YM}}\Bigr )^{3}\quad\ra
\eeq
\bq
\ra\quad\lambda^3_{YM}=\Bigl (\langle\lym^{(\rm br1)}\rangle\Bigr
)^3=\la^3\Bigl(\frac{m_Q}{\la}\Bigr )^{\frac{\nt-\no}{N_c-\no}}\Bigl
(\frac{\la}{\mph}\Bigr
)^{\frac{\no}{N_c-\no}}=\langle S\rangle_{\rm br1}. \label{(36.2.3)}
\eq
as it should be, see \eqref{(36.1.1)},\eqref{(36.1.8)}.

Because our main purpose in this section  is to calculate the mass spectra in
the direct theory in cases with quarks in the strong coupling regime 
$a(\mu)\gg 1$, we consider below this case $m_{Q,2}^{\rm pole}\gg
\Lambda^\prime_Q$ only. This requires then, see \eqref{(36.1.9)},
\bq
\frac{\Lambda^\prime_Q}{m_{Q,2}^{\rm pole}}\ll 1\quad\ra\quad \mph\ll
{\mu}_{\Phi,3}=\la\Bigl (\frac{\la}{m_Q}\Bigr
)^{\frac{3N_c-2N_F+\no}{2\no}\,<\,1}<\frac{\la^2}{m_Q}\,,\quad
\no>3N_c-2N_F\,.\label{(36.2.4)}
\eq

We distinguish then three regions. -\\

{\bf I)} $6N_c/5\,<\,N_F\,<\,5N_c/4,\,\,(3N_c-2N_F)\,<\,\no
\,<\,N_F/2$. Here the  hierarchies look as:
$\,\mo\ll\mu_{\Phi,1}\ll\mu_{\Phi,2}\ll\mu_{\Phi,3}\ll\la^2/m_Q$, see
\eqref{(36.1.4)},\eqref{(36.1.12)},\eqref{(36.1.13)}, while
$1/2\,<\gamma_Q^{-}\,<1$.\\
{\bf a)}\, The region $\mo\ll\mph\ll\mu_{\Phi,1}$. There are two complete
additional generations of all $N^2_F$ $\Phi$-particles, $\Phi_{1}^1,
\Phi_{1}^2,\Phi_{2}^1$ are relevant in the range of scales $\mu_3^{\rm
pole}(\Phi^1_1)<\mu<\mu_2^{\rm pole}(\Phi^1_1)$, while $\Phi_{2}^2$ is relevant
at $\mu_3^{\rm pole}(\Phi^2_2)<\mu<\mu_2^{\rm pole}(\Phi^2_2),\,\,\mu_2^{\rm
pole}(\Phi^2_2)=\mu_2^{\rm
pole}(\Phi^1_1)=\mos=\la(\la/m_Q)^{\nd/(5N_c-3N_F)}$,  see
\eqref{(36.1.4)},\eqref{(36.1.12)},\eqref{(36.1.13)}.\\
{\bf b)}\, The region $\mu_{\Phi,1}\ll\mph\ll\mu_{\Phi,2}$. $\,\,\,\Phi_{1}^1,
\Phi_{1}^2, \Phi_{2}^1$ are irrelevant at all scales $\mu<\mu^{\rm
pole}_1(\Phi)\sim\mph$. As for $\Phi_2^2$, there still will be two additional
generations of $\Phi_2^2$-particles with masses $m_{Q,2}^{\rm pole}<\mu^{\rm
pole}_{2} (\Phi_2^2)<\mgo$, see \eqref{(36.1.15)},\eqref{(36.1.16)}, and
$\mu^{\rm
pole}_{3}(\Phi_2^2)<m_{Q,2}^{\rm pole}$, see \eqref{(36.1.13)}.\\
{\bf c)}\, The region $\mu_{\Phi,2}\ll\mph\ll\mu_{\Phi,3}$. In this case all
$N^2_F$ $\Phi$-particles are irrelevant at all scales $\mu<\mu^{\rm
pole}_1(\Phi)\sim\mph$.\\

{\bf II)} $5N_c/4\,<\,N_F\,<\,4N_c/3,\,\, (5N_c-3N_F)/2\,<\,\no\,<\,N_F/2$.
Here the hierarchies look as:  $\mo\ll\mu_{\Phi,2}\ll\mu_{\Phi,1}\ll\mu_{\Phi,3}
\ll\la^2/m_Q$, while  $0\,<\gamma_Q^{-}\,<\,1/2$.\\
{\bf a)}\, The region $\mo\ll\mph\ll\mu_{\Phi,2}$. There will be 2-nd and 3-rd
generations of all $N_F^2\,\, \Phi^j_i$-particles with $\mu^{\rm
pole}_2(\Phi)=\mos=\la(\la/m_Q)^{\nd/(5N_c-3N_F)}$ and $\mu^{\rm
pole}_3(\Phi_1^1)$ in \eqref{(36.1.12)}, $\mu^{\rm
pole}_3(\Phi_2^2)<m_{Q,2}^{\rm  pole}$ in \eqref{(36.1.13)}.\\
{\bf b)}\, The region $\mu_{\Phi,2}\ll\mph\ll\mu_{\Phi,1}$. The difference with
"{a}"\, is that now there is no third generation of fions $\mu_3(\Phi_2^2)$.\\
{\bf c)}\, The region $\mu_{\Phi,1}\ll\mph\ll\mu_{\Phi,3}$. All fions are too
heavy and irrelevant in this case at all scales $\mu<\mu_1^{\rm
pole}\sim\mph\gg\la$. \\

{\bf III)} $4N_c/3\,<\,N_F\,<\,3N_c/2,\,\, (2N_c-N_F)\,<\,\no\,<\,N_F/2$.
Because $(-1/2)\,<\gamma_Q^{-}\,
<\,0$ and $z_{\Phi}^{-}=1$ in \eqref{(36.1.10)} now, then $\mu^{\rm
pole}_3(\Phi_2^2)\sim\mu^{\rm pole}_3(\Phi_1^1)$ \eqref{(36.1.12)} instead of
\eqref{(36.1.13)}, while the hierarchies look now as:
$\mo\ll\mu_{\Phi,3}\ll\mu_{\Phi,1}\ll\la^2/m_Q$. Because we consider only the
region $\mph<\mu_{\Phi,3}$ and $\mu_{\Phi,3}<\mu_{\Phi,1}$ now, there will be
2-nd and 3-rd generations of all $N_F^2\,\, \Phi^j_i$-particles with $\mu^{\rm
pole}_2(\Phi^j_i)=\mos=\la(\la/\mph)^{\nd/(5N_c-3N_F)}$ \eqref{(36.1.4)}, and
$\mu^{\rm pole}_3(\Phi^j_i)$ in \eqref{(36.1.12)}.\\

All formulas in special vacua with $\no=\nd,\, \nt=N_c$ can be obtained simply
substituting $\no=\nd$ in all expressions in this section 36.2.

\subsection{\quad  br2-vacua}

In these br2 vacua with $N_F/2<\nt<N_c$, all expressions for condensates and
corresponding masses in \eqref{(36.1.1)}-\eqref{(36.1.4)} are obtained
replacing \, $\no\leftrightarrow\nt$. Because we are interested in this article in
that the  direct theory stays in the strong coupling regime $a(\mu<\la)\gtrsim 1$, all
other formulas for br2-vacua can also be obtained from those for br1-vacua by
$\no\leftrightarrow\nt$, under corresponding restrictions.

\section{Broken flavor symmetry.\, Dual theory. }

\subsection{\quad br1 and special vacua with $1\leq n_1<{\rm
Min}\,(N_F/2,\,3N_c-2N_F)$}

This dual theory is mostly in the IR free logarithmic regime in this case and
all logarithmic effects of the RG evolution will be ignored below for
simplicity, as before. All $N_F^2$ fions $\Phi^j_i$ have very large masses
$\sim\mph\gg\la$ and can be integrated out from the beginning. The potentially
most important masses of dual quarks and gluons look here as follows,
\bq
\omp\sim\frac{\langle M_1\rangle_{\rm br1}=\langle\Qo\rangle_{\rm
br1}}{\la}\sim\frac{m_Q\mph}{\la}\gg\tmp\sim\frac{\langle M_2\rangle_{\rm
br1}}{\la} \sim\la\Bigl (\frac{m_Q}{\la}\Bigr )^{\frac{\nd-n_1}{N_c-n_1}}\Bigl
(\frac{\la}{\mph}\Bigr )^{\frac{n_1}{N_c-n_1}}\,,\label{(37.1.1)}
\eq
\bbq
\langle\lym^{(\rm br1)}\rangle^3\equiv\langle S\rangle_{\rm br1}=\frac{\langle
M_1\rangle\langle M_2\rangle}{\mph}\sim\la^3\Bigl (\frac{m_Q}{\la}\Bigr
)^{\frac{\nt-n_1}{N_c-n_1}}\Bigl (\frac{\la}{\mph}\Bigr
)^{\frac{n_1}{N_c-n_1}}\,,
\eeq
while, see \eqref{(37.1.1)},
\bq
({\ov\mu}^{\,\rm pole}_{gl,2})^2\sim\langle\qt\rangle_{\rm br1}=\frac{\langle
S\rangle_{\rm br1}\la}{\langle M_2\rangle_{\rm br1}}=\frac{\langle
M_1\rangle_{\rm br1}\la}{\mph}\sim m_Q\la\gg (\mgo)^2\,,\label{(37.1..2)}
\eq
\bq
\Bigl (\frac{{\ov\mu}^{\,\rm pole}_{gl,2}}{\tmp}\Bigr )^2\sim\Bigl
(\frac{m_Q}{\la}\Bigr )^{\frac{3N_c-2N_F+n_1}{N_c-n_1}}\Bigl
(\frac{\mph}{\la}\Bigr )^{\frac{2n_1}{N_c-n_1}}\ll 1\,,\label{(37.1.3)}
\eq
\bbq
\Bigl (\frac{{\ov\mu}^{\,\rm pole}_{gl,2}}{\omp}\Bigr
)^2\sim\frac{\la^3}{m_Q\mph^2}
<\frac{\la^3}{m_Q\mo^2}\sim\Bigl (\frac{m_Q}{\la}\Bigr
)^{\frac{3N_c-2N_F}{N_c}}\ll 1\,,
\eeq
\bbq
\Bigl (\frac{\langle\lym^{(\rm br1)\rangle}}{\tmp}\Bigr )^3\sim\Bigl
(\frac{{\ov\mu}^{\,\rm pole}_{gl,2}}{\tmp}\Bigr )^2\sim\Bigl
(\frac{m_Q}{\la}\Bigr )^{\frac{3N_c-2N_F+n_1}{N_c-n_1}}\Bigl
(\frac{\mph}{\la}\Bigr )^{\frac{2n_1}{N_c-n_1}}<\Bigl (\frac{m_Q}{\la}\Bigr
)^{\frac{3N_c-2N_F-n_1}{N_c-n_1}}\ll 1\,.
\eeq

Therefore,  the overall phase is $Hq_1-Hq_2$, all dual quarks are
not higgsed but confined, the confinement originates from the unbroken
$SU(\nd)\,\,{\cal N}=1$ SYM and so the scale of the confining string tension is
$\sqrt\sigma\sim\lym^{(\rm br1)}=\langle S\rangle^{1/3}_{\rm br1}$, see
\eqref{(37.1.1)}.

At scales $\mu<\omp$ all ${\ov q}^1, q_1$ quarks can be integrated out as heavy
ones and there remains $SU(\nd)$ gauge theory with $n_2$ flavors of quarks ${\ov
q}^2, q_2$ (and with $\bd^{\,\prime}=
2N_F-3N_c+n_1\,<0\,)$ and $N_F^2$ mions $M^i_j$. Integrating out then these
quarks as heavy ones at $\mu<\tmp$ and then all ${\cal N}=1\,\, SU(\nd)$
SYM-gluons at $\mu<\lym^{(\rm br1)}$ via the VY-procedure, the lower energy
Lagrangian of mions looks as
\bq
K\sim {\rm Tr}\,\frac{M^\dagger M}{\la^2}\,,\quad \quad{\cal W}={\cal
W}_M+{\cal W}_{\rm non-pert}\,,\quad
{\cal W}_{\rm non-pert}= -\nd\Bigl (\frac{\det M}{\la^{\rm\bo}}\Bigr
)^{1/\nd}\,,\label{(37.1.4)}
\eq
\bbq
{\cal W}_M=-\frac{1}{2\mph}\Bigl [{\rm Tr}\,(M^2)-\frac{1}{N_c}\Bigl ({\rm
Tr}\,M\Bigr )^2\Bigr ].
\eeq
From \eqref{(37.1.4)}, the masses of mions $M^i_j$ look as follows\,:\\
1)\,\, $n_2^2$ mions $M_2^2$ have masses
\bq
\mu^{\rm pole}(M_2^2)\sim\frac{\langle M_1\rangle_{\rm br1}}{\langle
M_2\rangle_{\rm br1}}\,\frac{\la^2}{\mph}\sim\la\Bigl (\frac{m_Q}{\la}\Bigr
)^{\frac{2N_c-N_F}{N_c-n_1}}\Bigl (\frac{\mph}{\la}\Bigr
)^{\frac{n_1}{N_c-n_1}}\,, \label{(37.1.5)}
\eq
\bbq
\Bigl (\frac{\mu^{\rm pole}(M_2^2)}{\lym^{(\rm br1)}}\Bigr )^{3/2}\sim\Bigl
(\frac{m_Q}{\la}\Bigr )^{\frac{3N_c-2N_F+n_1}{N_c-n_1}}\Bigl
(\frac{\mph}{\la}\Bigr )^{\frac{2n_1}{N_c-n_1}}<\Bigl (\frac{m_Q}{\la}\Bigr
)^{\frac{3N_c-2N_F-n_1}{N_c-n_1}}\ll 1\,,
\eeq
the main contribution to $\mu^{\rm pole}(M_2^2)$ originates from ${\cal W}_{\rm
non-pert}$ in \eqref{(37.1.4)}.

2)\,\, $n_1^2$ mions $M_1^1$ have masses
\bq
\mu^{\rm pole}(M_1^1)\sim\frac{\la^2}{\mph}\sim\frac{\langle M_2\rangle_{\rm
br1}}{\langle M_1\rangle_{\rm br1}}\,\mu^{\rm pole}(M_2^2)\sim\Bigl
(\frac{\mo}{\mph}\Bigr )^{\frac{N_c}{N_c-n_1}}\,\mu^{\rm pole}(M_2^2)\ll
\mu^{\rm pole}(M_2^2)\,,\label{(37.1.6)}
\eq
the main contribution to $\mu^{\rm pole}(M_1^1)$ originates from ${\cal W}_{M}$
in \eqref{(37.1.4)}.

3)\,\, $2n_1n_2$ mions $M_1^2, M_2^1$ are  Nambu-Goldstone particles and are
massless
\bq
\mu^{\rm pole}(M_1^2)= \mu^{\rm pole}(M_2^1)=0\,.\label{(37.1.7)}
\eq

On the whole, the mass spectrum looks in this case as follows. -\\
1) There is a large number of hadrons made from weakly interacting
non-relativistic dual quarks ${\ov q}^1, q_1$ with masses $\omp\sim
m_Q\mph/\la$\eqref{(37.1.1)}, and similarly for hybrids made from $({\ov
q}^{\,1}
q_2)$ or $({\ov q}^{\,2} q_1)$.\\
2) A large number of hadrons made from weakly interacting non-relativistic
quarks ${\ov q}^2, q_2$
with masses $\tmp\sim(\mo/\mph)^{N_c/(N_c-n_1)}\omp\ll\omp$, see
\eqref{(37.1.1)}.
All quarks are weakly confined, i.e. the tension of the confining string
originating from ${\cal N}=1\,\, SU(\nd)$ SYM is $\sqrt\sigma\sim\lym^{(\rm
br1)}\ll\tmp\ll\omp$.\\
3) A large number of gluonia from ${\cal N}=1\,\, SU(\nd)$ SYM, the scale of
their masses is $\lym^{(\rm br1)}\sim(\langle M_1\rangle_{\rm br1}\langle
M_2\rangle_{\rm br1}/\mph)^{1/3}$, see e.g. \eqref{(37.1.1)}.\\
4) $n_2^2$ mions $M_2^2$ with masses $\mu^{\rm pole}(M_2^2)\ll\lym^{(\rm
br1)}$,  see \eqref{(37.1.5)}.\\
5) $n_1^2$ mions $M_1^1$ with masses $\mu^{\rm pole}(M_1^1)\ll\mu^{\rm
pole}(M_2^2)$, see \eqref{(37.1.6)}.\\
6) $2n_1n_2$ mions $M_1^2, M_2^1$ are massless.

It is worth noting also that the scales $\mu_{\Phi,1}$ and $\mu_{\Phi,2}$,\,\,
$\mu_{\Phi,2}\gg\mu_{\Phi,1}\gg\mo$, see
\eqref{(36.1.4)},\eqref{(36.1.13)},\eqref{(36.2.4)} specific for the direct
theory play
no role in this dual theory at $1\leq n_1<{\rm Min}\,(N_F/2,\,3N_c-2N_F)$.\\

All formulas in special vacua with $\no=\nd,\, \nt=N_c$ can be obtained simply
substituting $\no=\nd$ in all expressions in this section 37.1.

\subsection{\quad  br1 and special vacua with $3N_c-2N_F < n_1< N_F/2$}

The masses of dual quarks $\omp$ and $\tmp$ are the same (with a logarithmic
accuracy), see \eqref{(37.1.1)}. The difference is that the dual coupling ${\ov
a}(\mu)$ still decreased logarithmically with diminished scale at
$\tmp<\mu<\omp$ at $1\leq n_1<{\rm Min}\,(N_F/2,\,3N_c-2N_F)$, while now
$\bd^{\,\prime}=3\nd-n_2=n_1-
(3N_c-2N_F)>0$ and ${\ov a}(\mu)$ increases logarithmically with diminishing
scale at $\mu<\omp$ until the dual theory enters the conformal regime at
$\mu<{\tilde\Lambda}_Q$ (if nothing prevents). ${\tilde\Lambda}_Q$ is
determinedfrom the matching of dual couplings at $\mu=\omp$, see
\eqref{(37.1.1)},\eqref{(36.2.2)},
\bq
({\tilde\Lambda}_Q )^{3\nd-n_2}=\la^{3\nd-N_F}(\omp)^{n_1}\quad\ra\quad
{\tilde\Lambda}_Q=\la\Bigl (\frac{m_Q\mph}{\la^2}\Bigr
)^{\frac{n_1}{n_1-(3N_c-2N_F)}\,>\,
0}=\Lambda^\prime_Q\ll\la\,.\label{(37.2.1)}\eq

Therefore, we need $\tmp>\Lambda^\prime_Q$ in order not to enter 
the conformal  regime, this requires, see \eqref{(37.2.1)},\eqref{(37.1.1)},
\bq
\frac{\Lambda^\prime_Q}{\tmp}=\Bigl (\frac{\mph}{\mu_{\Phi,3}}\Bigr
)^{\frac{n_1}
{(n_1-3N_c+2N_F)}\frac{2\nd}{(N_c-n_1)}}\ll 1\,\ra\, \mph\,\ll\,\mu_{\Phi,3}
\label{(37.2.2)}
\eq
as in the direct theory, see \eqref{(36.2.4)}. The masses $\mu^{\rm
pole}(M_2^2)$
and $\mu^{\rm pole}(M_1^1)\ll\mu^{\rm pole}(M_2^2)$ remain as in
\eqref{(37.1.5)},\eqref{(37.1.6)} and besides, see \eqref{(37.1.5)},
\bq
\frac{\mu^{\rm pole}(M_2^2)}{\lym^{(\rm br1)}}\sim \Bigl
(\frac{\mph}{\mu_{\Phi,3}}\Bigr )
^{\frac{4n_1}{3(N_c-n_1)}}\ll 1\quad{\rm at}\quad \mph\ll \mu_{\Phi,3}=\la\Bigl
(\frac{\la}{m_Q}\Bigr
)^{\frac{3N_c-2N_F+n_1}{2n_1}\,<\,1}<\frac{\la^2}{m_Q}\,.\label{(37.2.3)}
\eq

All formulas in special vacua with $\no=\nd,\, \nt=N_c$ can be obtained simply
substituting $\no=\nd$ in all expressions in this section 37.2.

\subsection{\quad  br2 - vacua}

In dual br2 vacua with $N_F/2<\nt<N_c$, all formulae can also be obtained from
those for dual br1-vacua by $\no\leftrightarrow\nt$, under corresponding
restrictions.

\section{\bf \large Conclusions to Part II} 

\hspace*{6mm} This Part IIc continues previous study in \cite{ch13,ch19}
of \, ${\cal N}=1$ SQCD-type theories with additional colorless but flavored
fields. But considered this time is the region $N_c<N_F<3N_c/2$. In this
region,\,  the UV free direct $SU(N_c)$ theory with light quarks enters {\it
smoothly} at  $\mu<\la$ to the (very) strongly coupled regime with the coupling
$a(\mu\ll\la)\gg 1$ (see Introduction, section 7 in \cite{ch1} and Conclusions
in \cite{Session} for additional arguments). The mass spectra were calculated
in \cite{ch16} in numerous different vacua. The calculations were performed
within  the dynamical scenario introduced in \cite{ch3}. This scenario assumes 
that  quarks
in such ${\cal N}=1$ SQCD-type theories can be in two {\it standard} phases
only\,: these are either the HQ (heavy quark) phase where they are confined, 
or  the Higgs phase. The word {\it standard} implies here also that, in such
theories without elementary colored adjoint scalars, no {\it additional}
\footnote{\,
i.e. in addition to Nambu-Goldstone particles due to spontaneously broken
global flavor symmetry
}
parametrically light solitons (e.g. magnetic monopoles or dyons) are formed at
those lower scales $\mu\ll\la$ where quarks decouple as heavy or are higgsed.
\footnote{
Besides, it is worth noting that the appearance of additional parametrically
light composite solitons will influence the 't Hooft triangles.
}

Similarly to previous studies in \cite{ch13,ch19} of these theories
within the conformal window at $3N_c/2<N_F<3N_c$, it is shown here that, due 
to the  strong powerlike RG evolution at scales $\mu<\la$ in the direct theory, the
seemingly heavy and dynamically irrelevant fields $\Phi^j_i$ can become light
and there appear then two additional generations of light $\Phi$-particles with
$\mu_{2,3}^{\rm pole}(\Phi)\ll\la$.

In parallel, were calculated the mass spectra of IR free and logarithmically
weakly coupled at $\mu<\la$ dual $\,SU(\nd=N_F-N_c)$ theory  proposed by 
Seiberg. \\

On the whole for the Part II, the comparison have shown that the  mass spectra of the 
direct and  dual theories are {\it parametrically different, in general, so that these 
two  theories  are not equivalent}.

As it is seen from the results in the whole Part II, the use of the dynamical scenario 
from  \cite{ch3} leads to the results for the mass spectra which look
self-consistent.  In other words, no internal inconsistences appeared in all 
cases considered and no contradictions were found with all previously proven results. 
It is worth to recall also that this dynamical scenario used 
in this article satisfies all those tests which were used as checks of the
Seiberg  proposal about the equivalence of the direct and dual theories. 
The parametric  difference of mass spectra of direct and dual theories shows, 
in particular,  that  all these tests, although necessary, may well be insufficient.

\section{ \bf   \large  Appendices 1 and 2}   

\addcontentsline{toc}{section}
{\bf  Appendix  1\,.\,  Anomalous dimensions in the strong coupling regime}

\begin{center}
{\bf  \large Appendix  1\,.\,  Anomalous dimensions in the strong coupling regime} 
\end{center}

\numberwithin{equation}{section}

The purpose of this Appendix is to find the anomalous dimensions of quarks and
colorless but flavored fields $\Phi^j_i$ in the theory  \eqref{(31.1.5)},\eqref{(31.1.6)}
in the strong coupling regime in the range of scales where they are dynamically
relevant, i.e. their running masses at the scale $\mu$ are smaller than $\mu$.

Note first that there is the first generation of all $N_F^2$ fions $\Phi_i^j$
with masses $\mu_1^{\rm pole}(\Phi)\sim\mph\gg\la$. As shown in section 5 of
\cite{ch19}, the anomalous dimension $\gamma^{(+)}_{\Phi}$ of fions in the
rangeof scales $\mos<\mu<\la$ where they are still heavy and dynamically
irrelevant  is:\,
$\gamma^{(+)}_{\Phi}=\,-\,2\gamma^{(+)}_Q,\,\,\gamma^{(+)}_Q=(2N_c-N_F)/\nd$.

Besides, there is also the second generation of all $N_F^2$ fions with masses,
see section 17 and \eqref{(31.1.7)},
\bq
{\hspace*{-0.5cm}}\mu_2^{\rm pole}(\Phi)\sim\mos\sim\la\Bigl
(\frac{\la}{\mph}\Bigr )^{\frac{1}{2\gamma^{(+)}_Q-1}}=\la\Bigl
(\frac{\la}{\mph}\Bigr )^{\frac{\nd}{5N_c-3N_F}}\ll\la, \label{(39.1)}
\eq
and all $N_F^2$ "heavy" fions become effectively massless and dynamically
relevant in the direct theory \eqref{(31.1.5)},\eqref{(31.1.6)} at $\mu<\mos$.\\

{\bf a}) Now, as for the anomalous dimension  ${\tilde\gamma}_Q^{(+)}$ of strongly 
coupled quarks  in the range  $m_{Q,1}^{\rm
pole}<\mu<\mos$ where the fions became relevant. We use first the approach
\cite{ch1} (see section 7 therein). For this, we have to introduce the dual
theory which has the same 't Hooft triangles (in a given range of scales) as
the \, direct theory.

In the range $\mos\ll\mu\ll\la$ where the all fields $\Phi^j_i$  are too heavy 
and  dynamically irrelevant in both the direct and dual theories ,  the dual theory 
can be taken as \eqref{(14.2.2)}, see section 14.2. The effectively massless particles in 
this  range \, of  scales in the direct theory are all quarks and gluons, while in the dual
one  these are the dual quarks and gluons and mions $M^i_j$. As
shown in \cite{S2}, all 't Hooft triangles will be the same in these two
theories.

Now, in the approach \cite{ch1}, we equate two NSVZ $\,{\widehat\beta}_{ext}$ -
functions of the external baryon and $SU(N_F)_{L}$ - flavor vector fields in
the \, direct and dual theories,
\bq
\frac{d}{d\,\ln \mu}\,\frac{2\pi}{\alpha_{ext}}={\widehat\beta}_{ext}=
-\frac{2\pi}{\alpha^2_{ext}}\,\beta_{ext}= \sum_i T_i\,\bigl (1+\gamma_i\bigr
)\,,\label{(39.2)}
\eq
where the sum runs over all fields which are effectively massless at scales
considered, the unity in the brackets is due to one-loop contributions while
the \, anomalous dimensions $\gamma_i$ of fields represent all higher-loop effects,
$T_i$ are the coefficients. It is worth noting that these general NSVZ forms
\eqref{(39.2)} of the external "flavored" $\widehat\beta_{ext}$-functions are
independent of the kind of massless perturbative regime of the internal gauge
theory, i.e. whether it is conformal, or the strong coupling or IR free
one.  For  the baryon charge \eqref{(39.2)} looks as
\bq
N_F N_c\,\Bigl ( B_Q=1 \Bigr )^2\,(1+\gamma^{(+)}_Q)=N_F \nd \,\Bigl (
B_q=\frac{N_c}{\nd}\Bigr )^2\,(1+\gamma_q)\,.\label{(39.3)}
\eq

In \eqref{(39.3)}: the left-hand side is from the direct theory while the
right-hand side is from the dual one. The dual theory is in the
IR free logarithmic regime at $\mu\ll\la$ with the dual coupling ${\ov
a}(\mu)\ll 1$ and $\gamma_q\sim {\ov a}\ra 0$. Then from \eqref{(39.3)}, see
section 7 in \cite{ch1}
\bq
\gamma^{(+)}_Q=\frac{2N_c-N_F}{N_F-N_c},\quad\frac{d\,a(\mu\ll\la)}
{d\log\mu}=\beta_{NSVZ}(a \gg 1)= \frac{a^2(\mu)}{a(\mu)-1}\,\frac{\bo-N_F
\gamma_Q^{(+)}}{N_c} \ra  -\,\nu^{(+)}\,a(\mu),\label{(39.4)}
\eq
\bbq
a(\mos\ll\mu\ll\la)\sim (\la/\mu)^{\nu^{(+)}}\gg 1,
\,\,\,\,\,\, \nu^{(+)}=\frac{3N_c-2N_F}{N_F-N_c}\,>\,0,\,\,\,  \bo=3N_c-N_F\,.
\eeq

For the flavor charge \eqref{(39.2)} looks as
\bq
N_c(1+\gamma^{(+)}_Q)=\nd(1+\gamma_q)+N_F(1+\gamma_{M})\,.\label{(39.5)}
\eq
Both $\gamma_q$ and $\gamma_{M}$ are logarithmically small at $\mu\ll\la$ and
\eqref{(39.5)} is incompatible with \eqref{(39.4)} (and with the NSVZ
$\beta$-function, see section 7 in \cite{ch1}). Therefore,  \eqref{(39.2)}  for the 
flavor charge will not be used  (here and below).  (But within the conformal 
window $3N_c/2 < N_F < 2N_c$ both \eqref{(39.3)} and \eqref{(39.5)} are fulfilled).\\

{\bf b}) Now, for the range $m_{Q,1}^{\rm pole}\ll\mu\ll\mos$ where the fions
in \, the direct theory became relevant in the strong coupling regime $a(\mu)\gg 1$,
\, the dual theory with the same 't Hooft triangles  (now
with effectively massless fions $\Phi$ in the direct theory) can be taken as \eqref{(3.0.1)} but
{\it without the mion fields} $ M^i_j$ (i.e. only massless dual quarks ${\ov
q}^{\,j}, q_i$ and $SU(\nd)$ dual gluons). But because the baryon charge of fions
$\Phi^j_i$  and mions $M^i_j$ is zero, \eqref{(39.3)} 
\eqref{(39.4)} remain the same, i.e. the anomalous dimension
${\wt\gamma}^{\,(+)}_Q$ of quarks ${\ov Q}, Q$ remains the same after fions
became relevant, ${\wt\gamma}^{\,(+)}_Q=\gamma^{(+)}_Q=(2N_c-N_F)/\nd$.\\

{\bf c}) Because \eqref{(39.2)} for the flavor charge is not fulfilled, the
above \, method does not allow to find the anomalous dimension
${\wt\gamma}^{\,(+)}_{\Phi}$ at $\mu<\mos$ where the fions become effectively
massless and dynamically relevant. Therefore, on the example of the strongly
coupled theory \eqref{(31.1.5)},\eqref{(31.1.6)}, we present now independent
way \, to connect the
anomalous dimensions of quarks and $\Phi$, $\,{\wt\gamma}^{\,(+)}_Q$ and
${\wt\gamma}^{\,(+)}_{\Phi}$, at those scales where $\Phi$ became relevant,  i.e. \, 
at  $\mu<\mos$. It follows from the internal self consistency when the quarks
\, ${\ov Q}_1, Q^{1}$ decouple as heavy ones at scales $\mu<m_{Q,1}^{\rm pole}$ in
the direct theory, while all $N_F^2$ fions $\Phi^j_i$ still remain
effectively massless in the large range of scales $\mu^{\rm pole}_{3}
(\Phi^1_1)<\mu<\mos,\,\, \mu^{\rm pole}_{3}(\Phi^1_1)\ll m^{\rm
pole}_{Q,1}\ll\mos$.
 
The Lagrangian of the direct theory, which is in the HQ (heavy
quark) phase with $m_{Q,1}^{\rm pole}\gg m_{Q,2}^{\rm pole}$,
has the form at the scale $\mu=m_{Q,1}^{\rm pole}\ll\la$
\bbq
K=K_{\Phi}+K_{Q}=z^{(+)}_{\Phi}{\rm Tr}\,(\Phi^\dagger\Phi)+z^{(+)}_Q{\rm
Tr}\,(Q^\dagger Q+Q\ra {\ov Q})=
{\rm Tr}\,(\delta{\Phi_R}^{\dagger}\delta{\Phi_R})+{\rm
Tr}\,({Q_R}^\dagger{Q_R}+Q_R\ra {\ov Q}_R),
\eeq
\bbq
z^{(+)}_Q=z^{(+)}_Q(\la,m_{Q,1}^{\rm pole})=z^{(+)}_Q(\la,\mos)\,{\wt
z}^{\,\,(+)}_Q(\mos,m_{Q,1}^{\rm pole})=\Bigl (\frac{\mos}{\la}\Bigr
)^{\gamma^{(+)}_Q}\Bigl (\frac{m_{Q,1}^{\rm pole}}{\mos}\Bigr
)^{{\wt\gamma}^{\,(+)}_Q}\,,
\eeq
\bq
\quad z^{(+)}_{\Phi}=z^{(+)}_{\Phi}(\la,m_{Q,1}^{\rm
pole})=z^{(+)}_{\Phi}(\la,\mos)\,{\wt z}^{\,\,(+)}_
{\Phi}(\mos,m_{Q,1}^{\rm pole})=\Bigl (\frac{\la}{\mos}\Bigr
)^{2\gamma^{(+)}_Q}\Bigl (\frac{m_{Q,1}^{\rm pole}}{\mos}\Bigr
)^{{\wt\gamma}^{\,(+)}_{\Phi}}\,, \label{(39.6)}
\eq
\bq
m_{Q,1}^{\rm pole}=\frac{\langle m_{Q,1}^{\rm
tot}\rangle}{z^{(+)}_Q(\la,m_{Q,1}^{\rm pole})},
\,\,\langle m_{Q,1}^{\rm tot}\rangle=\frac{\langle\Qt\rangle}
{\mph}\sim m_Q,\,\, m_Q^{\rm tot}=
m_Q-\Phi=\langle m_Q^{\rm tot}\rangle-\delta\Phi,\,\,
\langle\delta\Phi\rangle=0, \,\,\,\,\label{(39.7)}
\eq
where $\Phi_R$ and $Q_R$ are the canonically normalized fields and
$\delta\Phi_R$ is a pure quantum fluctuation,
\bq
{\cal W}_{\rm matter}={\cal W}_{\Phi}+{\rm Tr}\,({\ov Q}\,m_Q^{\rm tot}
Q)\,,\label{(39.8)}
\eq
 see \cite{ch16} for all details.

At $\mu<m_{Q,1}^{\rm pole}$ all quarks $Q^1, {\ov Q}_1$ decouple and can be
integrated out as heavy ones. As a result, there will appear a series of higher
dimension operators in D-terms, e.g.
\bq
O_n\sim a_f^{(1)} \delta \Phi_R^\dagger \delta \Phi_R \Biggl [
a^{(1)}_f\frac{\delta \Phi_R^\dagger \delta \Phi_R}
{\Bigl (m_{Q,1}^{\rm pole}\Bigr )^2}\,\Biggr ]^{\rm n}\,,\quad
a^{(1)}_f=\frac{a_f(\mu=\la)\sim 1}{z^{(+)}_{\Phi}\,(z^{(+)}_Q)^2}\,, 
\label{(39.9)}
\eq
where $a^{(1)}_f=a_f(\mu=m_{Q,1}^{\rm pole})$ is the Yukawa coupling at the
scale $\mu=m_{Q,1}^{\rm pole}$. These terms originate e.g. from the expansion in
powers of $\delta{{\cal M}_R}$ of the heavy quark loop integrated over the
non-parametric interval of momenta $p\sim m_{Q,1}^{\rm pole}$,
\bq
\Delta_{\Phi}\sim a^{(1)}_f\delta \Phi_R^\dagger \delta \Phi_R\int d^4 p\,[\, p^2+
{\cal M}^\dagger {\cal M}\, ]^{-2}, \,\,   {\cal M}=m_{Q,1}^{\rm  pole}\Biggl (1-
\sqrt{ a^{(1)}_f }\frac{\delta {\Phi_R}}{ m_{Q,1 }^{\rm pole}}\Biggr ), 
\,\, \langle  {\cal M}\rangle=m_{Q,1}^{\rm pole}.
\label{(39.10)}
\eq

When propagating in the loop with the momentum "k", the canonical fields
$\delta{\Phi_R}\sim k$. Therefore, in order that heavy quarks ${\ov Q}_1, Q^1$
really completely decouple as it should be, the contributions from
$O_{\rm n}$ in \eqref{(39.9)} have to be small at momenta $k < m_{Q,1}^{\rm
pole}$ in comparison with the canonical Kahler term
$\delta{\Phi_R}^{\dagger}\delta{\Phi_R}$. Then, because the expansion
\eqref{(39.9)} of \eqref{(39.10)} breaks down at larger momenta $> m_{Q,1}^{\rm
pole}$ (in any case, the running quark mass diminishes very quickly with
increasing momentum), all terms $O_{\rm n}/[\delta{\Phi_R}^{\dagger}
\delta{\Phi_R}]$ will be $O(1)$ at  momenta $\sim m_{Q,1}^{\rm pole}$. 
For  this, it should be parametrically $a^{(1)}_f\sim  1/[z^{(+)}_{\Phi}(z^{(+)}_Q)^2]
\sim 1$, see  \eqref{(39.6)},\eqref{(39.7)},\eqref{(39.9)},
\bq
a^{(1)}_f=O(1)\,\,\ra\,\, z^{(+)}_{\Phi}\,\Bigl ( z^{(+)}_Q\,\Bigr )^2=
z^{(+)}_{\Phi}(\mos,m_{Q,1}^{\rm pole})\,\Bigl ( z^{(+)}_Q(\mos,m_{Q,1}^{\rm
pole})\,\Bigr )^2\sim  1\,\,\ra\,\,
{\wt\gamma}^{\,(+)}_{\Phi}=-2\,{\wt\gamma}^{\,(+)}_Q\,.\label{(39.11)}
\eq
\eqref{(39.11)} ensures then that corrections from $O_{\rm n}$ will be small at
momenta $k<m_{Q,1}^{\rm pole}$, so that quarks ${\ov Q}_1, Q^1$ really decouple
completely.

Therefore, $\gamma_{\Phi}=-2\gamma_Q$ not only at $\mos<\mu<\la$ where  
fions \, $\Phi^j_i$ were irrelevant, but also  ${\wt\gamma}_{\Phi}=-2\,{\wt\gamma}_Q$ \, 
at  lower scales $\mu<\mos$ where they became relevant.\\

{\bf d}) And finally, we present also independent reasonings leading to the
same \, results.  First, we point out that the gauge coupling $a(\mu\ll\la)\gg 1$ entered \, 
already  into a  strong coupling regime, while
$\gamma^{(+)}_{\Phi}=-2\gamma^{(+)}_Q$ at $\mos<\mu<\la$ where the fions are
still irrelevant \cite{ch1}. Therefore, the gauge and Yukawa couplings look at
$\mu=\mos\ll\la$ as, see \eqref{(39.4)},
\bq
a(\mu=\mos)=(\la/\mos)^{\nu^{(+)}\,>\,0}\gg 1,\quad
a_f(\mu=\mos)=\frac{a_f(\mu=\la)= 1}{z^{(+)}_{\Phi}
(\la,\mos)[\,z^{(+)}_Q(\la,\mos)\,]^2}\sim 1\,.\label{(39.12)}
\eq

Consider now the Feynman diagrams in the strongly coupled direct theory
contributing to the renormalization factors $z_{\Phi}(\mu)$ and
$z_{Q}(\mu)$ at $m^{\rm pole}_{Q,1}\ll\mu\ll\mos$ where both the quarks and
fions $\Phi$ are effectively massless. Order by order in the perturbation theory\,
the extra loop with the exchange of the field $\Phi$ is $a_f(\mos)/a(\mos)\sim
(\mos/\la)^{\nu^{(+)}\,>\,0}\ll 1$ times smaller than the extra loop but with
the exchange of gluon.

In all cases with a resummation of perturbative series, a standard assumption
is\,
 that the leading contribution to the sum originates from summation of leading
terms at each order. With this assumption, we can neglect all exchanges of
$\Phi$ in comparison with those of gluons, order by order. Therefore, the fact
that the fields $\Phi^j_i$ became relevant at $\mu<\mos$ is really of no
importance for the RG-evolution, so that both ${\wt\gamma}^{\,\rm (+)}_{Q}$ and
${\wt\gamma}^{\,(+)}_{\Phi}$ {\it remain the same} at $m^{\rm
pole}_{Q,1}\ll\mu\ll\mos$ as they were at $\mos\ll\mu\ll\la$ when fions were
irrelevant:\,
${\wt\gamma}^{\,\rm(+)}_{Q}=\gamma^{\rm(+)}_{Q},\,\,{\wt\gamma}^{\,(+)}_{\Phi}
=\gamma^{(+)}_{\Phi}=
\,-2\,\gamma^{\rm (+)}_{Q}$ (i.e. the Yukawa coupling $a_f(\mu)$ still stays
intact at $a_f(\mu)\sim 1$ also at $m^{\rm pole}_{Q,1}\ll\mu\ll\mos$, as it was
at $\mos\ll\mu\ll\la$). The equality ${\wt\gamma}^{\,\rm (+)}_{Q}=\gamma^{\rm
(+)}_{Q}$ agrees with "${\bf b}$" above, while
${\wt\gamma}^{\,(+)}_{\Phi}=\,-2\,{\wt\gamma}^{\,\rm (+)}_{Q}$ agrees with
"${\bf c}$" above.\\

It is also worth to remind that, once the fions $\Phi^j_i$ become effectively
massless and dynamically relevant with respect to internal interactions, they
simultaneously begin to contribute to the 't Hooft triangles

The gluon contributions into the RG evolution of quarks at $m_{Q,1}^{\rm pole}
<\mu<\la$ were already accounted in \eqref{(39.6)}. Besides, because quarks 
$Q^1,{\ov Q}_1$  are confined, they form a set of color neutral strongly coupled
\, hadrons with  the characteristic scale of hadron masses ${\ov M}_{\rm hadr}\sim
m_{Q,1}^{\rm pole}$. Therefore, gluons with momenta $\lesssim m_{Q,1}^{\rm
pole}$ effectively decouple from these quarks, and for this reason gluons are
absent in \eqref{(39.10)}. On the other hand, there are no reasons to decouple
for the colorless fions $\Phi^j_i$. \\

\vspace*{1mm}

\addcontentsline{toc}{section}
{\bf  Appendix 2\,.\, \,Broken $\mathbf{\mathcal{N}=2}$ SQCD}

\begin{center}
{\bf \large Appendix  2\,. \,Broken $\mathcal{N}=2$ SQCD}
\end{center}

\hspace*{3mm} We consider in this Appendix  2\,\, ${\cal N}=2$ SQCD with 
$SU(N_c)$ colors,  $1 \leq N_F\ < 2N_c$
flavors of light quarks, the scale factor $\Lambda_2$ of the gauge coupling,
and  with ${\cal N}=2$ broken down to ${\cal N}=1$ by the large mass parameter
$\mx\gg \Lambda_2$ of the adjoint field $X=X^A T^A,\, {\rm Tr}\,(T^A
T^B)=\delta^{AB}/2$. At large $\mu\gg\mx$ the Lagrangian looks as
(the exponents with gluons are implied in the Kahler term K)
\bq
K=\frac{1}{g_2^2(\mu,\lm)}{\rm Tr}\,(X^\dagger X)+{\rm
Tr}\,({\textbf{Q}}^\dagger \textbf{Q}+\textbf{Q}\ra \ov{ \textbf{Q}})\,,\quad
   g_2^2=4\pi\alpha_2\,.\label{(39.13)}
\eq
\bbq
\cw=\frac{2\pi}{\alpha_2(\mu,\Lambda_2)}S+\mx {\rm Tr}\,(X^2)+\sqrt{2}\,{\rm
Tr}\,(\ov {\textbf{Q}}X\textbf{Q})+ m\,{\rm Tr}\,(\ov {\textbf{Q}}\textbf{Q}).
\eeq

The Konishi anomalies look here as
\bbq
\langle X^A\frac{\partial \cw}{\partial X^A}\rangle=\mx\langle X^A X^A
\rangle+{\rm Tr}\langle J^A X^A \rangle=2N_c\langle S\rangle,\quad
J^{A,i}_j={\sqrt 2}\,(\ov{\textbf{Q}}_j T^A \textbf{Q}^i)\,,\quad {\rm
Tr}\,\Bigl (T^A T^B \Bigr )=\frac{1}{2}\,\delta^{AB}\,,
\eeq
\bbq
\langle\ov{\textbf{Q}}_i\frac{\partial \cw}{\partial
\ov{\textbf{Q}}_i}\rangle=\langle J^{A,i}_i X^A\rangle+m\langle\ov
{\textbf{Q}}_i \textbf{Q}^i \rangle=\langle S\rangle\,, \quad \rm {no \,\,\,
summation\,\,\, over\,\,\, i}\,.
\eeq
From these
\bbq
2\langle{\rm Tr}\,X^2 \rangle=\langle X^A
X^A\rangle=\frac{1}{\mx}\Bigl [(2N_c-N_F)
\langle S\rangle+m\langle{\rm Tr}\,\ov{\textbf{Q}} \textbf{Q}\rangle \Bigr ]\,.
\eeq

The running mass of $X$ is $\mx(\mu)=g_{2}^{2}(\mu)\mx$, so that at scales
$\mu<\mu^{\rm pole}_{\rm x}=g_{2}^{2}(\mu^{\rm pole}_{\rm x})\mx$ the field $X$ decouples 
from  the dynamics and the RG evolution becomes those of ${\cal N}=1$ SQCD. The
matching of ${\cal N}=2$ and ${\cal N}=1$ couplings at $\mu=\mu^{\rm pole}_{\rm x}$
looks as ($\Lambda_2$ and $\la$ are the scale factors of ${\cal N}=2$ and
${\cal N}=1$ gauge couplings, $\la$ is held fixed when $\mx\gg\la$ is varied,
${\rm b}_2=2N_c-N_F\,,\,\bo=3N_c-N_F$)
\bbq
\frac{2\pi}{\alpha_2(\mu=\mu^{\rm  pole}_{\rm x},\Lambda_2)}=
\frac{2\pi}{\alpha(\mu=\mu^{\rm pole}_{\rm x},\la)}\,,
\quad \mx\gg\la\gg\Lambda_2\,,
\eeq
\bq
{\rm b}_2\ln\frac{\mu^{\rm pole}_{\rm x}}{\Lambda_2}=\bo\ln\frac{\mu^{\rm
pole}_{\rm x}}{\la}-N_F\ln z_Q(\la,\mu^{\rm pole}_{\rm x})+N_c\ln\frac{1}
{ g^2(\mu=\mu^{\rm  pole}_{\rm x})}\,,\label{(39.14)}
\eq
\bbq
\la^{\bo}=\frac{\Lambda^{{\rm b}_2}_2\mu^{N_c}_{\rm x}}{z^{N_F}_Q (\la,\mu^{\rm
pole}_{\rm x})},  \quad z_Q(\la,\mu=\mu^{\rm pole}_{\rm x})
\sim\Bigl (\ln\frac{\mu^{\rm pole}_{\rm x}}{\la}\Bigr )^{\frac{N_c}{\bo}}\gg 1.
\eeq

Although the field $X$ becomes too heavy and does not propagate any more at
$\mu<\mu^{\rm pole}_{\rm x}$, the loops of light quarks and gluons are still
active at $\la<\mu<\mu^{\rm pole}_{\rm x}$ if the next largest physical mass $\mu_H$
is below $\la$. These loops induce then to the field $X$ at $\mu=\la\ll\mu^{\rm
pole}_{\rm x}$ a non-trivial logarithmic renormalization factor $z_X(\mu^{\rm
pole}_{\rm x}, \la)=z^{-1}_X(\la,\mu^{\rm pole}_{\rm x})\ll 1$.

Therefore, finally, at scales $\la\ll\mu\ll\mu^{\rm pole}_{\rm x}$  the Lagrangian of the
broken  ${\cal N}=2$ - theory with $0<N_F<2N_c$ can
be written as
\bq
K=\frac{z_X(\mu^{\rm pole}_{\rm x},\mu)}{g^2(\mu^{\rm pole}_{\rm x})}\,{\rm Tr}\,(X^\dagger
X)+z_Q(\mu^{\rm pole}_{\rm x},\mu)\,{\rm Tr}\,({\textbf{Q}}
^\dagger \textbf{Q}+\textbf{Q}\ra \ov{\textbf{Q}})\,,\label{(39.15)}
\eq
\bbq
\cw=\frac{2\pi}{\alpha(\mu,\la)}S+\mx {\rm Tr}\,(X^2)+\sqrt{2}\,{\rm
Tr}\,(\ov{\textbf{Q}} X \textbf{Q})+ m\,{\rm
Tr}\,(\ov{\textbf{Q}}\textbf{Q})\,.\eeq
\bq
z_X(\mu^{\rm pole}_{\rm x},\mu)\sim\Biggl (\,\frac{\ln\, (\mu/\la)}{\ln\, (\mu^{\rm
pole}_{\rm x}/\la)}\,\Biggr)^{{\rm b}_2/\bo}\ll 1\,, \label{(39.16)}
\eq
\bbq
z_Q(\mu^{\rm pole}_{\rm x},\mu)=z_Q(\mu^{\rm pole}_{\rm x},\la)z_Q(\la,\mu),\quad
z_Q(\la,\mu)\sim \Bigl (\ln\frac{\mu}{\la}\Bigr)^{N_c/\bo}\gg 1\,.
\eeq

In all cases when the field $X$ remains too heavy and dynamically irrelevant,
it  can be integrated out in \eqref{(39.15)} and one obtains
\bq
K=z_Q(\mu^{\rm pole}_{\rm x},\mu)\,{\rm Tr}\,({\textbf{Q}}^\dagger
\textbf{Q}+\textbf{Q}\ra \ov{\textbf{Q}})\,,\label{(39.17)}
\eq
\bbq
\cw_Q=\frac{2\pi}{\alpha(\mu,\la)}S+ m\,{\rm Tr}(\ov{\textbf{Q}}
\textbf{Q})-\frac{1}{2\mx}\Biggl ({\rm
Tr}\,(\ov{\textbf{Q}}\textbf{Q})^2-\frac{1}{N_c}\Bigl({\rm Tr}\,\ov{\textbf{Q}}
\textbf{Q} \Bigr)^2 \Biggr ).
\eeq

Now we redefine the normalization of  quarks fields
\bq
\textbf{Q}=\frac{1}{z^{1/2}_Q(\mu^{\rm pole}_{\rm x},\la)}\,Q\,,\quad
\ov{\textbf{Q}}=\frac{1}{z^{1/2}_Q(\mu^{\rm pole}_{\rm x},\la)}\,{\ov
Q}\,,\label{(39.18)}
\eq
\bq
K=z_Q(\la,\mu){\rm Tr}\Bigl (\,Q^\dagger Q+(Q\ra {\ov Q})\,\Bigr ),\quad
\cw=\frac{2\pi}{\alpha(\mu,\la)}S+W_Q\,,\label{(39.19)}
\eq
\bq
\cw_Q=\frac{m}{z_Q(\mu^{\rm pole}_{\rm x},\la)}\,{\rm Tr}({\ov Q} Q)-\frac{1}
{2  z^2_Q(\mu^{\rm pole}_{\rm x},\la)\mx}
\Biggl ({\rm Tr}\,({\ov Q}Q)^2-\frac{1}{N_c}\Bigl({\rm Tr}\,{\ov Q} Q \Bigr)^2
\Biggr ).\label{(39.20)}
\eq
Comparing this with \eqref{(14.1.3)} and choosing
\bq
\frac{m}{z_Q(\mu^{\rm pole}_{\rm x},\la)}=m_Q\ll\la\,, \quad z^2_Q
(\mu^{\rm  pole}_{\rm x},\la)\mx=\mph\gg\la \label{(39.21)}
\eq
it is seen that with this matching the $\Phi$ - theory and the broken ${\cal
N}=2$ SQCD, both normalized canonically at $\mu=\la$,  will be equivalent.

Therefore, at large $\mx\gg\Lambda_2$ and until both $X$ and $\Phi$ fields
remain dynamically irrelevant, all results obtained above for the $\Phi$ -
theory will be applicable to the broken ${\cal N}=2$ SQCD as well. Besides, the
$\Phi$ and $X$ fields remain dynamically irrelevant in the same region of
parameters, i.e. at $N_F<N_c$ and at $\mph>\mo$ or $\mu_H>\mu_o$ at 
$\mph<\mo$ if $N_F>N_c$.

Moreover, some general properties of both theories such as {\it the
multiplicity  of vacua with unbroken or broken flavor symmetry and the values of
vacuum condensates of corresponding chiral superfields} (i.e. $\langle{\ov Q}_j
Q^i\rangle$ and $\langle S\rangle$)  {\it are the same in these
two theories, independently of whether the fields $\Phi$ and $X$ are irrelevant
or relevant}.

Nevertheless, once the fields $\Phi$ and $X$ become relevant (e.g. at
$\mx\ll\Lambda_2$), the phase states, the RG evolution, the mass spectra
etc., \, {\it become very different in these two theories}. The properties of the
$\Phi$ - theory were described in detail above in the text. In general, if $X$ is
sufficiently light and dynamically relevant, the dynamics of the softly broken
${\cal N}=2$ SQCD becomes complicated and is described in Part III.\\

Finally, we trace now a transition to the slightly broken ${\cal N}=2$ theory
with small $\mx\ll\Lambda_2$ and fixed $\Lambda_2$. For this, we write first 
the appropriate form of the
effective superpotential obtained from \eqref{(39.19)},\eqref{(39.20)}
\bbq
\cw^{\rm eff}_Q=\nd S+\frac{m}{z_Q(\mu^{\rm pole}_{\rm x},\la)}\,{\rm Tr}({\ov Q}
Q)-\frac{1}{2 z^2_Q(\mu^{\rm pole}_{\rm x},\la)\mx}\Biggl ({\rm Tr}\,({\ov
Q}Q)^2-\frac{1}{N_c}\Bigl({\rm Tr}\,{\ov Q} Q \Bigr)^2 \Biggr ),
\eeq
\bq
S=\Biggl (\,\,\frac{\det {\ov Q}Q}{\la^{\bo}}\,\,\Biggr )^{1/\nd}\,,\quad
\la^{\bo}=z^{N_F}_Q(\mu^{\rm pole}_{\rm x},\la)\Lambda^{b_2}_2\mu^{N_c}_{\rm x}
\label{(39.22)}
\eq
and restore now the original normalization of the quark fields
$\ov{\textbf{Q}},\textbf{Q}$ appropriate for the slightly broken ${\cal N}=2$
theory with varying  $\mx\ll\Lambda_2$ and fixed $\Lambda_2$, see
\eqref{(39.18)},
\bq
\hspace*{-1cm} \cw^{\rm eff}_Q=\nd S+m\,{\rm Tr}(\ov{\textbf{Q}}
\textbf{Q})-\frac{1}{2 \mx}\Biggl ({\rm
Tr}\,(\ov{\textbf{Q}}\textbf{Q})^2-\frac{1}{N_c}\Bigl({\rm Tr}\,\ov{\textbf{Q}}
\textbf{Q} \Bigr)^2 \Biggr ),\quad S=\Bigl (\frac{\det
\ov{\textbf{Q}}\textbf{Q}}{\Lambda^{{\rm b}_2}_2\mu^{N_c}_{\rm x}} \Bigr )^{1/\nd}.
\label{(39.23)}
\eq

One can obtain now from \eqref{(39.23)} the values of the quark condensates
$\langle\ov{\textbf{Q}}_j \textbf{Q}^i\rangle$ at fixed $\Lambda_2$ and small
$\mx\ll\Lambda_2$. Clearly, in comparison with values of   $\langle{\ov Q}_j Q^i
\rangle$  in the direct $\Phi$-theory,  the results for $\langle\ov{\textbf{Q}}_j 
\textbf{Q}^i \rangle$   in \eqref{(39.23)} are obtained by the replacements\,:\, 
$m_Q\ra m\,,\,  \mph\ra\mx,\,\,  \la^{\bo}=\Lambda^{b_2}_2\mu^{N_c}_{\rm x}$ 
\, in the effective  superpotential,  while multiplicities of vacua 
are the same.  From \eqref{(39.23)}, the dependence of  $\langle\ov{\textbf{Q}}_j 
\textbf{Q}^i\rangle$ and $\langle S\rangle$ on small  
$\mx \ll\lm$ \, at fixed $\lm$ is trivial in all vacua, $\sim \mx$.

It is worth only to recall that, as the effective
superpotential without the 4-quark term in the ${\cal N}=1$ SQCD,
\eqref{(39.23)} is {\it not} a genuine low energy superpotential, {\it it can \, 
be used only for finding the values of mean vacuum values
$\langle{\ov Q}_j Q^i\rangle_{N_c}$ and $\langle S\rangle_{N_c}$}. The
genuine low energy superpotentials in each vacuum are given below in
Part III of the text. \\

With the above replacements, the expressions for $\langle{\ov Q}_j Q^i\rangle$
in the region $\la\ll\mph\ll\mo$  for the direct $\Phi$-theory correspond here
 to the hierarchy
$m\ll\Lambda_2$, while those in the region $\mph\gg\mo$ correspond here to
$m\gg\Lambda_2$. In the language of \cite{APS} used in \cite{CKM} (see sections
6-9 therein), the correspondence between the $r$ - vacua \cite{APS,CKM} of the
slightly broken ${\cal N}=2$ theory with $0< \mx/\Lambda_2\ll 1,\,\, 0<
m/\Lambda_2\ll 1$ and our vacua in section 41 looks as
\footnote{\,
This correspondence is based on comparison of multiplicities of our vacua at
$\mph\ll\mo$ described in section 41 and those of $r$ - vacua at $m\ll\Lambda_2$
and $\mx\ll\Lambda_2$ as these last are given in \cite{CKM}.
}
\,: \,a)\, \,$r=n_1$,\, b)\, our L - vacua with the unbroken or the L - type
ones with spontaneously broken flavor symmetry correspond, respectively, to the
first group of vacua of the non-baryonic branches with $r=0$ and $r\geq 1,\,
r\neq\nd$ in \cite{CKM}\,,\,\, c)\, our S - vacua with the unbroken flavor
symmetry and $\rm br2$ - vacua with the spontaneously broken flavor symmetry
correspond to the first type from the second group of vacua of the baryonic
branches with, respectively, $r=0$ and $1\leq r<\nd$ in \cite{CKM},\,\, d)\,
our special vacua with $n_1=\nd,\, n_2=N_c$ correspond to the second type of
vacua from this group, see \cite{CKM}.

\addcontentsline{toc}{section}
{\bf \Large  Part III.  Softly broken $\mathbf {{\cal N}=2\ra {\cal N}=1}$ SQCD.\,\, \\
Mass spectra in vacua with unbroken  $\mathbf {Z_{\bb} }$  symmetry}

\begin{center}
\bf \Large    Part III.  Softly broken $\mathbf {\cal N}=2\ra {\cal N}=1$  SQCD.\,
Mass spectra in vacua with unbroken  $\mathbf {Z_{\bb}}$  symmetry   
\end{center}

\section{Introduction}

\hspace*{4mm} Considered are ${\cal N}=2\,\, SU(N_c)$ or $U(N_c)$ SQCD with
$N_F<2N_c$ equal mass quark flavors. ${\cal N}=2$ supersymmetry is
softly broken down to ${\cal N}=1$ by the mass term $\mx{\rm
Tr}\,(X^2)$ of colored adjoint scalar partners of gluons, $\mx
\ll\Lambda_2$ (\,$\Lambda_2$ is the scale factor of the $SU(N_c)$
gauge coupling).

There is a large number of different types of vacua in these theories
with both unbroken and spontaneously broken global flavor symmetry,
$U(N_F)\rightarrow U({\rm n}_1)\times U({\rm n}_2)$. We consider in
this paper the large subset of these vacua with the unbroken
non-trivial $Z_{2N_c-N_F\geq 2}$ discrete symmetry, at different
hierarchies between the Lagrangian parameters $m\gtrless\Lambda_2,\,
\,\mx\gtrless m$. The forms of low energy Lagrangians, quantum
numbers of light particles and mass spectra are described in the main
text for all these vacua.

The calculations of power corrections to the leading terms of the low
energy quark and dyon condensates are presented in two important
Appendices. The results agree with also presented in these Appendices
{\it independent} calculations of these condensates using roots of
the \, Seiberg-Witten spectral curve. {\it This agreement confirms in a
non-trivial way a self-consistency of the whole approach}.

Our results differ essentially from corresponding results in e.g.
related papers \cite{SY1}, \cite{SY2} and \cite{SY6}  of M.Shifman 
and A.Yung (and in a number of their  previous numerous papers on 
this subject), and we explain in the text below  the reasons for these 
differences.   (See also the extended critique of
a number of results of these authors in section 8.3 of \cite{ch13}.\\

\vspace*{2mm}

\hspace*{4mm} The masses and quantum numbers of BPS particles in
${\cal N}=2\,\, SU(2)$ SYM and $SU(2)$ SQCD with $N_F=1...4$ quark
flavors at small $\mx\neq 0$ were found in seminal papers of N.Seiberg \, 
and  E.Witten \cite{SW1,SW2}. They presented in particular the
corresponding spectral curves from which the masses and quantum
numbers of these BPS particles can be calculated at $\mx\ra 0$. The
forms of these curves have been generalized then, in particular, to
${\cal N}=2\,\, SU(N_c)$ SYM \cite{DS} and ${\cal N}=2\,\, SU(N_c)$
SQCD with $1\leq N_F\leq 2N_c$ quark flavors \cite{KL,AF,HO,Shap,APS}.
In what follows we deal mainly with ${\cal N}=2\,\, SU(N_c)$ SQCD
with \, $0 <  N_F<2N_c-1$ flavors of equal mass quarks $Q^i_a,\,{\ov
Q}_i^{\,a},\,i=1...N_F,\,\,a=1...N_c$. The Lagrangian of this UV free
theory, broken down to ${\cal N}=1$ by the mass term $\mx{\rm
Tr}\,(X^{\rm adj}_{SU(N_c)})^2$ of adjoint scalars, can be written at
sufficiently high scale $\mu\gg\max\{\lm, m\}$ as (the gluon
exponents in Kahler terms are implied in \eqref{(40.1)} and 
everywhere below)
\bbq
K=\frac{2}{g^2(\mu)}{\rm Tr}\,\Bigl [(X^{\rm adj}_{SU(N_c)})^\dagger
X^{\rm adj}_{SU(N_c)}\Bigr ]_{N_c}+{\rm Tr}\,(Q^\dagger Q+{\ov
Q}^{\,\dagger} {\ov Q})_{N_c}\,, \quad X^{\rm adj}_{SU(N_c)}=T^A
X^A\,,\quad A=1,\,...,N_c^2-1\,,
\eeq
\bq
{\cal W}_{\rm matter}=\mx{\rm Tr}\,(X^{\rm adj}_{SU(N_c)})_{N_c}^2
+{\rm Tr}\,\Bigl (m\,{\ov Q} Q-{\ov Q}\sqrt{2} X^{\rm adj}_{SU(N_c)} Q\Bigr
)_{N_c}\,,\quad {\rm Tr}\, (T^{A_1}
T^{A_2})=\frac{1}{2}\,\delta^{A_1 A_2}\,,\label{(40.1)}
\eq
with the scale factor $\lm$ of the gauge coupling $g(\mu)$, and ${\rm
Tr}$ in \eqref{(40.1)} is over all colors and flavors. The "softly
broken"\, ${\cal N}=2$ theory means $0<\mx\ll\lm$. Considering $\mx$
and $m$ as a soft background fields, there is the unbroken $U(1)$ 
R-symmetry with the charges of fields and parameters: $r_{\lambda}
=r_{\theta}=r_{Q}=r_{\ov Q}=1,\, r_X=r_m=r_{\lm}=0,\, r_{\mx}=2$.

The spectral curve corresponding to \eqref{(40.1)} can be written at
$N_F < 2N_c-1$ e.g. in the form \cite{APS}
\bq
y^2=\prod_{i=1}^{N_c}(z+\phi_i)^2-4\lm^{\bb}(z+m)^{N_F}\,,\quad
\sum_{i=1}^{N_c}\phi_i=0{\rm\,\,\, in\,\,\, SU(N_c)}\,, \label{(40.2)}
\eq
where $\{\phi_i\}$ is a set of gauge invariant co-ordinates on the
moduli space. Note that, as it is, this curve can be used for the
calculation of {\it \,only BPS particle masses at $\mx\ra 0$ only}.

In principle, it is possible to find at $\mx\ra 0$ from the curve
\eqref{(40.2)} the quantum numbers and masses of all massive and
massless BPS particles in different multiple vacua of this theory. 
But in \, practice it is very difficult to do this for general values of
$N_c$ and $N_F$, especially as for quantum numbers . The important
property of the $SU(N_c)$ curve \eqref{(40.2)} which will be used in
the text below is that there are maximum $N_c-1$ double roots. 
And the \, vacua we
will deal with below at $\mx\neq 0$ are just these vacua. In
other words, {\it in all $SU(N_c)$ vacua we will deal with, the
curve}\eqref{(40.2)} {\it has $N_c-1$ double roots and two single roots}.

The first detailed attempt has been made in \cite{APS} (only the case
$m=0$ was considered) to classify the vacua of \eqref{(40.1)} for
general values of $N_c$ and $N_F$, to find quantum numbers of light
BPS particles (massless at $\mx\ra 0$) in these vacua and forms of low \, 
energy  Lagrangians. It was found in particular that in all vacua the
$SU(N_c)$ gauge symmetry is broken spontaneously at the scale
$\mu\sim\lm$ by higgsed adjoint scalars, $\langle X^{\rm
adj}_{SU(N_c)}\rangle\sim\lm$.

That this should happen can be understood qualitatively as follows,
see e.g. \cite{ch16}. The perturbative NSVZ $\beta$-function of the
(effectively) massless unbroken ${\cal N}=2$ SQCD is exactly one loop
\cite{NSVZ-1,NSVZ-2}. The theory \eqref{(40.1)} is UV free at
$(2N_c-N_F)>0$. At small $\mx$ and $m\ll\lm$, and if the whole matrix
$\langle X^{adj}_{SU(N_c)}\rangle\ll\lm$ (i.e. in the effectively
massless at the scale $\mu\sim\lm$ unbroken $\,{\cal N}=2$ SQCD), the
coupling $g^2(\mu)$ is well defined at $\mu\gg\lm$, has a pole at
$\mu=\lm$ and becomes negative at $\mu<\lm$. To avoid this unphysical
behavior, {\it the field $X$ is necessarily higgsed breaking the
$SU(N_c)$ group, with (at least some) components $\langle
X^A\rangle\sim\lm$}. (This becomes especially clear in the limit of
the unbroken ${\cal N}=2$ theory at $\mx\ra 0$ and small $m\ll\lm$,
when all chiral quark condensates and the gluino condensate, i.e.
their corresponding mean vacuum values $\langle...\rangle$, approach zero, 
so \, that all quarks are not higgsed in any case, and all particles are
very light at $\langle X^{\rm adj}_{SU(N_c)}\rangle\ll\lm$).

Remind that exact values of $\langle X^A\rangle$ account for possible
nonperturbative instanton contributions. And it is worth also to note
that, in the limit of unbroken ${\cal N}=2$ SQCD with small $m$ and
$\mx\ra 0$, if there were no some components $\langle
X^A\rangle\sim\lm$ (i.e. all $\langle X^A\rangle\ll\lm$), {\it the
nonperturbative instanton contributions also would be non operative 
at \, the scale $\sim \lm$ without both the corresponding fermion masses
$\sim\lm$ and the infrared cut off $\,\,\rho\lesssim\, 1/\lm$ supplied \, 
by  some $\langle X^A\rangle\sim\lm$. Therefore, the problem with
$g^2(\mu<\lm)<0$ will survive in this case. Moreover, for the same
reasons, if there remains the non-Abelian subgroup unbroken at the
scale $\lm$, \,\, it \, has to be IR free (or at least conformal)}.

It was found in \cite{APS} that (at $m=0$) and $\mx\ra 0$ there are
two qualitatively different branches of vacua. On the baryonic branch
the lower energy gauge group at $\mu<\lm$ is $SU(N_F-N_c)\times
U^{\bb}(1)$, while on non-baryonic branches these are $SU(n_1)\times
U^{N_c-n_1}(1),\,\, n_1\leq N_F/2$.

Besides, at $m\neq 0$ and small $\mx\neq 0$, the global flavor
symmetry $U(N_F)$ of \eqref{(40.1)} is unbroken or broken spontaneously \, 
as  $U(N_F)\ra U(\no)\times U(\nt),\, 1\leq\no\leq [N_f/2]$ in various
vacua \cite{CKM}.

As was also first pointed out in \cite{APS}, there is the residual
$Z_{2N_c-N_F}$ discrete R-symmetry in theory \eqref{(40.1)}. {\it This
symmetry will play a crucial role in what follows, as well as a
knowledge of multiplicities of all different types of vacua}, see e.g.
 \cite{ch13, ch19}. The charges of \, fields and
parameters in the superpotential of \eqref{(40.1)} under
$Z_{\bb}=\exp\{i\pi/(\bb)\}$ transformation are: $q_{\lambda}=
q_{\rm\theta}=1,\,\, q_{X}=q_{\rm m}=2,\,\, q_Q=q_{\,\ov
Q}=q_{\lm}=0,\,\, q_{\mx}=-2$. If this non-trivial at $\bb\geq 2$
discrete $Z_{\bb}$ symmetry is broken spontaneously in some vacua,
there will appear then the factor $\bb$ in their multiplicity.
Therefore, if multiplicities of different types of vacua are known, we \, 
can see explicitly whether this $Z_{\bb}$ symmetry is broken
spontaneously or not in these vacua. As a result, it is seen the
following at $0<N_F<2N_c-1,\, m\ll\lm$ (in the language of
\cite{ch13,ch19}).

{\bf I)}.\, In $SU(N_c)$ theories at $m\ll\lm$ ($\,\langle ({\ov Q}
Q)_{1,2}\rangle_{N_c}$ and $\langle S\rangle_{N_c}=\langle
\lambda\lambda/32\pi^2\rangle$ are the quark and gluino condensates
summed over all their colors\,). The discrete $Z_{\bb\geq 2}$ symmetry \ 
is   broken spontaneously.-

1) In L (large) vacua with the unbroken global flavor symmetry
$U(N_F)$, the multiplicity $2N_c-N_F$ and with
$\langle\QQ\rangle_{N_c}\sim\mx\lm,\,\,\langle
S\rangle_{N_c}\sim\mx\lm^2,\,\,\langle{\rm Tr}
(X^{adj}_{SU(N_c)})^2\rangle\sim\lm^2$.\\
2) In Lt (\,L-type\,) vacua with the spontaneously broken flavor
symmetry, $U(N_F)\ra U(\no)\times U(\nt),\,\,1\leq\no\leq N_F/2$, the
multiplicity $(2N_c-N_F)C^{\,\no}_{N_F},\,C^{\,\no}_{N_F}=(N_F!/\no!\,\nt!)\,,$ 
and \, with $\langle\Qo\rangle_{N_c}\sim\langle\Qt\rangle_{N_c}\sim\mx
\lm,\,\,\langle S\rangle_{N_c}\sim\mx\lm^2,\,\, \langle{\rm Tr}
(X^{adj}_{SU(N_c)})^2\rangle\sim\lm^2$.\\
3) In special vacua with $U(N_F)\ra U(\no=N_F-N_c)\times U(\nt=N_c)$,
the multiplicity $(2N_c-N_F)C^{\,N_F-N_c}_{N_F}$, and with $\langle\Qo
\rangle_{N_c}\sim\mx m,\,\,
\langle\Qt\rangle_{N_c}=\mx\lm,\,\, \langle S\rangle_{N_c}\sim\mx
m\lm,\,\, \langle{\rm Tr} (X^{adj}_{SU(N_c)})^2\rangle\sim m\lm$.

While $Z_{\bb\geq 2}$ is unbroken. - \\
4) In br2 (br=breaking) vacua with $U(N_F)\ra U(\no)\times U(\nt),\,\,
\nt>N_c$, the multiplicity $(N_F-N_c-\no)C^{\,\no}_{N_F}$, and with
the dominant condensate $\langle\Qt\rangle_{N_c}\sim\mx m\gg
\langle\Qo\rangle_{N_c}\sim\mx m (m/\lm)^{\frac{2N_c-N_F}{\nt-N_c}}$,
$\langle S\rangle_{N_c}\sim\mx m^2
(m/\lm)^{(2N_c-N_F)/(\nt-N_c)},\,\,\langle{\rm Tr}
(X^{adj}_{SU(N_c)})^2\rangle\sim m^2$.\\
5) In S (small) vacua with unbroken $U(N_F)$, the multiplicity
$N_F-N_c$ and with $\langle\QQ\rangle_{N_c}
\sim\mx m,\,\, \langle S\rangle_{N_c}\sim\mx m^2
(m/\lm)^{(2N_c-N_F)/(N_F-N_c)},\,\, \langle{\rm Tr}
(X^{adj}_{SU(N_c)})^2\rangle\sim m^2$ (see  
\cite{ch13, ch19}  for all details).

{\bf II)}. $Z_{\bb\geq 2}$ is unbroken in all multiple vacua of $SU(N_c)$
theories at $m\gg\lm$ (see Part III.2  in the table of contents).

{\bf III)}. There are additional vs (very special)  vacua with e.g. 
$N_c+1\leq N_F\leq 2N_c-1$, with the spontaneously broken global flavor 
symmetry  $U(N_F)\ra U(\no=N_F-N_c)\times U(\nt=N_c),\,\,
\langle\Qt\rangle_{N_c}=\mx m,\,\,
\langle\Qo\rangle_{N_c}=0,\,\,\,\langle S\rangle_{N_c}=0,\,\,
\langle{\rm Tr} (X^{adj}_{SU(N_c)})^2\rangle=0$, and the multiplicity
$C^{N_F-N_c}_{N_F}$ in $U(N_c)$ theories at all $m\gtrless\lm$ (such
vacua are absent in $SU(N_c)$ theories).
$Z_{\bb\geq 2}$ is unbroken in these vs-vacua at all $m\gtrless\lm$.\\

It was stated in \cite{APS} that {\it in all vacua} of $SU(N_c)$
\eqref{(40.1)}, \, baryonic and non-baryonic, all light particles in low
energy Lagrangians with $m=0$ and small $\mx$, i.e. quarks, gluons 
and scalars in non-Abelian sectors and corresponding particles in 
Abelian ones, are pure magnetic.

Really, the final purpose of \cite{APS} was an attempt to derive the
Seiberg duality for ${\cal N}=1$ SQCD \cite{S2,IS}. The idea was as\, 
follows. On the one hand, the electric ${\cal N}=2$ SQCD in
\eqref{(40.1)} with small both $\mx$ and $m$ flows clearly to the
standard ${\cal N}=1$ electric SQCD with $SU(N_c)$ gauge group, $N_F$
flavors of light electric quarks and without adjoint scalars 
$X^{\rm adj}$ at \,$m={\rm const},\,\,\mx\ra \infty$ and fixed
$\la^{3N_c-N_F}=\lm^{2N_c-N_F}\mx^{N_c}$. On the other hand, the
authors of \cite{APS} expected that, starting in \eqref{(40.1)} at
$\mx\ll\lm$ from vacua of the baryonic branch with the $SU(\nd=N_F-N)$ \, 
low  energy \, gauge group {\it by itself} and increasing then
$\mx\gg\lm$, they will obtain Seiberg's dual ${\cal N}=1$\, {pure magnetic}
SQCD with $SU(\nd=N_F-N_c)$ gauge group and $N_F$ flavors of massless
dual magnetic quarks. (But, in any case, $N^2_F$ light Seiberg's
mesons $M^i_j\ra ({\ov Q}_j Q^i)$ of dual ${\cal N}=1\,\, SU(\nd)$
SQCD \cite{S2} were missing in this approach, see section 8 in
\cite{ch13} for a more detailed critique of \cite{APS}\,).

The results from \cite{APS} were generalized then in \cite{CKM} to
nonzero mass quarks, with emphasis on properties of flavor symmetry
breaking in different vacua. Besides, multiplicities of various vacua
were discussed in this paper in some details. (A complete detailed
description of different vacua multiplicities and values of quark and
gluino condensates, $\langle{\ov Q}_j Q^i\rangle_{N_c}$ and $\langle
S\rangle_{N_c}$, in various vacua and at different hierarchies between \, 
$\mx,\,m$  and $\lm$ (or $\la^{3N_c-N_F}=\lm^{2N_c-N_F}
\mx^{N_c}$) see in \cite{ch13,ch16,ch19}\,).

But further detailed investigations of light particles quantum numbers \  
in theory  \eqref{(40.1)}, in some simplest examples with small values
of $N_c$ and $N_F$, did not confirm the statement \cite{APS} that all
these particles, and in particular light quarks, are pure magnetic.

Specifically, the $SU(N_c=3)$ vacua of \eqref{(40.1)}, with $N_F=4,\,\,\no=2,\,
\nt=N_F-\no=2$, (these are Lt - vacua with the gluino condensate 
$\langle S\rangle_{N_c}\sim\mx\lm^2$
and spontaneously broken non-trivial $Z_{\bb=2}$ symmetry at
$m\ll\lm$),  and vacua with $N_F=5,\,\, \no=2,\, \nt=3$ (these 
are special vacua with $\langle S\rangle_{N_c}\sim m\, \mx\lm$, 
 $Z_{\bb=1}$ symmetry acts trivially in this case and gives \, no
restrictions on the form of $\langle X^{\rm adj}_{SU(3)}\rangle$),
were considered in details at $\mx\ra 0$ and $m\lessgtr\lm$ in
\cite{MY}. Besides, the vs (very special) $U(N_c=3)=SU(N_c=3)\times
U^{(0)}(1)$ vacua with $N_F=5,\,\, \no=2,\, \nt=3,\, \langle
S\rangle_{N_c}=0$ were considered earlier in \cite{SY3}. 
In these vacua, the light charged particles are not magnetic 
monopoles, but dyons $D^{\,i}_a,\,{\ov D}^{\,a}_i,\,\, a=1,\,2,\,
\,i=1...4 (5)$ (massless  at $\mx\ra 0$)   which form $SU(2)$ doublet   
of dyons with the $SU(3)$  fundamental electric charges and with 
the hybrid magnetic charges  which are fundamental with respect 
to $SU(2)$ but adjoint with respect  to  $SU(3)$.

We can propose e.g. the following picture for these UV free
$SU(N_c=3)$ theories with $N_F=4$ or $N_F=5$ at $m\ll\lm,\,\mx
\ra 0$. To avoid $g^2(\mu<\lm)<0$ and without restrictions from the
non-trivial unbroken $Z_{2N_c-N_F}$ symmetry, the electric gauge 
group \, $SU(3)$ is \, higgsed in these vacua by $\langle X^{\rm
adj}_{SU(3)}\rangle\sim diag (\lm, \lm, -2\lm)$, i.e. $SU(3)\ra
SU(2)\times U(1)$, so that all original pure electrically charged
quarks acquire masses $\sim\lm$. But, due to strong coupling
$a(\mu\sim\lm)=N_c g^2(\mu\sim\lm)/8\pi^2\sim 1$ and very specific
properties of the enhanced ${\cal N}=2$ SUSY, the heavy flavored
electric quarks $Q^{\,i}_a,\,{\ov Q}^{\,a}_i,\,\, a=1,\,2,\,\,i=1...4
(5)$ (i.e. $SU(2)$ color doublet) with equal masses $\sim\lm$ combined \, 
with two  heavy unflavored pure magnetic monopoles with the same equal
masses $\sim\lm$ and with above described hybrid magnetic charges, and \, 
formed  the \, $SU(2)$ doublet of light flavored BPS dyons
$D^{\,i}_a,\,{\ov D}^{\,a}_i$, massless at $\mx\ra 0$.

These massless at $\mx\ra 0$ quark-like dyons $D^{\,i}_a,\,{\ov  D}^{\,a}_i,\,
\, a,i=1,\,2$ coupled then with massless $SU(2)$ adjoint gluons and
scalars (which transformed from pure electric to dyonic ones, and this \, 
is a most  surprising phenomenon) and formed the dyonic ${\cal N}=2\,
SU(2)$ non-Abelian group (conformal at $N_F=4$ and IR free at
$N_F=5$ at lower energies $\mu < \lm$.

As follows from the curve \eqref{(40.2)} and 
 from \cite{SY5}, the $SU(2)$ dyons are higgsed
simultaneously at small $\mx\neq 0$ e.g. as $\langle{\ov
D}^{\,a}_i\rangle=\langle D_{\,a}^i\rangle\sim \delta^i_a
(\mx\lm)^{1/2},\, a,i=1,2$ (and for this to be possible in the ground
state they have to be massless at $\mx\ra 0$ and mutually local). As a
result, the whole color $SU(2)$ group is broken and $(\no=2)^2=4$ long\, 
${\cal  N}=2$ multiplets of massive gluons with masses $\sim g_D
(\mx\lm)^{1/2}$ are formed. Besides, the global flavor $U(N_F)$ symmetry 
is broken spontaneously as $U(N_F)\ra U(\no=2)\times U(N_F-2)$ at 
$\mx\neq 0, \,\,\delta_{ij}=(m_i-m_j)=0$,   and
massless ${\cal  N}=1$ Nambu-Goldstone multiplets 
appear   (in essence,  these are not higgsed dyons).

These higgsed $SU(2)$ quark-like dyons $D^{\,i}_a, {\ov D}^{\,a}_i,\, i,a=1,2$
are mutually non-local with all heavy original quarks with $SU(3)$
colors and with the $SU(2)$ doublet of pure magnetic monopoles, all
with masses $\sim\lm$. Therefore, when these dyons $D^{\,i}_a, {\ov
D}^a_i$ are higgsed at small $\mx\neq 0$ as $\langle
D^{\,i}_a\rangle=\langle{\ov  D}_i^{\,a}\rangle\sim\delta^i_a\,
(\mx\lm)^{1/2}$, this results in a
weak confinement of these heavy charged particles, the string tension
is $\sigma^{1/2}\sim (\mx\lm)^{1/2}\ll\lm$. (The confinement is weak
in a sense that the string tension is much smaller than particle
masses). These heavy particles form hadrons with masses $\sim\lm$. 
In  the case of $U(N_c=3)$ vs-vacua with $N_F=5$ and $\langle
S\rangle_{N_c}=0$ from \cite{SY3}, there are in addition the 
massless  at $\mx\ra 0$\, $SU(2)$  singlet BPS dyons $D_3, {\ov
D}_{3}$ with $\langle{\ov D}_{3}\rangle=\langle D_3\rangle\sim
(\mx\lm)^{1/2}$ at $\mx\neq 0$. Now the additional long ${\cal
N}=2\,\, U(1)$ multiplet of massive gluon with the mass $\sim g_D
(\mx\lm)^{1/2}$ is formed.\\

Besides, it was shown explicitly in \cite{SY3,MY} at $\mx\ra 0$ on
examples considered, how vacua with some massless at $m\gg\lm$ and
$\mx\ra 0$ pure electric original quarks $Q^i$ can evolve at $m\ll\lm$ \, 
into  vacua \, with only massless dyons $D^i$. E.g., in the vicinity of
the Argyres-Douglas point $m=m_0\sim\lm$, there are vacua in which
{\it additional} corresponding particles become light. At $m=m_0$ the
vacua \, collide, all these particles become massless, and the lower energy
regime at $\mu<\lm$ is conformal. In particular, both some flavored
pure electric quarks $Q^i$ and flavored dyonic quarks $D^i$ are
massless at this point (as well as some pure magnetic monopoles). The
masses of these quarks with {\it fixed quantum numbers} behave
continuously but non-analytically in $m/\lm$ at this special point $m=m_0$.
E.g., in selected at $m\lessgtr m_0$ vacua colliding at $m=m_0$
with other vacua, the mass of some light pure electric quarks $Q^i$
stays at zero at $m\geq m_0$, then begins to grow at $m<m_0$ and
becomes $\sim\lm$ at $m\ll\lm$, while the mass of light composite
dyonic quarks $D^i$ is zero in these vacua at $m\leq m_0$, then begins \, 
to grow  at \, $m>m_0$ and becomes large at $m\gg m_0$. {\it
Equivalently}, tracing e.g. the evolution with $m/\lm$ in the small
vicinity of $m=m_0$ of {\it quantum numbers} of, separately, massless
or nearly massless flavored quark-like particles at $m\lessgtr m_0$,
one can say that their quantum numbers jump at the collision point
$m=m_0$. I.e.: a) the massless particles at $m > m_0$ are pure
electric quarks $Q^i$, while they are dyonic quarks $D^i$ at $m <
m_0$; b) the very light (but not massless) flavored quark-like
particles are the dyonic quarks $D^i$ at $m > m_0$, while they are
pure electric quarks $Q^i$ at $m < m_0$. These two views of the
non-analytic behavior at the collision point $m= m_0$ when traversing
the small vicinity of the collision point, i.e. either looking at the
non-analytic but continuous evolution of small masses of particles
with fixed quantum numbers, or looking on the jumps of quantum numbers \, 
of,  separately, massless particles and of nearly massless particles,
are clearly {\it equivalent and indistinguishable} because all these
particles simultaneously become massless at the collision point
$m=m_0$. These are two different projections (or two sides) of the
one phenomenon.

The examples of these conformal regimes for the ${\cal N}=2\,\,SU(2)$
theory were described earlier e.g. in \cite{BF,GVY}. But it is clear
beforehand that to avoid unphysical $g^2(\mu<\lm)<0$ in UV free
${\cal  N}=2\,\,SU(2)$ with $N_F=1,2,3$ and $m\ll\lm$, the whole gauge group
is higgsed necessarily at the scale $\mu\sim\lm$ by $\langle X^{\rm
adj}_{SU(2)}\rangle\sim\lm,\, SU(2)\ra U(1)$ \cite{SW1,SW2}, so that
all original pure electrically charged particles are indeed heavy,
with masses~ $\sim\lm$.\\

The properties of light particles with masses $\ll\lm$ in theory
\eqref{(40.1)} (really, in the $U(N_c)$ theory with especially added
$U^{(0)}(1)$ multiplet) for general values of $N_c,\, N_F$ and
possible forms of low energy Lagrangians at scales $\mu<\lm$ in
different vacua were considered later by M.Shifman and A.Yung in a
series of papers, see e.g. the recent papers \cite{SY1,SY2,SY6} and
references therein. At $m\ll\lm$, for vacua of baryonic branch with
the lower energy ${\cal N}=2\,\,\, SU(\nd=N_F-N_c)\times
U^{2N_c-N_F+1}(1)$ gauge group at $\mu<\lm$ \cite{APS}, it was first
proposed naturally by these authors in \cite{SY1} that all light
charged particles of ${\cal N}=2\,\, SU(\nd)$ are the original pure
electric particles, as they have not received masses $\sim\lm$
directly from higgsed $\langle X^{\rm adj}_{SU(N_c)}\rangle$ (this is
forbidden in these vacua by the unbroken non-trivial $Z_{\bb\geq 2}$
symmetry \cite{APS}). But they changed later their mind to the
opposite
\footnote{\,
Because not $SU(N_c)$ but $U(N_c)$ group was considered in
\cite{SY1,SY2}, this their statement concerns really br2 and very
special vacua of $U(N_c)$, see table of contents,
(these are respectively zero vacua and $r=N_c$ vacua in the 
language  of \cite{SY1,SY2}). \label{(f48)}
}
and, extrapolating freely by analogy their previous results for
$r=N_c=3,\, N_F=5,\, \no=2$ very special vacua in \cite{SY3} (see
also\cite{SY4}) to all br2 and very special vacua, stated in \cite{SY2}
that this $SU(\nd)$ group is not pure electric but dyonic. This means
that at $m\ll\lm$ all original pure electrically charged particles of
$SU(\nd)$ have received large masses $\sim\lm$, but now not directly
from $\langle X^{\rm adj}_{SU(N_c)}\rangle\sim\lm$ (this is forbidden
by the non-trivial unbroken $Z_{\bb\geq 2}$ symmetry), but from some
mysterious "outside" sources and decoupled at $\mu<\lm$, while the
same gauge group $SU(\nd)$ of light composite dyonic solitons was
formed. But it was overlooked in \cite{SY2} that, in
distinction with their example with $U(N_c=3),\,\,N_F=5,\,\,\no=2$ in
\cite{SY3} with the trivial $Z_{2N_c-N_F=1}$ symmetry giving no
restrictions on the form of $\langle X^{adj}_{SU(N_c)}\rangle$, the
appearance e.g. in the superpotential at $\mu\sim\lm$ of mass terms of \, 
original  $SU(\nd)$ quarks like $\sim\lm{\rm Tr}\,({\ov Q}Q)_{\nd}$ or
terms like $\sim\lm{\rm Tr\,}(X^2_{SU(\nd)})$ for scalars is forbidden \, in 
vacua  with the non-trivial unbroken $Z_{2N_c-N_F\geq 2}$ symmetry,
independently of whether these terms originated directly from
$\langle  X^{adj}_{SU(N_c)}\rangle$ or from unrecognized "outside".

The authors of \cite{APS} understood that, at $m\ll\lm$ and small
$\mx$, the quarks $Q^i$ from $SU(\nd)$ cannot acquire heavy masses
$\sim\lm$ in vacua with unbroken non-trivial $Z_{2N_c-N_F\geq 2}$
symmetry and remain light (i.e. with masses $\ll\lm$). But they argued \, 
then that  a number of vacua with light (massless at $\mx\ra 0$) pure
electric quarks at $m\gg\lm$, when evolved to $m\ll\lm$, have only
light quarks with nonzero pure magnetic charges. And they considered 
this  as a literal evidence that phases with light pure electrically and
pure magnetically charged quarks are not different and can transform
freely into each other by some unidentified mechanism. And they
supposed that in vacua of the baryonic branch {\it just this
mysterious mechanism transformed at $m\ll\lm$ all light pure electric
particles from $SU(\nd)$ into light pure magnetic ones}.

Remind that, as described above on the examples of vacua with
spontaneously broken or trivial $Z_{2N_c-N_F}$ symmetry, even in such
vacua at $m\ll\lm$ the electric quarks do not literally transform by
themselves into light dyonic ones. The mechanism is different, see
above. Therefore, we will consider in what follows that such
mysterious mechanism which transforms literally at $m\ll\lm$ light
electric quarks {\it in vacua with unbroken non-trivial
$Z_{2N_c-N_F\geq 2}$ symmetry} into light magnetic (or dyonic) ones
does not exist really.~
\footnote{\,
It is clear from \cite{SY1,SY2} that these authors also disagree with
pure magnetic $SU(\nd)$ quarks from \cite{APS}.
}
\vspace*{2mm}

Our purpose in this paper is similar to \cite{SY1,SY2}, but we limit
ourselves at $m\ll\lm$ only to vacua of baryonic branch with the
non-trivial unbroken discrete $Z_{\bb\geq 2}$ symmetry. As will be
seen from the text below, this $Z_{\bb}$ symmetry is strong enough 
and \, helps  greatly.

We introduce also the two following assumptions of general
character.\,-

{\bf Assumption A}: {\it in unbroken ${\cal N}=2$ (i.e. at $\mx\ra 0$) \, 
and at  least in considered vacua with the non-trivial unbroken
$Z_{\bb\geq 2}$ symmetry, the original pure electrically charged
particles receive additional contributions to their masses only from
higgsed} $\langle X\rangle\neq 0$ (remind that $\langle X\rangle$
includes in general all nonperturbative instanton contributions). In
other words, at $\mx\ra 0$ {\it they are BPS particles} in vacua
considered. For instance, if light quarks $Q^i$ with $m\ll\lm$ do not
receive contributions to their masses from some $\langle
X^{A}\rangle\neq 0$, they remain as they were, i.e. the light electric \, 
quarks  with masses $m$. The opposite is not true in this very special
${\cal N}=2$ theory in the strong coupling (and nonperturbative)
regime $a(\mu\sim\lm)=N_c g^2(\mu\sim\lm)/8\pi^2=O(1)$. The original
pure electric quarks which have received masses $\sim\lm$ from some
$\langle X^{A}\rangle\sim\lm$, can combine e.g. with some heavy
composite pure magnetic monopoles, also with masses $\sim\lm$, and
form massless at $\mx\ra 0$ composite BPS dyons. And all that.\\

{\it This assumption A about   BPS properties of original electric 
particles is checked in the text  below  by two 
independent calculations of quark and gluino condensates}.\\

{\bf Assumption B}: {\it   there are no extra  massless particles  (i.e.  in addition 
to standard   Nambu-Goldstone particles due to spontaneous breaking of global 
$U(N_F)\ra U)\no)\times U(\nt)$  at $m_i=m$ and small $\mx \neq 0$)  in 
considered  vacua  of the $SU(N_c)$ theory} \, \eqref{(40.1)}.  This more particular  
assumption  will  help us to clarify the quantum numbers of $(2N_c-N_F)$  
massless  at $\mx\ra 0$  charged BPS dyons $D_j$, see section 41.1. \\ 

And, as a result, {\it  there are  no massless particles at all in 
considered vacua of $SU(N_c)$   theories at $\mx\neq 0,\,\,  
m_i\neq 0,\,\,(\delta m)_{ij}\neq 0$}.\\ 

Besides, to calculate mass spectra in some ${\cal N}=1$ SQCD theories
obtained from ${\cal N}=2$ after increasing masses $\sim\mx$ of adjoint
scalar multiplets $X^{\rm adj}$ and decoupling them as heavy ones at scales
$\mu<\mx^{\rm pole}$, we use the dynamical scenario introduced in
\cite{ch3}. {\it But it is really needed in this paper only for those
few cases with $3N_c/2<N_F<2N_c-1$ when so obtained ${\cal N}=1$ 
SQCD  enters at lower energies  $\mu < \la$ the strong coupling 
conformal regime with}  $a_*=O(1)$, see section 41.4 below. 

Remind   that this scenario \cite{ch3} assumes for these cases that at 
those even lower scales  where this conformal regime is broken 
when quarks are either  higgsed  or decouple as heavy,  
{\it no additional  parametrically light
composite solitons are formed at these scales in this ${\cal N}=1$
SQCD} without elementary  colored scalar fields $X^{\rm adj}$.
(`Additional' means here in addition to the standards 
Nambu-Golstone particles).
\footnote{\,
It is worth noting that the appearance of additional light solitons
will influence the 't Hooft triangles of this ${\cal N}=1$ theory.
\label{(f50)}
}

Let us emphasize that {\it this dynamical scenario satisfies all 
those  checks of Seiberg's duality hypothesis  for ${\cal N}=1$ 
SQCD which were used in \cite{S2}}, see e.g. 
\cite{ch1,ch21,ch19,ch13,ch16}. And it allows to calculate mass 
spectra of the standard ${\cal N}=1$ SQCD and SQCD-type 
.theories. And it was shown in these papers that the mass spectra 
of the direct and Seiberg's dual theories are parametrically different.

And this shows that \,  {\it all those  checks of  ${\cal N}=1$ duality 
from   \cite{S1,S2},   although necessary, may well be insufficient}.

Besides,  to the best  of our knowledge, 
{the results obtained  in the framework of 
this scenario look self-consistent,  satisfy a large number of 
independent checks, and do not contradict 
to any proven results}.

In other cases, if \, this  ${\cal N}=1$ SQCD stays in 
the IR free weak coupling logarithmic
regime, its dynamics is simple and clear and there is no need for
any  additional dynamical assumptions.\\

Below  are presented our results (see the table of
contents) for quantum numbers of light particles, the forms of 
genuine low  energy 
Lagrangians and mass spectra in various vacua of \eqref{(40.1)}
with the unbroken discrete $Z_{\bb\geq 2}$ symmetry, at different
hierarchies between lagrangian parameters $m,\,\mx$ and $\lm$.

For further convenience, we present here also the equations for the
quark condensates (with account of Konishi anomalies \cite{Konishi})
in the $SU(N_c)$ theory \eqref{(40.1)} in vacua with the spontaneously
broken global flavor symmetry $U(N_F)\ra U(\no)\times U(\nt)$,
following from the well known effective superpotential $\w^{\,\rm
eff}_{\rm tot}$ of ${\cal N}=1$ SQCD which contains {\it only} quark
bilinears $\Pi^i_j=({\ov Q}_j Q^i)_{N_c}$. Because, at all
$\mx\lessgtr\lm$, the quark and gluino condensates,
$\langle (\qq)_{1,2}\rangle_{N_c}$ and $\langle S\rangle_{N_c}$,
considered as functions of $\lm,\, m$ and $\mx$, depend trivially on
$\mx$, i.e. $\sim\mx$, we can take first sufficiently large  $\mx\gg\lm$ 
with fixed $\la$ and integrate out all heavy scalars in \eqref{(40.1)},
obtaining the superpotential of (modified only by addition of the
4-quark term) ${\cal N}=1$ $\,SU(N_c)$ SQCD. Proceeding then 
as in   \cite{ch3} (see also section 4 in \cite{ch19} or section 8 in
\cite{ch13} for the corresponding effective superpotential in 
\,${\cal  N}=1\,\,SU(N_c)$ SQCD), $\w^{\,\rm eff}_{\rm tot}$ looks as
($N_F=(\no+\nt)< 2 N_c,\,\,1\leq\no\leq [N_F/2],\,\, \nt\geq\no$)
\bq
\w^{\,\rm eff}_{\rm tot}(\Pi)=m\,{\rm Tr}\,({\ov Q}
Q)_{N_c}-\frac{1}{2\mx}\Biggl [ \,\sum_{i,j=1}^{N_F} ({\ov Q}_j
Q^i)_{N_c}({\ov Q}_{\,i} Q^j)_{N_c}-\frac{1}{N_c}\Bigl ({\rm
Tr}\,({\ov Q} Q)_{N_c}\Bigr )^2\Biggr ]-\nd S_{N_c} \,, \label{(40.3)}
\eq
\bbq
S_{N_c}=\Bigl(\frac{\det
(\QQ)_{N_c}}{\la^{3N_c-N_F}=\lm^{2N_c-N_F}\mx^{N_c}}\Bigr
)^{\frac{1}{N_F-N_c}}\,,\quad \langle S\rangle_{N_c}
=\Bigl(\frac{\langle\det\QQ\rangle=\langle\Qo\rangle^{\no}\langle\Qt
\rangle^{\nt}}{\lm^{2N_c-N_F}\mx^{N_c}}\Bigr
)^{\frac{1}{N_F-N_c}}_{N_c}\,.
\eeq

Indeed, in the case when the four-quark terms in \eqref{(40.3)} are
small and can be neglected (i.e. at $\lm \ll (m={\rm const})\ll
\mx,\,\,\mx\gg (\la={\rm
const}),\,\,\lm=\la(\la/\mx)^{N_c/(2N_c-N_F)}\ra 0$\,), this
effective \, superpotential reproduces the well known standard effective
superpotential of the ${\cal N}=1$ SQCD with the scale factor $\la$.
But, because as it is, it is holomorphic in $\mx$, it is valid also
e.g. both at small $\mx,\,\lm={\rm const}$, and in all other vacua of
${\cal N}=2$ SQCD. It is worth only to recall that, as the effective
superpotential without the 4-quark term in the ${\cal N}=1$ SQCD,
\eqref{(40.3)} is {\it not} a genuine low energy superpotential, it can \, 
be used  {\it only} for finding the values of mean vacuum values
$\langle{\ov Q}_j Q^i\rangle_{N_c}$ and $\langle S\rangle_{N_c}$. The
genuine low energy superpotentials in each vacuum are given below in
the text.

From  \eqref{(40.3)} (and also directly from \eqref{(40.1)}) and from 
Konishi anomalies \cite{Konishi}
\bq
\langle \Qo+\Qt-\frac{1}{N_c}{\rm Tr}\,(\QQ)\rangle_{N_c}=
m \mx\,,\quad \langle S \rangle_{N_c}=\frac{\langle
\Qo\rangle_{N_c}\langle\Qt\rangle_{N_c}}{\mx}\,,\label{(40.4)}
\eq
\bbq
\langle\,\sum_{a=1}^{N_c}{\ov Q}^{\,a}_j
Q^i_a\,\rangle=\delta^i_j\,\langle\Qo\rangle_{N_c}\,,\,\,i,j=
1\,...\,\no,\,\, \langle\,\sum_{a=1}^{N_c}{\ov Q}^{\,a}_j
Q^i_a\,\rangle=\delta^i_j\,
\langle\Qt\rangle_{N_c}\,,\,\,i,j=\no+1\,...\,N_F\,,
\eeq
\bbq
\langle {\rm Tr}\,({\ov Q} Q
\rangle_{N_c}=\no\langle\Qo\rangle_{N_c}+\nt\langle\Qt
\rangle_{N_c}\,,\eeq
\bbq
m_Q\langle\sum_{a=1}^{N_c}{\ov Q}^a_i
Q^i_a\rangle-\sum_{A=1}^{N_c^2-1}\sum_{a,b=1}^{N_c}\langle
{\ov Q}^b_i \sqrt{2} X^A (T^A)^a_b Q^i_a\rangle=\langle  S\rangle_
{N_c}\,,\quad  \rm{no\,\, summation\,\, over\,\, flavor\,\, here}
\eeq
\bbq
\mx\langle {\rm Tr}\, (\sqrt{2}\,X^{\rm
adj}_{SU(N_c)})^2\rangle=(2N_c-N_F)\langle S \rangle_{N_c}+m\,\langle
{\rm Tr}\,({\ov Q} Q) \rangle_{N_c}=2\langle\w^{\,\rm eff}_{\rm
tot}\rangle\,,
\eeq
where $\langle S\rangle_{N_c}=
\langle\lambda\lambda/32\pi^2\rangle_{N_c}$ is the
gluino condensate summed over all $N_c^2-1$ colors.\\

The organization of this paper is as follows (see also the table of
contents). In $SU(N_c)$ gauge theories with $m\ll\lm$, the br2 vacua
with the broken global flavor symmetry, $U(N_F)\ra U(\no)\times
U(\nt),\, \nt>N_c$, and the unbroken non-trivial discrete symmetry
$Z_{\bb\geq 2}$ are considered in section 41.1\,. We discuss in detail
in this section  the quantum numbers of massless at $\mx\ra 0$ BPS
particles in these vacua, the low energy Lagrangians and mass spectra
at smallest $0<\mx\ll\Bigl (\Lambda^{SU(\nd-{\rm n}_1)}_{{\cal
N}=2\,\, SYM}\Bigr )^2/\lm$. The mass spectra in these br2 vacua at
larger values of $\mx$ are described in sections 41.3 and 41.4. These
results serve then as a basis for a description of similar regimes in
next sections. The br2 vacua of the $U(N_c)$ theory at smallest
$0<\mx\ll\Bigl (\Lambda^{SU(\nd-\no)}_
{{\cal N}=2\,\, SYM}\Bigr )^2/\lm$ are considered in section 41.2.
Besides, the S vacua with the unbroken flavor symmetry $U(N_F)$ (i.e.
$\no=0$) in the $SU(N_c)$ theory at $m\ll\lm$ are considered in
section 42.

The light particle quantum numbers, the low energy Lagrangians and
mass spectra in a large number of vacua of $SU(N_c)$ or $U(N_c)$
theories with $m\gg\lm$ are considered in sections 43 and 44.

Mass spectra in specific additional vs (very special) vacua with
$\langle S\rangle_{N_c}=0,\, N_c+1\leq N_F< 2N_c-1$ present in $U(N_c)$
theory (but absent in $SU(N_c)$,  are
considered in section 45 at $m\lessgtr\lm$ and various values of
$0<\mx\ll\lm$.

The mass spectra are calculated in section 46 in br2 vacua of $SU(N_c) \, 
$theory  with $U(N_F)\ra U(\no)\times U(\nt),\, \nt<N_c,\,\,m\gg\lm$ at
various values of $\mx$.

And finally, the mass spectra are described in section 47 for vacua
with $0<N_F< N_c-1$ and $m\gg\lm\,$.\\

Calculations of power corrections to the leading terms of the low\,
energy quark and dyon condensates are presented in two important\,
Appendices. The results agree with also presented in these Appendices\,
{\it independent} calculations of these condensates using roots of\,
the \,\, Seiberg-Witten spectral curve. This agreement confirms in a\,
non-trivial way a self-consistency of the whole approach.\\

\addcontentsline{toc}{section}
{\bf \large Part III.1.\,\, Small quark masses,
$\Large\mathbf{m\ll\lm}$ }
\vspace*{1cm}
\begin{center}{\hspace*{1cm}\bf\large Part III.1.\,\, Small quark masses,
$\Large\mathbf{m\ll\lm}$ } 
\end{center}

\section{ Broken flavor symmetry,\,\, br2 vacua}
\numberwithin{equation}{section}

\hspace{4mm} From \eqref{(40.3)},\eqref{(40.4)} for the $SU(N_c)$
theory, the quark condensates (summed over all $N_c$ colors) at
$N_c+1<N_F<2N_c-1$ in the considered br2 vacua (i.e. breaking 2, with
the dominant condensate $\langle({\ov Q}Q)_2\rangle_{N_c}$) with
$\nt>N_c\,,\,1\leq\no<\nd=(N_F-N_c)$ in theory \eqref{(40.1)} (these
are vacua of the baryonic branch in the language \cite{APS} or zero
vacua in \cite{SY1,SY2}\,) look as, see 
section 4 in \cite{ch19}, (the leading terms only,
\footnote{\,
Here and everywhere below\,: $A\approx B$ has to be understood as an
equality neglecting smaller power corrections, and $A\ll B$ has to be
understood as $|A|\ll |B|$. \label{(f4)}
}
\bq
\langle({\ov Q}Q)_2\rangle_{N_c}=\mx\,
{\hat m}+\frac{N_c-\no}{\nt-N_c}\langle({\ov Q}Q)_1\rangle_{N_c}\approx
\ha\, \mx,\,\, \langle({\ov Q}Q)_1\rangle_{N_c}\approx\mx \ha\Bigl
(\frac{\ha}{\lm}\Bigr)^{\frac{\bb}{{\rm n}_2-N_c}}, \,\,\label{(41.1)}
\eq
\bbq
\frac{\langle ({\ov Q}Q)_1\rangle_{N_c}}{\langle({\ov
Q}Q)_2\rangle_{N_c}}\approx\Bigl (\frac{\ha}{\lm}
\Bigr)^{\frac{2N_c-N_F}{{\rm n}_2-N_c}}\ll 1\,,\,\, N_c+1 < N_F <
2N_c-1\,,\,\, \ha=\frac{N_c}{N_c-n_2}\,m\,,\,\, \nd\equiv N_F-N_c,
\eeq
\bbq
\langle
S\rangle_{N_c}=\frac{\langle\Qo\rangle_{N_c}\langle\Qt\rangle_{N_c}}
{\mx}\approx\mx\, \ha^2\Bigl (\frac{\ha}{\lm}\Bigr)^{\frac{\bb}{{\rm n}_2
-N_c}}\ll\mx m^2,\quad \nt > N_c\,,
\eeq
and this shows that {\it the multiplicity of these vacua} is $N_{\rm
br2}=(\nt-N_c)\, C^{\,\nt}_{N_F}=(\nd-n_1)\,C^{\,\no}_{N_F}$, the
factor $C^{\,\no}_{N_F}=N_F!/(\no!\,\nt!)$ originates from the
spontaneous breaking $U(N_F)\ra U(\no)\times U(\nt)$ of the global
flavor symmetry of \eqref{(40.1)}.

Because the multiplicity $N_{\rm br2}$ of these br2 vacua in
\eqref{(41.1)} does not contain the factor $\bb$, this shows that the
non-trivial at $\bb\geq 2$ discrete symmetry $Z_{\bb}$ {\it is not
broken spontaneously in these vacua}.

\numberwithin{equation}{subsection}
\subsection{$SU(N_c)$, smallest $\mx$}

\hspace{4mm} We describe first the overall qualitative picture of
various stages (in order of decreasing energy scale) of the gauge and
flavor symmetry breaking in these br2 vacua. As will be shown below,
it leads in a practically unique way to a right multiplicity of these
vacua with unbroken discrete $Z_{\bb\geq 2}$ symmetry, and it is
non-trivial to achieve this. The value of $\mx$ is taken in this
section to be very small, $0<\mx<\langle\Lambda^{SU(\nd-\no)}_{{\cal
N}=2\,\, SYM}\rangle^2/\lm$, see \eqref{(41.1.4)} below.\\

{\bf 1)}\, As was explained in Introduction, at small $m$ and $\mx$,
the adjoint field $X^{\rm adj}_
{SU(N_c)}$ higgses necessarily the UV free $SU(N_c)$ group at the
largest scale $\mu\sim\lm$ to avoid $g^2(\mu<\lm)<0$. In the case
considered this first stage looks as: $\,SU(N_c)\ra SU(N_F-N_c)\times
U^{(1)}(1)\times U(1)^{\bb-1}$ \cite{APS}, while all other components
of $\langle X^A \rangle\lesssim m$ are much smaller, see below
\eqref{(41.1.3)},\eqref{(41.1.6)},
\bq
\langle\sqrt{2}\, X^{\rm adj}_{SU(\bb)}\,\rangle =C_{\bb}\lm\,{\rm
diag}(\,\underbrace{\,0}_{\nd};
\underbrace{\,\omega^0,\,\,\omega^1,\,...,\omega^{\bb-1}}_{\bb}\,)\,,
\,\, \omega=\exp\{\frac{2\pi i}{\bb}\},\,\, \,\,\, \label{(41.1.1)}
\eq
\bbq
\sqrt{2}\, X^{(1)}_{U(1)}=a_{1}\,{\rm
diag}(\,\underbrace{\,1}_{\nd}\,;\, \underbrace{\,c_1}_{\bb}\,),\quad
c_1=-\,\frac{\nd}{\bb}\,,\quad \langle a_{1}\rangle=\frac{1}{c_1}\,
m\,,\quad C_{\bb}=O(1)\,.
\eeq
This pattern of symmetry breaking is required by the unbroken
$Z_{\bb\geq 2}$ discrete symmetry. Remind that charges under
$Z_{\bb}=\exp\{\pi i/(\bb)\}$ transformation are:
$q_{\lambda}=q_{\theta}=1,\,\,q_{X}=q_{\rm m}=2,\,\,q_Q=q_{\ov
Q}=q_{\lm}=0,\,\,\mx=-2$, so that $\langle a_{1}\rangle$ respects
$Z_{\bb}$ symmetry, while the form of $\langle\sqrt{2}\, X^{\rm
adj}_{SU(\bb)}\,\rangle$ ensures the right behavior under $Z_{\bb }$
transformation up to interchanging terms in \eqref{(41.1.1)} 
(the Weyl  symmetry).

As a result, all original $SU(2N_c-N_F)$ charged electric particles
acquire large masses $\sim\lm$ and decouple at lower energies
$\mu<\lm/(\rm several)$. But due to strong coupling,
$a(\mu\sim\lm)=N_c g^2(\mu\sim\lm)/2\pi\sim 1$ and the enhanced 
${\cal  N}=2$ \,SUSY, the light composite BPS dyons $D_j, {\ov D}_j,\,\,
j=1...2N_c-N_F$ (massless at $\mx\ra 0$) are formed in this sector at
the scale $\mu\sim\lm$, their number is required by the unbroken
$Z_{\bb\geq 2}$ symmetry which operates interchanging them among one
another. This is seen also from the spectral curve \eqref{(40.2)}: it
has $\bb$ unequal double roots $e_j\approx \omega^{\,j-1}\lm,\,\,
j=1...\bb,$ corresponding to these $\bb$ massless BPS solitons
\cite{APS,CKM}. The charges of these dyons $D_j$ are discussed below
in detail. They are all flavor singlets, their $U^{(1)}(1)$ and
$SU(N_c)$ baryon charges which are $SU(\bb)$ singlets are pure
electric, while they are coupled with the whole set of $U^{\bb-1}(1)$
independent light Abelian (massless at $\mx\ra 0$) multiplets in
\eqref{(41.1.2)}.

{\it All original electric particles with $SU(\nd)$ colors remain
light}, i.e. with masses $\ll\lm$. This is a consequence of the
non-trivial unbroken $Z_{\bb\geq 2}$ symmetry and their BPS
properties. As for quarks with $SU(\nd)$ colors, the appearance in
the \, superpotential of the mass terms like $\sim\lm {\rm Tr\,}({\ov Q} C_Q
Q)_{\nd}$ with the constants $C^i_Q$ such that this term will be
$Z_{\bb\geq 2}$ invariant is impossible, independently of the source
of such terms. Therefore, the unbroken $Z_{\bb\geq 2}$ is sufficient
by itself to forbid $SU(\nd)$ quark masses $\sim\lm$. Similarly, it
also forbids (as well as the unbroken $U(1)$ R-symmetry) the
"outside"contributions in the superpotential of mass terms of adjoint 
scalars  \,like $\lm (X^{A}_{SU(N_c)})^2$ or $m (X^{A}_{SU(N_c)})^2$.

But we will need below somewhat stronger assumption ${\bf "A"}$
formulated in Introduction that all $SU(N_c)$ quarks are BPS
particles, i.e. they receive additional contributions to their masses
{\it only directly} from the couplings with $\langle
X^{(adj)}_{SU(N_c)}\rangle$, see below. This forbids also the
appearance of {\it any "outside"} (i.e. originating not from the
direct couplings with $\langle X^{(adj)}_{SU(N_c)}\rangle$ )
contributions to their masses, e.g. the additional contributions in
the superpotential $c_m m {\rm Tr\,}({\ov Q} Q)_{\nd},\, c_m=O(1)$,
although these are not forbidden by the unbroken $Z_{\bb\geq 2}$.
Besides, this BPS assumption ${\bf "A"}$ concerns all original
electric particles of $SU(N_c)$. On the other hand, it is natural to
expect that all original particles of the $SU(N_c)$ theory have the
same properties, in the sense that in considered isolated vacua at
$\mx\ra 0$ either they all are BPS or they all are not BPS. And
because the "outside"\, contributions $\sim\lm$ to masses in the
superpotential are forbiden for $SU(\nd)$ quarks and scalars, and
$\sim m$ are forbidden for scalars, then all "outside"\, contributions \, to 
masses  are forbidden for all original electric particles, i.e. they \, all 
are indeed the  BPS particles. (The absence of such additional
contributions is confirmed by {\it two independent} calculations of
$SU(\nd)$ quark condensates, see \eqref{(41.1.14)} and the text just
below it).\\

The $SU(\nd)$ gauge symmetry can still be considered as unbroken at
scales $m\ll\mu<\lm$ (it is broken only at the scale $\sim m$ due to
higgsed $X^{\rm adj}_{SU(\nd)}$, see below). Therefore, after
integrating out all heavy particles with masses $\sim\lm$, the general \, 
form of  the superpotential at the scale $\mu_{\rm cut}=\lm/{(\rm
several)}$ can be written as (the dots in \eqref{(41.1.2)} denote
smaller power corrections)
\bq
\wh{\w}_{\rm
matter}=\w_{SU(\nd)}+\w_D+\w_{a_1}+\dots\,,\label{(41.1.2)}
\eq
\bbq
\w_{SU(\nd)}={\rm Tr}\,\Bigl [{\ov Q}\Bigl (m-a_1-\sqrt{2}X^{\rm
adj}_{SU(\nd)}\Bigr ) Q\Bigr ]_{\nd}+\wmu {\rm Tr}\,(X^{\rm
adj}_{SU(\nd)})^2\,,\quad \wmu=\mx(1+\delta_2)\,,
\eeq
\bbq
\w_{D}=(m-c_1 a_1)\sum_{j=1}^{\bb}{\ov D}_j D_j-\sum_{j=1}^{\bb}
a_{D,j}{\ov D}_j D_j-\mx\lm\sum_{j=1}^{\bb}\omega^{j-1}a_{D,j}+
\eeq
\bbq
+\mx L \sum_{j=1}^{\bb} a_{D,j}+O\Bigl (\mx a_D^2 \Bigr )\,,
\eeq
\bbq
\w_{a_1}=\frac{\mx}{2}(1+\delta_1)\frac{\nd N_c}{\bb}\,a^2_1+\mx N_c
\delta_3\,a_1(m-c_1 a_1)+\mx N_c\delta_4 (m-c_1 a_1)^2\,,
\eeq
where $a_{D,j}$ are the light neutral scalars, $L$ is the Lagrange
multiplier, $\langle L\rangle=O(m)$, while all $\delta_i=O(1)$.

These terms with $\delta_i$ in the superpotential \eqref{(41.1.2)}
originate finally from {\it the combined effect} of\,: a) the quantum
loops with heavy particles with masses $\sim\lm$, integrated over the
energy range $[\mu_{\rm cut}=\lm/(\rm several)]<\mu<(\rm several)
\lm$,\,  in the  strong coupling regime $a(\mu\sim\lm)\sim 1$ 
and in the background of lighter fields;\, b) the
unsuppressed\, non-perturbative instanton contributions from 
the broken color
subgroup $SU(\bb)\ra U^{\bb-1}(1)$ which are operating at the scale
$\mu\sim\lm$. Without instantons, different pure perturbative loop
contributions to the superpotential cancel in the sum, but the
additional instanton contributions spoil this cancelation. See also
the footnote \ref{(f53)}. Compare with the absence of such terms in
sections 43.1,\,43.2, see \eqref{(43.1.4)},\eqref{(43.2.3)}, when
$\langle  X^{adj}_{SU(N_c)}\rangle\sim m\gg\lm$ is higgsed in the weak 
coupling  regime, resulting in $SU(N_c)\ra SU(\no)\times U(1)\times 
SU(N_c-\no)$ \, with
unbroken at this scale the non-Abelian ${\cal N}=2\,\,
SU(N_c-\no)$ SYM part. The instanton contributions are then power
suppressed. They originate only from this ${\cal N}=2$ SYM part and
operate in the lower energy strong coupling region at the scale
$\mu\sim\langle\Lambda^{SU(N_{c}-{\rm n}_1)}_{{\cal
N}=2\,\,SYM}\rangle\ll m$, after the color breaking $SU(N_c-\no)\ra
U^{N_c-\no-1}(1)$ by higgsed $\langle
X^{adj}_{SU(N_c-\no)}\rangle\sim\langle\Lambda^{SU(N_{c}-{\rm
n}_1)}_{{\cal N}=2\,\,SYM}\rangle$, see \eqref{(41.1.2)}.

The way the parameter "m" enters \eqref{(41.1.2)} can be understood as
follows. This parameter "m" in the original superpotential $\w_{\rm
matter}$ in \eqref{(40.1)} can be considered as a soft scalar
$U^{(0)}(1)$ background field coupled universally to the $SU(N_c)$
singlet scalar electric baryon current, $\delta\w_B=m J_B$, and this
current looks at the scale $\mu\gg\lm$ as $J_B={\rm Tr}\,({\ov Q}
Q)_{N_c}$. But at the lower energy scale $\mu\ll\lm$ only lighter
particles with masses $\ll\lm$ and with a nonzero electric baryon
charge contribute to $J_B$, these are ${\rm Tr}\,({\ov Q} Q)_{\nd}$
and $\sum_{j=1}^{\bb}{\ov D}_j D_j$ (see bellow). Moreover, the field
$a_{1}$ plays a similar role, but its couplings are different for two
different color sectors of $SU(N_c)$, see
\eqref{(41.1.1)},\eqref{(41.1.2)}.

The terms with $\delta_i,\,\, i=1\,...\,4\,,\,\,\delta_i=O(1)$, arose
in \eqref{(41.1.2)} from integrating out heaviest original fields with
masses $\sim \lm$ in the soft background of lighter $X^{\rm
adj}_{SU(\nd)},\,\, a_1$ and $(m-c_1 a_1)$ fields. These heavy fields
are\,: charged $SU(\bb)$ adjoints with masses
$ (\Lambda_j-\Lambda_i),\,\, \Lambda_j\sim\omega^{\,j-1}\lm$, the
heavy quarks with $SU(\bb)$ colors and masses $\Lambda_j-(m-c_1 a_1)$, 
\, and heavy  adjoint hybrids $SU(N_c)/[SU(\nd)\times SU(\bb)\times
U^{(1)}(1)]$ with masses $\Lambda_j-[\sqrt{2}X^{\rm
adj}_{SU(\nd)}+(1-c_1)a_1]$. Note that terms like $\mx\lm^2,\,\,\mx
a_1\lm$ and $\mx (m-c_1 a_1)\lm$ cannot appear in \eqref{(41.1.2)},
they are forbidden by the unbroken $Z_{\bb\geq 2}$ symmetry. \\

The above described picture is more or less typical, the only really
non-trivial point concerns the quantum numbers of $2N_c-N_F$ dyons
$D_j$. Therefore, we describe now some necessary requirements to them.\\

{\bf a)}\, First, the $SU(\nd)$ part of \eqref{(41.1.2)}, {\it due to its own
internal dynamics}, ensures finally the right pattern of the
spontaneous flavor symmetry breaking $U(N_F)\ra U(\no)\times U(\nt)$
and the right multiplicity of vacua, $N_{\rm br2}=(\nd-\no)
C^{\,\no}_{N_F}$, see below. Therefore, the quantum numbers of these
$\bb$ dyons $D_j$ should not spoil the right dynamics in the $SU(\nd)$ \, 
sector.

{\bf b)}\, These quantum numbers should be such that the lower energy
Lagrangian respects the discrete $Z_{\bb\geq 2}$ symmetry. This fixes,\, 
in  particular, the number $\bb$ of these dyons because $Z_{\bb\geq
2}$ \, transformations interchange them among one another. Notice also 
that  the quantum numbers of these dyons $D_i$ formed at the scale
$\mu\sim\lm$ cannot know about properties of further color or flavor
breaking in the $SU(\nd)$ color part at much lower scales $\mu\sim
m\ll\lm$ or $\mu\sim (\mx m)^{1/2}\ll m$, i.e. they do not know the
number $\no$.

{\bf c)}\, According to M.Shifman and A.Yung, see e.g. \cite{SY6}, 
the massive \, diagonal  quarks ${\ov Q}^i_i,\, Q^i_i,\, i=1...2N_c-N_F$ with
$SU(2N_c-N_F)$ colors and masses $\sim\lm$ combine at the scale
$\mu\sim\lm$ with $SU(2N_c-N_F)$ appropriate massive flavorless
magnetic (anti)monopoles and form $2N_c-N_F$ massless dyons ${\ov
D}_i, D_i$ in \eqref{(41.1.2)}. But for quarks with equal own masses
$"m"$ the non-Abelian flavor symmetry $SU(N_F)$ is {\it not broken} by \, 
higgsed  $\langle X_{SU(\bb)}\rangle\sim\lm$ at the scale $\mu\sim\lm$, \,
and {\it  flavorless monopoles cannot distinguish between quarks with
different $SU(N_F)$ flavors. Therefore, such dyons will really have
the same flavor as quarks, i.e. they will form the (anti)fundamental
representation of global $SU(N_F)$}.

But these $\bb$ dyons ${\ov D}_i, D_i$ in \eqref{(41.1.2)} have to be
$SU(N_F)$ flavor singlets. Otherwise, because they all are higgsed at
the scale $\mu\sim (\mx\lm)^{1/2}\neq 0$, this will result in a wrong
pattern of the spontaneous non-Abelian flavor symmetry breaking
$U(N_F)\ra U(2N_c-N_F)\times U(2N_F-2N_c)$ instead of the right
one \, $U(N_F)\ra U(\no)\times U(\nt)$. Besides, there will be \,
$N_F (\bb)$ \,  these  dyons $D$, this is clearly too much.

In particular, the arguments in \cite{SY6} are based on considering 
the  unequal mass quarks with $\delta m\neq 0$, the
``outside''\, region $\delta m\gg\lm$ and ``inside''\, region 
$\delta m\ll\lm$ and the path between them going through 
Argyres-Douglas point.  

We would like to imphasize that these ``outside''\,  and ``inside''\, 
regions are {\it completely separated} by so-called ``wall crossing curves''.
And there is no such path in the complex plane $\delta m/\lm$
which connects the ``outside''\,  and ``inside''\,  regions and does 
not cross this ``wall crossing curve''.  And the path going through
Argyres-Douglas point is not the exclusion. 

The quantum numbers  of light particles jump generically when
the generic path crosses the ``wall crossing curve'' and have to be 
carefully considered separately in the ``outside''\,  and ``inside''\, 
regions.  And this questions the statements of M. Shifman and
A.Yung about quantum numbers of light particles.

{\bf d)}\, Their magnetic quantum numbers should be such that they are
mutually local with respect to all original pure electric particles in \, 
the  $SU(\nd)$ sector. Otherwise, in particular, because all these
dyons and all original pure electric quarks from $SU(\nd)$ have
nonzero $U^{(1)}(1)$ charges, and because all dyons are higgsed,
$\langle{\ov D}_j\rangle=\langle D_j\rangle\sim (\mx\lm)^{1/2}$, if
the magnetic $U^{(1)}(1)$ charge of these dyons were nonzero, this
would lead to confinement of all quarks from the $SU(\nd)$ sector with \, 
the  string \, tension $\sigma^{1/2}\sim (\mx\lm)^{1/2}$.  E.g., see below, 
\, all $N_F$  flavors of  confined light electric quarks with $SU(\no)$
colors would decouple then in any case at the scale
$\mu<(\mx\lm)^{1/2}$, the flavor symmetry $U(N_F)$ would remain
unbroken in the whole $SU(\nd)$ sector, while the remained ${\cal
N}=2\,\,SU(\no)$ SYM would give the additional wrong factor $\no$ in 
the  multiplicity. \, All this is clearly unacceptable. This excludes the
variant with the nonzero $U^{(1)}(1)$ magnetic charge (and also with
non-zero $SU(\nd)$ magnetic charges) of these dyons. Similarly, when
considering the $U(N_c)=SU(N_c)\times U^{(0)}(1)$ theory in section
41.2 below, this also requires the $SU(N_c)$ baryon charge of these
dyons to be pure electric. For the same reasons, the magnetic charges
of these dyons cannot be $SU(N_c)/[SU(\nd)\times SU(\bb)$ hybrids. On
the whole, to be mutually local with the whole $SU(\nd)$ part, the
magnetic parts of all charges of these dyons have to be $SU(\bb)$
root-like (i.e. $SU(\bb)$ adjoints most naturally).

{\bf e)}\, If the  $U^{(1)}(1)$ charge of these $\bb$ dyons $D_j$ were zero 
this would  correspond then to the first term of $\w_D$ in
\eqref{(41.1.2)},\eqref{(41.1.5)} of the form $m\sum_{j}^{\bb}{\ov D}_j
D_j$. They would not be coupled then with the $U^{(1)}(1)$ multiplet
\eqref{(41.1.1)} and this would lead to the internal inconsistency, see below.

The case with zero $SU(N_c)$ baryon charge of dyons, i.e. with the
first term of $\w_{D}$ of the form $[\,-c_1 a_1\sum_{j=1}^{\bb}{\ov
D}_j D_j\,]$ in \eqref{(41.1.2)}, \eqref{(41.1.5)}, will lead to wrong
values of quark condensates, see page 166 below

Finally, about the case when this first term of $\w_{D}$ in
\eqref{(41.1.2)} equals zero. The whole dyon sector will be then
completely decoupled from the $SU(\nd)\times U^{(1)}(1)$ sector, so
that $\langle{\ov D}_j D_j\rangle$ will not include corrections
containing $\langle\Lambda^{SU(\nd-\no)}_{{\cal N}=2\,\, SYM}\,
\rangle$. \, And this  will be wrong, because the values of $\langle{\ov D}_j
D_j\rangle$ obtained {\it independently} from the roots of the curve
\eqref{(40.2)} contain such corrections, see \eqref{(B.15)}. 

Moreover, counting degrees
of freedom in the $SU(\nd)\times U^{(1)}(1)$ sector (see below), it \,
is \, seen that  one ${\cal N}=1$  photon multiplet will remain exactly
massless in this $SU(\nd)$ sector at  small $\mx\neq 0$ 
(while its ${\cal N}=2$ scalar partner   will have very small mass  $\sim\mx$ 
only due to breaking ${\cal N}=2\ra {\cal N}=1$ at small  $\mx\neq 0$).

Besides,  when even one out of $U^{2N_c-N_F-1}(1)$ Abelian photon \, 
multiplets \,  is not  coupled with the whole set of $2N_c-N_F$ dyons then,
due to unbroken $Z_{2N_c-N_F}$ discrete symmetry, all $2N_c-N_F-1$
Abelian multiplets will not be  coupled. In any case (not even speaking 
about other problems), at \, least all  $U^{2N_c-N_F-1}(1)$ photons 
multiplets will  remain then  massless even  at $\mx\neq  0$. 

According to our assumption ``B'' (see  Introduction), all variants with 
extra  (i.e. in addition to standard massless  Nambu-Goldston particles 
at  $m_i= m$  and  $\mx\neq 0$,\, in $SU(N_c)$ theories  
are excluded.

{\bf f)}\, The regime in the $SU(\bb)$ color sector at low energies is IR
free. As follows from the curve \eqref{(40.2)}, see \eqref{(B.15)}),
all $\bb$ these BPS dyons are higgsed simultaneously in the ground
state at small $\mx\neq 0$. And for this to be possible, {\it they all \, 
have to  be \,  massless at $\mx\ra 0$ and to be mutually local} (so\, 
that  $\langle a_{D,j}\rangle=0$, see \eqref{(41.1.5)},\eqref{(41.1.6)}
below).

{\bf g)}\, No one of $U^{\bb-1}(1)$ charges of these light BPS dyons $D_j$ 
can \,  be {\it  pure electric}. Indeed, since all $SU(\bb)$ adjoint $\langle
a_j\rangle\sim\lm$ in \eqref{(41.1.1)}, \, all BPS solitons with some pure \, 
electric  $U^{\bb-1}(1)$ charges will have large masses $\sim\lm$ in
this case.

Therefore, as a result, these mutually local $\bb$ BPS dyons $D_j,
{\ov D}_j$, massless at $\mx\ra 0$, should have nonzero: i) {\it pure
electric} both $SU(N_c)$ baryon and $U^{(1)}(1)$ charges; ii) on the
whole, all $\bb-1$ nonzero independent Abelian charges of the broken
$SU(\bb)$ subgroup, with their $SU(\bb)$ adjoint magnetic parts. Then
they will be mutually local with the whole $SU(\nd)$ sector, and
coupled with $"m"$ and with the electric $U^{(1)}(1)$ multiplet, and
with all $U^{\bb-1}(1)$ Abelian multiplets,
see\eqref{(41.1.1)},\eqref{(41.1.2)}. There will be no massless
particles at $\mx\neq 0,\, m_i=m\neq 0, \,\,\delta_{ij}\neq 0$ 
and the multiplicity  and the pattern of  flavor 
symmetry breaking will be right, see below.\\

{\bf 2)}\, The second stage is the breaking $SU(\nd)\ra SU(n_1)\times
U^{(2)}(1)\times SU(\nd-n_1)$ at the lower scale $\mu\sim m\ll\lm$
{\it in the weak coupling regime}, see  \eqref{(41.1.5)},\eqref{(41.1.6)} 
below, $1\leq  \no<\nd=N_F-N_c$, (this stage is qualitatively similar 
to those in  section 43.1)
\bq
\sqrt{2}\, X^{(2)}_{U(1)}=a_{2}\,{\rm
diag}(\,\underbrace{\,1}_{\no}\,;\,
\underbrace{\,c_2}_{\nd-\no}\,;\,\underbrace{0}_{\bb}\,),\quad
c_2=-\,\frac{\no}{\nd-\no}\,,\quad \langle
a_{2}\rangle=\frac{N_c}{\nd}\, m\,.\label{(41.1.3)}
\eq

As a result, all original electric quarks with $SU(\nd-\no)$ colors
and $N_F$ flavors have masses $m_2=\langle m- a_1-c_2
a_2\rangle=m\,N_c/(\nd-\no)$, and all original electric
$SU(\nd)$ adjoint hybrids $SU(\nd)/[SU(\no)\times SU(\nd-\no)\times
U(1)]$ also have masses $(1-c_2)\langle a_2\rangle= m_2$, and so all
these particles decouple as heavy at scales $\mu\lesssim m$. But the
original electric quarks with $SU(\no)$ colors and $N_F$ flavors have
masses $\langle m- a_1- a_2\rangle=0$, they remain massless at $\mx\ra0$ 
and survive at $\mu\lesssim m$. Therefore, there remain two lighter \, 
non-Abelian  sectors $SU(\no)\times SU(\nd-\no)$. As for the first one,\, 
it is the IR free  ${\cal N}=2\,\, SU(\no)$ SQCD with $N_F$ flavors of
original electric quarks massless at $\mx\ra 0\,,\,\, N_F>2\no,\,\,
1\leq\no<\nd\,$. As for the second one, it is the ${\cal N}=2\,\,
SU(\nd-\no)$ SYM with the scale factor $\Lambda^{SU(\nd-{\rm
n}_1)}_{{\cal N}=2\,\, SYM}$ of its gauge coupling, compare with
$\langle S\rangle_{N_c}$ in \eqref{(41.1)},
\bq
\Bigl (\Lambda^{SU(\nd-\no)}_{{\cal N}=2\,\, SYM}\Bigr
)^{2(\nd-\no)}=\frac{\Lambda_{SU(\nd)}^{2\nd-N_F}
(m-a_1-c_2 a_2)^{N_F}}{[\,(1-c_2)a_2\,]^{\,2\no}}\,,\label{(41.1.4)}
\eq
\bbq
\langle\Lambda^{SU(\nd-\no)}_{{\cal N}=2\,\, SYM}\rangle^2=m^2_2\Bigl
(\frac{m_2}{\Lambda_{SU(\nd)}}\Bigr )^{\frac{2N_c-N_F}{\nd-\no}}\ll
m^2\,,\quad\langle
S\rangle_{\nd-\no}=\mx(1+\delta_2)\langle\Lambda^{SU(\nd-\no)}_{{\cal
N}=2\, SYM}\rangle^2\,,
\eeq
\bbq
m_2=\langle m-a_1-c_2 a_2\rangle=(1-c_2)\langle
a_2\rangle=\frac{N_c}{\nt-N_c}\,m= - \ha\,,
\eeq
where $\Lambda_{SU(\nd)}$ is the scale factor of the $SU(\nd)$ gauge
coupling at the scale $\mu=\lm/(\rm several)$ in \eqref{(41.1.2)},
after integrating out heaviest particles with masses $\sim\lm$ (it
will be determined below in \eqref{(41.1.15)}).\\

{\bf 3)}\, At the third stage, to avoid unphysical
$g^2(\mu<\langle\Lambda^{SU(\nd-\no)}_{{\cal N}=2\,\, SYM}\, 
\rangle)<0$  of UV free $SU(\nd-\no)\,\, {\cal N}=2$ SYM, the field
$X^{adj}_{SU(\nd-\no)}$ is higgsed, $\langle
X^{adj}_{SU(\nd-\no)}\rangle\sim \langle\Lambda^{SU(\nd-{\rm
n}_1)}_{{\cal N}=2\,\, SYM}\rangle$, breaking $SU(\nd-\no)$ in a well
known way \cite{DS}, $SU(\nd-\no)\ra U^{\nd-\no-1}(1)$. All original
pure electrically charged adjoint gluons and scalars of ${\cal
N}=2\,\,SU(\nd-\no)$ SYM acquire masses
$\sim\langle\Lambda^{SU(\nd-{\rm n}_1)}_{{\cal N}=2\,\, SYM}\rangle$
and decouple at lower energies $\mu<\langle\Lambda^{SU(\nd-{\rm
n}_1)}_{{\cal N}=2\,\, SYM}\rangle$. Instead, $\nd-\no-1$ light
composite pure magnetic monopoles (massless at $\mx\ra 0$),
$M_i,\,{\ov M}_i,\,\, i=1,...,\nd-\no-1$, with their $SU(\nd-\no)$
adjoint magnetic charges are formed at this scale
$\sim\langle\Lambda^{SU(\nd-{\rm n}_1)}_{{\cal N}=2\,\, SYM}\rangle$.
The factor $\nd-\no$ in the overall multiplicity of considered br2
vacua originates from the multiplicity of vacua of this ${\cal
N}=2\,\,SU(\nd-\no)$ SYM. Besides, the appearance of two single roots
with $(e^{+}-e^{-})\sim \langle\Lambda^{SU(\nd-\no)}_{{\cal N}=2\,\,
SYM}\rangle$, of the curve \eqref{(40.2)} in these br2 vacua is
connected just with this ${\cal N}=2\,\,SU(\nd-\no)$ SYM \cite{DS}.
Other $\nd-\no-1$ double roots originating from this ${\cal N}=2$ SYM
sector are unequal double roots corresponding to $\nd-\no-1$ pure
magnetic monopoles $M_{\rm n}$, massless at $\mx\ra 0$.\\

{\bf 4)}\, All $\bb$ dyons ${\ov D}_j, D_j$ are higgsed at the scale
$\sim (\mx\lm)^{1/2}$ in the weak coupling regime: $\,\,\langle
D_j\rangle=\langle {\ov D}_j\rangle\sim (\mx\lm)^{1/2}\gg (\mx
m)^{1/2}$. As a result, $\bb$ long ${\cal N}=2\,\, U(1)$ multiplets of \, 
massive  photons are formed (including $U^{(1)}(1)$ with its scalar
$a_1$), all with masses $[\,g_D(\mu\sim(\mx\lm)^{1/2})\ll
1\,](\mx\lm)^{1/2}$ (for simplicity, we ignore below logarithmically
small factors $\sim g$ in similar cases, these are implied where
needed). No massless particles remain in this sector at $\mx\neq 0$.
And there remain no lighter particles  in this sector at lower energies.
The $(\bb)$-set of these higgsed dyons with the $SU(2N_c-N_F)$ adjoint \, 
magnetic  charges is mutually non-local with all original
$SU(2N_c-N_F)$ pure electrically charged particles with largest
masses$\sim\lm$. \, Therefore, all these heaviest electric particles are
weakly \,, confined, the string
tension is $\sigma^{1/2}_D\sim (\mx\lm)^{1/2}$.
This confinement is weak in the sense that the tension of the
confining string is much smaller than particle masses,
$\sigma^{1/2}_{D}\sim (\mx\lm)^{1/2}\ll\lm\,$. All these heaviest
confined particles form a large number of hadrons with masses
$\sim\lm$.\\

{\bf 5)}\, $\no$ out of $N_F$ electric quarks of IR free ${\cal
N}=2\,\, SU(\no)$ SQCD are higgsed at the scale $\mu\sim (\mx\, 
m)^{1/2}$ in the weak coupling region, with $\langle{\ov Q}_k^{\,a}
\rangle=\langle Q^k_a\rangle\sim \delta^k_a\,(m\, \mx )^{1/2},\,\,
a=1,...,\no,\,\, k=1,...,N_F$. As a result, the whole electric color group
$SU(\no)\times U^{(2)}(1)$ is broken and $\no^2$ s;ightly broken 
long ${\cal N}=2$  multiplets of
massive electric gluons are formed (including $U^{(2)}(1)$ with its
scalar $a_2$), all with masses $\sim (m\, \mx)^{1/2}\ll (\mx\lm)^{1/2}$. \, 
The  global \, flavor symmetry is broken spontaneously {\it only at this
stage}, $U(N_F)\ra U(\no)\times U(\nt)$, and $2\no\nt$ (complex)
massless Nambu-Goldstone multiplets remain in this sector at lower
energies (in essence, these are quarks $Q^k_a,\, {\ov
Q}^{\,a}_{k},\,a=1,...,\no,\,\, k=\no+1,...,N_F$ ). This is a reason
for the origin of the factor $C^{\,\no}_{N_F}$ in multiplicity of
these br2 vacua. (See also the text somewhat below \eqref{(41.1.6)}
about $(\delta m)_{ij}$).\\

{\bf 6)}\, $\nd-\no-1$ magnetic monopoles are higgsed at the lowest
scale, $\langle{\ov M}_{\rm n}\rangle=\langle M_{\rm n}\rangle\sim
(\mx\langle\Lambda^{SU(\nd-{\rm n}_1)}_{{\cal N}=2\,\,
SYM}\rangle)_{,}^{1/2}$ and $\nd-{\rm n}_1-1$ long ${\cal N}=2$
multiplets of massive $U^{\nd-\no-1}(1)$ photons are formed, all with
masses $\sim (\mx\langle\Lambda^{SU(\nd-{\rm n}_1)}_{{\cal N}=2\,\,
SYM}\rangle)^{1/2}\ll (\mx m)^{1/2}$. All original pure electrically
charged particles with non-singlet $SU(\nd-{\rm n}_1)$ charges and
masses either $\sim m$ or $\sim\langle\Lambda^{SU(\nd-{\rm
n}_1)}_{{\cal N}=2\,\, SYM}\rangle$ are weakly confined, the string
tension is $\sigma^{1/2}_{\rm SYM}\sim \,(\mx\langle\Lambda^
{SU(\nd-{\rm  n}_1)}_{{\cal N}=2\,\, SYM}\rangle)^{1/2}$. No
massless particles  remain in this sector at $\mx\neq 0,\, m\neq 0$.\\

As a result of all described above, the lowest energy superpotential
at the scale $\mu=\langle\Lambda^  {SU(\nd-\no)}_{{\cal N}=2\,\, SYM}\,
\rangle/ \\ (\rm several)$, and e.g. at  smallest $0<\mx\ll\langle\Lambda^
{SU(\nd-\no)}_{{\cal N}=2\,\, SYM}\rangle^2/\lm$, can be written in
these br2 vacua as
\bq
\w^{\,\rm low}_{\rm tot}=\w^{\,(SYM)}_{SU(\nd-\no)}+\w^{\,\rm
low}_{\rm matter}+\dots\,,\quad
\w^{\,\rm low}_{\rm
matter}=\w_{SU(\no)}+\w_{D}+\w_{a_1,\,a_2}\,,\label{(41.1.5)}
\eq
\bbq
\w^{\,(SYM)}_{SU(\nd-\no)}=(\nd-\no)\wmu\Bigl (\Lambda^{SU(\nd-{\rm
n}_1)}_{{\cal N}=2\,\, SYM}\Bigr )^2+\w^{\,(M)}_{SU(\nd-\no)}\,,\quad
\wmu=\mx(1+\delta_2)\,,
\eeq
\bbq
\w^{\,(M)}_{SU(\nd-\no)}= - \sum_{n=1}^{\nd-\no-1} a_{M, \rm n}\Biggl
[\, {\ov M}_{\rm n} M_{\rm n}+\wmu\Lambda^{SU(\nd-\no)}_{{\cal
N}=2\,\,SYM}\Biggl (1+O\Bigl
(\frac{\langle\Lambda^{SU(\nd-\no)}_{{\cal
N}=2\,\,SYM}\rangle}{m}\Bigr )\Biggr ) d_{\rm n}\,\Biggr ]\,,
\eeq
\bbq
\w_{SU(\no)}=(m-a_1-a_2)\,{\rm Tr}\,({\ov Q} Q)_{\no}-\,{\rm
Tr}\,({\ov Q}\sqrt{2} X^{\rm adj}_{SU(\no )} Q)_{\no}+\wmu{\rm
Tr}\,(X^{\rm adj}_{SU(\no)})^2\,,
\eeq
\bbq
\w_{D}=(m-c_1 a_1)\sum_{j=1}^{\bb}{\ov D}_j D_j-\sum_{j=1}^{\bb}
a_{D,j}\,{\ov D}_j D_j\,-
\,\mx\lm\sum_{j=1}^{\bb}\omega^{j-1}\,a_{D,j}+\mx L \sum_{j=1}^{\bb}
a_{D,j}\,,
\eeq
\bbq
\w_{a_1,\,a_2}=\frac{\mx}{2}(1+\delta_1)\frac{\nd N_c}{\bb}
a_1^2+\frac{\wmu}{2}\frac{\no \nd}{\nd-\no} a_2^2+\mx N_c \delta_3
a_1(m-c_1 a_1)+\mx N_c\delta_4 (m-c_1 a_1)^2+O\Bigl (\mx a_{D,j}^2
\Bigr ),
\eeq
where coefficients $d_i=O(1)$ in \eqref{(41.1.5)} are known from
\cite{DS}, and dots in \eqref{(41.1.5)} denote smaller power
suppressed corrections (these are always implied and dots are omitted
below in the text), . It is useful to notice that the unbroken
$Z_{2N_c-N_F\geq 2}$ symmetry restricts strongly their possible
values. The matter is that e.g. power suppressed corrections in
$\w_{a_1}$ like $\sim\mx a_1^2(a_1/\lm)^{N}$ originate from
$SU(N_c)\ra SU(\nd)\times U^{(1)}(1)\times SU(2N_c-N_F)$ at the scale
$\sim\lm$ and they do not know the number $\no$ originating only at
much lower scales $\mu\sim m\ll\lm$. Then the unbroken
$Z_{2N_c-N_F\geq 2}$ symmetry requires that $N=(2N_c-N_F) k,\,
k=1,2,3...$.

Remind that charges of fields and parameters entering \eqref{(41.1.5)}
under $Z_{\bb}=\exp\{i\pi/(2N_c-N_F)\}$ transformation are\,:
$q_{\lambda}=q_{\theta}=1,\,\, q_{X_{SU(\no)}^{\rm
adj}}=q_{a_1}=q_{a_2}=q_{a_{D,j}}=q_{a_{M,\rm n}}=q_{\rm
m}=q_{L}=2,\,\,q_{Q}=q_{\ov Q}=q_{D_j}=q_{{\ov D}_j}=q_{M_{\rm
n}}=q_{{\ov M}_{\rm n}}=q_{\lm}=0,\,\, q_{\mx}=-2$. The non-trivial
$Z_{\bb\geq 2}$ transformations change only numerations of dual
scalars $a_{D,j}$ and dyons $D_j, {\ov D}_j$ in \eqref{(41.1.5)}, so
that $\int d^2\theta\,\w_{\rm tot}^{\,\rm low}$ is
$Z_{\bb}$-invariant.

All (massless at $\mx\ra 0$) $\bt=2N_c-N_F$ dyons $D_j$, $\no<\nd$
quarks $Q^k$ and $\nd-\no-1$ monopoles $M_{\rm n}$ in \eqref{(41.1.5)}
are higgsed at $\mx\neq 0$. As a result, we obtain from
\eqref{(41.1.5)} (neglecting power corrections)
\bq
\langle a_1\rangle=\frac{1}{c_1}\,m=-\frac{\bb}{\nd}\,m\,,\,\,
\langle  a_2\rangle=\langle m-a_1\rangle=
\frac{N_c}{\nd}\,m\,,\,\, \langle X^{\rm adj}_{SU({\rm
n}_1)}\rangle=\langle a_{D,\,j}\rangle=\langle a_
{M,\rm n}\rangle=0\,,\label{(41.1.6)}
\eq
\bbq
\langle{\ov M}_{\rm n} M_{\rm n}\rangle=\langle{\ov M}_{\rm
n}\rangle\langle M_{\rm n}\rangle\approx -
\wmu\langle\Lambda^{SU(\nd-\no)}_{{\cal N}=2\,\, SYM}\rangle d_{\rm
n}\,,\,\, d_{\rm n}=O(1),\quad \langle{\ov D}_j\,
D_j\rangle=\langle{\ov  D}_j\rangle\langle D_j\rangle\approx
-\mx\lm\,\omega^{j-1}\,,
\eeq
\bbq
\langle\Qo\rangle_{\no}=\langle{\ov Q}^1_1\rangle\langle
Q^1_1\rangle\approx\wmu\frac{\nd}{\nd-\no}\langle a_2\rangle\approx
\wmu\frac{N_c}{\nd-\no}\,m,\,\,
\langle\Qt\rangle_{\no}=\sum_{a=1}^{\no}\langle{\ov
Q}^{\,a}_2\rangle\langle Q^2_a\rangle=0,\,\,\langle S\rangle_{\no}=0
\eeq
(the dyon condensates are dominated by terms $\sim\mx\lm$ plus smaller \, 
terms  $\sim\mx m$, see Appendix B).\\

We have to explain at this point the meaning of notations used e.g.
in\, \eqref{(41.1.5)},\eqref{(41.1.6)} for the superpotential and mean 
values \, written  for the theory with the upper cutoff $(\mx
m)^{1/2}\ll\mu_{uv}=\langle\Lambda^{SU(\nd-\no)}_{{\cal N}=2\,\,
SYM}\rangle/(\rm several)\ll m$. By definition, for any operator
$O,\,\,\langle O\rangle$ denotes its {\it total mean vacuum value}
integrated from sufficiently high energies (formally, from the very
high UV cutoff $M_{UV}$ for the elementary fields and from $\mu_{\rm
sol}$ for composite solitons formed at the scale $\sim\mu_{\rm sol}$)
down to $\mu^{\rm lowest}_{\rm cut}=0$, i.e. $\langle
O\rangle\equiv\langle O\rangle^{M_{UV}\,{\rm or}\,\mu_{\rm
sol}}_{\mu^{\rm lowest}_{\rm cut}=0}$. Therefore, we have, strictly
speaking, to write e.g. for the lower energy superpotential in
\eqref{(41.1.5)}: $\,a_{1,2}=\langle a_{1,2}\rangle^{M_{UV}}_{\mu^{\rm
low}_{\rm cut}=\mu_{uv}}+(\hat a_{1,2})^{\mu_{uv}}$, with $\mu^{\rm
low}_{\rm cut}=\mu_{uv}=\langle\Lambda^{SU(\nd-\no)}_{{\cal N}=2\,\,
SYM}\rangle/(\rm several)$, and the remaining soft parts $(\hat
a_{1,2})^{\mu_{uv}}$ of operators $a_{1,2}$ with the upper energy
cutoff $\mu_{uv}\ll m$. But in the case considered, with higgsed
$\langle a_{1,2}\rangle\sim m\ll\lm$, these their total mean values
{\it originate and saturate} in the range of scales $m/(\rm
several)<\mu<(\rm several) m$, because otherwise all $SU(\nd)$ quarks
and all dyons would have masses $\sim m$. They all will decouple then
as heavy at the scale $\mu<m/(\rm several)$ and this will be clearly
wrong for these br2-vacua (the multiplicity will be wrong, the flavor
symmetry will remain unbroken, the number of charged BPS particles
massless at $\mx\ra 0$ will be wrong, etc.). Therefore, $\langle
a_{1,2}\rangle^{M_{UV}}_{\mu_{uv}}=\langle
a_{1,2}\rangle^{M_{UV}}_{\mu^{\rm lowest}_{\rm cut}=0}\equiv\langle
a_{1,2}\rangle=\langle a_{1,2}\rangle^{(\rm several) m}_{m/(\rm
several)}$, and $\langle{\hat a}\rangle^{\mu_{uv}}_{\mu^{\rm
lowest}_{\rm cut}=0}=0$. For this reason, only in order to simplify
all notations (here and in most cases below in the text) all is
written as in \eqref{(41.1.5)},\eqref{(41.1.6)}. We hope that the
meaning of operators and mean values entering all formulas will \,
be  clear in each case from the accompanying text.\\

Now, in a few words, it is not difficult to check that the variant of
\eqref{(41.1.5)} with the first term of $\w_D$ of the form
$m\sum_{j=1}^{\bb}{\ov D}_j D_j$ (i.e. with zero $U^{(1)}(1)$ charge
of dyons) results in $\langle\partial\w^{\,\rm low}_{\rm
matter}/\partial L\rangle=\mx m (\bb)/2\neq 0$, and this is the
internal inconsistency. Another way, with such a form of the first
term of $\w_D$, one obtains from $\langle{\partial\w_D}/{\partial{\ov
D}_j}\rangle=0$ : $\langle a_{D,j}\rangle=m,\,\, \sum_{j} \langle
a_{D,j}\rangle=(\bb) m\neq 0$, and this is wrong. \\

The $\bb$ unequal double roots $e^{(D)}_j\approx\omega^{j-1}\lm$ of
the curve \eqref{(40.2)} correspond to BPS dyons formed at the scale $\mu
\sim\lm$.  For slightly unequal mass quarks with $(\delta m)_{ij}\neq 0$ 
(see just below),  together with $\no$ unequal double roots $e^{(Q)}_k=
 - m_k$  of original pure electric quarks from
${\cal N}=2 \,\, SU(\no)$ SQCD, and $\nd-\no-1$ unequal double roots
of $SU(\nd-\no)$ of adjoint pure magnetic monopoles from ${\cal N}=2
\,\,SU(\nd-\no)$ SYM (formed at the scale
$\mu\sim\langle\Lambda^{SU(\nd-\no)}_{{\cal N}=2\,\, SYM}\rangle$),
they constitute the total set of $N_c-1$ double roots of the curve
\eqref{(40.2)} at $\mx\ra 0$.. There are no {\it additional}  massless at 
$\mx\ra 0$  charged  BPS solitons because there will be then too \, 
many  double roots of the curve \eqref{(40.2)}.

The short additional explanations will be useful at this point. As was \, 
pointed  out in Introduction, the spectral curve \eqref{(40.2)} can be
used, in general, to determine  multiplicities and charges of massless 
charged  BPS particles  in the formal limit $\mx\ra 0$ only.  But in cases 
with  equal mass quarks as in br2 vacua considered 
in  this  section (and in many other vacua), some additional infrared
regularization is needed to really have only $N_c-1$ exactly massless
at $\mx\ra 0$ charged BPS particles and corresponding them $N_c-1$
well defined {\it unequal double roots} of the $SU(N_c)$ curve
\eqref{(40.2)}. We have in mind that all charged particles of the
${\cal N}=2\,\, SU(\no)$ subgroup will be massless at $\mx\ra 0$ for
equal mass quarks, this corresponds to $\no$ {\it equal double roots}
$e^{(Q)}_k= - m,\,\, k=1...\no$ of the curve \eqref{(40.2)}. To have
really only $\no$ exactly massless at $\mx\ra 0$ charged BPS particles \, 
in this  $SU(\no)$ sector, we have to split slightly the quark masses.
For instance, in br2 vacua of this section it is sufficient to split
slightly the masses of quarks from color $SU(\no)$ with the first $\no$
flavors. I.e.\,: $m_k= m+{\delta m}_k,\,\, k=1...N_F,\,\,
{\delta m}_k\sim {\delta m}_n, \,\,
\sum_{k=1}^{\no}{\delta m}_k=0$, for $k, n=1...\no$\,, while ${\delta
m}_k=0$ for $k=\no+1\, ...\, N_F$. The mass splittings ${\delta
m}_k,\,\, k=1...\no$ are arbitrary small but fixed to be unequal and
nonzero,  $(\delta m)_{ij}=(\delta m_i-\delta m_j)\neq 0$ at $i\neq j$. 
In this case, $\langle X^{\rm adj}_{SU({\rm n}_1)}\rangle$
in \, \eqref{(41.1.6)} will be nonzero, $\langle{\sqrt 2} X^{\rm
adj}_{SU({\rm n}_1)}\rangle={\rm diag}(\,{\delta m}_{1}\,...\,{\delta
m}_{\no}\,)$, and $SU(N_F)\ra U^{N_F-1}(1)$ due to not spontaneous 
but explicit breaking of the global $SU(N_F)$ flavor symmetry by
 $(\delta m)_{ij}\neq 0$.  The massless charged
BPS particles at $\mx\ra 0$ will be only $\no$ quarks $Q^{i=a}_a,\, {\ov
Q}^{\,a}_{j=a},\, a=1...\no$ (higgsed at $0<\mx\ll (\delta m)_{ij}$),
while all other charged BPS particles of color $SU(\no)$
will acquire nonzero  masses $O((\delta m)_{ij})$. 
And the $SU(N_c)$ curve \eqref{(40.2)} will  have just $\no$ slightly 
split quark double roots $e^{(Q)}_k= - m_k,\,\, k=1...\no$.   And this
shows now most clearly that {\it there are no other charged BPS
solitons massless at $\mx\ra 0$}. This infrared regularization will be \, 
always  implied  when we will speak about only   $N_c-1$ \, 
double roots of  the $SU(N_c)$ curve \eqref{(40.2)} at $\mx\ra 0$ 
(or  about $N_c$ double roots of the $U(N_c)$ curve \eqref{(40.2)} 
in vs-vacua of section 45).\\

Our purpose now is to calculate $\delta_2$ in order to find the
leading terms of quark condensates $\langle{\rm Tr}\,{\ov Q}
Q\rangle_{\no}$ and the monopole condensates $\langle M_i\rangle$ in
\eqref{(41.1.6)}, and to calculate $\Lambda_{SU(\nd)}$ in
 \eqref{(41.1.2)},\eqref{(41.1.4)}. For this, on account of leading terms $\sim\mx m^2$
and the leading power correction $\sim\langle S\rangle_{SU(\nd-\no)}$,\, 
we write,  see \eqref{(40.1)},\eqref{(41.1.5)},\eqref{(41.1.4)},
\bq
\langle\w^{\,\rm low}_{\rm
tot}\rangle=\langle\w^{\,(SYM)}_{SU(\nd-\no)}\rangle+\langle\w^{\,\rm
low}_{\rm matter}\rangle\,,\quad
\wmu=\mx(1+\delta_2)\,,\label{(41.1.7)}
\eq
\bbq
\langle\w^{\,(SYM)}_{SU(\nd-\no)}\rangle=\wmu\langle{\rm
Tr\,}(X^{adj}_{SU(\nd-\no)})^2\rangle=(\nd-\no)
\langle
S\rangle_{SU(\nd-\no)}=(\nd-\no)\wmu\langle\Lambda^
{SU(\nd-\no)}_{{\cal  N}=2\,\, SYM}\rangle^2\,,
\eeq
where $\langle\w^{\,(SYM)}_{SU(\nd-\no)}\rangle$ is the contribution
of the ${\cal N}=2\,\,SU(\nd-\no)$ SYM.

First, to determine $\delta_{1,2}=O(1)$ in \eqref{(41.1.5)}, it will
be \, sufficient \, to keep only the leading term $\langle\w^{\,\rm low}_{\rm
matter}\rangle\sim\mx m^2$ in \eqref{(41.1.7)},\eqref{(41.1.5)} and to
neglect even the leading power correction
$\langle\w^{\,(SYM)}_{SU(\nd-\no)}\rangle$,
\bq
\langle\w^{\,\rm low}_{\rm tot}\rangle\approx\langle\w^{\,\rm
low}_{\rm matter}\rangle=\frac{\mx}{2}(1+\delta_1)\frac{\nd
N_c}{\bt}\Bigl [ \langle \, a_1\rangle^2=\frac{\bt^2}{\nd^{\,2}}\,m^2
\Bigr ]+\frac{\mx}{2}(1+\delta_2)\frac{{\rm n}_1\nd}{\nd-{\rm n}_1}
\Bigl [\langle a_2\rangle^2=\frac{N_c^{\,2}}{\nd^{\,2}}\,m^2 \Bigr
].\quad\,\,\label{(41.1.8)}
\eq

On the other hand, the exact total effective superpotential $\w^{\,\rm
eff}_{\rm  tot}$ which accounts explicitly for all anomalies and
contains {\it only} quark bilinears $\Pi^i_j=({\ov Q}_j Q^i)_{N_c}$
looks as, see \cite{ch13,ch16,ch19} and \eqref{(40.3)},\eqref{(40.4)},
\bq
\w^{\,\rm eff}_{\rm tot}(\Pi)=m\,{\rm Tr}\,({\ov Q}
Q)_{N_c}-\frac{1}{2\mx}\Biggl [ \,\sum_{i,j=1}^{N_F} ({\ov Q}_j
Q^i)_{N_c}({\ov Q}_{\,i} Q^j)_{N_c}-\frac{1}{N_c}\Bigl ({\rm
Tr}\,({\ov Q} Q)_{N_c}\Bigr )^2\Biggr ]-\nd S_{N_c} \,,
\label{(41.1.9)}
\eq
\bbq
\langle
S\rangle_{N_c}=\Bigl(\frac{\langle\det\QQ\rangle=\langle\Qo
\rangle^{\no}\langle\Qt\rangle^{\nt}}{\lm^{2N_c-N_F}\mx^
{N_c}}\Bigr )^{\frac{1}{N_F-N_c}}_{N_c}\,.
\eeq

Kipping only the leading terms $\sim\mx m^2$ in \eqref{(41.1.9)} and
using \eqref{(41.1)} obtained from \eqref{(41.1.9)},\eqref{(40.4)},
\bq
\langle\w^{\,\rm eff}_{\rm tot}\rangle\approx\frac{1}{2}\frac{\nt
N_c}{N_c-\nt}\mx m^2\approx -\,\frac{1}{2}\frac{\nt N_c}{\nd-\no}\mx
m^2\,.\label{(41.1.10)}
\eq

From $\langle\w^{\,\rm low}_{\rm tot}\rangle=\langle\w^{\,\rm
eff}_{\rm tot}\rangle$ in \eqref{(41.1.8)},\eqref{(41.1.10)},
\bq
-2\nd
N_c=\nd\bt\delta_1+\no(N_c\delta_2-\bt\delta_1)\,.\label{(41.1.11)}
\eq
But all coefficients $\delta_{i}=O(1),\,\, i=1...4,$ in
\eqref{(41.1.2)} originate from the large scale $\sim\lm$, from
integrating out the heaviest fields with masses $\sim\lm$, and they do \, 
not know  the number $\no$ which originates only from the behavior of
$X^{\rm adj}_{SU(\nd)}$ in \eqref{(41.1.2)} at the much lower scale
$\sim\langle a_2\rangle\sim m\ll\lm$, so that
\bq
N_c\delta_2=\bt\delta_1\,,\quad \delta_1=-\,\frac{2N_c}{\bt}\,,\quad
\delta_2= -2\,,\quad \bt=2N_c-N_F\,.\label{(41.1.12)}
\eq
(Besides, it is shown in Appendix B that $\delta_3=0$ in
\eqref{(41.1.2)},\eqref{(41.1.5)}, see \eqref{(B.4)}).\\

Therefore, we obtain from $\langle\partial\w^{\,\rm low}_{\rm
matter}/\partial a_2\rangle=0$, see  \eqref{(41.1.4)}, \eqref{(41.1.5)},
\eqref{(41.1.6)},\eqref{(41.1.12)}, for the leading term \, 
of  $\langle\Qo\rangle_{\no}\rangle$ in br2 vacua, compare with
\eqref{(41.1)},
\bq
\langle{\rm Tr}\,({\ov Q}
Q)\rangle_{\no}=\no\langle\Qo\rangle_{\no}+\nt\langle\Qt
\rangle_{\no}=\no\langle\Qo\rangle_{\no}\approx \,
\frac{{\rm n}_1\, N_c}{N_c-\nt}\,\mx \,
m\approx\no\mx \ha\,,\label{(41.1.13)}
\eq
\bbq
\langle\Qo\rangle^{SU(N_c)}_{\no}=\sum_{a=1}^{\no}\langle{\ov
Q}_1^{\,a} Q^1_a\rangle=\langle{\ov Q}_1^{\,1} \rangle\langle
Q^1_1\rangle\approx\mx \,
\ha\approx\langle\Qt\rangle^{SU(N_c)}_{N_c},
\eeq
\bbq
\langle\Qt\rangle_{\no}=\sum_{a=1}^{\no}\Bigl [\langle{\ov Q}_2^{\,a}
\rangle\langle Q^{(2)}_a\rangle=0\Bigr ]+\frac{1}{|\langle\Qo\rangle_{\no}|}
\Bigl [\langle{\Pi}_{2}^{1}\rangle\,\langle{\Pi}_{1}^{2}\rangle=0\Bigr ],\quad
\eeq
\bbq
{\Pi}_{2}^{1}=\sum_{a=1}^{\no}{\ov Q}^{\,a}_2\langle Q^{(1)}_{a}\rangle,\quad
{\Pi}_{1}^{2}=\sum_{a=1}^{\no}\langle{\ov Q}^{\,a}_{1}\rangle Q_{a}^{\,(2)},
\eeq

The whole group $SU(\no)\times U^{(2)}(1)$ is higgsed by $\no$
electric quarks. The power correction from the SYM part 
is accounted for in \eqref{(41.1.14)}. This power  correction 
is obtained either directly from  \eqref{(41.1.28)}, \eqref{(41.1.32)}, 
\eqref{(41.4.5)},  or independently from the Seiberg-Witten curve 
\eqref{(A.24)}),
\bq
\sum_{a=1}^{\no}\langle{\ov Q}_k^{\,a} Q^{(k)}_a\rangle\approx \delta
^k_a\mx \ha \Bigl [\,1+\frac{\bb}{\nt-N_c}\Bigl (\frac{\ha}{\lm}\Bigr
)^{\frac{2N_c-N_F}{\nt-N_c}}\,\Bigr ],\,\, \wmu\langle
S\rangle_{\no}=\langle\Qo\rangle_{\no}\langle\Qt\rangle_
{\no}=0,\quad  \,\,\label{(41.1.14)}
\eq
\bbq
a=1...\no\,,\quad k=1...N_F\,,\quad \ha=\frac{N_c}{N_c-\nt}\, m\,,
\eeq
and $\no^2$ long ${\cal N}=2$ multiplets of massive gluons with masses \, 
$\sim  (m \mx)^{1/2}$ are formed (including $U^{(2)}(1)$ with its
scalar $a_2$). It is worth to remind that at sufficiently small $\mx$
the ${\cal N}=2$ SUSY remains unbroken at the level of particle
masses \, $O(\sqrt{\mx})$, and only small corrections $O(\mx)$ to masses 
of  corresponding scalar superfields break ${\cal N}=2$ to ${\cal N}=1$,
see e.g. \cite{VY2}.
\footnote{\,
In comparison with the standard ${\cal N}=1$ masses $\sim (\mx\, 
\ha)^{1/2}$ of $\no^2$ gluons and $\no^2$ of their ${\cal N}=1$ (real)\,  
scalar  superpartners originating from D-terms, the same masses of
additional $2\no^2$ (complex) scalar superpartners of long ${\cal
N}=2$ gluon multiplets originate here from F-terms in
\eqref{(41.1.5)}:$\sim\sum_{a,b=1}^{\no} X^a_b \Pi^b_a$, where $\no^2$ 
of   $X^a_b,\,\langle X^a_b\rangle=0$ at $m_i=m$, include $(\no^2-1)$ of 
$X^{adj}_{SU(\no)}$  and one ${\hat a}_2=a_2-\langle a_2\rangle$, and $\no^2$ 
colored "pions"  are  $\Pi^b_a=\sum_{i=1}^{\no}({\ov Q}^{\,b}_i Q_a^i),\,
\langle\Pi^b_a\rangle\approx \delta^b_a\,\mx\,  \ha,\,\, a,b=1...\no$. 
The Kahler terms  of \, all  $\no^2$ pions $\Pi^b_a$ look as $K_{\Pi}=2\,{\rm
Tr}\,\sqrt{\Pi^\dagger \Pi}\,$. \label{(f52)}
}

The value of the quark condensate in \eqref{(41.1.14)} (see also \eqref{(41.1.30)}) 
is verified by  two independent calculations: either directly from the corresponding 
Lagrangian  and Konishi anomalies,  or using the
values of double roots of the curve \eqref{(40.2)} and
\eqref{(41.2.10)}, see also \eqref{(A.31)},\eqref{(B.14)}, and {\it these
calculations with roots of the curve \eqref{(40.2)} are valid only for
charged ${\cal N}=2$ BPS particles massless at $\mx\ra 0$}. This right \, 
value in  \eqref{(41.1.14)} is of real importance because \, {\it  it is a direct check
of the main assumption "A" formulated in Introduction, i.e. the BPS
properties of original quarks in \eqref{(40.1)} at 
sufficiently small $\mx$}. Indeed,
if original quarks were not BPS particles, then their masses 
would receive e.g. the {\it additional \, contributions}  $O(\mx^{1/2})$ from
additional terms $c_{a_1} a_1 {\rm Tr}(\QQ)_{\nd},\,\, c_{a_1}=O(1)$
and/or $c_m m{\rm Tr}(\QQ)_{\no},\,\, c_m=O(1)$ in \eqref{(41.1.2)},
\eqref{(41.1.5)}  from integrated out
heavier particles (or from elsewhere). The value of $\langle
a_2\rangle$ in \eqref{(41.1.6)} will be then changed, and this will
spoil the correct value of the quark condensate in \eqref{(41.1.14)}.
(And the same for the $U(N_c)$ theory in section 41.2 below, see
\eqref{(41.2.5)},\eqref{(41.2.8)},\eqref{(A.27)},\eqref{(A.29)}). On the \, other
hand, the additional contributions $\sim\mx\delta_i$ into masses \, of 
adjoint  scalars in \eqref{(41.1.2)} do not contradict the ${\cal
N}=2$ BPS properties of these scalars as these additional
contributions to their masses are only $O(\mx)$.\\

As it should be, for equal mass quarks, there remain $2\no\nt$ {\it exactly 
massless}  Nambu-Goldstone multiplets $({\ov Q}_2 \langle Q^1\rangle_{\no}), 
(\langle{\ov Q}_1\rangle_{\no}  Q^2)$. These are in essence the quarks 
${\ov Q}$ and $Q$ with
$\nt=N_F-\no$ flavors and $SU(\no)$ colors). The factor $\nd-\no$ in
the multiplicity $N_{\rm br2}=(\nd-\no)C^{\,\no}_{N_F}$ of these
vacua \, originates \, from $SU(\nd-\no)$ SYM, while $C^{\,\no}_{N_F}$ is \, 
due to  the spontaneous global flavor symmetry breaking 
 $U(N_F)\ra U(\no(\times U(\nt)$  in the  $SU(\no)$ sector.\\

Clearly, higgsed flavorless scalars $\langle
X^{adj}_{SU(\bb)}\rangle^{(\rm several)\lm}_{\lm/(\rm
several)}\sim\lm$ in \eqref{(41.1.1)} {\it operating at  the scale
$\mu\sim\lm$ do not break by themselves the global flavor $SU(N_F)$
symmetry}. In particular, largest $|\rm masses|\sim\lm$ of all $N_F$
quarks with $SU(\bb)$ colors are the same. At the same time, e.g. the
non-Abelian color $SU(\bb)$ is broken at this scale by this higgsed
$X^{adj}_{SU(\bb)}$ down to Abelian one, and masses $\sim\lm$ of
heavy  flavorless  $SU(\bb)$ adjoint gluons (and scalars) differ greatly. 
{\it  There   is no  color-flavor locking in this $SU(\bb)$ color sector}. We
would like to emphasize (see below in this section) that the global
$SU(N_F)$ is really broken spontaneously {\it only} at lower energies
$\sim (\mx m)^{1/2}\ll\lm$ in the $SU(\no)$ color sector due to
higgsed light quarks with $N_F$ flavors and $SU(\no)$ colors, see
\eqref{(41.1.13)},\eqref{(41.1.14)}.\\

Now, in short about the variant of \eqref{(41.1.5)} with the first term \, 
of $\w_D$  of the form $[\,-c_1 a_1\sum_{j=1}^{\bt}{\ov D}_j D_j\,]$,
i.e. with zero $SU(N_c)$ baryon charge of dyons $D_j$. With such a
form of the first term of $\w_D$, one obtains from
$\langle{\partial\w_D}/{\partial{\ov D}_j}\rangle=0$ : $\langle
a_{D,j}\rangle= -c_1\langle a_1\rangle,\,\, \sum_{j} \langle
a_{D,j}\rangle=0=(\bb)( -c_1\langle a_1\rangle)\ra \langle
a_1\rangle=0,\, \langle a_2\rangle= m$. Then, instead of
\eqref{(41.1.12)},\eqref{(41.1.13)} we will obtain
\bbq
(1+\delta_2)=-\frac{N^2_c}{\nd^{\,2}}\,\, \ra\,\,
\langle\Qo\rangle^{SU(N_c)}_{\no}\approx\mx(1+\delta_2)
\frac{\nd}{\nd-\no}\langle a_2\rangle\approx  -\frac{N_c^2}
{\nd(\nd-\no)}\mx\, m\approx\frac{N_c}{\nd}\mx\, \ha\,,
\eeq
and this disagrees with the {\it independent} calculation of
$\langle\Qo\rangle^{SU(N_c)}_{\no}\approx \mx\, \ha$ in this 
$SU(N_c)$  theory using the roots of the curve \eqref{(40.2)}, 
see  \eqref{(A.31)},\eqref{(B.14)}. \\

And finally, to determine $\Lambda_{SU(\nd)}$ in \eqref{(41.1.4)}, we
account now for the leading power corrections $\langle\,\delta
\w^{\,\rm eff}_{\rm tot}\rangle\sim\langle
S\rangle_{N_c}\sim\langle\,\delta \w_{\rm tot}^{\,\rm
low}\rangle\sim\langle S\rangle_{\nd-\no}$, \,
see  \eqref{(41.1)},\eqref{(41.1.4)},\eqref{(41.1.9)},\eqref{(41.1.12)},
\bbq
\langle\,\delta \w^{\,\rm eff}_{\rm tot}\rangle=(N_c-\nt)\langle\,
S\rangle_{N_c}\approx (N_c-\nt)\mx\, \ha^2\Bigl
(\frac{\ha}{\Lambda_2}\Bigr )^{\frac{2 N_c-N_F}{\nt-N_c}}\,,
\quad  \ha=\frac{N_c}{N_c-\nt}\,m\,,
\eeq
\bbq
\langle\,\delta \w^{\,\rm low}_{\rm  tot}\rangle=
\langle\w^{\,(SYM)}_{SU(\nd-\no)}\rangle=
\, {\wmu} \,\langle\,{\rm  Tr\,} 
(X^{adj}_{SU(\nd-\no)} )^2\,\rangle=(\nd-\no)\langle
S\rangle_{SU(\nd-\no)}= (\nd-\no)\wmu\langle\Lambda
^{SU(\nd-\no)}_{{\cal N}=2\,\, SYM}\rangle^2
\eeq
\bbq
\approx(N_c-\nt)\mx m^{2}_2\Bigl (\frac{m_2}{\Lambda_{SU(\nd)}}\Bigr
)^{\frac{2N_c-N_F}{\nt-N_c}},\quad \wmu=\mx(1+\delta_2)=-\mx\,,\quad
m_2=\frac{N_c}
{\nd-\no}\,m = - \ha\,,
\eeq
\bq
\langle\,\delta \w^{\,\rm eff}_{\rm tot}\rangle=\langle\,\,
\delta  \w^{\,\rm low}_{\rm tot}\rangle\quad\ra\quad
\Lambda_{SU(\nd)}=-\lm\,.\label{(41.1.15)}
\eq

The whole spectrum of masses in the br2 vacua considered and, in
particular, the spectrum of nonzero masses arising from
\eqref{(41.1.5)} at small $\mx\ll\langle\Lambda^{SU(\nd-\no)}_{{\cal
N}=2\,\, SYM}\rangle$ was described above in this section. It is in
accordance with $N_c-1$ double roots of the curve \eqref{(40.2)} in
these br2-vacua at $\mx\ra 0$. Of them: $\,2N_c-N_F$ unequal roots
corresponding to $2N_c-N_F$ dyons $D_j$, then $\no$ equal roots
corresponding to $\no$ pure electric quarks $Q^k$ of $SU(\no)$
(higgsed at $\mx\neq 0$), and finally $\nd-\no-1$ unequal roots
corresponding to $\nd-\no-1$ pure magnetic monopoles $M_i$ of
$SU(\nd-\no)\,\, {\cal N}=2$ SYM. Remind that two single roots
$(e^{+}-e^{-})\sim\langle\Lambda^{SU(\nd-
{\rm n}_1)}_{{\cal N}=2\,\, SYM}\rangle$ of the curve \eqref{(40.2)}
originate in these br2 vacua from the ${\cal N}=2\,\, SU(\nd-\no)$\,
 SYM \,  sector.

As it is seen from the above, {\it the only exactly massless at $m\neq \, 0,\,
\mx\neq 0$ particles} in the Lagrangian \eqref{(41.1.5)} are $2\no\nt$ 
(complex) Nambu-Goldstone multiplets originating from the
spontaneous breaking of the global flavor symmetry, $U(N_F)\ra
U(\no)\times U(\nt)$ in the $SU(\no)$ color sector.

The calculation of the leading power correction
$\sim\mx\langle\Lambda^{SU(\nd-{\rm n}_1)}_{{\cal N}=2\,\,
SYM}\rangle^2/m$ to the value of $\langle\Qo\rangle^{SU(N_c)}_{\no}$
in \eqref{(41.1.14)} is presented in Appendices A and B, and power
corrections to the dyon condensates in \eqref{(41.1.6)} are calculated
in Appendix B. The results agree with also presented in these
Appendices {\it independent} calculations of these condensates using
roots of the Seiberg-Witten curve. And these last calculations with
roots are valid {\it only for BPS particles. This agreement confirms
in a non-trivial way a self-consistency of the whole approach and BPS
properties of dyons and original quarks $Q^i, {\ov Q}_j$}.\\

According to reasonings given above in this section, the quarks with
all $N_F$ flavors, $SU(2N_c-N_F)$ colors, and masses $\sim\lm$ are 
too \, heavy and too short ranged and cannot form a coherent condensate 
for  this reason. In other words, they are {\it not higgsed}, i.e.
$\langle  Q^i_a\rangle=\langle{\ov Q}^{\,a}_j\rangle=0,\,\, a=
\nd+1...N_c$.  Indeed, in the range from high energies down to
$\mu^{\rm low}_{\rm cut}=(\mx\lm)^{1/2}$, {\it for independently
moving} heavy quarks ${\ov Q}^{\,a}_j$ and $Q_a^i$ with $SU(\bb)$
colors, all their physical (i.e. path-dependent) phases induced by
interactions with the {\it light} $\,U^{(1)}(1)\times U^{\bb-1}(1)$
photons still fluctuate freely, so that the mean values of quark
fields integrated over this energy range, as well as the mean values
of factorizable parts of local quarks bilinears, look as (the
non-factorizable parts are also zero in this energy range, see below)
\bq
\langle{\ov Q}^{\,a}_j\rangle_{\mu^{\rm low}_{\rm cut}}=\langle
Q_a^i\rangle_{\mu^{\rm low}_{\rm cut}}=\langle{\ov Q}^{\,a}_j
Q_a^i\rangle^{\rm factor}_{\mu^{\rm low}_{\rm cut}}=0\,,\quad
a=\nd+1...N_c\,,\quad \mu^{\rm low}_{\rm
cut}=(\mx\lm)^{1/2}\,.\,\,\label{(41.1.16)}
\eq

For this reason, at the scale $\mu_{\rm cut}=\lm/(\rm several)$ where
all particles with masses $\sim\lm$ {\it decoupled already as heavy,
there are no any contributions to particle masses in the Lagrangian
\eqref{(41.1.2)} from mean values of heavy quark fields}, $\langle{\ov
Q}^{\,a}_j\rangle_{\mu_{\rm cut}}=\langle Q_a^i\rangle_{\mu_{\rm
cut}}=0,\,\,\mu_{\rm cut}=\lm/(\rm several)$. And after all heavy
particles decoupled at $\mu<\mu_{\rm cut}$, resulting in the
Lagrangian \eqref{(41.1.2)}, {\it they themselves do not affect then
the behavior of lower energy theory}. (Really, because $\nt>N_c,\,
\no<\nd$ in these br2 vacua, it is clear beforehand that the quarks
$Q^2, {\ov Q}_2$ in $\nt>N_c$ equal condensates
$\langle\Qt\rangle_{N_c}$ are definitely not higgsed at all due to the \, 
rank  restrictions, because otherwise this will result in a wrong
pattern of flavor symmetry breaking. Higgsed are the light quarks
$Q^i_a, {\ov Q}^{\,a}_i,\, \, i,a=1...\no$ only, see \eqref{(41.1.14)}
and the end of section 41.4).\\

It is worth emphasizing also the following. The equality
$\langle\,\partial \w/\partial \phi\rangle=0$ {\it is valid only when
all nonperturbative contributions are accounted for in the
superpotential} $\w$. If nonperturbative contributions are smaller
than the perturbative ones, they can be neglected and then the
equality $\langle\,\partial \w_{\rm pert}/\partial \phi\rangle\approx
0$ will be approximately right. If nonperturbative contributions are
of the same order as perturbative ones, then ignoring them will give
at best only the order of magnitude estimate. Remind that, by
definition for any operator $O=A-B$, the equality $\langle O\rangle=0 \, 
$denotes  the {\it total mean value} integrated from the appropriate
high energy down to $\mu^{\rm lowest}_{\rm cut}=0$. And following\,  
from \, it equality of {\it total mean values} of two operators, $\langle
A\rangle=\langle B\rangle$, {\it does not mean in general that the
numerical values of $\langle A\rangle$ and $\langle B\rangle$
originate from the same energy regions}. The equality of {\it such
total mean values} $\langle A\rangle=\langle B\rangle$ means only
that, when integrated from the appropriate high energies down to zero,\, 
{\it  their \, numerical values} will be equal. In general, {\it the
numerical values of $\langle A\rangle$ and $\langle B\rangle$ can
originate from different energy regions}. This is the case with e.g.
\eqref{(41.1.23)}, see below. And similarly in many cases in the text
below.\\

As it is seen from \eqref{(41.1)},\eqref{(41.4.1)}, the $SU(\bb)$
singlet parts of heavy quark {\it total mean values}, $\langle{\ov
Q}_j Q^i\rangle_{\bb}=\sum_{a=\nd+1}^{N_c}\langle{\ov Q}^{\,a}_j
Q^i_{a}\rangle\sim\,\delta^i_j\,\mx m\ll\mx\lm$ are of the same size
as condensates of light higgsed quarks
$\langle\Qo\rangle_{\nd}\approx\langle\Qo\rangle_{\no}=\langle{\ov
Q}^1_1\rangle\langle Q^1_1\rangle\sim\mx m$ (the total mean values
$\langle{\ov Q}_i Q^i\rangle_{\nd}$ of all flavors $i=1...N_F$ of
quarks are holomorphic in $\mx$ and their values in
\eqref{(41.4.1)},\eqref{(41.4.5)},\eqref{(41.4.6)} are valid at smaller
$\mx\ll m$ as well),
\bq
\langle\Qt\rangle_{\bb}=\langle\Qt\rangle_{N_c}-\langle\Qt
\rangle_{\nd}\approx\mx\,  \ha\Biggl [1+\frac{\bb}{\nt-N_c}\Bigl
(\frac{\ha}{\lm}\Bigr )^{\frac{\bb}{\nt-N_c}}\Biggr ) \Biggr
]\,,\label{(41.1.17)}
\eq
\bbq
\langle\Qo\rangle_{\bb}=\langle\Qo\rangle_{N_c}-\langle\Qo\rangle
_{\nd}\approx \,- \mx\, \ha\Biggl [1+\frac{\bb}{\nt-N_c}\Bigl (\frac{\ha}
{\lm}\Bigr )^{\frac{\bb}{\nt-N_c}}\Biggr ) \Biggr ]\,,
\eeq
\bbq
\langle{\rm
Tr\,}\QQ\rangle_{\bb}=\nt\langle\Qt\rangle_{2N_c-N_F}+
\no\langle  \Qo\rangle_{2N_c-N_F}\approx
(\nt-\no)\mx\,  \ha\Biggl [1+\frac{\bb}{\nt-N_c}\Bigl
(\frac{\ha}{\lm}\Bigr )^{\frac{\bb}{\nt-N_c}}\Biggr ) \Biggr].
\eeq

Besides, the total mean value of the flavor singlet part $\langle{\rm
Tr\,}\QQ\rangle_{\bb}$ in \eqref{(41.1.17)} can be connected with the
$\langle\Sigma_D\rangle$ part of the dyon condensate, $\Sigma_D
=\sum_{j=1}^{\bb}{\ov D}_j D_j$. Considering 'm' as the background
field, from \eqref{(40.1)},\eqref{(41.1.2)} and using $\delta_3=0$ from
\eqref{(B.4)},
\bq
\langle\,\frac{\partial}{\partial m}\w_{\rm
matter}\rangle=\langle\,\frac{\partial}{\partial m}
{\widehat\w}_{\rm matter}\rangle \ra\langle{\rm
Tr\,}\QQ\rangle_{\bb}=\langle\Sigma_D\rangle
\equiv\sum_{j=1}^{\bb}\langle{\ov D}_j
D_j\rangle=\sum_{j=1}^{\bb}\langle{\ov D}_j\rangle
\langle D_j\rangle.  \quad\,\,\,\,\label{(41.1.18)}
\eq

As for the value of $\langle\Sigma_D\rangle$ in br2 vacua of the
$SU(N_c)$ theory, it is obtained {\it independently} in Appendix B
using \eqref{(41.2.10)} and the values of roots of the curve
\eqref{(40.2)}, see \eqref{(B.7)},\eqref{(B.15)}
\bq
\langle\Sigma_D\rangle=\sum_{j=1}^{\bb}\langle{\ov D}_j\rangle\langle
D_j\rangle\approx (\nt-\no)\mx\,  \ha \Biggl [1+\frac{\bb}{\nt-N_c}\Bigl
(\frac{\ha}{\lm}\Bigr )^{\frac{\bb}{\nt-N_c}}\Biggr ) \Biggr]\,,\quad
\ha=\frac{N_c}{N_c-\nt}\,,\quad \label{(41.1.19)}
\eq
this agrees with \eqref{(41.1.17)},\eqref{(41.1.18)}. \\

As an example, the {\it operator} ${\rm Tr\,}(\QQ)_{\bb}$ of heavy
quarks with masses $\sim\lm$ in \eqref{(41.1.17)}, when integrated over \, 
the  interval $(\rm several) m < \mu<(\rm several)\lm$, transforms in
general into a linear combination of three {\it operators} of light
fields, see \eqref{(41.1.1)}
\bq
\hspace*{-2mm} [\,{\rm Tr\,}(\QQ)_{\bb}\,]^{(\rm several)\lm}_{(\rm
several) m}=\Biggl [f_1 N_c\mx (m-c_1 a_1)+f_2 \Sigma_D+ f_3 {\rm
Tr\,}(\QQ)_{\nd}\Bigr ]_{,}^{(\rm several) m}\,\,\,\,
f_{1,2,3}=O(1).\quad \label{(41.1.20)}
\eq

This {\it operator expansion} relates then {\it the numerical total
mean values} of the left and right hand sides of \eqref{(41.1.20)}. But \, 
the total  mean values of three operators in the right hand side of
\eqref{(41.1.20)} originate and saturate at parametrically different
energies. $\langle a_1\rangle$ originates and saturates at $\mu\sim
m$, as otherwise all dyons would have masses $\sim m\gg
(\mx\lm)^{1/2}$. They will not be higgsed then but will decouple as
heavy at $\mu<m$. $\bb$ unequal double roots of the curve
\eqref{(40.2)} will be absent, while e.g. all $U^{(1)}(1)\times
U^{\bb}(1)$ photons will remain massless, etc.
$\langle\Sigma_D\rangle$ originates and saturates at $\mu\sim
(\mx\lm)^{1/2}\ll\langle\Lambda^{SU(\nd-\no)}_{{\cal N}=2\,\,
SYM}\rangle\ll m$. And $\langle{\rm Tr\,}(\QQ)_{\nd}\rangle$
originates and saturates from the sum of three regions, $\mu\sim
m\ll\lm\,,\,\, \mu\sim\langle\Lambda^{SU(\nd-\no)}_{{\cal N}=2\,\,
SYM}\rangle\ll m$, and $\mu\sim (\mx m)^{1/2}\ll\langle
\Lambda^{SU(\nd-\no)}_{{\cal N}=2\,\, SYM}\rangle$, see below
in this section. Besides, the {\it numerical total mean value} of the
right hand side in \eqref{(41.1.20)} equals $\langle\Sigma\rangle$,
see\eqref{(41.1.18)}. And because $\langle m-c_1 a_1\rangle=0$ from
\eqref{(41.1.1)}, we obtain: $f_2=1,\, f_3=0$.

The contributions to the heavy quark total mean values $\langle
\QQ_{1,2}\rangle_{\bb}$ also originate on the whole: a) from the
quantum loop effects at the scale $\mu\sim\lm$ in the strong coupling
(and nonperturbative) regime, transforming on the first "preliminary"
stage such bilinear {\it operators} of heavy quark fields with masses
$\sim\lm$ into the appropriate {\it operators} of light fields, i.e.
$\mx a_1$ and bilinear {\it operators} of light $SU(\nd)$ quarks and
dyons, see \eqref{(41.1.20)} and \eqref{(41.1.22)}-\eqref{(41.1.25)}
below; \, b) finally, from formed at lower energies $\mu\ll\lm$ the
genuine condensates of these light higgsed fields.

In connection with \eqref{(41.1.17)}-\eqref{(41.1.19)}, it is worth
asking the following question. Because the difference between 
$\w_{\rm  matter}$in \eqref{(40.1)} and $\widehat\w_{\rm matter}$ 
in  \eqref{(41.1.2)} and formation of dyons originate from the color
symmetry breaking $SU(N_c)\ra SU(\nd)\times U^{\bb}(1)$ in the high
energy region $\mu\sim\lm$, how e.g. the condensate
$\langle\Sigma_D\rangle$ of these dyons knows about the number $\no$
which originates only at much lower energies $\sim m$ ? (Put attention \, 
that even  the leading term $\sim\mx m$ of $\langle\Sigma_D\rangle$ in
\eqref{(41.1.20)} is much smaller than the separate terms $\langle{\ov
D}_j\rangle\langle D_j\rangle\sim\mx\lm$, but these largest terms
$\sim\mx\lm$ cancel in the sum over j). And similarly, because the
color breaking $\langle X^{adj}_{SU(\bb)}\rangle\sim\lm$ does not
break {\it by itself} the global flavor symmetry, why the condensates
of heavy quarks $\langle\Qt\rangle_{\bb}$ and
$\langle\Qo\rangle_{\bb}$ are different ?

First, about $\langle\Sigma_D\rangle$. The answer is that, indeed,
both {\it operators} ${\rm Tr\,}(\QQ)_{2N_c-N_F}$ and $\Sigma_D$ in
\eqref{(41.1.18)} are by themselves $U(N_F)_{\rm flavor}\times
U(\nd)_{\rm color}$ singlets. The explicit dependence on $\no$ appears \, 
in  $\langle\Sigma_D\rangle$ in \eqref{(41.1.19)} (and in left hand
sides of \eqref{(41.1.17)},\eqref{(41.1.25)},\eqref{(41.1.26)}) {\it only \, from the
region of much lower energies, after taking the corresponding \, total 
vacuum   averages $\langle...\rangle$ in br2 vacua}. Recall that
in these vacua. -

1) $\langle a_1\rangle\sim m$ is higgsed and operates at the scale
$\sim m$. It "eats" the masses $"m"$ of all dyons in \eqref{(41.1.5)},
as otherwise all these dyons would not be higgsed but would decouple
as heavy at $\mu<m$. $\bb$ unequal double roots of the curve
\eqref{(40.2)} will be absent, while $\bb$ photons will remain
massless.

2) $SU(\nd)$ is broken also at the scale $\sim m$ to $SU(\no)\times
U^{(2)}(1)\times SU(\nd-\no)$ by higgsed $\langle
X^{adj}_{SU(\nd)}\rangle\sim\langle a_2\rangle\sim m$. Together with
$\langle a_1\rangle$, this $\langle a_2\rangle$ "eats" masses "m" of
all $SU(\no)$ quarks, $\langle(m-a_1-a_2)\rangle=0$. Otherwise all
these $SU(\no)$ quarks will decouple as heavy at $\mu<m$, there will
remain $SU(\no)\,\,{\cal N}=2$ SYM at $\mu<m$, the multiplicity will
be wrong and $U(N_F)$ will remain unbroken. At the same time, all
quarks in the $SU(\nd-\no)$ color sector remain with masses $\sim m$.

3) To avoid $g^2(\mu<\langle\Lambda^{SU(\nd-\no)}_{{\cal N}=2\,\,
SYM}\rangle )< 0$ in ${\cal N}=2\,\, SU(\nd-\no)$ SYM, $\langle
X^{adj}_{SU(\nd-\no)}\rangle$ is higgsed at the scale
$\sim\langle\Lambda^{SU(\nd-\no)}_{{\cal N}=2\,\, SYM}\rangle,\,\,
SU(\nd-\no)\ra U^{\nd-\no-1}(1)$.

The (non-factorizable) mean values $\langle (\QQ)_{1,2}
\rangle_{N_c-\no}$ of quarks with masses $\sim m$ in the
$SU(N_c-\no)$ SYM sector are determined by the one-loop Konishi
anomaly for these massive non-higgsed quarks. I.e., on the first
"preliminary" {\it operator stage} at the scale $\mu\sim m$ in the
weak coupling regime, the one-loop diagrams with scalar quarks and
their fermionic superpartners with masses $\sim m$ inside, transform
the quark {\it operators} $[\,(\QQ)_{1,2}\,]_{N_c-\no}$ into the {\it
operator} $S_{N_c-\no}/m$ of lighter $SU(N_c-\no)$ gluino (plus
operators with total superderivatives). And the mean value of
$\langle  S\rangle_{\nd-\no}=\wmu\langle\Lambda^{SU(\nd-\no)}_{{\cal
N}=2\,\,SYM}\rangle^2$, originates  only at much lower energies
$\mu\lesssim\langle\Lambda^{SU(\nd-\no)}_{{\cal N}=2\,\,SYM}\rangle$,
see \eqref{(41.1.7)}.

4) The global $U(N_F)$ symmetry is broken to $U(\no)\times U(\nt)$ by
higgsed quarks in the $SU(\no)$ color sector at the scale $\sim (\mx
m)^{1/2}\ll\langle\Lambda^{SU(\nd-\no)}_{{\cal N}=2\,\, SYM}\rangle\ll
m\ll\lm$.

5) The {\it total mean value} of $\langle\Sigma_{D}\rangle\sim \mx
m\ll\mx\lm$ in \eqref{(41.1.19)} originates and saturates at energies
$\sim (\mx\lm)^{1/2}$. The reason is that $D_j$ and ${\ov D}_j$ {\it
move independently} in the whole energy interval $[\mu^{\rm low}_{\rm
cut}=(\mx\lm)^{1/2}]< \mu <\lm/(\rm several)$, and their physical
(i.e. path-dependent) phases induced by interactions with the light
$U^{\bb-1}(1)\times U^{(1)}(1)$ photons {\it still fluctuate freely},
so that $\langle{\ov D}_j\rangle_{\mu^{\rm low}_{\rm cut}}=\langle
D_j\rangle_{\mu^{\rm low}_{\rm cut}}=\langle\Sigma_D\rangle_{\mu^{\rm
low}_{\rm cut}}=0$. The nonzero mean values of all light dyon fields
are formed only at the lower scale $\sim (\mx\lm)^{1/2}\ll\lm$, in the \, 
weak  coupling regime, from {\it coherent condensates} of these {\it
higgsed light dyons} (resulting in the appearance of $\sim
(\mx\lm)^{1/2}$ masses of all $U^{\bb-1}(1)\times U^{(1)}(1)\, 
\,{\cal  N}=2$  multiplets): $\langle{\ov D}_j D_j\rangle =\langle{\ov
D}_j\rangle\langle D_j\rangle= -
\mx\lm\,\omega^{j-1}+(\bb)^{-1}\langle\Sigma_D\rangle$, see
\eqref{(41.1.6)},\eqref{(B.8)},\eqref{(B.15)}, and \eqref{(41.1.19)}
for \, $\langle\Sigma_{D}\rangle$. (Clearly, the leading terms $\langle
D_j\rangle=\langle {\ov D}_j\rangle\sim (\mx\lm)^{1/2},\, j=1...\bb$
cancel in $\langle\Sigma_D\rangle$).

{\it All particles} in this dyonic sector acquire masses $\sim
(\mx\lm)^{1/2}$ and decouple as heavy at lower energies. No
lighter particles  at all remain in this sector at lower energies.\\

The number $\no$ penetrates then into $\langle\Sigma_D\rangle$ from
the {\it numerical} relation between the {\it total vacuum averages}
in br2 vacua following from \eqref{(41.1.5)},\eqref{(41.1.12)} (see
also\eqref{(41.4.6)},\eqref{(A.23)},\eqref{(B.4)})
\bq
\langle\frac{\partial\w^{\,\rm low}_{\rm tot}}{\partial
a_{1}}\rangle=0\ra \frac{\nd}{\bb}\langle\Sigma_D
\rangle=  \langle{\rm Tr\,}\QQ\rangle_{\nd}
-\Bigl [\langle\frac{\partial\w_{\rm a_{1}}}{\partial a_{1}}\rangle=
-\mx\frac{\nd N_F N_c}{(\bb)^2}\langle a_1\rangle\,\Bigr
]\,,\label{(41.1.21)}
\eq
\bbq
\langle{\rm Tr\,}\QQ\rangle_{\nd}=\langle{\rm
Tr\,}\QQ\rangle_{\no}+\langle{\rm Tr\,}\QQ\rangle_{\nd-\no}\,,
\quad  \langle{\rm Tr\,}\QQ\rangle_{\nd-\no}=
-\langle\frac{\partial\w^{(SYM)}_{SU(\nd-\no)}}{\partial
a_{1}}\rangle\,,
\eeq
\bbq
\langle\Sigma_D\rangle\approx (\nt-\no)\mx\, \ha \Biggl
[1+\frac{\bb}{\nt-N_c}\Bigl (\frac{\ha}{\lm}\Bigr
)^{\frac{\bb}{\nt-N_c}}\Biggr ) \Biggr]\,,
\eeq
because all $N_F$ flavors of quarks from the color $SU(\nd)$ and all
dyons interact with $a_1$. \eqref{(41.1.21)} agrees with
\eqref{(41.1.17)}-\eqref{(41.1.19)}.
Remind that, by definition, $\langle O\rangle$ means the total mean
value of any operator $O$ integrated from very high energies down to
$\mu^{\rm lowest}_{\rm cut}=0$. The equation \eqref{(41.1.21)} is the
example of {\it numerical} relations between total mean values where
each term originates and saturates at different energies.

Now, about $\langle\Qo\rangle_{\bb}$ and $\langle\Qt\rangle_{\bb}$.
The heavy quark operators in \eqref{(41.1.17)} include both the
$SU(N_F)$ adjoint and singlet in flavor parts. As for the singlet
part, its mean value is given in \eqref{(41.1.18)},\eqref{(41.1.19)}. 
As \, for the  adjoint parts, they look as, see \eqref{(41.1.18)}
\bbq
\langle{\ov Q}_j Q^i -\delta^i_j\frac{1}{N_F}{\rm
Tr}\,(\QQ\rangle_{\bb}=A\langle{\ov Q}_j Q^i
-\delta^i_j\frac{1}{N_F}{\rm Tr\,}\QQ\rangle_{\nd}\,,
\eeq
\bq
\langle\Qt-\frac{1}{N_F}{\rm
Tr\,}\QQ\rangle_{\bb}=A\langle\Qt-\frac{1}{N_F}{\rm
Tr\,}\QQ\rangle_{\nd}, \label{(41.1.22)}
\eq
\bq
\langle\Qt\rangle_{\bb}=A\langle\Qt-\frac{1}{N_F}{\rm
Tr\,}\QQ\rangle_{\nd}+
\frac{1}{N_F}\langle\Sigma_D\rangle.\label{(41.1.23)}
\eq
\bq
\langle\Qo-\frac{1}{N_F}{\rm
Tr\,}\QQ\rangle_{\bb}=A\langle\Qo-\frac{1}{N_F}{\rm
Tr\,}\QQ\rangle_{\nd} \label{(41.1.24)}
\eq
\bq
\langle\Qo\rangle_{\bb}=A\langle\Qo-\frac{1}{N_F}{\rm
Tr\,}\QQ\rangle_{\nd}+
\frac{1}{N_F}\langle\Sigma_D\rangle\,,\label{(41.1.25)}
\eq
where $A=O(1)$ is some constant originating from integrating out the
loop (and nonperturbative) effects from the Lagrangian \eqref{(40.1)}
over the energy region $\mu\sim\lm$. It is seen from
\eqref{(41.1.22)},\eqref{(41.1.24)}, \\ \eqref{(41.4.1)},\eqref{(41.4.6)} that \, 
nonzero  contributions to the "non-genuine" (i.e. non-factorizable)
adjoint in $U(N_F)$ bilinear condensates of heavy non-higgsed quarks
are induced in two stages. a) On the first "preliminary" operator
stage from the energy range $\lm/(\rm several)<\mu<(\rm several)
\lm$,  transforming the heavy quarks {\it operators} in left parts of
\eqref{(41.1.22)},\eqref{(41.1.24)} into {\it operators} of light $SU(\nd)$ 
quarks, this  is the operator expansion. b) Finally, at the second 
numerical stage at lower energies $\sim (\mx m)^{1/2}$, by the 
"genuine" (i.e.  coherent) factorizable condensate
$\langle\Qo\rangle_{\no}=\langle{\ov Q}^1_1\rangle
\langle Q^1_1\rangle\approx\,  \ha\mx$ \eqref{(41.1.14)} of
light higgsed quarks which break spontaneously $U(N_F)\ra U(\no)\times
U(\nt)$. \, Unfortunately, we cannot calculate directly the coefficient 'A'
originating from the scale $\mu\sim\lm$, but we can claim that it does \, 
not know  the number $\no$ appearing only at much lower scale $\mu
\sim  m\ll\lm$. Besides, we can find 'A' e.g. from \eqref{(41.1.23)} using
\eqref{(41.4.1)},\eqref{(41.4.6)},\eqref{(41.1.19)} for the right hand
part, and then predict $\langle\Qo\rangle_{\bb}$ in \eqref{(41.1.25)}.
As a result,
\bq
A= -2\,,\quad \langle\Qo\rangle_{\bb}\approx - \mx\,  \ha\Biggl
[1+\frac{\bb}{\nt-N_c}\Bigl (\frac{\ha}{\lm}\Bigr
)^{\frac{\bb}{\nt-N_c}} \Biggr]\,, \label{(41.1.26)}
\eq
this agrees with \eqref{(41.1.17)}, and $A=\,-2$ is independent of
$\no$ as it should be.\\

On the whole, the total decomposition of quark condensates $\langle
\QQ_{1,2}\rangle_{N_c}$ over their separate color parts look in
these br2 vacua as follows.\\
I) The condensate $\langle\Qo\rangle_{N_c}$.\\
a) From \eqref{(41.1.14)}, the (factorizable) condensate of higgsed
quarks in the $SU(\no)$ part
\bq
\langle\Qo\rangle_{\no}=\langle{\ov Q}^1_1\rangle\langle
Q^1_1\rangle\approx\mx \ha \Bigl [\, 1+\frac{\bb}{\nt-N_c}\Bigl
(\frac{\ha}{\lm}\Bigr )^{\frac{2N_c-N_F}{\nt-N_c}}\,\Bigr
]\,.\label{(41.1.27)}
\eq
b) The (non-factorizable) condensate $\langle\Qo\rangle_{\nd-\no}$ is
determined by the one-loop Konishi anomaly for the heavy not higgsed
quarks with the mass $m_2$ in the $SU(\nd-\no)$ SYM sector, see
\eqref{(41.1.4)},\eqref{(41.1.15)}
\bq
\langle\Qo\rangle_{\nd-\no}=\frac{\langle
S\rangle_{\nd-\no}}{m_2}\approx
\frac{\wmu\langle\Lambda^{SU(\nd-\no)}_{{\cal N}=2\,\,
SYM}\rangle^2}{m_2}\approx\mx \ha\Bigl (\frac{\ha}{\lm}\Bigr
)^{\frac{\bb}{\nt-N_c}}\,.\label{(41.1.28)}
\eq
c) From \eqref{(41.1.17)}, the (non-factorizable) condensate of
 heavy  non-higgsed quarks with masses $\sim\lm$
\bq
\langle\Qo\rangle_{\bb}\approx - \mx\, \ha\Biggl
[1+\frac{\bb}{\nt-N_c}\Bigl (\frac{\ha}{\lm}\Bigr
)^{\frac{\bb}{\nt-N_c}} \Biggr ]\,.\label{(41.1.29)}
\eq
Therefore, on the whole
\bq
\langle\Qo\rangle_{N_c}=\langle\Qo\rangle_{\no}+\langle
\Qo\rangle_{\nd-\no}+\langle\Qo\rangle_{\bb}\approx
\mx\, \ha\Bigl (\frac{\ha}{\lm}\Bigr
)^{\frac{\bb}{\nt-N_c}}\,,\,\,\quad\label{(41.1.30)}
\eq
as it should be, see \eqref{(41.1)}.\\
II) The condensate $\langle\Qt\rangle_{N_c}$.\\
a) From \eqref{(41.1.6)} the (factorizable) condensate of the
non-higgsed massless Nambu-Goldstone particles in the
 $SU(\no)$ part
\bq
\langle\Qt\rangle_{\no}=\sum_{a=1}^{\no}\langle{\ov
Q}^{\,a}_2\rangle\langle Q^2_{a}\rangle=0\,.\label{(41.1.31)}
\eq
b) The (non-factorizable) condensate $\langle\Qt\rangle_{\nd-\no}$ 
is  determined by the same one-loop Konishi anomaly for the heavy
non-higgsed quarks with the mass $m_2$ in the $SU(\nd-\no)$ SYM
sector,
\bq
\langle\Qt\rangle_{\nd-\no}=\frac{\langle
S\rangle_{\nd-\no}}{m_2}\approx\mx\,  \ha\Bigl (\frac{\ha}{\lm}\Bigr
)^{\frac{\bb}{\nt-N_c}}\,.\label{(41.1.32)}
\eq
c) From \eqref{(41.1.19)}, the (non-factorizable) condensate of \,
heavy  non-higgsed quarks with masses $\sim\lm$
\bq
\langle\Qt\rangle_{\bb}\approx \mx\,  \ha\Biggl
[1+\frac{\bb}{\nt-N_c}\Bigl (\frac{\ha}{\lm}\Bigr
)^{\frac{\bb}{\nt-N_c}} \Biggr ]\,.\label{(41.1.33)}
\eq
Therefore, on the whole
\bq
\langle\Qt\rangle_{N_c}=\langle\Qt\rangle_{\no}+\langle\Qt\rangle
_{\nd-\no}+\langle\Qt\rangle_{\bb}\approx
\mx\,  \ha \Biggl [1+\frac{N_c-\no}{\nt-N_c}\Bigl (\frac{\ha}{\lm}\Bigr
)^{\frac{\bb}{\nt-N_c}} \Biggr ],\,\,\,\quad \label{(41.1.34)}
\eq
as it should be, see \eqref{(41.1)}.\\

And finally, remind that the mass spectra in these br2 vacua depend
essentially on the value of $m/\lm$ and all these vacua with $\nt>N_c$ \, 
evolve at  $m\gg\lm$ to br1 vacua of section 4.3 below, see e.g.
section 4 in \cite{ch19}.

\numberwithin{equation}{subsection}
\subsection{$U(N_c)$\,,\,\,smallest $\mx$}

$U(N_c)$ theory is obtained by adding one $SU(N_c)$ singlet
$U^{(0)}(1)$ ${\cal N}=2$ multiplet, with its scalar field $\sqrt{2}
X^{(0)}=a_0 I$, where $I$ is the unit $N_c\times N_c$ matrix. 
Instead of \eqref{(40.1)}, the superpotential looks now as
\bq
{\cal W}_{\rm matter}=\frac{\mu_{0}}{2} N_c a^2_0+\mx{\rm
Tr}\,(X^{\rm  adj}_{SU(N_c)})^2 +{\rm Tr}\,\Bigl [ (m-a_0)\,{\ov Q} Q-{\ov
Q}\sqrt{2} X^{\rm adj}_{SU(N_c)} Q \Bigr ]_{N_c}\,,\quad
\mu_0=\mx\,,\label{(41.2.1)}
\eq
and we consider in this paper only the case with $\mu_0=\mx$. The only \,
change in  the Konishi anomalies \eqref{(40.4)} and in the value of
$\langle\Qt\rangle_{N_c}$ in \eqref{(41.1)} will be that now
\bq
\langle \Qo+\Qt\rangle_{N_c}=\mx\, m\,,\quad
\langle\Qt\rangle_{N_c}\approx \mx\, m\gg
\langle\Qo\rangle_{N_c}\approx\mx\, m\Bigl (\frac{m}{\lm}
\Bigr)^{\frac{\bb}{{\rm n}_2-N_c}},  \label{(41.2.2)}
\eq
\bbq
\langle
S\rangle_{N_c}=\frac{\langle\Qo\rangle_{N_c}\langle\Qt\rangle_{N_c}}
{\mx}\approx\mx\, 
m^2\Bigl (\frac{m}{\lm}\Bigr)^{\frac{\bb}{{\rm n}_2-N_c}},
\eeq
while in br2 vacua the small ratio $\langle
\Qo\rangle_{N_c}/\langle\Qt\rangle_{N_c}\ll 1$ and small $\langle
S\rangle_{N_c}\ll\mx\, m^2$ in \eqref{(41.1)} remain parametrically the
same. Besides, from \eqref{(41.2.1)},\eqref{(41.2.2)} (neglecting power
corrections)
\bq
\langle a_0\rangle=\frac{\langle\,{\rm
Tr}\,(\QQ)\rangle_{N_c}}{N_c\mx}=\frac{\no
\langle\Qo\rangle_{N_c}+\nt\langle\Qt\rangle_{N_c}}{N_c
\mx}\approx\frac{\nt}{N_c}\,m,\,\,
\langle m-a_0\rangle\approx
-\,\frac{\nd-\no}{N_c}\,m.\,\,\,\,\label{(41.2.3)}
\eq

The changes in \eqref{(41.1.2)} and \eqref{(41.1.5)} are also very
simple\,: a)\, the term "$\mx N_c a^2_{0}/2$"\, is added,\,\, b)
"$m$"\, is replaced by "$m-a_0$" in all other terms. So, instead of
\eqref{(41.1.5)},\eqref{(41.1.6)}, we have now
\bq
\w^{\,\rm low}_{\rm tot}=\w^{\,(SYM)}_{SU(\nd-\no)}+\w^{\,\rm
low}_{\rm matter}+\dots\,,\quad
\w^{\,\rm low}_{\rm matter}=\w_{SU(\no)}+\w_{D}+\w_{a}\,,
\label{(41.2.4)}
\eq
\bbq
\w^{\,(SYM)}_{SU(\nd-\no)}=(\nd-\no)\,\wmu\Bigl
(\Lambda^{SU(\nd-\no)}_{{\cal N}=2\,\, SYM}\Bigr
)^2+\w^{\,(M)}_{SU(\nd-\no)}\,,\quad\langle
\Lambda^{SU(\nd-\no)}_{{\cal N}=2\,\,SYM}\rangle^2
\approx m^2\Bigl (\frac{m}{\lm}\Bigr )^{\frac{2N_c-N_F}{\nt-N_c}}\,,
\eeq
\bbq
\w_{SU(\no)}=(m-a_0-a_1-a_2)\,{\rm Tr}\,({\ov Q} Q)_{\no}-\,{\rm
Tr}\,({\ov Q}\sqrt{2} X^{\rm adj}_{SU(\no )} Q)_{\no}+\wmu{\rm Tr}
\,(X^{\rm adj}_{SU(\no)})^2\,,
\eeq
\bbq
\w_{D}=(m-a_0-c_1 a_1)\sum_{j=1}^{\bb}{\ov D}_j D_j-\sum_{j=1}^{\bb}
a_{D,j}\,{\ov D}_j D_j\,-\,\mx\lm\sum_{j=1}^{\bb}
\omega^{j-1}\,a_{D,j}+\mx L \sum_{j=1}^{\bb}  a_{D,j}\,,
\eeq
\bbq
\w_{a}=\frac{\mx}{2}\Biggl [N_c a^2_0+(1+\delta_1)\frac{\nd N_c}{\bt}
a_1^2+(1+\delta_2)\frac{\no \nd}{\nd-\no} a_2^2+2 N_c \delta_3
a_1(m-a_0-c_1 a_1)+2 N_c \delta_4(m-a_0-c_1 a_1)^2\Biggr ].
\eeq

From \eqref{(41.2.3)},\eqref{(41.2.4)} (the leading terms only)
\bq
\langle a_1\rangle= -\,\frac{\bt}{\nd}\langle m-a_0\rangle\approx
\frac{\bt(\nd-\no)}{\nd N_c}\,m\,,\quad \langle a_2\rangle=\langle
m-a_0-a_1\rangle\approx -\,\frac{\nd-\no}{\nd}\,m\,, \label{(41.2.5)}
\eq
while, according to reasonings given in section 41.1, all $\delta_i$
should remain the same. Of course, this can be checked directly.
Instead of \eqref{(41.1.8)} we have now for the leading terms
\bq
\langle\w^{\,\rm low}_{\rm tot}\rangle\approx\langle\w^{\,\rm
low}_{\rm matter}\rangle\approx\frac{\mx}{2}\Biggl [\,N_c\langle
a_0\rangle^2+\frac{\nd N_c}{\bt}(1+\delta_1)\langle
a_1\rangle^2+\frac{{\rm n}_1\nd}{\nd-{\rm n}_1}(1+\delta_2)\langle
a_2\rangle^2\,\Biggr ]\,, \label{(41.2.6)}
\eq
where $\langle a_i\rangle$ are given in \,\eqref{(41.2.3)},\eqref{(41.2.5)}. 
Instead of \, \eqref{(41.1.9)},\eqref{(41.1.10)} 
we have now (with the same accuracy)
\bq
\w^{\,\rm eff}_{\rm tot}(\Pi)=m\,{\rm Tr}\,({\ov Q}
Q)_{N_c}-\frac{1}{2\mx}\Biggl [ \,\sum_{i,j=1}^{N_F} ({\ov Q}_j
Q^i)_{N_c}({\ov Q}_{\,i} Q^j)_{N_c}\Biggr ]-\nd S_{N_c}\ra
\langle\w^{\,\rm eff}_{\rm tot}\rangle\approx\frac{1}{2}\,\nt\,\mx
m^2.\,\,\,\label{(41.2.7)}
\eq
It is not difficult to check that values of $\delta_{1,2}$ obtained
from \eqref{(41.2.6)}=\eqref{(41.2.7)} are the same as in
~\eqref{(41.1.12)}.~
\footnote{\, Besides, one can check that \eqref{(41.2.4)} with all
$\delta_{i}=0,\,i=1...4$, will result in the wrong value:\, $\langle
a_0\rangle\approx [\,m\,(2 N_c-\nt)/N_c\,]$, this differs from
\eqref{(41.2.3)}.  \label{(f53)}
}

Therefore, instead of \eqref{(41.1.13)}, we obtain now finally for the
leading term, see \eqref{(41.2.4)},\eqref{(41.2.5)},
\bbq
\langle{\rm Tr}\,({\ov Q}
Q)\rangle_{\no}=\no\langle\Qo\rangle_{\no}+\nt\langle\Qt\rangle
_{\no}=\no\langle\Qo\rangle_{\no}\approx
\wmu \,\frac{{\rm
n}_1\nd}{\nd-\no}\langle a_2\rangle\approx\no \mx m\,,
\eeq
\bq
\langle\Qo\rangle^{U(N_c)}_{\no}=\sum_{a=1}^{\no}\langle{\ov
Q}_1^{\,a}Q^1_a\rangle=\langle{\ov Q}^1_1\rangle\langle
Q^1_1\rangle\approx\mx m\approx
\langle\Qt\rangle^{U(N_c)}_{N_c},\,\,
\langle\Qt\rangle_{\no}=0,\,\, \langle S\rangle_{\no}=0,
\,\,\,\,\,\label{(41.2.8)}
\eq
while as before in $SU(N_c)$
\bq
\langle{\ov D}_j\rangle\langle D_j\rangle= -\mx\lm
\omega^{j-1}+O(\mx \, m)\,. \label{(41.2.9)}
\eq

It is worth to remind that the number of double roots of the curve
\eqref{(40.2)} in this case is still $N_c-1$, as two single roots with
$e^{+}=-e^{-}\approx 2\langle\Lambda^{SU(\nd-{\rm n}_1)}_{{\cal
N}=2\,\, SYM}\rangle$ still originate here from $SU(\nd-\no)\,\, {\cal  N}=2$
SYM. \, In other words, the number of charged particles massless at \, 
$\mx\ra 0$  (and  higgsed at small $\mx\neq 0$) remains $N_c-1$ as in
the $SU(N_c)$ theory. But now, in comparison with the $SU(N_c)$
theory, one extra $U^{(0)}(1)$ gauge multiplet is added. As a result,
in addition to $2\no\nt$ massless Nambu-Goldstone multiplets, there
remains now one exactly massless ${\cal N}=1$ photon multiplet, while
the corresponding scalar ${\cal N}=1$ multiplet has now smallest
nonzero mass $\sim\mx$ due to breaking ${\cal N}=2\ra {\cal N}=1$.

In so far as we know from \eqref{(41.1.5)},\eqref{(41.2.4)} the charges
and numbers of all particles massless at $\mx\ra 0$, we can establish
the definite correspondence with the roots of the curve \, \eqref{(40.2)}:\, 
$2N_c-N_F$ \, unequal double roots $e_j\sim\lm,\, j=1...2N_c-N_F$
correspond to our dyons $D_j$,\, $\nd-\no-1$ unequal double roots
$e_i\sim\langle\Lambda^{SU(\nd-{\rm n}_1)}_{{\cal N}=2\,\,
SYM}\rangle,\, i=1...N_c-\no-1$ correspond to pure magnetic monopoles
from $SU(\nd-\no)$ SYM, and $\no$ equal double roots $e_k\sim m,\,
k=1...\no$ correspond to $\no$ higgsed original pure electric quarks
from $SU(\no)$. Besides, we know from \eqref{(41.2.8)},\eqref{(41.2.9)}
the leading terms of their condensates (see \eqref{(41.1.6)} for the
monopole condensates). Then, as a check, we can compare the values of
condensates in \eqref{(41.2.8)},\eqref{(41.2.9)} with the formulas
proposed in \cite{SY5} where these condensates are expressed through
the roots of the $U(N_c)$ curve \eqref{(40.2)}.
\footnote{\,
Remind that these formulas were derived in \cite{SY5} from and checked \, 
then on a \, few simplest examples only, and proposed after this to be
universal.

Besides, in general, the main problem with the use of the expressions
like \eqref{(41.2.10)} from \cite{SY5} for finding the condensates of
definite light BPS particles (massless at $\mx\ra 0$) is that one
needs to find first the values of all roots of the curve
\eqref{(40.2)}. But for this, in practice, one has to understand, at
least, the main properties of the color and flavor symmetry breaking
in different vacua, the corresponding mass hierarchies, and
multiplicities of each type roots. \label{(f54)}
}

Specifically, these roots look in these $U(N_c)$ br2 vacua as:\,  1) the
quark double roots $e^{(Q)}_k= - m,\,\, k=1...\no$,\,\, 2) the dyon
double roots $e^{(D)}_j
\approx \omega^{j-1}\lm,\,\,j=1...(2N_c-N_F)$,\,\, 3) two single
roots \, \cite{DS,CIV,CSW} which we know originate from $SU(\nd-\no)\,\,{\cal
N}=2$ SYM,\,\, $e^{+}=-e^{-}\approx 2\langle\Lambda^{SU(\nd-{\rm
n}_1)}_{{\cal N}=2\,\, SYM}\rangle$. Therefore, with this knowledge,
the formulas from \cite{SY5} look as
\bq
\langle{\ov Q}_k Q^k\rangle^{U(N_c)}_{\no}= -
\mx\sqrt{(e^{(Q)}_k-e^+)(e^{(Q)}_k-e^-)}\approx -\mx
e^{(Q)}_k\approx\mx m\,,\quad k=1...\no, \label{(41.2.10)}
\eq
\bbq
\langle{\ov D}_j D_j\rangle= -
\mx\sqrt{(e^{(D)}_j-e^+)(e^{(D)}_j-e^-)}\approx -\mx e^{(D)}
_j\approx \, -\mx\omega^{j-1}\lm\,,\quad j=1...2N_c-N_F\,.
\eeq
( Both $e^{\pm}$ are neglected here because they are non-leading). It
is seen that the leading terms of \eqref{(41.2.8)},\eqref{(41.2.9)} and
\eqref{(41.2.10)} agree, this is non-trivial as the results were
obtained very different methods. The calculations of power corrections \, 
to  \eqref{(41.2.8)},\eqref{(41.2.9)}, and \eqref{(41.2.10)} are presented \, 
in important \, Appendices A and~ B.

\subsection{$SU(N_c)$\,,\,\,larger
$\mx\,,\,\,\,\langle\Lambda^{SU(\nd-\no)}_{{\cal N}=2\,\,
SYM}\rangle\ll\mx\ll\, m$}

Significant changes for such values of $\mx$ occurs only in the ${\cal
N}=2\,\,SU(\nd-\no)$ SYM sector. All $SU(\nd-\no)$ adjoint scalars
$X^{adj}_{SU(\nd-\no)}$ are {\it too heavy and too short ranged} now,
their physical (i.e. path dependent) $SU(\nd-\no)$ phases induced by
interactions with the lighter $SU(\nd-\no)$ gluons fluctuate freely at \, 
all  scales \, $\mu\gtrsim\langle\Lambda^{SU(\nd-\no)}_{{\cal
N}=1\,\,SYM}\rangle$ in this case, and they are not higgsed, i.e.
$\langle X^{adj}_{SU(\nd-\no)}\rangle_{\mx^{\rm pole}/(\rm
several)}=0$. Instead, they all decouple as heavy already at the
scale \, $\langle\Lambda^{SU(\nd-\no)}_{{\cal
N}=1\,\,SYM}\rangle\ll\mu<[\mx^{\rm pole}={\it g}^2\mx]/(\rm
several)\ll m$ {\it in the weak coupling region} and do not affect
then by themselves the lower energy dynamics. There remains ${\cal
N}=1\,\,SU(\nd-\no)$ SYM with the scale factor of its gauge coupling
$\langle\Lambda^{SU(\nd-\no)}_{{\cal N}=1\, SYM}\rangle=[\,\wmu
\langle\Lambda^{SU(\nd-\no)}_{{\cal
N}=2\,\,SYM}\rangle^2\,]^{1/3},\\\langle\Lambda^
{SU(\nd-\no)}_{{\cal  N}=2\,\, SYM}\rangle\ll
\langle\Lambda^{SU(\nd-\no)}_{{\cal N}=1\,\, SYM}\rangle
\ll\mx\ll~ m$, \, see  \eqref{(41.1.4)},\eqref{(41.1.15)}. The small nonzero
(non-factorizable) value $\langle {\,\rm
Tr\,}(X^{adj}_{SU(\nd-\no)})^2\rangle=(\nd-\no)\langle
S\rangle_{\nd-\no}/\wmu=(\nd-\no)\langle
\Lambda^{SU(\nd-\no)}_{{\cal  N}=2\,\, SYM}\rangle^2 \ll
\langle\Lambda^{SU(\nd-\no)}_{{\cal N}=1\,\, SYM}\rangle^2$ arises
here not because $X^{adj}_{SU(\nd-\no)}$ are higgsed, but only due to
the Konishi anomaly. I.e., on the first "preliminary" operator stage,
from one-loop diagrams with heavy scalars $X^{adj}_{SU(\nd-\no)}$ and
their fermionic superpartners with masses $\sim\mx$ inside,
transforming (in the weak coupling regime at the scale $\sim\mx$) this\, 
\, {\it  operator} ${\,\rm Tr\,}(X^{adj}_{SU(\nd-\no)})^2$ of heavy
scalars into the bilinear {\it operator} $\sim ({\rm
Tr\,}\lambda\lambda)_{\nd-\no}/\mx$ of lighter gluinos of ${\cal
N}=1\,\, SU(\nd-\no)$ SYM. But the mean vacuum value of this latter
originates and saturates only in the strong coupling (and
non-perturbative) region at much smaller energies
$\mu\sim\langle\Lambda^{SU(\nd-\no)}_{{\cal N}=1\,\,
SYM}\rangle\ll\mx$.

The multiplicity of vacua of this ${\cal N}=1 \,\, SU(\nd-\no)$ SYM is \, 
also  $\nd-\no$, as it should be.. There appears now a large number 
of  strongly coupled ${\cal N}=1$ gluonia with the mass scale
$\sim\langle\Lambda^{SU(\nd-\no)}_{{\cal N}=1\,\,SYM}\rangle$. All
heavier $SU(\nd-\no)$ charged original pure electric quarks and
hybrids with masses $\sim m$ and scalars $X^{adj}_{SU(\nd-\no)}$ with
masses $\sim\mx\gg\langle\Lambda^{SU(\nd-\no)}_{{\cal
N}=1\,\,SYM}\rangle$ are still weakly confined, but the tension of the \, 
confining  string is larger now,
$\sigma^{1/2}_{SU(\nd-\no)}\sim\langle\Lambda^{SU(\nd-\no)}_{{\cal
N}=1\,\,SYM}\rangle\ll\mx\ll m$.

At the same time, in this range $\langle\Lambda^{SU(\nd-\no)}_{{\cal
N}=1\,\,SYM}\rangle\ll\mx\ll m$, nothing has happened yet with
$\langle a_2\rangle\sim m$ {\it still higgsed} at the scale $\mu\sim
m\gg\mx$, resulting in $SU(\nd)\ra SU(\no)\times U^{(2)}(1)\times
SU(\nd-\no)$, and in the $SU(\no)$ sector with its quarks higgsed at
$\mu\sim (\mx m)^{1/2}\gg\mx$.

\subsection{$SU(N_c)$\,,\,\, even larger $\mx\,,\,\,\,m\ll\mx\ll\lm$}

\hspace*{4mm} The case with $m\ll\mx\ll\lm$ and $N_c+1<N_F<3N_c/2$ is
described in section 8.1 of \cite{ch13}. Therefore, we consider here in \, 
addition  the region $3N_c/2<N_F<2N_c-1$.

Remind, that the whole dyonic $\bt=2N_c-N_F$ sector, including the
$U^{(1)}(1)\,\, {\cal N}=2$ photon multiplet with its scalar partner
$a_1$, acquires masses $\sim (\mx\lm)^{1/2}\gg\mx$ and decouples at
lower energies. But finally, the trace of the heavier scalar $a_1$
remains in the lower energy ${\cal W}_{SU(\nd)}$ superpotential at
scales $\mu<(\mx\lm)^{1/2},\,\, \mx\ll(\mx\lm)^{1/2}\ll\lm$, in the
form: $\,m\ra{\wt m}= (m-\langle a_1\rangle)=m N_c/\nd\,$. 
Remind also \, that, see  \eqref{(41.1.2)},\eqref{(41.1.12)},
$\mx\ra\wmu=(1+\delta_2)\mx= - \mx$.

Therefore, after integrating out all particles with masses $\sim
(\mx\lm)^{1/2}$, the lower energy superpotential at the scale $\mu=
(\mx\lm)^{1/2}/{(\rm several)}\gg\mx$ looks as, see also
\eqref{(41.4.5)},\eqref{(41.4.6)},
\bq
\w^{({\cal N}=2,\,\nd)}_{\rm matter}={\rm Tr}\,\Bigl [\,{\ov
Q}\,(\tm-\sqrt{2}X^{\rm adj}_{SU(\nd)})\,Q\,
\Bigr ]_{\nd}+\wmu{\rm Tr}\,(X^{\rm adj}_{SU(\nd)})^2,\,\,
\tm=\frac{N_c}{\nd}m,\,\, \wmu=-\mx\,,\label{(41.4.1)}
\eq
\bbq
\langle\Qo\rangle_{\nd}= [\,\frac{\nd\,\wmu \tm}{\nd-\no}=\wmu
m_2=\mx\,\ha ]+\frac{N_c-\no}{\nd-\no}\langle\Qt\rangle_{\nd}
\approx  \mx \,\ha   \approx\langle\Qt\rangle_{N_c}\,,
\eeq
\bbq
\langle\Qt\rangle_{\nd}\approx\mx \ha\Bigl
(\frac{\ha}{\lm}\Bigr)^{\frac{\bb}{{\rm
n}_2-N_c}}\approx\langle\Qo\rangle_{N_c}\,,\quad m_2= -
\ha=\frac{N_c}{\nt-N_c}\,m\,,
\eeq
\bbq
\langle{\rm Tr}(\sqrt{2}X^{\rm
adj}_{SU(\nd)})^2\rangle=\frac{1}{\wmu}\Bigl [(2\nd-N_F)\langle
S\rangle_{\nd}+\tm\langle{\rm Tr}({\ov Q} Q)\rangle_{\nd}\Bigr
]\approx \frac{\no\nd}{\nd-\no}\tm^2\,.
\eeq

All $\nd^{\,2}-1\,\,X^{\rm adj}_{SU(\nd)}$ {\it are not higgsed and
decouple as heavy} at scales $\mu<\mx^{\rm pole}/(\rm several)={\it
g}^2(\mu=\mx^{\rm pole})\mx/(\rm several)$ {\it in the weak coupling
region}, and can be integrated out. Therefore, the ${\cal N}=1\,\,
SU(\nd)$ SQCD with $N_F$ flavors of light quarks with masses $\ll\mx$
emerges already at the scale $\mu=\mx^{\rm pole}/(\rm several)$. 
The  scale factor $\lt$ of its gauge coupling is, see \eqref{(41.1.15)},
\bq
(\,\lt\,)^{3\nd-N_F}=\Bigl [\Lambda_{SU(\nd)}= -\lm\Bigr
]^{2\nd-N_F}\,\wmu^{\,\nd}\,,\quad \frac{\lt}{\wmu}=\Bigl
(\frac{\mx}{\lm}\Bigr )^{\frac{2N_c-N_F}{2N_F-3N_c}}\ll 1
\,.\label{(41.4.2)}
\eq

This ${\cal N}=1$ theory is not IR free at $3N_c/2<N_F<2N_c-1$ and so
its logarithmically small coupling $g^2(\mu\sim\mx)$ begins to grow
logarithmically at $\mu<\mx$. There is a number of variants of the
mass spectrum. To deal with them it will be convenient to introduce
colorless but flavored auxiliary (i.e. sufficiently heavy, with
masses \, $\sim\mx$ \, and dynamically irrelevant at $\mu<\mx$) fields 
$\Phi_i^j$  \cite{ch19}, so that the Lagrangian at the scale 
$\mu=\mx^{\rm  pole}/(\rm several)$ instead of
\bq
K={\rm Tr}\,(Q^\dagger Q+{\ov Q}^\dagger {\,\ov Q}),\,\, \w^{\,({\cal
N}=1,\,\nd)}_{\rm matter}=\tm\, {\rm Tr}\,({\ov Q}
Q)_{\nd}-\frac{1}{2\wmu}\Biggl ( {\rm Tr}\,({\ov Q}
Q)_{\nd}^2-\frac{1}{\nd}\Bigl ({\rm Tr}({\ov Q} Q)_{\nd}\Bigr )^2
\Biggr ) \label{(41.4.3)}
\eq
can be rewritten as
\footnote{\,
all factors of the ${\cal N}=1$ RG evolution in Kahler terms are
ignored below for simplicity if they are logarithmic only, as well as
factors of $g$ if $g$ is either constant or logarithmically small
\label{(f8)}
}
\bq
K={\rm Tr}\,(\Phi^\dagger \Phi)+{\rm Tr}\,(Q^\dagger Q+
{\ov Q}^\dagger{\,\ov  Q})\,,\quad
\w^{\,({\cal N}=1,\,\nd)}_{\rm matter}=\w_{\Phi}+{\rm Tr}\,
\,\Bigl ({\ov  Q}\,\tm^{\rm tot} Q \Bigr )\,,\label{(41.4.4)}
\eq
\bbq
\w_{\Phi}=\frac{\wmu}{2}  \Bigl ( {\rm Tr}\,(\Phi^2)-\frac{1}{N_c}\, 
({\,Tr}\,\Phi)^2 \Bigr )\,,\quad {\tm}
=\frac{N_c}{\nd}\,m\,,\quad (\tm^{\rm tot})^j_i=
\tm\,\delta^j_i -\Phi^j_i\,,\quad \wmu= - \mx\,,
\eeq
and {\it its further evolution at lower energies is determined now by
the dynamics of the genuinely ${\cal N}=1$ theory \eqref{(41.4.4)}}.

The Konishi anomalies for \eqref{(41.4.4)} look as, see
\eqref{(41.4.2)}, compare with \eqref{(40.3)},\eqref{(40.4)},
\bq
\langle\Qo+\Qt-\frac{1}{\nd}{\rm Tr}\,\QQ\rangle_{\nd}=\wmu
\tm\,,\label{(41.4.5)}
\eq
\bbq
\langle
S\rangle_{\nd}=\frac{\langle\Qo\rangle_{\nd}\langle\Qt\rangle
_{\nd}}{\wmu}=\Biggl(\frac{\langle  \det\QQ\rangle_{\nd}}{\lt^
{3\nd-N_F}}\Biggr)^{1/N_c}=\Biggl(\,\frac{[\langle\Qo\rangle_
{\,\nd}]^{\no}[\langle\Qt\rangle_{\,\nd}]^{\nt}}{\Lambda_
{SU(\nd)}^{2\nd-N_F}\,\wmu^{\,\nd}}\,\Biggr )^{1/N_c}\,,
\eeq
\bbq
\langle\tm^{\rm
tot}_1=\tm-\Phi^1_1\rangle=\frac{\langle\Qt\rangle_{\nd}}
{\wmu}\,,\quad  \langle\tm^{\rm
tot}_2=\tm-\Phi^2_2\rangle=\frac{\langle\Qo\rangle_{\nd}}
{\wmu}\,.
\eeq
\bbq
\langle\Phi^i_j\rangle=\frac{1}{\wmu}\Biggl [\langle{\ov Q}_{j}
Q^i\rangle_{\nd}-\delta^i_j\,\frac{1}{\nd}\langle{\rm
Tr\,}\QQ\rangle_{\nd} \,\Biggr ]\,.
\eeq

From \eqref{(41.4.5)}, see \eqref{(41.4.1)},\eqref{(41.1.15)} and section \,
8.1.1 in \,  \cite{ch13} (neglecting all power corrections for simplicity),
\bq
\langle\Qo\rangle_{\nd}= [\,\frac{\nd\,\wmu \tm}{\nd-\no}=\wmu
m_2= \mx \ha]+\frac{N_c-\no}{\nd-\no}
\langle\Qt\rangle_{\nd}\approx\mx
\ha\approx\langle\Qt\rangle_{N_c}\,,\label{(41.4.6)}
\eq
\bbq
\langle\Qt\rangle_{\nd}\approx\wmu m_2\Bigl
(\frac{m_2}{\Lambda_{SU(\nd)}}\Bigr)^{\frac{\bb}{\nt-N_c}}
\approx\mx\ha\Bigl (\frac{\ha}{\lm}\Bigr)^{\frac{\bb}{{\rm
n}_2-N_c}}\approx\langle\Qo\rangle_{N_c}\,,
\quad \Lambda_{SU(\nd)}= -\lm\,,
\eeq
\bbq
\langle S\rangle_{\nd}=\langle S\rangle_{\,SU(\nd-\no)}
^{\,(SYM)}=  -\langle S\rangle_{N_c}\,,\quad
m_2=\frac{N_c}{\nt-N_c}\,m= -\ha\,.
\eeq

Besides, it is not difficult to check that accounting for both the
leading terms $\sim \mx m^2$ and the leading power corrections 
$\sim \langle S\rangle_{N_c}= -\langle S\rangle_{\nd}$, see
\eqref{(41.4.3)}-\eqref{(41.4.6)},\,
\eqref{(41.1)},\eqref{(41.1.9)},\eqref{(41.1.12)}
\bq
\Bigl [\langle\w^{\,({\cal N}=1,\,\nd)}_{\rm
tot}\rangle=\langle\w^{\,({\cal N}=1,\,\nd)}_{\rm
matter}\rangle-N_c\langle S\rangle_{\nd}\Bigr ]+\Bigl
[\langle\w_{a_1}\rangle=\frac{\mx}{2}(1+\delta_1)\frac{\nd
N_c}{\bt}\langle a_1\rangle^2\Bigr ]=\langle \w^{\,\rm eff}_{\rm
tot}\rangle\,,\label{41.4.7)}
\eq
as it should be.\\

{\bf A)} If $m$ is not too small so that $(\mx m)^{1/2}\gg\lt$, see
\eqref{(41.4.2)}. Then $\no<\nd$ quarks are higgsed still {\it in the
weak coupling regime} at the scale $\mu\sim(\mx m)^{1/2}\gg\lt,\,\,
SU(\nd)\ra SU(\nd-\no)$, and $\no(2\nd-\no)$ gluons acquire masses
$\mu_{\rm gl,1}\sim (\mx m)^{1/2}$. There remains at lower energies
${\cal N}=1$ SQCD with the unbroken $SU(\nd-\no)$ gauge group,
$\nt>N_c$ flavors of still active lighter quarks $Q^2,\,{\ov Q}_2$
with $SU(\nd-\no)$ colors, $\no^2$ pions $\Pi^1_{1^\prime}\ra
\sum_{a=1}^{\nd}({\ov Q}^{\,a}_{i^\prime}
Q^i_a),\,\,i^\prime,i=1,\,...,\,\no,\,\,
\langle\Pi^1_1\rangle=\langle\Qo\rangle_{\nd}\sim \mx m$, and 
$2\no\nt$ hybrids $\Pi^2_1,\,\Pi^1_2$(these are in essence the 
quarks $Q^2,\,{\ov Q}_2$ with  broken
colors). The scale factor $\widehat\Lambda$ of the gauge 
coupling is,  see \eqref{(41.4.2)},
\bq
{\wh\Lambda}^{\,\wh{\rm b}_{\rm o}}=\frac{\lt^{3\nd-N_F}}{\det
\Pi^1_1}\,,\quad
\Bigl (\frac{{\langle\widehat\Lambda}\rangle^{2}}{\mx m}\Bigr
)^{\wh{\rm b}_{\rm o}}\sim\Bigl (\frac{\lt^{2}}{\mx m} \Bigr
)^{2N_F-3N_c}\ll 1\,.\,\quad \widehat{\rm b}_{\rm o}=
3(\nd-\no)-\nt\,.\label{(41.4.8)}
\eq

{\bf a1)} If $\wh{\rm b}_{\rm o}<0$. The $SU(\nd-\no)$ theory at
$\mu<\mu_{\rm gl,1}$ is then IR free, its coupling is small and still
decreases logarithmically with diminishing energy. The quarks
$Q^2,\,{\ov Q}_2$ with $\nd-\no$ still active colors and $\nt>N_c$
flavors have masses $m^{\rm pole}_{Q,2}\sim m$ and decouple as 
heavy  at $\mu<\,m$. Notice that {\it all this occurs in the weak 
coupling  regime} in this case, so that {\it there is no need to use 
the proposed dynamical  scenario from} \cite{ch3}.

There remains at lower energies ${\cal N}=1\,\, SU(\nd-\no)$ 
SYM with  the scale factor of its coupling
\bq
\Bigl (\Lambda^{SU(\nd-\no)}_{{\cal N}=1\,\,SYM}\Bigr )^{3(\nd-\no)}
=\frac{{\lt}^{\,3\nd-N_F}\,\det\tm^{\rm tot}_2}{\det \Pi^1_1},\quad
(\tm^{\rm
tot}_2)^{2^\prime}_2=\tm\,\delta^{2^\prime}_2-\Phi^{2^\prime}_2\,,
\,\, \frac{\langle\Lambda^{SU(\nd-\no)}_{{\cal
N}=1\,\,SYM}\rangle}{m}\ll 1.\label{(41.4.9)}
\eq

After integrating this ${\cal N}=1$ SYM via the VY-procedure
\cite{VY}, the Lagrangian looks as, see
\eqref{(41.1.15)},\eqref{(41.4.9)},\, \eqref{(41.4.2)},
\bbq
K={\rm Tr}\,(\Phi^\dagger\Phi)+ {\rm Tr}\,\Biggl
[2\sqrt{(\Pi^1_1)^{\dagger}\Pi^1_1}+\Pi_2^1\frac{1}
{\sqrt{(\Pi^1_1)^{\dagger}\Pi^1_1}}\,(\Pi_2^1)^{\dagger}+
(\Pi_1^2)^{\dagger}\frac{1}{\sqrt{
(\Pi^1_1)^{\dagger}\Pi^1_1}}\,\Pi_1^2 \Biggr ]\,,
\eeq
\bq
\w^{\,(\nd)}=\w_{\rm non-pert}+\w_{\Phi}+{\rm Tr}\,\Bigl 
(\tm^{\rm \,tot}_1\,\Pi^1_1\Bigr )+\w_{\rm hybr}\,, 
\label{(41.4.10)}
\eq
\bbq
\w_{\rm non-pert}=(\nd-\no)\Bigl (\Lambda^{SU(\nd-\no)}_{{\cal
N}=1\,\,SYM}\Bigr )^3= (\nd-\no)\Biggl  [\frac{(\Lambda_{SU(\nd)}=
\,- \lm)^{2\nd-N_F}\wmu^{\,\nd}\,\det\tm^{\rm
tot}_2}{\det \Pi^1_1}\Biggr ]^{1/(\nd-\no)}\,,
\eeq
\bbq
\w_{\rm hybr}={\rm Tr}\,\Bigl (\tm^{\rm
tot}_2\,\Pi_2^1\frac{1}{\Pi^1_1}\Pi_1^2-\Phi_1^2
\Pi_2^1-  \Phi_2^1\Pi_1^2\Bigr )\,.
\eeq

The masses of $\no^{\,2}$ pions $\Pi^1_1$ and hybrids from
\eqref{(41.4.10)} are, see \eqref{(41.4.4)} for $\w_{\Phi}$,
\bq
\mu(\Pi^1_1)\sim\frac{\langle\Pi^1_1\rangle=\langle\Qo
\rangle_{\nd}}{\mx}\sim  m\sim m^{\rm
pole}_{Q,2}\,,\quad\langle\Pi^1_{1^\prime}\rangle=\langle({\ov
Q}_{1^\prime} Q^1)\rangle_{\nd}\approx - \delta^1_{1^\prime}
\frac{N_c}{\nd-\no}\,\mx m\,,\label{(41.4.11)}
\eq
\bbq
\mu(\Pi^1_2)=\mu(\Pi^2_1)=0\,,
\eeq
(the main contribution to $\mu(\Pi^1_1)$ gives the term $\sim
(\Pi^1_1)^2/\mx$ originating from ${\rm Tr}\, (\tm^{\rm
tot}_1\,\Pi^1_1 )$ in \eqref{(41.4.10)} after integrating out 
heavier  $\Phi^1_1$ with masses $\sim\mx\gg m$ ).\\

On the whole, the mass spectrum of this ${\cal N}=1\,\,SU(\nd)$ 
theory at $\mx\neq 0$ looks in  this case as follows.\\
1) There are $\no(2\nd-\no)\,\,{\cal N}=1$ multiplets of massive
gluons, $\mu_{\rm gl,1}\sim (\mx m)^{1/2}$.\\
2) A large number of hadrons made from non-relativistic and weakly
confined ${\ov Q}_2, Q^2$ quarks with $\nt>N_c$ flavors and $\nd-\no$
colors, their mass scale is $\mu_H\sim m,\,\,
\Lambda^{SU(\nd-\no)}_{{\cal N}=1\,\,SYM}\ll \mu_H\ll\mu_{\rm gl,1}$
(the tension of the confining string is
$\sigma^{1/2}\sim\Lambda^{SU(\nd-\no)}_{{\cal N}=1\,\,SYM}\ll m$).\\
3) $\no^2$ pions $\Pi^1_1$ with masses $\sim m$.\\
4) A large number of ${\cal N}=1\,\,SU(\nd-\no)$ SYM strongly coupled
gluonia with the mass scale $\sim\Lambda^{SU(\nd-\no)}_{{\cal
N}=1\,\,SYM}$.\\
5) $2\no\nt$ massless (complex) Nambu-Goldstone ${\cal N}=1$
multiplets $\Pi^1_2,\,\Pi^2_1$.\\
6) All $N_F^2$ fields $\Phi$ have masses $\mu(\Phi)\sim\mx\gg m$. 
They \, are dynamically irrelevant and not observable as real particles 
at  scales $\mu<\mx$.\\
Comparing with the mass spectrum at $N_c+1<N_F<3N_c/2$ in 
\,"$\bf A$" \,  of \, section  8.1.1 in \cite{ch13} it is seen that it is the 
same (up to  different logarithmic factors).

The multiplicity of these vacua is $N_{\rm
br2}=(\nd-\no)C^{\,\no}_{N_F}$, as it should be. The factor $\nd-\no$
originates from ${\cal N}=1\,\,SU(\nd-\no)$ SYM and the factor
$C^{\,\no}_{N_F}$ - from spontaneous flavor symmetry breaking
$U(N_F)\ra U(\no)\times U(\nt)$ due to higgsing of $\no$ quarks
$\,Q^1, {\ov Q}_1$.\\

{\bf a2)} If $\wh{\rm b}_{\rm o}>0$. Then, because $\wh\Lambda\ll m$,
the quarks $Q^2,\,{\ov Q}_2$ decouple as heavy at $\mu\sim m$, still
in the weak coupling region. As a result, the mass spectrum will be
the same as in "{\bf a1}" above (up to different logarithmic
factors).\\

{\bf B)} If $(\mx m)^{1/2}\ll\lt$. Then ${\cal N}=1$ SQCD with
$SU(\nd)$ colors and $3N_c/2<N_F<2N_c-1$ quark flavors enters 
\, first at \, $\mu<\lt$
the strongly coupled conformal regime with the frozen gauge
coupling $a_*=\nd g^2_*/2\pi=O(1)$.

And only now we use  {{\it for the first time} the dynamical scenario
proposed in \cite{ch3} {\it to calculate the mass spectra}. Really,
what is only assumed in this scenario in the case of the standard
${\cal N}=1$ SQCD conformal regime is the following.

In this ${\cal N}=1$ theory without elementary colored adjoint scalars, 
(unlike  the very special ${\cal N}=2$ theory with its additional colored
scalar fields $X^{\rm adj}$ and enhanced supersymmetry),\, {\it no
additional  parametrically lighter solitons,  i.e. \,  in addition} to the 
ordinary  mass  spectrum \,  with Nambu-Goldstone particles
described below)  are formed at those scales where the
standard ${\cal N}=1$ conformal regime is broken explicitly by 
nonzero \,  particle  masses (see also the footnote \ref{(f50)}).

The potentially competing masses look then as
\bbq
{\wt m}_{Q,2}^{\rm pole}\sim\frac{m}{z_Q(\lt,m_{Q,2}^{\rm
pole})}\sim\lt\Bigl (\frac{m}{\lt}\Bigr )^{N_F/3\nd},\,\,
z_Q(\lt,\mu\ll\lt)=\Bigl (\frac{\mu}{|\lt|}\Bigr )^{\gamma_Q}\ll
1,\,\, 0<\gamma_Q=\frac{3\nd-N_F}{N_F}<\frac{1}{2}\,,
\eeq
\bq
\mu^2_{\rm gl,1}\sim z_Q(\lt,\mu_{\rm gl,1})\Bigl
[\langle\Qo\rangle_{\nd}\sim\mx m\Bigr ]\,,\quad \mu_{\rm
gl,1}\sim\lt\Bigl (\frac{\mx m}{\lt^2}\Bigr )^{N_F/3N_c}\ll\lt\,,
\label{(41.4.12)}
\eq
\bbq
m\ll\mu_{\rm gl,1}\ll\lt\ll\mx\ll\lm\,,\quad \frac{{\wt m}_
{Q,2}^{\rm \,pole}}{\mu_{\rm gl,1}}\ll 1\,.
\eeq
Therefore, the quarks $Q^i, {\ov Q}_j,\,\, i,j=1...\no,$ are higgsed
at $\mu=\mu_{\rm gl,1}$, the flavor symmetry is broken spontaneously,
$U(N_F)\ra U(\no)\times U(\nt)$, and there remains at lower energy
${\cal N}=1$ SQCD with $SU(\nd-\no)$ colors and $\nt$ flavors of still \, 
active  quarks $Q^2, {\ov Q}_2\,$ with $SU(\nd-\no)$ colors.

{\bf b1)} If $\wh{\rm b}_{\rm o}<0$. Then this theory is IR free and
the gauge coupling becomes logarithmically small at
$\Lambda^{SU(\nd-\no)}_{{\cal N}=1\,\,SYM}\ll\mu\ll\mu_{\rm gl,1}$.
The RG evolution at $\Lambda^{SU(\nd-\no)}_{{\cal
N}=1\,\,SYM}<\mu<\mu_{\rm gl,1}$ is logarithmic only (and ignored 
for simplicity). The pole mass of $Q^2, {\ov Q}_2$ quarks with still
active unbroken $\nd-\no$ colors looks really as
\bq
m_{Q,2}^{\rm pole}\sim \frac{m}{z_Q(\lt,\mu_{\rm gl,1})}\sim\lm\Bigl
(\frac{\mx}{\lm}\Bigr )^{1/3}\Bigl (\frac{m}{\lm} \Bigr
)^{\frac{2(3N_c-N_F)}{3N_c}}\ll\mu_{\rm gl,1}\,.\label{(41.4.13)}
\eq
They decouple then as heavy at $\mu<m_{Q,2}^{\rm pole}$ {\it in the
weak coupling region} and there remains ${\cal N}=1\,\, SU(\nd-\no)$
SYM with $\Lambda^{SU(\nd-\no)}_{{\cal N}=1\,\,SYM}\ll m_{Q,2}^{\rm
pole}$, see \eqref{(41.4.9)}. Integrating it via the VY procedure
\cite{VY}, the lower energy Kahler terms looks now as \cite{ch19},
\bbq
K=z_{\Phi}(\lt,\mu_{\rm gl,1}){\rm Tr} (\Phi^\dagger\Phi
)+z_Q(\lt,\mu_{\rm gl,1})
{\rm Tr}\,\Biggl
[\,2\,\sqrt{(\Pi^1_1)^{\dagger}\Pi^1_1}+\Pi_2^1\frac{1}
{\sqrt{(\Pi^1_1)^{\dagger}\Pi^1_1}}\,(\Pi_2^1)^{\dagger}+(\Pi_1^2)
^{\dagger}\frac{1}{\sqrt{
(\Pi^1_1)^{\dagger}\Pi^1_1}}\,\Pi_1^2 \Biggr ]\,,
\eeq
\bq
z_{\Phi}(\lt,\mu_{\rm gl,1})= \Bigl (\frac{\mu_{\rm
gl,1}}{|\lt|}\Bigr)^{\gamma_{\Phi}}\,,\quad
-1<\gamma_{\Phi}=-2\gamma_{Q}<0\,,\quad z_{\Phi}(\lt,\mu_{\rm
gl,1})=1/z^2_{Q}(\lt,\mu_{\rm gl,1})\gg 1\,,\label{(41.4.14)}
\eq
while the superpotential is as in \eqref{(41.4.10)}. Therefore, in
comparison with \eqref{(41.4.11)}, only the pion masses have
 changed  and are now
\bq
\mu(\Pi^1_1)\sim\frac{\langle\Pi^1_1\rangle=\langle\Qo\rangle_{\nd}}
{\mx}\frac{1}{z_Q(\lt,\mu_{\rm
gl,1})}\sim\lm\Bigl (\frac{\mx}{\lm}\Bigr )^{1/3}\Bigl (\frac{m}{\lm}
\Bigr )^{\frac{2(3N_c-N_F)}{3N_c}}\sim m_{Q,2}^{\rm pole}\,,
\label{(41.4.15)}
\eq
while, because $0<\gamma_Q<1/2$, all $N_F^2$ fions $\Phi$ remain too
heavy, dynamically irrelevant and not observable as real particles at
$\mu<\mx^{\rm pole}$.\\

{\bf b2)} If $\wh{\rm b}_{\rm o}>0$. Then the ${\cal N}=1\,\,
SU(\nd-\no)$ theory remains in the conformal window at $\mu<\mu_{\rm
gl,1}$ with the frozen gauge coupling $a^{*}=O(1)$.

The anomalous dimensions at $\mu<\mu_{\rm gl,1}$ look now as, see
\eqref{(41.4.8)},
\bq
0<{\wh\gamma}_Q=\frac{{\rm {\wh b}}_{\rm
o}}{\nt}<\gamma_Q<\frac{1}{2}\,,\quad
{\wh\gamma}_{\Phi}=-2{\wh\gamma}_Q\,,\quad {\wh z}_Q(\mu_{\rm
gl,1},\mu\ll\mu_{\rm gl,1})=\Bigl (\frac{\mu}{\mu_{\rm gl,1}}\Bigr
)^{{\wh\gamma}_Q}\ll 1\,.\label{(41.4.16)}
\eq
The pole mass of $Q^2, {\ov Q}_2$ quarks with unbroken colors looks
now as, see \eqref{(41.4.9)}, \eqref{(41.4.12)},
\bq
{\wh m}_{Q,2}^{\rm pole}\sim\frac{m}{z_Q(\lt,\mu_{\rm
gl,1})}\frac{1}{{\wh z}_Q(\mu_{\rm gl,1},m_{Q,2}^{\rm
pole})}\sim\lm\Bigl (\frac{\mx}{\lm}\Bigr )^{1/3}\Bigl
(\frac{m}{\lm}\Bigr )^{\frac{\nt-\no}{3(\nd-\no)}}\sim (\rm
several)\,\Lambda^{SU(\nd-\no)}_{{\cal N}=1\,\,SYM}\,,\label{(2.4.17)}
\eq
while the masses of $\no^2$ pions $\Pi^1_1$ remain the same 
as in  \eqref{(41.4.15)} and all $N_F^2\,\, \Phi^j_i$ also remain 
dynamically irrelevant. The overall hierarchies of
nonzero masses look in this case as
\bq
\mu(\Pi^1_1)\ll\Lambda^{SU(\nd-\no)}_{{\cal N}=1\,\,SYM}\sim {\wh
m}_{Q,2}^{\rm pole}\ll\mu_{\rm
gl,1}\ll\lt\ll\mx\ll\lm\,.\label{(41.4.18)}
\eq

On the whole for this section 41.4 with $m\ll\mx\ll\lm$. -

In all cases considered the overall phase $\rm\mathbf{Higgs_1-HQ_2}$
(HQ=heavy quark) remains the same in the non-Abelian $SU(\nd)$ sector. \, 
At scales \, $\mx\ll\mu\ll\lm$ it behaves as the effectively massless IR
free ${\cal N}=2$ theory. At the scale $\mu\sim \mx^{\rm pole}=g^2\mx$\, 
all \, $X^{adj}_{SU(\nd)}$ decouple as heavy and its dynamics becomes
those of the ${\cal N}=1$ theory. $2\no^2$ quarks $Q^1, {\ov Q}_1$
with $SU(\no)$ colors and $\no$ flavors are higgsed while $2\nt
(\nd-\no)$ quarks $Q^2, {\ov Q}_2$ with $SU(\nd-\no)$ colors and $\nt$ \, 
flavors  are in the HQ phase and confined by the ${\cal N}=1\,
SU(\nd-\no)$ SYM. $2\no\nt$ massless Nambu-Goldstone particles in all
cases are in essence the hybrid quarks with $\nt$ flavors and broken
$SU(\no)$ colors. And the overall qualitative picture is also the same \, 
in all  variants considered, see e.g. the text after \eqref{(41.4.11)}.
Changes only the character of the ${\cal N}=1$ RG evolution, i.e. it
is logarithmic or power-like, and this influences the values of
nonzero masses and mass hierarchies.\\

Finally, we think it will be useful to make the following short
additional comments.\\
In this section, the largest masses $\mx^{\rm pole}=g^2(\mu=\mx^{\rm
pole})\mx\gg (\mx m)^{1/2}\gg m$ in the whole $SU(\nd)$ sector have
$\nd^{\,2}-1$ adjoint scalars $X^{\rm adj}_{SU(\nd)}$. They all are
{\it too heavy and too short ranged} now, and in the whole energy
region $\mu > (\rm several) (\mx m)^{1/2}$ at least
their physical (i.e. path dependent) $SU(\nd)$ phases induced by
interactions with the {\it lighter} (and effectively massless at such
energies) $SU(\nd)$ gluons fluctuate freely. Therefore, the mean\, 
value \, of \, $X^{\rm  adj}_{SU(\nd)}$, integrated not only over the 
interval from \, the high  energy down
to $\mx^{\rm pole}/({\rm several})$ as in the
Lagrangian \eqref{(41.4.3)},\eqref{(41.4.4)}, but even down to much
lower energies $\mu_{\rm cut}^{\rm low}= (\rm several) (\mx
m)^{1/2}\ll\mx$ is zero,
\bbq
\langle\sqrt{2} X^{\rm adj}_{SU(\nd)}\rangle_{\mu^{\rm low}_{\rm
cut}}=2\sum_{i=1}^{N_F}\sum_{A=1}^{\nd^{\,2}-1}T^{A}\langle{\rm
Tr\,}{\ov Q}_i  T^{A} Q^i\rangle_{\mu^{\rm low}_{\rm cut}}/\wmu=
\eeq
\bq
= 2\sum_{i=1}^{\no}\sum_{A=1}^{\nd^{\,2}-1}T^{A} {\rm Tr}\Biggl
[\langle{\ov Q}_i\rangle_{\mu^{\rm low}_{\rm cut}} T^{A} \langle
Q^i\rangle_{\mu^{\rm low}_{\rm cut}}\Biggr ]/\wmu=0\,,\,
\label{(41.4.19)}
\eq
because all light quark fields $Q^i_a$ and ${\ov Q}^{\,a}_i,\,
a=1...\nd,\, i=1...N_F$ (and all light $SU(\nd)$ gluons) {\it still
fluctuate independently and freely} in this higher energy region.

But its formal {\it total mean value} $\langle X^{\rm
adj}_{SU(\nd)}\rangle$, integrated by definition down to $\mu^{\rm
lowest}_{\rm cut}=0$, contains the small but nonzero, $\sim m\ll (\mx
m)^{1/2}$, contribution originating only in the lower energy region
$\mu\sim (\mx m)^{1/2}\ll\mx$\, from the classical coupling of heavy
$X^{\rm adj}_{SU(\nd)}$ with {\it lighter higgsed quarks} ${\ov
Q}^{\,a}_i, Q^i_a,\, a,i=1...\no$, see e.g. \eqref{(41.4.1)} or
\eqref{(41.1.14)},
\bbq
\langle\sqrt{2}X^{\rm \,adj}_{SU(\nd)}\rangle=\langle\sqrt{2}
\sum_{A=1}^{\nd^{\,2}-1} X^{A} \,T^{A}\rangle\approx
2\sum_{i=1}^{N_F}\sum_{A=1}^{\nd^{\,2}-1}T^{A}
\langle{\rm Tr\,}{\ov Q}_i T^{A} Q^i\rangle/\wmu \approx
\eeq
\bq
\approx 2\sum_{i=1}^{\no}\sum_{A=1}^{\nd^{\,2}-1}T^{A}{\rm
Tr\,}[\,\langle{\ov Q}\rangle_i T^{A} \langle
Q^i\rangle\,]/\wmu\approx m_2\,{\rm
diag}\,(\,\underbrace{\,1-\frac{\no}{\nd}}_{\no}\,;\,
\underbrace{\,-\frac{\no}{\nd}}_{\nd-\no}\,)\,.\label{(41.4.20)}
\eq
This leads to the main contribution $\sim m^2$ to ${\rm Tr\,}[\langle
(\sqrt{2} X^{\rm adj}_{SU(\nd)})^2\rangle ]$, which in the case
considered looks as \, ${\rm Tr\,}[\langle (\sqrt{2} X^{\rm adj}
_{SU(\nd)})^2\rangle ]\approx{\rm Tr\,}[\langle \sqrt{2} X^{\rm
adj}_{SU(\nd)}\rangle\langle \sqrt{2} X^{\rm adj}_{SU(\nd)}
\rangle]$, \, see \eqref{(41.4.20)},
\bq
{\rm Tr\,} [\langle (\sqrt{2} X^{\rm adj}_{SU(\nd)})^2\rangle ]\approx\Bigl
[\frac{\no (\nd-\no)}{\nd}\, m_2^2=\frac{\no\nd}{\nd-\no}\,\tm^2\,\Bigr ], 
\,\, \tm=\frac{N_c}{\nd}m,\,\, m_2=\frac{N_c}{\nd-\no} m\,, \,\,\,\,
\label{(41.4.21)}
\eq
this agrees with the exact relation in the last line of
\eqref{(41.4.1)} following from the Konishi anomaly (up to small
nonperturbative power corrections $\sim\langle S\rangle_{\nd}/\mx\sim
\ha^2 (\ha/\lm)^{(\bb)/(\nt-N_c)}$ originating finally at even lower
energies from the SYM part), and explains the origin of a smooth
holomorphic behavior of $\langle{\rm Tr\,}(\sqrt{2}X^{\rm
adj}_{SU(\nd)})^2\rangle$ when $\mx$ is increased from 
$\mx\ll m$ in  section 41.1 to $\mx\gg m$ in this section.

But, in the case considered, this small formal nonzero total mean
value of $\langle X^{\rm adj}_{SU(\nd)}  \rangle\sim m
\ll (\mx m)^{1/2}$ does not mean that these heavy adjoint \, 
scalars  are {\it higgsed}, in the sense that they give by themselves
{\it the additional contributions} $\sim m$ {\it to particle masses}
(those of hybrid gluons and $SU(\nd)$ quarks in this case), as it
really occurs at $\mx\ll m$ in section 41.1. (Finally, the physical
reason that the fields $X^{\rm adj}_{SU(\nd)}$ do not give by
themselves originating from the scale $\mu\sim\mx$ contributions to
particle masses is that they are too heavy now and so too short ranged \, 
and\,  cannot \, form the coherent condensate).

Clearly, {\it there are no any additional contributions to particle
masses in the Lagrangian \eqref{(41.4.3)},\eqref{(41.4.4)} at the scale
$\mu=\mu_{\rm x}^{\rm pole}/({\rm several})$ from $\langle X^{\rm
adj}_{SU(\nd)}\rangle_{\mu_{\rm x}^{\rm pole}/(\rm several)}=0$}, see
\eqref{(41.4.19)}, because all $\nd^{\,2}-1$ fields $X^{\rm
adj}_{SU(\nd)}$ with masses $\mx^{\rm pole}$ {\it decoupled already as \, 
heavy} at  $\mu=\mx^{\rm pole}/(\rm several)$ where \eqref{(41.4.19)} is \, 
definitely valid,  and they {\it do not affect by themselves the
dynamics of the lower energy ${\cal N}=1$ theory at $\mu < \mx^{\rm
pole}/(\rm several)$}. This is the reason for a qualitative difference \, in
patterns of color symmetry breaking and mass spectra at $\mx\ll m$
in section 41.1 and at $\mx\gg m$ in this section. Remind that the
color breaking (and corresponding mass splitting) occurs at the scale
$\mu\sim m\gg (\mx m)^{1/2}\gg\mx$ and the unbroken gauge symmetry
looks as $SU(\nd)\ra SU(\no)\times U^{(2)}(1)\times SU(\nd-\no)$ in
section 41.1, while it occurs at the scale $m\ll\mu\sim (\mx
m)^{1/2}\ll\mx$ and the unbroken gauge symmetry looks as 
$SU(\nd)\ra  SU(\nd-\no)$ in this section.

As it is seen from \eqref{(41.4.3)}-\eqref{(41.4.5)} and the whole
content of this section, the whole contributions $\sim (\mx m)^{1/2}$
to the masses of $SU(\no)$ and hybrid gluons  originate only at
 lower energies $\mu\sim (\mx
m)^{1/2}\ll\mx$ in D-terms of the genuine ${\cal N}=1$ SQCD
\eqref{(41.4.3)},\eqref{(41.4.4)} from higgsed quarks only. And the
whole additional F-term contributions $\sim m$ to the quark masses
originate really from the quark self-interaction term $\sim{\rm
Tr\,}(\QQ)^2/\mx$ with higgsed quarks
$\langle\Qo\rangle_{\nd}=\langle{\ov Q}^1_1\rangle\langle
Q^1_1\rangle$ in \eqref{(41.4.3)} or, what is the same, from
$\langle\Phi_{1,2}\rangle\sim m$ in \eqref{(41.4.4)}, this last also
originating finally from these higgsed quarks, forming the coherent
condensate and giving masses to corresponding gluons, see the last
line of \eqref{(41.4.5)}. And the additional difference between the
section 41.1 with higgsed light $\langle \, X^{adj}_{SU(\nd)}
\rangle\sim\langle a_2\rangle\sim m$ at $\mx\ll m$
in \, \eqref{(41.1.3)},\eqref{(41.1.6)} contributing to quark masses, and
replacing it heavier $\langle\Phi^j_i\rangle\sim m$ with the mass 
$\sim\mx\gg  m$ in this 
section, see \eqref{(41.4.4)},\eqref{(41.4.5)}, is that $
X^{adj}_{SU(\nd)}$ acts in the color space, while $\Phi^j_i$ acts in
the flavor space.\\

\section{Unbroken flavor symmetry, \,\,S-vacua,\,\,$\mathbf{SU(N_c)}$}
\numberwithin{equation}{section}

As in the br2 vacua in section 41.1, the non-trivial discrete
$Z_{\bb\geq 2}$ symmetry is also unbroken in these vacua, i.e. they
also belong to the baryonic branch in the language \cite{APS}, this
case corresponds to $\no=0$. From \eqref{(40.3)},\eqref{(41.1.9)}, the
quark condensates $\langle\QQ\rangle_{N_c}$ in these vacua with the
multiplicity $\nd$ look as (neglecting smaller power corrections), see \, 
section 4 in \cite{ch19},
\bq
\langle\QQ\rangle_{N_c}\approx -\,\frac{N_c}{\nd}\,\mx m =
-\mx\tm,\,\, \langle S\rangle_{N_c}=\Bigl (\frac
{\det\langle\QQ\rangle_{N_c}}{\lm^{\bb}\mx^{N_c}}\Bigr
)^{1/\nd}\approx\mx \tm^2\Bigl (\frac{-\tm}{\lm}\Bigr
)^{(\bb)/\nd}\ll\mx m^2,\,\,\, \label{(42.1)}
\eq
\bbq
\langle{\rm Tr\,}(\sqrt{2}\,X^{adj}_{SU(N_c)})^2\rangle=\Bigl
[(2N_c-N_F)\langle S\rangle_{N_c}+m\langle{\rm
Tr\,}\QQ\rangle_{N_c}\Bigr ]\approx m\,(-N_F\mx\tm)\,,\quad
\tm=\frac{N_c}{\nd} m\,.
\eeq

As before, the scalar $X^{\rm adj}_{SU(N_c)}$ higgses the $SU(N_c)$
group at the scale $\mu\sim\lm$ as $SU(N_c)\ra SU(\nd)\times
U^{(1)}(1)\times U^{\bt-1}(1)$, see \eqref{(41.1.1)},\eqref{(41.1.16)},
\bq
\langle X \rangle=\langle\,X^{adj}_{SU(\nd)}+ X^{(1)}_{U(1)}+
X^{adj}_{\bt}\,\rangle\,,\quad \nd=N_F-N_c\,,\quad
\bt=2N_c-N_F\,,\label{(42.2)}
\eq
\bbq
\langle\sqrt{2}\,X^{adj}_{\bt}\,\rangle=C_{\bt}\lm\,{\rm
diag}(\,\underbrace{\,0}_{\nd}\,; \underbrace
{\,\omega^0,\,\,\omega^1,\,...\,,\,\omega^{\bt-1}}_{\bt}\,)\,,
\quad \omega=\exp\{\frac{2\pi i}{\,\bt}\,\}\,,
\eeq
\bbq
\sqrt{2}\,X^{(1)}_{U(1)}=a_1\,{\rm
diag}(\,\underbrace{\,1}_{\nd}\,;\,\underbrace{\,c_1}_{\bt}\,),
\quad \, c_1=-\,\frac{\nd}{\bt}\,,\quad
\langle a_1\rangle=\frac{1}{c_1}m=\,-\frac{\bt}{\nd} m\,.
\eeq

I.e., as in br2 vacua, {\it the same} $\bt=2N_c-N_F\geq 2$ dyons
$D_j$, massless in the limit $\mx\ra 0$ (this number is required by
the unbroken $Z_{2N_c-N_F\geq 2}$ discrete symmetry) are formed at 
the \, scale  $\mu\sim\lm$ and at sufficiently small $\mx$ the 
superpotential  $\wh\w$ at $m\ll\mu\ll\lm$ is as in \eqref{(41.1.2)}.

The difference is in the behavior of the $SU(\nd)$ part. In these
vacua with $\no=0$, there is no analog of the additional gauge
symmetry breaking $SU(\nd)\ra SU(\no)\times U^{(2)}(1)\times
SU(\nd-\no)$ in br2 vacua at the scale $\mu\sim m$.

{\bf A)}\,\, At $\mx\ll\Lambda^{SU(\nd)}_{{\cal N}=2\,\,SYM}\ll m$
and \, $N_c<N_F<2N_c-1$, {\it all quarks} in the $SU(\nd)$ part 
in these S  vacua have now 
masses $\tm=\langle m-a_1\rangle=m N_c/\nd$ and
decouple as heavy {\it in the weak coupling regime} at scales
$\mu\,<\, \tm/(\rm several)$, there remains ${\cal N}=2\,\,\,SU(\nd)$
SYM with the scale factor $\Lambda^{SU(\nd)}_{{\cal N}=2\,\, SYM}$ of
its gauge coupling, see \eqref{(41.1.15)},
\bq
\langle\Lambda^{SU(\nd)}_{{\cal N}=2\,\, SYM}\rangle^{2\nd}=\Bigl
(\Lambda_{SU(\nd)}= -\lm\Bigr )^{2\nd-N_F}\,\tm^{N_F}\,,\quad
\tm=m\,\frac{N_c}{\nd}\,,\quad \langle\Lambda^{SU(\nd)}_{{\cal
N}=2\,\, SYM}\rangle\ll \,m\,.\label{(42.3)}
\eq

And the field $X^{adj}_{SU(\nd)}$ higgses this UV free 
${\cal \, N}=2\,\,\, SU(\nd)$ SYM at the scale
$\sim\langle\Lambda^{SU(\nd)}_{{\cal N}=2\,\, SYM}\rangle$ in a
standard way, $SU(\nd)\ra U^{\nd-1}(1)$ \cite{DS}. This results in the \, 
right  multiplicity $N_s=\nd=N_F-N_c$ of these S vacua. The two single
roots with $(e^{+}-e^{-})\sim \Lambda^{SU(\nd)}_{{\cal N}=2\,\,SYM}$
of the curve \eqref{(40.2)} originate here from this ${\cal N}=2\,\,\,
SU(\nd)$ SYM. Other $\nd-1$ unequal double roots originating from this \, 
$SU(\nd)$  sector correspond to massless at $\mx\ra 0\,\,\nd-1$ pure
magnetic monopoles $M_{\rm n}$ with the $SU(\nd)$ adjoint charges.

As a result of all described above, the low energy superpotential at
$\mx\ll (\langle\Lambda^{SU(\nd)}_{{\cal N}=2\,\, SYM}\rangle)^2/\lm$
and at the scale
$\mu<\langle\Lambda^{SU(\nd)}_{{\cal N}=2\,\, SYM}\rangle$
can be written in these S-vacua as (the coefficients $f_n=O(1)$ in
\eqref{(42.4)} are known from \cite{DS})
\bq
\w^{\,\rm low}_{\rm tot}=\w^{\,(SYM)}_{SU(\nd)}+\w^
{\,\rm low}_{\rm  matter}+\dots\,,\quad
\w^{\,\rm low}_{\rm matter}=\w_{D}+\w_{a_1}\,,\label{(42.4)}
\eq
\bbq
\w^{\,(SYM)}_{SU(\nd)}=\nd \mx(1+\delta_2)\Bigl
(\Lambda^{SU(\nd)}_{{\cal N}=2\,\, SYM}\Bigr
)^2+\w^{\,(M)}_{SU(\nd)}\,,
\eeq
\bbq
\w^{\,(M)}_{SU(\nd)}= - \sum_{n=1}^{\nd-1} {\tilde a}_{M,\rm n}\Biggl
[\, {\ov M}_{\rm n} M_{\rm n}+\mx(1+\delta_2)\Lambda^{SU(\nd)}_{{\cal
N}=2\,\,SYM}\Biggl (1+O\Bigl (\frac{\langle\Lambda^{SU(\nd)}_{{\cal
N}=2\,\,SYM}\rangle}{m}\Bigr )\Biggr ) f_{\rm n}\,\Biggr ]\,,
\eeq
\bbq
\w_{D}=(m-c_1 a_1)\sum_{j=1}^{\bb}{\ov D}_j D_j-\sum_{j=1}^{\bb}
a_{D,j}\,{\ov D}_j D_j\,- \,\mx\lm\sum_{j=1}^{\bb}\omega^{j-1}\,a_{D,j}
+\mx \, L_S \, \sum_{j=1}^{\bb} a_{D,j}\,,
\eeq
\bbq
\w_{a_1}=\frac{\mx}{2}(1+\delta_1)\frac{\nd N_c}{\bt} a_1^2+\mx N_c
\delta_3 a_1(m-c_1 a_1)+\mx N_c\delta_4 (m-c_1 a_1)^2\,,
\eeq
where $\delta_{1,2}$ are given in \eqref{(41.1.12)}, while
$\delta_3=0$, see \eqref{(B.4)}.

From \eqref{(42.4)}:
\bq
\langle a_1\rangle=\frac{1}{c_1} m= - \frac{2N_c-N_F}{\nd}\, 
m\,,\quad \langle a_{D,j}\rangle=\langle {\tilde a}_{M,\rm
n}\rangle=0\,,\label{(42.5)}
\eq
\bbq
\langle{\ov M}_{\rm n} M_{\rm n}\rangle=\langle{\ov M}_{\rm
n}\rangle\langle M_{\rm n}\rangle\approx
\mx\langle\Lambda^{SU(\nd}_{{\cal N}=2\,\,SYM}\rangle f_n\,,
\quad \, (2N_c-N_F)\mx\langle L_S\rangle=\langle\Sigma_D
\rangle=\sum_{i=1}^{N_F}\langle{\ov Q}_i Q^i\rangle_{N_c}\,,
\eeq
\bbq
\langle{\ov D}_j D_j\rangle=\langle{\ov D}_j\rangle\langle
D_j\rangle\approx \Bigl [\,-\mx\lm\omega^{j-1}- \frac{N_c N_F}{\nd
(\bb)}\mx m\,\Bigr ]\,,\quad \Sigma_D=\sum_{j=1}^{\bb}\langle{\ov
D}_j\rangle\langle D_j\rangle\,.
\eeq

And the qualitative situation with the condensates of $N_F$ flavors of \, 
heavy not higgsed quarks with $SU(2N_c-N_F)$ colors and masses
$\sim\lm$ in these S-vacua with $\no=0$ is the same as those in br2
vacua described at the end of section 41.1, see
\eqref{(41.1.16)}-\eqref{(41.1.26)}. The mean vacuum values of all
lighter non-higgsed quarks with $N_F$ flavors, $\nd$ colors and masses \, 
$\tm$ in  these S-vacua are power suppressed, see \eqref{(41.4.1)} with
$\no=0,\, \nt=N_F$ or (8.2.2) in \cite {ch13}, their small nonzero
values originate in this case only from the Konishi anomaly for quarks \, 
in the  $SU(\nd)$ sector, see \eqref{(42.3)},
$\,\langle\QQ\rangle_{\nd}=\langle S\rangle_{\nd}/{\tm}= -
\mx\langle\Lambda^{SU(\nd)}_{{\cal N}=2\,\,  SYM}\rangle^{2}/
{\tm}\sim\mx m\Bigl (m/\lm\Bigr)^{(\bb)/\nd}\ll\mx  m\,$.

Besides, from \eqref{(42.1)},\eqref{(42.5)} (the leading terms only for
simplicity, compare with \eqref{(41.1.18)}\,)
\bq
\langle\Sigma_D\rangle=\langle{\rm
Tr\,}\QQ\rangle_{N_c}\approx\langle{\rm Tr\,}\QQ\rangle_{\bb}
\approx -\frac{N_c \,N_F}{\nd}\mx m\,. \label{(42.6)}
\eq

On the whole for the mass spectrum in these $\nd$ S-vacua with the
unbroken flavor symmetry at $\mx\ll(\langle\Lambda^{SU(\nd)}_{{\cal
N}=2\,\,SYM}\rangle)^2/\lm$. - \\
1) All original electric quarks $Q, {\ov Q}$ are not higgsed but
confined. The masses of all original electric particles charged under
$SU(\bb)$ are the largest ones, $\sim\lm\,$, and they all are weakly
confined because all $\bb$ dyons $D_j$ are higgsed, $\langle{\ov
D}_j\rangle=\langle D_j\rangle\sim (\mx\lm)^{1/2}$, the string tension \, 
in this   sector is $\sigma^{1/2}_{SU(\bb)}\sim \langle D_j\rangle\sim
(\mx\lm)^{1/2}\ll\lm\,$. These confined particles form a large number
of hadrons with masses $\sim\lm$.\\
2) The next mass scale is $\tm$, these are masses of original electric \, 
quarks  with $SU(\nd)$ colors and $N_F$ flavors, they are also weakly
confined due to higgsing of $\nd-1$ pure magnetic monopoles $M_{\rm
n}$ of ${\cal N}=2\,\,SU(\nd)$\,\,SYM, the string tension here is much \, 
smaller,  $\sigma^{1/2}_{SU(\nd)}\sim \langle M_{\rm
n}\rangle_{SU(\nd)}\sim (\mx\langle\Lambda^{SU(\nd)}_{{\cal
N}=2\,\,SYM}\rangle)^{1/2}
\ll (\mx\lm)^{1/2}$. They form hadrons with masses $\sim m$.\\
3) The next mass scale is $\langle\Lambda^{SU(\nd)}_{{\cal
N}=2\,\,SYM}\rangle\ll m$, see \eqref{(42.3)}, these are masses of
charged $SU(\nd)$ electric gluons and scalars. They are also weakly confined
due to higgsing of $\nd-1$ magnetic monopoles $M_{\rm n}$, the tension \, 
of the  confining string is the same, $\sigma^{1/2}_{SU
(\nd)}\sim \langle M_{\rm n}\rangle_{SU(\nd)}\sim
(\mx\langle\Lambda^{SU(\nd)}_{{\cal N}=2\,\,SYM}\rangle)^{1/2}$. 
But  the mass scale of these hadrons is
$\sim\langle\Lambda^{SU(\nd)}_{{\cal N}=2\,\,SYM}\rangle\,$.\\
4) The other mass scale is $\sim (\mx\lm)^{1/2}$ due to higgsing of
$\bt$ dyons $D_j$. As a result, $\bt=2N_c-N_F$ long ${\cal N}=2$
multiplets of massive photons with masses $\sim (\mx\lm)^{1/2}$ are
formed (including $U^{(1)}(1)$ with its scalar $a_1$).\\
5) The lightest are $\nd-1$ long ${\cal N}=2$ multiplets of massive
dual photons with masses $\sim (\mx\langle\Lambda^{SU(\nd)}_{{\cal
N}=2\,\,SYM}\rangle)^{1/2}$.\\
6) Clearly, there are no massless Nambu-Goldstone particles because
the global flavor symmetry remains unbroken. And there are no massless \, 
particles  at all at $\mx\neq 0,\,\, m\neq 0$.

The corresponding scalar multiplets have additional small
contributions $\sim\mx$ to their masses, these small corrections 
break \, ${\cal  N}=2$ down to ${\cal N}=1$.\\

{\bf B)}\,\, At larger $\langle\Lambda^{SU(\nd)}_{{\cal
N}=2\,\,SYM}\rangle\ll\mx\ll\, m$, the difference is that all
$SU(\nd)$ adjoint scalars $X^{adj}_{SU(\nd)}$ are now {\it too heavy}, \, 
their  physical $SU(\nd)$ phases induced by interactions with light
$SU(\nd)$ gluons fluctuate freely at all scales  $\mu\gtrsim
\langle\Lambda^{SU(\nd)}_{{\cal N}=1\,\,SYM}\rangle$, and
they are not higgsed. Instead, they decouple as heavy much before,
already at the scale $\mu<[\,\mx^{\,\rm pole}=|g^2\mx|\,]/{\rm
several)}\ll m$, and there remains ${\cal N}=1 \,\, SU(\nd)$ SYM 
with  the scale factor $\langle\Lambda^{SU(\nd)}_{{\cal
N}=1\,\,SYM}\rangle=[\,\wmu \langle\Lambda^{SU(\nd)}_{{\cal
N}=2\,\,SYM}\rangle^2\,]^{1/3}$ of its gauge coupling,
$\,\,\langle\Lambda^{SU(\nd)}_{{\cal
N}=2\,\,SYM}\rangle\ll\langle\Lambda^{SU(\nd)}_{{\cal
N}=1\,\,SYM}\rangle\ll\mx\ll\, m$. As in section 41.3, the small
nonzero value $\langle {\rm Tr\,}
(X^{adj}_{SU(\nd)})^2\rangle=\nd\langle
S\rangle_{\nd}/\wmu=\nd\langle\Lambda^{SU(\nd)}_{{\cal N}=2\,\,
SYM}\rangle^2\ll\langle\Lambda^{SU(\nd)}_{{\cal N}=1\,\,
SYM}\rangle^2$, see \eqref{(42.3)}, arises here only due to the Konishi \, 
anomaly,  i.e. from one-loop Feynman diagrams with heavy scalars
$X^{adj}_{SU(\nd)}$ and their fermionic superpartners with masses
$\sim\mx$ inside, not because $X^{adj}_{SU(\nd)}$ are higgsed.

The multiplicity of vacua of this ${\cal N}=1 \,\, SU(\nd)$ SYM is
also $\nd$ as it should be. There appears now a large number of
strongly coupled gluonia with the mass scale
$\sim\langle\Lambda^{SU(\nd)}_{{\cal N}=1\,\,SYM}\rangle$. All
$SU(\nd)$ electrically charged particles are still confined, the
tension of the confining string is larger now,
$\sigma^{1/2}_{SU(\nd)}\sim\langle\Lambda^{SU(\nd)}_{{\cal
N}=1\,\,SYM}\rangle\ll\mx\ll\, m$.

The case with $m\ll\mx\ll\lm$ is described in section 8.2 of
\cite{ch13}.

\addcontentsline{toc}{section}
{\bf \large Part III.2.\,\, Large quark masses,
$\Large\mathbf{m\gg\lm}$ }}

\begin{center}
\hspace*{1cm}\bf\Large {Part III.2.\,\,Large quark
masses}, $\Large \mathbf{m\gg\lm}$ 
\end{center}

\section{Broken flavor symmetry}

\hspace*{4mm} We present below in this section the mass spectra at
$m\gg\lm$ of the direct (electric) $SU(N_c)$ theory \eqref{(40.1)} in
vacua with spontaneously broken flavor symmetry, $U(N_F)\ra
U(\no)\times U(\nt)$,\, $N_c+1<N_F<2N_c-1$. There are the br1-vacua
(br=breaking) with $1\leq\no< N_F/2\,,\,\, N_F=\no+\nt\,,$ in which
$\langle\Qo\rangle_{N_c}\gg\langle\Qt\rangle_{N_c}$, see 
\eqref{(40.3)},\eqref{(40.4)}. The quark and gluino
condensates look in these br1 vacua of $SU(N_c)$ as (the leading 
terms \, only), 
\bq
\langle\Qo\rangle_{N_c}\approx\mx m_3\,,\quad
\langle\Qt\rangle_{N_c}\approx\mx m_3\Bigl (\frac{\lm}{m_3}\Bigr
)^{\frac{2N_c-N_F}{N_c-\no}} \,,\quad m_3=\frac{N_c}{N_c-\no}\, m\,,
\label{(43.1)}
\eq
\bbq
\langle
S\rangle_{N_c}=\frac{\langle\Qo\rangle_{N_c}\langle\Qt\rangle_{N_c}}
{\mx}\approx\mx \, m_3^2\Bigl (\frac{\lm}{m_3}\Bigr
)^{\frac{2N_c-N_F}{N_c-\no}}\,,\quad\frac{\langle\Qt\rangle_{N_c}}
{\langle\Qo\rangle_{N_c}}\approx\Bigl
(\frac{\lm}{m_3}\Bigr)^{\frac{2N_c-N_F}{N_c-\no}}\ll 1\,,
\eeq
while $\no\leftrightarrow\nt$ in \eqref{(46.1.1)} in br2-vacua, but with
$N_F/2<\nt<N_c$ in this case. The multiplicity of these br1 and br2
vacua is correspondingly $N_{\rm br1}=(N_c-\no)C_{N_F}^{\,\no}$ 
and  $N_{\rm br2}=(N_c-\nt)C_{N_F}^{\,\nt},\,\,
C_{N_F}^{\,\no}=C_{N_F}^{\,\nt}=[\,N_{F} !/\no!\,\nt!\,],\,\,\, 1\leq
\no< N_F/2,\,\, \nt> N_F/2$. It is seen from \eqref{(43.1)} that the
discrete $Z_{\bb}$ symmetry is unbroken at $m\gg\lm$ in all these br1
and br2 vacua with $\no\neq\nd$. In contrast, $Z_{\bb}$ symmetry is
(pure formally) broken spontaneously in special vacua with
$\no=\nd,\,\,\nt=N_c$ and the multiplicity $N_{spec}=(\bb)
C_{N_F}^{\,\nd}$ (as it is really broken spontaneously in these
special vacua at $m\ll\lm$).

Out of them, see  section 4 in  \cite{ch19}, $(\nd-\no)C_{N_F}^{\,\no}$ 
part of br1-vacua with
$1\leq\no <\nd$ evolves at $m\ll\lm$ into the br2-vacua of section 
41.1 \,   with  $\langle\Qt\rangle^{(\rm br2)}_{N_c}\sim\mx
m\gg\langle\Qo\rangle^{(\rm br2)}_{N_c}$, while the other part of
br1-vacua and br2-vacua evolves at $m\ll\lm$ into $Lt$ vacua with
$\langle\Qo\rangle^{(\rm Lt)}_{N_c}\sim\langle\Qt\rangle^{(\rm
Lt)}_{N_c}\sim\mx\lm,\, ,\, \langle S\rangle_{Lt}\sim\mx\lm^2$.

\subsection{$SU(N_c),$ br1 vacua, smaller
$\mx,\,\,\,\mx\ll\Lambda^{SU(N_{c}-\no)}_{{\cal N}=2\,\,SYM}$}
\numberwithin{equation}{subsection}

In the case considered, $X^{\rm adj}_{SU(N_c)}$ breaks the $SU(N_c)$
group {\it in the weak coupling regime} at the largest scale $\mu\sim
m\gg\lm$ as\,: $\,SU(N_c)\ra SU(\no)\times U(1)\times SU(N_c-\no)$,
see \eqref{(43.1.4)} below,
\bq
\langle X^{\rm adj}_{SU(N_c)}\rangle= \langle\, X^{adj}_{SU(\no)}+
X_{U(1)}+ X^{adj}_{SU(N_c-\no)}\,\rangle\,,\label{(43.1.1)}
\eq
\bbq
\sqrt{2}\, X_{U(1)}=a \,{\rm
diag}\Bigl(\underbrace{\,1}_{\no}\,;\,\underbrace{\,c}_{N_c-\no}\,
\Bigr )\,,\quad c=-\frac{\no}{N_c-\no}\,,\quad a\equiv \langle
a\rangle+{\hat a}\,,\quad \langle a \rangle=m\,,
\quad \langle{\hat a}\rangle\equiv 0\,.
\eeq
As a result, all quarks $Q^i_a,\,{\ov Q}_j^{\,a}$ with flavors $i,
j=1...N_F$ and colors $a=(\no+1)...N_c$ have large masses
$m_3=m-c\langle a\rangle=m N_c/(N_c-\no)$, the hybrid gluons and
hybrid $X$ also have masses $m_3\gg\lm$, and they all decouple at
scales $\mu\lesssim\, m_3$. The lower energy theory at $\mu<\, m_3$
consists of ${\cal N}=2\,\, SU(N_c-\no)$\,\, SYM, then ${\cal N}=2\,\,
SU(\no)$  SQCD with $N_F$ flavors of massless at $\mx\ra 0$ quarks
$Q^{i}_{\rm b},\, {\ov Q}^{\,\rm b}_j$ with $\no$ colors, ${\rm
b}=1\,...\,\no$, and finally one ${\cal N}=2\,\, U(1)$ photon
multiplet with its scalar superpartner $X_{U(1)}$. The scale factor of \, 
the  ${\cal  N}=2\,\,\, SU(N_{c}-\no)$ SYM gauge coupling is
$\langle\Lambda^{SU(N_{c}-{\rm n}_1)}_{{\cal N}=2\,\,SYM}\rangle\ll
m$. The field $X^{\rm adj}_{SU(N_c-\no)}$ in \eqref{(43.1.1)}) breaks
this ${\cal N}=2\,\, SU(N_{c}-\no)$ SYM at the scale
$\mu\sim\langle\Lambda^{SU(N_{c}-{\rm n}_1)}_{{\cal
N}=2\,\,SYM}\rangle$ in a standard way, $SU(N_c-\no)\ra
U(1)^{N_c-\no-1}$ \cite{DS}, see also \eqref{(43.1)},
\bq
\langle\sqrt{2}\,
X^{adj}_{SU(N_c-\no)}\rangle\sim\langle\Lambda^
{SU(N_{c}-\no)}_{{\cal  N}=2\,\,SYM}\rangle
\,{\rm diag}\Bigl (\,\underbrace{\,0}_{\no}\,; k_1,k_2,\,...\,,
k_{N_c-\no}\Bigr )\,,\quad k_i=O(1)\,,\label{(43.1.2)}
\eq
\bbq
\langle\Lambda^{SU(N_{c}-\no)}_{{\cal
N}=2\,\,SYM}\rangle^2\approx\Bigl
(\frac{\lm^{\bb}{(m_3)}^{N_F}}{{(m_3)}^{2\no}}\Bigr
)^{\frac{1}{N_c-\no}}\approx
m^2_3\Bigl (\frac{\lm}{m_3}\Bigr )^{\frac{2N_c-N_F}
{N_c-\no}}\ll  m^2\,,
\eeq
\bbq
\langle S\rangle_{N_c-\no}=\mx\langle\Lambda^{SU(N_{c}-\no)}
_{{\cal\, N}=2\,\,SYM}\rangle^2=\langle S\rangle_{N_c}\,,
\eeq
where $k_i$ are known numbers \cite{DS}. We note that
$\Lambda^{SU(N_{c}-\no)}_{{\cal N}=2\,\,SYM}$ in \eqref{(43.1.2)} and
$\langle a\rangle=m$ in \eqref{(43.1.1)} have the same charge 2 under
the discrete $Z_{2N_c-N_F}$ transformations as $X$ itself, this is due\, 
to  unbroken $Z_{\bb}$ symmetry. There are $N_c-\no$ physically
equivalent vacua and $N_c-\no-1$ light pure magnetic monopoles ${\ov
M}_n,\, M_n,\,\, n=1... N_c-\no-1$ with the $SU(N_{c}-\no)$ adjoint
charges (massless at $\mx\ra 0$) in each of these ${\cal N}=2$ SYM
vacua. The relevant part of the low energy superpotential of this SYM
part looks as
\bq
{\cal W}^{\,\rm low}_{\rm tot}={\cal W}^{(M)}_{SYM}+{\cal
W}_{\no}+\dots\,,\label{(43.1.3)}
\eq
\bbq
{\cal W}^{(M)}_{SYM}= - \sum_{n=1}^{N_c-\no-1} {\tilde a}_{M,
\rm \, n}\Biggl [\, {\ov M}_{\rm n} M_{\rm \,
n}+\mx\langle\Lambda^{SU(N_{c}-{\rm \, n}_1)}_{{\cal
N}=2\,\,SYM}\rangle\Biggl (1+O\Bigl
(\frac{\langle\Lambda^{SU(N_{c}-{\rm n}_1)}_{{\cal
\, N}=2\,\,SYM}\rangle}{m}\Bigr )\Biggr ) z_{\rm n}\, \,\Biggr ]\,,
\quad  \,\,z_{\rm  n}=O(1),
\eeq
where $z_{\rm n}$ are known numbers \cite{DS} and dots denote 
as  always smaller power corrections.

All monopoles in \eqref{(43.1.3)} are higgsed at $\mx\ne 0$ with
$\langle M_{\rm n}\rangle=\langle{\ov M}_{\rm n} \rangle\sim
[\,\mx\langle\Lambda^{SU(N_{c}-{\rm n}_1)}_{{\cal
N}=2\,\,SYM}\rangle\,]^{1/2}$, so that all light particles of this
${\cal N}=2\,\, U(1)^{N_c-\no-1}$ Abelian magnetic part form
$N_c-\no-1$ long ${\cal N}=2$ multiplets of massive dual photons 
with  masses $\sim (\mx\langle\Lambda^{SU(N_{c}-\no}_{{\cal
N}=2\,\,SYM})^{1/2}\rangle$ .

Besides, for this reason, all heavier original $SU(N_{c}-\no)$
electrically charged particles with masses $\sim m$ or
$\sim\langle\Lambda^{SU(N_{c}-{\rm n}_1)}_{{\cal N}=2\,\,SYM}\rangle$
are weakly confined (the confinement is weak in the sense that the
tension of the confining string is much smaller than their masses,
$\sigma^{1/2}_{SYM}\sim (\mx\langle\Lambda^{SU(N_{c}-{\rm
n}_1)}_{{\cal \, N}=2\,\,SYM}\rangle)^{1/2}\ll\langle
\Lambda^{SU(N_{c}-{\rm \, n}_1)}_{{\cal N}=2\,\,SYM}\rangle\ll m$).\\

And finally, as for the (independent) ${\cal N}=2\,\,SU(\no)\times
U(1)$ part with $N_F$ flavors of original electric quarks ${\ov
Q}^b_j, Q^{\,i}_b,\,\, i,j=1...N_F,\,\, b=1...\no$. Its superpotential \, 
at the  low \, scale looks as
\bq
{\cal W}_{\no}=(\, m-a\,)\,{\rm Tr}\,({\ov Q} Q)_{\no} - {\rm
Tr}\,({\ov Q}\sqrt{2} X^{adj}_{SU(\no)} Q) +\mx {\rm Tr}\,(
X^{adj}_{SU(\no)})^2+\, \label{(43.1.4)}
\eq
\bbq
+ \,\frac{\mx}{2}\,\frac{\no N_c}{N_c-\no}\,a^2+(N_c-\no)\mx\Biggl
[\,\Bigl (\Lambda^{SU(N_{c}-{\rm n}_1)}_{{\cal N}=2\,\,SYM}\Bigr
)^2\approx\Bigl (\frac{\lm^{2N_c-N_F}(m_3-c\,{\hat
a})^{N_F}}{[\,m_3+(1-c){\hat a}\,]^{2\no}}\Bigr
)^{\frac{1}{N_c-\no}}\,\Biggr ]\,,\quad a\equiv \langle a
\rangle+{\hat  a}\,.
\eeq

Because the type of color breaking $SU(N_c)\ra SU(\no)\times
U(1)\times SU(N_c-\no)$ in \eqref{(43.1.1)} is qualitatively different
from $SU(N_c)\ra SU(N_F-N_c)\times U^{(1)}(1)\times U(1)^{\bb-1}$ 
in  section 41.1 and occurs here {\it in the weak coupling regime}
$a(\mu\sim m)=N_c g^2(\mu\sim m)/8\pi^2\ll 1$, the nonperturbative
instanton contributions do not operate in this high energy region.
They originate and operate in these vacua only at much lower energies
$\sim\langle\Lambda^{SU(N_{c}-{\rm n}_1)}_{{\cal N}=2\,\,SYM}\rangle$
from the SYM part, see \eqref{(43.1.3)}. For this reason, there are no
in \eqref{(43.1.4)} analogs of terms $\sim\delta_i$ in \eqref{(41.1.2)}.
From \eqref{(43.1.4)} (neglecting here small power corrections in the
quark condensate $\langle\Qo\rangle
_{\no}$ originating from the SYM part, see \eqref{(A.4)}\,), compare
also with \eqref{(43.1)},
\bq
\langle a\rangle=m\,,\quad \langle
X^{adj}_{SU(\no)}\rangle=0\,,\quad\langle{\rm Tr}\,({\ov Q}
Q)_{\no}\rangle=\no\langle\Qo\rangle_{\no}+\nt\langle\Qt
\rangle_{\no}\approx\no \mx\, m_3\,,\label{(43.1.5)}
\eq
\bbq
\langle\Qo\rangle^{SU(N_c)}_{\no}=\langle{\ov
Q}^{1}_{1}\rangle\langle  Q^{1}_{1}\rangle\approx\mx
m_3\approx\langle\Qo\rangle^{SU(N_c)}_{N_c}\,,\quad
\langle\Qt\rangle_{\no}=\sum_{a=1}^{\no}\langle{\ov
Q}^{\,a}_2\rangle\langle Q^2_a\rangle=0\,,\quad \langle
S\rangle_{\no}=0\,,
\eeq
\bbq
\langle Q^i_b\rangle=\langle {\ov
Q}_i^{\,b}\rangle\approx\delta^i_b\,(\mx m_3)^{1/2},\,
\, b=1... \no,\,\,i=1...N_F\,,\,\, m_3=\frac{N_c}{N_c-\no}\,.
\eeq

On the whole, all this results in the right multiplicity of these br1
vacua, $N_{\rm br1}=(N_c-\no) C_{N_F}^{\,\no}$, the factor $N_c-\no$
originates from ${\cal N}=2\,\,\, SU(N_{c}-\no)$ SYM, while
$C_{N_F}^{\,\no}$ is due to the spontaneous breaking $U(N_F)\ra
U(\no)\times U(\nt)$ by $\no$ higgsed quarks in the $SU(\no)
\times  U(1)$ sector.

It is worth emphasizing that the value $\langle a\rangle=m$ in
\eqref{(43.1.5)} is {\it exact} (i.e. there is no any small
correction). This is seen from the following. At very small $\mx\ra
0$, the possible correction $\delta m\neq 0$ is independent of $\mx$,
and so $\delta m\gg\mx$. Then quarks $Q^i_b, {\ov Q}_i^{\,b},\,\,
i=1...N_F,\,\, b=1...\no$ will acquire masses $\delta m$, much larger
than the scale of their potentially possible coherent condensate,
$\delta m\gg (\mx m)^{1/2},\,\, \mx\ra 0$. In this case, these quarks
will be not higgsed but will decouple as heavy ones at scales
$\mu<\delta m$ and the flavor symmetry will remain unbroken. 
There  will remain ${\cal N}=2\,\, SU(\no)\times U(1)$ SYM at scales
$\mu<\delta m$. This lowest energy ${\cal N}=2\,\, SU(\no)$ SYM 
will  give then its own multiplicity factor $\no$, so that the overall
multiplicity of vacua will be $\no (N_c-\no)$, while the right
multiplicity is $(N_c-\no)C^{\,\no}_{N_F}$, see \eqref{(43.1)}. 
To have this right multiplicity and spontaneous flavor symmetry
breaking all $\no$ quarks have to be higgsed at arbitrary small 
$\mx$, \, and this  is only possible when they are {\it exactly massless} 
in the  limit $\mx\ra 0$, i.e. at $\delta m=0$.~
~\footnote{\, These considerations clearly concern also all other
similar cases. \label{(56)}
}

As a result of higgsing of $\no$ out of $N_F$ flavors of pure
electric original quarks  in
$SU(\no)\times U(1)$ in \eqref{(43.1.4)}, the flavor symmetry is
broken \, spontaneously as $U(N_F)\ra U(\no)\times U(\nt)$, the 
quarks  $Q^{\,l}_b, {\ov Q}^{\,b}_l$ with flavors $l=(\no+1)...N_F$ and
colors \, $b=1...\no$ will be the massless Nambu-Goldstone particles
($\,2\no\nt$ complex degrees of freedom), while all other particles
in\eqref{(43.1.4)} will acquire masses $\sim (\mx m)^{1/2}$ and will 
form \, $\no^2$  long ${\cal N}=2$ multiplets of massive gluons.

On the whole, the hierarchies of nonzero masses look in this case as
\bq
\langle M_{\rm n}\rangle\sim (\mx\langle\Lambda^{SU(N_{c}-{\rm
n}_1)}_{{\cal N}=2\,\,SYM}\rangle)^{1/2}\ll\langle
Q^i_{b=i}\rangle\sim (\mx m)^{1/2}\,,\quad
\lm\ll\langle\Lambda^{SU(N_{c}-\no)}_{{\cal N}=2\,\,SYM}\rangle\ll
m\,,\label{(43.1.6)}
\eq
where ${\rm n}=1...N_{c}-\no-1\,,\,\,\, i=1...\no$\,. There are no
massless particles, except for the $\,2\no\nt$ Nambu-Goldstone
multiplets.

The curve \eqref{(40.2)} has $N_c-1$ double roots in these vacua.
$N_c-\no-1$ unequal roots correspond to pure magnetic monopoles
$M_{\rm n}$ in \eqref{(43.1.3)} and $\no$ equal ones correspond to
higgsed quarks from $SU(\no)$. The two single roots with
$(e^{+}-e^{-})\sim\langle\Lambda^{SU(N_{c}-\no)}_{{\cal
N}=2\,\,SYM}\rangle$ originate from SYM.

The calculation of leading power corrections to $\langle{\cal
W}^{\,\rm low}_{\rm tot}\rangle$ is presented in Appendix A, 
and those  to $\langle\Qo\rangle^{SU(N_c)}_{\no}$ are presented 
in Appendices A  and B. These latter are also compared therein 
with the {\it  independent} calculations of $\langle\Qo\rangle^
{SU(N_c)}_{\no}$ using \, the roots  of the curve \eqref{(40.2)}.\\

Now, let us recall some additional explanations about notations in
\eqref{(43.1.4)},\eqref{(43.1.5)}. As for {\it higgsed} $\langle
a\rangle$, it looks as $\langle a\rangle=\langle a\rangle^{(\rm
several) m}_{m/(\rm several)}=m$. In other words, its total mean
value \, originates \, and saturates at $\mu\sim m$. If not, all quarks will
decouple as heavy at $\mu=m/(\rm several)$, the multiplicity of these
br1 vacua will be wrong and $U(N_F)$ will remain unbroken. As for
$\langle Q_a\rangle,\,\, a=1...\no$, of light quarks with $SU(\no)$
colors. - At scales $[\mu^{\rm low}_{\rm cut}=(\mx \, m_3)^{1/2}]<
\mu<m/(\rm several)$, all $SU(\no)$ gluons are effectively \, massless
and all light quarks with $SU(\no)$ colors move freely and
{\it independently}. Therefore, the physical (i.e. path dependent)
$SU(\no)$ phases of these quarks induced by their interactions with
$SU(\no)$ gluons fluctuate {\it freely} in this higher energy region,
so that
\bq
\langle{\ov Q}^{\,b}_j Q^i_a\rangle_{\mu^{\rm low}_{\rm
cut}}=\langle{\ov Q}^{\,b}_j\rangle_{\mu^{\rm low}_{\rm cut}}\langle
Q^i_a\rangle_{\mu^{\rm low}_{\rm cut}}=0\,,\quad a,b=1...\no\,,\quad
i,j=1...N_F\,\quad \mu^{\rm low}_{\rm cut}=(\mx  \, 
m_3)^{1/2}\,.\label{(43.1.7)}
\eq
Therefore, in the whole region $0\leq \mu<m/(\rm several)$, the
nonzero quark mean values in \eqref{(43.1.5)}, i.e.
$[\langle\Qo\rangle_{\no}=\langle{\ov Q}^{1}_{1}\rangle\langle
Q^{1}_{1}\rangle\,]/\mx\approx 
[\langle a\rangle=m]N_c/(N_c-\no)]$ (the approximation
here is controllable and consists in neglecting small power
corrections originating from the SYM part, see
\eqref{(43.1.3)},\eqref{(43.1.4)} and Appendix A) originate and saturate \, 
{\it  only} \, in the region $g (\mx m_3)^{1/2}/(\rm several)<\mu< (\rm
several){\it g} (\mx\,  m_3)^{1/2}$, after $\no$ out of $N_F$ quarks
are higgsed (i.e. form the coherent condensate), {\it giving masses}
$g (\mx m)^{1/2}$ to themselves and all $SU(\no)\times U(1)$ 
gluons. \, And there  remain only $2\no\no$ massless Nambu-
Goldstone   multiplets in  this sector at  lower  energies.

And there is no contradiction in the relation following from
\eqref{(43.1.4)}
\bbq
\langle\Qo\rangle_{\no}=\Bigl (\langle{\ov Q}^1_1\rangle\langle
Q^1_1\rangle\Bigr )^{(\rm several){\it g}_Q (\mx m_3)^{1/2}}_{g_Q
(m_3 \mx )^{1/2}/(\rm several)}\approx \Bigl [\mx\frac{N_c}{N_c-\no}\Bigl
(\langle a\rangle^{(\rm several) m}_{m/(\rm several)}=m\,\Bigr )=\mx
m_3\,\Bigr ],
\eeq
in that the {\it total mean values} of $\langle a\rangle$ and
$\langle  Q^1_1\rangle$ originate and saturate in different energy regions,
because this is {\it not} the equality between mean values of two
equal to each other operators, but only the numerical equality between \, 
two total  mean values of two different operators.\\

On the whole, the total decomposition of quark condensates $\langle
(\QQ)_{1,2}\rangle_{N_c}$ over their separate color parts look in
these br1 vacua as follows.\\
I) The condensate $\langle\Qo\rangle_{N_c}$.\\
a) From \eqref{(43.1.5)} and accounting for the leading power
correction, see \eqref{(A.5)}, the (factorizable) condensate of
higgsed quarks in the $SU(\no)$ part
\bq
\langle\Qo\rangle_{\no}\approx \mx\, m_3
\Biggl [1-\frac{2N_c-N_F}{N_c-\no}
\Bigl (\frac{\lm}{m_3}\Bigr )^{\frac{2N_c-N_F}{N_c-\no}}\Biggr ], \,\, 
m_3=\frac{N_c}{N_c-\no} \, m\gg \langle\Lambda
^{SU(N_{c}-\no)}_{{\cal N}=2\,\,SYM}\rangle\gg\lm.
\,\,\,\,\,\label{(43.1.8)}
\eq
b) The (non-factorizable) condensate $\langle\Qo\rangle_{N_c-\no}$ is
determined by the one-loop Konishi anomaly for the heavy not higgsed
quarks with the mass $m_3$ in the $SU(N_c-\no)$ SYM sector. I.e., on
the first "preliminary" stage, from the one-loop diagrams with heavy
scalar quarks and their fermionic superpartners inside, transforming
in the weak coupling regime at the scale $\mu\sim m$ the heavy quark
{\it operator} $(\Qo)_{N_c-\no}$ into the {\it operator} $\sim
(\lambda\lambda)_{N_c-\no}/m_3$ of lighter $SU(N_c-\no)$ gluino. And
the mean value of this latter originates only at much lower energies
$\mu\lesssim\langle\Lambda^{SU(N_{c}-\no)}_{{\cal
N}=2\,\,SYM}\rangle$, see \eqref{(41.1.2)},
\bq
\langle\Qo\rangle_{N_c-\no}=\frac{\langle
S\rangle_{N_c-\no}}{m_3}=\frac{\mx\langle\Lambda^{SU(N_{c}-\no)}
_{{\cal N}=2\,\,SYM}\rangle^2}{m_3}\approx\mx m_3\Bigl
(\frac{\lm}{m_3}\Bigr )^{\frac{\bb}{N_c-\no}}\,.\label{(43.1.9)}
\eq

Therefore, on the whole
\bq
\langle\Qo\rangle_{N_c}=\langle\Qo\rangle_{\no}+\langle\Qo
\rangle_{N_c-\no}  \approx \, \mx m_3\Biggl [ 1+\frac{\nt-N_c}
{N_c-\no}\Bigl (\frac{\lm}{m_3}\Bigr
)^{\frac{\bb}{\nt-N_c}}\Biggr ]\,,\,\,\quad \label{(43.1.10)}
\eq
as it should be, see \eqref{(A.1)}.\\
II) The condensate $\langle\Qt\rangle_{N_c}$.\\ a) From 
\eqref{(43.1.5)}, the (factorizable) condensate of non-higgsed
massless Nambu-Goldstone particles in the $SU(\no)$ part
\bq
\langle\Qt\rangle_{\no}=\sum_{a=1}^{\no}\langle{\ov
Q}^{\,a}_2\rangle\langle  Q^2_{a}\rangle=0\,.\label{(43.1.11)}
\eq
b) The (non-factorizable) condensate $\langle\Qt\rangle_{N_c-\no}$ is
determined by the same one-loop Konishi anomaly for the heavy
non-higgsed quarks with the mass $m_3$ in the $SU(\nd-\no)$ SYM
sector,
\bq
\langle\Qt\rangle_{N_c-\no}=\frac{\langle
S\rangle_{N_c-\no}}{m_3}\approx\mx m_3\Bigl (\frac{\lm}{m_3}\Bigr
)^{\frac{\bb}{N_c-\no}}\,.\label{(43.1.12)}
\eq

Therefore, on the whole
\bq
\langle\Qt\rangle_{N_c}=\langle\Qt\rangle_{\no}+\langle\Qt\rangle_
{N_c-\no}\approx \mx\,  m_3\Bigl (\frac{\lm}{m_3}\Bigr )^{\frac{\bb}
{N_c-\no}}\,,  \label{(43.1.13)}
\eq
as it should be, see \eqref{(A.1)}.\\

\subsection{$U(N_c)$, br1 vacua, smaller $\mx,\,\,
\mx\ll\Lambda^{SU(N_{c}-\no)}_{{\cal N}=2\,\,SYM}$}

When the additional $U^{(0)}(1)$ is introduced, the Konishi anomalies
look as in \eqref{(41.2.2)} but $\langle\Qo\rangle_{N_c}$ is dominant
in br1 vacua, i.e. (the leading terms
only)
\bq
\langle\Qo+\Qt\rangle_{N_c}=\mx m\,,\quad \langle\Qo\rangle_{N_c}=\mx
m-\langle\Qt\rangle_{N_c}\approx\mx m\,,\label{(43.2.1)}
\eq
\bbq
\langle\Qt\rangle_{N_c}\approx\mx m\Bigl (\frac{\lm}{m}\Bigr
)^{\frac{2N_c-N_F}{N_c-\no}}\approx
\mx m\,\frac{\langle\Lambda^{SU(N_{c}-\no)}_{{\cal
N}=2\,\,SYM}\rangle^2}{m^2}\,,
\eeq
while
\bq
\frac{\langle a_0\rangle}{m}=\frac{1}{\mx m\, N_c}\langle{\rm
Tr}\,\QQ\rangle_{N_c}\approx\Bigl [\,\frac{\no}{N_c}
+\frac{\nt-\no}{N_c}\Bigl (\frac{\lm}{m}\Bigr
)^{\frac{2N_c-N_F}{N_c-\no}}\,\Bigr ]
\approx\frac{\no}{N_c}\,.\label{(43.2.2)}
\eq
(Put attention that $\langle\Lambda^{SU(N_{c}-\no)}_{{\cal
N}=2\,\,SYM}\rangle$ in this section is different from those in
section 43.1, see \eqref{(43.1.2)}).

Instead of \eqref{(43.1.4)} the superpotential of the $SU(\no)
\times  U^{(0)}(1)\times U(1)$ part looks now as
\bbq
{\cal W}^{\,\rm low}_{\rm matter}=\w_{\no}+\w_{a_0,a}+\dots\,,
\eeq
\bq
{\cal W}_{\no}=( m-a_0-a){\rm Tr}({\ov Q} Q)_{\no}-{\rm Tr}({\ov
Q}\sqrt{2} X^{adj}_{SU(\no)} Q)+\mx {\rm Tr}(
X^{adj}_{SU(\no)})^2,\,\,\,\label{(43.2.3)}
\eq
\bbq
\w_{a_0,a}=\frac{\mx}{2}N_c a^2_0+\frac{\mx}{2}
\frac{\no N_c}{N_c-\no}a^2\,,
\eeq
where dots denote smaller power corrections.

From \eqref{(43.2.3)} (for the leading terms)
\bbq
\langle a_0\rangle\approx\frac{\no}{N_c} m\,,\,\,\,\langle
a\rangle=\langle m-a_0\rangle\approx\frac{N_c-
\no}{N_c} m\,,\,\,\, \langle{\rm Tr}\,({\ov Q}
Q)\rangle_{\no}\approx\mx N_c\langle a_0\rangle
\approx\mx\frac{\no N_c}{N_c-\no}\langle a\rangle
\approx\no \mx \,  m\,,
\eeq
\bq
\langle
X^{adj}_{SU(\no)}\rangle=0,\,\,\langle\Qo\rangle^{U(N_c)}_{\no}=
\sum_{a=1}^{\no}\langle{\ov
Q}^{\,a}_1 Q^1_a\rangle=\langle{\ov Q}^{\,a}_1\rangle\langle
Q^1_a\rangle\approx\mx m\approx\langle\Qo\rangle^{U(N_c)}_
{N_c},\,\, \langle\Qt\rangle_{\no}=0.\quad\,\,\,\label{(43.2.4)}
\eq

The qualitative difference with section 4.1 is that one extra ${\cal
N}=1$ photon multiplet remains massless now in this $U(N_c)$ theory
while corresponding scalar multiplet has the smallest mass $\sim\mx$
because ${\cal N}=2$ is broken down to ${\cal N}=1$ at the level
$O(\mx)$\,.

The calculation of leading power corrections to $\langle{\cal
W}^{\,\rm low}_{\rm tot}\rangle$ in these $U(N_c)$ br1 vacua is
presented in Appendix A, and those to
$\langle\Qo\rangle^{U(N_c)}_{\no}$ are presented in Appendices A and
B. These latter are also compared therein with the {\it independent}
calculations of $\langle\Qo\rangle^{U(N_c)}_{\no}$ using the roots of
the curve \eqref{(40.2)}.\\

\subsection{$SU(N_c)$, br1 vacua, larger $\mx,\,
\Lambda^{SU(N_{c}-\no)}_{{\cal N}=2\,\,SYM}\ll\mx\ll m$}

Nothing changes significantly in this case with the $SU(\no)\times
U(1)$ part in \eqref{(43.1.4)}. The mean value $\langle
a\rangle=m\gg\mx$ of the field $\langle\sqrt{2}\,X_{U(1)}\rangle$ in
\eqref{(43.1.4)} stays intact as far as $\mx\ll m$, and so the
contributions $\sim (\mx m)^{1/2}$ to the masses of $\no^2-1$ long
${\cal N}=2\,\,\,SU(\no)$ multiplets of massive gluons and to the mass \, of 
the  long ${\cal N}=2\,\,\,U(1)$ multiplet of the massive photon due \, to 
higgsing   of  $\no$ quarks remain dominant. At this level the ${\cal
N}=2$ SUSY remains unbroken in this sector. It breaks down here to
${\cal N}=1$ only at the level $\sim\mx$ by the additional smaller
contributions $\sim\mx\ll (\mx m)^{1/2}$ to the masses of
corresponding scalar multiplets.

The situation with $SU(N_c-\no)$ SYM in
\eqref{(43.1.2)},\eqref{(43.1.3)} is different. At $\langle\Lambda^
{SU(N_c-\no)}_{{\cal N}=2\,\,SYM}\rangle\ll\mx\ll m$, the adjoint
scalars $X^{adj}_{SU(N_c-\no)}$ of the $SU(N_c-\no)$ subgroup in
\eqref{(43.1.2)} become too heavy and too short ranged. Their light
$SU(N_c-\no)$ physical (i.e. path dependent) phases induced by
interactions with effectively massless at all scales
$\mu\gtrsim\langle\Lambda^{SU(N_c-\no)}_{{\cal
N}=1\,\,SYM}\rangle\,\,SU(N_c-\no)$ gluons fluctuate freely at 
the scale  $\mu>\mx^{\rm
pole}/(\rm several)={\it g}^2\mx/(\rm several)$, and they are 
{\it not \, higgsed},  i.e. $\langle X^{adj}_{SU(N_c-\no)}\rangle_{\mx^{\rm
pole}/(\rm several)}=0$. Instead, they decouple as heavy already at
the scale $\mu<\mx^{\rm pole}/(\rm several)\ll m$ in the weak coupling \, 
region,  can all be integrated out and do not affect by themselves the
lower energy dynamics in this $SU(N_c-\no)$ sector. There remains
${\cal N}=1 \,\, SU(N_c-\no)$ SYM with the scale factor
$\langle\Lambda^{SU(N_c-\no)}_{{\cal N}=1\,\,SYM}\rangle=[\,\mx
\langle\Lambda^{SU(N_c-\no)}_{{\cal N}=2\,\,SYM}\rangle^2\,]^{1/3}$ 
of \, its gauge  coupling, $\,\,\langle\Lambda^{SU(N_c-\no)}_{{\cal
N}=2\,\,SYM}\rangle\ll\langle\Lambda^{SU(N_c-\no)}_{{\cal
N}=1\,\,SYM}\rangle\ll\mx\ll\, m$,
\bq
\langle S\rangle_{N_c-\no}=\langle\Lambda^{SU(N_c-\no)}_{{\cal N}=
1\,\,SYM}\rangle^3=\mx m_3^2\Bigl (\frac{\lm}{m_3}\Bigr
)^{\frac{2N_c-N_F}{N_c-\no}},\quad
\frac{\langle\Lambda^{SU(N_c-\no)}_{{\cal
N}=1\,\,SYM}\rangle}{\langle\Lambda^{SU(N_{c}-{\rm n}_1)}_{{\cal
N}=2\,\,SYM}\rangle}\sim\Biggl
(\frac{\mx}{\langle\Lambda^{SU(N_{c}-{\rm n}_1)}_{{\cal
N}=2\,\,SYM}\rangle}\Biggr )^{1/3}\gg 1\,. \label{(43.3.1)}
\eq
The small (non-factorizable) value $\langle{\rm Tr\,}
(X^{adj}_{SU(N_c-\no)})^2\rangle=
(N_c-\no)\langle
S\rangle_{N_c-\no}/\mx=(N_c-\no)\langle\Lambda^{SU(N_c-\no)}_{{\cal
N}=2\,\, SYM}\rangle^2 \\ \ll \langle\Lambda^{SU(N_c-\no)}_{{\cal
N}=1\,\, SYM}\rangle^2$ arises here not because
$X^{adj}_{SU(N_c-\no)}$ are higgsed, but only due to the Konishi
anomaly. I.e. from one-loop diagrams with heavy scalars
$X^{adj}_{SU(N_c-\no)}$ and their fermionic superpartners with masses
$\sim\mx$ inside. On the first "preliminary" stage these loop effects
transform, in the weak coupling regime at the scale $\sim \mx$, the
heavy field {\it operator} ${\rm Tr\,} (X^{adj}_{SU(N_c-\no)})^2$ into \, the
{\it
operator} $\sim {\rm Tr\,}(\lambda\lambda)_{SU(N_c-\no)}/\mx$ \, of lighter 
${\cal\, N}=1\,\, SU(N_c-\no)$ SYM gluinos. And the nonzero
mean value of this latter originates and saturates only in the strong
coupling and non-perturbative regime at much lower energies
$\sim\langle\Lambda^{SU(N_c-\no)}_{{\cal N}=1\,\, SYM}\rangle\ll\mx$ .
There is now a large number of strongly coupled gluonia with the mass
scale $\sim\langle\Lambda^
{SU(N_c-\no)}_{{\cal N}=1\,\,SYM}\rangle$ in this ${\cal N}=1$ SYM.
The multiplicity of vacua in this sector remains equal $N_c-\no$ as it \, 
should  be. \,  All heavier original electric particles with masses $\sim
m$ or $\sim\mx$ charged with respect to $SU(N_c-\no)$ remain weakly
confined, but the string tension is larger now, $\sigma^{1/2}_{{\cal
N}=2}\ll\sigma^{1/2}_{{\cal  N}=1}\sim\langle\Lambda^{SU(N_c-\no)}
_{{\cal  N}=1\,\,SYM}\rangle\ll\mx\ll m$.

\subsection{$SU(N_c)$\,, \, special vacua}

The values of quark condensates in these special vacua with
$\no=\nd,\, \nt=N_c$, see \cite{ch13} and \eqref{(43.1)},
\bq
\langle\Qo+\Qt-\frac{1}{N_c}{\rm
Tr}\,(\qq)\rangle_{N_c}=(1-\frac{\no=\nd}{N_c})\langle\Qo
\rangle_{N_c}=\mx \, m\,,\label{(43.4.1)}
\eq
\bbq
\langle\Qo\rangle_{N_c}=\frac{N_c}{\bb}\mx m=\mx m_3\,, 
\quad \langle\Qt\rangle_{N_c}=\mx\lm\,,\quad
\langle S\rangle_{N_c}=\frac{\langle\Qo\rangle_{N_c}
\langle\Qt \rangle_{N_c}}{\mx}=
\eeq
\bbq
=\Bigl (\frac{\det \langle\qq\rangle_{N_c}}{\lm^{2N_c-N_F}
\mx^{N_c}}\Bigr )^{1/\nd}=\langle
S\rangle_{2N_c-N_F}=\mx\langle\Lambda^{SU(\bb)}_{{\cal
N}=2\,\,SYM}\rangle^2=\mx m_3\lm\,,
\eeq
are exact and valid at any values $m\gtrless\lm$. Therefore, the
discrete $Z_{\bb}$ symmetry is formally broken spontaneously therein
at large $m\gg\lm$ also, as it is really broken at $m\ll\lm$. But
practically they behave at $m\gg\lm$ as the br1 vacua described above
in section (43.1) with $\no=\nd\,$. I.e., the factor $\bb$ in their
multiplicity $N_{\rm spec}=(\bb) C^{\,\nd}_{N_F}$ originates at
$m\gg\lm$ from the multiplicity $\bb$ of the $SU(\bb)\,\,{\cal N}=2$
SYM at $\mx\ll\langle\Lambda^{SU(\bb)}_{{\cal N}=2\,\,SYM}\rangle$ 
or  ${\cal N}=1$ SYM at $\langle\Lambda^{SU(\bb)}_{{\cal
N}=1\,\,SYM}\rangle\ll\mx\ll m$, after the breaking $SU(N_c)\ra
SU(\nd)\times U(1)\times SU(\bb)$ by $\langle X_{U(1)}\rangle$ at 
the  scale $\mu= m_3\gg\lm$, see \eqref{(43.4.1)}. The factor
$C^{\,\nd}_{N_F}$ originates due to higgsing of $\no=\nd$ original
electric quarks in the $SU(\nd)$ sector,
$\langle\Qo\rangle_{\nd}=\langle{\ov Q}^{\,1}_1
\rangle\langle Q^1_1\rangle=\mx \, m_3$, see \eqref{(43.4.1)}, and
spontaneous breaking $U(N_F)\ra U(\nd)\times U(N_c)$. The two single
roots with $( e^{+}-e^{-})\sim\langle\Lambda^{SU(\bb)}_{{\cal
N}=2\,\,SYM}\rangle$ of the curve \eqref{(40.2)} at small $\mx$
originate from the ${\cal N}=2\,\,SU(\bb)$ SYM. There are no massless
particles at $\mx\neq 0$, except for the standard $2\no\nt$ 
Nambu-Goldstone  multiplets for equal mass quarks.

\section{Unbroken flavor symmetry,\,\, SYM vacua,\,\, $\mathbf{SU(N_c)}$}

These SYM vacua of the $SU(N_c)$ theory have the multiplicity $N_c$. \,  
$\langle  X^{\rm adj}\rangle_{N_c}\sim
\langle\Lambda^{SU(N_c)}_{{\cal N}=2\, SYM}\rangle\ll m$ in these
vacua, so that all quarks have here large masses
$m\gg\langle\Lambda^{SU(N_c)}_{{\cal N}=2\, SYM}\rangle$ and 
decouple  in the weak coupling regime
at $\mu<m/(\rm several)$. There remains at this scale the ${\cal
N}=2\,\, SU(N_c)$ SYM with the scale factor of its gauge coupling:
$\langle\Lambda^{SU(N_c)}_{{\cal N}=2\, SYM}\rangle^2=(\lm^{\bb} m^
{N_F})^{1/N_c}=m^2\,
(\lm/m)^{(\bb)/N_c},\,\,\,\lm\ll\langle\Lambda^{SU(N_c)}_{{\cal
N}=2\,SYM}\rangle\ll m$.

If $\mx\ll\langle\Lambda^{SU(N_c)}_{{\cal N}=2\, SYM}\rangle$, then
the field $X^{\rm adj}_{SU(N_c)}$ is higgsed at the scale
$\mu\sim\langle\Lambda^{SU(N_c)}_{{\cal N}=2\, SYM}\rangle$ in a
standard way for the ${\cal N}=2$ SYM, $\langle\sqrt{2} X^{\rm
adj}_{SU(N_c)}\rangle\sim\langle\Lambda^{SU(N_c)}_{{\cal N}=2\,
SYM}\rangle\,{\rm diag}\,(k_1,k_2,\,...\,k_{N_c}),\,\, k_i=O(1)$
\cite{DS}, $SU(N_c)\ra U^{N_c-1}(1)$. Note that the value
$\sim\langle\Lambda^{SU(N_c)}_{{\cal N}=2\, SYM}\rangle$ of $\langle
X^{\rm adj}_{SU(N_c)}\rangle$ is consistent with the unbroken
$Z_{\bb}$ discrete symmetry.

All original electrically charged gluons and scalars acquire masses
$\sim\langle\Lambda^{SU(N_c)}_{{\cal N}=2\, SYM}\rangle$ and $N_c-1$
lighter Abelian pure magnetic monopoles $M_k$ are formed. These are
higgsed at $\mx\neq 0$ and $N_c-1$ long ${\cal N}=2$ multiplets of
massive dual photons are formed, with masses $\sim
(\mx\langle\Lambda^{SU(N_c)}_{{\cal N}=2\,
SYM}\rangle)^{1/2}\ll\langle\Lambda^{SU(N_c)}_{{\cal N}=2\,
SYM}\rangle$ (there are non-leading contributions $\sim\mx\ll
(\mx\langle\Lambda^{SU(N_c)}_{{\cal N}=2\, SYM}\rangle)^{1/2}$ to the
masses of corresponding scalars, these break slightly ${\cal N}=2$
down to ${\cal N}=1$ ). As a result, all original electrically charged \, 
quarks,  gluons and scalars $X$ are weakly confined (i.e. the tension
of the confining string, $\sigma^{1/2}_2\sim
(\mx\langle\Lambda^{SU(N_c)}_{{\cal N}=2\, SYM}\rangle)^{1/2}$, is
much smaller than their masses $\sim m$ or
$\sim\langle\Lambda^{SU(N_c)}_{{\cal N}=2\, SYM}\rangle\,$).

All $N_c-1$ double roots of the curve \eqref{(40.2)} correspond in this \, 
case to  $N_c-1$ pure magnetic monopoles (massless at $\mx\ra 0$),
while two single roots with $(e^+ -
e^-)\sim\langle\Lambda^{SU(N_c)}_{{\cal N}=2\, SYM}\rangle$ also
originate from this ${\cal N}=2 \,\,SU(N_c)$ SYM.\\

The mass spectrum is different if $\langle\Lambda^{SU(N_c)}_{{\cal
N}=2\, SYM}\ll\mx\ll m$. All scalar fields $X^{\rm adj}_{SU(N_c)}$ are \, 
then too  heavy and not higgsed. Instead, they decouple as heavy at
scales $\mu<\mx^{\rm pole}/(\rm several)$, still in the weak coupling
regime, and there remains at lower energies ${\cal N}=1\,\, SU(N_c)$
SYM with its $N_c$ vacua and with the scale factor
$\langle\Lambda^{SU(N_c)}_{{\cal N}=1\, SYM}\rangle$ of its gauge
coupling, $\langle\Lambda^{SU(N_c)}_{{\cal N}=1\,
SYM}\rangle=[\,\mx(\langle\Lambda^{SU(N_c)}_{{\cal N}=2\,
SYM}\rangle)^{\,2}\,]^{1/3},\,\, \langle\Lambda^{SU(N_c)}_{{\cal
N}=2\, SYM}\rangle\ll\langle\Lambda^{SU(N_c)}_{{\cal N}=1\,
SYM}\rangle\ll\mx\ll m$. A large number of strongly coupled gluonia
with the mass scale $\sim\langle\Lambda^{SU(N_c)}_{{\cal N}=1\,
SYM}\rangle$ is formed in this ${\cal N}=1$ SYM theory, while all
heavier original charged particles are weakly confined (the tension of \, 
the  confining string,
$\sigma^{1/2}_1\sim\langle\Lambda^{SU(N_c)}_{{\cal N}=1\,
SYM}\rangle$, is much smaller than the quark masses $\sim m$ or the
scalar masses $\sim\mx$).

\section{Very special vacua with $\mathbf{\langle S\rangle_{N_c}=0}$ in
$\mathbf{U(N_c)}$ gauge theory}
\numberwithin{equation}{section}

\hspace{4mm} We consider in this section the vs (very special) vacua
with $\no=\nd,\,\nt=N_c,\, N_c < N_F<2 N_c-1$\,, $\langle
S\rangle_{N_c}=0$ (\,as always, $\langle S\rangle_{N_c}$ is the
bilinear gluino condensate summed over all its colors) and the
multiplicity $N_{vs}=C^{\nd}_{N_F}=C^{N_c}_{N_F}$ in the
$U(N_c)=SU(N_c)\times U^{(0)}(1)$ theory, when the additional Abelian
$U^{(0)}(1)$ with $\mu_0=\mx$ is added to the $SU(N_c)$ theory (see
e.g. \cite{SY2} and references therein, these are called $r=N_c$ vacua \, 
in  \cite{SY2}). Note that vacua with $\langle S\rangle_{N_c}=0$ are
absent in the $SU(N_c)$ theory with $m\neq 0,\,\, \mx\neq 0$, see
Appendix B in \cite{ch19}.

The superpotential of this $U(N_c)$ theory looks at high energies as
\bq
{\cal W}_{\rm matter}=\mx\Bigl [{\rm Tr}\,(X_0)^2=\frac{1}{2}N_c
a_0^2\Bigr ]+\mx{\rm Tr}\,(X^{\rm adj}_{SU(N_c)})^2+{\rm Tr}\,\Biggl
[\,(m-a_0)\,{\ov Q} Q-{\ov Q}\sqrt{2} X^{adj}_{SU(N_c)} Q \Biggr
]_{N_c}\,,\label{(45.1)}
\eq
\bbq
\sqrt{2} X_{0}=a_0\, {\rm diag}(\,\underbrace{\,1}_{N_c}\,)\,,\quad
X^{\rm adj}_{SU(N_c)}=X^A T^A\,,\,\, {\rm Tr}\,(T^A
T^B)=\frac{1}{2}\,\delta^{AB}\,, \quad A, B=1,\,...,\,N_c^2-1\,.
\eeq

From \eqref{(45.1)} and Konishi  anomalies
\bq
\langle a_0\rangle=\frac{\langle {\rm Tr}({\ov Q}\,
Q)\rangle_{N_c}=\nd\langle\Qo\rangle_{N_c}
+N_c\langle\Qt\rangle_{N_c}}{\mx N_c}\,,\label{(45.2)}
\eq
\bbq
\langle m-a_0\rangle\sum_{a=1}^{N_c}\langle{\ov Q}_i^a
Q^i_a\rangle-\sum_{A=1}^{N^2_c-1}\sum_{a,b=1}^{N_c}\langle{\ov
Q}_i^b\sqrt{2} X^A (T^{A})^a_b Q^i_a\rangle=\langle
S\rangle_{N_c},\,\, i=1...N_F,\,\, \rm {no\,\,summation\,\,over\,\,
flavor},
\eeq
\bbq
\mx\langle{\rm Tr}\,(\sqrt{2} X^{\rm
adj}_{SU(N_c)})^2\rangle=(2N_c-N_F)\langle S\rangle_{N_c}+\langle
m-a_0\rangle\langle{\rm Tr}\,(\qq)\rangle_{N_c}\,.
\eeq

Note that, as a consequence of a unique property $\langle
S\rangle_{N_c}=0$ of these $U(N_c)$ vs -vacua, the curve \eqref{(40.2)} \,
at  $\mx\ra0$ for theory \eqref{(45.1)} has not $N_c-1$ but $N_c$
double roots: $\,\nd$ equal double roots $e_k= - m$ corresponding to
low energy non-Abelian $SU(\nd)$, and $2N_c-N_F$ unequal double roots
$e_j= -m+\omega^{j-1}\lm,\, j=1...\bb$, see e.g. \cite{Bo}. There are
no single roots of the curve \eqref{(40.2)} in these vacua.

As previously in section 40, taking $\mx$ sufficiently large
\footnote{\,
The scale factor $\Lambda_0$ of the Abelian coupling
$g_0(\Lambda_0/\mu)$ is taken sufficiently large, the theory (45.1) is
considered only at scales $\mu\ll\Lambda_0$ where $g_0$ is small, the
large $\mx$ means here $\lm\ll m\ll\mx\ll\Lambda_0$ if $m\gg\lm$ (and
only here, while $\mx\ll\lm$ in all other cases).
}
and integrating out all $N^2_c$ scalars $X$ as heavy, the
superpotential of the $U(N_c)\,\,\, {\cal N}=1$ SQCD looks then as
\bq
{\cal W}_{\rm matter}=m\,{\rm Tr}\, (\,{\ov Q}
Q)_{N_c}-\frac{1}{2\mx}\,\sum_{i,j=1}^{N_F}\,({\ov Q}_j
Q^i)_{N_c}({\ov Q}_i Q^j)_{N_c}\,,\quad ({\ov Q}_j
Q^i)_{N_c}=\sum_{a=1}^{N_c}({\ov Q}_j^{\,a} Q^i_a)\,,\label{(45.3)}
\eq
while, instead of \eqref{(40.3)}, $\w^{\,\rm eff}_{\rm tot}$ looks now
as
\bq
\w^{\,\rm eff}_{\rm tot}={\cal W}_{\rm matter}-\nd S_{N_c}\,,\quad
S_{N_c}=\Bigl(\frac{\det (\QQ)_{N_c}}{\lm^{2N_c-N_F}\mx^{N_c}}\Bigr
)^{1/\nd}\,,\quad \langle S\rangle_{N_c}=\Bigl
(\frac{\langle\Qo\rangle^{\nd}\langle\Qt\rangle^{N_c}}
{\lm^{2N_c-N_F}\mx^{N_c}}\Bigr )^{1/\nd}_{N_c}.\,\,\label{(45.4)}
\eq
From \eqref{(45.4)}, in these specific vs-vacua with $\langle
S\rangle_{N_c}=0$, the equations for determining quark 
condensates look as
\bq
\langle{\ov Q}_k\frac{\partial\w^{\rm eff}_{\rm tot}}{\partial{\ov
Q}_k}\rangle= m\langle (\QQ)_k\rangle_
{N_c}-\frac{\langle (\QQ)_k\rangle^2_{N_c}}{\mx}-\langle
S\rangle_{N_c}=0\,,\quad k=1,\,2\,,\label{(45.5)}
\eq
\bbq
\langle\Qo+\Qt\rangle_{N_c}=\mx m\,,\quad\langle
S\rangle_{N_c}=\langle\Qo\rangle_{N_c}\langle\Qt\rangle_{N_c}/\mx\,.
\eeq
The only self-consistent solution for these vs-vacua with $\no=\nd,\,
\nt=N_c,\,\, \langle S\rangle_{N_c}=0$ looks as, see
\eqref{(45.1)}-\eqref{(45.3)},
\bq
\langle\Qt\rangle_{N_c}=\mx m,\,\, \langle\Qo\rangle_{N_c}=0,\,\,
\langle
S\rangle_{N_c}=\frac{\langle\Qo\rangle_{N_c}\langle\Qt\rangle_
{N_c}}{\mx}=0\,,\label{(45.6)}
\eq
\bbq
\sum_{A=1}^{N^2_c-1}\sum_{a,b=1}^{N_c}\langle{\ov Q}_i^b\sqrt{2} 
X^A  (T^{A})^a_b Q^i_a\rangle=0\,, \quad i=1...N_F\,,\quad \rm{no\,\,
summation\,\, over\,\, flavor},
\eeq
\bbq
\langle a_0\rangle=\frac{\langle\Qt\rangle_{N_c}}{\mx} = m\,,\quad
\langle{\rm Tr}\,(\sqrt{2} X^{\rm adj}_{SU(N_c)})^2\rangle=0\,,
\quad\langle\w^{\,\rm eff}_{\rm tot}\rangle=
\langle{\cal W}_{\rm matter}\rangle=\frac{\mx}{2}N_c \langle
a_0\rangle^2=\frac{\mx}{2}N_c m^2\,.
\eeq
(The variant with $\langle\Qo\rangle_{N_c}=\mx
m,\,\,\langle\Qt\rangle_{N_c}=0$ and $\langle a_0\rangle=\nd m/N_c$
from \eqref{(45.2)}, will result at $m\gg\lm$ in all heavy not higgsed
quarks with masses ${\hat m}=\langle m-a_0\rangle=(\bb) m/N_c\gg\lm$,
the lower energy $SU(N_c)$ \,\,${\cal N}=2$ SYM, the multiplicity of
vacua equal $N_c$, $\,\langle S\rangle_{N_c}=\mx (\lm^{\bb}{\hat
m}^{N_F})^{1/N_c}\neq 0$, the unbroken flavor symmetry, and the
massless photon. There will be not $N_c$ but $N_c-1$ unequal ${\cal
N}=2$ SYM double roots at $\mx\ra 0$. This is incompatible with
$\langle S\rangle_{N_c}=\langle\Qo\rangle_{N_c}\langle\Qt
\rangle_{N_c}/\mx = 0$ from \eqref{(45.5)}, and the multiplicity is
wrong. While at $m\ll\lm$ this variant with $\langle a_0\rangle=\nd
m/N_c$ from \eqref{(45.2)} will be at least incompatible with
\eqref{(45.2.2)},\eqref{(45.2.3)}).

\subsection {$m\gg\lm\,,\,\, \mx\gg\lm^2/m \,,\,\, N_c < N_F < 2
N_c-1$}
\numberwithin{equation}{subsection}

\hspace*{4mm} Consider first this simplest case with
$0<\mx\ll\lm,\,\,m\gg\lm$ and $\mx m\gg\lm^2$. Note that $\langle
a_0\rangle=m$ "eats" all quark masses $m$ in \eqref{(45.1)} in the
range of scales $m/(\rm several)<\mu<(\rm several) m$, see
\eqref{(45.6)} (as otherwise all quarks will decouple as heavy at
$\mu<m/(\rm several)$ resulting in the wrong lower energy theory). In
this region of parameters, $\nt=N_c$ quarks are higgsed at $\mu\sim
(\mx m)^{1/2}\gg\lm$ {\it in the weak coupling regime}, e.g. (compare
with \eqref{(41.4.19)},\eqref{(41.4.20)}),
\bq
\langle{\ov Q}_{i=b+\nd}^{\, b}\rangle=\langle Q^{i=b+\nd}_b\rangle=
(\mx m)^{1/2}\gg\lm\,,\quad b=1...N_c\,,\quad i=\nd+1...N_F\,,
\label{(45.1.1)}
\eq
\bbq
\langle X^{adj,A}_{SU(N_c)}\rangle={\sqrt 2}\,{\rm Tr\,}\Bigl
[\langle{\ov Q}\rangle T^{A}\langle Q\rangle\Bigr ]/\mx =0\,,\quad
A=1...N_c^2-1\,,\quad \langle a_0\rangle=m\,,\quad \langle
S\rangle_{N_c}=0\,.
\eeq

Higgsed quarks with $N_c$ colors and $\nt=N_c$ flavors give in this
case large masses $\sim (\mx m)^{1/2}\gg\lm$ to themselves and to all
gluons and scalars, and simultaneously {\it prevent} all adjoint
scalars (which are light by themselves, in the sense $\mx\ll\lm$) from \, 
higgsing,  see second line in \eqref{(45.2)}, $\langle a_0\rangle=m,\,
\langle S\rangle_{N_c}=0$.

From \eqref{(45.3)},\eqref{(45.1.1)}, the multiplicity of these vs-vacua \, 
is  $N_{vs}=C^{N_c}_{N_F}$, the factor $C^{N_c}_{N_F}$ originates in
this phase from the spontaneous flavor symmetry breaking, $U(N_F)\ra
U(N_c)\times U(\nd)$, due to higgsing of $\nt=N_c$ out of $N_F$
quarks. This multiplicity shows that the non-trivial at $2N_c-N_F\geq
2$ discrete $Z_{\bb}$ symmetry is {\it unbroken}.

As far as $\mx$ remains sufficiently larger than $\lm^2/m$, the mass
spectrum includes in this phase: $N_c^2$ long ${\cal N}=2$ multiplets
of massive gluons and their ${\cal N}=2$ superpartners with masses
$\mu_{\rm gl}\sim (\mx m)^{1/2}\gg\lm$ and $2\no\nt=2\nd N_c$
massless \, Nambu-Goldstone multiplets (these are remained original 
electric  quarks with $\nd$ flavors and $N_c$ colors). There are no 
heavier  particles with masses $~\sim m$.

\subsection{$m\gg\lm\,,\,\, \mx\ll\lm^2/m\,,\,\, N_c < N_F < 2
N_c-1$}\numberwithin{equation}{subsection}

\hspace*{4mm} Consider now the case of smaller $\mx$, such that $(\mx
m)^{1/2}\ll\lm$, while $m$ remains large, $m\gg\lm$. The quark
condensates still look as in \eqref{(45.6)}, and $\,\langle
a_0\rangle=m$ {\it stays intact} and still "eats" all quark masses
$"m"$ in \eqref{(45.1)}. Therefore, because the $SU(N_c)$ theory is UV
free, its gauge coupling grows logarithmically with diminished energy
and, if nothing prevents, it will become $g^2(\mu)<0$ at $\mu<\lm$.
Clearly, even if quarks were higgsed as above in section 45.1, i.e.
$\langle Q^2\rangle=\langle{\ov Q}_2\rangle\sim (\mx m)^{1/2}$ 
(while  $\langle X^{adj}_{SU(N_c)}\rangle
\ll\lm$), see \eqref{(45.1.1)}, this would result only in appearance of \, 
small  particle masses $\sim (\mx m)^{1/2}\ll\lm$, so that {\it all
particles would remain effectively massless at the scale
$\mu\sim\lm\gg (\mx m)^{1/2}$} (this is especially clear in the
unbroken ${\cal N}=2$ theory at $\mx\ra 0$). Therefore, this will not
help and the problem with $g^2(\mu<\lm)<0$ cannot be solved in this
way (these quarks ${\ov Q}_2, Q^2$ are really not higgsed now at all,
i.e. $\langle Q^i_b\rangle=\langle{\ov Q}_i^{\,b}\rangle=0,\,
i=\nd+1...N_F,\, b=1...N_c$, see below).

As explained in Introduction, to really avoid the unphysical
$g^2(\mu<\lm)<0$, the field $X^{adj}_{SU(N_c)}$ is higgsed \,
necessarily \, in this case at $\mu\sim\lm$. Because the non-trivial 
$Z_{\bb\geq 2}$
discrete symmetry is {\it unbroken} in these vs-vacua, this results,
as in section 41.1, in $SU(N_c)\ra SU(\nd)\times U^{(1)}(1)\times
U^{\bb-1}(1)$, and main contributions $\sim\lm$ to particle masses
originate now from this higgsing of $X^{adj}_{SU(N_c)}$, see
\eqref{(41.1.1)},\eqref{(41.1.16)},
\bq
\langle\, X^{adj}_{SU(N_c)}\rangle=\langle
X^{adj}_{SU(\nd)}+X^{(1)}_{U(1)}+
X^{adj}_{SU(\bb)}\rangle,\label{(45.2.1)}
\eq
\bbq
\langle \sqrt{2} X^{adj}_{SU(\bb)}\rangle=C_{\bb} \lm\,{\rm
diag}\,\Bigl (\,\underbrace{0}_{\nd}\,;\,\underbrace{\omega^0,\,
\omega^1,\,...\,\omega^{\bb-1}}_{\bb} \,\Bigr )\,,\quad
\omega=\exp\{\frac{2\pi i}{\bb}\}\,,
\eeq
\bbq
\sqrt{2}\, X^{(1)}_{U(1)}=a_{1}\,{\rm
diag}(\,\underbrace{\,1}_{\nd}\,;\,\underbrace{\,c_1}
_{\bb}),\, c_1=-\,\frac{\nd}{\bb},\,\, \langle
Q^i_k\rangle=\langle{\ov Q}_i^{\,k}\rangle=0,\,\,k=
\nd+1...\,N_c,\,  i=1...\,N_F.
\eeq

Now, vice versa, higgsed $\langle X^{A}_{SU(\bb)}\rangle\sim\lm$\, 
{\it  prevent} quarks $Q^i_k,\, {\ov Q}^{\,k}_i,\, k=\nd+1...\,N_c,\,
i=1...\,N_F$ from higgsing, i.e. $\langle Q^i_k\rangle=\langle{\ov
Q}_i^{\,k}\rangle=0$, see second line in \eqref{(45.2)}, $\langle
a_0\rangle=m,\, \langle S\rangle_{N_c}=0$. The physical reason is 
that \, these  quarks acquire now masses $\sim\lm$, much larger than 
a  potentially possible scale $\sim (\mx m)^{1/2}$ of their coherent
condensate, see \eqref{(45.1.1)}. These quarks are too heavy and too
short ranged now and it looks unrealistic that they can form the
coherent condensate with $\langle Q^i_k\rangle=
(\mx\,  m)^{1/2}$  as before  in  section 45.1.

Besides, for independently moving heavy ${\ov Q}$ and $Q$, their
physical (i.e. path dependent) phases induced by interactions with the \,
{\it  light} $U^{2N_c-N_F}(1)$ photons fluctuate {\it independently and \,  
freely} at  least at all scales $\mu > \mu^{\rm low}_{\rm cut}=(\rm
several){\rm max}\{(\mx m)^{1/2},\, (\mx\lm)^{1/2} \}$. I.e., in any
case, $\langle Q^i_k\rangle_{\mu^{\rm low}_{\rm cut}}=\langle{\ov
Q}_i^{\,k}\rangle_{\mu^{\rm low}_{\rm cut}}=0$. Clearly, this concerns \, 
also all  other heavy charged particles with masses $\sim\lm$.
Therefore, after integrating out the theory \eqref{(45.1)} over the
interval from the high energy down to the scale $\mu^{\rm low}_{\rm
cut}$, all particles with masses $\sim\lm$ {\it decouple} as heavy
already at the scale $\mu<\lm/(\rm several)$, i.e. much above
$\mu^{\rm low}_{\rm cut}$, and definitely {\it the heavy quarks by
themselves do not give any contributions to masses of remaining
lighter degrees of freedom in the Lagrangian \eqref{(45.2.2)}} because
$\langle Q^i_k\rangle_{\mu^{\rm low}_{\rm cut}}=\langle{\ov
Q}_i^{\,k}\rangle_{\mu^{\rm low}_{\rm cut}}=0$.
And after the heavy particles decoupled at $\mu<\lm/(\rm several)$,
they themselves do not affect the lower energy theory. This is exactly \, 
the same  situation as with heavy quarks with masses $\sim\lm$
discussed in detail in section 41.1 (and qualitatively the same as
with \, $X^{adj}_{SU(\nd)}$ with masses $\sim\mx\gg (\mx m)^{1/2}$ in 
section  41.4), and only the parameter $m\gg\lm$ in this section while 
$m\ll\lm$ \, in  sections 41.1 and 41.4, and always $\mx\ll\lm$.

The lower energy original electric ${\cal N}=2\,\,SU(\nd)$ theory at
the scale $\mu=\mu^{\rm low}_{\rm cut}\ll\lm$, with $2\nd<N_F<\,
2N_c-1$  flavors of remained light quarks $Q^i_a, {\ov Q}_j^{\,a},\,\,
a=1...\nd,\,\, i,j=1...N_F$, is IR free and weakly coupled. The whole
matter superpotential has the same form as \eqref{(41.1.5)}, but with
omitted $a_2$ and $SU(\nd-\no)$ SYM part, $\no=\nd$, $m\ra (m-a_0)$,
and with the addition of $\sim\mx a_0^2\,$,
\bq
\w_{N_c}=\w_{\,\nd}+\w_{D}+\w_{a_0,\,a_1}\,,\label{(45.2.2)}
\eq
\bbq
\w_{\,\nd} =(m-a_0-a_1){\rm Tr}\,({\ov Q} Q)_{\nd}-{\rm Tr}\,\Bigl
({\ov Q}\sqrt{2}X_{SU(\nd)}^{\rm adj} Q\Bigr )_{\nd}+\mx
(1+\delta_2){\rm Tr}\,(X^{\rm adj}_{SU(\nd)})^2\,,
\eeq
\bbq
\w_{D}=\Bigl ( m-a_0-c_1 a_1 \Bigr )\sum_{j=1}^{\bb}{\ov D}_j
D_j\,-\sum_{j=1}^{\bb} a_{D,j}{\ov D}_j D_j
-\,\mx\lm \sum_{j=1}^{\bb}\omega^{j-1} a_{D,j}+\mx L\,\Bigl
(\,\sum_{j=1}^{\bb} a_{D,j}\Bigr )\,,
\eeq
\bbq
\w_{a_0,\,a_1}=\frac{\mx}{2}N_c\,a^2_0+\frac{\mx}{2}\frac{\nd
N_c}{\bt}\,(1+\delta_1)\,a^2_1+
\mx N_c\delta_3\, a_1(m-a_0-c_1 a_1)+\mx N_c\delta_4\, 
(m-a_0-c_1  a_1)^2\,,
\eeq
where $\delta_1$ and $\delta_2$ are the same as in \eqref{(41.1.12)}
(and $\delta_3=0$, see \eqref{(B.4)}).

Proceeding now similarly to section 41.1, we obtain from
\eqref{(45.2.2)}
\bq
\langle a_0\rangle=m\,,\quad \langle a_1\rangle=0\,,\quad 
\langle  a_{D,j}\rangle=0\,, \quad 
\langle X_{SU(\nd)}^{\rm adj}\rangle=0\,,\label{(45.2.3)}
\eq
\bbq
\langle{\ov D}_j D_j\rangle=\langle{\ov D}_j\rangle\langle D_j
\rangle=\,-\mx\lm\,\omega^{j-1}+\mx \langle L\rangle,\,\, 
\langle  L\rangle=m,\,\, \langle\Sigma_D\rangle=\sum_{j=1}
^{\bb}\langle{\ov  D}_j\rangle\langle D_j \rangle=(\bb)\mx\, m,
\eeq
\bbq
\langle\Qo\rangle_{\nd}=\langle{\ov Q}^1_1\rangle\langle
Q^1_1\rangle=\mx m=\langle\Qt\rangle_{N_c},\,\,
\langle\Qt\rangle_{\nd}=\sum_{a=1}^{\nd}\langle{\ov
Q}^{\,a}_2\rangle\langle Q^2_a\rangle=0=\langle\Qo
\rangle_{N_c}\,,
\eeq
\bbq
\langle{\rm Tr} ({\ov Q}
Q)\rangle_{\nd}=\nd\langle\Qo\rangle_{\nd}+N_c\langle\Qt
\rangle_{\nd}=\nd\mx  \, m\,, \quad \wmu\langle
S\rangle_{\nd}=\langle\Qo\rangle_{\nd}\langle\Qt
\rangle_{\nd}=0.\,\,\,
\eeq

From \eqref{(45.2.3)}, the multiplicity of these vs -vacua in this
phase is $N_{vs}=C^{\nd}_{N_F}=C^{N_c}_{N_F}$ as it should be, the
factor $C^{\nd}_{N_F}$ originates now from the spontaneous flavor
symmetry breaking, $U(N_F)\ra U(\nd)\times U(N_c)$, due to higgsing 
of \,   $\nd$  quarks in the $SU(\nd)$ color sector.

On the whole, the mass spectrum looks now as follows.\\
1) Due to higgsing of $X^{adj}_{SU(\bb)}$ at the scale $\sim\lm$,
$\,\,SU(N_c)\ra SU(\nd)\times U^{(1)}(1)\times U^{\bb-1}(1)$, there 
is \,   a large  number of original pure electrically $SU(\bb)$ charged
particles with the largest masses $\sim\lm$. They all are weakly
confined due to higgsing at the scale $\mu\sim (\mx m)^{1/2}$ of
mutually non-local with them $\bb$ BPS dyons $D_j, \ov D_j$ with the
nonzero $SU(\bb)$ adjoint magnetic charges, the string tension is
$\sigma^{1/2}_D\sim \langle{\ov D}_j D_j\rangle^{1/2}\sim (\mx
m)^{1/2}\ll\lm$. These original electrically charged particles form
hadrons with the mass scale $\sim\lm\,$. Clearly, higgsed $\langle
X^{adj}_{SU(\bb)}\rangle\sim\lm$ does not break {\it by itself} the
global flavor symmetry. All heavy quarks with $SU(\bb)$ colors have
equal masses $|C_{\bb}\lm|$ and are in the (anti)fundamental
representation of the non-Abelian $SU(N_F)$, they form e.g. a number
of $SU(N_F)$ adjoint hadrons with masses $\sim\lm$. Original $SU(\bb)$
\, adjoint  heavy gluons and scalars are flavorless and have different
masses $|\Lambda_j-\Lambda_k|,\, \Lambda_j=C_{\bb}\lm\omega^{j-1},\,
\omega=\exp\{2\pi i/(\bb)\}$. {\it There is no color-flavor locking in \, 
this  heaviest sector}\,:\, the global $U(N_F)$ is unbroken while color
$SU(\bb)\ra U^{\bb-1}(1)$. \\
2) Due to higgsing of these dyons $D_j, {\ov D}_j$, there are $\bb$
long ${\cal N}=2\,\, U(1)$ multiplets of massive photons, all with
masses $\sim \langle{\ov D}_j D_j\rangle^{1/2}\sim (\mx m)^{1/2}$.\\
3) There are $\nd^{\,2}$ long ${\cal N}=2$ multiplets of massive
gluons with masses $\sim \langle\Qo\rangle^{1/2}_{\nd}\sim (\mx
m)^{1/2}\ll\lm$ due to higgsing of $\nd$ flavors of original electric
quarks with $SU(\nd)$ colors.\\
4) $2 \no\nt=2\nd N_c$ (complex) Nambu-Goldstone multiplets are
massless (in essence, these are remained original electric quarks
with \, $\nt=N_c=N_F-\nd$ flavors and $SU(\nd)$ colors). There are 
no other  massless particles at $\mx\neq 0,\, m\neq 0$.\\
5) There are no particles with masses $\sim m\gg\lm$.

The $U(N_c)$ curve \eqref{(40.2)} has in this phase at $\mx\ra 0$\,\,
$\nd$ equal double roots $e_k= - m$ corresponding to $\nd$ out of
$N_F$ higgsed original electric quarks with $SU(\nd)$ colors, and
$\bb$ unequal double roots $e_j= -m+\omega^{j-1}\lm$ 
corresponding   to  dyons $D_j$.

It is worth recalling once more that higgsed $\langle\,
X^{adj}_{SU(\bb)}\rangle\sim\lm$ in \eqref{(45.2.1)} {\it does not
break by itself the global flavor symmetry, it is really broken only
in the $SU(\nd)$ color sector by higgsed $\nd$ quarks with $N_F$
flavors and $SU(\nd)$ colors}.\\

In essence, the situation here with the heavy not higgsed quarks
$Q^i_k, {\ov Q}^{\,k}_i,\, i=1...N_F,\, k=\nd+1...N_c$ with masses
$\sim\lm$ and with their chiral bilinear condensates (really, it is
more adequate to speak about mean values, using the word 
"condensate" \, only for "genuine", i.e. coherent, condensates of 
higgsed fields) {\it  is the  same} as in br2 vacua of section 41.1, see
\eqref{(41.1.16)}-\eqref{(41.1.26)} and accompanying text. In the lower
energy phase of this section at $\mx\ll\lm^2/m$, the nonzero mean
values of {\it non-factorizable} local bilinears
$\langle(\QQ)_{1.2}\rangle_{\bb}$ of these heavy quarks with masses
$\sim\lm$ originate from: a) the quantum loop effects in the strong
coupling (and non-perturbative) regime at the scale $\mu\sim\lm$,
transforming these bilinears {\it operators} of heavy quark fields
into bilinears {\it operators} of light $SU(\nd)$ quark fields and the \,
{\it  operator} $\Sigma_D$; b) and finally, from formed at much lower
scales $\mu\sim (\mx m)^{1/2}\ll\lm$ the genuine (i.e. coherent)
condensates of light higgsed quarks and dyons. I.e., the equations
\eqref{(41.1.17)}-\eqref{(41.1.19)} and
\eqref{(41.1.22)}-\eqref{(41.1.26)} (with $A=-2$) remain the same and
only the values of condensates are different in br2 and vs vacua, see
\eqref{(45.2.3)}.

In this section, the corresponding values of condensates of heavy
quarks with $SU(\bb)$ colors, $N_F$ flavors, and masses $\sim\lm$ in
these vs-vacua with $\nt=N_c,\, \no=\nd$, and $\langle
S\rangle_{N_c}=0$ differ from those in the br2 vacua with $\nt>N_c,\,
\no<\nd$ and $\langle S\rangle_{N_c}\neq 0$ in section 41.1 only by:\,
 1)\, $\ha\ra  m$  due to $SU(N_c)\ra U(N_c)$,\,\, 2) the absence of
non-leading power corrections originating in br2 vacua from the SYM
part, see \eqref{(45.3)},\eqref{(45.2.3)}, compare with
\eqref{(45.1.1)},\eqref{(41.1.17)},\eqref{(41.1.19)},
\bq
\langle\Qt\rangle_{N_c}=[\langle\Qt\rangle_{\bb}=\mx
m\,]+[\langle\Qt\rangle_{\nd}=\sum_{a=1}^{\nd}\langle
{\ov Q}^{\,a}_2\rangle\langle Q^2_a\rangle=0\,]=\mx m\,,
\label{(45.2.4)}
\eq
\bbq
\langle\Qo\rangle_{N_c}=[\langle\Qo\rangle_{\nd}=\langle{\ov
Q}^{\,1}_1\rangle\langle Q^1_1\rangle=\mx
m\,]+[\langle\Qo\rangle_{\bb}= - \mx m\,]=0\,,
\eeq
\bbq
\langle{\rm
Tr\,}\QQ\rangle_{\bb}=N_c\langle\Qt\rangle_{\bb}+\nd\langle\Qo
\rangle_{\bb}=\langle  \Sigma_D\rangle=(2N_c-N_F)\mx m\,,
\eeq
while $\langle \sqrt{2} X^{adj}_{SU(\bb)}\rangle$ in \eqref{(45.2.1)}
is the same as \eqref{(41.1.1)} and differs from \eqref{(45.1.1)}.\\

There is one important qualitative difference between the vs vacua
and \, br2  vacua. Because \eqref{(45.3)},
\eqref{(45.4)},\eqref{(45.2.2)},\eqref{(45.2.3)} remain the same at
$m\lessgtr\lm$, and $\langle a_0\rangle=m$ still "eats" all quark
masses $"m"$ in \eqref{(45.1)} at all $m\lessgtr\lm$, the whole physics \, 
in vs-vacua is qualitatively independent separately of the ratio
$m/\lm$ and {\it depends only on a competition between $(\mx m)^{1/2}$ \, 
and  $\lm$}. For this reason, the lower energy phase with $m\gg\lm,\,
(\mx m)^{1/2}\ll\lm$ in vs-vacua of this section continues smoothly
into the region $\mx\ll m\ll\lm$. This is in contrast with the br2
vacua with $\mx\ll m\ll\lm$ in section 41.1. In the latter, all
properties depend essentially on the ratio $m/\lm$ and they evolve
into the qualitatively different br1 vacua of section 43 at
$m\gg\lm$.\\

Comparing the mass spectra at $(\mx m)^{1/2}\gg\lm$ in section 45.1
with those in this section at $(\mx m)^{1/2}\ll\lm$ one can see the
qualitative difference between these two cases. {\it The lower energy
phase with $\mx\ll\lm^2/m$ in this section is, in essence, the phase
of the unbroken ${\cal N}=2$ theory because it continues smoothly 
down \, to $\mx\ra0$. While the higher energy phase of section 45.1 with
$\mx\gg\lm^2/m$, because it cannot be continued as it is down to
$\mx\ra 0$, is, in essence, the phase of the strongly broken ${\cal
N}=2$ theory (i.e. already the genuine ${\cal N}=1$ theory in this
sense)}. Remind that the applicability of the whole machinery with
the \, use of roots of the curve \eqref{(40.2)} is guaranteed only at 
$\mx\ra 0$. I.e., it is applicable only to the lower energy phase of this
section, and clearly not applicable to the higher energy phase of
section 45.1, in spite of that $\mx\ll\lm$ therein also.

And these two phases at $\mx\gtrless\lm^2/m$ are qualitatively
different, compare e.g. \eqref{(45.1.1)} and \eqref{(45.2.1)}. (In the
higher energy phase with
$\langle{\ov Q}^{\,b}_{i=b+\nd}\rangle=\langle
Q^{i=b+\nd}_b\rangle=(\mx m)^{1/2},\,\,b=1...N_c,\,\,i=\nd+1...N_F$
and $\langle m - a_0\rangle=0,\,\langle S\rangle_{N_c}=0$, the mean
value $\langle X^{\rm adj}_{SU(N_c)}\rangle\neq 0$ {\it is
incompatible with the Konishi anomaly in}
\eqref{(45.2)},\eqref{(45.6)}). Besides, in this higher energy phase in
section 45.1 all charged and neutral gluons from $SU(\bb)$ have {\it
exactly equal masses $| g (\mx m)^{1/2}|$ in the whole wide region
from $(\mx m)^{1/2}\gg\lm$ down to $(\mx m)^{1/2} >(\rm several)\lm$},\, 
see  \eqref{(45.1.1)}. On the other hand, in the lower energy phase of
this section, the masses of charged $SU(\bb)$ gluons differ greatly at
$ \, \mx\ra  0:\, \mu_{\rm gl}^{ij}\sim |(\omega^{i}-\omega^{j})\lm
|\,,\,\, i\neq j,\,\,i,j=0...\bb-1$, see \eqref{(45.2.1)}, and differ
from masses of neutral gluons. This shows that the dependence of
masses of these gluons on $\mx m$ is non-analytic. The mean values 
of  $\langle X^{adj}_{SU(N_c)}\rangle$ and $\langle
Q^i_k\rangle=\langle{\ov Q}_i^{\,k}\rangle,\, k=\nd+1...\,N_c,\,
i=1...\,N_F$ also behave non-analytically, compare \eqref{(45.1.1)}
and\eqref{(45.2.1)}. In this sense, we can say that there is a phase
transition somewhere at $(\mx m)^{1/2}\sim\lm$. This is in
contradiction with a widespread opinion that there are no phase
transitions in the supersymmetric theories because the gauge 
invariant \, mean values of chiral fields ( e.g. $\langle
(\QQ)_{1,2}\rangle_{N_c},\,\langle S\rangle_{N_c},\, \,\langle{\rm
Tr\,}(X^{adj}_{SU(N_c)})^2\rangle$ in \eqref{(45.2)}-\eqref{(45.4)})
depend holomorphically on the chiral superpotential parameters.

This phenomenon with the phase transition of the type described above
occurring at very small $\mx\ll\lm$ from the nearly unbroken ${\cal
N}=2$ at $\mx\ll\lm^2/m$ to the strongly broken ${\cal N}=2$ at
$\mx\gg\lm^2/m$, i.e. ${\cal N}=2\ra {\cal N}=1$, where the use of
roots of the curve \eqref{(40.2)} becomes not legitimate, appears
clearly {\it only} at $m\gg\lm$. Let us start e.g. with the nearly
unbroken ${\cal N}=2$ theory at {\it sufficiently small} $\mx$ where,
in particular, the whole machinery with the use of roots of the 
curve  \eqref{(40.2)} is still legitimate (for independent of $\mx$
quantities). In this section, when $\mx$ begins to increase, the
phase \, transition \, occurs at $\mx\sim (\lm^2/m)\ll\lm$. But in 
other cases,
see sections 46 and 47 below, it occurs even at parametrically smaller
values of $\mx$ than in this section. In br2 vacua of section 46 it
occurs at $\mx\sim \langle\Lambda_{SU(\nt)}\rangle^2/m,\,
\langle\Lambda_  {SU(\nt)}\rangle\ll\lm$, see \eqref{(46.1.6)}, and 
in   S- vacua of  section 47 it occurs at $\mx\sim
\langle\Lambda_{SU(N_F)}\rangle^2/m,\,\langle\Lambda_{SU(N_F)
}\rangle\ll\lm$,  see \eqref{(47.4)}. \\

Let us compare now the corresponding parts of unbroken non-Abelian
global symmetries of two phases, those in section 45.1 and here. It is
$SU(N_c)_{C+F}\times SU(\nd)_F$ of original pure electric particles in \, 
section  45.1, where "C+F" denotes the color-flavor locking. The higgsed \, 
massive electric  quarks $Q^{i=b+\nd}_b, {\ov Q}^{\,b}_{i=b+\nd},\,\,
b=1...N_c,\, i=\nd+1...N_F$, with masses $m_Q\sim (\mx
m)^{1/2}\gg\lm$, together with massive electric $SU(N_c)$ gluons and
$X^{adj}_{SU(N_c)}$ scalars form the long ${\cal N}=2$ multiplet in
the adjoint representation of $SU(N_c)_{C+F}$. The $2\nd N_c$
massless \, Nambu-Goldstone multiplets are bifundamental, these are
 original  not higgsed quarks $Q^i_a$ and ${\ov Q}^a_i$ with $\nd$ 
 flavors and  $N_c$ colors.

The corresponding non-Abelian global symmetry in this section is
$SU(\nd)_{C+F}\times SU(N_c)_F$ in the lighter sector, and higgsed
massive electric quarks $Q^{i=a}_a, {\ov Q}^{\,a}_{i=a},\,\,
a,i=1...\nd$, with masses $m_Q\sim (\mx m)^{1/2}\ll\lm$, together with \,  
$SU(\nd)$  gluons and $X^{adj}_{SU(\nd)}$ scalars form the long ${\cal
N}=2$ multiplet in the adjoint representation of $SU(\nd)_{C+F}$, see
the point "3" above. The $2\nd N_c$ massless Nambu-Goldstone
multiplets are also bifundamental. But, in addition, {\it all $N_F$
flavors of original electric quarks with $SU(\bb)$ colors have equal
large masses $\sim\lm$, so that the global $SU(N_F)$ is unbroken in
this heavy sector. In the limit $\mx\ra 0$ these heavy quarks are not
confined and form the (anti)fundamental representation of $SU(N_F)$}.
All original unconfined at $\mx\ra 0$ $\,SU(\bb)$-adjoint and hybrid
gluons and scalars with the large masses $\sim\lm$ are clearly flavor
singlets. At small $\mx\neq 0$ all these heavy pure electrically
charged particles are weakly confined and form hadrons with masses
$\sim \lm$, but this confinement does not break {\it by itself}
$SU(N_F)$. In particular, the heavy quarks form a number of $SU(N_F)$
adjoint hadrons.

Besides, it is worth noting the following. It is erroneous to imagine
that, because $\langle\Qt\rangle_{N_c}=\mx\, m$ still in the $\mx\, 
m\ll\lm^2$ phase of this section, then all quarks with $n_2 = N_c$
flavors are still higgsed in this lower energy phase as previously in
section 45.1 at $\mx m\gg\lm^2$, i.e. $\langle{\ov
Q}^{\,b}_{i=b+\nd}\rangle=\langle Q^{i=b+\nd}_b\rangle=(\mx\, 
m)^{1/2},\,\,b=1...N_c,\,\,i=\nd+1...N_F$, and so give corresponding
contributions to particle masses, but because $(\mx m)^{1/2}\ll\lm$
now, this will be of little importance. It is sufficient to notice
that, with $\langle X^{\rm adj}_{SU(\bb)}\rangle\sim\lm\neq 0$ now,
see \eqref{(45.2.1)}, this variant with still higgsed quarks with all
$N_c$ colors is incompatible with the Konishi anomalies in
\eqref{(45.2)},\eqref{(45.6)}.\\

The difference of adjoint representations
$SU(N_c)_{C+F}\leftrightarrow SU(\nd)_{C+F}$ at $\mx m\gtrless\lm^2$
was considered e.g. in \cite{SY4,SY2} as an evidence that all
$SU(\nd)$ particles cannot be the same original pure electric
particles and, by analogy with \cite{SY3}, it was finally claimed that \, 
they all  are dyons. E.e., the dyonic quarks ${\ov d}^{\,b}_i,\, d^i_b$\, 
with  $N_F$ flavors of this $SU(\nd)$ are composites of original
flavored quarks with $SU(\nd)$ colors and corresponding flavorless
magnetic monopoles.

This line of reasonings is in a clear contradiction with the picture
described above in this section. The reason for different adjoint
representations is that the phase and mass spectrum are determined by
$N_c$ out of $N_F$ higgsed quarks of $SU(N_c)$ at $\mx\,  m\gg\lm^2$
(and$X^{adj}_{SU(N_c)}$ is not higgsed,
$\langle X^{adj}_{SU(N_c)}\rangle=0$, see \eqref{(45.1.1)}), while they \, 
are  determined by higgsed both $X^{\rm adj}_{SU(2N_c-N_F)}$ and 
$\nd$  out of $N_F$ quarks of $SU(\nd)$ at $\mx m\ll\lm^2$, see
\eqref{(45.2.1)},\eqref{(45.2.3)}, (both phases in section 45.1 and here
at $m\gg\lm$, and the case $\mx\ll m\ll\lm$ clearly corresponds to 
the \, regime  $\mx m <\lm^2$, see section 45.3 below).

Besides, the above described in this section picture with the {\it
phase transition} at $\mx m\gtrless\lm^2$ (to avoid entering the
unphysical regime with $g^2(\mu<\lm)<0$ without higgsed $\langle
X^{adj}_{SU(\bb)}\rangle\sim\lm$), is in a clear contradiction with
the proposed by M.Shifman and A.Yung {\it a smooth analytic 
crossover} \,  from the region $\mx m\gg\lm^2$ to the 
"instead-of-confinement" regime \, in the region $\mx\, 
m\ll\lm^2$, with {\it still unbroken the
non-Abelian} $SU(N_c)_{C+F}$ global symmetry as it was 
in section 45.1 above, see e.g. \cite{SY4,SY2}. And the 
smooth analytic crossover  implies that
$N_c$ of $N_F$ original $SU(N_c)$ quarks are {\it still higgsed}, 
see  \eqref{(45.1.1)}, and only the value of their condensate 
is smaller  now,
\bq
\langle{\ov Q}_{i=b+\nd}^{\, b} Q^{i=b+\nd}_b\rangle=\langle{\ov
Q}_{i=b+\nd}^{\, b}\rangle\langle Q^{i=b+\nd}_b\rangle= (\mx
m)^{1/2}\ll\lm\,,\quad b=1...N_c\,,\quad
i=\nd+1...N_F\,.\label{(45.2.5)}
\eq

\vspace*{2mm}

At $\mx m\ll\lm^2$, according to M.Shifman and A.Yung,  see e.g.
\cite{SY6} and refs therein to their previous related papers. -

a) There is a large number of light particles with masses $\ll\lm$
which all are dyonic solitons. The first part of this set is the
$SU(\nd)$ gauge theory with $N_F$ flavors of above dyonic quarks
$d^i_b,\,{\ov d}^{\,b}_i,\, i=1...N_F,\, b=1...\nd$. $\,\nd$ them of
$N_F$ are higgsed at small $\mx\neq 0$ and, together with dyonic
$SU(\nd)$ gluons and scalars, form the adjoint representation of the
unbroken $SU(\nd)_{C+F}$ global symmetry. The second part of this set
includes $2 N_c\nd$ massless Nambu-Goldstone particles which are
remained non-higgsed dyonic quarks $d^i_b,\,{\ov d}^{\,b}_i,\,
i=\nd+1...N_F,\, b=1...\nd$. The third part includes $U^{\bb}(1)$
photon multiplets and $\bb$ dyons $\ov{\textsf{D}}_n,\,
{\textsf{D}}_n$ which are composites of diagonal original quarks
${\ov  Q}^{\,n}_n,\, Q^n_n,\, n=1 ... \bb$ and corresponding $SU(\bb)$
magnetic monopoles\,: ${\textsf{D}}_n=( Q^n_n+M_n)$. All quark-like
dyons (except for Nambu-Goldstone particles) are higgsed at 
$\mx\neq 0$.

As was emphasized in section 41.1, as a result of this picture, {\it
the dyons} ${\textsf{D}}^i_n,\, n=1...\bb,\, i=1...N_F$ {\it will
really realize the fundamental representation $N_F$ of the flavor}
$SU(N_F)$ and this will result in a wrong pattern of the flavor
symmetry breaking at $\mx\neq 0$. Besides, the number of such 
dyons  ${\textsf{D}}^i_n$ will be too large.

b) At the same time, in the region $\mx m\ll\lm^2$ of this section
{\it all original electric particles from the unbroken adjoint
representation of $SU(N_c)_{C+F}$ in section 45.1 mysteriously 
acquire  large equal masses} $\sim\lm$ (in addition to equal 
contributions  $\sim (\mx m)^{1/2}$ to their masses from still 
higgsed quarks, see  \eqref{(45.2.5)}).

It is not even attempted to recognize the dynamical mechanism
responsible for the origin of these large equal masses $\sim\lm$ of
all original particles forming the long ${\cal N}=2$ adjoint
representation of $SU(N_c)_{C+F}$.

Let us recall that in the whole weak coupling region $\mx\,  m\gg\lm^2$,
with quarks higgsed as in \eqref{(45.1.1)}, it follows from the Konishi \, 
anomalies  \eqref{(45.2)},\eqref{(45.6)} (valid for each flavor
separately) that all $\langle X^A\rangle=0,\, A=1...N_c^2-1$. If,
according to M.Shifman and A.Yung, the transition from $\mx m\gg
\lm^2$to $\mx\,  m\ll\lm^2$ is a smooth analytic crossover, then 
all $\langle  X^A\rangle=0$ at $\mx m\ll\lm^2$ also and quarks are 
higgsed still as in \eqref{(45.1.1)}. From where then appear large 
masses $\sim\lm$ of  all original particles forming the adjoint 
representation of  $SU(N_c)_{C+F}$ ? This contradicts the BPS 
properties of original  particles.

And it is not attempted to answer the question: are the properties of
this mysterious mechanism responsible for the appearance of these
large masses $\sim\lm$ compatible with the smooth analytic crossover
and with the unbroken global $SU(N_c)_{C+F}$ symmetry ? Recall that
the described above in this section concrete dynamical mechanism
responsible for the appearance of large masses $\sim\lm$ of all
original {\it charged} particles with $SU(\bb)$ colors, i.e. higgsed
$\langle X^{adj}_{SU(\bb)}\rangle\sim\lm$ \eqref{(45.2.1)}, is
incompatible with the unbroken global $SU(N_c)_{C+F}$ requiring 
equal  masses of all $N_c^2-1$ $\,SU(N_c)_{C+F}$ adjoint particles. 
E.g.,  charged $SU(\bb)$ adjoint gluons have different masses.

Besides, as explained in detail in section 41.1, e.g. the quarks
$Q^i_a,\,{\ov Q}^a_i,\, a=1...\bb,\, i=1...N_F$ with masses $\sim\lm$
at $(\mx m)^{1/2}\ll\lm$ are too heavy and too short ranged and 
cannot  form the coherent condensate with $\langle Q\rangle_
{\lm/(\rm  several)}=\langle{\ov Q}\rangle_{\lm/(\rm several)}\sim (\mx
\, m)^{1/2}$, because they decouple already at the scale $\sim\lm/(\rm
several)\gg (\mx\,  m)^{1/2}$ where $\langle Q\rangle_{\lm/(\rm
several)}=\langle{\ov Q}\rangle_{\lm/(\rm several)}=0$ and do not
affect by themselves further lower energy evolution of remained
lighter fields. In other words, as opposed to \eqref{(45.2.5)}, the
nonzero total mean values of heavy quarks bilinears
$\langle\Qt\rangle_{\bb}$ and $\langle\Qo\rangle_{\bb}$ in
\eqref{(45.2.4)} are non-factorizable. As explained in detail in
section 41.1, they are induced by nonzero genuine condensates of light
higgsed quarks with $SU(\nd)$ colors and light dyons ${\ov D}_j, D_j$.
Moreover, e.g. the appearance at $(\mx m)^{1/2}\ll\lm$ of the quark
mass terms $\sim\lm \sum_{a,i=1}^{N_c}{\ov Q}^a_i Q^i_a$ in the
superpotential contradicts the unbroken in these vs-vacua $Z_{\bb}$
symmetry, and also the mass terms $\sim\lm{\rm Tr}
(X^{adj}_{SU(N_c)})^2$ of all $N_c^2-1$ scalars in the superpotential
contradict both the unbroken $U_{R}(1)$ and $Z_{\bb}$ symmetries. The
very idea that original quarks acquired their masses $\sim\lm$ at
$\mx\ra 0$ not from higgsed $\langle X^{adj}_{SU(\bb)}\rangle\sim\lm$
but from some (unrecognized) outside sources contradicts their BPS
properties. Recall that in similar br2 vacua in section 41.1, with
higgsed $\langle X^{\rm adj}_{SU(\bb)}\rangle\sim\lm$ in
\eqref{(41.1.1)}, the condensates of light higgsed original quarks
$\langle\Qo\rangle=\langle{\ov Q}^1_1\rangle\langle Q^1_1\rangle$ 
in  \eqref{(41.1.4)} agree with {\it independently} calculated their
values from roots of the curve \eqref{(40.2)} (and the same in
vs-vacua \, considered \, here with $\no=\nd$). And because the 
values calculated  from roots are valid only for BPS particles,
{\it this confirms the BPS properties of original quarks}.

c) In addition, in this variant with the smooth analytic crossover,
all dyons $d^k_b,\, k,b=1...\nd$, ${\textsf{D}}_n,\, n=1...\bb$ {\it
and still quarks} $Q^i_a,\, a=1...N_c,\, i=\nd+1...N_F$ are {\it
higgsed simultaneously} with $\langle d\rangle\sim \langle
Q\rangle\sim (\mx m)^{1/2}\,,\, \langle{\textsf{D}}\rangle\sim \max
[(\mx\lm)^{1/2},\, (\mx m)^{1/2}]$, see e.g. \eqref{(45.2.5)}. But, in
any case, these quarks $Q^i_a$ are definitely mutually non-local with
respect to these dyons. How then can it be?

Comparing properties of proposed in \cite{SY3, SY4,SY2,SY6} the smooth
analytic crossover between regions $\mx m\gtrless\lm^2$ (with the
"instead-of-confinement" regime in the region $\mx \, m\ll\lm^2$), and
those with the phase transition described above in this section, it is \, seen 
that  these properties are, so to say, "orthogonal" to each other. \, And, 
on account of critical remarks presented above, we consider the
proposal of the smooth analytic crossover and the
"instead-of-confinement" regime as not self-consistent. \\

In addition, we would like to especially emphasize (and this concerns
the whole content of this paper, not only those of this section) that, \, for 
{\it  higgsed fields giving nonzero contributions to particle masses}, 
{\it the mean values of fields themselves should be nonzero},\,
i.e.  $\langle Q\rangle=\langle{\ov Q}\rangle\neq 0$ or $\langle
X\rangle\neq 0$, {\it not only} something like $\langle Q^{\dagger}
Q\rangle^{1/2}\neq 0$ or $\langle X^{\dagger} X\rangle^{1/2}\neq 0$
for the D-terms, or $\langle{\ov Q} Q\rangle^{1/2}\neq 0,\, \langle
X^2\rangle^{1/2}\neq 0$ for the F-terms (which are, as a rule, nonzero
\, even for  those fields which are really not higgsed).

This is especially clearly seen e.g. from the mass terms of fermionic
superpartners. For the D-terms of {\it higgsed} quarks or $X$ these
look as:
\footnote{\,
Really $\langle ...\rangle$ in \eqref{(45.2.6)},\eqref{(45.2.7)} have to  be
understood as $\langle ...\rangle_{\mu^{low}_{cut}=M/(\rm
several)}$, where $M$ is the pole mass of $Q,\, {\ov Q}$ or $X$}.
\label{(f58)}
\bq
{\rm Tr\,}\Bigl [\,\langle Q^{\,\dagger}\rangle\lambda
\chi+{\ov\chi}\lambda\langle{\ov Q}^{\,\dagger}\rangle\,\,
\Bigr ] \quad{\rm \, or}  \quad {\rm Tr\,}\Bigl (\,\langle X^
{\dagger}\rangle\,[\psi,  \lambda]\,\Bigr )\,,\label{(45.2.6)}
\eq
and for the F-terms they look as
\bq
{\rm Tr\,}\Bigl [\,{\ov\chi}\psi\langle Q\rangle+\langle{\ov
Q}\rangle\psi\chi\,\Bigr ]\quad {\rm or}\quad {\rm Tr\,}\Bigl [\,{\ov
\chi}\langle X\rangle \chi\,\Bigr ]\,.\label{(45.2.7)}
\eq

\subsection{$m\ll\lm$}

\hspace*{4mm} Comparing with the case $\mx\ll\lm,\,\, m\gg\lm$ in
sections 45.1-45.2, the case $\mx\ll m\ll\lm$ corresponds clearly the
regime $\mx\ll\lm^2/m$. The quark condensates remain the same as in
\eqref{(45.2.3)}, and $\langle a_0\rangle=m$ remains the same as in
\eqref{(45.2.3)} and still "eats" all quark masses $"m"$. {\it The
phase remains the same as in section} 45.2, and at $m\ll\lm$ the
difference with the case $m\gg\lm,\,\, \mx\ll\lm^2/m$ in section 45.2
is only quantitative, not qualitative, see \eqref{(45.2.3)}. The main
difference is that $\langle D_j\rangle\sim (\mx m)^{1/2}\gg
(\mx\lm)^{1/2}$ at $m\gg\lm$, while $\langle D_j\rangle\sim
(\mx\lm)^{1/2}\gg (\mx m)^{1/2}$ at $m\ll\lm$. Still, there are no
particles with masses $\sim m$, the only particle masses 
independent  of $\mx$ are $\sim\lm$.

\subsection{$N_F=N_c$}

Now, in short about the vs vacuum with $N=N_F=N_c=\nt,\, \nd=0=\no$
at \, $m\lessgtr\lm$ and $\mx\ll\lm$. There is only one such vacuum. And
really, there is nothing extraordinary with this case.

\subsubsection{Higher energy phase at $(\mx m)^{1/2}\gg\lm$}

As always in all vs vacua, see \eqref{(45.3)},\eqref{(45.4)},
$\langle\QQ\rangle_{N_c}=\mx m,\, \langle S\rangle_{N_c}=0$ in this
vacuum, and $\langle a_0\rangle=m$ "eats" all quark masses. The whole
color group $U(N_c)$ is broken in this higher energy phase as before
in vs-vacua of section 45.1 by higgsed quarks at the scale $\mu\sim
(\mx m)^{1/2}\gg\lm$ in the weak coupling regime, while $\langle
X^{adj}_{SU(N_c)}\rangle=0$ is not higgsed, see \eqref{(45.1.1)}. {\it
All $SU(N_c)$ gluons have equal masses $\sim (\mx m)^{1/2}$ in this
phase in the whole wide region from $(\mx m)^{1/2}\gg\lm$ down to
$(\mx m)^{1/2} >(\rm several)\lm$}. There is the residual global
non-Abelian $SU(N)_{C+F}$ symmetry (i.e. color-flavor locking).
$N^2-1$ long ${\cal N}=2$ multiplets of massive gluons are formed in
the adjoint representation of this remained unbroken global symmetry
$SU(N)_{C+F}$. There are no particles with masses $\sim m$.

The global $SU(N)_{F}$ is substituted by global $SU(N)_{C+F}$, so that \, 
there are  no massless Nambu-Goldstone particles in this vacuum. 
And  there are no massless particles at all in this phase.

\subsubsection{Lower energy phase at $(\mx m)^{1/2}\ll\lm$}

As in all vs-vacua, the total mean value of higgsed $a_0=\langle
a_0\rangle^{(\rm several) m}_{m/(\rm several)}+[\,{\hat a}_0^{\rm
soft}\,]^{m/(\rm several)}$ originates and saturates at $\mu\sim m$
and "eats" masses "m" of all quarks
\bq
\langle a_0\rangle=\frac{\langle{\,\rm Tr\,}\QQ\rangle_{N_c}}{N_c\mx
}=m\,,\quad \langle a_0\rangle=[\,\langle a_0\rangle^{(\rm several)\,  
m}_{m/(\rm  several)}=m\,]+[\,\langle {\hat a}_0^{\rm \, soft}
\rangle^{m/(\rm several)}_{\mu_{\rm cut}^{\rm lowest}=0}=0
\,]\,.\label{(45.4.1)}
\eq
In this phase, at all $m\gtrless\lm$, to avoid entering the unphysical \, 
regime  $g^2(\mu<\lm)<0$ in this UV free theory, the non-Abelian color
$SU(N_c)$ is also broken in the strong coupling and non-perturbative
regime at the scale $\mu\sim \lm$ by higgsed $\langle
X^{adj}_{SU(N_c)}\rangle\sim\lm,\, SU(N_c)\ra U^{N_c-1}(1)$, 
see  \eqref{(41.1.1)},\eqref{(45.2.1)},
\bq
X^{adj}_{SU(N_c)}=\langle X^{adj}_{SU(N_c)}\rangle^{(\rm\,
several)\lm}_{\lm/(\rm several)}+[\, {\hat X}^{adj,\rm\,
soft}_{SU(N_c)}\,]^{\lm/(\rm\, several)}\,, \label{(45.4.2)}
\eq
\bbq
\langle\sqrt{2} X^{adj}_{SU(N_c)}\rangle^{(\rm several)\lm}_
{\lm/(\rm \, several)}=C_N\lm\,{\rm\,
diag}\,\Bigl(\,\rho^0,\,\rho^1,\,...,\,\rho^{N_c-1}\,\Bigr ),\quad
\rho=\exp\{\,\frac{2\pi i}{N_c}\,\}\,,\quad \langle{\hat X}^{adj,\rm\,
soft}_{SU(N_c)}\rangle^{\lm/(\rm several)}_{\mu_{\rm\, cut}^{\rm\,
lowest}=0}=0\,. \eeq

The unbroken non-trivial discrete symmetry is $Z_{2N_c-N_F}=Z_{N\geq
2}$ in this vacuum, and it determines the standard pattern of this
color symmetry breaking by higgsed $\langle X^{adj}_{SU(N_c)}\rangle$
giving the largest contributions $\sim\lm$ to particle masses. All
original charged particles acquire masses $\sim\lm$, while $N_c=N$
light flavorless BPS dyons $D_j$ (massless at $\mx\ra 0$) are formed
at the scale $\mu\sim\lm$.

Clearly, higgsed flavorless $X^{adj}_{SU(N_c)}$ does not break by
itself the flavor symmetry $SU(N_F)$, i.e. {\it there is no
color-flavor locking}. The global non-Abelian flavor symmetry
$SU(N_F)=SU(N)_F$ remains unbroken, while the non-Abelian color
$SU(N_c)=SU(N)_C$ is broken down to Abelian one.

As before in section 45.2 for heavy quarks with $SU(\bb)$ colors, {\it
all quarks have equal masses}, but they are too heavy and too short
ranged now and cannot form the coherent condensate. I.e., they are
{\it not higgsed} in this phase. This is in accord with the Konishi
anomaly \eqref{(45.6)} (valid for each flavor separately) which
requires $\langle Q^i_i\rangle=\langle{\ov Q}^{\,i}_i\rangle=0,\,
i=1...N$ when $X^{adj}_{SU(N_c)}$ is higgsed as in \eqref{(45.4.2)}.

All equal mass quarks are clearly {\it in the (anti)fundamental
representation} of unbroken global $SU(N)_F$, while {\it the
non-Abelian color $SU(N)_C$ is broken at the scale 
$\mu\sim\lm$ down  to Abelian $U^{N-1}(1)$, 
and masses of all charged flavorless gluons  (and scalars) 
differ greatly in this lower energy phase at $\mx\ra
0:\, \mu_{\rm gl}^{ij}\sim |(\rho^{i}-\rho^{j})\lm |\,,\,\, i\neq
j,\,\,i,j=1...N$}, see \eqref{(45.4.2)}. There is the phase transition
at $(\mx m)^{1/2}\sim\lm$.

After all heavy particles with masses $\sim\lm$ decoupled at
$\mu=\lm/(\rm several)$, the low energy superpotential looks as
\bq
\w_{N_F}=\w_{D}+\w_{a}\,, \label{(45.4.3)}
\eq
\bbq
\w_{D}= (m-a_0)\sum_{j=1}^{N}{\ov D}_j D_j -\sum_{j=1}^{N}
a_{D,\,j}{\ov D}_j D_j -\mx\lm\sum_{j=1}^{N}\rho^{\,{j-1}}
a_{D,\,j}+\mx L\,\Bigl (\sum_{j=1}^{N} a_{D,\,j}\Bigr )\,
\eeq
\bbq
\w_{a}=\frac{\mx}{2} N \,a_0^2+\mx N\,{\hat\delta}_4\, (m-a_0)^2\,,
\quad N=2N_c-N_F=N_c=N_F\,.
\eeq

We would like to emphasize that all dyons in
\eqref{(41.1.5)},\eqref{(41.2.4)},\eqref{(42.4)},\eqref{(45.2.2)} 
and  \eqref{(45.4.3)} (and in \eqref{(46.2.2)},\eqref{(47.6)} 
below) are {\it the same} particles, and only
corresponding values of $2N^\prime_c-N^\prime_F$ in 
different   vacua  may be different.

From \eqref{(45.4.3)}
\bbq
\langle a_0\rangle=m\,,\quad \langle a_{D,\,j}\rangle=0\,,\quad
\langle{\ov D}_j\rangle\langle D_j \rangle= -\mx\lm\rho^{\,{j-1}}
+\mx\,  m\,,
\eeq
\bq
\langle\Sigma_D\rangle=\sum_{j=1}^{N}\langle{\ov D}_j\rangle\langle
D_j\rangle=N\mx\, m=\langle{\rm Tr\,}\QQ\rangle_{N_c}\,.\label{(45.4.4)}
\eq

All $N$ dyons in \eqref{(45.4.3)} are higgsed at small $\mx$ at the
scale $\mu_D\sim\langle{\ov D}_j D_j\rangle^{1/2},\, \mu_D\sim\max [
(\mx m)^{1/2},\\(\mx\lm)^{1/2}]$ (at $m\gg\lm$ below in this section
for definiteness), in the weak coupling regime $g_D(\mu_D)\ll 1$, and
$N$ long ${\cal N}=2$ multiplets of massive photons are formed with
masses $\sim \mu_D\ll\lm$. All original heavy charged particles with
masses $\sim\lm$ are weakly confined, the string tension is
$\sigma^{1/2}\sim\langle{\ov D}_j D_j\rangle^{1/2}\ll\lm$. In
particular, the heavy flavored quarks form a number of $SU(N_F)$
adjoint hadrons with masses $\sim\lm$. There are no massless
Nambu-Goldstone particles because the global flavor symmetry 
$U(N_F)$  is unbroken.\\

As for non-factorizable condensates of heavy non-higgsed in this lower \, 
energy  phase quarks with masses $\sim\lm$, \eqref{(41.1.16)} looks
similarly here, with $(\bb)\ra N_c$, as well as \eqref{(41.1.18)}.
Instead of \eqref{(41.1.20)} we have here
\bq
[\,{\rm Tr\,}(\QQ)_{N_c}\,]^{(\rm several)\lm}_{(\rm
several)(\mx\lm)^{1/2}}=\Bigl [\,{\hat d}_1 N_c
\mx {\hat a}^{\rm soft}_0+\Sigma_D\,\Bigr ]^{(\rm
several)(\mx\lm)^{1/2}},\quad
{\hat d}_1=O(1)\,.\,\label{(45.4.5)}
\eq
This operator expansion leads to the numerical equality
\bq
\langle {\rm Tr\,}(\QQ)_{N_c}\rangle^{(\rm\, several)\lm}_
{\mu_{\rm \,cut}^{\rm\,  lowest}=0}=
\langle\Sigma_D\rangle=\langle\Sigma_D\rangle^{(\rm\,
several) (\mx\, m)^{1/2}}_{\mu_{\rm\, cut}^{\rm\, lowest}=0}\,,\quad
\langle {\hat a}^{\rm soft}_0\rangle^{\lm/(\rm\, several)}_{\mu_
{\rm\,  cut}^{\rm\, lowest}=0}=0\,,\quad m\gg\lm \label{(45.4.6)}
\eq
as it should be, see \eqref{(45.4.1)},\eqref{(45.4.4)}. As before in
section 45.2 for $\langle{\rm Tr\,}\QQ\rangle_{\bb}$ at $\nd\geq 1$,
the nonzero total mean value of the heavy quark non-factorizable
bilinear operator originates and saturates not at the scale
$\mu\sim\lm$, but only at much lower energies
$\mu_D\sim (\mx m)^{1/2}$.\\

Moreover, at $\mx\ra 0$, because $\langle a_0\rangle=m$ "eats" all
quark masses, all charged solitons will have masses either $\sim\lm$
or zero. But the $U(N_c)$ curve \eqref{(40.2)} has only $N_c$ unequal
double roots $e^{D}_j= - m +\rho^{j-1}\lm,\, j=1...N_c$ in this
vacuum \, corresponding to $N_c=N$ massless BPS dyons $D_j$, 
and   shows that  
there are no {\it additional charged ${\cal N}=2$ BPS solitons
massless at $\mx\ra 0$}. In particular, there are no massless pure
magnetic monopoles or other additional dyons, they all have masses
$\sim\lm$. And there are no massless particles at all at small
$\mx\neq 0$. Besides, there are no particles with masses $\sim
m\gg\lm$, the largest masses are $\sim\lm$.\\

According to e.g. \cite{SY6} (and refs therein to their previous
related papers), the transition from $(\mx\, m)^{1/2}\gg\lm$ to 
$(\mx\, m)^{1/2}\ll\lm$ is a smooth analytical crossover with {\it still
higgsed quarks}, $\langle Q^i_a\rangle=\delta^i_a (\mx\, m)^{1/2},\,
i,a=1...N$, {\it and still unbroken $SU(N)_{C+F}$ global symmetry}.
And all original particles from the long ${\cal N}=2\,\, SU(N)_{C+F}$
adjoint multiplet mysteriously acquire equal masses $\sim\lm$ at
$(m \mx)^{1/2}<\lm$. Besides, e.g. each heavy original quark $Q^i$ 
with the  mysteriously 
appeared mass $\sim\lm$, decays into the (flavored  dyon-
flavorless antimonopole) pair $({\cal D}^i+{\ov{\cal M}})$ (both
particles with masses $\sim\lm$) in the "instead-of-confinement"
regime at $(\mx\, m)^{1/2} < \lm$. Due to higgsed at small $\mx\neq 0$
{\it light} dyons $\ov{\textsf{D}}_n,\, \textsf{D}_n,\, n=1...N$
(composed, according to \cite{SY6}, from diagonal quarks $Q^n_n$ and
flavorless magnetic monopoles $M_n:\, \textsf{D}_n=(Q^n_n+M_n)$\,),
these heavy ${\cal D}^i$ and ${\ov{\cal M}}$ are confined in this
"instead-of-confinement" regime, and this pair to which quark decays
forms a "stringy meson" with the mass $\sim\lm$. This is, so to say,
"internal decay", so that, as a whole, each pair $({\cal
D}^i+{\ov{\cal M}})$ remains the same quark $Q^i$.

Let us recall that, in any case, the considered theory is not at the
arbitrary point on the formal moduli space existing at the
mathematical point $\mx\equiv 0$ only, but stays in definite isolated
vacua - those at $\mx\neq 0$, even in the formal limit $\mx\ra 0$.  And\, 
e.g.  massive BPS quarks are at least marginally stable at $\mx\ra 0$
in vacua considered in this paper and can in this case only formally
dissolve "emitting" massless quanta.

And, in any case, similarly to section 45.2, it remains completely
unclear in the variant with the smooth crossover transition from
$(m \mx)^{1/2}\gg\lm$ to $(m \mx)^{1/2}\ll\lm$ proposed in \cite{SY6}\,:\, 
a)  from where mysteriously appeared large equal masses $\sim\lm$ of 
all  particles from still unbroken long ${\cal N}=2\,\, SU(N)_{C+F}$
adjoint multiplet (because, with still higgsed quarks $\langle
Q^i_a\rangle=\delta^i_a (\mx m)^{1/2},\, i,a=1...N$, the Konishi
anomaly \eqref{(45.6)} requires $\langle X^{adj}_{SU(N)}\rangle=0,\,
A=1...N^2-1$);\, b) because all $N^2-1\,\, SU(N)_{C+F}$ adjoint gluons \,
acquired  masses $\sim\lm$, from where then mysteriously appeared
$U^{N-1}(1)$ massless at $\mx\ra 0$ photon multiplets;\, c) how their
dyons $\textsf{D}_n=(Q^n_n+M_n)$ and mutually non-local with them
quarks $Q^n_n$ can be higgsed simultaneously at small $\mx\neq 0$.

\subsection{$m\gg\lm\,,\,\, 2N_c-N_F=1$}
\numberwithin{equation}{subsection}

At $\mx m\gg\lm^2$ the behavior of all vs-vacua is the same and is
described in section 45.1. Therefore, we consider below only the case
$\mx\,  m\ll\lm^2$ when the vs-vacua with $\bb=1$ and $\bb\geq 2$ 
behave  differently.

The pattern of flavor symmetry breaking remains fixed,
$\no=\nd=N_c-1\,,\,\, \nt=N_c\,,\,\, U(N_F)\ra U(\no)\times U(\nt)$,
as well as the multiplicity
$N_{vs}=C^{\,\nt=N_c}_{N_F}=C^{\,\no=\nd}_{N_F}$. And $\langle
a_0\rangle=m$ also remains the same in all vs -vacua (really, at all
$m\gtrless\lm$) and still "eats" all quark masses $"m"$. But the
discrete symmetry $Z_{2N_c-N_F}$ becomes trivial at $\bb=1$ and gives
no restrictions on the form of $\langle X^{\rm
adj}_{SU(N_c)}\rangle\sim\lm$ (remind that $\langle X^{\rm
adj}_{SU(N_c)}\rangle\sim\lm$ is higgsed necessarily at $\mx
m\ll\lm^2$ to avoid $g^2(\mu<\lm)<0$ ). As above in section 45.2, the
right flavor symmetry breaking implies the unbroken $SU(\nd)$ lower
energy group and a presence of $N_F$ flavors of quark-like particles
massless at $\mx\ra 0$, then $\no=\nd$ of them will be higgsed at
small $\mx\neq 0$ and there will remain $2\nd N_c$ 
Nambu-Goldstone  multiplets.

However, it is not difficult to see that a literal picture with {\it
light all BPS original electric particles} of remained unbroken at the 
scale  $\mu\sim\lm$ electric subgroup $SU(\nd)$ cannot be right at
$\bb=1$. Indeed, $X^{\rm adj}_{SU(N_c)}$ is higgsed now as 
$\langle  X^{\rm adj}_{SU(N_c)}\rangle\sim\lm\, {\rm diag}(\,1... 1\,;\,
-\nd\,),\,\,\nd=N_c-1$. And all original quarks $Q^i,\,{\ov Q}_i$ 
with \, $SU(N_c)$  colors acquire masses $\sim\lm$ and 
decouple at  $\mu\lesssim\lm$.
\footnote{\,
Clearly, at $m\ll\lm,\, \mx\ra 0$ and $\bb=1$, this argument concerns
also $SU(N_c)$ br2 vacua with $0<\no<\nd$ and the multiplicity
$(\nd-\no)C^{\no}_{N_F}$, and S vacua with the multiplicity $\nd$,
both with $\langle S\rangle_{N_c}\neq 0$ and with the $SU(\nd=N_c-1)$
lower energy gauge group at energy $\mu<\lm/(\rm several)$.
\label{(59)}
}

Therefore, the vs-vacua with $2N_c-N_F=1$ are not typical vacua, they
are really exceptional. The absence of the non-trivial unbroken
$Z_{2N_c-N_F}$ symmetry is crucial.
\footnote{\,
Considered in \cite{SY3} example of $U(N_c)$ vs-vacua ($r=N_c$ 
vacua  in the language of \cite{SY3}) with $N_c=3,\, N_F=5,\,
\no=\nd=N_c-1=2$ and $\langle S\rangle_{N_c}=0$ belongs just 
to this  exceptional type of vs-vacua with $2N_c-N_F=1$.
\label{(60)}
}

The possible exceptional properties of vacua with $\bb=1$ at $\mx\ra
0$ are indicated also by the curve \eqref{(40.2)}, its form is changed
at $\bb=1$\,:\, $m\ra m+(\lm/N_c)$ \cite{APS}.

For this reason we do not deal with such vacua in this paper.

\section{$\mathbf{SU(N_c),\,\,m\gg\lm,\,\, U(N_F)\ra U(\no)\times
U(\nt)},$\,\, br2 vacua}

\subsection{Larger $\mx$}

\hspace*{4mm} The quark condensates in these br2 vacua with $\no\geq
1,\, \nt<N_c$ are obtained from \eqref{(43.1)},\eqref{(43.1.1)} by the
replacement $\no\leftrightarrow\nt$, i.e.\,:
$\langle\Qt\rangle_{N_c}\approx\mx \ha\gg\langle\Qo\rangle_{N_c},\,
\ha=m \,N_c/(N_c-\nt)$. The  discrete $Z_{\bb\geq 2}$ 
symmetry is also unbroken in these vacua and
the multiplicity is $N_{\rm br2}=(N_c-\nt) C^{\,\nt}_{N_F}$. They
evolve at $m\ll\lm$ to Lt-vacua with  $\langle\Qo\rangle_{N_c}
\sim\langle\Qt\rangle_{N_c}\sim\mx\lm$.

Similarly to br1 vacua, the scalars $X^{adj}$ are also higgsed at the
largest scale $\mu\sim m\gg\lm$ in the weak coupling region, now as
$SU(N_c)\ra SU(\nt)\times U(1)\times SU(N_c-\nt),\,\, N_F/2<\nt<N_c$,
so that (the leading terms only)
\bq
\langle\Qt\rangle_{N_c}\approx\mx \ha\,,\quad
\langle\Qo\rangle_{N_c}\approx\mx\,  \ha\Bigl (\frac{\lm}{\ha}\Bigr
)^{\frac{2N_c-N_F}{N_c-\nt}} \,,\quad \ha=\frac{N_c}{N_c-\nt}\, m\,,
\label{(46.1.1)}
\eq
\bbq
\langle
S\rangle_{N_c}=\frac{\langle\Qo\rangle_{N_c}\langle\Qt\rangle_{N_c}}{\mx}
\approx\mx \, \ha^2\Bigl (\frac{\lm}{\ha}\Bigr )^{\frac{2N_c-N_F}{N_c-\nt}}
\ll\mx \, m^2\,,
\quad\frac{\langle\Qo\rangle_{N_c}}{\langle\Qt\rangle_{N_c}}\approx\Bigl
(\frac{\lm}{\ha}\Bigr)^{\frac{2N_c-N_F}{N_c-\nt}}\ll 1\,,
\eeq
\bq
\langle X^{\rm adj}_{SU(N_c)}\rangle\equiv \langle\,
X^{adj}_{SU(\nt)}+ X_{U(1)}+
X^{adj}_{SU(N_c-\nt)}\,\rangle\,,\label{(46.1.2)}
\eq
\bbq
\sqrt{2}\, X_{U(1)}=a \,{\rm diag}\Bigl
(\underbrace{\,1}_{\nt}\,;\,\underbrace{\,\wh{c}}_{N_c-\nt}\, \Bigr
)\,,\quad \wh{c}=-\frac{\nt}{N_c-\nt}\,,\quad \langle a \rangle= m\,,
\eeq
\bbq
\mx\langle{\rm Tr}\,(\sqrt{2} X^{\rm
adj}_{SU(N_c)})^2\rangle=(2N_c-N_F)\langle S\rangle_{N_c}+ m
\langle{\rm Tr}\,(\QQ)\rangle_{N_c}\approx m (\nt\mx \ha)\,.
\eeq
As a result, all quarks charged under $SU(N_c-\nt)$ and hybrids
$SU(N_c)/[\,SU(\nt)\times SU(N_c-\nt)\times U(1)\,]$ acquire large
masses $\ha=m-{\wh c}\,\langle a\rangle=m N_c/(N_c-\nt)$ and decouple
at scales $\mu<\,\ha$, there remains ${\cal N}=2\,\, SU(N_c-\nt)$\,
SYM with the scale factor $\Lambda^{SU(N_c-\nt)}_
{{\cal N}=2\, SYM}$ of its gauge coupling, see also \eqref{(46.1.1)},
\bq
\langle\Lambda^{SU(N_c-\nt)}_{{\cal N}=2\, SYM}\rangle^2=\Bigl
(\frac{\lm^{\bb}{(\ha)}^{N_F}}{{(\ha)}^{2\nt}
}\Bigr )^{\frac{1}{(N_c-\nt)}}={\ha^2}\Bigl (\frac{\lm}{{\ha}}\Bigr
)^{\frac{\bb}{(N_c-\nt)}}\ll\lm^2\,,\quad \ha=\frac{N_c}{N_c-\nt}
m\,,\label{(46.1.3)}
\eq
\bbq
\langle S\rangle_{N_c-\nt}=\mx\langle\Lambda^{SU(N_c-\nt)}_{{\cal
N}=2\, SYM}\rangle^2\approx\langle S\rangle_{N_c}\,.
\eeq

At $\mx\ll\langle\Lambda^{SU(N_c-\nt)}_{{\cal N}=2\, SYM}\rangle$ it
is higgsed in a standard way \cite{DS}, $SU(N_c-\nt)\ra
U^{N_c-\nt-1}(1)$, due to higgsing of $X^{adj}_{SU(N_c-\nt)}$
\bq
\langle {\sqrt 2}
X^{adj}_{SU(N_c-\nt)}\rangle\sim\langle\Lambda^{SU(N_c-\nt)}_{{\cal
N}=2\, SYM}\rangle\, {\rm diag} \Bigl
(\underbrace{0}_{\nt}\,;\,\underbrace{k_1,\,...\,k_{N_c-\nt}}_{N_c-\nt}\Bigr
)\,,\quad k_i=O(1)\,.\label{(46.1.4)}
\eq
Note that the value $\langle\Lambda^{SU(N_c)}_{{\cal N}=2\,
SYM}\rangle$ of $\langle X^{adj}_{SU(N_c-\nt)}\rangle$ in
\eqref{(46.1.3)},\eqref{(46.1.4)} is consistent with the unbroken
$Z_{\bb}$ discrete symmetry.

As a result, $\,N_c-\nt-1$ pure magnetic monopoles $M_{\rm n}$
(massless at $\mx\ra 0$) with the $SU(N_c-\nt)$ adjoint charges are
formed at the scale $\mu\sim\langle\Lambda^{SU(N_c-\nt)}_{{\cal
N}=2\,SYM}\rangle$ in this SYM sector. These correspond to $N_c-\nt-1$
unequal double roots of the curve \eqref{(40.2)} connected with this
SYM sector. Note also that two single roots with $(e^{+} -e^{-})
\sim\langle\Lambda^{SU(N_c)}_{{\cal N}=2\, SYM}\rangle$ of the curve
\eqref{(40.2)} also originate from this SYM sector. Other $\nt$ double
roots of the curve at $\mx\ra 0$ originate in these vacua from the
$SU(\nt)$ sector, see below.

These $N_c-\nt-1$ monopoles are all higgsed with $\langle M_{\rm
n}\rangle=\langle {\ov M}_{\rm n}\rangle\sim \,(\mx\langle
\Lambda^{SU(N_c-\nt)}_{{\cal N}=2\, SYM}\rangle)^{1/2}$, so \, that
$N_c-\nt-1$ long ${\cal N}=2$ multiplets of massive photons
appear, with masses $\sim (\mx\langle\Lambda^{SU(N_c-\nt)}_{{\cal
N}=2\, SYM}\rangle)^{1/2}$. Besides, this leads to a weak confinement
of all heavier original $SU(N_c-\nt)$ electrically charged particles,
i.e. quarks and all hybrids with masses $\sim m$, and all ${\cal
N}=2\,\, SU(N_c-\nt)$ SYM adjoint charged particles with masses
$\sim\langle\Lambda^{SU(N_c-\nt)}_{{\cal N}=2\, SYM}\rangle$,\, 
the  tension of the confining string is $\sigma^{1/2}_2\sim
(\mx\langle\Lambda^{SU(N_c-\nt)}_{{\cal N}=2\,
SYM}\rangle)^{1/2}\ll\langle\Lambda^
{SU(N_c-\nt)}_{{\cal N}=2\, SYM}\rangle\ll\lm\ll m$. The factor
$N_c-\nt$ in the multiplicity of these br2 vacua arises just from 
this \, ${\cal  N}=2\,\, SU(N_c-\nt)$\, SYM part.

At larger $\mx$ in the range $\langle\Lambda^{SU(N_c-\nt)}_{{\cal
N}=2\, SYM}\rangle\ll\mx\ll m$, the scalars $X^{adj}_{SU(N_c-\nt)}$
become too heavy, their light $SU(N_c-\nt)$ physical phases fluctuate
then freely at the scale $\sim\mx^{\rm pole}=g^{2}(\mu=\mx^{\rm
pole})\mx$ and they are not higgsed. Instead, they decouple as heavy
in the weak coupling region at scales $\mu<\mx^{\rm pole}/(\rm
several)$. There remains then ${\cal N}=1\,\, SU(N_c-\nt)$ SYM with
the scale factor $\Lambda^{SU(N_c-\nt)}_{{\cal N}=1\, SYM}$ of its
gauge coupling
\bq
\langle S\rangle_{{\cal N}=1\,
SYM}=\langle\Lambda^{SU(N_c-\nt)}_{{\cal N}=1\, SYM}\rangle^3
=\mx\langle\Lambda^{SU(N_c-\nt)}_{{\cal N}=2\, SYM}\rangle ^2,\,\,
\,\, \langle\Lambda^{SU(N_c-\nt)}_{{\cal N}=2\,
SYM}\rangle\ll\langle\Lambda^{SU(N_c-\nt)}_{{\cal N}=1\,
SYM}\rangle\ll\mx\ll m\,.\quad\label{(46.1.5)}
\eq
Therefore, there will be in this case a large number of strongly
coupled $SU(N_c-\nt)$ gluonia with the mass scale  $\sim\langle
\Lambda^{SU(N_c-\nt)}_{{\cal N}=1\, SYM}\rangle$, while
all original heavier charged particles with $SU(N_c-\nt)$ 
electric  charges and masses either $\sim
m\gg\langle\Lambda^{SU(N_c-\nt)}_{{\cal N}=1\, SYM}\rangle$ 
or $\sim  g^{2}\mx\gg\langle\Lambda^{SU(N_c-\nt)}_{{\cal N}=1\, 
SYM}\rangle$  still will be weakly confined, the tension of the 
confining string is  larger in this case,
$\sigma^{1/2}_1\sim\langle\Lambda^{SU(N_c-\nt)}_{{\cal N}=1\,
SYM}\rangle$. This ${\cal N}=1\,\,SU(N_c-\nt)\,$ SYM gives the 
same  factor $N_c-\nt$ in the multiplicity of these vacua.\\

Now, about the electric $SU(\nt)$ part decoupled from 
$SU(N_c-\nt)$  SYM. The scale factor of its gauge coupling is
\bq
\langle\Lambda_{SU(\nt)}\rangle=\Bigl
(\,\frac{\lm^{\bb}}{(\ha)^{2(N_c-\nt)}}\,\Bigr
)^{\frac{1}{2\nt-N_F}}=\ha\Bigl (\frac{\lm}{\ha} \Bigr
)^{\frac{\bb}{2\nt-N_F}}\ll\lm\,,\quad \ha=\frac{N_c}{N_c-\nt}
m\,.\label{(46.1.6)}
\eq
Note that the value of $\langle\Lambda_{SU(\nt)}\rangle$ in
\eqref{(46.1.6)} is also consistent with the unbroken $Z_{\bb}$
discrete symmetry.

What {\it qualitatively differs} this $SU(\nt)$ part in br2 vacua at
scales $\mu<\, m$ from its analog $SU(\no)$ in br1 vacua of section
43.1 is that $SU(\no)$ with $N_F>2\no$ is IR free, while $SU(\nt)$
with \, $N_F<2\nt$ \, is UV free, and its small ${\cal N}=2$ gauge coupling 
at  the scale $\mu\sim m\gg\lm$ begins to grow logarithmically with
diminished energy at $\mu<m$. Therefore, {\it if nothing prevents},
$\,\,X^{adj}_{SU(\nt)}$ will be higgsed necessarily at the scale
$\sim\langle\Lambda_{SU(\nt)}\rangle$, with $\langle
X^{adj}_{SU(\nt)}\rangle\sim\langle\Lambda_{SU(\nt)}\rangle$, to
avoid \, unphysical \, $g^2(\mu<\langle\Lambda_{SU(\nt)}\rangle)<0$
 (see section  46.2 below).

But if $(\mx m)^{1/2}\gg\langle\Lambda_{SU(\nt)}\rangle$, {\it the
phase will be different}. The leading effect in this case will be the
breaking of the whole $SU(\nt)$ group due to higgsing of $\nt<N_c$ 
out \, of $N_F$  quarks $Q^i,\,{\ov Q}_i$ in the weak coupling region, 
with  e.g. $\langle Q^{i=b+\no}_b\rangle\sim \delta^{i}_b (\mx
m)^{1/2},\,\,b=1...\nt,\,\, i=\no+1... N_F$. This will give the 
additional  factor  $C^{\,\nt}_{N_F}$ 
in the multiplicity of these br2 vacua due to  spontaneous 
flavor symmetry breaking $U(N_F)\ra U(\no)\times U(\nt)$,
so that the overall multiplicity will be $(N_c-\nt)C^{\,\nt}_{N_F}$,
as it should be. The corresponding parts of the superpotential will be \,  
as in  \eqref{(43.1.3)},\eqref{(43.1.4)}, with a replacement $\no\ra\nt$,
\bq
{\cal W}_{\nt}=(\, m-a\,)\,{\rm Tr}\,({\ov Q} Q)_{\nt} - {\rm
Tr}\,({\ov Q}\sqrt{2} X^{\rm adj}_{SU(\nt)} Q)_{\nt} +\mx 
{\rm Tr}\,( X^{\rm adj}_{SU(\nt)})^2+\frac{\mx}{2}\,\frac{\nt
N_c}{N_c-\nt}\,a^2\,. \label{(46.1.7)}
\eq

From \eqref{(46.1.7)} in the case considered (no summation 
over $j$ in  \eqref{(46.1.8)}\,)
\bbq
\langle a\rangle=m,\,\, \langle X^{adj}_{SU(\nt)}
\rangle=0,\,\,\langle\Qt\rangle_{\nt}=\sum_{b=1}^{\nt}
\langle{\ov Q}_2^{\,b} Q^2_b\rangle=\langle{\ov
Q}_2^{\,2}\rangle\langle Q^2_2\rangle\approx\frac{N_c}
{N_c-\nt}\mx\, m=\mx\, \ha\,,
\eeq
\bq
\langle \Qo\rangle_{\nt}=\sum_{b=1}^{\nt}\langle{\ov
Q}_j^{\,b}\rangle\langle Q^j_b\rangle=0\,,\,\, j=1...\no,\,\,\langle
{S}\rangle_{\nt}=\frac{1}{\mx}\langle \Qt\rangle_{\nt}\langle
\Qo\rangle_{\nt}=0\,.\quad\label{(46.1.8)}
\eq

$2\no\nt$ massless Nambu-Goldstone multiplets are formed as a result
of the spontaneous flavor symmetry breaking, $\Pi_j^i= ({\ov Q}_{j}
Q^i)_{\nt}\,\ra\,({\ov Q}_{j}\langle Q^i\rangle)_{\nt},\,\,\Pi_i^j=
({\ov Q}_{i} Q^j)_{\nt}\,\ra\,(\langle{\ov Q}_{i}\rangle
Q^j)_{\nt},\,\, i=\no+1...N_F,\,\, j=1...\no,\,\,
\langle\Pi^i_j\rangle=\langle \Pi^j_i\rangle=0$ (in essence, these
are \, not higgsed quarks with $\no$ flavors and $SU(\nt)$ colors). 
Besides,  there are $\nt^2$ long ${\cal N}=2$ multiplets of massive 
gluons with  masses $\sim (\mx m)^{1/2}$.

We emphasize that higgsing of $\nt$ quarks at the higher scale 
$(\mx\, m)^{1/2}\gg\langle\Lambda_{SU(\nt)}
\rangle$, see \eqref{(46.1.7)}, \eqref{(46.1.8)}, {\it prevents
$X^{adj}_{SU(\nt)}$ from higgsing}, i.e. $\langle
X^{adj}_{SU(\nt)}\rangle_{\nt}$ is not simply smaller, but 
exactly  zero (see also section 45.1 for a similar regime).

On the whole for these br2 vacua in the case considered, i.e. at
$(m \mx)^{1/2}\gg\langle\Lambda_{SU(\nt)}
\rangle$, all qualitative properties are similar to those of br1 
vacua \, in  section \, 43.1.

\subsection{Smaller $\mx$}

\hspace*{4mm} Consider now the behavior of the $SU(\nt)\times U(1)$
part in the opposite case of smaller
$\mx\ll\langle\Lambda_{SU(\nt)}\rangle^2/m,\,
\langle\Lambda_{SU(\nt)}\rangle\ll\lm\ll m$. As will be seen below, in \, 
this case  even all qualitative properties of mass spectra in this part \, 
will be quite  different (the behavior of the $SU(N_c-\nt)$ SYM part
decoupled from $SU(\nt)\times U(1)$ was described above in section 46.1).

Now, at smaller $\mx$, instead of quarks, the scalar field
$X^{adj}_{SU(\nt)}$ is higgsed first at the scale
$\sim\langle\Lambda_{SU(\nt)}\rangle$ to avoid unphysical
$g^2(\mu<\langle\Lambda_{SU(\nt)}\rangle)<0$ of UV free ${\cal
N}=2\,\, SU(\nt)$. Although $\langle\Lambda_{SU(\nt)}\rangle$ respects \, 
the  original $Z_{\bb}$ discrete symmetry of the UV free $SU(N_c)$
theory, see \eqref{(46.1.6)}, because the $SU(N_c-\nt)$ SYM part is
decoupled, there is now the analog of $Z_{\bb}$ in the $SU(\nt)$ 
group \, with  $N_F$ \, flavors of quarks, this is the unbroken non-trivial
$Z_{2\nt-N_F}=Z_{\nt-\no}$ discrete symmetry with $\nt-\no\geq 2$,
\,${\rm b}_2=(\bb)\ra {\rm b}^{\prime}_2=(2\nt-N_F)=(\nt-\no)$.
Therefore, the field $X^{adj}_{SU(\nt)}$ is higgsed at $\nt-\no\geq 2$ \, 
at the  scale $\mu\sim\langle\Lambda_{SU(\nt)}\rangle$ (in the strong
coupling and nonperturbative regime), qualitatively similarly to
\eqref{(41.1.1)},\eqref{(41.1.2)} in section 41.1\,: $SU(\nt)\ra SU(\no)\times
U^{(1)}(1)\times U^{\nt-\no-1}(1)$. There will be similar dyons $D_j$
(massless at $\mx\ra 0$) etc., and a whole qualitative picture will be \, 
similar  to \, those in section 41.1, with evident replacements of
parameters. The only qualitative difference is that the additional SYM \, 
part is  absent here because $\no$ quark flavors are higgsed now in
$SU(\no)$, see \eqref{(41.1.1)},\eqref{(41.1.16)},
\bq
\langle X^{adj}_{SU(\nt)}\rangle=\langle
X^{adj}_{SU(\no)}+X^{(1)}_{U(1)}+
X^{adj}_{SU(\nt-\no)}\rangle,\label{(46.2.1)}
\eq
\bbq
\langle\sqrt{2}
X^{adj}_{SU(\nt-\no)}\rangle=C_{\nt-\no}\langle\Lambda_{SU(\nt)}
\rangle\, {\rm\, diag}\,\Bigl(\,\underbrace{0}_{\no}\,;\,\underbrace{\,
\tau^0,\,\tau^1,\,..., \,\tau^{\nt-\no-1}}_{\nt-\no}\,; \underbrace{0}_
{N_c-\nt}\,\Bigr )\,,\quad \tau=\exp\{\frac{2\pi i}{\nt-\no}\}\,,
\eeq
\bbq
\sqrt{2}\, X^{(1)}_{U(1)}=a_{1}\,{\rm
diag}\,(\,\underbrace{\,1}_{\no}\,;\, \underbrace{\, 
\wh{c}_1}_{\nt-\no}\,;\underbrace{0}_{N_c-\nt}\,),\quad
\wh{c}_1=-\,\frac{\no}{\nt-\no}\,.\quad
\eeq
Therefore, similarly to \eqref{(41.1.2)},\eqref{(45.2.2)}, in this case
the low energy superpotential of this $SU(\nt)\times U(1)$ sector 
can  be written as
\footnote{\,
In other words, in this case the $SU(N^\prime_c=\nt)\times U(1)$ part
of $SU(N_c)$ with $m\gg\lm$ and with decoupled $SU(N_c-\nt)$ 
SYM part, \, {\it is in \, its own vs-vacuum} of section 45.2, with
$\langle\,{S}\,\rangle_{N^\prime_c}=0$ and with $a$ of $U(1)$ 
in \eqref{(46.2.2)} playing a  role of $a_0$ in \eqref{(45.2.2)}, etc.
\label{(f61)}}
\bq
\w_{\nt}=\w_{\no}+\w_{D}+\w_{a,\,a_1}+\dots\,,\label{(46.2.2)}
\eq
\bbq
\w_{\no} =(m-a-a_1){\rm Tr}\,({\ov Q} Q)_{\no}-{\rm Tr}\,
\Bigl ({\ov Q}\sqrt{2}X_{SU(\no)}^{\rm adj} Q\Bigr )_{\no}+\mx
(1+\wh{\delta}_2){\rm Tr}\,(X^{\rm adj}_{SU(\no)})^2\,,
\eeq
\bbq
\w_{D}=\Bigl ( m-a-\wh{c}_1 a_1 \Bigr )\sum_{j=1}^{\nt-\no}
{\ov D}_j D_j -\sum_{j=1}^{\nt-\no} a_{D,j}{\ov D}_j D_j
-\mx\Lambda_{SU(\nt)}\sum_{j=1}^{\nt-\no}\tau^{j-1} a_{D,j}
+\mx{\wh  L}\,\Bigl (\sum_{j=1}^{\nt-\no} a_{D,j}\Bigr ),
\eeq
\bbq
\w_{a,\,a_1}=\frac{\mx}{2}\frac{\nt N_c}{N_c-\nt}\,a^2+\frac{\mx}{2}
\frac{\no\nt}{\nt-\no}(1+\wh{\delta}_1)\,a^2_1+
\mx\nt\wh{\delta}_3a_1(m-a-\wh{c}_1  a_1)+\mx\nt\wh{\delta}_4\, 
(m-a-\wh{c}_1 a_1)^2\,,
\eeq
where $\wh{L}$ is the Lagrange multiplier field,
$\langle\wh{L}\rangle=O(m)$ and dots denote small power corrections.
The additional terms with $\wh{\delta}_{1,2,3,4}$, as previously in
br2 vacua in section 41.1 and in vs -vacua in section 45.2, originate
from integrating out all heavier fields with masses
$\sim\Lambda_{SU(\nt)}$ in the soft background. Here these soft
background fields are $(m-a-\wh{c}_1 a_1)$ and $[(1-\wh{c}_1)
a_1+\sqrt{2}X^{\rm adj}_{SU(\no)}]$ (\,while $a_1$ plays here a role
of $a_1$ in \eqref{(41.1.1)}\,). These constants $\wh\delta_i$ differ
from \eqref{(41.1.12)} only because the number of colors is different,
$N_c\ra N^\prime_c=\nt$, i.e. $\wh\delta_1=[\,-
2N^\prime_c/(2N^\prime_c-N_F)]
=[\,-2\nt/(\nt-\no)\,],\,\, \wh\delta_2=-2$ (while $\wh
\delta_3=0$ as  in br2 vacua, see \eqref{(47.4)}).

The charges of fields and parameters entering \eqref{(46.2.2)} 
under  $Z_{\nt-\no}=\exp\{i\pi/(\nt-\no)\}$ transformation are\,:
$q_{\lambda}=q_{\rm\theta}=1,\,\, q_{X_{SU(\no)}^{\rm
adj}}=q_{a}=q_{a_1}=q_{a_{D,j}}=q_{\rm
m}=q_{\wh{L}}=2,\,\,q_{Q}=q_{\ov Q}=q_{D_j}=q_{{\ov
D}_j}=q_{\Lambda_{SU(\nt)}}=0,\,\, q_{\mx}=-2$. The non-trivial
$Z_{\nt-\no\geq 2}$ transformations change only numerations 
of dual  scalars $a_{D,j}$ and dyons in \eqref{(46.2.2)}, so that $\int
d^2\theta\,\w_{\nt}$ is $Z_{\nt-\no}$-invariant.

From \eqref{(46.2.2)}:
\bq
\langle a\rangle=m\,,\quad\langle a_1\rangle=0\,,\quad 
\langle  a_{D,j}\rangle=0\,,
\quad \langle X_{SU(\no)}^{\rm adj}\rangle=0\,,\quad 
\langle  S\rangle_{\no}=0\,,\label{(46.2.3)}
\eq
\bbq
\langle{\ov D}_j D_j\rangle=\langle{\ov D}_j\rangle\langle D_j
\rangle\approx -\mx\langle\Lambda_{SU(\nt)}\rangle\tau^{j-1}
+\mx  \langle{\wh L}\rangle\,,\quad
\langle\Sigma_D\rangle=\sum_{j=1}^{\nt-\no}\langle{\ov  D}_j
\rangle\langle D_j \rangle=(\nt-\no)\mx \langle{\wh L}\rangle,
\eeq
\bbq
\langle{\wh L}\rangle\approx\frac{N_c}{N_c-\nt}\, m=\ha\,, \quad
\langle{\rm Tr}\,({\ov Q}
Q)\rangle_{\no}=\no\langle\Qo\rangle_{\no}+\nt\langle\Qt
\rangle_{\no}\approx\no\mx\, \ha\,,
\eeq
\bbq
\langle\Qo\rangle_{\no}=\langle{\ov Q}^{\,1}_1\rangle\langle
Q^1_1\rangle\approx\frac{N_c}{N_c-\nt}\,\mx m\approx\mx  \ha
\approx\langle\Qt\rangle_{N_c}\,,\quad
\langle\Qt\rangle_{\no}=\sum_{a=1}^{\no}\langle{\ov
Q}^{\,a}_2\rangle\langle Q^2_a\rangle=0\,.
\eeq

The multiplicity of these br2 vacua in the case considered is $N_{\rm
br2}=(N_c-\nt)C^{\,\no}_{N_F}=(N_c-\nt)C^{\,\nt}_{N_F}$, as it should
be. The factor $(N_c-\nt)$ originates from the $SU(N_c-\nt)$ SYM. The
factor $C^{\,\no}_{N_F}$ is finally a consequence of the color
breaking $SU(\nt)\ra SU(\no)\times U(1)^{\nt-\no}$ with the unbroken
$Z_{2\nt-N_F}=Z_{\nt-\no}$ discrete symmetry, and spontaneous breaking \, 
of flavor  symmetry due to higgsing of $\no$ quarks flavors (massless
at $\mx\ra 0$) in the $SU(\no)$ color subgroup of $SU(\nt)$, $\langle
Q^i_a\rangle=\langle{\ov Q}^{\,a}_i\rangle\approx \delta_a^i (\mx
\ha)^{1/2}$, $a=1...\no,\,\,i=1...N_F$.

Note also that if the non-trivial discrete symmetry $Z_{\nt-\no\geq
2}$ were broken spontaneously in the $SU(\nt)$ subgroup at the scale
$\mu\sim\langle\Lambda_{SU(\nt)}\rangle\gg(\mx m)^{1/2}$, this 
would  lead then
to the factor $\nt-\no$ in the multiplicity of these vacua, and this
is a wrong factor.

Due to higgsing of $\nt-\no$ dyons with the nonzero $SU(\nt-\no)$
adjoint magnetic charges, $\langle{\ov D_j}\rangle\langle
D_j\rangle\sim \mx m\gg\mx\lm\gg\mx\langle\Lambda_{SU(\nt)}\rangle$,
all original pure electrically charged particles with $SU(\nt-\no)$
colors and masses either $\sim m$ or
$\sim\langle\Lambda_{SU(\nt)}\rangle$ are weakly confined, the 
string  tension is $\sigma^{1/2}\sim (\mx
m)^{1/2}\ll\langle\Lambda_{SU(\nt)}\rangle\ll\lm\ll m$. Besides,
$\nt-\no$ long ${\cal N}=2$ multiplets of massive photons with masses
$\sim (\mx m)^{1/2}$ are formed. Remind that, due to higgsing of
$N_c-\nt-1$ magnetic monopoles $M_{\rm n}$ from $SU(N_c-\nt)$ 
SYM at  $\mu\sim (\mx\langle\Lambda^{SU(N_c-\nt)}_{{\cal N}=2\,
SYM}\rangle)^{1/2}$, all original pure electrically charged particles
with $SU(N_c-\nt)$ charges and masses either $\sim m$ or
$\sim\langle\Lambda^{SU(N_c-\nt)}_{{\cal N}=2\, SYM}\rangle$ are also
weakly confined, but this string tension $\sigma^{1/2}_2\sim
(\mx\langle\Lambda^{SU(N_c-\nt)}_{{\cal N}=2\, SYM}\rangle)^{1/2}
\ll  (\mx m)^{1/2}$ is much smaller.

Due to higgsing of $\no$ flavors of original electric quarks in the IR \, 
free  color \, $SU(\no)$ with $N_F>2\no$ flavors, there are $\no^2$ long
${\cal N}=2$ multiplets of massive gluons with masses $\sim (\mx
m)^{1/2}\ll\langle\Lambda_{SU(\nt)}\rangle$, while the original
non-higgsed quarks with $\nt$ flavors and $SU(\no)$ colors form
$2\no\nt$ massless Nambu-Goldstone multiplets. There are no other
massless particles at $\mx\neq 0$. And there are no particles with
masses $\sim m\gg\lm$ in this $SU(\nt)\times U(1)$ sector.

Remind that the curve \eqref{(40.2)} has $N_c-1$ double roots in these
br2 vacua at $\mx\ra 0$. Of them: $\nt-\no$ unequal roots correspond
to $\nt-\no$ dyons $D_j,\,\, \no$ equal roots correspond to $\no$
higgsed quarks $Q^i$ of $SU(\no)$, the remaining unequal $N_c-\nt-1$
double roots correspond to $N_c-\nt-1$ pure magnetic monopoles 
$M_{\rm  n}$ from$SU(N_c-\nt)\,\, {\cal N}=2$ SYM. Two single 
roots with  $(e^{+}-e^
{-})\sim\langle\Lambda^{SU(N_c-\nt)}_{{\cal N}=2\, SYM}\rangle$
originate from this ${\cal N}=2$ SYM.

Besides, similarly to vs-vacua in section 45, there is a phase
transition in these br2 vacua in the region
$\mx\sim\langle\Lambda_{SU(\nt)}\rangle^2/m$. E.g., all $SU(\nt-\no)$
charged gluons have equal masses in the whole wide range $(\rm
several)\langle\Lambda_{SU(\nt)}\rangle < (\mx m)^{1/2}\ll m$ of the
higher energy phase of section 46.1, while masses of these gluons
differ greatly at $\mx\ra 0$ in the lower energy phase of this
section, $\mu^{ij}_{gl}\sim
|(\tau^i-\tau^j)\langle\Lambda_{SU(\nt)}\rangle|,\,i\neq j,\,
i,j=0...\nt-\no-1$. And $\langle X^{adj}_{SU(\nt-\no)}\rangle$,
$\,\langle\Qt\rangle_{\nt}$ also behave non-analytically at
$\mx\gtrless\langle\Lambda_{SU(\nt)}\rangle^2/m$. But, unlike the
phase transition at $\mx\sim\lm^2/m$ in
vs-vacua of section 45, the phase transition in these br2 vacua is 
at  much smaller $\mx\sim\langle\Lambda_
{SU(\nt)}\rangle^2/m\,,\,\,\langle\Lambda_{SU(\nt)}\rangle\ll
\lm$,  see \, \eqref{(46.1.6)}.\\

And finally, remind that the mass spectra in these br2 vacua with
$\nt<N_c$ and $m\gg\lm$ depend essentially on the value of $m/\lm$,
and all these vacua evolve at $m\ll\lm$ to Lt-vacua with the
spontaneously broken discrete $Z_{\bb}$ symmetry, see 
section 4 in \cite{ch19}.

\section{\bf Large quark masses $\mathbf{m\gg\lm,\,\,\, SU(N_c),\,\,\, 
1 \leq N_F<N_c}$}

\numberwithin{equation}{section}

\hspace*{4mm} The vacua with the unbroken $U(N_F)$ flavor symmetry
(here and in what follows at not too small $N_c$ and $N_F$) are in
this case the SYM-vacua with the multiplicity $N_c$ and S-vacua with
the multiplicity $N_c-N_F$, see  section 3
in \cite{ch19}. The vacua with the broken $U(N_F)\ra U(\no)\times
U(\nt)$ symmetry are the br1 and br2 vacua with multiplicities
$(N_c-{\rm n}_i)C^{\,{\rm n}_i}_{N_F},\, i=1,\,2$.

Except for S-vacua, all formulae for this case $N_F<N_c,\,\, m\gg\lm$
are the same as described above for br1 vacua in section 43, for SYM
vacua in section 44 and for br2 vacua in section 46 (and only $N_F<N_c$
now). Therefore, we describe below only the mass spectra in S-vacua
with $m\gg\lm,\,\mx\ll\lm\,$.

From
\bbq
\w^{\,\rm eff}_{\rm tot}(\Pi)=m\,{\rm Tr}\,({\ov Q}
Q)_{N_c}-\frac{1}{2\mx}\Biggl [ \,\sum_{i,j=1}^{N_F} ({\ov Q}_j
Q^i)_{N_c}({\ov Q}_{\,i} Q^j)_{N_c}-\frac{1}{N_c}\Bigl ({\rm
Tr}\,({\ov Q} Q)_{N_c}\Bigr )^2\Biggr ]+(N_c-N_F) S_{N_c} \,,
\eeq
\bbq
S_{N_c}=\Bigl(\frac{\lm^{2N_c-N_F}\mx^{N_c}}{\det (\QQ)_{N_c}}
\Bigr )^{\frac{1}{N_c-N_F}}\,,
\eeq
the quark and gluino condensates look in these $SU(N_c)$ S-vacua 
with  $\no=0,\, \nt=N_F$ as,
\bq
\langle\QQ\rangle_{N_c}\approx\mx\,\ha\Bigl
[1-\frac{N_c}{N_c-N_F}\Bigl(\frac{\lm}{\ha}\Bigr )^{\frac{2N_c-N_F}
{N_c-N_F}} \,\Bigr ],\,\, \ha=\frac{N_c}{N_c-N_F}\,m,\,\,\,
\langle S\rangle_{N_c}\approx\mx\, \ha^2\Bigl (\frac{\lm} 
{\ha}\Bigr )^{\frac{2N_c-N_F}{N_c-N_F}}.\,\,\, \label{(47.1)}
\eq
It is seen from \eqref{(47.1)} that the non-trivial discrete symmetry
$Z_{2N_c-N_F\geq 2}$ is not broken.

The scalars $ X^{adj}_{SU(N_c)}$ are higgsed at the highest scale
$\sim m\gg\lm$ {\it in the weak coupling regime}, $SU(N_c)\ra
SU(N_F)\times U(1)\times SU(N_c-N_f)$, see \eqref{(47.7)} below,
\bq
\langle X^{adj}_{SU(N_c)}\rangle=\langle X^{adj}_{SU(N_F)}+X_{U(1)}+
X^{adj}_{SU(N_c-N_F)}\rangle,\label{(47.2)}
\eq
\bbq
\sqrt{2}\, X_{U(1)}=a\,{\rm diag}\,(\,\underbrace{\,1}_{N_F}\,;\,
\underbrace{\,\wh{c}}
_{N_c-N_F}\,),\quad \wh{c}=-\,\frac{N_F}{N_c-N_F}\,,\quad 
\langle  a\rangle= m\,.
\eeq

The quarks in the $SU(N_c-N_F)$ sector and $SU(N_c)/[SU(N_F)\times
SU(N_c-N_F)\times U(1)]$ hybrids have large masses $m_Q=
\langle m-{\wh{c}}\,a\rangle=N_c m/(N_c-N_F)=\ha$. 

After they are integrated out, the  scale factor of  the $SU(N_c-N_F)$ 
SYM gauge coupling is, see also  \eqref{(47.1)},
\bq
\langle\Lambda^{SU(N_c-N_F)}_{{\cal N}=2\, SYM}\rangle^2=\Bigl
(\frac{\lm^{\bb}{(\ha)}^{N_F}}{(\ha)^{2 N_F}}\Bigr
)^{\frac{1}{N_c-N_F}}={\ha^2}\Bigl (\frac{\lm}{{\ha}}\Bigr
)^{\frac{\bb}{N_c-N_F}}\ll\lm^2\,,\quad \ha=\frac{N_c}{N_c-N_F}
m\,,\label{(47.3)}
\eq
\bbq
\langle S\rangle_{N_c-N_F}=\mx\langle\Lambda^{SU(N_c-N_F)}_
{{\cal  N}=2\, SYM}\rangle^2=\langle S\rangle_{N_c}\,,
\eeq
while those of $SU(N_F)$ SQCD is
\bq
\langle\Lambda_{SU(N_F)}\rangle=\Bigl
(\,\frac{\lm^{\bb}}{(\ha)^{2(N_c-N_F)}}\,\Bigr
)^{\frac{1}{N_F}}=\ha\Bigl (\frac{\lm}{\ha} \Bigr
)^{\frac{\bb}{N_F}}\ll\lm\,.\label{(47.4)}
\eq

If $\mx\ll\langle\Lambda^{SU(N_c-N_F)}_{{\cal N}=2\, SYM}\rangle$, the \, 
scalars  $X^{adj}_{SU(N_c-N_F)}$ of ${\cal N}=2$ SYM are higgsed at
$\mu\sim\langle\Lambda^{SU(N_c-N_F)}_{{\cal N}=2\, SYM}\rangle$
\eqref{(47.3)} in a known way, $SU(N_c-N_F)\ra U^{N_c-N_F-1}(1)$
\cite{DS}, giving $N_c-N_F$ vacua and $N_c-N_F-1$ massless at $\mx\ra
0$ magnetic monopoles ${\ov M}_{\rm n}, M_{\rm n}$. These all are
higgsed at $\mx\neq 0$ giving $N_c-N_F-1$ $\,{\cal N}=2$ multiplets of \, 
dual  massive photons with masses $\sim
(\mx\langle\Lambda^{SU(N_c-N_F)}_{{\cal N}=2\, SYM}\rangle)^{1/2}$.
All heavy original particles with masses $\sim m$ and all charged
$SU(N_c-N_F)$ adjoints with masses
$\sim\langle\Lambda^{SU(N_c-N_F)}_{{\cal N}=2\, SYM}\rangle$ are
weakly confined, the string tension in this sector is
$\sigma^{1/2}_2\sim (\mx\langle\Lambda^{SU(N_c-N_F)}_
{{\cal N}=2\, SYM}\rangle)^{1/2}\ll\langle\Lambda^{SU(N_c-N_F)}_
{{\cal  N}=2\,  SYM}\rangle$.

On the other hand, if e.g. $\langle\Lambda^{SU(N_c-N_F)}_{{\cal
N}=2\,SYM}\rangle\ll\mx\ll m$, see \eqref{(47.3)}, all scalars
$X^{adj}_{SU(N_c-N_F)}$ are too heavy and not higgsed, they all
decouple in the weak coupling regime and can be integrated out at
$\mu<g^2\mx/(\rm several)$, there remains then ${\cal N}=1\,\,
SU(N_c-N_F)$ SYM with $\langle\Lambda^{SU(N_c-N_F)}_
{{\cal N}=1\,  SYM}\rangle=[\,\mx\langle
\Lambda^{SU(N_c-N_F)}_{{\cal N}=2\,
SYM}\rangle^2\,]^{1/3}\gg\langle\Lambda^{SU(N_c-N_F)}_
{{\cal N}=2\,  SYM}\rangle$. There will be a large number of 
strongly coupled  $SU(N_c-N_F)$ gluonia with the mass scale
$\sim\langle\Lambda^{SU(N_c-N_F)}_{{\cal N}=1\, SYM}\rangle$. All
original heavier $SU(N_c-N_F)$ charged particles with masses $\sim m$
or $g^2\mx$ are still weakly confined, but the string tension in this
sector is larger now, $\langle\Lambda^{SU(N_c-N_F)}_{{\cal N}=2\,
SYM}\rangle\ll\sigma^{1/2}_1 \sim\langle
\Lambda^{SU(N_c-N_F)}_{{\cal N}=1\, SYM}\rangle\ll\mx$.\\

Now, as for the $SU(N_F)\times U(1)$ sector. At the scale $\mu= m/(\rm
several)$ \, this is the UV free ${\cal N}=2$ SQCD with $N_F$ flavors of
massless quarks ${\ov Q}^{\,a}_j, Q^i_a,\,\,i,a=1...N_F,\,\,
\langle m-a\rangle=0$, its gauge coupling grows logarithmically 
with  diminished energy. The dynamics of this part is the same as in
vs-vacua of section 45.4. When
$\langle\Lambda_{SU(N_F)}\rangle^2/m\ll\mx\ll m$, see \eqref{(47.4)},
then the low energy superpotential in this sector will be as e.g. in
\eqref{(46.1.7)} with $\nt=N_F$. I.e., all quarks in this $SU(N_F)$
sector will be higgsed in the weak coupling regime in this higher
energy phase (while $\langle X^{adj}_{SU(N_F)}\rangle=0$), and
$[(N_F^2-1)+1]$ long ${\cal N}=2$ multiplets of massive gluons
(including $U(1)$ with its scalar "a") with masses $\sim
\langle\QQ\rangle^{1/2}_{N_F}\sim (\mx m)^{1/2}$ will be formed, see
\eqref{(46.1.8)}. The global symmetry $SU(N_F)_{C+F}$ is unbroken in
this higher energy phase, and $(N_F^2-1)\,\, {\cal N}=2$ long
multiplets form the $SU(N_F)_{C+F}$ adjoint representation.
There are no particles with mass $\sim m$ and there are no massless
particles. The multiplicity of vacua in this phase is determined by
the multiplicity $N_c-N_F$ of $SU(N_c-N_F)$ SYM.

But at $(\mx m)^{1/2}\ll\langle\Lambda_{SU(N_F)}\rangle$ the lower
energy phase is different. To avoid
$g^2(\mu<\langle\Lambda_{SU(N_F)}\rangle)<0$, the main contributions
to masses originate now in this UV free $SU(N_F)$ sector from higgsed
$\langle X^{adj}_{SU(N_F)}\rangle\sim\langle\Lambda_{SU(N_F)}\rangle$, \, 
in the  strong coupling and nonperturbative regime. The situation in
this phase is also qualitatively similar to those in section 45.4 (or
in 46.2 with $\no\geq 1$, the difference is that $\no=0,\, \nt=N_F$ now \, 
as in  section 45.4, so that the whole $SU(N_F)$ color group will be
broken at the scale $\sim\langle\Lambda_{SU(\nt=N_F)}\rangle$, see
below). Because $SU(N_c-N_F)$ sector is decoupled, there is the
residual unbroken discrete symmetry $Z_{2N_F-N_F\geq\, 2}=Z_{N_F
\geq\,2}$.  Therefore, $X^{adj}_{SU(N_F)}$ are higgsed as, see
\eqref{(41.1.1)},\eqref{(41.1.16)},
\bq
\langle\sqrt{2}
X^{adj}_{SU(N_F)}\rangle=C_{N_F}\langle
\Lambda_{SU(N_F)}\rangle\,{\rm  diag}\,\Bigl(\,
\underbrace{{\tilde\rho}^{\,0},\,{\tilde\rho}^{\,
1},\,...,\,{\tilde\rho}^{\, N_F-1}}_{N_F}\,;\underbrace{0}
_{N_c-N_F}\,\Bigr ),\quad \tilde\rho=\exp\{\,\frac{2\pi
i}{N_F}\,\}\,,\label{(47.5)}
\eq
and the whole $SU(N_F)_C$ color group is broken, $SU(N_F)_C\ra
U^{N_F-1}(1)$. All quarks with $SU(N_F)_C$ colors acquire now masses
$m_Q\sim\langle\Lambda_{SU(N_F)}\rangle$ which are large in comparison \, 
with the  potentially possible scale of their coherent condensate,
$m_Q\gg (\mx m)^{1/2}$, so that {\it they are not higgsed but decouple \, 
as   heavy  at $\mu<\langle\Lambda_{SU(N_F)}\rangle/(\rm several)$}, as
well as all charged $SU(N_F)_C$ adjoints. Instead, $N_F$ massless at
$\mx\ra 0$ dyons ${\ov D}_j, D_j$ are formed at the scale
$\sim\langle\Lambda_{SU(N_F)}\rangle$.

The relevant part of the low energy superpotential of this sector is
in this case qualitatively the same as in vs-vacua of section 45.4,
see\eqref{(45.4.3)}, and only $\w_{SYM}$ is added
\bq
\w^{\,(N_F)}_{\rm low}=\w_{D}+\w_{a}+\w_{SYM}\,, \label{(47.6)}
\eq
\bbq
\w_{D}= (m-a)\sum_{j=1}^{N_F}{\ov D}_j D_j -\sum_{j=1}^{N_F}
a_{D,\,j}{\ov D}_j D_j
-\mx\Lambda_{SU(N_F)}\sum_{j=1}^{N_F}{\tilde\rho}^{\,{j-1}}
a_{D,\,j}+\mx{\,\tilde L}\,
\Bigl (\sum_{j=1}^{N_F} a_{D,\,j}\Bigr )\,,
\eeq
\bbq
\w_{a}=\frac{\mx}{2}\frac{N_c N_F}{N_c-N_F}\,a^2+\mx
N_F\,{\tilde\delta}_4\, (m-a)^2\,,
\eeq
\bbq
\w_{SYM}= (N_c-N_F)\mx \Bigl (\Lambda^{SU(N_{c}-N_F)}_{{\cal N}=
2\,\,SYM}\Bigr )^2= (N_c-N_F)\mx\Biggl (\frac{\lm^{2N_c-N_F}(m-{\hat
c}\,a)^{N_F}}{[\,(1-{\hat c}) a\,]^{2 N_F}}\Biggr
)^{\frac{1}{N_c-N_F}}\approx
\eeq
\bbq
\approx \,\mx\langle\Lambda^{SU(N_c-N_F)}_{{\cal
N}=2\,\,SYM}\rangle^2\Biggl [(N_c-N_F) -N_F
\frac{(2N_c-N_F)}{N_c-N_F}\,\frac{\hat a}{\ha}\,\Biggr ]\,,\quad
\ha=\frac{N_c}{N_c-N_F} m\,,\quad a=\langle a\rangle+{\hat a}\,.
\eeq

From \eqref{(47.6)}
\bbq
\langle a\rangle=m\,,\quad \langle a_{D,\,j}\rangle=0\,,\quad
\langle{\ov D}_j\rangle\langle D_j \rangle=
-\mx\langle\Lambda_{SU(N_F)}\rangle{\tilde\rho}^{\,{j-1}}+\mx
\langle{\tilde L}\rangle\,,\quad
\langle\Sigma_D\rangle=\sum_{j=1}^{N_F}\langle{\ov D}_j\rangle
\langle  D_j\rangle=N_F\mx \langle{\tilde L}\rangle,
\eeq
\bq
\langle\Sigma_D\rangle\approx N_F\mx\,  \ha\Bigl
[\,1-\frac{2N_c-N_F}{N_c-N_F}\frac{\langle
\Lambda^{SU(N_c-N_F)}_{{\cal N}=2\,\,SYM}\rangle^2}{\ha^2}\,
\Bigr ]\approx N_F\mx \ha\Bigl [\,1-\frac{2N_c-N_F}{N_c-N_F}\Bigl
(\frac{\lm}{{\ha}}\Bigr )^{\frac{\bb}{N_c-N_F}}\Bigr ]\,,\label{(47.7)}
\eq
\bbq
\langle S\rangle_{N_F}=0,\quad\langle{\ov D}_j\rangle\langle
D_j\rangle\approx \mx\, \ha \Bigl
[\,1-\frac{2N_c-N_F}{N_c-N_F}\frac{\langle\Lambda^
{SU(N_c-N_F)}_{{\cal  N}=2\,\,SYM}\rangle^2}{\ha^2}\,\Bigr
]-\mx\langle\Lambda_{SU(N_F)}\rangle{\tilde\rho}^{\,{j-1}}\,,
\quad  j=1...N_F\,.
\eeq

As an example, we present also the analog of \eqref{(41.1.18)} in 
theseS-vacua.  From \eqref{(40.1)},\eqref{(47.6)},\eqref{(47.7)}
\bq
\langle\,\frac{\partial}{\partial m}\w_{\rm
matter}\rangle=\langle\,\frac{\partial}{\partial m}
{\w}^{\,(N_F)}_{\rm low}\rangle\ra\langle{\rm
Tr\,}\QQ\rangle_{N_c}=\langle\Sigma_D\rangle+
\langle\,\frac{\partial}{\partial m}{\w_{SYM}}\rangle\approx 
\label{(47.8)}
\eq
\bbq
\approx N_F\mx \, \ha\Bigl [\,1-\frac{N_c}{N_c-N_F}\Bigl
(\frac{\lm}{{\ha}}\Bigr )^{\frac{\bb}{N_c-N_F}}\Bigr ]\,,
\eeq
this agrees with \eqref{(47.1)}.\\

The curve \eqref{(40.2)} has $N_c-1$ double roots in these $N_c-N_F$
S-vacua. From these, $N_c-N_F-1$ unequal double roots correspond to
massless at $\mx\ra 0$ magnetic monopoles $M_{\rm n}$ from
$SU(N_c-N_F)\,\, {\cal N}=2$ SYM, and remaining $N_F$ unequal double
roots correspond to massless at $\mx\ra 0$ dyons $D_j$. Two single
roots with $(e^{+}-e^{-})\sim \langle\Lambda^{SU(N_c-N_F)}_{{\cal
N}=2\, SYM}\rangle$ \eqref{(47.3)} originate at $\mx\ra 0$ from the
$SU(N_c-N_F)\,\,{\cal N}=2$ SYM sector. The global flavor symmetry
$U(N_F)$ is unbroken and there are no massless Nambu-Goldstone
particles. And there are no massless particles at all at $\mx\neq 0$.
At the same time, as in section 45.4.2, the color is broken in this
sector, $SU(N_F)_C\ra U^{N_F-1}(1)$, and masses of charged $SU(N_F)_C$ \, 
adjoints  are different, see \eqref{(47.5)}. There is no color-flavor
locking in this lower energy phase.

All original charged electric particles are confined in the lower
energy phase at $\mx<\langle\Lambda_
{SU(N_F)}\rangle^2/\ha$, see \eqref{(47.4)}. But the phase changes 
in  the $SU(N_F)$ sector already at very small $\mx
:\,\langle\Lambda_{SU(N_F)}\rangle^2/\ha<\mx\ll\langle
\Lambda_{SU(N_F)}\rangle\ll\lm\ll  m$. Instead of higgsed $\langle
X^{adj}_{SU(N_F)}\rangle\sim\langle\Lambda_{SU(N_F)}\rangle,\,\,
SU(N_F)\ra U^{N_F-1}(1)$ and subsequent condensation of $N_F$ 
dyons  $D_j$ \eqref{(47.7)}, all $N_F$ flavors of original electric quarks
${\ov Q}^{\,a}_i,\,Q^i_a,\,\,a,i=1...N_F$ are then higgsed as
$\langle{\ov Q}^{\,a}_i
\rangle=\langle Q_a^i\rangle\approx\delta^i_a (\mx\,
\ha)^{1/2}>\langle\Lambda_{SU(N_F)}\rangle$ (and clearly 
not confined  but screened), forming $N_F^2$ long ${\cal N}=2$ 
multiplets of massive \, gluons.

Remind that, at $N_F<N_c$ and $m\gg\lm$, the mass spectra in these 
S vacua with the multiplicity $N_c-N_F$, as well as in the SYM vacua
with the multiplicity $N_c$ in section 44, depend essentially on the
value of $m/\lm$, and all these vacua with the unbroken global flavor
symmetry $U(N_F)$ evolve at $\mx\ll m\ll\lm$ to L-vacua with the
multiplicity $2N_c-N_F$ and with the spontaneously broken discrete
$Z_{\bb}$ symmetry, see sections 3.1 and 12 in \cite{ch19}.

\section{Conclusions to Part III}

\hspace*{4mm} We presented above in this Part III the calculations of
mass spectra of softly broken ${\cal N}=2\,\, SU(N_c)$ or $U(N_c)$
gauge theories in a large number of their various vacua with the
unbroken non-trivial $Z_{\bb\geq 2}$ discrete symmetry. The quantum
numbers of light particles in each vacuum were determined, as well as
forms of corresponding low energy Lagrangians and mass spectra at
different hierarchies between the Lagrangian parameters
$m,\,\mx,\,\lm$. A crucial role in obtaining these results was played
by the use\,: a) the unbroken $Z_{\bb\geq 2}$ symmetry,\, b) the
pattern of spontaneous flavor symmetry breaking, $U(N_F)\ra
U(\no)\times U(\nt),\, 0\leq \no<N_F/2,\,$ c) the knowledge of
multiplicities of various vacua. Besides, the knowledge of the quark
and gluino condensates
$\langle\Qo\rangle_{N_c},\,\langle\Qt\rangle_{N_c}$,  
and $\langle S\rangle_{N_c}$ (obtained from the effective superpotential 
$\w_{\rm  tot}^{\rm eff}$) was also of great importance. In addition, two
dynamical {\it assumptions} "A" and "B\," of general character
formulated in section 40 were used. They concern the BPS properties
of original particles and the absence of extra non-BPS fields massless 
at  $\mx\neq 0$. All this appeared to be sufficient to
calculate mass spectra in vacua considered at small but nonzero 
values \, of $\mx$,  in particular at $0<\mx<M_{\rm min}$, where $M_
{\rm min}$ is \ the independent of  $\mx$ mass scale appropriate for 
each given vacuum, \, such that the phase and mass
spectrum continue smoothly down to
$\mx\ra 0$. In other words, the theory stays at $\mx<M_{\rm min}$ 
in  the same regime and all hierarchies in the mass spectrum are the 
same  as in the ${\cal N}=2$ theory at $\mx\ra 0$.

Within this framework we considered first in detail in section 41.1 the \, br2
vacua  of $SU(N_c)$ theory with $N_c < N_F< 2 N_c-1$ (these are vacua \, 
of the baryonic branch in \cite{APS} or zero vacua in \cite{SY1,SY2}), \,
at \, $m\ll\lm$ and $0<\mx\ll M^{(\rm br2)}_{\rm min}=
\langle\Lambda^{SU(\nd-{\rm n}_1)}_{{\cal N}=2\,\,SYM}
\rangle^2/\lm\ll\langle\Lambda^{SU(\nd-{\rm n}_1)}_{{\cal
N}=2\,\,SYM}\rangle\ll m\ll\lm$. The original color symmetry is
broken \, spontaneously in these vacua  by higgsed  $\langle  X^{adj}
_{SU(N_c)}\rangle\neq 0$ in three stages. The first stage 
is at \,   the  highest scale $\mu\sim\lm$, $\,\langle
X^{adj}_{SU(2N_c-N_F)}\rangle\sim\lm$, $\,\,SU(N_c)\ra
SU(N_F-N_c)\times U^{(1)}(1)\times U^{2N_c-N_F-1}(1)$, this pattern is \, 
required  by the unbroken discrete $Z_{2N_c-N_F\geq 2}$ symmetry and is \, 
necessary to avoid  $g^2(\mu<\lm) < 0$ in this effectively massless at the
scale $\mu\sim\lm$ $\,\,{\cal N}=2$ UV free theory. The second stage
is in the $SU(N_F-N_c)$ color sector at the scale $\mu\sim m$,
$\,\langle X^{adj}_{SU(N_F-N_c)}\rangle\sim m\ll\lm$,
$\,SU(N_F-N_c)\ra SU(\no)\times U^{(2)}(1)\times SU(N_F-N_c-\no)$. And \, 
the third  stage is in the $SU(N_F-N_c-\no)$ color sector  at the scale
 $\mu\sim\langle\Lambda^
{SU(\nd-{\rm n}_1)}_{{\cal N}=2\,\,SYM}\rangle\ll m$,
$\,SU(N_F-N_c-\no)\ra U^{N_F-N_c-\no-1}(1)$.

The global flavor symmetry $U(N_F)$ is broken spontaneously at
$\mx\neq 0$ in these br2-vacua at the scale $\mu\sim (\mx m)^{1/2}$ 
in \, the  $SU(\no)$ color sector by higgsed original pure electric quarks,
$\,\langle{\ov Q}^{\,a}_i\rangle=\langle Q^i_a\rangle\sim
\delta_a^i\,(\mx m)^{1/2}\,,\,\, a=1...\no,\,\, i=1...N_F$, as\,:
$U(N_F)\ra U(\no)\times U(\nt),\, 1\leq\no< N_F-N_c$.

It was shown that the lightest charged BPS particles (massless at
$\mx\ra 0$) are the following.\,-\\
1) $\bb$ flavorless mutually local dyons $D_j, {\ov D}_j$ (this number \, 
$\bb$ is  required by the unbroken $Z_{\bb\geq 2}$ discrete symmetry
which operates interchanging them among each other), with the two
nonzero pure electric charges, the $SU(N_c)$ baryon charge and
$U^{(1)}(1)$ one. The nonzero magnetic charges of all these dyons are
corresponding $SU(\bb)$ adjoints. The whole $\bb$ set of these dyons
is coupled with $\bb$ Abelian ${\cal N}=2$ photon multiplets,
$U^{\bb-1}(1)$ magnetic and $U^{(1)}(1)$ electric. These dyons are
formed at the scale $\mu\sim\lm$, at the first stage of the color
breaking, and are mutually local with respect to all particles of the
$SU(N_F-N_c)$ sector and between themselves.\\
2) All original electric particles of remaining unbroken at $\mx\ra
0$$\,SU(\no)$ subgroup of original $SU(N_c)$.\\
3) $\nd-\no-1$ pure magnetic monopoles with the $SU(\nd-\no)$ 
adjoint  charges. These last are formed at the low scale
$\mu\sim\langle\Lambda^{SU(\nd-{\rm n}_1)}_{{\cal
N}=2\,\,SYM}\rangle\ll m\ll\lm$ in the ${\cal N}=2\,\,SU(\nd-\no)$ 
SYM \, sector, with the scale factor $\langle\Lambda^{SU(\nd-{\rm
n}_1)}_{{\cal N}=2\,\,SYM}\rangle$ of its gauge coupling. These
magnetic monopoles are coupled with $U^{\nd-\no-1}(1)$
${\cal N}=2$ Abelian magnetic photon multiplets.\\
4) In addition to original electric ${\cal N}=2\,\, SU(\no)$ adjoints, \, 
there are  magnetic $U^{\nd-\no-1}(1)$ and $U^{\bb-1}(1)$, and one
$U^{(1)}(1)$ electric ${\cal N}=2$ Abelian photon multiplets, all
massless at $\mx\ra 0$.

Massless at $\mx\ra 0$ $\,2N_c-N_F$ dyons $D_j,\,\no$ original pure
electric quarks $Q^i_a, {\ov Q}^a_i,\, a=1...\no,\, i=1...N_F$ from
$SU(\no)$, and $\nd-\no-1$ pure magnetic monopoles are all higgsed 
at  small $\mx\neq 0$, and no massless particles remains (except for
$2\no\nt$ Nambu-Goldstone multiplets due to spontaneous breaking 
of  global flavor symmetry, $U(N_F)\ra U(\no)\times U(\nt)$) in the
$SU(\no)$ color sector due to higgsed original pure electric quarks
with $N_F$ flavors). The mass spectrum was described in these
br2-vacua at smallest $0<\mx\ll\langle\Lambda^{SU(\nd-{\rm
n}_1)}_{{\cal N}=2\,\,SYM}\rangle^2/\lm$.

The material of this section 41.1 served then as a basis for similar
regimes in sections 42,\,45,\,46, and 47.

The mass parameter $\mx$ was increased then in two different
stages\,:\, 1) $\langle\Lambda^{SU(\nd-{\rm n}_1)}_{{\cal
N}=2\,\,SYM}\rangle\ll\mx\ll m$ in section 41.3\,;\,\,2)
$m\ll\mx\ll\lm$ in section 41.4, and changes in the mass spectra were
described. In all those cases when the corresponding ${\cal N}=1$ SQCD \, 
lower  energy theories were weakly coupled, we needed no additional
dynamical assumptions. And only in those few cases in section 41.4,
when the corresponding ${\cal N}=1$ SQCD theories were in the strongly \, 
coupled  conformal regime at $3N_c/2<N_F<2N_c$, we used additionally
the assumption of the dynamical scenario introduced in \cite{ch3}. 

In  essence, this assumption from \cite{ch3} looks here as follows\,: \,
{\it  no  additional parametrically light composite solitons are formed in
this ${\cal N}=1\,\, SU(\nd)$ SQCD without decoupled colored adjoint
scalars $X^{\rm adj}_{SU(\nd)}$ at those even lower scales where this
${\cal N}=1$ conformal regime is broken by nonzero particle masses}
(see also the footnote \ref{(f50)}). As was shown in
\cite{ch3,ch19,ch16}, {\it this dynamical scenario does not contradict to any
\,proven properties of ${\cal N}=1$ SQCD and satisfies all those checks of
Seiberg's duality hypothesis for ${\cal N}=1$ SQCD which were used in }
\cite{S2}, and allows to calculate the corresponding mass spectra.

Many other vacua with the broken or unbroken flavor symmetry and at
different hierarchies between the Lagrangian parameters $m,\, \mx,\,
\lm$ were considered in sections 42-47 and corresponding mass spectra
were calculated within the above described framework. Besides,
calculations of power corrections to the leading terms of the low
energy condensates of original quarks $Q,\,{\ov Q}$ and dyons in a
number of vacua are presented in two important Appendices A and B.  
The results \, 
agree  with \, also presented in these Appendices {\it independent}
calculations of these quark and dyon condensates using roots of the
Seiberg-Witten spectral curve. And because these last calculations
with roots are valid {\it only for BPS particles, these agreements
serve as an independent confirmation of the main assumption "A" from
Introduction about BPS properties of original quarks,  and serve as 
numerous  checks of a self-consistency of the whole approach}.

We consider that there is no need to repeat in detail in these
conclusions all results obtained in this paper (see the table of
contents). But at least two points have to be emphasized. 

{\bf 1)} As shown e.g. in section 45.2 (and similarly in sections 45.4, 46 and 47), 
the  widespread opinion that the holomorphic dependence of gauge invariant
chiral condensates (e.g. $\langle (\QQ)_{1,2}\rangle_{N_c},\, \langle
S\rangle_{N_c},\, \\ 
\langle {\rm Tr\,}(X^{adj}_{SU(N_c)})^2\rangle\,)$  on chiral parameters 
of the superpotential implies the absence of phase
transitions in supersymmetric theories is not right. 

{\bf 2)} In contradiction with the smooth analytical crossover transition from 
the \,  higher  energy region to lower energy one (in sections 45,\, 46 and 47) 
and  the emergence of the "instead-of-confinement" regime proclaimed in 
a  number of papers by M.Shifman and A.Yung, see e.g. the latest paper
\cite{SY6} and references therein to their previous papers with this
crossover, we have found that this transition is not an analytic  crossover 
but a \,  phase  transition. See the critique of the M.Shifman and A.Yung
proposal in e.g. sections 45.2, 45.4 and in section 8 of \cite{ch13}.

In particular, the arguments in \cite{SY3} in favour of crossover are based 
on considering  the  unequal mass quarks with $\Delta m\neq 0$, the
``outside''\, region $\Delta m\gg\lm$ and ``inside''\, region 
$\Delta m\ll\lm$ and the path between them going through 
Argyres-Douglas point.  

We would like to emphasize that these ``outside''\,  and ``inside''\, 
regions are {\it completely separated} by so-called ``wall crossing curves''.
And there is no such pass in the complex plane $\Delta m/\lm$
which connects the ``outside''\,  and ``inside''\,  regions and does 
not cross this ``wall crossing curve''.  And the path going through 
Argyres-Douglas point is not the exception.

The quantum numbers  of light particles jump generically when
the generic path crosses the ``wall crossing curve'' and have to be 
carefully considered separately in the ``outside''\,  and ``inside''\, 
regions.  

And besides, as for  the example considered in \cite{SY3} in support of 
crossover between the confinement and higgs regimes (and 
proposal of  ``instead of confinement''\,  regime),  i.e.   $N_c=3,\, 
N_F=5,\, 2N_c-N_F=1$. It is the example  with the trivial  $Z_{2N_c-N_F=1}$ 
symmetry which gives no  restrictions. It belongs 
to another type of vacua, see section 45.5.
And this  clearly  invalidates  all statements of M. Shifman and
A.Yung about quantum numbers of light particles, about crossover 
and  about ``instead of confinement'' regime in vacua with 
non-trivial unbroken $Z_{2N_c-N_F\geq 2}$ symmetry. \\

{\appendix{\Large \bf Appendices.\, Quark and dyon condensates}}\\

\hspace*{4mm} As examples, we present below in two important 
Appendices the
calculation of leading power corrections to the low energy quark
condensates $\langle\Qo\rangle_{\no}$ in br1 vacua of $SU(N_c)$ and
$U(N_c)$ in sections 43.1 and 43.2, and in br2 vacua of $SU(N_c)$ and
$U(N_c)$ in sections 41.1 and 41.2, and to the dyon condensates in
sections 41.1 and 41.2. The results agree with also presented in these
Appendices {\it independent} calculations of these condensates using
roots of the Seiberg-Witten spectral curve. {\t These agreements serve as
the independent numerous checks of a few used assumptions and 
of  self-consistency of the  whole approach}.

\section{Calculations of quark condensates}

{\bf 1) br1 vacua of $SU(N_c),\,\, m\gg\lm$ in section 43.1}\\

From \eqref{(40.3)},\eqref{(40.4)}, in these vacua (up to even smaller
power corrections)
\bq
\langle\Qo\rangle_{N_c}=\mx
m_3-\frac{N_c-\nt}{N_c-\no}\langle\Qt\rangle_{N_c}\,,\quad
\langle\Qt\rangle_{N_c}\approx\mx m_3\Bigl (\frac{\lm}{m_3}\Bigr
)^{\frac{2N_c-N_F}{N_c-\no}}\,,\label{(A.1)}
\eq
\bbq
\langle
S\rangle_{N_c}=\frac{\langle\Qo\rangle_{N_c}\langle\Qt\rangle_{N_c}}
{\mx}\approx \,m_3\langle\Qt\rangle_{N_c}\,,\quad
m_3=\frac{N_c}{N_c-\no} m\,.
\eeq
From \eqref{(41.1.9)},\eqref{(A.1)}, the leading power correction to
$\langle\w^{\,\rm eff}_{\rm tot}\rangle$ is
\bq
\langle\delta \w^{\,\rm eff}_{\rm tot}\rangle\approx (N_c-\no)\mx
m^2_3\Bigl (\frac{\lm}{m_3}\Bigr )^{\frac{2N_c-N_F}{N_c-\no}}
\approx  (N_c-\no)\langle S\rangle_{N_c}\,.\label{(A.2)}
\eq
On the other hand, this leading power correction to $\langle\w^{\,\rm
low}_{\rm tot}\rangle$ originates from the $SU(N_c-\no)$ SYM part,
see \, \eqref{(43.1.2)},\eqref{(43.1.4)},
\bbq
\langle\delta \w^{\,\rm low}_{\rm tot}\rangle\approx\mx\langle\,
{\rm  Tr\,}\Bigl (X^{adj}_{SU(N_c-\no)}\Bigr
)^2\,\rangle\approx(N_c-\no)\langle S\rangle_{N_{c}-\no}
\approx (N_c-\no)\mx\langle\Lambda^{SU(N_{c}-\no)}_{{\cal
N}=2\,\,SYM}\rangle^2\approx
\eeq
\bq
\approx (N_c-\no)\mx m_3^2\Bigl (\frac{\lm}{m_3}\Bigr
)^{\frac{2N_c-N_F}{(N_c-\no)}}\,,\quad \ra\quad \langle\delta
\w^{\,\rm low}_{\rm tot}\rangle=\langle\delta \w^
{\,\rm eff}_{\rm  tot}\rangle\,,\label{(A.3)}
\eq
as it should be.

To calculate the leading power correction
$\delta\langle\Qo\rangle_{\no}\sim\mx\langle\Lambda^
{SU(N_{c}-\no)}_{{\cal N}=2\,\,SYM}\rangle^2/m$ to the low energy
quark condensate $\langle\Qo\rangle_{\no}$, we have to account for 
the \, terms  with \, the first power of quantum fluctuation ${\hat a}$ 
of the  field $a=\langle a\rangle+{\hat a},\,\,\langle{\hat a}\rangle=0$, 
in  the $SU(N_c-\no)$ SYM contribution to $\delta \w^{\,\rm low}_{\rm
tot}$, see \eqref{(43.1.2)}-\eqref{(43.1.4)},
\bbq
\delta \w^{\,\rm low}_{\rm tot}\approx (N_c-\no)\mx\, \delta\, \Bigl
(\Lambda^{SU(N_{c}-\no)}_{{\cal N}=2\,\,SYM}\Bigr )^2\approx
(N_c-\no)\mx\,\delta\,\Bigl (\frac{\lm^{2N_c-N_F}(m_3-c\,{\hat
a})^{N_F}}{[\,m_3+(1-c){\hat a}\,]^{2\no}}\Bigr
)^{\frac{1}{N_c-\no}}\approx
\eeq
\bq
\approx \,\mx\langle\Lambda^{SU(N_{c}-\no)}_{{\cal
N}=2\,\,SYM}\rangle^2\,\Bigl [\,-\no
\frac{(2N_c-N_F)}{N_c-\no}\,\Bigr]\,\frac{{\hat a}}{m_3}\,,
\quad  m_3=\frac{N_c}{N_c-\no}  m\,.\label{(A.4)}
\eq
As a result, from \eqref{(43.1.4)},\eqref{(A.4)},\eqref{(A.1)},
\bbq
\delta\langle\Qo\rangle_{\no}\approx\frac{1}{\no}\langle
\frac{\partial}{\partial{\hat
a}}\delta \w^{\,\rm low}_{\rm tot}\rangle\approx -
\frac{(2N_c-N_F)}{N_c-\no}\,\mx\frac{\langle\Lambda^
{SU(N_{c}-\no)}_{{\cal  N}=2\,\,SYM} \rangle^2}{m_3}\,,
\eeq
\bq
\langle\Qo\rangle^{SU(N_c)}_{\no}=\langle{\ov Q}^1_1\rangle
\langle  Q^1_1\rangle\approx \mx m_3\Biggl [1-\frac{2N_c-N_F}
{N_c-\no}  \,\frac{\langle\Lambda^{SU(N_{c}-\no)}_{{\cal N}=2\,\,
SYM}  \rangle^2}{m^2_3}\Biggr ]\approx \label{(A.5)}
\eq
\bbq
\approx\mx m_3\Biggl [1-\frac{2N_c-N_F}{N_c-\no}\Bigl
(\frac{\lm}{m_3}\Bigr )^{\frac{2N_c-N_F}{N_c-\no}} \Biggr
]\,,\quad\langle\Qo\rangle^{SU(N_c)}
_{N_c}\approx\mx m_3\Biggl [1-\frac{N_c-\nt}{N_c-\no}\Bigl
(\frac{\lm}{m_3}\Bigr )^{\frac{2N_c-N_F}{N_c-\no}} \Biggr ]\,.
\eeq
It is seen from \eqref{(A.5)} that, although the leading terms are 
the \,  same in  $\langle\Qo\rangle_{\no}$ and $\langle\Qo\rangle_
{N_c}$, the  leading power corrections are different.\\

{\bf 2) br1 vacua of $U(N_c),\,\, m\gg\lm$ in section 43.2}

From \eqref{(40.4)},\eqref{(43.2.1)},\eqref{(43.2.2)},\eqref{(43.2.3)}
(keeping now leading power corrections)
\bq
\langle\Qo\rangle_{N_c}=\mx m-\langle\Qt\rangle_{N_c}\,,
\label{(A.6)}
\eq
\bbq
\langle\Qt\rangle_{N_c}=\mx\lm^{\frac{\bb}{N_c-\no}}\Bigl
(\frac{\langle\Qo\rangle_{N_c}}{\mx}\Bigr
)^{\frac{\nd-\no}{N_c-\no}}\approx\mx m\Bigl (\frac{\lm}{m}
\Bigr  )^{\frac{2N_c-N_F}{N_c-\no}}
\Biggl (1+O\Bigl (\frac{\langle\Lambda^{SU(N_{c}-\no)}_
{{\cal  N}=2\,\,SYM}\rangle^2}{m^2} \Bigr )\Biggr )\,,
\eeq
\bbq
\langle
S\rangle_{N_c}=\frac{\langle\Qo\rangle_{N_c}\langle\Qt\rangle_
{N_c}}{\mx}\approx\mx  m^2\Bigl (\frac{\lm}{m}\Bigr )^{\frac
{2N_c-N_F}{N_c-\no}}\Biggl  (1+O\Bigl (\frac{\langle\Lambda^{
SU(N_{c}-\no)}_{{\cal  N}=2\,\,SYM}\rangle^2}{m^2} \Bigr )\Biggr )\,,
\eeq
\bq
\frac{\langle a_0\rangle}{m}=\frac{1}{N_c\mx m}\langle{\rm
Tr}\,\qq\rangle_{N_c}\approx\frac{\no}{N_c} +\frac{\nt-\no}{N_c}
\Bigl  (\frac{\lm}{m}\Bigr )^{\frac{2N_c-N_F}{N_c-\no}}\,,\quad 
\langle  a\rangle=\langle m-a_0\rangle\,, \label{(A.7)}
\eq
\bq
\langle\w^{\,\rm low}_{\rm tot}\rangle\approx \Bigl
[\langle\w_{\no}\rangle=0\,\Bigr ]+\Biggl [
\langle\w_{a_0,a}\rangle=\frac{\mx}{2}\Bigl (\,N_c\langle
a_0\rangle^2+\frac{\no N_c}{N_c-\no}\langle a\rangle^2
\Bigr )\Biggr
]+\langle\w_{SU(N_c-\no)}^{\,SYM}\rangle\,,\label{(A.8)}
\eq
\bbq
\langle\w_{SU(N_c-\no)}^{\,SYM}\rangle=\mx\langle{\,\rm Tr\,}
\Bigl ( X^{adj}_{SU(N_c-\no)}\Bigr )^2\rangle= (N_c-\no)\langle
S\rangle_{SU(N_c-\no)}=(N_c-\no)\mx\langle\Lambda^
{SU(N_{c}-\no)}_{{\cal  N}=2\,\,SYM}\rangle^2\approx
\eeq
\bbq
\approx (N_c-\no)\mx\langle\Biggl (\frac{\lm^{2N_c-N_F}
(m-a_0-c  \,a)^{N_F}}{[\,(1-c) a \,]^{2\no}} \Biggr
)^{\frac{1}{N_c-\no}}\rangle\approx (N_c-\no)\mx\, 
m^2\Bigl  (\frac{\lm}{m}\Bigr )^{\frac{2N_c-N_F}{N_c-\no}}\,.
\eeq
From \eqref{(A.6)}-\eqref{(A.8)}, the leading power correction
$\langle\,\delta\,\w^{\,\rm low}_{\rm
tot}\rangle\sim\mx\langle\Lambda^{SU(N_{c}-\no)}_{{\cal
N}=2\,\,SYM}\rangle^2$ to $\langle\w^{\,\rm low}
_{\rm tot}\rangle$ looks as
\bbq
\langle\,\delta\, \w^{\,\rm low}_{\rm tot}\rangle=\Bigl
(\langle\,\delta\,\w_{a_0,a}\rangle=0\Bigr )+
\langle\,\delta\,\w_{SU(N_c-\no)}^{\,SYM}\rangle\approx
\eeq
\bq
\approx (N_c-\no)\mx\langle\Lambda^{SU(N_{c}-\no)}_{{\cal
N}=2\,\,SYM}\rangle^2\approx
(N_c-\no)\,m\langle\Qt\rangle_{N_c}\approx (N_c-\no)\mx m^2
\Bigl  (\frac{\lm}{m}\Bigr )^{\frac{2N_c-N_F}{N_c-\no}}\,.\label{(A.9)}
\eq

Instead of \eqref{(41.1.9)},\eqref{(41.1.10)} we have now, see
\eqref{(41.2.7)},
\bq
\w^{\,\rm eff}_{\rm tot}(\Pi)=m\,{\rm Tr}\,({\ov Q}
Q)_{N_c}-\frac{1}{2\mx}\Biggl [ \,\sum_{i,j=1}^{N_F} ({\ov Q}_j
Q^i)_{N_c}({\ov Q}_{\,i} Q^j)_{N_c}\Biggr ]-\nd \Biggl
[\,S_{N_c}=\frac{\langle\Qo\rangle_{N_c}\langle\Qt\rangle_{N_c}}
{\mx}\,\Biggr ],\,\,\,\,\,\label{(A.10)}
\eq
and using \eqref{(A.6)} (with the same accuracy)
\bbq
\langle\delta\w^{\,\rm eff}_{\rm tot}(\Pi)\rangle\approx
(N_c-\no)m\langle\Qt\rangle_{N_c}\approx (N_c-\no)\mx\, 
m^2\Bigl  (\frac{\lm}{m}\Bigr
)^{\frac{2N_c-N_F}{N_c-\no}}\approx\langle\delta\w^
{\,\rm\, low}  _{\rm tot}\rangle\,,
\eeq
as it should be.

As for the leading power correction $\delta\langle\Qo\rangle_
{\no}\sim\mx \, m\Bigl(\lm/m\Bigr )^
{\frac{2N_c-N_F}{N_c-\no}}$ to the quark condensate, we 
have now  instead of \eqref{(A.4)}
\bbq
\delta\w^{\,\rm low}_{\rm tot}=\delta \w_{a_0,a}+\delta
\w_{SU(N_c-\no)}^{\,SYM}\,,\quad a_0=\langle a_0\rangle +
{\hat  a}_0\,,\quad a=\langle a\rangle +{\hat a}\,,
\eeq
\bbq
\delta \w_{SU(N_c-\no)}^{\,SYM}=(N_c-\no)\mx\, \delta\, \Bigl
(\Lambda^{SU(N_{c}-\no)}_{{\cal N}=2\,\,SYM}\Bigr )^2\approx
(N_c-\no)\mx\,\delta\,\Bigl (\frac{\lm^{2N_c-N_F}(m-{\hat
a}_0-c\,{\hat a})^{N_F}}{[\,m+(1-c){\hat a}\,]^{2\no}}\Bigr
)^{\frac{1}{N_c-\no}}\approx
\eeq
\bq
\approx \Bigl [\,-N_F{\hat a_0}-\frac{\no(2N_c-N_F)}{N_c-\no}\,
{\hat  a}\, \Bigr ) \mx\frac{\langle\Lambda^{SU(N_{c}-\no)}_
{{\cal  N}=2\,\,SYM}\rangle^2}{m}\,,  \label{(A.11)}
\eq
while from \eqref{(43.2.3)},\eqref{(A.7)}
\bq
\delta \w_{a_0,a}\approx (\nt-\no)\Bigl [\,{\hat
a}_0-\frac{\no}{N_c-\no}\,{\hat a}\,\Bigr
]\mx\frac{\langle\Lambda^{SU(N_{c}-\no)}_{{\cal
N}=2\,\,SYM}\rangle^2}{m}\,.\label{(A.12)}
\eq

Instead of \eqref{(A.5)} (with the same accuracy), on the 
one hand,  see \eqref{(43.2.3)},\eqref{(43.2.4)},\eqref{(A.6)},
\bbq
\langle{\,\rm Tr\,}\qq\rangle_{\no}=\no\mx m+\delta
\langle{\,\rm  Tr\,}\qq\rangle_{\no}=\no\Bigl (\mx
m+\delta\,\langle\Qo\rangle_{\no}\Bigr )\,,
\eeq
\bq
\delta\langle{\,\rm  Tr\,}\Qo\rangle_{\no}=
\frac{1}{\no}\langle\frac{\partial}{\partial
{\hat a}_0} \Bigl (\delta \w_{SU(N_c-\no)}^{\,SYM}+
\delta  \w_{a_0,a}\Bigr )\rangle\approx -2
\mx\frac{\langle\Lambda^{SU(N_{c}-\no)}_{{\cal
N}=2\,\,SYM}\rangle^2}{m}\,,\label{(A.13)}
\eq
while on the other hand,
\bbq
\delta\langle{\,\rm  Tr\,}\Qo\rangle_{\no}=
\frac{1}{\no}\langle\,\frac{\partial}{\partial
{\hat a}}\Bigl (\delta \w_{SU(N_c-\no)}^{\,SYM}+\delta 
\w_{a_0,a}\Bigr)\approx
-2 \mx\frac{\langle\Lambda^{SU(N_{c}-\no)}_{{\cal
N}=2\,\,SYM}\rangle^2}{m} \,,
\eeq
as it should be.

Therefore, on the whole for these $U(N_c)$ br1 vacua, see
\eqref{(A.6)} for $\langle\Qo\rangle^{U(N_c)}_{N_c}$,
\bq
\langle\Qo\rangle^{U(N_c)}_{\no}=\langle{\ov Q}^1_1\rangle
\langle  Q^1_1\rangle\approx\mx m\Bigl
[\,1-2\frac{\langle\Lambda^{SU(N_{c}-\no)}_{{\cal
N}=2\,SYM}\rangle^2}{m^2}\,\Bigr ]\approx\mx m \Bigl 
[\, 1-2\Bigl  (\frac{\lm}{m}\Bigr )^{\frac{2N_c-N_F}{N_c-\no}}\,
\Bigr  ]\,,\label{(A.14)}
\eq
\bbq
\langle\Qo\rangle^{U(N_c)}_{N_c}\approx\mx m\Bigl
[\,1-\frac{\langle\Lambda^{SU(N_{c}-\no)}_{{\cal
N}=2\,\,SYM}\rangle^2}{m^2}\,\Bigr ].
\eeq

We can compare also the result for $\langle\Qo\rangle^{U(N_c)}_
{\no}$  in \eqref{(A.14)} with those from \eqref{(41.2.10)}.
\footnote{\,
In these simplest br1 vacua of $U(N_c)$ the charges and multiplicities \, 
of  particles massless at $\mx\ra 0$ are evident, these are original
electric particles from $SU(\no)\times U^{(0)}(1)$ and $N_c-\no-1$
magnetic monopoles from $SU(N_c-\no)$ SYM.
}
To obtain definite predictions from \eqref{(41.2.10)} one has to find
first the values of roots entering \eqref{(41.2.10)}. In the case
considered, to obtain definite values of all roots of the curve
\eqref{(40.2)} in the br1 vacua of $U(N_c)$ theory one needs only one
additional relation \cite{CIV,CSW} for the two single roots $e^{\pm}$
of the curve \eqref{(40.2)}
\bq
e^{\pm}=\pm 2\sqrt{\langle S\rangle_{N_c}/\mx}\,,\quad
e^{\pm}_c=\frac{1}{2} (e^{+}+e^{-})=0\,.\label{(A.15)}
\eq
One can obtain then from the $U(N_c)$ curve \eqref{(40.2)} the values
of $\no$ equal double roots $e^{(Q)}_i$ of original electric quarks
from $SU(\no)$ and $N_c-\no-1$ unequal double roots $e^{(M)}_k$ of
magnetic monopoles from $SU(N_c-\no)$ SYM. We find from
\eqref{(40.2)},\eqref{(A.15)} with our accuracy, see
\eqref{(A.6)},\eqref{(A.8)},
\bbq
e^{+}=-e^{-}\approx 2 m\Bigl (\frac{\lm}{m}\Bigr
)^{\frac{2N_c-N_F}{2(N_c-\no)}}\Biggl (1+O\Bigl
(\frac{\langle\Lambda^{SU(N_{c}-\no)}_{{\cal
N}=2\,\,SYM}\rangle^2}{m^2} \Bigr ) \Biggr )
\approx 2\langle\Lambda^{SU(N_{c}-\no)}_{{\cal
N}=2\,\,SYM}\rangle,\quad e^{(Q)}_i= - m,\,\, i=1...\no\,,
\eeq
\bq
e^{(M)}_k\approx 2\cos (\frac{\pi
k}{N_c-\no})\langle\Lambda^{SU(N_{c}-\no)}
_{{\cal
N}=2\,\,SYM}\rangle-\frac{\nt-\no}{N_c-\no-1}\frac{\langle
\Lambda^{SU(N_{c}-\no)}
_{{\cal N}=2\,\,SYM}\rangle^2}{m}\,,\quad
k=1...(N_c-\no-1)\,.\label{(A.16)}
\eq
Let us note that these roots satisfy the sum rule, see \eqref{(A.7)},
\bq
\sum_{n=1}^{N_c}\phi_n=\frac{1}{2}\sum_{n=1}^{2 N_c}(-e_n)=
N_c\langle  a_0\rangle\,, \label{(A.17)}
\eq
as it should be.

Therefore, from \eqref{(41.2.10)},\eqref{(A.16)}
\bbq
\langle\Qo\rangle^{U(N_c)}_{\no}= -
\mx\sqrt{(e^{(Q)}_i-e^+)(e^{(Q)}_i-e^-)}\approx
\eeq
\bq
\approx\mx m \Bigl (1-2\frac{\langle\Lambda^{SU(N_{c}-
\no)}_{{\cal  N}=2\,\,SYM}\rangle^2}{m^2} \Bigr)\approx
\mx m \Bigl [\, 1-2\Bigl (\frac{\lm}{m}\Bigr
)^{\frac{2N_c-N_F}{N_c-\no}}\,\Bigr ]\,,\quad
i=1...\no\,,\,\label{(A.18)}
\eq
this agrees with \eqref{(A.14)}.\\

Besides, we can use the knowledge of roots of the curve \eqref{(40.2)}
for the $U(N_c)$ theory and of $\langle a_0\rangle$, see
\eqref{(A.7)},\eqref{(A.16)}, to obtain the values of
$\langle\Qo\rangle^{SU(N_c)}_{\no}$ condensates in the $SU(N_c)$
theory. For this, the curve \eqref{(40.2)} for the $U(N_c)$ theory
\bq
y^2=\prod_{i=1}^{N_c}(z+\phi_i)^2-4\lm^{\bb}(z+m)^{N_F}\,,\quad
\sum_{i=1}^{N_c}\phi_i=\frac{1}{2}\sum_{n=1}^{2 N_c}(-e_n)=N_c
\langle  a_0\rangle{\rm\,\,\, in\,\,\, U(N_c)}\,,\label{(A.19)}
\eq
we rewrite in the form of the $SU(N_c)$ theory
\bq
y^2=\prod_{i=1}^{N_c}({\hat z}+{\hat \phi}_i)^2-4\lm^{\bb}({\hat
z}+{\hat m})^{N_F}\,,\,\,
\sum_{i=1}^{N_c}{\hat\phi}_i=\frac{1}{2}\sum_{n=1}^{2 N_c}(-{\hat
e}_n)=0\,,\,\, {\hat m}=m\, (1-m^{-1}\langle a_0\rangle)\,.\,\label{(A.20)}
\eq

But clearly, it remained the same $U(N_c)$ theory with the same
condensates \eqref{(A.18)}. Therefore, for the $SU(N_c)$ curve
\eqref{(40.2)} the quark condensates will be as in \eqref{(A.18)} with
$m$ replaced by  ${m^\prime}=m/(1-m^{-1}\langle a_0\rangle)$, 
both in the leading terms \, and in  all \, power corrections, i.e., 
see \eqref{(A.7)},
\bbq
\frac{\langle a_0\rangle}{m}\,\,\ra\,\,\approx \Biggl
(\,\frac{\no}{N_c}+\frac{\nt-\no}{N_c}\Bigl (\frac{\lm}{m_3}
\Bigr )^{\frac{2N_c-N_F}{N_c-\no}}\,\Biggr )\,,
\eeq
\bq
m\ra {m^\prime}\approx m_3\Biggl (1+\frac{\nt-\no}{N_c-\no}
\Bigl  (\frac{\lm}{m_3}\Bigr )^{\frac{\bb}{N_c-\no}}\Biggr )\,,\quad
m_3=\frac{N_c}{N_c-\no}\,m\,.\label{(A.21)}
\eq
Then, from \eqref{(A.18)},\eqref{(A.21)},
\bq
\langle\Qo\rangle^{SU(N_c)}_{\no}\approx \mx m_3 \Biggl
(1+\frac{\nt-\no}{N_c-\no}\Bigl (\frac{\lm}{m_3}\Bigr
)^{\frac{\bb}{N_c-\no}} \Biggr )\Bigl [\, 1-2\Bigl
(\frac{\lm}{m_3}\Bigr )^{\frac{2N_c-N_F}{N_c-\no}}\,\Bigr ]
\approx  \label{(A.22)}
\eq
\bbq
\approx \mx m_3 \Bigl [\, 1-\frac{\bb}{N_c-\no}\Bigl
(\frac{\lm}{m_3}\Bigr )^{\frac{2N_c-N_F}{N_c-\no}}\,\Bigr ]\,,
\eeq
this agrees with \eqref{(A.5)}.\\

{\bf 3) br2 vacua of $SU(N_c),\,\, m\ll\lm$ in section 41.1}

The calculations of leading power corrections to $\langle\,\delta
\w_{\rm tot}^{\,\rm low}\rangle$ and $\langle\,\delta \w^{\,\rm
eff}_{\rm tot}\rangle$ have been presented in \eqref{(41.1.15)}.
Therefore, we present here the calculation of the leading power
correction $\delta\,\langle\Qo\rangle_{\no}\sim \mx
\langle\Lambda^{SU(N_{c}-\no)}_{{\cal N}=2\,\,SYM}\rangle^2/m$ to 
the  quark condensate. For this, see \eqref{(41.1.4)},\eqref{(41.1.15)}, we
write ($a_{1,2}=\langle a_{1,2}\rangle+{\hat a}_{1,2}$)
\bbq
\delta \w^{(SYM)}_{SU(\nd-\no)}\approx (\nd-\no)\wmu\, \delta\,\Bigl
(\Lambda^{SU(\nd-\no)}_{{\cal N}=2\,\,SYM}\Bigr )^2\approx
(\nd-\no)\wmu\,\delta\,\Biggl
(\frac{\Lambda_{SU(\nd)}^{2\nd-N_F}(m_2-{\hat a}_1-c_2\,{\hat
a}_2)^{N_F}}{[\,m_2+(1-c_2)\,{\hat a}_2\,]^{\,2\no}}\Biggr
)^{\frac{1}{\nt-N_c}}\approx
\eeq
\bq
\approx \wmu\frac{\langle\Lambda^{SU(N_{c}-\no)}_{{\cal
N}=2\,\,SYM}\rangle^2}{\ha}\,\Bigl [\, N_F{\hat
a}_1+\frac{\no(2N_c-N_F)}{N_c-\nt}\,{\hat a}_2 \Bigr ],
\,\, \ha= \frac{N_c}{N_c-\nt}m= -\, m_2,\,\,
\wmu=-\,\mx,\label{(A.23)}
\eq
\bbq
\langle{\,\rm Tr\,}\qq\rangle_{\no}=\no\mx\, \ha +\delta
\langle{\,\rm  Tr\,}\qq\rangle_{\no}=\no\Bigl (\mx\,
\ha+\delta\,\langle\Qo\rangle_{\no}\Bigr )\,,\quad
\Lambda_{SU(\nd)}=-\lm\,,
\eeq
From this, see \eqref{(41.1.5)}, and \eqref{(41.1)} for
$\langle\Qt\rangle^{SU(N_c)}_{N_c}$,
\bbq
\delta\,\langle\Qo\rangle_{\no}=\frac{1}{\no}\langle
\frac{\partial}{\partial  {\hat
a}_2}\w^{(SYM)}_{SU(\nd-\no)}\rangle\approx\wmu\, \ha 
\Bigl  [\,\frac{2N_c-N_F}{N_c-\nt}\frac{\langle\Lambda^
{SU(\nd-\no)}_{{\cal N}=2\,\,SYM}\rangle^2}{\ha^2}\,\Bigr
],\,\,\frac{\langle\Lambda^{SU(\nd-\no)}_{{\cal
N}=2\,\,SYM}\rangle^2}{\ha^2}\approx\Bigl (\frac{\ha}
{\lm}\Bigr )^{\frac{2N_c-N_F}{\nt-N_c}},
\eeq
\bq
\langle\Qo\rangle^{SU(N_c)}_{\no}=\langle{\ov Q}^1_1\rangle
\langle  Q^1_1\rangle\approx\mx\,  \ha \Bigl [\,1+\frac{2N_c-N_F}
{\nt-N_c}\,  \frac{\langle\Lambda^{SU(\nd-\no)}_{{\cal
N}=2\,\,SYM}\rangle^2}{\ha^2}\,\Bigr ]\approx \label{(A.24)}
\eq
\bbq
\approx\mx\,  \ha\Bigl [\,1+\frac{2N_c-N_F}{\nt-N_c}\,\Bigl
(\frac{\ha}{\lm}\Bigr )^{\frac{2N_c-N_F}{\nt-N_c}}\Bigr ]\,,
\quad  \ha=\frac{N_c}{N_c-\nt}m\,,
\eeq
\bbq
\langle\Qt\rangle^{SU(N_c)}_{N_c}=\mx\,
\ha-\frac{N_c-\no}{N_c-\nt}\,\langle\Qo\rangle_{N_c}
\approx\mx\,  \ha\Bigl [\,
1+\frac{N_c-\no}{\nt-N_c}\frac{\langle\Lambda^{SU(\nd-\no)}_
{{\cal  N}=2\,\,SYM}\rangle^2}{\ha^2}\,\Bigr ]\,.
\eeq
It is seen that, as in \eqref{(A.5)}, the leading corrections to
$\langle\Qo\rangle_{\no}$ and $\langle\Qt\rangle_{N_c}$ 
are different.\\

{\bf 4) br2 vacua of $U(N_c),\,\, m\ll\lm$ in section 41.2}

Keeping the leading order corrections
$\sim\langle\Lambda^{SU(\nd-\no)}_{{\cal N}=2\,\, SYM}\rangle^2/m$,
from \eqref{(41.2.2)},\eqref{(41.2.4)},\eqref{(41.1.15)} (put attention
that the quark and gluino condensates, and
$\Lambda^{SU(\nd-\no)}_{{\cal N}=2\,\, SYM}$ are different in br2
vacua of $SU(N_c)$ or $U(N_c)$ theories, see
\eqref{(A.24)},\eqref{(A.25)},
\bq
\langle\Qt\rangle_{N_c}=\mx m-\langle\Qo\rangle_{N_c}\,,\quad
\langle\Qo\rangle_{N_c}\approx\mx m\Bigl ( \frac{m}{\lm}\Bigr
)^{\frac{2N_c-N_F}{\nt-N_c}}\approx
\mx\frac{\langle\Lambda^{SU(\nd-\no)}_{{\cal N}=2\,\,
SYM}\rangle^2}{m} \,,\label{(A.25)}
\eq
\bbq
\frac{\langle a_0\rangle}{m}=\frac{1}{N_c\mx m}\langle{\rm
Tr\,}\qq\rangle_{N_c}\approx \frac{\nt}{N_c}+\frac{\no-\nt}{N_c}\,
\Bigl ( \frac{m}{\lm}\Bigr )^{\frac{2N_c-N_F}{\nt-N_c}}\,.
\eeq
\bq
\delta \w_{SU(\nd-\no)}^{\,SYM}=(\nd-\no)\wmu\,\delta\,\Biggl [\Bigl
(\Lambda^{SU(\nd-\no)}_{{\cal N}=2\,\, SYM}\Bigr )^{2}\approx\Biggl
(\frac{\Lambda_{SU(\nd)}^{2\nd-N_F}(-m-{\hat a}_0-{\hat
a}_1-c_2\,{\hat a}_2)^{N_F}}
{[\,-m+(1-c_2)\,{\hat a}_2\,]^{\,2\no}}\Biggr
)^{\frac{1}{\nt-N_c}}\Biggr ]\,\,\,\label{(A.26)}
\eq
\bbq
\approx\wmu\,\frac{\langle\Lambda^{SU(\nd-\no)}_{{\cal N}=2\,\,
SYM}\rangle^2}{m}\, \Bigl [\,N_F({\hat a}_0+{\hat a}_1)-\no\frac{
2N_c-N_F}{\nt-N_c}\,{\hat a}_2\,\Bigr ],\,\,
\frac{\langle\Lambda^{SU(\nd-\no)}_{{\cal N}=2\,\,SYM}\rangle^2}
{m^2}=\Bigl  (\frac{m}{\lm}\Bigr )^{\frac{2N_c-N_F}{\nt-N_c}},\,\,
\Lambda_{SU(\nd)}=-\,\lm\,,
\eeq
\bbq
\delta \w_{a}\approx\mx\frac{\langle\Lambda^{SU(\nd-\no)}_{{\cal
N}=2\,\, SYM}\rangle^2}{m}\,\Bigl [(\no-\nt) {\hat
a}_0-\frac{N_F}{2N_c-N_F} {\hat
a}_1-\no\frac{\nt-\no}{\nt-N_c}\,{\hat  a}_2\,\Bigr ]\,,\quad a_i=
\langle  a_i\rangle+{\hat a}_i,\,
i=0,\,1,\,2\,,\,
\eeq
\bbq
\delta\langle\Qo\rangle^{(U(N_c)}_{\no}=\frac{1}{\no}\langle
\frac{\partial}{\partial
{\hat a}_2}\Bigl (\delta\w_{SU(\nd-\no)}^{\,SYM}+\delta \w_{a}\Bigr
)\rangle \approx - 2\mx m \frac{\langle\Lambda^{SU(\nd-\no)}_{{\cal
N}=2\,\, SYM}\rangle^2}{m^2}\approx - 2\mx m\Bigl (\frac{m}{\lm}\Bigr
)^{\frac{2N_c-N_F}{\nt-N_c}}\,.
\eeq
Therefore, on the whole for these $U(N_c)$ br2 vacua after accounting
for the leading power corrections, see \eqref{(A.25)},\eqref{(A.26)},
\bq
\langle\Qo\rangle^{U(N_c)}_{\no}=\langle{\ov Q}^1_1\rangle\langle
Q^1_1\rangle\approx\mx m\Bigl [1-2\Bigl (\frac{m}{\lm}\Bigr
)^{\frac{2N_c-N_F}{\nt-N_c}}
\Bigr ],\,\, \langle\Qt\rangle^{U(N_c)}_{N_c}\approx\mx m\Bigl
[1-\Bigl (\frac{m}{\lm}\Bigr )^{\frac{2N_c-N_F}{\nt-N_c}}\Bigr
].\,\,\,\,\label{(A.27)}
\eq

It is seen that, as above in $SU(N_c)$ br2 vacua \eqref{(A.24)}, the
non-leading terms in $\langle\Qo\rangle_{\no}$ and
$\langle\Qt\rangle_{N_c}$ are different.\\

We can compare now the value of $\langle\Qo\rangle
^{U(N_c)}_{\no}$  from \eqref{(A.27)} with those from
\eqref{(41.2.10)} (see however the footnote \ref{(f7)}, in these br2
vacua the charges of massless at $\mx\ra 0$ particles are non-trivial
and not obvious beforehand, as well as their multiplicities). But we
know the charges and multiplicities of massless at $\mx\ra 0$
particles from sections 41.1 and 41.2. To obtain definite predictions
from \eqref{(41.2.10)} for the quark condensates
$\langle\Qo\rangle_{\no}$ we need the values of $\no$ double roots
$e^{(Q)}_k$ corresponding to original electric quarks from $SU(\no)$
and the values of two single roots $e^{\pm}$. Similarly to
\eqref{(A.15)},\eqref{(A.16)}, these look here as
\bq
e^{(Q)}_k= - m\,,\quad k=1...\no\,,\quad e^{+}=-e^{-}\approx
2\langle\Lambda^{SU(\nd-\no)}_{{\cal N}=2\,\, SYM}\rangle\,,
\quad  \frac{\langle\Lambda^{SU(\nd-\no)}_{{\cal N}=2\,\,
SYM}\rangle^2}{m^2}\approx\Bigl (\frac{m}{\lm}\Bigr
)^{\frac{2N_c-N_F}{\nt-N_c}}\ll 1\,.\,\,\,\label{(A.28)}
\eq
From this
\bq
\langle\Qo\rangle^{U(N_c)}_{\no}= -
\mx\sqrt{(e^{(Q)}_k-e^+)(e^{(Q)}_k-e^-)}\approx\mx m \Bigl
[\,1-2\Bigl(\frac{m}{\lm}\Bigr )^{\frac{2N_c-N_F}{\nt-N_c}}\,\Bigr
]\,,\label{(A.29)}
\eq
this agrees with \eqref{(A.27)}.

Besides, proceeding similarly to br1 vacua in
\eqref{(A.19)}-\eqref{(A.22)}, we obtain for
$\langle\Qo\rangle^{SU(N_c)}_{\no}$ in br2 vacua 
of $SU(N_c)$, see  \eqref{(A.25)},
\bbq
\frac{\langle a_0\rangle}{m}\,\,\ra\,\,\approx \Biggl
(\,\frac{\nt}{N_c}+\frac{\no-\nt}{N_c}\Bigl (\frac{\ha}{\lm}
\Bigr )^{\frac{2N_c-N_F}{\nt-N_c}}\,\Biggr )\,,
\eeq
\bq
m\ra {m^\prime}\approx \ha\Biggl (1+\frac{\no-\nt}{N_c-\nt}
\Bigl (\frac{\ha}{\lm}\Bigr )^{\frac{\bb}{\nt-N_c}} \Biggr )\,,\quad
\ha=\frac{N_c}{N_c-\nt}\,m\,,\label{(A.30)}
\eq
and from \eqref{(A.29)},\eqref{(A.30)}
\bq
\langle\Qo\rangle^{SU(N_c)}_{\no}\approx \mx\,  \ha \Biggl
(1+\frac{\no-\nt}{N_c-\nt}\Bigl (\frac{\ha}{\lm}\Bigr
)^{\frac{\bb}{\nt-N_c}}\Biggr )\Bigl [\, 1-2\Bigl
(\frac{\ha}{\lm}\Bigr )^{\frac{2N_c-N_F}{\nt-N_c}}\,\Bigr ]
\approx  \label{(A.31)}
\eq
\bbq
\approx \mx\,  \ha \Bigl [\, 1+\frac{\bb}{\nt-N_c}\Bigl
(\frac{\ha}{\lm}\Bigr )^{\frac{2N_c-N_F}{\nt-N_c}}\,\Bigr ]
\approx\, \ha  \mx\Biggl[\,1+\frac
{2N_c-N_F}{\nt-N_c}\frac{\langle\Lambda^{SU(\nd-\no)}_
{{\cal  N}=2\,\,SYM}\rangle^2}{\ha^2}\,\Biggr ]\,,
\eeq
this agrees with \eqref{(A.24)}.

\section{Calculations of dyon condensates}

Consider first the br2 vacua of $\mathbf{U(N_c)}$ with $m\ll\lm$ in
section 41.2. Because, unlike the br1 vacua of $U(N_c)$ theory in
section 43.2, there are now in addition massless dyons in br2 vacua, to \, 
obtain  with our accuracy the definite values of roots from the
$U(N_c)$ curve \eqref{(40.2)}, one needs to use not only
\eqref{(A.15)}, but also the sum rule \eqref{(A.17)}. The sum rule
\eqref{(A.17)} looks in these br2 vacua as, see \eqref{(A.25)},
\bq
N_c\langle a_0\rangle\approx \Bigl (\nt
m+(\no-\nt)\frac{\langle\Lambda^{SU(\nd-\no)}_{{\cal N}=2\,\,
SYM}\rangle^2}{m} \Bigr )\approx \Biggl (\nt m+(\no-\nt) m\Bigl
(\frac{m}{\lm}\Bigr )^{\frac{\bb}{\nt-N_c}}\Biggr )\approx
\label{(B.1)}
\eq
\bbq
\approx\Bigl (\no
m+(\nd-\no-1)A^{(M)}_c\frac{\langle\Lambda^{SU(\nd-\no)}
_{{\cal  N}=2\,\,  SYM}\rangle^2}
{m}-\frac{1}{2}(e^{+}+e^{-}=0)+(2N_c-N_F)C^{(D)}_c \Bigr)\,,
\eeq
where $\no(-e^{(Q)}_i)=\no m$ is the contribution of $\no$ equal
double roots of original electric quarks from $SU(\no)$, the term
with \, $A^{(M)}_c$ is the contribution from the center of $\nd-\no-1$
unequal \, double \, roots of magnetic monopoles, and the term with 
$C^{(D)}_c$ is  the contribution from the center of $2N_c-N_F$ unequal 
double roots of \, dyons. We obtain then from the curve \eqref{(40.2)}
togetherwith \eqref{(B.1)} (with our accuracy)
\bq
A^{(M)}_c=\frac{\nt-\no}{\nd-\no-1}\,,\quad
C^{(D)}_c=\frac{\nt-\no}{2N_c-N_F} m \Bigl ( 1-2
\frac{\langle\Lambda^{SU(\nd-\no)}_{{\cal N}=2\,\, SYM}\rangle^2}
{m^2}\Bigr )\,,\,\,\,\label{(B.2)}
\eq
\bbq
e^{(M)}_k\approx 2\cos (\frac{\pi
k}{\nd-\no})\langle\Lambda^{SU(\nd-\no)}
_{{\cal
N}=2\,\,SYM}\rangle-\frac{\nt-\no}{\nd-\no-1}\frac{\langle\Lambda^
{SU(\nd-\no)}_{{\cal N}=2\,\,SYM}\rangle^2}{m}\,,\quad k=1...(\nd-\no-1)\,,
\eeq
\bbq
e^{(D)}_j\approx \omega^{j-1}\lm\Biggl (1+O\Bigl (\frac{m}{\lm}
\Bigr )^{\bb}\Biggr )-\frac{\nt-\no}{2N_c-N_F}\,m\Biggl (1-2\Bigl
(\frac{m}{\lm}\Bigr )^{\frac{\bb}{\nt-N_c}}\Biggr )\,,\,\,
j=1...(2N_c-N_F)\,,
\eeq
while $e^{\pm}$ are given in \eqref{(A.28)}.

From \eqref{(41.2.10)},\eqref{(B.2)} we obtain now for the dyon
condensates in these br2 vacua of $U(N_c)$ at $m\ll\lm$ 
(with the sameaccuracy)
\bq
\langle{\ov D}_j D_j\rangle_{U(N_c)}=\langle{\ov D}_j\rangle\langle
D_j\rangle\approx \mx\Biggl [\,-\,
\omega^{j-1}\lm+\frac{\nt-\no}{2N_c-N_F}\,m
\Biggl (1-2\Bigl (\frac{m}{\lm}\Bigr )^{\frac{\bb}{\nt-N_c}}\Biggr )
\Biggr ]\,, \label{(B.3)}
\eq
\bbq
\langle\Sigma^{U(N_c)}_D\rangle\equiv\sum_{j=1}^{2N_c-N_F}
\langle{\ov  D}_j D_j\rangle_{U(N_c)}=\sum_{j=1}^{2N_c-N_F}
\langle{\ov  D}_j\rangle\langle D_j\rangle
\approx \mx m\,(\nt-\no)\Biggl (1-2\Bigl (\frac{m}{\lm}\Bigr
)^{\frac{\bb}{\nt-N_c}}\Biggr ), \,\,\, j=1...\bb\,.
\eeq

Now, using only the leading terms $\sim \mx m$ of
\eqref{(B.3)},\eqref{(41.2.3)},\eqref{(41.2.5)},
\eqref{(41.2.8)}, we can find the value of $\delta_3=O(1)$ in
\eqref{(41.1.2)},\eqref{(41.1.5)},\eqref{(41.2.4)},\eqref{(45.2.3)}. 
From  \eqref{(41.2.4)}
\bq
\langle\frac{\partial}{\partial a_0}\w^{\,\rm low}_{\rm tot}\rangle=0=-
\langle{\rm
Tr\,}\qq\rangle_{\no}-\langle\Sigma^{U(N_c)}_D\rangle+\mx N_c\langle
a_0\rangle-\mx N_c\delta_3\langle a_1\rangle= - \mx
N_c\delta_3\langle  a_1\rangle\ra
{\boldsymbol\delta}{\mathbf{_3=0}}.\,\,\quad\label{(B.4)}
\eq
(The same result follows from $\langle\partial \w^{\,\rm low}_{\rm
tot}/{\partial a_1}\rangle=0$ in \eqref{(41.2.4)}\,).

As for the non-leading terms
$\delta\langle\Sigma^{U(N_c)}_D\rangle\sim \mx m
(m/\lm)^{\frac{\bb}{\nt-N_c}}$ of $\langle\Sigma^{U(N_c)}_D\rangle$
in\eqref{(B.3)}, with $\delta_3=0$ from \eqref{(B.4)}, these can now be
also calculated {\it independently} from
\eqref{(41.2.4)},\eqref{(A.25)},\eqref{(A.26)}, i.e.
\bbq
\hspace*{-3mm}\langle\frac{\partial}{\partial {\hat a}_0}\delta\w^{\,\rm
low}_{\rm
tot}\rangle=0\ra \delta\langle\Sigma^{U(N_c)}_D\rangle\approx\Bigl
[-\delta\langle{\rm Tr}\qq\rangle_{\no}=2\no\mx m\Bigl
(\frac{m}{\lm}\Bigr )^{\frac{\bb}{\nt-N_c}}\Bigr ]_Q+\Bigl
[(\no-\nt)\mx m\Bigl (\frac{m}{\lm}\Bigr )^{\frac{\bb}{\nt-N_c}}\Bigr
]_{\delta\w_{a_0}}\,\,\,
\eeq
\bq
+\Bigl [\langle\frac{\partial}{\partial {\hat a}_0}\delta
\w_{SU(\nd-\no)}^{\,SYM}\rangle= - N_F\mx m\Bigl (\frac{m}{\lm}
\Bigr )^{\frac{\bb}{\nt-N_c}}\Bigr ]_{\delta\w_{SYM}}\approx -2\mx
m\,(\nt-\no) \Bigl (\frac{m}{\lm}\Bigr )^{\frac{\bb}{\nt-N_c}}\,,\label{(B.5)}
\eq
this agrees with \eqref{(B.3)}.\\

As for the value of $\langle\Sigma^{SU(N_c)}_D\rangle$ in br2 vacua
of \, $\mathbf{SU(N_c)}$, with $\delta_3=0$ from \eqref{(B.4)}, it can be
found now directly from
\eqref{(41.1.5)},\eqref{(41.1.12)},\eqref{(A.23)},\eqref{(A.24)}
\bq
\langle\frac{\partial}{\partial {\hat a}_1}\w^{\,\rm low}_{\rm
tot}\rangle=0\ra \langle\Sigma^{SU(N_c)}_D\rangle\approx 
\,\mx\, \ha(\nt-\no)\Biggl [1+\frac{\bb}
{\nt-N_c}\Bigl (\frac{\ha}{\lm}\Bigr )^{\frac{\bb}{\nt-N_c}} \Biggr
]\,,\,\, \ha=\frac{N_c}{N_c-\nt}\,m\,.\,\,\,\label{(B.6)}
\eq
(Really, the power suppressed term in \eqref{(B.6)} can be found {\it
independently of the value of $\delta_3$} because it is parametrically \, 
different  and, unlike \eqref{(B.5)} in $U(N_c)$ theory, there are no
power corrections to $\langle a_1\rangle$ in $SU(N_c)$ theory, see
\eqref{(41.1.6)}).

It can also be obtained from \eqref{(B.3)} in br2 vacua of $U(N_c)$
proceeding similarly to \eqref{(A.30)},\eqref{(A.31)} for the quark
condensates. We obtain from \eqref{(A.30)} and \eqref{(B.3)}
\bbq
\langle\Sigma^{SU(N_c)}_D\rangle=\sum_{j=1}^{2N_c-N_F}
\langle{\ov D}_j\rangle\langle D_j\rangle\approx \mx\,  \ha\Biggl
(1+\frac{\no-\nt}{N_c-\nt}\Bigl (\frac{\ha}{\lm}\Bigr
)^{\frac{\bb}{\nt-N_c}}\Biggr )(\nt-\no)\Bigl [\, 1-2\Bigl
(\frac{\ha}{\lm}\Bigr )^{\frac{2N_c-N_F}{\nt-N_c}}\,\Bigr ]\approx
\eeq
\bq
\approx\mx\,  \ha(\nt-\no)\Biggl [1+\frac{\bb}{\nt-N_c}\Bigl
(\frac{\ha}{\lm}\Bigr )^{\frac{\bb}{\nt-N_c}} \Biggr ]\,,\label{(B.7)}
\eq
this agrees with \eqref{(B.6)}.

On the whole, the dyon condensates in br2 vacua of $SU(N_c)$ look at
$m\ll\lm$ as, $j=1,...,\bb$,
\bq
\hspace*{-4mm}\langle{\ov D}_j D_j\rangle^{SU(N_c)}\approx \mx\Biggl
[-\omega^{j-1}\lm\Biggl (1+O\Bigl (\frac{m}{\lm}\Bigr )^{\bb}\Biggr
)+\frac{\nt-\no}{2N_c-N_F}\, \ha\Biggl (1+\frac{\bb}{\nt-N_c}\Bigl
(\frac{\ha}{\lm}\Bigr )^{\frac{\bb}{\nt-N_c}}\Biggr ) \Biggr
].\,\,\,\label{(B.8)}
\eq
\vspace*{2mm}

Finally, not going into details, we present below the values of roots
of the $\mathbf{SU(N_c)}$ spectral curve \eqref{(40.2)} in br1 vacua 
of \, section 43.1 and br2 vacua of section 41.1.

{\bf 1) br1 vacua}. There are at $\mx\ra 0$ $\,\no$ equal double
roots \, $e^{(Q)}_i$ of massless original electric quarks $Q_i$, then
$N_c-\no-1$ unequal double roots $e^{(M)}_k$ of massless pure 
magnetic \, monopoles
$M_k$ and two single roots $e^{\pm}$ from ${\cal N}=2\,\,
SU(N_c-\no)$ SYM. They look (with our accuracy) as, see \eqref{(A.3)},
\bbq
e^{(Q)}_i= - m\,,\,\,i=1...\no\,,\quad e^{\pm}\approx e^{\pm}_{c}\pm
2\langle\Lambda^{SU(N_{c}-\no)}_{{\cal N}=2\,\,SYM}\rangle,\quad
\langle\Lambda^{SU(N_{c}-\no)}_{{\cal N}=2\,\,SYM}\rangle\approx
m_3\Bigl (\frac{\lm}{m_3}\Bigr )^{\frac{2N_c-N_F}{2(N_c-\no)}},
\eeq
\bq
e^{\pm}_{c}=\frac{1}{2}(e^{+}+e^{-})\approx m_3\Biggl
[\,\frac{\no}{N_c}
+\frac{\nt-\no}{N_c-\no}\frac{\langle\Lambda^{SU(N_{c}-\no)}_{{\cal
N}=2\,\,SYM}\rangle^2}{m^2_3} \,\Biggr ], \quad
m_3=\frac{N_c}{N_c-\no} m\,,\label{(B.9)}
\eq
\bbq
e^{(M)}_k\approx \Biggl
[\,e^{\pm}_{c}-\frac{\nt-\no}{N_c-\no-1}\frac{\langle\Lambda^{SU(N_{c}
-\no)}_{{\cal  N}=2\,\,SYM}\rangle^2}{m_3}\,\Biggr ]+ 2\cos (\frac{\pi
k}{N_c-\no})\langle\Lambda^{SU(N_{c}-\no)}
_{{\cal N}=2\,\,SYM}\rangle,\quad \frac{1}{2}\sum_{n=1}^{2
N_c}(-e_n)=0\,.
\eeq
It should be emphasized that instead of $e^{\pm}_{c}=0$ in
$\mathbf{U(N_c)}$ theory, see \eqref{(A.15)}, it looks in
$\mathbf{SU(N_c)}$ theory as, see \eqref{(A.21)},\eqref{(B.9)},
\bbq
e^{\pm}_{c}\approx \Bigl [\frac{\langle a_0\rangle}{m}\Bigr ]\Bigl
[\,m\,\Bigr ]\,\ra\,\Biggl [\,\frac{\langle
a_0\rangle}{m}\,\,\ra\,\,\, \approx\Biggl
(\,\frac{\no}{N_c}+\frac{\nt-\no}{N_c}\Bigl (\frac{\lm}{m_3}\Bigr
)^{\frac{2N_c-N_F}{N_c-\no}}\,\Biggr )\,\Biggr ]\Biggl [m\,\ra\,
{m^\prime}
\approx m_3\Biggl (\,1+\frac{\nt-\no}{N_c-\no}
\eeq
\bq
\cdot\Bigl (\frac{\lm}{m_3}\Bigr )^{\frac{\bb}{N_c-\no}}\Biggr
)\,\Biggr ]\approx m_3\Biggl
[\frac{\no}{N_c}+\frac{\nt-\no}{N_c-\no}\Bigl (\frac{\lm}{m_3}\Bigr
)^{\frac{2N_c-N_F}{N_c-\no}} \Biggr ]\approx m_3\Biggl
[\,\frac{\no}{N_c}+
\frac{\nt-\no}{N_c-\no}\frac{\langle\Lambda^{SU(N_{c}-\no)}_{{\cal
N}=2\,\,SYM}\rangle^2}{m^2_3} \,\Biggr ]\,.\label{(B.10)}
\eq

{\bf 2) br2 vacua}. There are at $\mx\ra 0$ $\,\no$ equal double
roots \, $e^{(Q)}_i$ of massless original electric quarks $Q_i$, then
$\nd-\no-1$ unequal double roots $e^{(M)}_k$ of massless pure 
magnetic \, monopoles
$M_k$ and two single roots $e^{\pm}$ from ${\cal N}=2\,\,
SU(\nd-\no)$ SYM, and $2N_c-N_F$ unequal double roots of 
massless  dyons $D_j$. They all look (with our accuracy) as, see
\eqref{(A.24)},\eqref{(A.25)},\eqref{(A.30)},
\bbq
e^{(Q)}_i= - m\,,\,\,i=1...\no\,,\quad e^{\pm}\approx e^{\pm}_{c}\pm
2\langle\Lambda^{SU(\nd-\no)}_{{\cal N}=2\,\,SYM}\rangle,\quad
\langle\Lambda^{SU(\nd-\no)}_{{\cal N}=2\,\,SYM}\rangle^2\approx
\,\ha^2\Bigl (\frac{\ha}{\lm}\Bigr )^{\frac{2N_c-N_F}{\nt-N_c}},
\eeq
\bbq
e^{\pm}_{c}=\frac{1}{2}(e^{+}+e^{-})\approx\Biggl [\,\frac{\langle
a_0\rangle}{m}\,\ra\,\, \approx \Biggl
(\,\frac{\nt}{N_c}+\frac{\no-\nt}{N_c}\Bigl (\frac{\ha}{\lm}\Bigr
)^{\frac{2N_c-N_F}{\nt-N_c}}\,\Biggr )\,\Biggr ]\Biggl
[\,m\,\ra\,\,{m^\prime}\approx\, \ha\Biggl (1+
\eeq
\bq
+\frac{\no-\nt}{N_c-\nt}\Bigl (\frac{\ha}{\lm}\Bigr
)^{\frac{\bb}{\nt-N_c}}\Biggr )\,\Biggr ]\approx\, \ha\Biggl
[\,\frac{\nt}{N_c}+\frac{\nt-\no}{\nt-N_c}
\frac{\langle\Lambda^{SU(\nd-\no)}_{{\cal
N}=2\,\,SYM}\rangle^2}{\ha^2} \,\Biggr ], \quad
\ha=\frac{N_c}{N_c-\nt} m\,,\label{(B.11)}
\eq
\bbq
e^{(M)}_k\approx \Biggl
[\,e^{\pm}_{c}-\frac{\nt-\no}{\nd-\no-1}\frac{\langle\Lambda^
{SU(\nd-\no)}_{{\cal
N}=2\,\,SYM}\rangle^2}{\ha}\,\Biggr ]+ 2\cos (\frac{\pi
k}{\nd-\no})\langle\Lambda^{SU(\nd-\no)}_{{\cal
N}=2\,\,SYM}\rangle,\quad k=1...(\nd-\no-1)\,,
\eeq
\bq
e^{(D)}_j\approx \omega^{j-1}\lm+\frac{N_F}{2N_c-N_F}\,m\,,
\quad  j=1...2 N_c-N_F\,,
\quad \frac{1}{2}\sum_{n=1}^{2 N_c}(-e_n)=0\,.\label{(B.12)}
\eq

Now, with \eqref{(B.9)}-\eqref{(B.12)}, we can perform checks of
possible applicability of \eqref{(41.2.10)} not only to softly broken
$\mathbf{U(N_c)}$ $\,{\cal N}=2$ SQCD, but also directly to softly
broken $\mathbf{SU(N_c)}$ $\,{\cal N}=2$ SQCD at appropriately small
$\mx$ (clearly, this applicability is far not evident beforehand, see
the footnote \ref{(f54)}).

{\bf a)} From \eqref{(B.9)} for the quark condensates in br1 vacua of
section 43.1
\bq
\langle\Qo\rangle^{SU(N_c)}_{\no}= -
\mx\sqrt{(e^{(Q)}_k-e^+)(e^{(Q)}_k-e^-)}\approx\mx\,  m_3\Biggl
[\,1-\frac{2N_c-N_F}{N_c-\no}
\frac{\langle\Lambda^{SU(N_{c}-\no)}_{{\cal
N}=2\,\,SYM}\rangle^2}{m^2_3}\, \Biggr ]\,,\label{(B.13)}
\eq
this agrees with the {\it independent} calculation in \eqref{(A.5)}.\\{\bf b)}
From \eqref{(B.11)} for the quark condensates in br2 vacua ofsection 41.1
\bq
\langle\Qo\rangle^{SU(N_c)}_{\no}= -
\mx\sqrt{(e^{(Q)}_k-e^+)(e^{(Q)}_k-e^-)}\approx\mx\,  \ha\Biggl
[\,1+\frac
{2N_c-N_F}{\nt-N_c}\frac{\langle\Lambda^{SU(\nd-\no)}_{{\cal
N}=2\,\,SYM}\rangle^2}{\ha^2}\,\Biggr ]\,,\label{(B.14)}
\eq
this agrees with the {\it independent} calculation in
\eqref{(A.24)}.\\

{\bf c)} From \eqref{(B.11)},\eqref{(B.12)} for the dyon condensates
in br2 vacua of section 41.1
\bbq
\langle{\ov D}_j D_j\rangle^{SU(N_c)}=\langle{\ov
D}_j\rangle^{SU(N_c)}\langle D_j\rangle^{SU(N_c)}= -
\mx\sqrt{(e^{(D)}_j-e^+)(e^{(D)}_j-e^-)}\approx
\eeq
\bq
\approx\mx\Biggl
[-\omega^{j-1}\lm+\frac{\nt-\no}{2N_c-N_F}\,\ha\Biggl(1+
\frac{\bb} {\nt-N_c}\frac{\langle\Lambda^{SU(\nd-\no)}_{{\cal
N}=2\,\,SYM}\rangle^2}{\ha^2}\Biggl )\,\Biggr ]\,,\quad j=1...2
N_c-N_F\,,  \label{(B.15)}
\eq
this agrees with the {\it independent} calculation in
\eqref{(B.8)}

{\addcontentsline{toc}{section}
{\large \bf References}}

\end{document}